\documentclass[11pt,a4paper]{article}
\pdfoutput=1

\usepackage{jcappub}
\usepackage{latexsym}
\usepackage{graphicx}
\usepackage{float}
\usepackage{afterpage}
\usepackage{mathrsfs}
\usepackage{amsmath}
\usepackage{amsfonts} 
\usepackage{amssymb}  
\usepackage[bf, sf, FIGTOPCAP, normal, tight]{subfigure}
\usepackage{multirow}
\usepackage{multicol}
\usepackage[numbers,sort&compress]{natbib}
\usepackage{array}
\usepackage{bigstrut}
\usepackage{rotating}
\usepackage{booktabs}
\usepackage{hhline}

\def\urltilda{\kern -.15em\lower .7ex\hbox{\~{}}\kern .04em}

\def\tab#1{Table~\ref{#1}}

\def\fig#1{Fig.~\ref{#1}}
\def\figs#1#2{Figs.~\ref{#1} and \ref{#2}}

\def\sec#1{Sec.~\ref{#1}}
\def\eq#1{Eqn.~\ref{#1}}
\def\eqs#1#2{Eqns.~\ref{#1} and \ref{#2}}

\definecolor{gold}{rgb}{0.85,.66,0}

\newcommand{\newc}{\newcommand}

\newc{\be}{\begin{equation}}
\newc{\ee}{\end{equation}}
\newc{\nn}{\nonumber}
\newc\ps{\mbox{ ps}}
\newc{\mev}{\mbox{ MeV}}
\newc{\gev}{\mbox{ GeV}}
\newc{\tev}{\mbox{ TeV}}
\newc{\GeV}{\gev}
\newc{\MeV}{\mev}
\newc{\TeV}{\tev}
\newc{\cl}{\text{CL}}
\newc\BR{BR}
\newc{\alphaemmz}{\alpha_{\text{em}}(m_Z)^{\overline{MS}}}
\newc{\alphas}{\alpha_s(m_Z)^{\overline{MS}}}
\newc\zetah{\zeta_h}
\newc\eg{{\rm {e.g.}}}
\newc\etal{{\rm {et al.}}}
\newc\ie{{\rm i.e.}}
\newc\etc{{\rm {etc}}}
\newc{\mhalf}{m_{1/2}}
\newc{\mzero}{m_0}
\newc{\tanb}{\tan\beta}
\newc{\azero}{A_0}
\newc{\sgn}{{\rm sgn}}
\newc{\deltaamususy}{\delta a_{\mu}^{\text{SUSY}}}
\newc{\bsg}{\bsgamma}
\newc\gmtwo{(g-2)_{\mu}}
\newc\deltaamu{\Delta a_{\mu}}
\newc{\abundchi}{\Omega_{\chi} h^2}
\newc{\msbar}{\overline{MS}}
\newc{\mtop}{m_t}
\newc{\mtpole}{m_t}
\newc{\hl}{h}
\newc{\mhl}{m_{\hl}}  
\newc{\mgut}{M_{\rm GUT}}
\newc{\mplanck}{M_{\rm P}}
\newc{\mpl}{M_{\text{Pl}}}
\newc{\msusy}{M_{\rm SUSY}}
\newc{\ms}{M_{\text{S}}}
\newc{\VEV}[1]{\langle #1 \rangle}
\newc{\sineff}{\sin^2 \theta_{\rm{eff}}}
\newc\MN{{\sf {MultiNest}}}

\newc\bsgamma{b\rightarrow s \gamma }
\newc\brbsgamma{\BR(\overline{B}\rightarrow X_s\gamma)}
\newc\bsmumu{\overline{B}_s\to\mu^+\mu^-}
\newc\brbsmumu{\BR(\overline{B}_s\to\mu^+\mu^-)}
\newc\bdmmumu{\overline{B}_d\to\mu^+\mu^-}
\newc\bbbarmix{\overline{B}_s\mbox{--}B_s}
\newc\delmbs{\Delta M_{B_s}}
\newc\brbtaunu{\BR(\overline{B}_u\to \tau \nu)}
\newc{\mbmbmsbar}{m_b(m_b)^{\msbar} }

\newc\AIPCP[3] {{AIP Conf. Proc.} {\bf #1} (#2) #3}
\newc\AJ[3] {{Astrophys. J.} {\bf #1} (#2) #3}
\newc\AMS[3] {{Ann. Math. Statist.} {\bf #1} (#2) #3}                
\newc\AP[3] {{Ann. Phys.} {\bf #1} (#2) #3}
\newc\APJ[3] {{Astropart. J.} {\bf #1} (#2) #3}
\newc\APP[3] {{Astropart. Phys.} {\bf #1} (#2) #3}
\newc\APS[3] {{Astrophys. J. Suppl.} {\bf #1} (#2) #3}
\newc\ARNPS[3] {{Ann. Rev. Nucl. Part. Sci.} {\bf C#1} (#2) #3}
\newc\BA[3] {{Bayesian Anal.} {\bf C#1} (#2) #3}              
\newc\CPC[3] {{Comput. Phys. Commun.} {\bf C#1} (#2) #3}
\newc\CP[3] {{Contemp. Phys.} {\bf #1} (#2) #3}                     
\newc\EPJ[3] {{Euro. Phys. Journ.} {\bf C#1} (#2) #3}
\newc\JCAP[3] {{JCAP} {\bf #1} (#2) #3}
\newc\JHEP[3] {{JHEP} {\bf #1} (#2) #3}
\newc\JPG[3] {{J. Phys.} {\bf G #1} (#2) #3}
\newc\IJMP[3] {{Int. J. Mod. Phys.} {\bf A #1} (#2) #3}
\newc\MNRAS[3] {{Mon. Not. Roy. Astron. Soc.} {\bf #1} (#2) #3}
\newc\MPL[3] {{Mod. Phys. Lett.} {\bf A #1} (#2) #3}
\newc\NAR[3] {{New Astron. Rev.} {\bf #1} (#2) #3}                  
\newc\NCA[3] {{Nuovo Cimento} {\bf #1} (#2) #3}
\newc\NIM[3] {{Nucl. Instrum. Methods} {\bf #1} (#2) #3}
\newc\NIMA[3] {{Nucl. Instrum. Methods} {\bf A #1} (#2) #3}
\newc\NAT[3] {{Nature} {\bf #1} (#2) #3}
\newc\NPB[3] {{Nucl. Phys.} {\bf B #1} (#2) #3}
\newc\NPA[3] {{Nucl. Phys.} {\bf A #1} (#2) #3}
\newc\NPPS[3] {{Nucl. Phys. Proc. Suppl.} {\bf #1} (#2) #3}                
\newc\PLB[3] {{Phys. Lett.} {\bf B #1} (#2) #3}
\newc\PR[3] {{Phys. Rep.} {\bf #1} (#2) #3}
\newc\PRL[3] {{Phys. Rev. Lett.} {\bf #1} (#2) #3}
\newc\PRD[3] {{Phys. Rev.} {\bf D #1} (#2) #3}
\newc\PRC[3] {{Phys. Rev.} {\bf C #1} (#2) #3}
\newc\PTP[3] {{Prog. Theor. Phys.} {\bf #1} (#2) #3}
\newc\RMP[3] {{Rev. Mod. Phys.} {\bf #1} (#2) #3 }
\newc\RPP[3] {{Rept. Prog. Phys.} {\bf #1} (#2) #3 }
\newc\SC[3] {{Science} {\bf #1} (#2) #3 }
\newc\ZPC[3] {{Z. Phys.} {\bf C #1} (#2) #3}
\newc\Err[3] {{Erratum-ibid.} {\bf #1} (#2) #3 }

\usepackage[normalem]{ulem}
\definecolor{DarkGreen}{rgb}{0.0,0.5,0.0}

\newcommand{\refeqn}[2][eqn:]{Eqn.~(\ref{#1#2})}

\newcommand{\reftab}[2][tab:]{Table~\ref{#1#2}}

\newcommand{\reffig}[2][fig:]{Figure~\ref{#1#2}}

\newcommand{\refsec}[2][sec:]{Section~\ref{#1#2}} 

\newcommand{\rmd}{\ensuremath{\mathrm{d}}}

\newcommand{\erf}{\mathop{\mathrm{erf}}}

\newcommand{\gae}{%
  \ensuremath{\lower 2pt \hbox{%
    $\, \buildrel {\scriptstyle >}\over {\scriptstyle \sim}\,$}%
    }%
  }
\newcommand{\lae}{%
  \ensuremath{\lower 2pt \hbox{%
    $\, \buildrel {\scriptstyle <}\over {\scriptstyle \sim}\,$}%
    }%
  }

\newcommand{\Enr}{\ensuremath{E_{nr}}}

\newcommand{\Rnr}{\ensuremath{R_{nr}}}
\newcommand{\dRdE}{\ensuremath{\frac{dR}{dE}}}

\newcommand{\dRdEi}{\ensuremath{\frac{dR_i}{dE_{\phantom{i}}}}}
\newcommand{\dRdEnr}{\ensuremath{\frac{d\Rnr}{d\Enr}}}
\newcommand{\dRSdE}{\ensuremath{\frac{dR_S}{dE_{\phantom{S}}}}}
\newcommand{\dRBdE}{\ensuremath{\frac{dR_B}{dE_{\phantom{B}}}}}
\newcommand{\dRtotdE}{\ensuremath{\frac{dR_{tot}}{dE_{\phantom{tot}}}}}

\newcommand{\Emin}{\ensuremath{E_{\mathrm{min}}}}
\newcommand{\Emax}{\ensuremath{E_{\mathrm{max}}}}

\newcommand{\rhoDM}{\ensuremath{\rho_{0}}}

\newcommand{\vmin}{\ensuremath{v_\mathrm{min}}}
\newcommand{\vmp}{\ensuremath{v_0}}
\newcommand{\vrot}{\ensuremath{v_\mathrm{rot}}}
\newcommand{\vobs}{\ensuremath{v_\mathrm{obs}}}
\newcommand{\bvobs}{\ensuremath{\mathbf{v}_\mathrm{obs}}}
\newcommand{\vesc}{\ensuremath{v_\mathrm{esc}}}
\newcommand{\Nesc}{\ensuremath{N_\mathrm{esc}}}
\newcommand{\bu}{\ensuremath{\mathbf{u}}}  
\newcommand{\bv}{\ensuremath{\mathbf{v}}}  
\newcommand{\bV}{\ensuremath{\mathbf{V}}}  

\newcommand{\rmu}{\ensuremath{\mathrm{u}}}
\newcommand{\rms}{\ensuremath{\mathrm{s}}}
\newcommand{\rmc}{\ensuremath{\mathrm{c}}}
\newcommand{\rmb}{\ensuremath{\mathrm{b}}}
\newcommand{\rmt}{\ensuremath{\mathrm{t}}}
\newcommand{\rmpp}{\ensuremath{\mathrm{p}}}
\newcommand{\rmnn}{\ensuremath{\mathrm{n}}}

\newcommand{\muu}{\ensuremath{m_{\mathrm{u}}}}
\newcommand{\mdd}{\ensuremath{m_{\mathrm{d}}}}
\newcommand{\mumd}{\ensuremath{\muu / \mdd}}
\newcommand{\mss}{\ensuremath{m_{\mathrm{s}}}}
\newcommand{\msmd}{\ensuremath{\mss / \mdd}}

\newcommand{\SigmapiN}{\ensuremath{\Sigma_{\pi\!{\scriptscriptstyle N}}}}

\newcommand{\athree}{\ensuremath{a_{3}^{\mathrm{(p)}}}}
\newcommand{\aeight}{\ensuremath{a_{8}^{\mathrm{(p)}}}}
\newcommand{\Deltaps}{\ensuremath{\Delta_{\rms}^{(\rmpp)}}}

\newcommand{\fpSI}{\ensuremath{f_{\mathrm{p}}}}
\newcommand{\fnSI}{\ensuremath{f_{\mathrm{n}}}}
\newcommand{\apSD}{\ensuremath{a_{\mathrm{p}}}}
\newcommand{\anSD}{\ensuremath{a_{\mathrm{n}}}}

\newcommand{\sigmaSI}{\ensuremath{\sigma^{\mathrm{SI}}}}
\newcommand{\sigmaSD}{\ensuremath{\sigma^{\mathrm{SD}}}}

\newcommand{\sigmaNSI}{\ensuremath{\sigma^{\mathrm{SI}}_{\mathrm{N}}}}
\newcommand{\sigmapSI}{\ensuremath{\sigma^{\mathrm{SI}}_{\mathrm{p}}}}
\newcommand{\sigmanSI}{\ensuremath{\sigma^{\mathrm{SI}}_{\mathrm{n}}}}
\newcommand{\sigmaNSD}{\ensuremath{\sigma^{\mathrm{SD}}_{\mathrm{N}}}}
\newcommand{\sigmapSD}{\ensuremath{\sigma^{\mathrm{SD}}_{\mathrm{p}}}}
\newcommand{\sigmanSD}{\ensuremath{\sigma^{\mathrm{SD}}_{\mathrm{n}}}}

\newcommand{\fTq}[1]{\ensuremath{f_{T_{#1}}}}
\newcommand{\fNTq}[1]{\ensuremath{f_{T_{#1}}^{(N)}}}

\newcommand{\DeltaNq}[1]{\ensuremath{\Delta_{#1}^{(N)}}}
\newcommand{\Deltapq}[1]{\ensuremath{\Delta_{#1}^{(\rmpp)}}}
\newcommand{\Deltanq}[1]{\ensuremath{\Delta_{#1}^{(\rmnn)}}}
\newcommand{\qqbar}[1]{\ensuremath{\bar{#1}#1}}

\newcommand{\smatrix}[2][\rmpp]{\ensuremath{\langle#1|#2|#1\rangle}}

\title{How well will ton-scale dark matter direct detection experiments constrain minimal supersymmetry?}

\author[a]{Yashar Akrami,}
\author[a]{Christopher Savage,}
\author[a,b]{Pat Scott,}
\author[a]{Jan Conrad}
\author[a]{and Joakim Edsj\"o}

\affiliation[a]{Oskar Klein Centre for Cosmoparticle Physics and Department of Physics,\\
Stockholm University,\\
AlbaNova University Centre, SE-10691 Stockholm, Sweden}
\affiliation[b]{Department of Physics, McGill University,\\
3600 rue University, Montr\'eal, QC, H3A 2T8, Canada}

\emailAdd{yashar@fysik.su.se}
\emailAdd{savage@fysik.su.se}
\emailAdd{patscott@physics.mcgill.ca}
\emailAdd{conrad@fysik.su.se}
\emailAdd{edsjo@fysik.su.se}

\abstract{Weakly interacting massive particles (WIMPs) are amongst the most interesting dark matter (DM) candidates. Many DM candidates naturally arise in theories beyond the standard model (SM) of particle physics, like weak-scale supersymmetry (SUSY). Experiments aim to detect WIMPs by scattering, annihilation or direct production, and thereby determine the underlying theory to which they belong, along with its parameters. Here we examine the prospects for further constraining the Constrained Minimal Supersymmetric Standard Model (CMSSM) with future ton-scale direct detection experiments. We consider ton-scale extrapolations of three current experiments: CDMS, XENON and COUPP, with 1000 kg-years of raw exposure each. We assume energy resolutions, energy ranges and efficiencies similar to the current versions of the experiments, and include backgrounds at target levels. Our analysis is based on full likelihood constructions for the experiments. We also take into account present uncertainties on hadronic matrix elements for neutralino-quark couplings, and on halo model parameters. We generate synthetic data based on four benchmark points and scan over the CMSSM parameter space using nested sampling. We construct both Bayesian posterior PDFs and frequentist profile likelihoods for the model parameters, as well as the mass and various cross-sections of the lightest neutralino. Future ton-scale experiments will help substantially in constraining supersymmetry, especially when results of experiments primarily targeting spin-dependent nuclear scattering are combined with those directed more toward spin-independent interactions.}

\keywords{dark matter theory, dark matter experiments, cosmology of theories beyond the SM, supersymmetry and cosmology}

\begin{document}

\maketitle

\section{Introduction} \label{sec:intro}

The Standard Model (SM) of particle physics (for an introduction, see ref.~\cite{SM:Burgess}) is superbly consistent with all existing laboratory experimental data. There are however various reasons to consider the SM as an effective theory, valid only up to certain energies. These stem from conceptually irritating problems existent in the theoretical structure of the SM itself, and its inability to explain certain celestial observations. One of the most subtle issues in the first category is the so-called ``gauge hierarchy problem''. The problem is that the Higgs mass should be subject to very large (but experimentally excluded) quantum corrections, whose absence can only be explained in the framework of the SM by requiring extremely fine-tuned cancellations between physically-unrelated components of the theory. From the second category, the standard $\Lambda$CDM model of cosmology (for an introduction, see refs.~\cite{cosmology:Weinberg,cosmology:Mukhanov}) is the strongest argument for new physics. The new physics called for ranges from a long-awaited reconciliation of Einstein's classical theory of gravity with the quantum-field-theoretical nature of the SM (required to describe very early moments of the Universe, and maybe also the properties of some extreme astrophysical objects) to the lack of any compelling explanation for dark matter (DM) or dark energy (see e.g. refs.~\cite{Komatsu:2008hk,Iocco:2008va,Riess:2004nr,Astier:2005qq,Cole:2005sx,Kessler:2009ys,Massey:2007wb}). If we add to this evidence the absence of any explanation for the origin of the gauge symmetries of the SM, the large differences in strengths of the gauge interactions, the number of matter generations, the origin of fermion masses, the strong CP problem, the specific scale and mechanism of electroweak symmetry breaking, and neutrino masses and mixings, going beyond the SM seems entirely unavoidable~\cite{SM:Burgess}.

Supersymmetry (SUSY), if realised at TeV energy scales, is arguably the most favoured theory beyond the SM. TeV-scale SUSY can solve many of the problems listed above, and pave the way for the resolution of many others in some broader theoretical framework (for an introduction, see refs.~\cite{Martin:9709356,SUSY:Aitchison,SUSY:Baer}). This weak-scale SUSY provides a natural solution to the hierarchy problem (by eliminating the quadratic divergences in the Higgs mass via the cancellation of the contributions from SM particles and corresponding superpartners), suggests a natural mechanism for electroweak symmetry breaking (via renormalisation group evolution) and contains viable DM candidates (such as the lightest neutralino, gravitino, sneutrino and axino). Gauge-coupling unification can be achieved in models with SUSY~\cite{Ellis:1990} and many of the questions about the mathematical structure of the SM can be gracefully answered by assuming some grand unified theory (GUT) above this unification scale, in which the SM gauge group is extended to some larger groups (e.g. $SU(5)$ or $SO(10)$). Finally, SUSY, either as incumbent in the framework of supergravity (SUGRA) or as the essential ingredient of most versions of string theory, delivers extensive scope for unifying gravity and other fundamental interactions.

Although it is not yet known how exactly SUSY is implemented in Nature (if at all), one can take a phenomenological approach and upgrade the SM to its SUSY extensions by simply adding new degrees of freedom and interactions to the SM Lagrangian. The most conventional framework for doing this is that of the so-called Minimal Supersymmetric Standard Model (MSSM), in which the particle content is increased by a minimum set of extra degrees of freedom necessary for supersymmetrising the model; i.e.~every SM particle gets a ``superpartner''. Were SUSY an exact symmetry of Nature, the structure of the MSSM would be highly restricted, including just one more parameter than the SM. This would however imply equal masses for SM particles and their superpartners, immediately excluding the theory by simple low-energy laboratory experiments. However, all the desirable features of SUSY remain entirely intact if it is broken around the TeV scales, in such a way that the gauge hierarchy remains stabilised, original gauge invariance of the model is preserved, and the renormalisability of the model is not violated. This is implemented in the MSSM by introducing the so-called ``soft SUSY-breaking'' terms to the (previously) SUSY-symmetric Lagrangian. 

The full MSSM Lagrangian contains more than a hundred free parameters to be determined experimentally. The model is phenomenologically rich, but its huge parameter space makes any comparison of the model with real data highly challenging (see e.g. refs.~\cite{Allanach:07050487,Trotta:08093792,Akrami:2009hp,Scott:2009jn,Akrami:2010} and references therein). There are several ways to approach this problem, so as to extract useful information about the structure of the model from experimental constraints. In general, one can pick out a particular SUSY-breaking mechanism (for some reviews, see e.g. refs.~\cite{Chung:0312378,Luty:0509029}) that relates or even unifies many of the model parameters at certain energies (usually much higher than the electroweak scale), and then study its phenomenological consequences at low energies by means of the renormalisation group equations (RGEs). Alternatively, one can impose experimentally-motivated relations directly on the parameters at low energy, so as to reduce their number. The latter proposal is justified by the fact that even small values of many of the soft parameters induce large flavour-changing neutral currents (FCNCs), or CP violation beyond that observed experimentally; these parameters can then be (more or less) reasonably set to zero.

As far as SUSY-breaking mechanisms are concerned, several schemes have been proposed in the literature. These include gravity or Plank-scale mediation (PMSB)~\cite{Chamseddine:1982jx,Barbieri:1982eh,Ohta:1982wn,Hall:1983iz,AlvarezGaume:1983gj,Nilles:1983ge}, gauge mediation (GMSB)~\cite{Dine:1981za,Dimopoulos:1981au,Nappi:1982hm,AlvarezGaume:1981wy,Dine:1994vc,Dine:1995ag} and extra-dimensional or anomaly mediation (AMSB)~\cite{Randall:1998uk,Giudice:1998xp}.  These schemes relate the model parameters in very different ways, giving rise to very different phenomenological predictions (for a comparison of some mediation models using existing data, see ref.~\cite{AbdusSalam:09060957}). A hybrid approach also exists, where one imposes phenomenological assumptions on model parameters at high energy scales, then studies low-energy predictions by solving the RGEs down to interesting scales. The Constrained MSSM (CMSSM)~\cite{CMSSM} is one of the most popular examples in this class of models, in which various boundary conditions are imposed at the GUT scale ($\sim$10$^{16}$\,GeV). These conditions are motivated by minimal supergravity (mSUGRA)~\cite{Chamseddine:1982jx} as the simplest PMSB scenario. Although the number of free parameters is dramatically reduced by going from the MSSM to the CMSSM, many phenomenologically interesting features of SUSY remain intact, making the CMSSM particularly interesting for phenomenologists trying to compare SUSY models with experimental data.

SUSY signatures can be probed in a variety of complementary ways. The most obvious way is to explore physics at TeV scales directly at particle colliders; the Large Hadron Collider (LHC) will play a central role in this endeavour. If supersymmetric particles show up at such experiments, any positive measurement of their mass spectrum can put strong constraints on SUSY models and their parameters~\cite{Roszkowski:2009ye,Bertone:2010rv}. Even if no positive signal is observed up to relatively high energies, one can still exclude extensive regions of many interesting SUSY scenarios using the resultant bounds on sparticle masses. 

There are however various other possibilities for testing SUSY. These are almost all related to the abilities of such models to provide DM candidates (for some general reviews, see e.g. refs.~\cite{DMBergstrom:2000,DMBertone:2005,DMBergstrom:2009,DMBertone:2010}, and for an introduction to supersymmetric DM in particular, see ref.~\cite{Jungman:1995df}). If DM consists of Weakly Interacting Massive Particles (WIMPs) thermally produced in the early Universe, these can be detected either directly, via interactions with SM particles in atomic nuclei of large-volume target materials on Earth, or indirectly by measuring the primary and secondary products of their self-annihilation in the sky.  These products may take the form of photons, neutrinos, antimatter or other cosmic rays. To these detection channels we should also add the strong constraint on the relic abundance of DM, garnered from various ground-based telescopes and space missions. WMAP measurements of the cosmic microwave background anisotropies~\cite{Komatsu:2008hk} have been perhaps the most influential of these.  Direct and indirect searches are crucial for establishing that a particular candidate WIMP is in fact stable on cosmological timescales, and constitutes the majority of DM; whilst the LHC will be able to produce and discover heavy neutral particles, it will not be able to confirm their stability beyond the time they take to exit the detector~\cite{Bertone:2010rv}.

Over the past few years, both direct and indirect detection experiments have seen huge boosts in both quality and quantity. In indirect detection (ID), gamma-ray telescopes and other particle detectors are currently observing potential products of DM annihilation from different astrophysical objects, through different annihilation channels. Some examples are: the \emph{Fermi} gamma-ray space telescope~\cite{Atwood:2009ez}, together with large air \v{C}erenkov gamma-ray telescopes (ACTs) such as VERITAS~\cite{Holder:2008ux}, MAGIC~\cite{magic} and H.E.S.S.~\cite{Aharonian:2008aa} for detection with photons, the PAMELA satellite~\cite{Adriani:2008zr} together with \emph{Fermi}, H.E.S.S. and some balloon missions such as ATIC~\cite{:2008zzr} for detection with electrons, positrons and antiprotons, and IceCube~\cite{Ahrens:2002dv} for detection with neutrinos. These experiments have already started touching upon interesting regions of SUSY parameter spaces (see e.g. refs.~\cite{Scott:2009jn,Ripken}), but in some cases this ability should improve significantly in the next few years. Future experiments such as the AMS-02 space shuttle mission~\cite{Choutko} and the GAPS long-duration balloon flight~\cite{Aramaki:2010zz} provide the possibility of detecting antideuterons, another promising ID channel. The major problem with ID searches is the large contamination of signals with backgrounds from astrophysical sources. This becomes especially problematic when one searches for a signal from the Galactic Centre.

The situation is even more promising for direct detection (DD) experiments. Compared to the ID case, background contamination is much better understood. DD experiments are also relatively cheap and far easier to construct than ID ones. These factors have motivated several experimental groups to build such experiments. Currently, a copious number exist, sometimes with very similar properties, something that is always good for error reduction. These experiments have already reached sufficient sensitivity to probe small parts of SUSY parameter spaces, but this sensitivity is expected to significantly improve in the near future. It is therefore very interesting for both SUSY phenomenologists and DD experimentalists to get some feeling for how strong future DD limits and detections will be for SUSY DM models. It is our intention to provide that feeling in this paper.

In this paper, we study the ability of future ton-scale DD experiments to constrain the parameter space of the CMSSM. For these purposes, we consider ton-scale versions of three current experiments (CDMS, XENON and COUPP), which are together expected to cover a large part of the remaining CMSSM parameter space. The analysis is performed using a state-of-the-art parameter estimation technique optimised for Bayesian statistical inference, and is based on full likelihood constructions for the experiments. We include typical energy resolutions, thresholds and efficiencies for such experiments, as well as backgrounds at target levels. We take into account different uncertainties in the values of Milky Way halo parameters, as well as the hadronic matrix elements for neutralino-nucleon couplings.

The paper is organised as follows: in~\sec{sec:analysis} we start with a brief review of the theory of direct detection of WIMPs, in particular the neutralino in the framework of SUSY DM. We describe our different assumptions regarding halo models and the calculation of neutralino-nucleon scattering cross-sections, as well as our treatment of different astrophysical and particle physics uncertainties. We then specialise the discussion to our chosen ton-scale experiments CDMS, XENON and COUPP, and present our likelihoods for the experiments. After discussing our benchmarks and generated data, we introduce our SUSY model and parameters, scanning strategy and statistical framework. Results are presented and discussed in~\sec{sec:results}, followed by concluding remarks and future prospects in~\sec{sec:concl}.

\section{Framework, data and analysis} \label{sec:analysis}
\subsection{Direct detection of supersymmetric WIMPs} \label{sec:DD}

WIMP direct detection experiments seek to measure the energy deposited
when a WIMP interacts with a nucleus in a detector
\cite{Goodman:1984dc,SmithLewin,Jungman:1995df}.
If a WIMP of mass $m$ scatters elastically from a nucleus of mass $M$,
it will deposit a recoil energy $\Enr = (\mu^2v^2/M)(1-\cos\theta)$,
where $\mu \equiv m M/ (m + M)$ is the reduced mass of the WIMP-nucleus
system, $v$ is the speed
of the WIMP relative to the nucleus, and $\theta$ is the scattering
angle in the centre of mass frame.  The differential recoil rate per
unit detector mass for a WIMP mass $m$, typically given in units of
cpd\,kg$^{-1}$\,keV$^{-1}$ (where cpd is counts per day), can be written as
\begin{equation} \label{eqn:dRdEnr}
  \dRdEnr = \frac{1}{2 m \mu^2}\, \sigma \, F^2(q)
            \, \rhoDM \, \eta(\Enr,t) \, ,
\end{equation}
where $q = \sqrt{2 M \Enr}$ is the nucleus recoil momentum,
$\sigma$ is the WIMP-nucleus cross-section in the limit of zero momentum transfer,
$F^2(q)$ is a form factor due to the finite size of the nucleus
(normalised to $F^2(0) = 1$),
$\rho$ is the local WIMP density,
and information about the WIMP velocity distribution is encoded into
the mean inverse speed $\eta(\Enr,t)$,
\begin{equation} \label{eqn:eta}  
  \eta(\Enr,t) = \int_{u > \vmin} d^3u \, \frac{f(\bu,t)}{u} \, .
\end{equation}
Here
\begin{equation} \label{eqn:vmin}
  \vmin = \sqrt{\frac{M \Enr}{2\mu^2}}
\end{equation}
represents the minimum WIMP velocity that can result in a recoil energy
$\Enr$ and $f({\bf u},t)$ is the (time-dependent) distribution of WIMP
velocities ${\bf u}$ relative to the detector.

Real experimental apparatuses
cannot determine the event energies with perfect precision.
The recoiling nucleus (or recoiling electron, in the case of some
backgrounds) will transfer its energy to either electrons,
which may be observed as \eg\ ionisation or scintillation in the
detector, or to other nuclei, producing phonons and heat.  Measurements
of these signals are then used to make an estimate $E$ of the recoil
energy for the event that produced those signals.  However, due to
intrinsic stochastic fluctuations in the chemical and physical processes
that produce and propagate the ionisation, scintillation, and phonons,
and due to technical limitations in measuring those signals, the
measured signal (and thus estimated recoil energy $E$) exhibits some
random variation for a given true recoil energy $\Enr$,
resulting in a finite (imperfect) energy resolution.
In addition, limitations in detection efficiency and data cuts
applied to reduce backgrounds mean that not all recoil events will be detected and/or accepted.
Thus, in a real experiment, the expected observed spectrum is
\begin{equation} \label{eqn:dRdE}
  \dRdE = \int_{0}^{\infty} d\Enr \; \phi(E,\Enr)
                            \, \dRdEnr(\Enr,t) \, ,
\end{equation}
where $\phi(E,\Enr) \: \Delta E$ is the probability of an event with
recoil energy $\Enr$ being observed with a measured energy between
$E$ and $E + \Delta E$ (in the limit $\Delta E \to 0$) after
accounting for efficiencies (data cuts) and the finite energy
resolution.
There are two important distinctions between \refeqn{dRdEnr} and
\refeqn{dRdE}.
(1) The latter is the \textit{observed} rate, which is smaller than
the former (nuclear recoil rate) as some events will fail to be
observed and/or fall into the analysis window.
(2) The latter equation is in terms of the measured energy $E$ rather
than the true recoil energy $\Enr$ and accounts for the random
fluctuations in the measurements.  In many cases, the fluctuations are
small and taking $E = \Enr$ is a reasonable approximation.  However,
for many experiments, the relative size of the fluctuations grows as
$\Enr$ becomes small, so the finite energy resolution should be taken
into account whenever low energy recoils become important (typically
the case for light WIMPs, as they can produce only low energy recoils).

The expected number of observed (signal) recoil events between $\Emin$
and $\Emax$ is given by
\begin{equation} \label{eqn:mu}
  \mu_S = M_{det} T \int_{\Emin}^{\Emax} dE \, \sum_i f_i \dRdEi(E) \, ,
\end{equation}
where $M_{det}$ is the detector mass, $T$ is the exposure time,
the sum is over the isotopes present in the detector material
(possibly involving multiple elements),
and $f_i$ is the mass fraction of each isotope.
Here, we have neglected the time dependence in $\dRdE$, the source of
which is described in \refsec{HaloModel}.  In principle, the time
dependence can be
properly accounted for by replacing $T$ in the equation above with an
integral over the time that the experiment is running.  However, the
time dependence is small and not relevant for the experiments we
consider in this paper, so we will use only the time averaged recoil
spectrum of \refeqn{dRdEnr}.

We use \textsf{DarkSUSY}~\cite{darksusy} to calculate the differential
recoil rate of \refeqn{dRdEnr}, albeit with some non-default
parameters used in calculating the scattering cross-sections and
describing the dark matter halo distribution.
In the following sections, we examine these cross-sections and halo models in more detail.

\subsubsection{\label{sec:CrossSection} Cross-section}

For an MSSM neutralino, the only neutralino-quark couplings that do not
vanish in the non-relativistic limit, and are thus not suppressed in
the scattering of the low velocity relic neutralinos with baryons (the
basis of direct detection), are the axial vector and scalar
couplings~\cite{Jungman:1995df,Falk:1998xj}:
\begin{equation} \label{eqn:Lagrangian}
  \mathcal{L} = \alpha_{2q} \bar{\chi} \gamma^\mu \gamma^5 \chi
                \bar{q} \gamma_{\mu} \gamma^{5}  q
                + \alpha_{3q} \bar{\chi} \chi \bar{q} q.
\end{equation}
The axial vector and scalar couplings give rise to spin-dependent (SD)
and spin-independent (SI) cross-sections for elastic scattering of a
neutralino with a nucleus, respectively. Different form factors apply in
each case, so that
\begin{equation} \label{eqn:CStot}
  \sigma F^2(q) \to \sigmaSD F_{\mathrm{SD}}^2(q)
                      + \sigmaSI F_{\mathrm{SI}}^2(q)
\end{equation}
in \refeqn{dRdEnr}, where $\sigmaSD$ and $\sigmaSI$ are the SD and
SI scattering cross-sections.
The two cross-sections are described below.

\paragraph{\label{sec:SDCS} Spin-dependent cross-section (SD).}
The SD WIMP-nucleus cross-section can be written in terms of effective
couplings to the proton ($\apSD$) and neutron ($\anSD$) as
\begin{equation} \label{eqn:sigmaSD}
  \sigmaSD = \frac{32}{\pi} G_{F}^{2} \mu^2 \Lambda^2 J(J+1) \; ,
\end{equation}
where $\mu$ is the neutralino-nucleus reduced mass, $J$ is the
spin of the nucleus,
\begin{equation} \label{eqn:Lambda}
  \Lambda \equiv \frac{1}{J} \left(
                 \apSD \langle S_{\rmpp} \rangle
                 + \anSD \langle S_{\rmnn} \rangle
                 \right) \; ,
\end{equation}
and\footnote{
  The effective spin-dependent couplings to the proton and neutron are
  sometimes instead given as $G_a^{\rmpp}$ and $G_a^{\rmnn}$,
  differing from the above couplings only by the normalisation:
  $G_a^{N} = 2\sqrt{2}G_F\,  a_{N}$ \cite{darksusy,Gondolo:1996qw}.
  }
\begin{equation} \label{eqn:aN}
  \apSD = \sum_{q} \frac{\alpha_{2q}}{\sqrt{2} G_{f}} \Deltapq{q} , \qquad
  \anSD = \sum_{q} \frac{\alpha_{2q}}{\sqrt{2} G_{f}} \Deltanq{q} \; .
\end{equation}
The factors $\DeltaNq{q}$ ($N \in \{\rmpp,\rmnn\}$) parametrise the
quark spin content of the
nucleon and are only significant for the light (u,d,s) quarks.
A combination of experimental and theoretical results tightly
constrain the linear combinations~\cite{Yao:2006px}
\begin{equation} \label{eqn:a3}
  \athree \equiv \Deltapq{\rmu} - \Deltapq{\rmd}
          = 1.2695 \pm 0.0029
\end{equation}
and~\cite{Goto:1999by,Leader:2002ni}
\begin{equation} \label{eqn:a8}
  \aeight \equiv \Deltapq{\rmu} + \Deltapq{\rmd} - 2 \Deltapq{\rms}
          = 0.585 \pm 0.025 ,
\end{equation}
while the strange quark contribution is given by the COMPASS
result~\cite{Alekseev:2007vi},
\begin{equation} \label{eqn:Deltaps}
  \Deltapq{\rms} = -0.09 \pm 0.01 \, \text{(stat)} \,
                         \pm 0.02 \, \text{(syst)}
                 \approx -0.09 \pm 0.03 \; ,
\end{equation}
where we have conservatively combined the statistical and systematic
uncertainties.
The proton and neutron spin contributions are related by an
interchange of $\Delta_{\rmu}$ and $\Delta_{\rmd}$
(\ie\ $\Deltanq{\rmu} = \Deltapq{\rmd}$,
$\Deltanq{\rmd} = \Deltapq{\rmu}$, and
$\Deltanq{\rms} = \Deltapq{\rms}$).
The SD couplings $\apSD$ and $\anSD$ can be given in terms of the SD
scattering cross-sections for a WIMP with the proton ($\sigmapSD$) and
neutron ($\sigmanSD$), respectively, via \refeqn{sigmaSD}.

The SD form factor $F_{\mathrm{SD}}^2(q)$ is dependent upon the spin
structure within each nucleus and is thus unique to individual nuclei.
We use the default form factors implemented in \textsf{DarkSUSY}, which are
taken from Ref.~\cite{Ressell:1997kx} for iodine and xenon and from
Ref.~\cite{Ressell:1993qm} for germanium (an interacting shell model
form factor calculation is not implemented in \textsf{DarkSUSY}
for fluorine, for which a simple exponential form factor is used).
These and additional SD form factors may be found in the review of
Ref.~\cite{Bednyakov:2006ux}.

\paragraph{\label{sec:SICS} Spin-independent cross-section (SI).}
The SI WIMP-nucleus cross-section can likewise be written in terms of
effective couplings to the proton ($\fpSI$) and neutron ($\fnSI$) as
\begin{equation} \label{eqn:sigmaSI}
  \sigmaSI = \frac{4}{\pi} \mu^2
             \left[ Z \fpSI + (A-Z) \fnSI  \right]^{2} \; ,
\end{equation}
where $\mu$ is again the reduced mass and $Z$ and $A-Z$ are the number
of protons and neutrons in the nucleus, respectively. 
For the scalar neutralino-quark couplings of
\refeqn{Lagrangian},\footnote{
  As with the spin-dependent case, the spin-independent couplings to
  the proton and neutron are sometimes given with an alternate
  normalisation: $G_s^{\rmpp} = 2\fpSI$ and $G_s^{\rmnn} = 2\fnSI$
  \cite{darksusy,Gondolo:1996qw}.
  }
\begin{equation} \label{eqn:fN}
  \frac{f_N}{m_N}
    = \sum_{q=\rmu,\rmd,\rms} \fNTq{q} \frac{\alpha_{3q}}{m_{q}}
      + \frac{2}{27} f_{TG}^{(N)}
        \sum_{q=\rmc,\rmb,\rmt} \frac{\alpha_{3q}}{m_q},
\end{equation}
with
\begin{equation} \label{eqn:fTq}
  m_N \fNTq{q}
  \equiv \smatrix[N]{m_q \qqbar{q}}
\end{equation}
and 
\begin{equation} \label{eqn:fTG}
  f_{TG}^{(N)} = 1 - \sum_{q=\rmu,\rmd,\rms} \fNTq{q} .
\end{equation}
Here $\fNTq{q}$ is the contribution of quark $q$ to the composition of nucleus $N$.  The contribution from the heavy quarks (c,b,t) arises
through an anomaly involving gluons; the factor $f_{TG}^{(N)}$ is fixed by this anomaly \cite{Shifman:1978zn,Vainshtein:1980ea}.
The matrix elements $\smatrix{\qqbar{q}}$ for the light quarks (u,d,s)
can be determined from estimates of the linear
combinations~\cite{Pavan:2001wz,Ellis:2005mb}
\begin{equation} \label{eqn:SigmapiN}
  \SigmapiN \equiv \frac{1}{2} (\muu + \mdd) \,
                   \smatrix{\left( \qqbar{\rmu} + \qqbar{\rmd} \right)}
            = 64 \pm 8 \, \mathrm{MeV}
\end{equation}
and~\cite{Borasoy:1996bx,Gasser:1990ce,Knecht:1999dp,Sainio:2001bq}
\begin{equation} \label{eqn:sigma0}
  \sigma_0 \equiv \frac{1}{2} (\muu + \mdd) \,
                   \smatrix{\left( \qqbar{\rmu} + \qqbar{\rmd}
                                   - 2 \qqbar{\rms} \right)}
           = 36 \pm 7 \, \mathrm{MeV} \, ,
\end{equation}
along with~\cite{Cheng:1988im,Ellis:2005mb}
\begin{equation} \label{eqn:z}
  z \equiv \frac{\smatrix{\left( \qqbar{\rmu} - \qqbar{\rms} \right)}}%
                {\smatrix{\left( \qqbar{\rmd} - \qqbar{\rms} \right)}}
    = 1.49
\end{equation}
and the quark mass ratios $\mumd = 0.553$ and $\msmd = 18.9$
\cite{Leutwyler:1996qg}.
The uncertainties in $z$ and the quark mass ratios are negligible
compared to the uncertainties in $\SigmapiN$ and $\sigma_0$.
The pion-nucleon sigma term $\SigmapiN$ parametrises the nucleon mass
shift from the chiral limit and is determined from pion-nucleon
scattering, while the combinations of matrix elements corresponding
to $\sigma_0$ and $z$ are determined from octet baryon mass differences.
The neutron matrix elements are related to the proton matrix elements
by an interchange of $\smatrix[N]{\qqbar{\rmu}}$ with
$\smatrix[N]{\qqbar{\rmd}}$
(\ie\ $\smatrix[\rmnn]{\qqbar{\rmu}} = \smatrix{\qqbar{\rmd}}$,
$\smatrix[\rmnn]{\qqbar{\rmd}} = \smatrix{\qqbar{\rmu}}$, and
$\smatrix[\rmnn]{\qqbar{\rms}} = \smatrix{\qqbar{\rms}}$).

As with the SD case, the SI couplings $\fpSI$ and $\fnSI$ can be given
in terms of the SI scattering cross-sections for a WIMP with the proton
($\sigmapSI$) and neutron ($\sigmanSI$), respectively, via
\refeqn{sigmaSI}.
For an MSSM neutralino, \refeqn{fN} typically yields $\fnSI \sim \fpSI$,
so that $\sigmanSI \approx \sigmapSI$; SI scattering is then typically
characterised by the single parameter $\sigmapSI$ (aside from the
neutralino mass).  MSSM models often predict $\sigmaNSD \gg \sigmaNSI$.
However, the contributions to the total SI cross-section of a nucleus
are a coherent sum over the individual proton and neutrons within,
which yields the $\sim A^2$ scaling of the SI cross-section apparent
in \refeqn{sigmaSI}.  On the other hand, the contribution from the
spins of individual nucleons tend to cancel in the SD case, so there
is no such scaling with nucleus size for $\sigmaSD$. As a result of
this scaling, heavier elements are often expected to have larger
SI cross-sections than SD cross-sections, even if the reverse is true
for a lone proton or neutron.  For this reason, most direct detection
experiments use heavy elements such as xenon, iodine, and germanium as
their target materials and they mainly aim to observe or place
constraints on SI interactions.

\textsf{DarkSUSY} implements a variety of models for the SI form factor
$F_{\mathrm{SI}}^2(q)$.  We use the default \textsf{DarkSUSY} form
factors, which is usually the commonly used Helm form
factor \cite{SmithLewin}, but is a Fourier-Bessel or sum-of-Gaussians
expansion for a limited number of isotopes in which these expansions
are available.  For the target elements used in our hypothetical
experiments (see \refsec{likes}), most of the most significant carbon
and germanium isotopes use Fourier-Bessel expansions (the exceptions
being $^{13}$C and $^{73}$Ge), while the remainder use a Helm form
factor.  See Ref.~\cite{Duda:2006uk} for a detailed discussion of these
form factors.
\begin{table}
  \begin{center}
  \addtolength{\tabcolsep}{1em}
  \begin{tabular}{lr@{$\,\pm\,$}lll}
    \toprule
    \textbf{Parameter}  & \multicolumn{2}{c}{\textbf{Mean/s.d.}\ \ \ }
                                  & \textbf{Distribution} & \textbf{Ref.} \\
    \toprule
    \Deltaps   & -0.09  & 0.03    & normal     &~\cite{Alekseev:2007vi} \\
    $\sigma_0$~(MeV)& 36     & 7   & log-normal &~\cite{Borasoy:1996bx} \\
    \SigmapiN~(MeV) & 64     & 8   & log-normal &~\cite{Pavan:2001wz,Ellis:2005mb} \\
    \midrule
    \rhoDM~(GeV/cm$^3$)    & 0.40   & 0.15
                                  & log-normal
      &~\cite{Caldwell:1981rj,Catena:2009mf,Weber:2009pt,Salucci:2010qr} \\
    \vrot~(km/s)     & 235    & 20 & normal
      &~\cite{Kerr:1986hz,Reid:2009nj,McMillan:2009yr,Bovy:2009dr} \\
    \vmp~(km/s)       & 235    & 20 & normal     & \\
    \vesc~(km/s)      & 550    & 35 & normal     &~\cite{Smith:2006ym} \\
    \bottomrule
  \end{tabular}
  \end{center}
  \caption[Direct detection parameters]{
    \footnotesize{Likelihood distributions for parameters used to calculate the
    scattering cross-sections and to define the halo model.
    The parameters are defined in the text.
    The likelihoods for $\rhoDM$ and $\vrot$ have been chosen based on
    a variety of estimates and do not represent any one particular
    estimate. The distribution in $\vmp$ was chosen so that the allowed values of
    $\vmp$ and $\vrot$ cover a range of reasonable smooth dark matter
    halo models.  Notably, we treat these two parameters as independent
    and do not make the common assumption that $\vmp = \vrot$.
    See the text for further discussion regarding the choice of
    distribution for each parameter.
    }
    }
  \label{tab:DDParameters}
\end{table}

\paragraph{\label{sec:CSLike} Parameters and likelihoods.}
We vary the quantities 
$\Deltaps$, $\sigma_0$, and $\SigmapiN$ in our scans with the
likelihood distributions given in \reftab{DDParameters}.
For the positive parameter $\SigmapiN$, we use a log-normal
distribution rather than a normal distribution as the latter would
allow for unphysical negative values.  In practice, though,
there is little difference between the two distributions for the
assumed mean and variance except in the tails of the distribution
(beyond the 2-3$\sigma$ level).  We also avoid (likely disfavoured)
small values of $\sigma_0$ by using a log-normal distribution as
these small values are suppressed in this distribution relative to a
normal distribution.\footnote{
  To first order in SU(3) breaking,
  $\sigma_0 \approx \frac{\hat{m}}{m_{\rms} - \hat{m}}
   \left( m_{\Xi} + m_{\Sigma} - 2 \, m_{N} \right)$, where
  $\hat{m} \equiv \frac{1}{2} \left( m_{\rmu} + m_{\rmd} \right)$
  \cite{Cheng:1988im,Sainio:2001bq}; a negative value of $\sigma_0$
  would thus suggest the proton and neutron are not the lightest
  baryons.  Even with higher order corrections, it would be difficult
  to rectify a negative or even small positive value of $\sigma_0$
  with observed baryon mass splittings.
  }
As with $\SigmapiN$,
there would be little difference in practice had we used here a normal
distribution for $\sigma_0$ as the two distributions are substantially
similar in this case, except at their tails.
We additionally require $\SigmapiN \ge \sigma_0$, to avoid the strange scalar
matrix element becoming (unphysically) negative.

The parameters $\athree$, $\aeight$, and $\Deltaps$ affect the
determination of the SD scattering cross-section $\sigmaSD$ for a
given SUSY model.  The uncertainties in these parameters as given in
\reftab{DDParameters} and in \refsec{SDCS} result in variations in the
calculated $\sigmaSD$ by $\sim \times$1.5, almost entirely due to the
uncertainties in $\Deltaps$.  As variations in $\athree$ and $\aeight$
would not significantly affect our results, we fix them to their
mean values and vary only $\Deltaps$.

Uncertainties in both $\sigma_0$ and $\SigmapiN$ yield significant
variations in the predicted SI scattering cross-section $\sigmaSI$
for a given SUSY model, by as much as an order of magnitude at the
2$\sigma$ level.  In many cases, the strange quark contribution in
\refeqn{fN} is the dominant contribution to $f_N$.\footnote{%
  Due to the quark mass factor in \refeqn{fTq}, $\fTq{\rms} \gg
  \fTq{\rmu}, \fTq{\rmd}$.  In the CMSSM and some other SUSY models,
  $\alpha_{3q} \propto m_q$, so that the $\frac{\alpha_{3q}}{m_q}$
  terms in \refeqn{fN} are not suppressed for heavier quarks by the
  explicit quark mass appearing there.
  Combined with the $\frac{2}{27}$ factor that suppresses the heavy
  quark contributions, the strange quark term will thus dominate unless
  $\alpha_{3\rms}$ is suppressed relative to other quark couplings.
  }
As $\smatrix{\qqbar{\rms}} \propto \SigmapiN - \sigma_0$,
$f_N \sim (\SigmapiN - \sigma_0)^2$ (approximately) in this case.
The uncertainties in $\sigma_0$ and $\SigmapiN$ gain additional
significance due to this dependence on the difference of the two
terms, which has a larger total uncertainty than the individual
parameters themselves.
Given the very strong dependence of $\sigmaSI$ on the values of
$\sigma_0$ and $\SigmapiN$, the uncertainties in these two parameters
are the most important to take into account when fitting a SUSY model
to a direct detection result.

We neglect uncertainties in parameters other than these three
(\eg\ quark mass ratios) as the resulting uncertainties in the
calculated SD and SI cross-sections are negligible in comparison.  A more detailed discussion
of how these and other parameters affect the elastic scattering
cross-sections may be found in Ref.~\cite{Ellis:2008hf}.

\subsubsection{\label{sec:HaloModel} Halo model}

The dark matter halo in the local neighbourhood is likely to be composed
mainly of a smooth, well mixed (virialised) component with an average
density $\rhoDM \sim 0.4$~GeV/cm$^3$.
The simplest model of this smooth component is the Standard Halo Model
(SHM) \cite{Freese:1987wu}.  The SHM is an isothermal sphere with an
isotropic, Maxwellian velocity distribution characterised by an rms
velocity dispersion $\sigma_v$.  Sufficiently high velocity WIMPs would
escape the Galaxy's potential well and will not be present in the halo
in any significant quantities; the lack of these high velocity WIMPs
can be accounted for by truncating the Maxwellian distribution at some
escape velocity $\vesc$, yielding
\begin{equation} \label{eqn:Maxwellian}
  \widetilde{f}(\bv) =
    \begin{cases}
      \frac{1}{\Nesc} \left( \pi \vmp^2 \right)^{-3/2}
        \, e^{-\bv^2\!/\vmp^2} , 
        & \textrm{for} \,\, |\bv| < \vesc  \\
      0 , & \textrm{otherwise}.
    \end{cases}
\end{equation}
Here
\begin{equation} \label{eqn:Nesc}
  \Nesc = \erf(z) - 2 z \exp(-z^2) / \pi^{1/2} ,   
\end{equation}
with $z \equiv \vesc/\vmp$, is a normalisation factor that corrects for
the truncation in the Maxwellian. Here we have also defined
\begin{equation} \label{eqn:vmp}
  \vmp = \sqrt{2/3} \, \sigma_v,
\end{equation}
the most probable speed (in the non-truncated Maxwellian).

The dark matter halo is essentially non-rotating, while the Sun
moves with the disk, which rotates about the centre of the Galaxy
at a speed $\vrot$.
The halo thus exhibits a bulk motion relative to Earth, so that
\begin{equation} \label{eqn:vdist}
  f(\bu,t) = \widetilde{f}(\bvobs(t) + \bu)
\end{equation}
is the velocity distribution as seen from Earth's frame,
where $\bvobs(t)$ is the motion of the observer (Earth) relative to the
rest frame of the dark matter halo described by \refeqn{Maxwellian}.
The motion relative to this stationary halo is
\begin{equation} \label{eqn:vobs}
  \bvobs(t) = \bv_{\mathrm{LSR}} + \bv_{\odot,\mathrm{pec}}
              + \bV_{\oplus}(t)
    \, ,
\end{equation}
where $\bv_{\mathrm{LSR}} = (0,\vrot,0)$ is the motion of the Local
Standard of Rest in Galactic coordinates,
$\bv_{\odot,\mathrm{pec}} = (10,13,7)$~km/s is the Sun's peculiar
velocity (see \eg\ Refs.~\cite{Mignard:2000aa,Schoenrich:2009bx} and
references therein),
and $\bV_{\oplus}(t)$ is the velocity of the Earth relative to the Sun.
This last term varies throughout the year as the Earth orbits the Sun,
leading to an annual modulation in the velocity distribution and,
thus, the recoil rate of \refeqn{dRdEnr} \cite{Drukier:1986tm,
Freese:1987wu}.
However, $\vrot \sim 200$-$300$~km/s, while the Earth's orbital speed
is 30~km/s, so $\bV_{\oplus}$ makes only a small contribution to
$\bvobs$.  In addition, the Earth's orbital plane is inclined to the
plane of the Galactic disk by 60$^{\circ}$ so that $\vobs$ changes by
$\sim \frac{1}{2} V_{\oplus}$ ($\sim$15~km/s) due to the Earth's orbit.
The result is that $\vobs$ fluctuates by $\lae 5\%$ throughout the year.
This results in a small time dependence of the velocity distribution
that yields a small annual modulation in the recoil spectrum of
\refeqn{dRdEnr};
observation of this modulation is the goal of experiments such as
DAMA/NaI and DAMA/LIBRA \cite{Bernabei:2003za,Bernabei:2008yi}.
This small fluctuation is not important for the direct detection
experiments and cases considered here; for simplicity, we drop the
small $\bV_{\oplus}(t)$ term in \refeqn{vobs} and use only the
yearly average for $\bvobs$.
Further discussion of the annual modulation effect,
along with analytical expressions for the integral in \refeqn{eta}
for this model,
may be found in Ref.~\cite{Savage:2006qr}.

The isotropic, Maxwellian velocity distribution of \refeqn{Maxwellian},
intended to describe a class of smooth spherical halo models, should be
considered only a first approximation of the local halo profile.
Oblate/prolate or triaxial halos would be expected to have an anisotropic
velocity distribution, though the assumption of isotropy is not
necessarily a poor one if the level of oblateness/triaxiality is
small, as is probably the case for the Milky Way.  In addition, the
formation of the Milky Way via merger events throughout its history
leads to significant structure in both the spatial and velocity
distribution of the dark matter halo.  In the inner portions of the
Galaxy, including the region containing the Sun, this structure is
thought to have been substantially mixed to the point that a smooth
distribution such as \refeqn{Maxwellian} is a reasonable approximation.
However, this does not preclude the possibility that a neutralino
population from a late merger event (such as a tidal stream from a
dwarf galaxy in the process of being absorbed by the Milky Way) might
contribute an appreciable amount to the local density, with a velocity
distribution significantly different from the well mixed portion.
However, the presence of such structure locally, while possible, is not
probable and we ignore it here (see \eg\ Ref.~\cite{Schneider:2010jr}).

We note that the choice of excluding high velocity WIMPs by truncating
the Maxwellian of \refeqn{Maxwellian} at the local escape velocity is
somewhat arbitrary.  An alternative ad-hoc method of removing the high
velocity WIMPs from the velocity distribution via subtraction instead
of truncation can be seen in \eg\ Refs.~\cite{Fairbairn:2008gz,
McCabe:2010zh}.  Both of these forms are attempts to remove the most
problematic part of the velocity profile ``by hand.''  
Chaudhury \etal\ \cite{Chaudhury:2010hj} use King models to more
self-consistently handle the finite size and mass of the Galaxy
in determining the velocity distribution;
in these models, the maximum WIMP velocity in the halo is, in
fact, somewhat below the escape velocity.  While a Maxwellian might
reasonably approximate the bulk of the WIMP velocity distribution, the
appropriate form of the tail is not so clear (but perhaps it will be
clarified by numerical simulations \cite{Kuhlen:2009vh}).
Fortunately, the tail of the velocity distribution is unimportant
for most direct detection experiments and WIMP candidates, the
exception being very light WIMPs ($\sim$10~GeV or less), which
we do not consider.\footnote{
  In practice, there is typically very little difference in direct
  detection constraints when using the two different methods for
  modifying the Maxwellian distribution (truncation and subtraction).
  For CDMS or XENON exclusion curves, one finds the low mass fall-off
  of the constraint in the cross-section vs.\ mass plane is shifted
  by $\sim$1/2~GeV \cite{Schwetz:2010pc}.
  }
The imposition of a cut-off at the escape velocity $\vesc$ above serves
to avoid unrealistic cases; however, any result that depends
significantly on the value of this cut-off should be viewed with
caution, as it likely also depends strongly on the 
(probably inaccurately) assumed form for the tail of the
velocity distribution.

\paragraph{\label{sec:rho0} Local dark matter density ($\rhoDM$).}
The dark matter halo density has long been understood on the basis of Galactic
modelling to have an average value of $\sim$0.3~GeV/cm$^3$ in the solar
neighbourhood, but with an uncertainty on the order of a
factor of 2 \cite{Caldwell:1981rj}.  This often-quoted value (though not the most recent or precise estimate) is typically taken as the fiducial case in determining direct detection constraints.
As one can always write \refeqn{dRdEnr} in terms of the quantity
$\rhoDM \sigma$, direct detection experiments are unable to separately
constrain the local density and scattering cross-section.
Direct detection cross-section constraints are given either explicitly
or implicitly in terms of the quantity $\xi\sigmapSI$ (and similar for
the SD scattering cross-sections), where
$\xi \equiv \rhoDM/$(0.3~GeV/cm$^3$).  This allows for easy rescaling
of constraints for other local density values, while the choice of a
common reference value allows for comparison of constraints for
different experiments.

For scans over SUSY parameters, however, knowledge of the true local
density is important.  If the assumed local density is too high or
too low, the scans will find SUSY models yielding scattering
cross-sections that are too low or too high, respectively.  Thus,
our limited knowledge of the true local density must be taken into account.

Estimates of the local density are often made by fitting a variety of
galactic observations to a particular model of the Galaxy that
includes both dark matter and baryons.  The reliability of these
estimates are limited not only by statistical and systematic errors
in the various observations, but in how well the assumed model matches the true Galaxy.
Catena and Ullio \cite{Catena:2009mf} have suggested that for a spherical halo model, the
local dark matter density can be very accurately determined as
$0.385 \pm 0.027$~GeV/cm$^3$ for an Einasto density profile and
$0.389 \pm 0.025$~GeV/cm$^3$ for a Navarro-Frank-White (NFW) profile.
On the other hand, Weber and Boer, using a different method, find the
local dark matter density is more weakly constrained, taking values of
0.2--0.4~GeV/cm$^3$ for spherical models and even higher for oblate
models \cite{Weber:2009pt}.
Salucci \etal\ have tried to estimate the local density using a more
model independent method, finding
$\rhoDM = 0.43 \pm 0.11 \pm 0.10$~GeV/cm$^3$ \cite{Salucci:2010qr};
see their paper for a description of the two uncertainties.
Pato \etal\ have shown that, when relaxing complete spherical symmetry
of the dark matter halo, the dark matter density near the stellar disk
can be 1--41\% larger than the spherically-shelled average at the same
distance to the Galactic Centre \cite{Pato:2010yq}.  This last issue
should affect the estimates of Catena and Ullio, so the local halo
density is probably not so precisely constrained as their stated
values imply, though their novel method for determining the local
density might also be applied to more complicated halo models.

\paragraph{\label{sec:vrot} Velocity distribution.}
The velocity distribution characterised by \refeqn{Maxwellian} is fixed
by the three parameters $\vrot$, $\vmp$, and $\vesc$.
Based upon a variety of measurements, earlier estimates of the local
disk circular velocity $\vrot$ placed it at $\sim$220~km/s
\cite{Kerr:1986hz}.  This 220~km/s has been the canonical value
recommended by the International Astronomical Union (IAU) and is the
most widely assumed value when determining direct detection
constraints.\footnote{
  Often in the literature, the halo velocity distribution is
  characterised by a solar velocity of 232-233~km/s,
  which is obtained from this rotation speed of 220~km/s after
  including the Sun's peculiar velocity.  See \refeqn{vobs} and the
  surrounding discussion.
  }
However, a recent estimate by Reid \etal, based on Galactic masers,
suggests a higher value of $254 \pm 16$~km/s \cite{Reid:2009nj}.
A reanalysis of the same data by McMillan and Binney find that $\vrot$
ranges from $200 \pm 20$~km/s to $279 \pm 33$~km/s, depending on how
the Galactic rotation curve is modelled and what parameters are included
in their fits, though the more extreme cases suggest a distance to the
Galactic Centre $R_0$ that is significantly lower or higher than
typically found using other techniques \cite{McMillan:2009yr}.  The two
most favoured models of McMillan and Binney yield $235 \pm 19$~km/s and
$258 \pm 32$~km/s, but a a major point of discussion is how estimates
of the Sun's peculiar velocity $\bv_{\odot,\mathrm{pec}}$ impact these fits.  Bovy \etal\ have also reanalysed the
same maser data to find $\vrot = 244 \pm 13$~km/s, or
$\vrot = 236 \pm 11$~km/s if other observations are included in the
fit \cite{Bovy:2009dr}.
See Ref.~\cite{Savage:2009mk} for an example of how different values of
$\vrot$ impact direct detection constraints.

The velocity dispersion $\sigma_v$, or most probable speed $\vmp$, of
the dark matter halo is not directly observable, but it is related to
the disk rotational velocity.  For a spherical dark matter halo,
$\vmp \sim \vrot$.  For specific halo models, the relation is more
concrete: for an isothermal spherical halo (the SHM), $\vmp = \vrot$,
while for an NFW profile, $\vmp$ would be $\sim$10\% smaller than
$\vrot$ \cite{Serpico:2010ae} (the exact relation depends on the
scales used in the NFW profile).

WIMPs travelling faster than the local escape velocity for the Milky Way
will escape the Galaxy and thus will not contribute to any direct
detection signals.  High velocity stars in the RAVE survey have been
used to estimate the local escape velocity to be $\vesc = 544$~km/s
(498--608~km/s at the 90\% confidence level), though this estimate
is based upon a disk rotation speed of 220~km/s and could change for
other values of $\vrot$ \cite{Smith:2006ym}.

\paragraph{\label{sec:HaloLike} Parameters and likelihoods.}
We assume a spherical halo model as described above and allow the
nuisance parameters $\rhoDM$, $\vrot$, $\vmp$, and $\vesc$ to vary in
our scans with likelihood distributions as given in \reftab{DDParameters}.
As noted previously, \refeqn{Maxwellian} is only a first approximation
to the dark matter halo velocity profile; more general models of
the halo would allow for triaxiality of the halo, anisotropy of the
velocity profile, and the possible presence of structure remaining from
recent merger events.  We continue to use the 
spherical model for its simplicity, and note that our results will still
be representative of how uncertainties in describing the smooth
halo background affect the ability of scans to constrain the
neutralino parameters.  In any case, the direct detection experiments
considered here are not likely to be able to distinguish between more
complicated models of the smooth halo and this first order approximation
for SUSY models producing less than $O(100)$ events in the
detectors.\footnote{
  The annual modulation signal and directional distribution of nuclear
  recoils searched for by experiments such as
  DAMA \cite{Bernabei:2003za, Bernabei:2008yi}
  and DRIFT \cite{Burgos:2007zz}, respectively, are more sensitive to
  the halo velocity profile than experiments such as CDMS, XENON,
  and CoGeNT.  These latter three experiments would be required to
  observe many events in order to distinguish the modest changes
  induced in \refeqn{eta} by the use of a more complicated halo model.
  }

While many of these four parameters are highly correlated for a
specific halo model (such as an isothermal sphere, where
$\vmp = \vrot$), we do not assume any particular halo model, so we
take these quantities to be independent.  This allows the scans to
cover a wider range of realistic smooth halo backgrounds, allowing our lack of
knowledge of the true halo profile to be taken into account in our reconstructions of CMSSM parameters.

The distribution for the local density $\rhoDM$
($0.40 \pm 0.15$~GeV/cm$^3$) has been chosen to be representative of
typical averages and uncertainties found in the literature, and does
not correspond to any one estimate
\cite{Catena:2009mf,Weber:2009pt,Salucci:2010qr,Pato:2010yq}.
A log-normal distribution is chosen to avoid inappropriately small
local densities and to provide what is most likely to be an asymmetric
distribution.
The disk rotation velocity of $\vrot = 235 \pm 20$~km/s has been chosen
to be similar to current estimates \cite{Reid:2009nj,Bovy:2009dr},
apart from the more extreme cases found in Ref.~\cite{McMillan:2009yr},
and is still consistent with the older, IAU recommended estimate of
$220$~km/s \cite{Kerr:1986hz}.
For the velocity dispersion, we expect $\vmp \sim \vrot$, so the
distribution in $\vmp$ should have a relative variance at least as
large as that of $\vrot$, but the distribution should also allow for
a $\sim$10\% relative variation between $\vmp$ and $\vrot$ that
correspond to a variety of reasonable smooth halo models.
We take $\vmp = 235 \pm 20$~km/s, which satisfies both requirements.
This choice for a central value induces a mild preference for an
isothermal spherical halo ($\vmp = \vrot$), but the variance still
allows for velocity profiles consistent with reasonable alternative
halo profiles (such as the NFW).
Note that, even if the s.d.\ for $\vrot$ had been much
smaller, we would still wish to maintain the s.d.\ for $\vmp$ of
$\sim$20~km/s in order to allow for a variety of halo models.
For the escape velocity, we use $\vesc = 550 \pm 35$~km/s, comparable
to the estimates made in Ref.~\cite{Smith:2006ym}.  We ignore the
asymmetry in the distribution apparent in Figure~7 of
Ref.~\cite{Smith:2006ym}.  We further note that this estimate was
based on models of the Galaxy that assumed $\vrot = 220$~km/s; we
ignore the fact that this $\vesc$ estimate should be adjusted for
the different values of $\vrot$ that we use.  Again, we apply this
distribution to indicate how uncertainties in halo parameters impact
the scans and do not restrict ourselves to any particular halo model used to estimate the various halo parameters.  In any
case, the value of the escape velocity is not particularly important
for the experiments we consider as signals in these experiments are
dominated by contributions from neutralinos in the bulk of the
velocity distribution rather than the tail (at least for the reasonably
massive neutralinos we consider here).

See Refs.~\cite{McCabe:2010zh,Serpico:2010ae,Green:2010gw} and
references therein for further discussions regarding dark matter
halo parameters and their impact on direct detection.
See also Refs.~\cite{Strigari:2009zb,Bertone:2010rv} for examinations
of the impacts of these astrophysical uncertainties on reconstruction
of WIMP parameters in direct detection results.

\subsection{Likelihoods for ton-scale CDMS, XENON and COUPP} \label{sec:likes}

\begin{table}
  \begin{center}
  \begin{tabular}{llllll}
    \toprule
    \textbf{Experiment} & \textbf{Target}  & \textbf{Exposure}  & \textbf{Energy}      & \multicolumn{2}{c}{\textbf{Event measurements:}} \\
               &         & \textbf{[kg-year]} & \textbf{range [keV]} & \textbf{Number?} & \textbf{Energies?} \\
    \toprule
    CDMS1T     & Ge      & 1000      & 10--100     & yes     & yes \\
    XENON1T    & Xe      & 1000      & 8--75       & yes     & yes \\
    COUPP1T    & CF$_3$I & 1000      & $>$10       & yes     & no  \\
    \bottomrule
  \end{tabular}
  \end{center}
  \caption[Direct detection experiments]{\footnotesize{Future ton-scale experiments considered in this work, based on
    extrapolations from current generation experiments.
    The efficiencies and energy resolution of each experiment are
    discussed in the text.}
    }
  \label{tab:DDExperiments}
\end{table}

To examine the prospects for constraining SUSY models with future
direct detection results, we will use ton-scale extrapolations of the
current experiments
CDMS \cite{Akerib:2005zy,Ahmed:2009zw},
XENON10/100 \cite{Aprile:2010bt,Angle:2007uj,Angle:2009xb,Aprile:2010um},
and COUPP \cite{Behnke:2008zza,Behnke:2010xt};
these (hypothetical) future detectors will be referred to as CDMS1T,
XENON1T, and COUPP1T, respectively.\footnote{
  Ton-scale versions of these experiments are in various stages of
  planning, though the characteristics of those detectors may differ
  from what we use for our hypothetical detectors here.}
In all three cases, we assume 1000 kg-years of raw exposure, with
efficiencies, energy resolutions, and energy ranges similar to the
present day versions of these experiments.  The primary characteristics
of the experiments are shown in \reftab{DDExperiments}.

We define the likelihood of seeing $N$ events with observed energies
$E_{1,..,N}$ in a direct detection experiment as
\begin{equation} \label{eqn:DDLike}
  \mathcal{L}_{DD}(N,E_{1,..,N})
    \ = \ P(N|\mu_{tot}) \prod_{i=1..N} \! f(E_i) \, ,
\end{equation}
where $P(N|\mu_{tot})$ is the probability of seeing $N$ events for a
Poisson distribution with average $\mu_{tot}$ and $f(E)$ is the
probability of an event having an observed energy $E$.  The latter is
normalised such that $\int_{\Emin}^{\Emax}dE\,f(E) = 1$, where
$\Emin$--$\Emax$ is the energy range for the experiment.
The quantity $\mu_{tot}$ and functional form of $f(E)$ are dependent
on the CMSSM parameters and cross-section nuisance parameters, whose
dependences arise only through the neutralino mass and couplings to
the nucleons $(m,\apSD,\anSD,\fpSI,\fnSI)$,
as well as the halo nuisance parameters. For COUPP1T, which cannot
determine event energies, only the number of events, we simply use the
Poisson probability for the observed number of events
\begin{equation} \label{eqn:DDLikeN}
  \mathcal{L}_{DD}(N) = P(N|\mu_{tot}) \, .
\end{equation}
This special case will be discussed below.

We briefly discuss here the use of \refeqn{DDLike}.
This unbinned likelihood cannot be used to directly compare
different experimental results (realisations) for a given set of
CMSSM and nuisance parameters as the number of spectrum factors is
$N$, which is not constant.  This means that this form has units of
keV$^{-N}$ and the likelihoods for two different experimental results
(with different $N$) do not even have the same units.
One way of comparing experimental realisations is to examine
the likelihood ratio
\begin{equation} \label{eqn:DDLikeRatio}
R = \frac{\mathcal{L}_{DD}(N,E_{1,..,N} | \boldsymbol{\theta})}{\mathcal{L}_{DD}(N,E_{1,..,N} | \boldsymbol{\hat{\theta}})} \, ,
\end{equation}   
where $\boldsymbol{\theta}$ represents the SUSY and nuisance parameters
and $\boldsymbol{\hat{\theta}}$ is the set of parameters that maximises
the likelihood for that set of experimental results
\cite{Feldman:1997qc} (see Ref.~\cite{Abrams:2002nb} for an example of
the usage of this ratio in analysing CDMS results).
However, \refeqn{DDLike} provides all the dependence of the likelihood
on the SUSY and nuisance parameters for a given experimental result
and is sufficient for our purposes here.

Cosmic rays, trace radioactive isotopes, and other sources of radiation
induce recoils in a detector, providing a number of background events
in addition to the WIMP-induced (signal) recoil events.  Many 
direct detection experiments aim to reduce the expected background
to $\sim$1 background event by applying a variety of
techniques to both reduce the amount of radiation entering the
detector, and to discriminate background from signal
events.  The former typically involves reducing 
radioactive contaminants during detector fabrication,
placing detectors deep underground to reduce the cosmic ray flux,
placing shielding around detectors to reduce the influx of external
radiation, etc.  Discriminating background from signal
is often done by observing two or more signatures from a recoil in
the detector, such as ionisation and phonons (CDMS) or ionisation and
scintillation (XENON).  This allows very efficient removal of
electron recoil events induced by $\beta$ and $\gamma$ radiation, as
recoiling electrons tend to produce the two
signatures in a different ratio than recoiling nuclei do (where WIMP
interactions induce only nuclear recoils, not electron recoils).
Though this discrimination is efficient (often rejecting 99.9\% or
more of the $\gamma$-induced recoil events), the number of such
background events can be in the 1000s, so some leakage into the signal
region can still occur.
Neutrons, on the other hand, induce nuclear recoils, which are
indistinguishable from WIMP-induced nuclear recoils, and are harder
to discriminate.  However, the neutron radiation level is considerably
smaller than $\gamma$-rays, and can be further reduced by removal of
radioactive contaminants and using a muon veto to detect cosmic ray
showers (which can produce neutrons).  In addition, neutrons can
scatter multiple times in a detector; observation of a multiple scatter
is a clear signature of a neutron, because the probability of a WIMP
scattering more than once is negligible. This allows all
multiple-scattering events to be excluded from further analysis.

Though the goal of reaching a background level of $\sim$1 event might
seem overly ambitious, it is not an unreasonable one.  The current
generation of detectors have been able to reach this target, such as
CDMSII \cite{Ahmed:2009zw} (2 background/signal events observed),
XENON100 \cite{Aprile:2010um} (0 events), and
COUPP \cite{Behnke:2010xt} (3 events).
Though extending this background level to much larger detectors is no
easy task, we note that the discrimination efficiency tends to increase
with the size of the detector.
The larger detector volume increases the likelihood of seeing neutrons
multiple-scattering within the detector and thus allows for better
rejection of neutrons.  A larger volume also allows
for self-shielding: external radiation is unlikely to penetrate far
into the active volume without interacting; ignoring recoil events in
the outer portion of the detector volume (referred to as a fiducial
volume cut) can significantly reduce the background level with only
a modest reduction in the expected WIMP signal, as WIMPs easily
penetrate the detector volume.

In addition to the WIMP-induced recoil spectrum, we assume a
background spectrum consisting of a component that is flat in energy and a component that decreases exponentially with increasing energy
\begin{equation} \label{eqn:dRBdE}
  \dRBdE = C_{f} + C_{e} e^{-E/E_0} \, ,
\end{equation}
where $C_{f}$ and $C_{e}$ are normalisation constants chosen
such that the expected number of background events from the flat and
exponential components are $\mu_{B,f}$ and $\mu_{B,e}$,
respectively.
The above form was chosen to reflect the two common energy dependences
of typical backgrounds: some are fairly independent of energy (flat
spectrum), while some backgrounds tend to increase rapidly at low
(and decreasing) energy.  The particular form of the background for
any given experiment depends on the detection technique and analysis
cuts; some experiments may expect to have backgrounds more consistent
with only one of the components, while some may have a background
spectrum not adequately approximated by \refeqn{dRBdE} (\eg\ there
may be a Gaussian peak due to a particular radioactive decay).
We take the above form for all the experiments we consider, as a representative example 
of what a background spectrum might be,
not as an indication of what the actual spectrum will be if/when
such experiments are built.
Indeed, while the expected number of background events $\mu_B$ can
often be found for direct detection experiments, the spectra of
these backgrounds can rarely be found in the literature even for
existing experiments.\footnote{
  The lack of background spectra may be the result of large systematic
  uncertainties that significantly affect its determination.
  Nevertheless, if/when a potential WIMP signal is detected in an
  experiment, characterising the background contributions will be of
  utmost importance in order to use the signal to constrain SUSY
  parameter spaces.
  }
We presume the future experiments reach their low background level goal
and take $\mu_{B,f} = \mu_{B,e} = 1.0$ in all three cases, with
$E_0$ = 10~keV.  The total expected number of (signal+background)
events is then
\begin{equation} \label{eqn:mutot}
  \mu_{tot} = \mu_S + \mu_B \, ,
\end{equation}
where $\mu_S$ is the expected WIMP signal of \refeqn{mu} and
$\mu_B \equiv \mu_{B,f} + \mu_{B,e}$ is the total expected background.
The expected (signal+background) energy spectrum is
\begin{equation} \label{eqn:dRtotdE}
  \dRtotdE = \dRSdE + \dRBdE \, ,
\end{equation}
where $\dRSdE$ is the WIMP spectrum of \refeqn{dRdE}.  With this
spectrum, the spectral factor in \refeqn{DDLike} is
\begin{equation} \label{eqn:fE}
  f(E) = \frac{1}{N_f} \dRtotdE \, , \qquad \textrm{with} \quad
  N_f \equiv \int_{\Emin}^{\Emax} dE \, \dRtotdE \, .
\end{equation}

We now describe the CDMS1T, XENON1T, and COUPP1T experiments we
use in our study.  We emphasise that, while their characteristics are
loosely based on their existing present-day versions, we make various
assumptions and approximations that do not necessarily reflect the true
behaviour of the current experiments or their future incarnations.
  Our intent is to define experiments that provide a
variety of targets and realistically account for such issues as finite
energy resolution.

\subsubsection{\label{sec:CDMS1T} CDMS 1-ton}

The Cryogenic Dark Matter Search (CDMS) experiments use germanium
detectors to observe both phonons and ionisation produced in recoils
\cite{Akerib:2005zy,Ahmed:2009zw}.
For our hypothetical future ton-scale CDMS-like experiment (CDMS1T),
we take a 1000~kg-year germanium exposure over energies of 10-100~keV.

The 10~keV threshold is typical of the CDMS low background analyses.
Though CDMS is capable of examining events at energies much lower than
10~keV, background discrimination rapidly becomes worse at lower
energies and the spectrum at these energies is significantly
contaminated with backgrounds \cite{Akerib:2010pv,Ahmed:2010wy}.  The 10~keV
threshold already allows for significant sensitivity to most WIMPs;
inclusion of lower energy events is generally only important when
searching for very light WIMPs (less than $\sim$ 10~GeV) incapable of
producing any events above threshold. Due to the increased
background contamination, addition of the low-energy spectrum to the
analysis would give only a modest or even negligible improvement to
the sensitivity to WIMPs heavier than $\sim$ 20~GeV.
As we do not focus on low mass WIMPs, the low-energy portion of the
spectrum is not important for our analysis and this 10~keV threshold
is sufficient for our purposes.
Due to the form factor and mean inverse velocity factor $\eta$ in
\refeqn{dRdEnr}, the recoil rate at energies of $\sim 100$~keV and
above is far lower than near threshold, so even though CDMS is capable of detecting such recoils, these high energy events
contribute very little to the overall rate (at most 5\%).
An upper limit on energies is imposed rather than including this small
high-energy signal, as backgrounds can easily produce a similar or greater number of events at these energies.

We now define the $\phi$ factor in \refeqn{dRdE} for CDMS1T.
The CDMS detectors allow for a very good energy resolution; we
use \cite{Ahmed:2009rh}
\begin{equation} \label{eqn:CDMSER}
  \sigma^2(\Enr) = (0.293)^2 + (0.056)^2 (\Enr/\mathrm{keV}) \ \mathrm{[keV^2]}
\end{equation}
to describe the normally distributed fluctuations of the observed
energy $E$ about the true recoil energy $\Enr$.
This energy resolution technically applies to low energy electron
recoil events, though we assume here that nuclear recoil events have
a similar resolution.  Though the efficiency $\varepsilon$ of the
various cuts is in principal energy dependent, we assume a constant
30\% efficiency, or\footnote{
  Depending on how the cuts are defined, the efficiency may
  alternatively be defined as a function of the observed energy $E$,
  \ie\ $\varepsilon (E)$, rather than as a function of the true
  nuclear recoil energy $\Enr$.
  }
\begin{equation} \label{eqn:CDMSEff}
  \varepsilon(\Enr) = 0.3 \, ,
\end{equation}
which is comparable to the 25-32\% efficiency over 10-100~keV in the
latest CDMS analysis \cite{Ahmed:2009zw}.  With this energy resolution
and efficiency,
\begin{equation} \label{eqn:CDMSphi}
  \phi(E,\Enr) = \varepsilon(\Enr) \frac{1}{\sqrt{2\pi \sigma^2(\Enr)}}
                 \, e^{-(E-\Enr)^2/2\sigma^2(\Enr)} \, .
\end{equation}

\subsubsection{\label{sec:XENON1T} XENON 1-ton}

The XENON10 \cite{Aprile:2010bt,Angle:2007uj,Angle:2009xb} and
XENON100 \cite{Aprile:2010um} experiments use a dual phase liquid/gas
xenon detector and observe both scintillation and ionisation from
recoil events.
For our hypothetical future ton-scale XENON-like experiment (XENON1T),
we take a 1000~kg-year xenon exposure over energies of 8-75~keV.
The XENON experiments have a poorer energy resolution than CDMS,
but the heavier target ($A \approx 131$ for Xe versus $A \approx 73$
for Ge) gives a somewhat better sensitivity to SI scattering due to
the $A^2$ scaling.  The same issues as with CDMS, regarding the energy range, also apply here.  Lower energy events can be examined, but to
reach much lower energies requires a different type of analysis, leading to greater background contamination \cite{Sorensen:2010idm}.
As with CDMS, ignoring lower energy events only significantly impacts
the XENON sensitivity to light WIMPs and is not important for
our purposes.

As with CDMS, we define a normally distributed fluctuation of $E$ about
$\Enr$, with \cite{Baudis:2008pc}
\begin{equation} \label{eqn:XENONER}
  \sigma(\Enr) = (0.579\,\mathrm{keV}) \sqrt{\Enr/\mathrm{keV}} + 0.021 \Enr \, ,
\end{equation}
and a constant efficiency
\begin{equation} \label{eqn:XENONEff}
  \varepsilon(\Enr) = 0.3 \, ,
\end{equation}
which is the approximate efficiency for XENON100 when including the
fiducial volume cut \cite{Aprile:2010um}.  With this energy resolution
and efficiency, the factor $\phi$ for XENON1T is given by
\refeqn{CDMSphi}.
This energy resolution is a fairly simplified form for the
XENON10 experiment and the resolution at low recoil energies is
actually much more complicated for XENON-like experiments
\cite{Sorensen:2010hq}.  However, the above form still describes an
experiment with poorer energy resolution than CDMS, and should
introduce the same type of relative limitations in constraining parameters as
would be expected from a XENON-like experiment.
We note that there are currently significant issues in determining the
energy calibration of XENON-like experiments (\ie\ estimating the
energy $E$ from a given scintillation signal), which affect
both the threshold and energy resolution determinations; see
Refs.~\cite{Savage:2010tg,Manalaysay:2010mb} and references therein for
further discussion.
We assume that these energy calibration issues will be resolved by the time a
ton-scale xenon experiment is built, and neglect any systematic errors
that would otherwise be introduced in the analysis of our XENON1T
experiment.

\subsubsection{\label{sec:COUPP1T} COUPP 1-ton}

The Chicagoland Observatory for Underground Particle Physics (COUPP)
is a bubble chamber experiment, different from CDMS and XENON in that it does not measure event energies, only the
number of events above a threshold nuclear recoil energy
\cite{Behnke:2008zza,Behnke:2010xt}.  The target material, CF$_3$I,
provides some sensitivity to SI scattering, primarily through the heavy
iodine nuclei, but the high number of nuclei with non-zero spins
(all the fluorine and iodine nuclei) allows for a significant
sensitivity to SD scattering.  The sensitivity to SD scattering is
likely to be much higher than in CDMS or XENON,
allowing COUPP to potentially constrain WIMP parameters in a
manner different than (and complementary to) the other two experiments.

The principle behind the operation of the COUPP bubble chamber begins
with its use of superheated liquid CF$_3$I within the chamber.
When sufficient energy is deposited in a small region, a
bubble forms in this liquid, undergoing runaway growth and causing the
liquid to boil.
However, if insufficient energy is deposited or if the energy is spread
out over a larger region, it is thermodynamically favourable for a
bubble to recollapse rather than expand, and the liquid
remains in its superheated state (such bubbles are too small to be seen).
Electron recoils, induced by $\beta$ or $\gamma$ radiation, deposit
energy over too large a region to induce runaway bubble growth, and
are thus not observed.  Nuclear recoils, on the other hand, deposit
their energy over a small region and, if above a certain minimum
recoil energy, will generate an observable bubble in the detector.
This minimum energy (threshold) is adjustable by changing the
operating temperature and pressure of the liquid.  Essentially
all nuclear recoil events above threshold produce bubbles, which are
indistinguishable and give no indication of their recoil energies (other
than that they exceeded the threshold).  Thus, COUPP provides no direct
measurement of the recoil energy spectrum. The number of observed bubbles is the only observation that can be used to
constrain WIMP properties, leading to the use of \refeqn{DDLikeN}
instead of \refeqn{DDLike} for the likelihood.
Alpha decays unfortunately also lead to bubble formation, and constitute
a source of background events.  These alpha decays produce a
different acoustic signature in the detector than nuclear recoils,
allowing for some background discrimination; see
Ref.~\cite{Behnke:2010xt} for details.

For our hypothetical ton-scale COUPP-like experiment (COUPP1T),
we take a 1000~kg-year CF$_3$I exposure with a threshold of 10~keV.
We take an efficiency of
\begin{equation} \label{eqn:COUPPEff}
  \varepsilon(\Enr) = 0.7 \, ,
\end{equation}
comparable to the efficiency of the most recent COUPP run after
fiducial volume and acoustic signature cuts \cite{Behnke:2010xt}.
As energies are not measured, there is no finite energy resolution
to take into account, and we take $E = \Enr$ by setting
\begin{equation} \label{eqn:COUPPphi}
  \phi(E,\Enr) = \varepsilon(\Enr) \, \delta(E-\Enr) \, .
\end{equation}
Carrying out the integration in \refeqn{dRdE} yields a spectrum
$\dRdE$ proportional to the true nuclear recoil spectrum of
\refeqn{dRdEnr}, though $\dRdE$ will only be used in \refeqn{mu}
and does not represent the ``observed'' spectrum here, as no energy spectrum
is actually observed.

In principle, some sensitivity to the energy spectrum can be gained
by dividing the COUPP exposure into multiple periods with different
thresholds.  The relative rates observed at different thresholds is
dependent upon the recoil spectrum; however, sensitivity to the
spectrum is far weaker via this technique than with
experiments like CDMS and XENON.  In
practice, each period at a different threshold would be treated as a
separate ``experiment'' with a likelihood defined by \refeqn{DDLikeN}.
We do not consider this case here.

\subsection{Benchmarks and generated data} \label{sec:bench}

\begin{figure}
\begin{center}
\subfigure{\includegraphics[scale=0.75, trim = 10 370 50 10, clip=true]{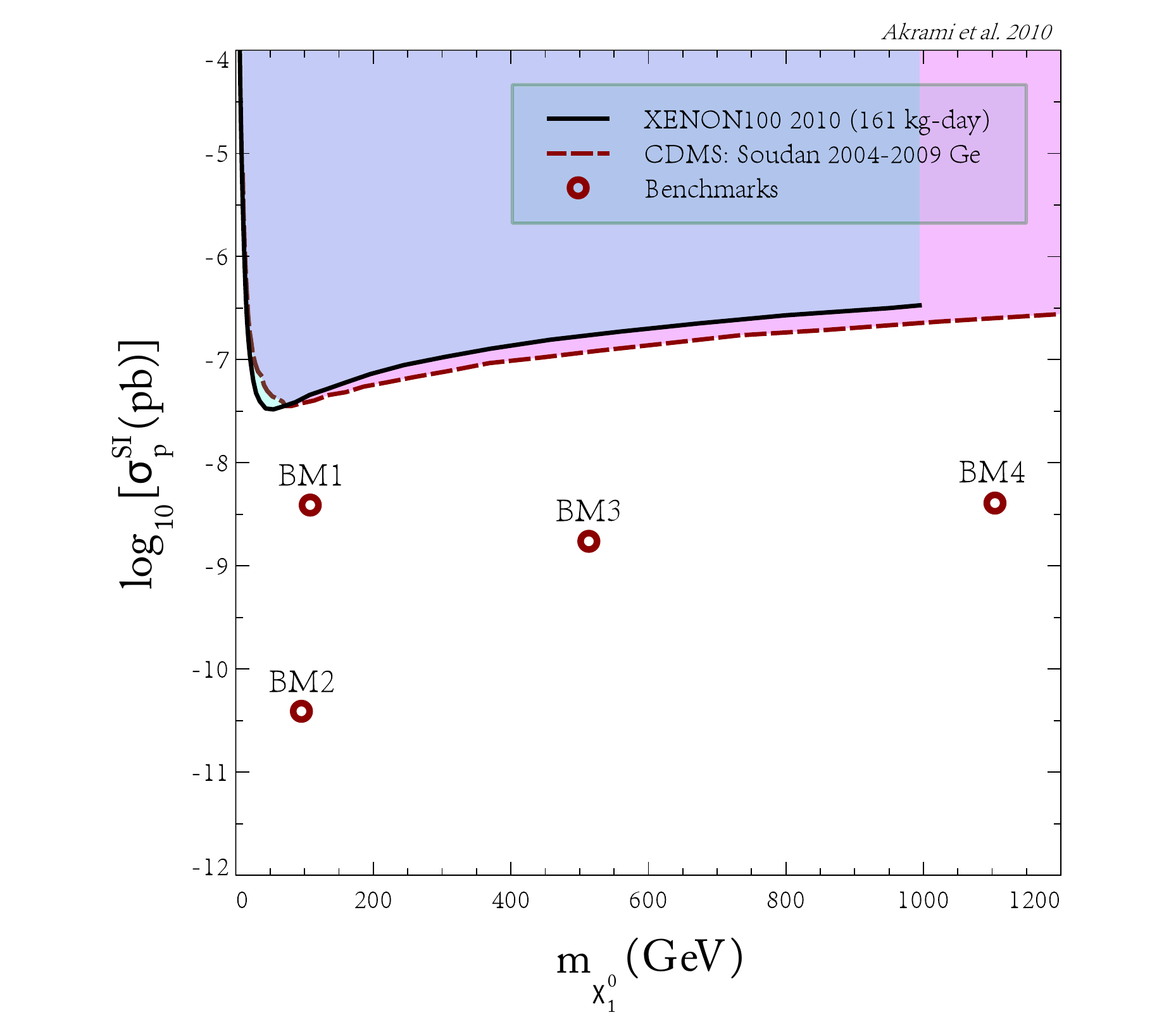}}\\
\caption[aa]{\footnotesize{Strongest experimental limits on the spin-independent scattering cross-section of the neutralino and a proton $\sigma^{SI}_p$, as a function of the neutralino mass, from CDMS and XENON experiments. All WIMPs above the curves are excluded at the $90\%$ confidence level if a standard local halo configuration is assumed. Red circles show the benchmark points chosen in our analysis. For corresponding benchmark values of spin-dependent cross-sections, see~\tab{tab:BMs}.}}\label{fig:currentlimits}
\end{center}
\end{figure}

In order to assess the ability of our hypothetical ton-scale experiments CDMS1T, XENON1T and COUPP1T to constrain the CMSSM, we first look at~\fig{fig:currentlimits}, where the strongest limits on WIMP properties from current direct detection (DD) experiments are given. The black and red curves, shown in the $m_{\tilde\chi^0_1}$-$\sigma^{SI}_p$ plane, are exclusion limits at the $90\%$ confidence level from the CDMS and XENON experiments under the assumption of a standard local halo configuration. We see that almost all WIMPs with $\sigma^{SI}_p > 10^{-7}$ pb are excluded.

So, what will our hypothetical ton-scale DD experiments add to our knowledge of WIMPs in the CMSSM? One way to examine this is to assume that such experiments do not see any signal events (i.e. no WIMPs exist with cross-sections within the reach of these experiments), and then put stronger exclusion limits on WIMP properties. For these types of analyses one can use the likelihoods we constructed for different experiments, and scan over the model parameter space by setting the observed number of signal events to zero. This will exclude regions of parameter space that predict some signal events. What we are interested in however is how well the experiments can zero-in on parts of the parameter space if they see an actual signal. For this purpose, we need to choose some hypothetical `benchmark' models, and test how well we can recover them in our hypothetical experiments.

We perform this investigation by selecting some benchmark points in the $m_{\tilde\chi^0_1}$-$\sigma^{SI}_p$ plane of~\fig{fig:currentlimits} in such a way that they represent WIMPs with interesting properties that are not yet ruled out. For each point we randomly generate some hypothetical DD data based on the likelihoods of~\sec{sec:likes}, and use that data in CMSSM scans to constrain the model parameter space. Our selected benchmark points are shown in~\fig{fig:currentlimits} with red circles. Values of the neutralino mass, SI and SD scattering cross-sections for our chosen points are explicitly listed in~\tab{tab:BMs}.

\begin{table}[t]
\begin{center}
{\small
\begin{tabular}{p{3.5cm} p{2.5cm} p{2.5cm} p{2.3cm} l}
\toprule
\textbf{Quantity}& \textbf{BM1}& \textbf{BM2}& \textbf{BM3}& \multicolumn{1}{c}{\textbf{BM4}} \\ \toprule
$\sigma^{SI}_p$ (pb) & 3.9$\times 10^{-9}$  & 3.9$\times 10^{-11}$  & 1.7$\times 10^{-9}$ & 4.1$\times 10^{-9}$ \\
$\sigma^{SD}_p$ (pb) & 2.8$\times 10^{-5}$  & 1.5$\times 10^{-8}$  & 4.8$\times 10^{-7}$ & 4.5$\times 10^{-7}$ \\
$\sigma^{SD}_n$ (pb) & 2.1$\times 10^{-5}$  & 2.8$\times 10^{-8}$  & 3.9$\times 10^{-7}$ & 3.4$\times 10^{-7}$ \\
$m_{\tilde\chi^0_1}$ (GeV)  &  $108.32$    & $95.51$ & $513.45$    & $1104.05$ \\ \midrule
$\mzero$ (GeV) & 3354.7  & 1773.8  & 2008.8 & 3905.5 \\
$\mhalf$ (GeV) & 271  & 239.7  & 1194.7 & 2490 \\
$\azero$ (GeV) & 3515.3  & 4340.2  & 2840.6 & 3609.5 \\
$\tanb$  &  40.2    & 22.3 & 39.5    & 22 \\ \bottomrule
\end{tabular}
} \caption[aa]{\footnotesize{(Top) Values of spin-independent (SI) and spin-dependent (SD)
cross-sections of the neutralino as well as the neutralino mass at our
benchmark points BM1, BM2, BM3 and BM4. (Bottom) CMSSM parameters for the benchmark points. These parameters were
arbitrarily chosen to produce the desired neutralino mass and SI
cross-section for each benchmark.  We note that there are other choices for
CMSSM parameters, sometimes with very different values, that can also
produce the same neutralino mass and cross-sections in each case.}} \label{tab:BMs}
\end{center}
\end{table}

We have chosen four benchmarks (shown as BM1, BM2, BM3 and BM4 in~\fig{fig:currentlimits}). With benchmarks 1, 3 and 4 we cover the most interesting regions in the $m_{\tilde\chi^0_1}$-$\sigma^{SI}_p$ plane with relatively high cross-sections. The points correspond to WIMPs with low, intermediate and high masses ($\sim 100$ GeV, $\sim 500$ GeV and $\sim 1100$ GeV, respectively). These are WIMPs for which we expect a significant number of signal events in ton-scale experiments. We also select a point at low mass ($\sim 100$ GeV) and relatively small cross-section ($\sim 4\times 10^{-11}$ pb), for which we expect only a couple of events above background. The cross-sections for this benchmark point approximately correspond to the sensitivity limits of our detectors; below this cross-section we do not expect them to be able to distinguish WIMPs from backgrounds. We do not look at higher masses with low cross-sections, because in those cases the expected number of events is so low that the detectors would see only backgrounds. For each of the points we have selected in \reffig{currentlimits},
we have chosen for the corresponding benchmark an arbitrary set of CMSSM
parameters that yields the desired neutralino mass and SI cross-section, as
shown in~\tab{tab:BMs}, in order to generate reasonable SD cross-sections.

The neutralino mass and cross-sections for each of our benchmarks were
determined from a particular set of CMSSM parameters.  However, the same
mass and cross-sections in each case can be approximately reproduced with
many different CMSSM models, in some cases with very different parameter
values, due to various degeneracies between and/or weak dependencies on
some of the CMSSM parameters.  Thus, direct detection experiments will only
be able to directly constrain the neutralino mass and cross-sections and
will be unable to significantly constrain all of the CMSSM parameters
($\mhalf$ is the exception as it is strongly related to the neutralino
mass).  We therefore focus our studies on how direct detection experiments
constrain the neutralino mass and cross-sections rather than on individual
CMSSM parameters, though we do include the constraints for some of the
latter.

Table~\ref{tab:eventsBMs} shows the total and fractional numbers of expected events for the benchmark points, as well as the actual numbers we found in our randomly-generated data, based on the likelihoods. Our synthetic data set also includes randomly-generated recoil energies for the events, according to the expected energy spectrum.

\begin{table}[t]
\begin{center}
{\small
\begin{tabular}{p{3.5cm} p{2.5cm} p{2.5cm} p{2.3cm} r}
\toprule
\textbf{}& \textbf{BM1}& \textbf{BM2}& \textbf{BM3}& \multicolumn{1}{c}{\textbf{BM4}} \\ \toprule
\multicolumn{5}{c}{\textbf{CDMS1T}} \\ \midrule
$\mu_{tot}$    & 135.2     & 3.4                   & 18.7                 & 21 \\
$\mu_S$        & 133.2     & 1.4                   & 16.7                 & 19 \\
$\mu_S^{SI}$   & 131.8       & 1.4                   & 16.7                 & 19 \\
$\mu_S^{SD}$   &  1.39     & 2.2$\times 10^{-3}$   & 6.9$\times 10^{-3}$  & 2.9$\times 10^{-3}$ \\
$\mu_B$        &  2        & 2                     & 2                    & 2 \\
$N_{obs}$      &  139      & 4                     & 18                   & 23 \\ \midrule
\multicolumn{5}{c}{\textbf{XENON1T}} \\ \midrule
$\mu_{tot}$    & 204.1     & 4.1                   & 24.6                 & 27.3 \\
$\mu_S$        & 202.1     & 2.1                   & 22.6                 & 25.3 \\
$\mu_S^{SI}$   & 195.3     & 2.1                   & 22.6                 & 25.3 \\
$\mu_S^{SD}$   &  6.7      & 1.1$\times 10^{-2}$   & 3.2$\times 10^{-2}$  & 1.3$\times 10^{-2}$ \\
$\mu_B$        &  2        & 2                     & 2                    & 2 \\
$N_{obs}$      &  213      & 4                     & 24                   & 28 \\ \midrule
\multicolumn{5}{c}{\textbf{COUPP1T}} \\ \midrule
$\mu_{tot}$    & 389.9     & 5                     & 34.4                 & 38 \\
$\mu_S$        & 387.9     & 3                     & 32.4                 & 36 \\
$\mu_S^{SI}$   & 268.6     & 2.9                   & 31.9                 & 35.7 \\
$\mu_S^{SD}$   &  119.3    & 6.6$\times 10^{-2}$   & 0.56                 & 0.25 \\
$\mu_B$        &  2        & 2                     & 2                    & 2 \\
$N_{obs}$      &  392      & 5                     & 35                   & 38 \\
\bottomrule
\end{tabular}
} \caption[aa]{\footnotesize{Expected and generated numbers of events for our hypothetical ton-scale direct detection experiments CDMS1T, XENON1T and COUPP1T at the four benchmark points BM1, BM2, BM3 and BM4 employed in our scans of the CMSSM parameter space. $\mu_{tot}$, $\mu_S$, $\mu_S^{SI}$, $\mu_S^{SD}$, $\mu_B$ and $N_{obs}$ denote the total expected numbers of (signal+background) events, expected numbers of signal events, expected numbers of spin-independent signal events, expected numbers of spin-dependent signal events, expected numbers of background events and the total generated numbers of events in our particular realisations, respectively.}} \label{tab:eventsBMs}
\end{center}
\end{table}

Given some scattering cross-sections and a neutralino mass, the expected and generated number of events, and their corresponding recoil energies, also depend on the assumed values for the halo parameters $\rhoDM$, $\vrot$, $\vmp$ and $\vesc$. In generating our synthetic data, we set these values to the means of the probability distribution functions given in~\sec{sec:HaloModel}; values are given in~\tab{tab:Halomeans}.

\begin{table}[t]
\begin{center}
{\small
\begin{tabular}{c c c c}
\toprule
\textbf{$\rhoDM$}& \textbf{$\vrot$}& \textbf{$\vmp$}& \multicolumn{1}{c}{\textbf{$\vesc$}} \\
\textbf{(GeV/cm$^3$)} & \textbf{(km/s)} & \textbf{(km/s)}& \multicolumn{1}{c}{\textbf{(km/s)}} \\ \midrule
0.4   & 235  & 235 & 550 \\ \bottomrule
\end{tabular}
} \caption[aa]{\footnotesize{Values of halo parameters used for generating direct detection data at our benchmarks; these are the same as the mean values for distributions given in~\sec{sec:HaloModel}.}} \label{tab:Halomeans}
\end{center}
\end{table}

\subsection{CMSSM scans} \label{sec:scans}
\subsubsection{Model, parameters and ranges} \label{sec:model}

The model of choice in our analysis is the CMSSM~\cite{CMSSM}.\footnote{An alternative analysis using a purely phenomenological setup has been performed in ref.~\cite{BertoneDD}. Results of that analysis are in generally good agreement with those we present in this paper.} Here one unifies all scalar, gaugino and trilinear coupling parameters of the MSSM to $\mzero$, $\mhalf$ and $\azero$, respectively. This unification happens at the gauge coupling unification (or GUT) scale ($\sim$10$^{16}$\,GeV). Assuming such boundary conditions, one obtains low-energy quantities such as the masses of the SM particles and their superpartners by solving the RGEs from the GUT scale down to scales where these quantities are defined. One key assumption in the CMSSM is that electroweak symmetry is broken at TeV scales; this requirement fixes the magnitude of the Higgs/higgsino mass parameter $\mu$ in the MSSM superpotential. The sign of $\mu$ remains as a free parameter to be determined by experiment. In addition to $\mzero$, $\mhalf$, $\azero$ and $\sgn\,\mu$, the CMSSM possesses one more free parameter: $\tanb$, the ratio of the up-type and down-type Higgs vacuum expectation values. Compared to the SM, the CMSSM has therefore five new free parameters in total, four continuous and one discrete. Throughout this paper we only work with positive $\mu$. We use \textsf{SOFTSUSY}~\cite{softsusy} (available from ref.~\cite{softsusyweb}) to solve the RGEs for every set of CMSSM parameters.

There are also some SM quantities whose values have not been fully determined by experiment. The ones that impact the CMSSM predictions most are $\mtpole$, the top quark mass, $m_b$, the bottom quark mass, $\alpha_{\mbox{em}}$, the electromagnetic coupling constant and $\alpha_s$, the strong coupling constant. We therefore add them to the set of free parameters in our scans. Here we evaluate the bottom quark mass $m_b$ at $m_b$ and denote it as $\mbmbmsbar$ while for the top quark mass we use its pole mass and show it simply as $\mtpole$. The electromagnetic and strong coupling constants are evaluated at the $Z$-boson pole mass $m_Z$ and are denoted as $\alphaemmz$ and $\alphas$, respectively. $\msbar$ means that these quantities are computed in the modified minimal subtraction renormalisation scheme. Values of the four SM quantities $\mtpole$, $\mbmbmsbar$, $\alphaemmz$ and $\alphas$ are strongly constrained by electroweak precision measurements and we therefore include them in our scans as nuisance parameters. We model the likelihoods for these parameters by normal distributions with means and standard deviations given in~\tab{tab:Nuisances}.

\begin{table}[t]
\begin{center}
{\small
\begin{tabular}{p{3cm} p{3cm} p{4cm} r} \toprule
\textbf{Observable} &   \textbf{Mean value} & \textbf{Standard deviation} & \textbf{Ref.} \\
\toprule
$\mtpole$ (GeV)           &  $172.6$    & $1.4$& \cite{topmass:mar08} \\
$m_b (m_b)^{\overline{MS}}$ (GeV) & $4.20$  & $0.07$ & \cite{Yao:2006px} \\
$\alphas$       &   $0.1176$   & $0.002$ & \cite{Yao:2006px}\\
$1/\alphaemmz$  & $127.955$ & $0.03$ & \cite{Hagiwara:2006jt}
\\ \bottomrule
\end{tabular}
} \caption[aa]{\footnotesize{Experimental constraints on SM nuisance parameters used in our analysis.}} \label{tab:Nuisances}
\end{center}
\end{table}

Given a point in the eight-dimensional parameter space of the CMSSM+SM nuisance parameters, one can theoretically predict properties of the neutralino, such as its mass and scattering cross-sections. However, as pointed out in~\sec{sec:CrossSection}, the cross-sections depend also on the values we assume for the neutralino-quark couplings through the hadronic matrix elements. As we discussed before, the uncertainties on three of these quantities, $\Deltaps$, $\sigma_0$ and $\SigmapiN$, can significantly impact predictions for the cross-sections. We therefore vary them in our scans as cross-section nuisance parameters with likelihoods given in~\sec{sec:CrossSection}.

Even though the neutralino mass and cross-sections can be calculated just from CMSSM parameters, SM nuisances and cross-section nuisances, the DD data observed by a detector depend also on the halo model and parameters (\eqs{eqn:dRdEnr}{eqn:eta}). As discussed in~\sec{sec:HaloModel}, we do not fix the values of our four halo parameters $\rhoDM$, $\vrot$, $\vmp$, and $\vesc$ in our scans; they form yet another set of nuisance parameters with likelihoods given in~\sec{sec:HaloModel}.

With four CMSSM parameters, four SM nuisances, three cross-section nuisances and four halo nuisances, the parameter space that we scan over becomes fifteen-dimensional. Ranges for all parameters used in our scans are listed in~\tab{tab:paramranges}.

\begin{table}[t]
\begin{center}
{\small
\begin{tabular}{p{3.5cm} p{3cm} r}
\toprule
\textbf{Parameter}& \textbf{Lower limit}& \multicolumn{1}{c}{\textbf{Upper limit}} \\ \toprule
$\mzero$ (GeV)           &  $60$    & $4000$ \\
$\mhalf$ (GeV)       &  $60$    & $4000$ \\
$\azero$ (GeV)          &  $-7000$    & $7000$ \\
$\tanb$     &  $2$    & $65$ \\ \midrule
$\mtpole$ (GeV)      &  $163.7$    & $178.1$ \\
$m_b (m_b)^{\overline{MS}}$ (GeV)& $3.92$  & $4.48$ \\
$\alphas$       &  $0.1096$ & $0.1256$ \\
$1/\alphaemmz$  & $127.846$ & $127.990$ \\ \midrule
\Deltaps     &  $-0.18$    & $0$ \\
$\sigma_0$ (MeV)& $20$  & $65$ \\
\SigmapiN (MeV) &  $40$ & $100$ \\ \midrule
\rhoDM (GeV/cm$^3$)      &  $0.1$    & $1.2$ \\
\vrot (km/s)& $175$  & $295$ \\
\vmp (km/s)     &  $175$ & $295$ \\
\vesc (km/s) & $445$ & $655$ \\ \bottomrule
\end{tabular}
} \caption[aa]{\footnotesize{Ranges of model and nuisance parameters employed in our scans.}} \label{tab:paramranges}
\end{center}
\end{table}

The full likelihood for our model is a composite likelihood comprised of the likelihoods for our ton-scale DD experiments and the likelihoods for the SM, cross-section and halo nuisance parameters. In addition, we require some physicality conditions at each sample point in the parameter space: (i) self-consistent solutions to the RGEs must exist, (ii) electroweak symmetry breaking (EWSB) conditions must be fulfilled, (iii) all particle masses must be non-tachyonic, and (iv) the neutralino must be the lightest supersymmetric particle (LSP). To ensure that all these conditions are satisfied, we disregard any points that do not fulfil these criteria simply by assigning them extremely small likelihoods.

\subsubsection{Statistical frameworks and scanning technique} \label{sec:stat}

In order to make any inference about the parameters of a theoretical model by comparison with experimental data, one has to first decide which statistical framework to use, Bayesian or frequentist statistics. SUSY parameter estimation is no exception; for a general introduction, see e.g. ref.~\cite{Cowan:1998}, and for some applications in SUSY phenomenology, see refs.~\cite{Trotta:08093792,Akrami:2009hp,Akrami:2010} and references therein.

In Bayesian statistics (for its applications in physics, see e.g. ref.~\cite{DAgostini:1995}, and for reviews of its applications in cosmology, see e.g. refs.~\cite{Trotta:2005,Trotta:2008,Liddle:2009,Hobson:2010}), probabilities are defined in a `subjective' way and consequently, one can assign probabilities to model parameters. The cornerstone of Bayesian inference is Bayes' Theorem through which our `prior' degree of belief about different values of model parameters is updated when new experimental data are taken into account. Our updated knowledge, given in terms of a multi-dimensional probability density function (PDF) and called the `posterior' PDF, is then used to make statistical statements about the model and its parameters, including the construction of intervals/regions in the parameter space with certain levels of confidence (`credible' intervals/regions). This construction can be easily performed by integrating (or `marginalising') the full posterior PDF over all unwanted parameters resulting in joint `marginal' PDFs for the parameters of interest.

In frequentist statistics, on the other hand, probability has an `objective' definition, so probabilities cannot be assigned to model parameters. Instead of posterior PDFs, frequentists work only with the likelihood and the idea of repeated experiments. One widely-used method for constructing `confidence' intervals/regions in the parameter space in the frequentist framework is the `profile likelihood' construction~\cite[and references therein]{profilelike}. In this method, in order to make inference about some parameters, we maximise (or `profile') the full multi-dimensional likelihood function along the unwanted parameters. For complex and poorly-understood models like the CMSSM, with non-Gaussian likelihoods, the profile likelihood method is only an approximation to the exact frequentist construction of confidence regions proposed by Neyman~\cite{Neyman}. One method that provides exact confidence regions is the so-called `confidence belt' construction~\cite{Feldman:1997qc}; this method is however harder to implement numerically.

Results of Bayesian and frequentist statistics, in particular the credible or confidence intervals/regions obtained, do not agree in general. One central ingredient of Bayesian inference is the set of prior assumptions one employs in the inference. This can significantly affect the final results, in particular when data are not constraining enough and/or the model at hand is too complex (for a detailed discussion of the impacts of priors for SUSY parameter estimation, see ref.~\cite{Trotta:08093792}). Bayesian statistics therefore provides a good measure of the robustness of a fit; if the fit depends strongly on priors, it is not to be considered definitive. On the other hand, results of frequentist inference including the profile likelihood framework are in principle independent of priors. This is however the case only if we have an accurate construction of the likelihood function. How true this is depends on how powerful the employed sampling algorithm is. Most existing scanning algorithms are optimised for Bayesian inference (for an example of frequentist scanning techniques, see e.g. ref.~\cite{Akrami:2009hp}), so any mapping of the likelihood function based on these scans can be strongly impacted by prior effects (for detailed discussions, see refs.~\cite{Akrami:2009hp,Akrami:2010}).

The other interesting distinction between marginal posteriors and profile likelihoods is that the marginalisation procedure makes the posterior sensitive only to the total posterior probability mass in the subspace that has been marginalised over.  Consequently, fine-tuned regions with spike-like likelihood surfaces cannot be probed. Profile likelihoods, however, are strongly sensitive to fine-tuned regions by construction, as confidence regions are constructed as iso-likelihood contours around the best-fit point. Both marginal posteriors and profile likelihoods should be considered in any scan, so as to acquire a complete picture of the favoured parameter regions; this is our strategy in this paper.

In order to map the posterior PDFs and profile likelihoods for our SUSY model, we employ nested sampling~\cite{SkillingNS1,SkillingNS2} to explore the parameter space. We use the \MN\ algorithm~\cite{MultiNest1,MultiNest2}, which is widely used in SUSY parameter estimation.  \MN\ is one scanning option in \textsf{SuperBayeS}~\cite{Trotta:08093792,deAustri:2006pe,Roszkowski:2007fd} (available from ref.~\cite{superbayes}), a package for constraining CMSSM parameters using different types of experimental data. Nested sampling is designed and optimised for computing the Bayesian evidence (and hence, posterior PDFs) in a relatively fast and efficient way. One other popular class of scanning algorithms are Markov-Chain Monte Carlos (MCMCs), which are also optimised for Bayesian inference. Results of nested sampling and MCMC scans are very similar in the CMSSM~\cite{Trotta:08093792}, although the former requires remarkably less computational power.

\section{Results and discussion} \label{sec:results}
\subsection{Bayesian credible and frequentist confidence regions} \label{sec:confreg}

\begin{figure}[t]
\subfigure{\includegraphics[scale=0.24, trim = 40 250 130 153, clip=true]{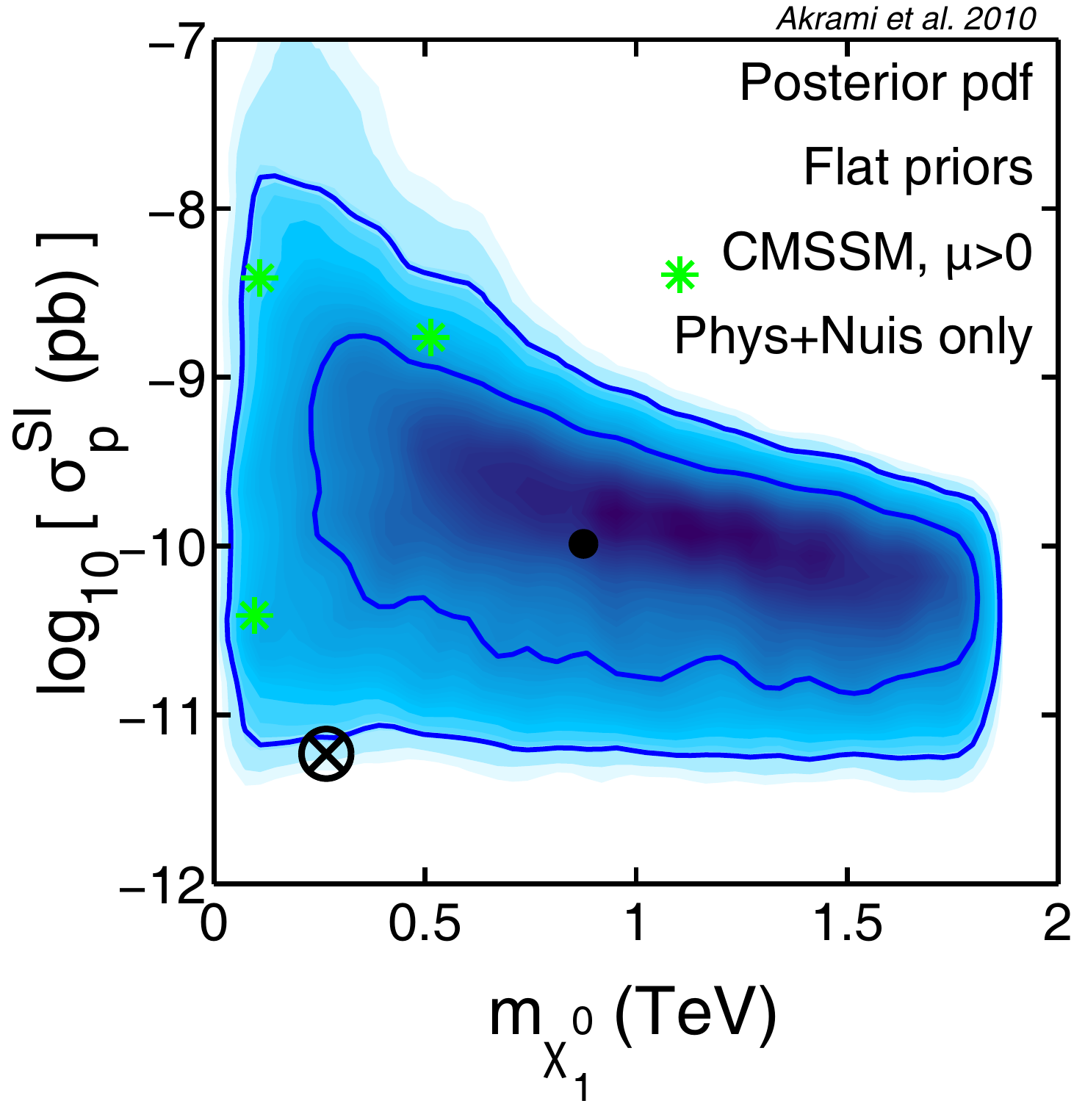}}
\subfigure{\includegraphics[scale=0.24, trim = 40 250 130 153, clip=true]{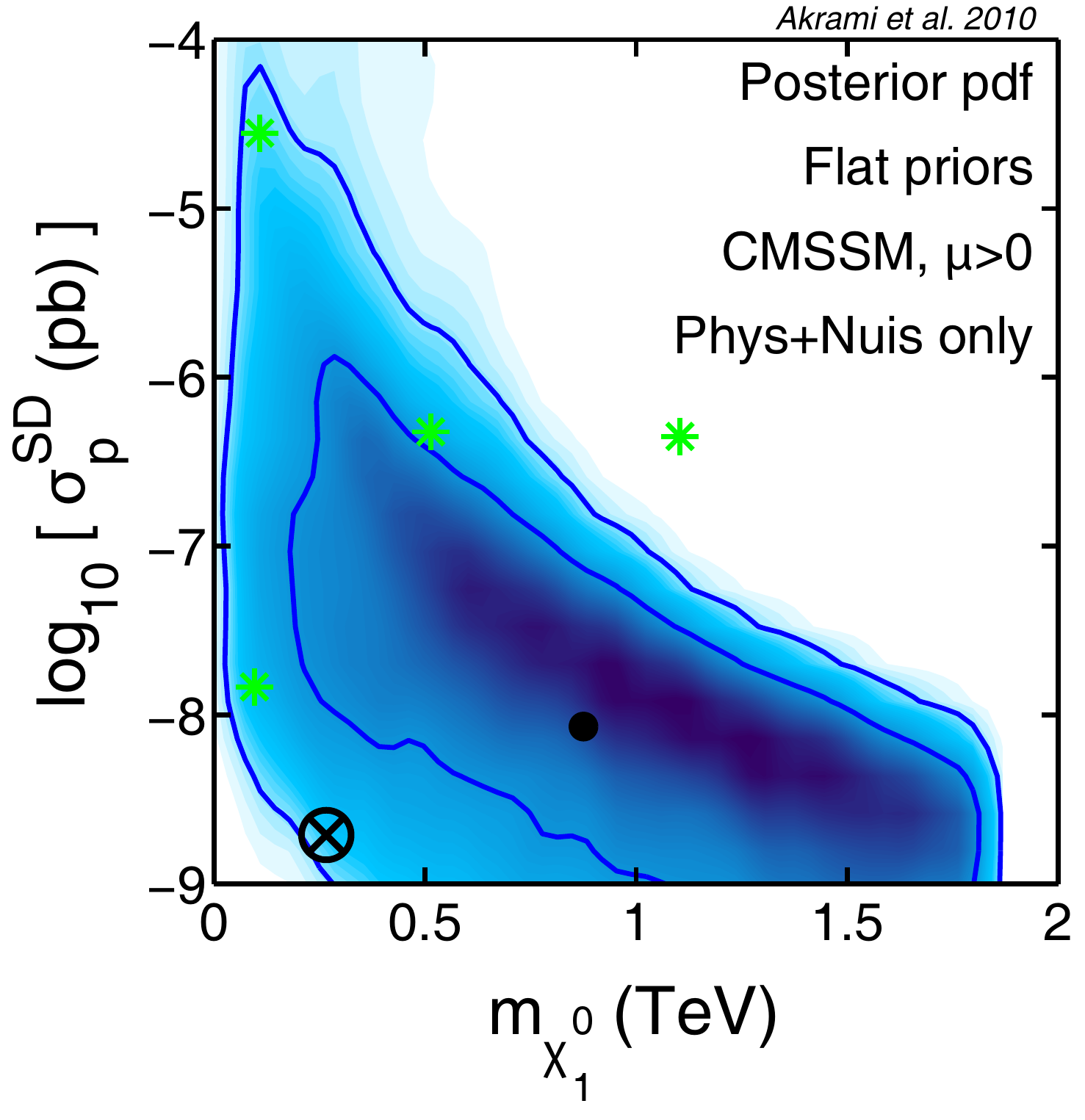}}
\subfigure{\includegraphics[scale=0.24, trim = 40 250 130 153, clip=true]{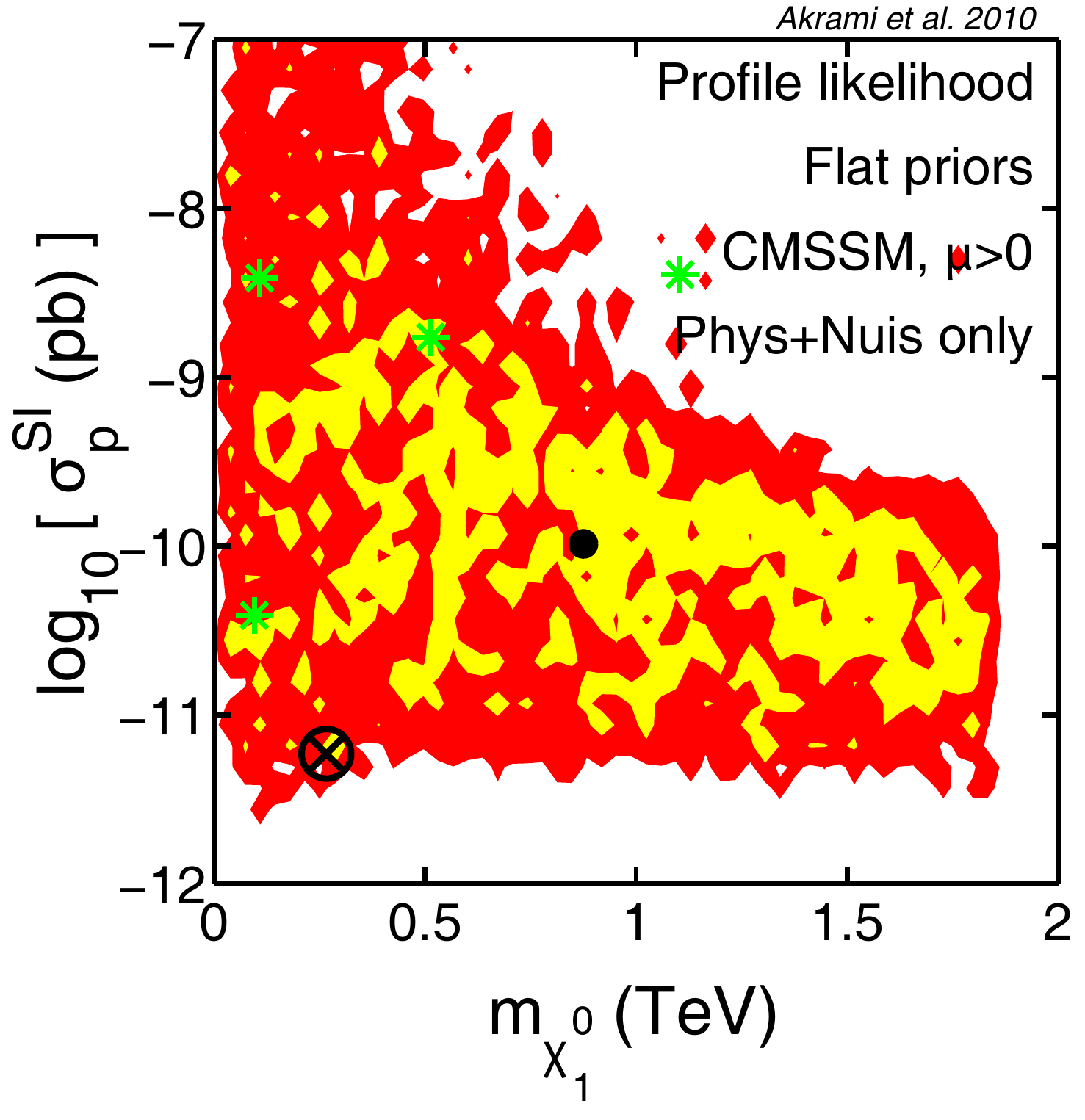}}
\subfigure{\includegraphics[scale=0.24, trim = 40 250 60 153, clip=true]{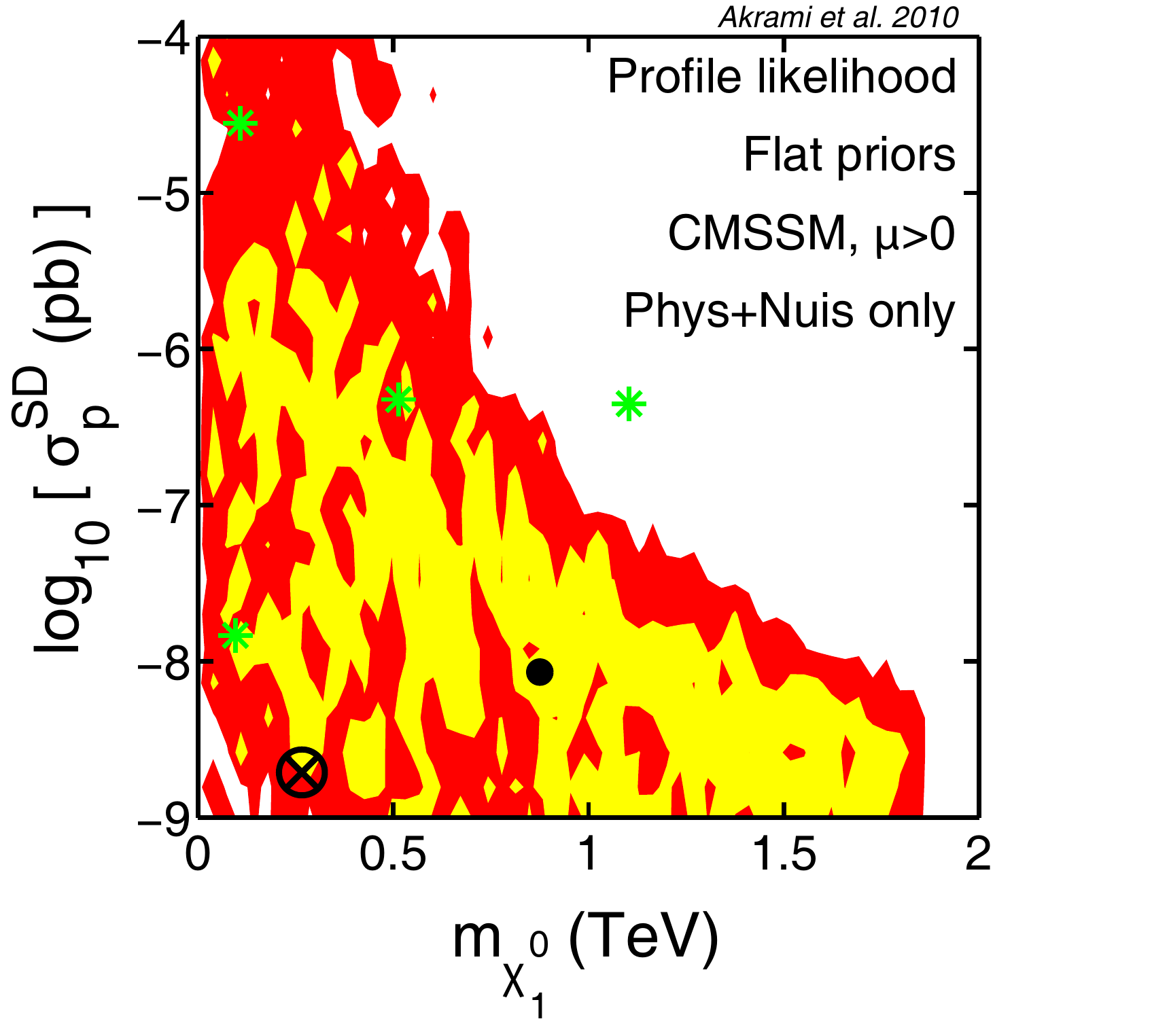}}\\
\subfigure{\includegraphics[scale=0.24, trim = 40 250 130 153, clip=true]{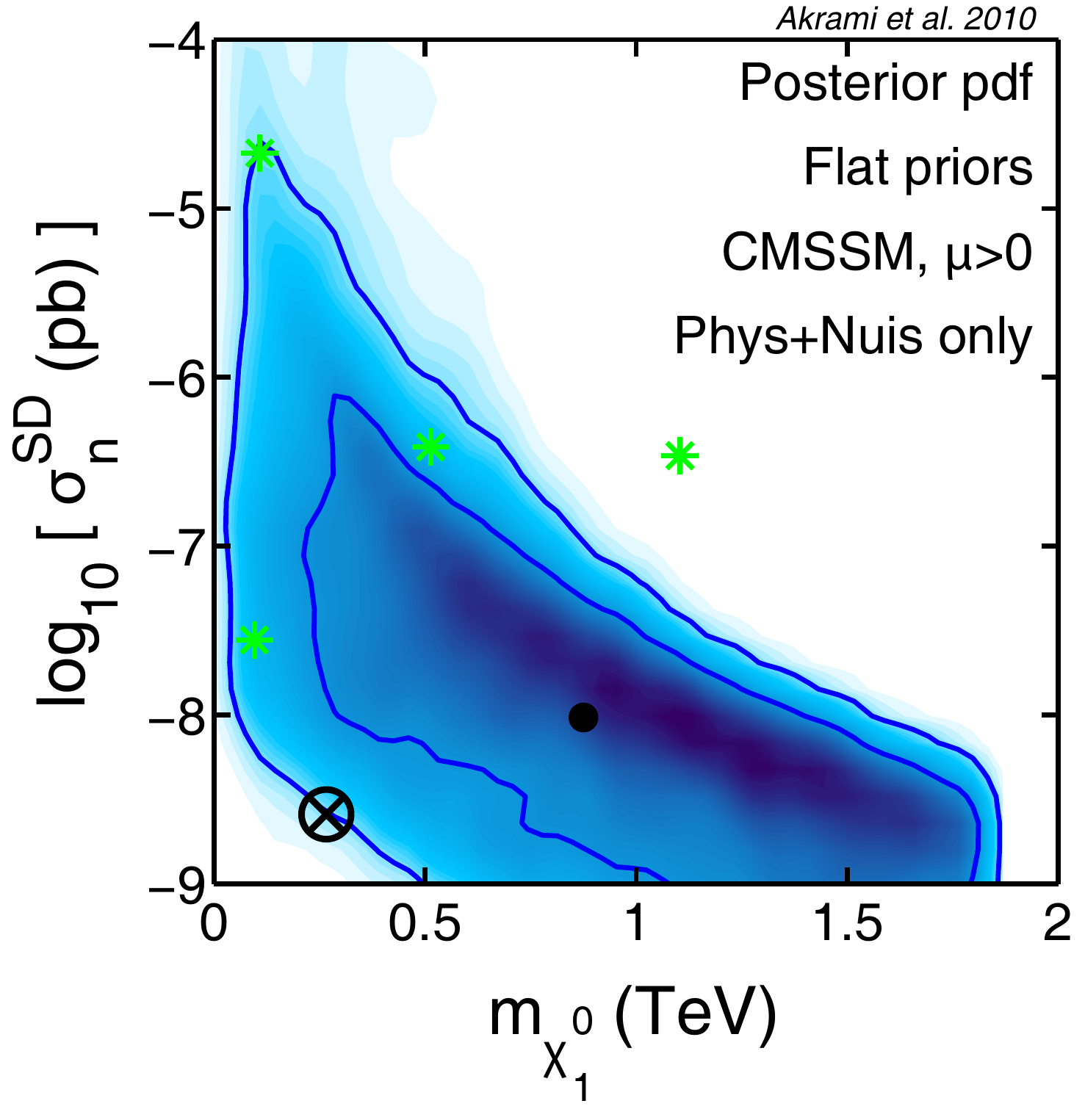}}
\subfigure{\includegraphics[scale=0.24, trim = 40 250 130 153, clip=true]{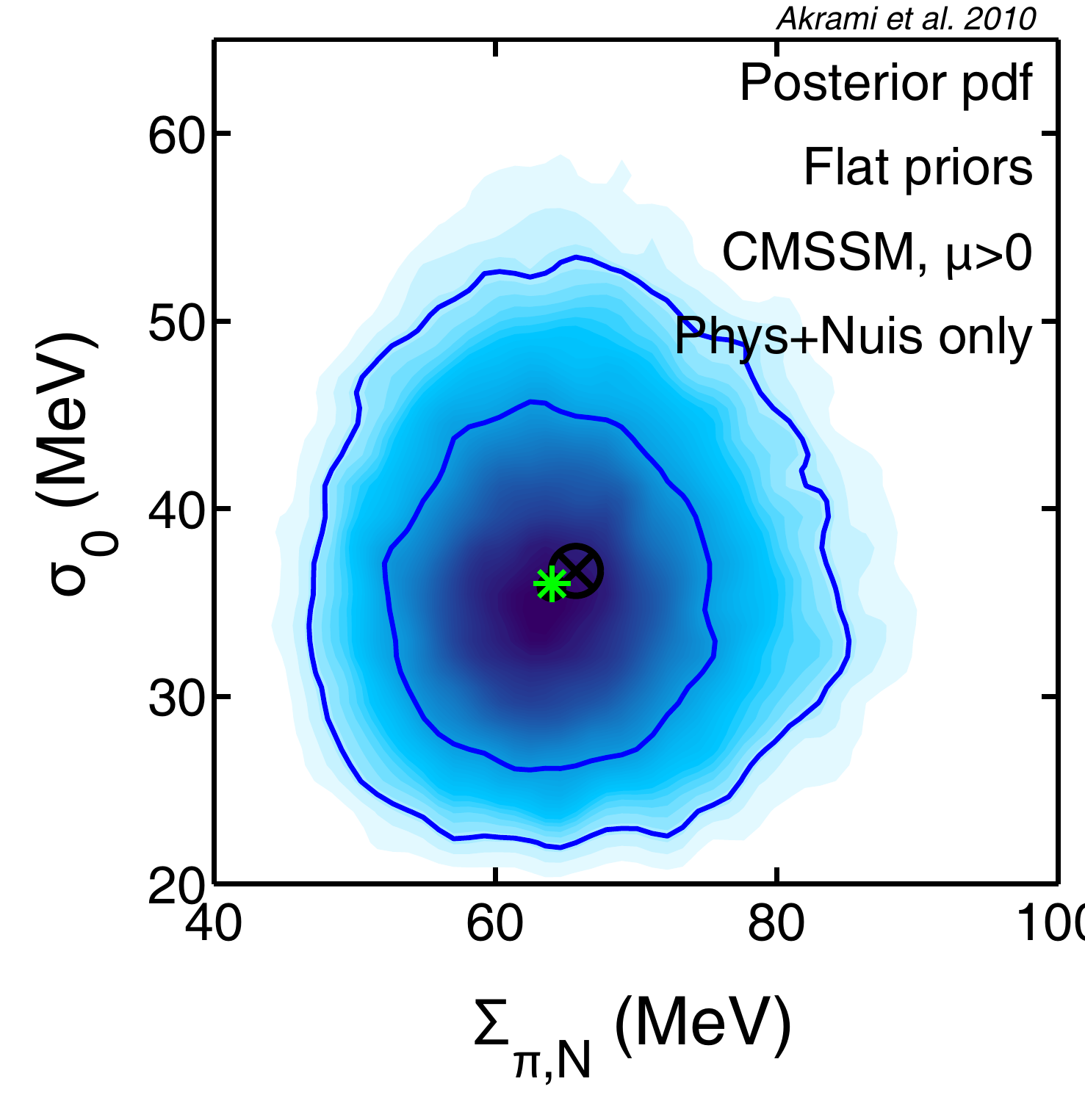}}
\subfigure{\includegraphics[scale=0.24, trim = 40 250 130 153, clip=true]{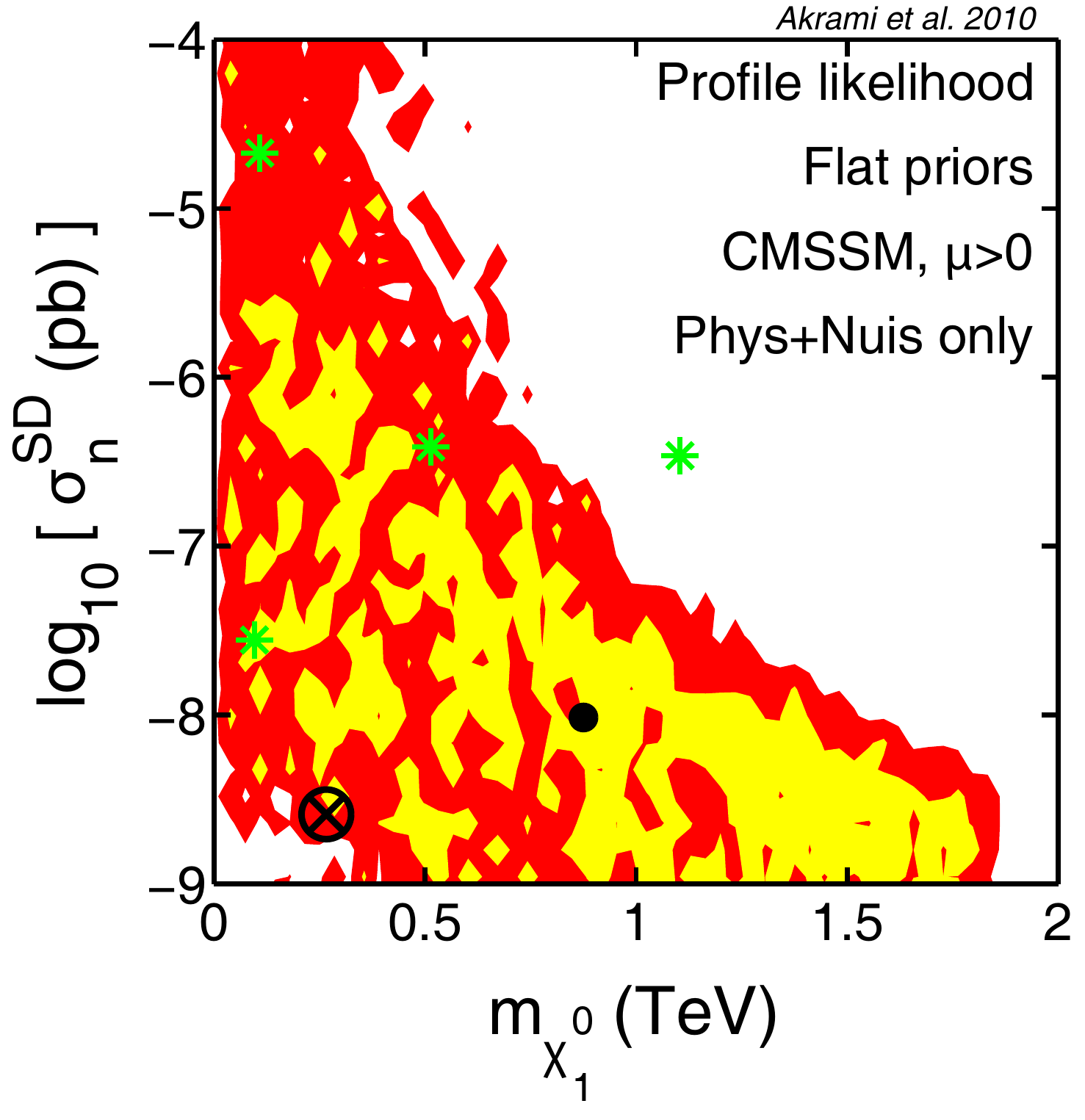}}
\subfigure{\includegraphics[scale=0.24, trim = 40 250 60 153, clip=true]{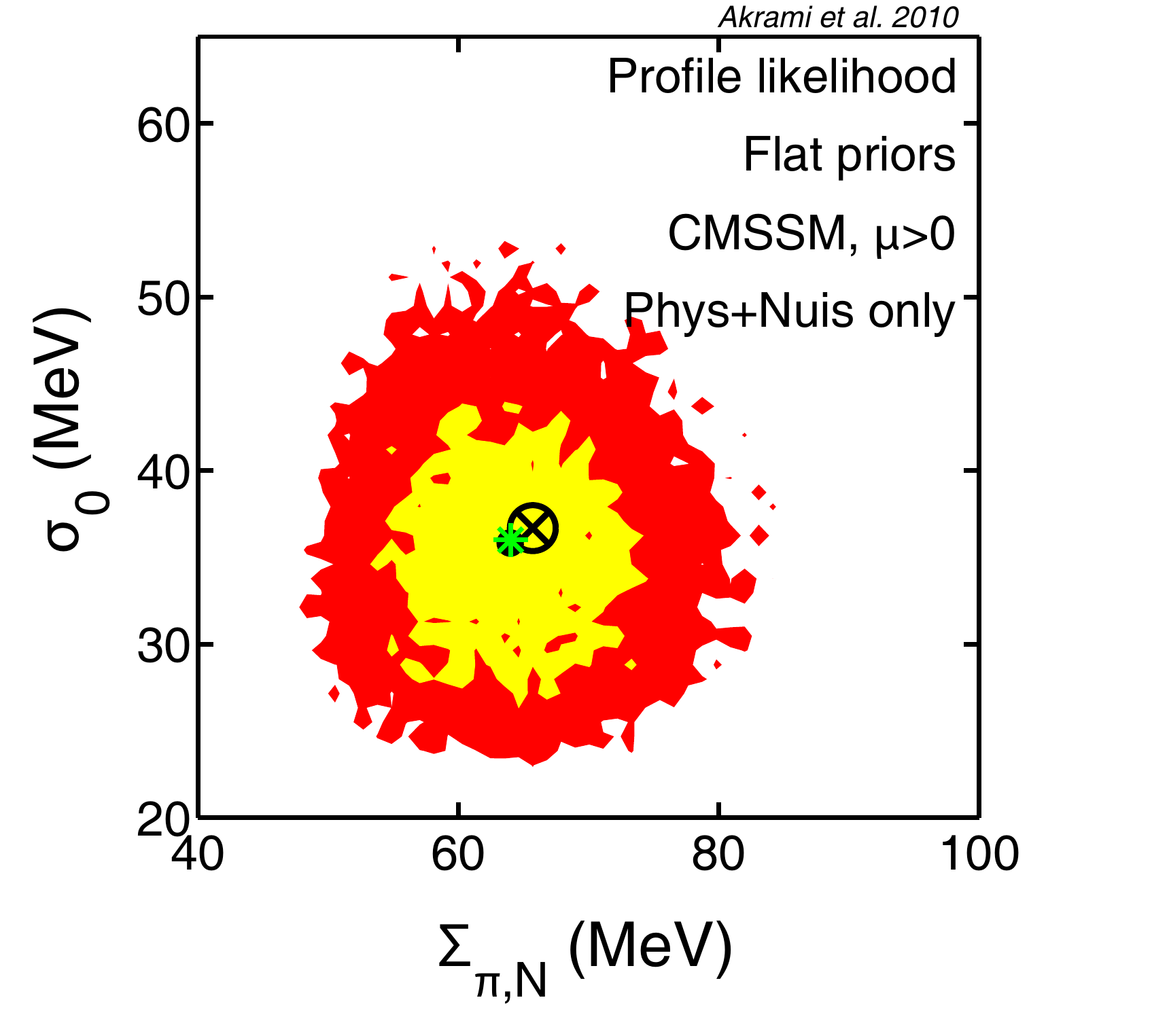}}\\
\subfigure{\includegraphics[scale=0.24, trim = 40 250 130 153, clip=true]{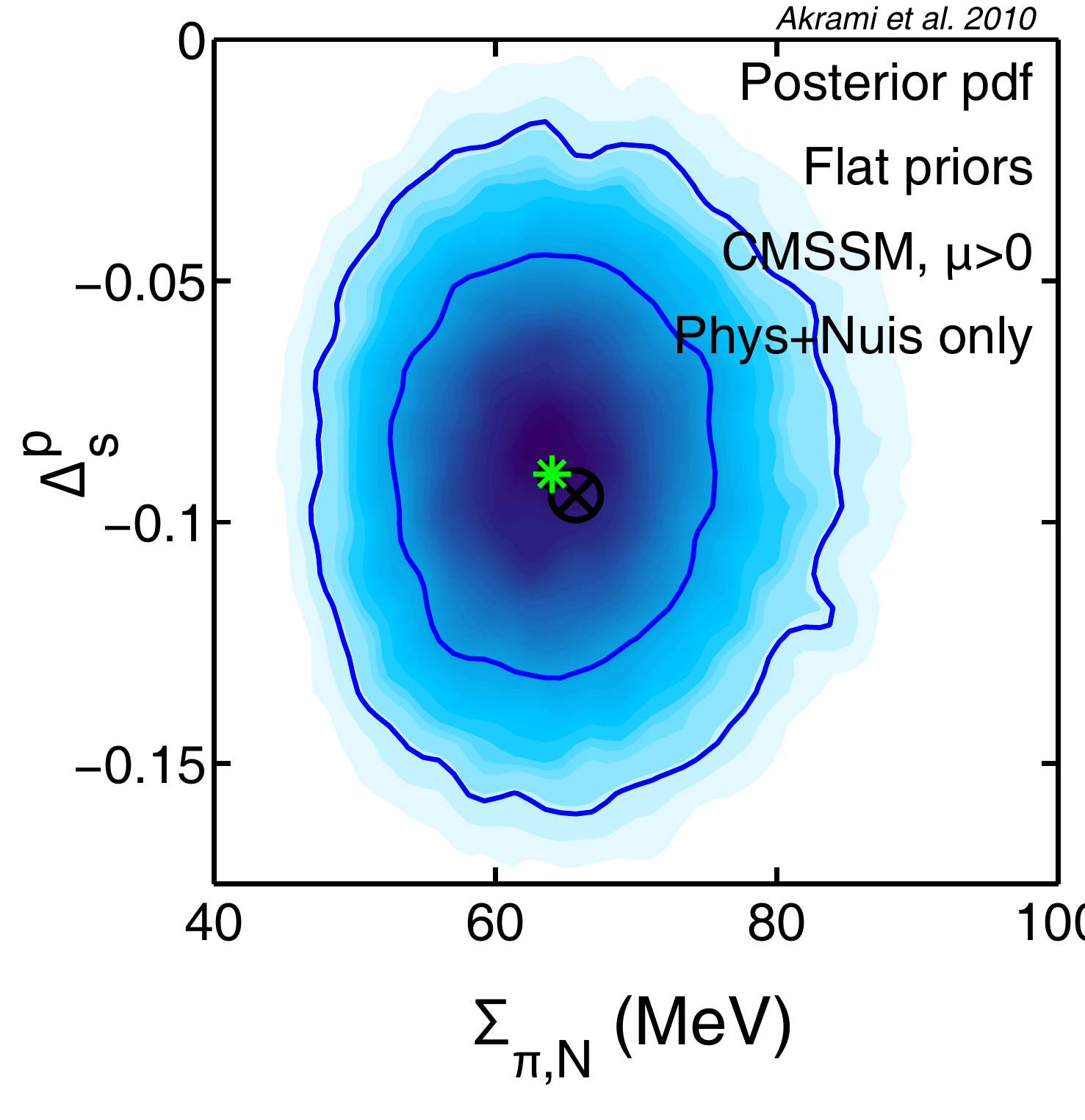}}
\subfigure{\includegraphics[scale=0.24, trim = 40 250 130 153, clip=true]{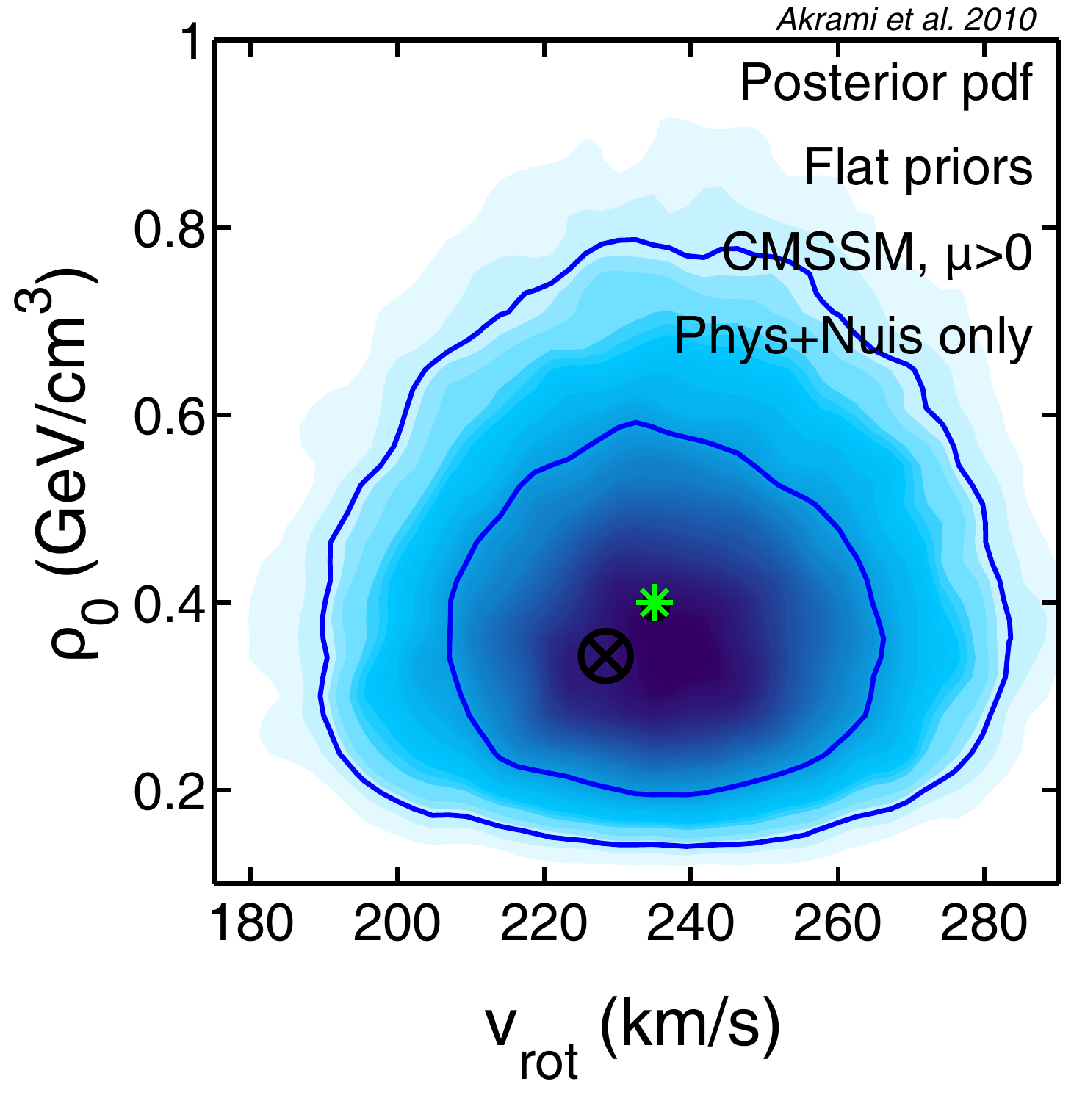}}
\subfigure{\includegraphics[scale=0.24, trim = 40 250 130 153, clip=true]{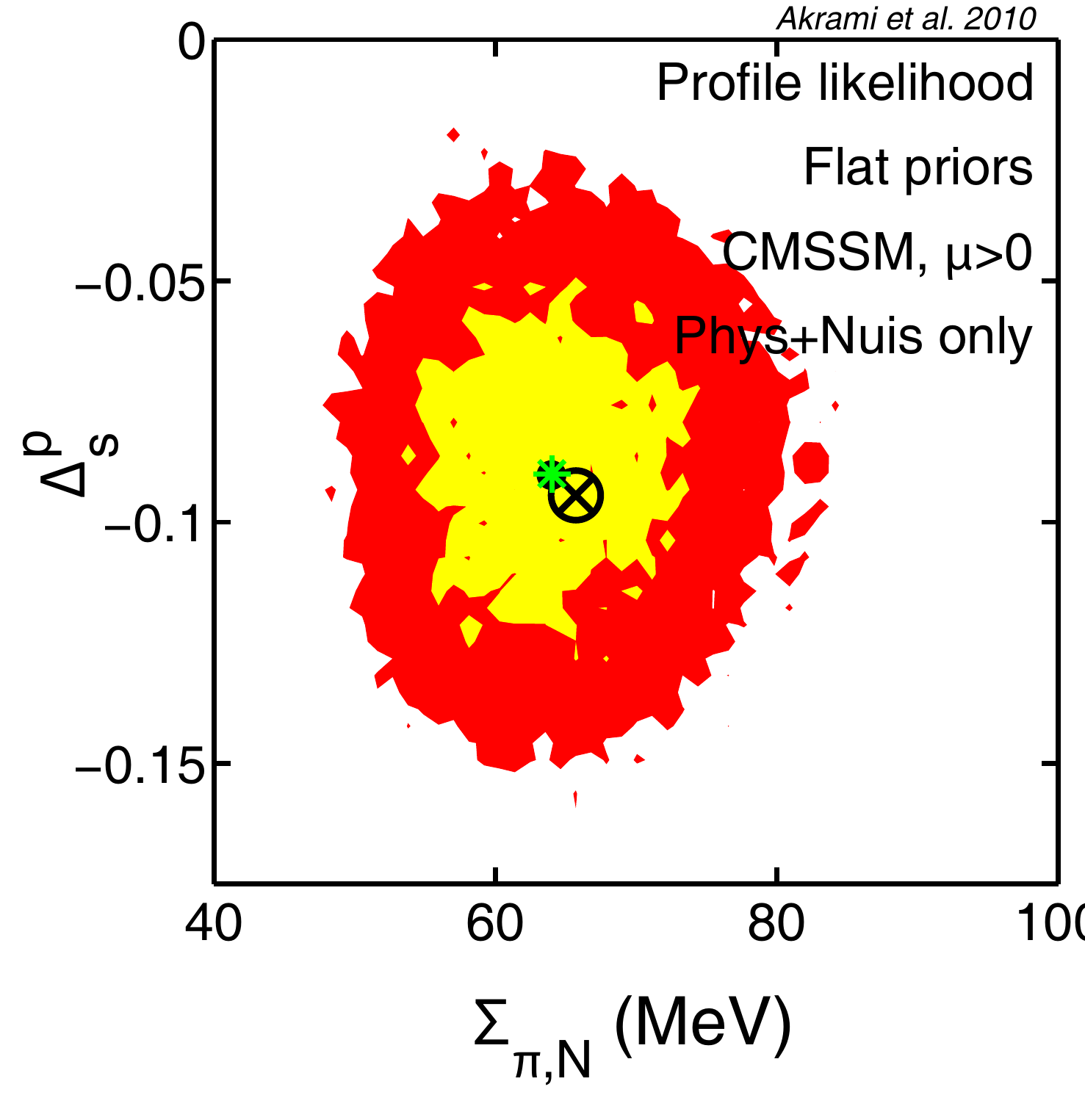}}
\subfigure{\includegraphics[scale=0.24, trim = 40 250 60 153, clip=true]{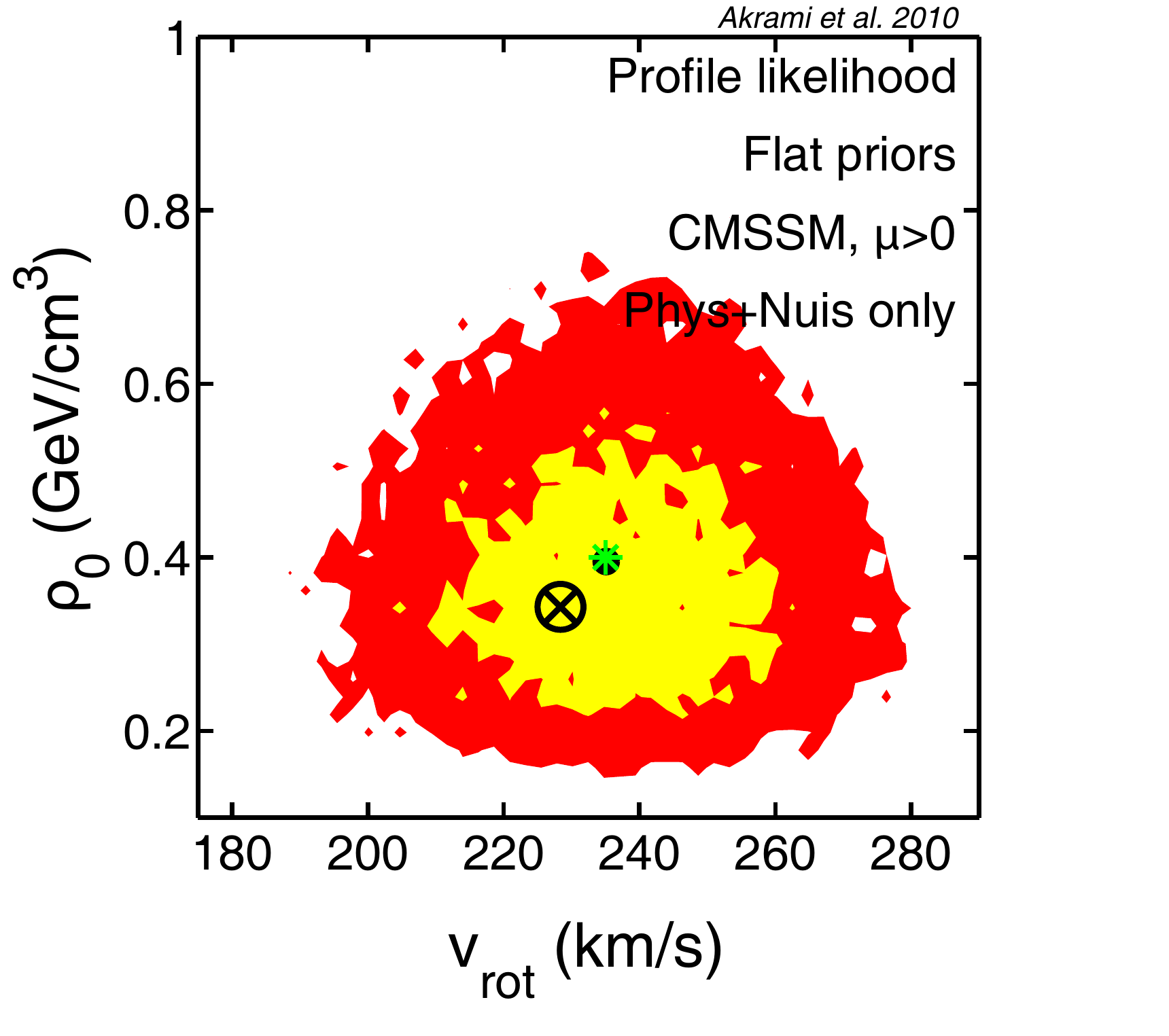}}\\
\subfigure{\includegraphics[scale=0.24, trim = 40 250 130 153, clip=true]{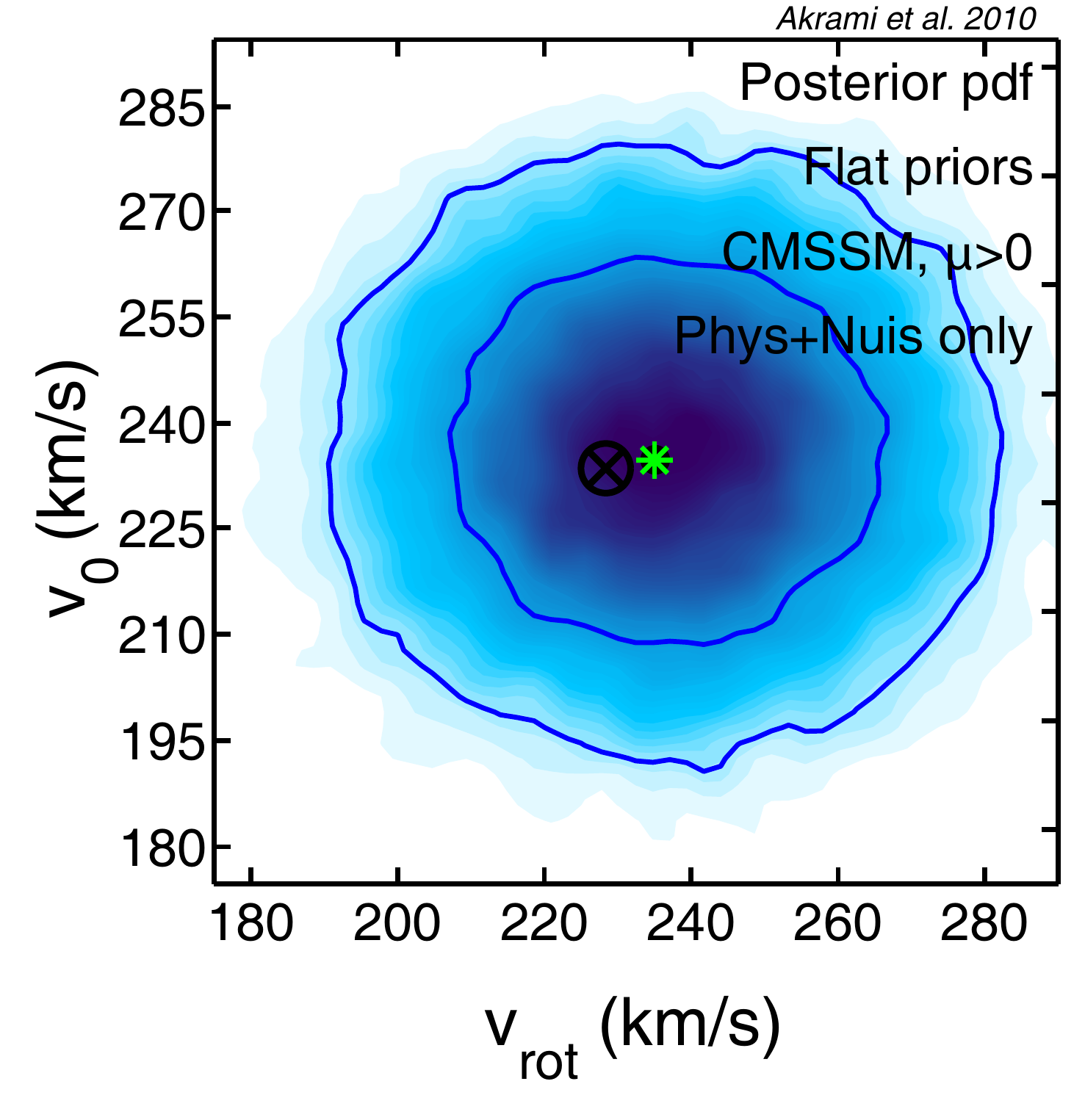}}
\subfigure{\includegraphics[scale=0.24, trim = 40 250 130 153, clip=true]{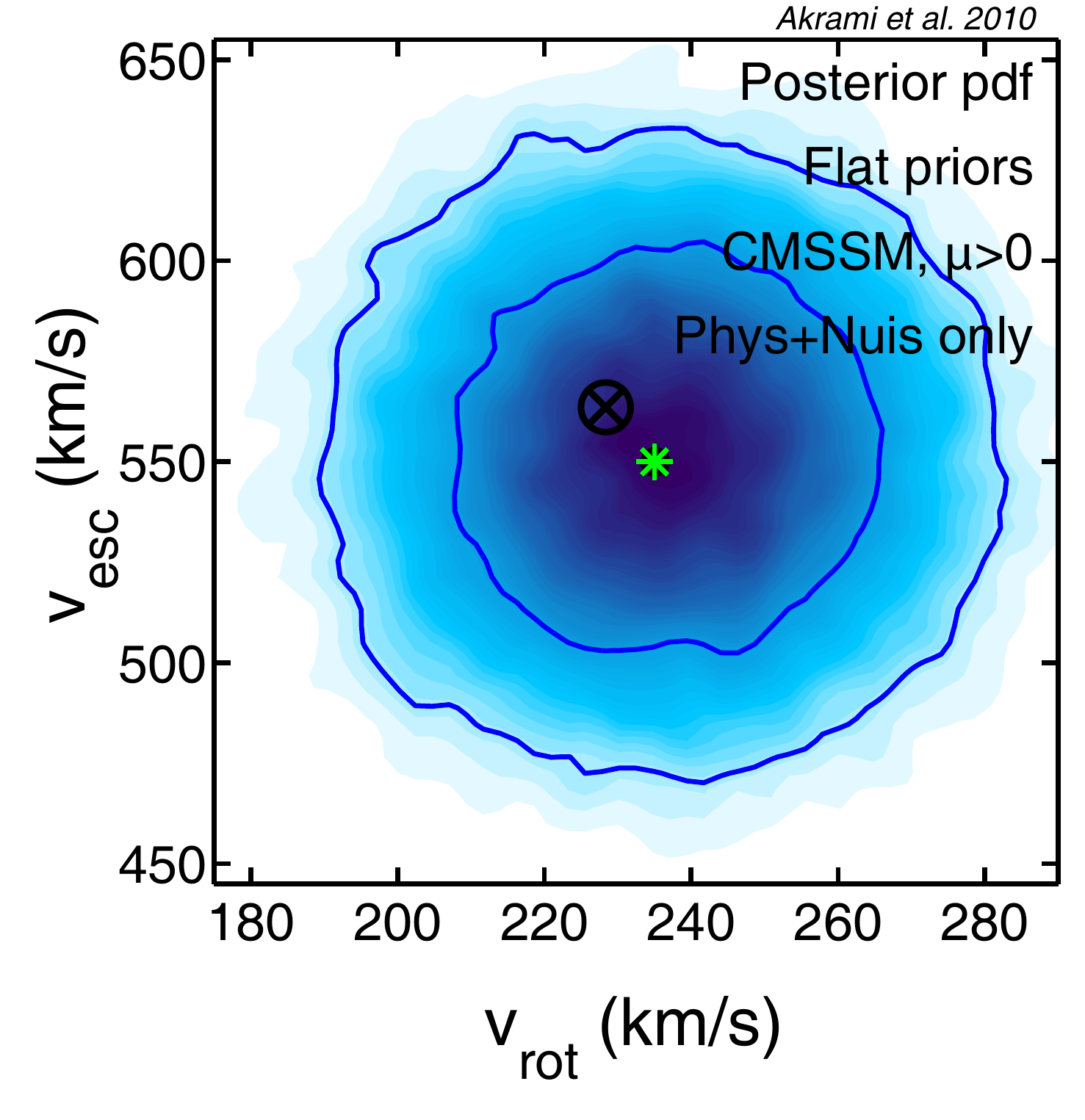}}
\subfigure{\includegraphics[scale=0.24, trim = 40 250 130 153, clip=true]{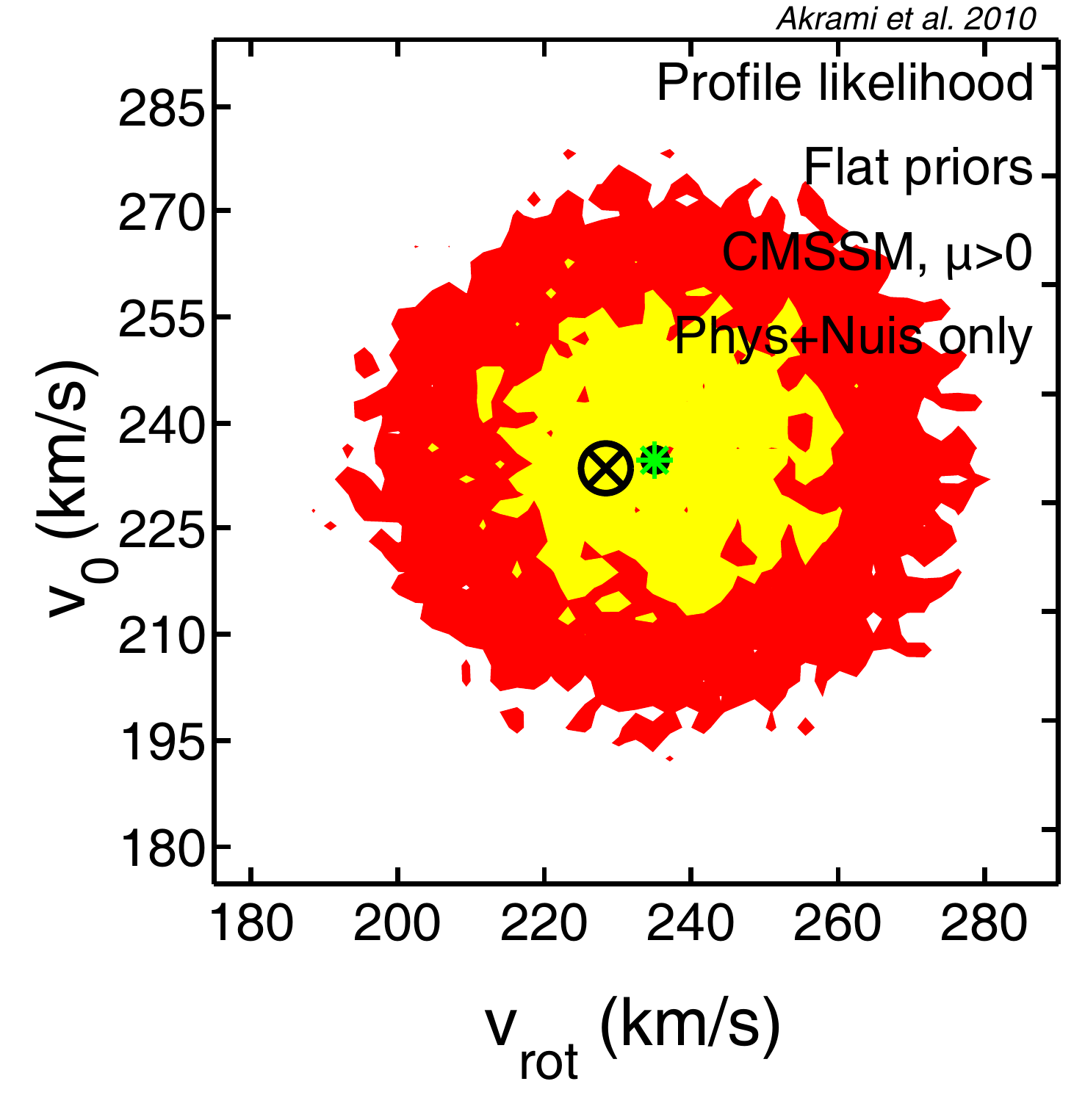}}
\subfigure{\includegraphics[scale=0.24, trim = 40 250 60 153, clip=true]{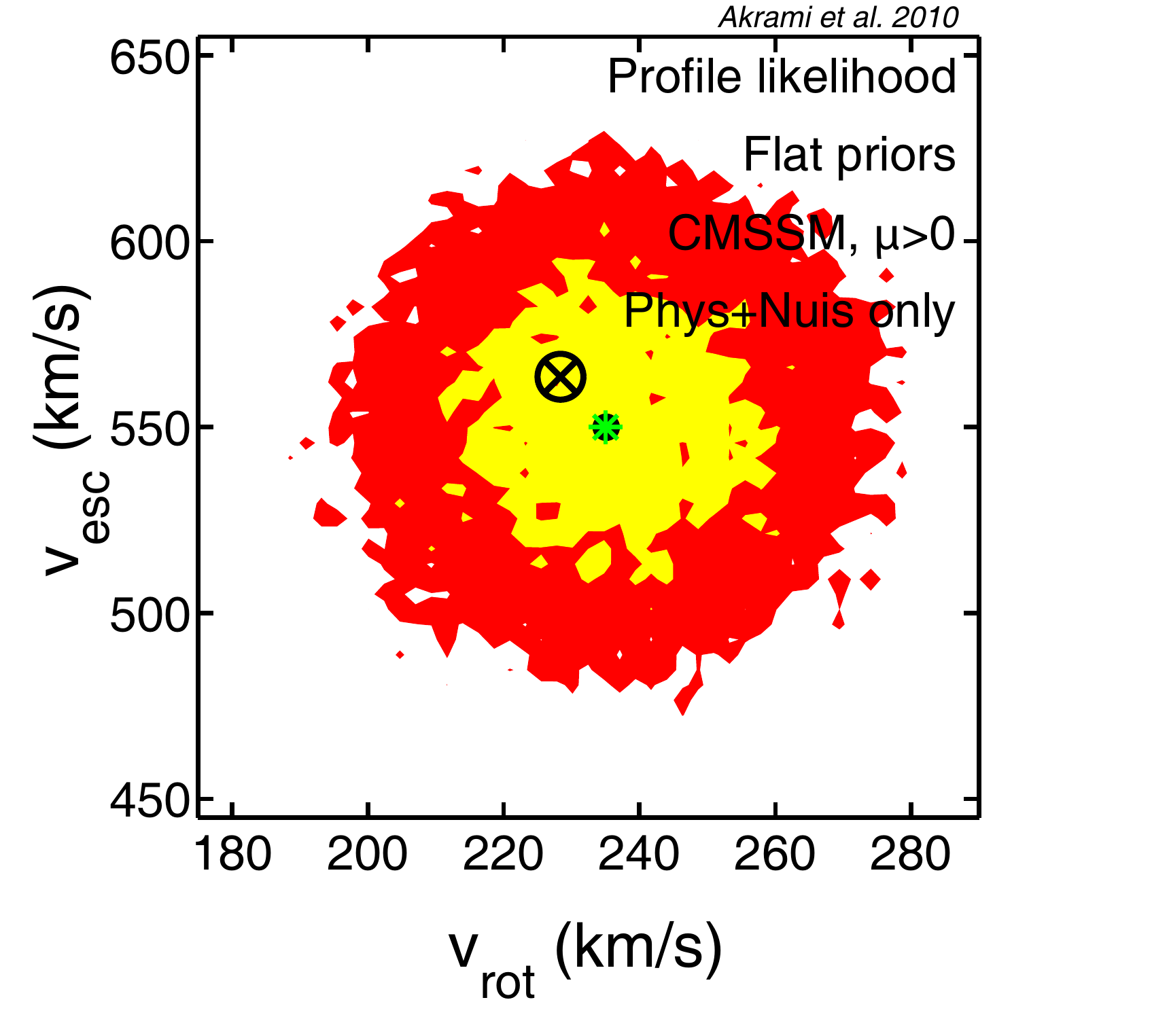}}\\
\subfigure{\includegraphics[scale=0.23, trim = 40 230 130 123, clip=true]{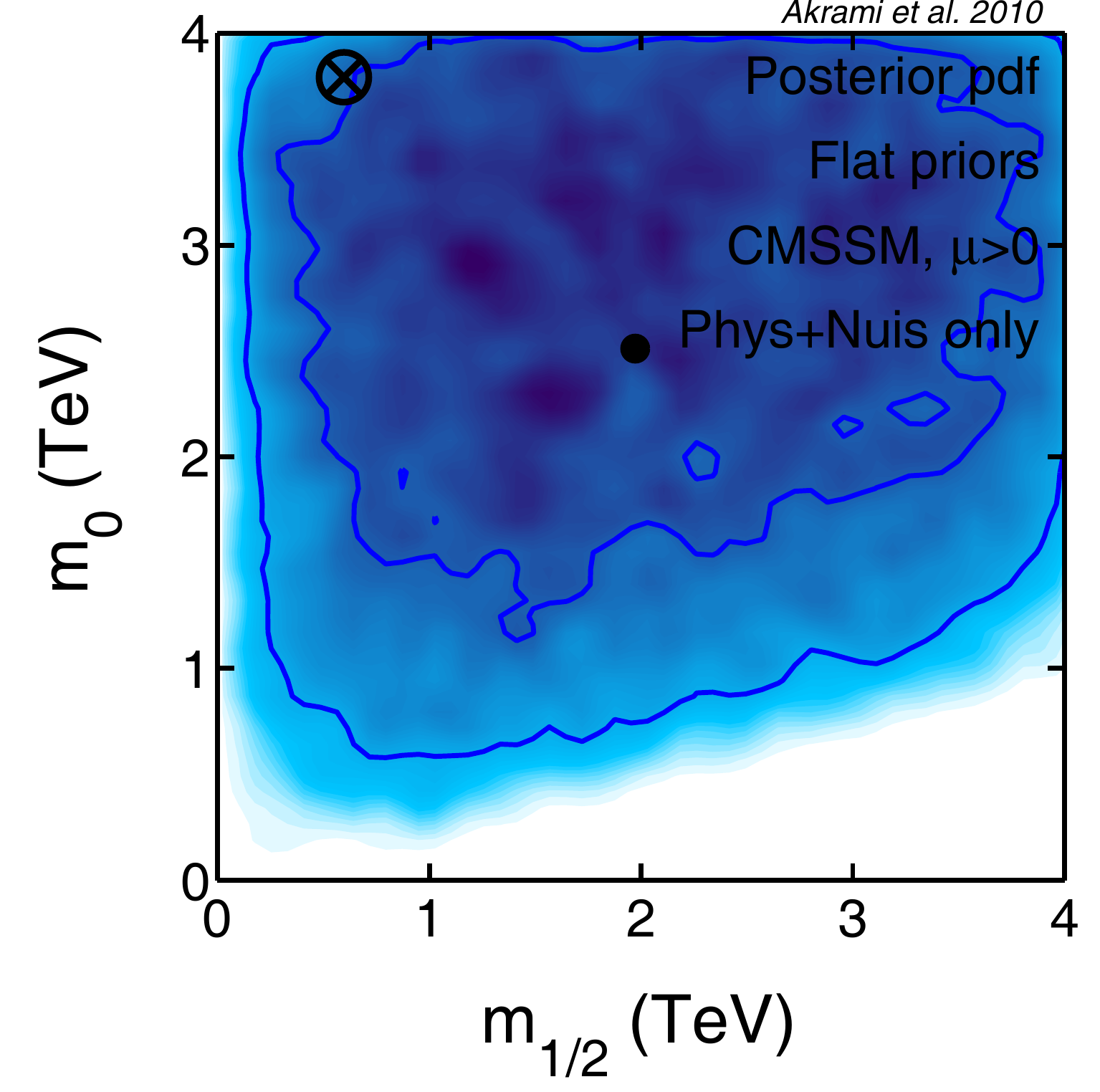}}
\subfigure{\includegraphics[scale=0.23, trim = 40 230 130 123, clip=true]{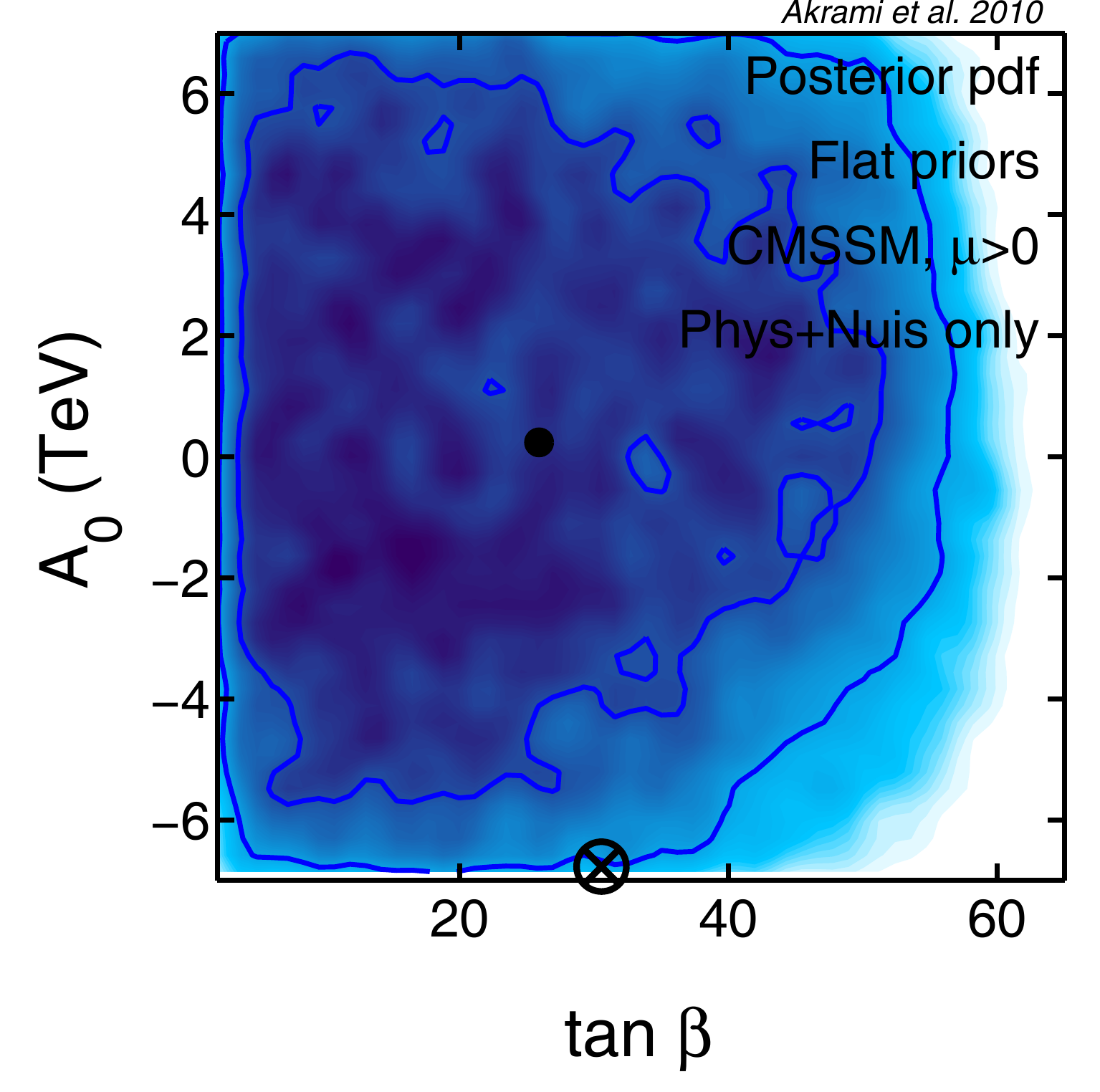}}
\subfigure{\includegraphics[scale=0.23, trim = 40 230 130 123, clip=true]{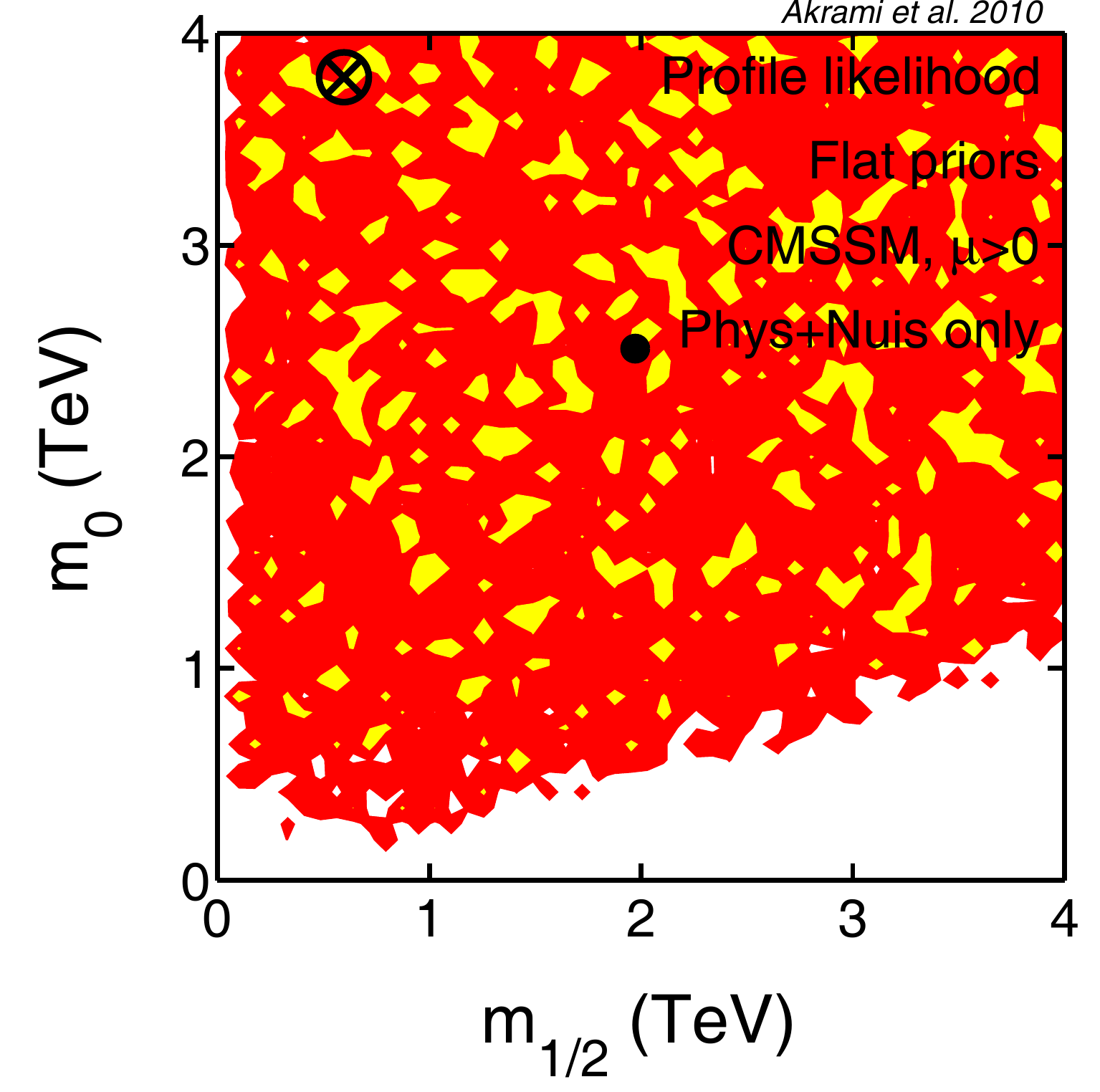}}
\subfigure{\includegraphics[scale=0.23, trim = 40 230 60 123, clip=true]{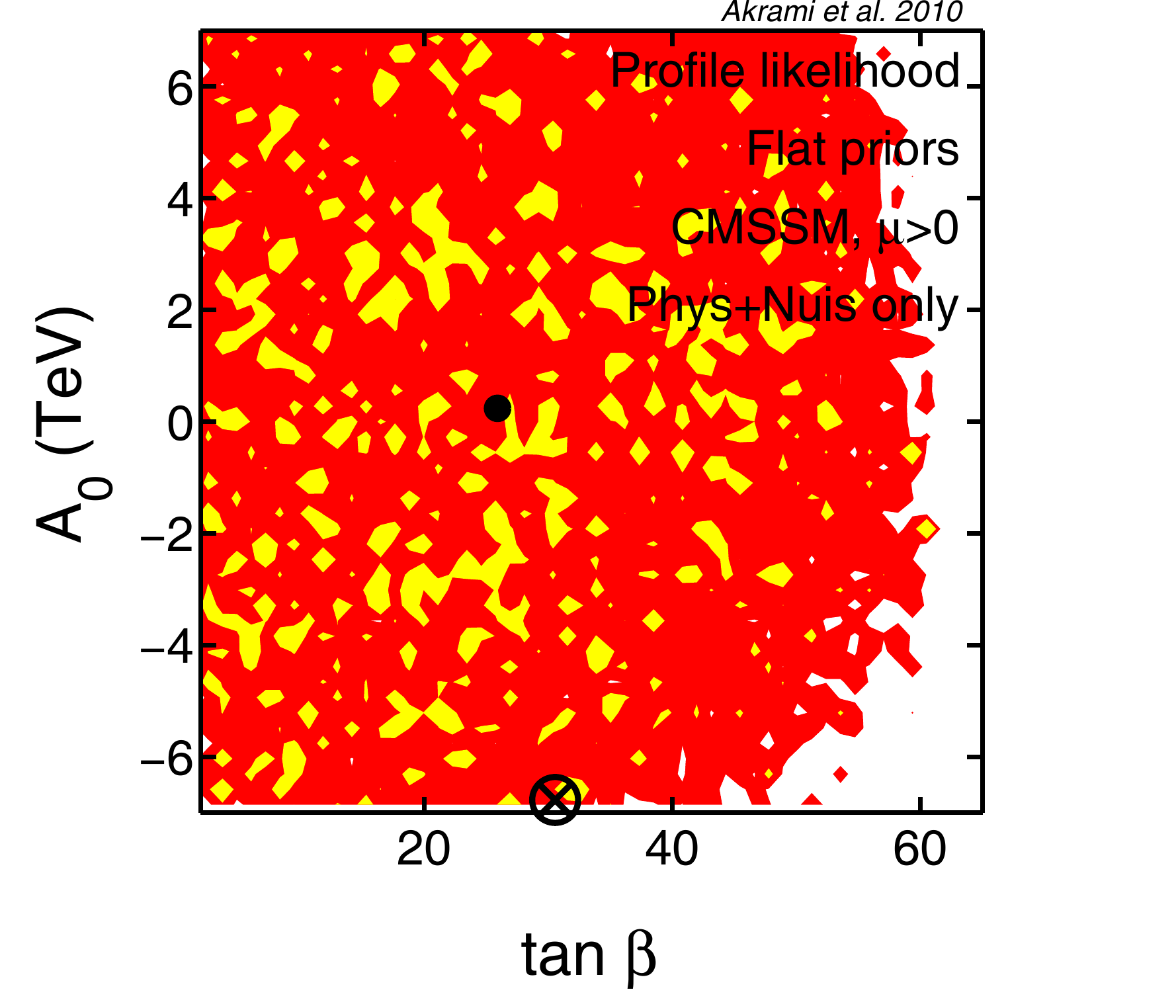}}\\
\caption[aa]{\footnotesize{Two-dimensional marginalised posterior PDFs and profile likelihoods for the mass and nuclear scattering cross-sections of the lightest neutralino, as well as the halo and cross-section nuisance parameters, when no direct detection data are included in the scans. The same plots are also shown for CMSSM parameters (last row). These plots essentially show the effects of explicit priors imposed on the parameters, as well as the physicality constraints. The inner and outer contours in each panel represent $68.3\%$ ($1\sigma$) and $95.4\%$ ($2\sigma$) confidence levels, respectively. Black dots and crosses show the posterior means and best-fit points, respectively. Benchmarks are marked with green stars.}}\label{fig:PriorsOnly}
\end{figure}

Before we present the results of our scans when DD data are employed, we give in~\fig{fig:PriorsOnly} the two-dimensional marginal posteriors and profile likelihoods when only the priors and physicality constraints are taken into account. Plots are in the SI and SD scattering cross-sections of the neutralino $\sigma^{SI}_p$, $\sigma^{SD}_p$ and $\sigma^{SD}_n$ versus the neutralino mass $m_{\tilde\chi^0_1}$, as well as the CMSSM parameter planes $\mzero$-$\mhalf$ and $\azero$-$\tanb$. We also show the same plots for the halo and hadronic mass matrix nuisance parameters $\rhoDM$, $\vrot$, $\vmp$, $\vesc$, $\Deltaps$, $\sigma_0$ and $\SigmapiN$. These show how our scanning algorithm samples the parameter space when no experimental data are provided, so can be thought of as the ``effective priors'' we place on the parameters by our choice of physicality constraints and scanning algorithm. The inner and outer contours in each panel (in dark and light blue for marginal posteriors and in yellow and red for profile likelihoods) represent $68.3\%$ ($1\sigma$) and $95.4\%$ ($2\sigma$) confidence levels, respectively. Black dots show the posterior means while black crosses show best-fit (highest likelihood) points, respectively.

\afterpage{\clearpage}

These types of plots are always necessary when one wants to examine how constraining certain sets of data (from DD experiments in our case) are. A combination of both posterior PDFs and profile likelihoods show that in our scans some regions of the parameter space are sampled better than others. This becomes clear when we examine the plots for the DD quantities $\sigma^{SI}_p$, $\sigma^{SD}_p$, $\sigma^{SD}_n$ and $m_{\tilde\chi^0_1}$ in particular. Obviously, the algorithm tends to sample low cross-sections and low neutralino masses at a higher resolution. For a detailed discussion of these sampling effects, their relationship to the choice of priors and implications for statistical inference, see the companion paper \cite{Akrami:2010}. For the halo and cross-section nuisance parameters, the plots are basically reflections of what we assumed for their likelihoods, i.e. normal and log-normal distributions.

\begin{figure}[t]
\subfigure{\includegraphics[scale=0.23, trim = 40 230 130 123, clip=true]{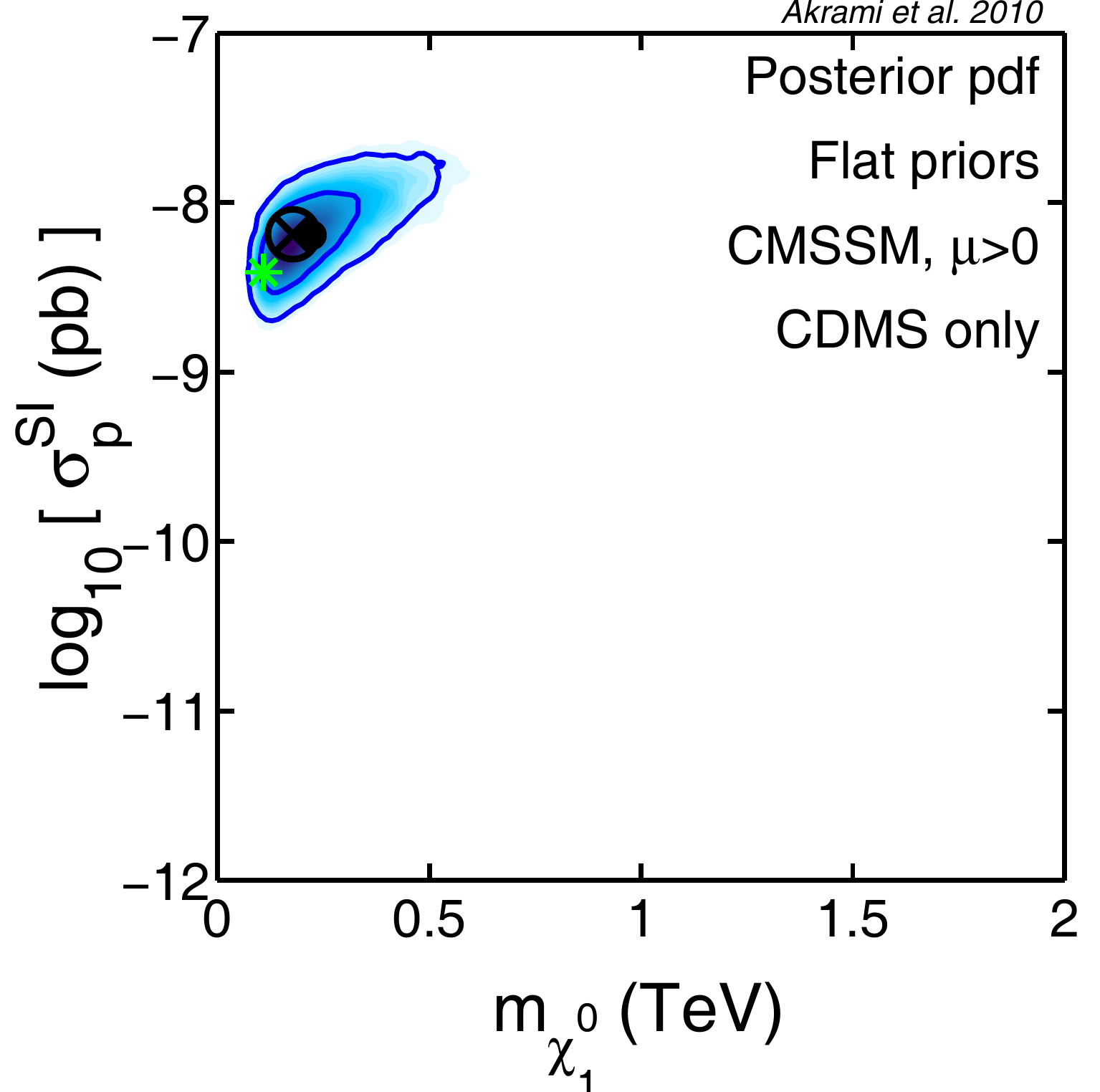}}
\subfigure{\includegraphics[scale=0.23, trim = 40 230 130 123, clip=true]{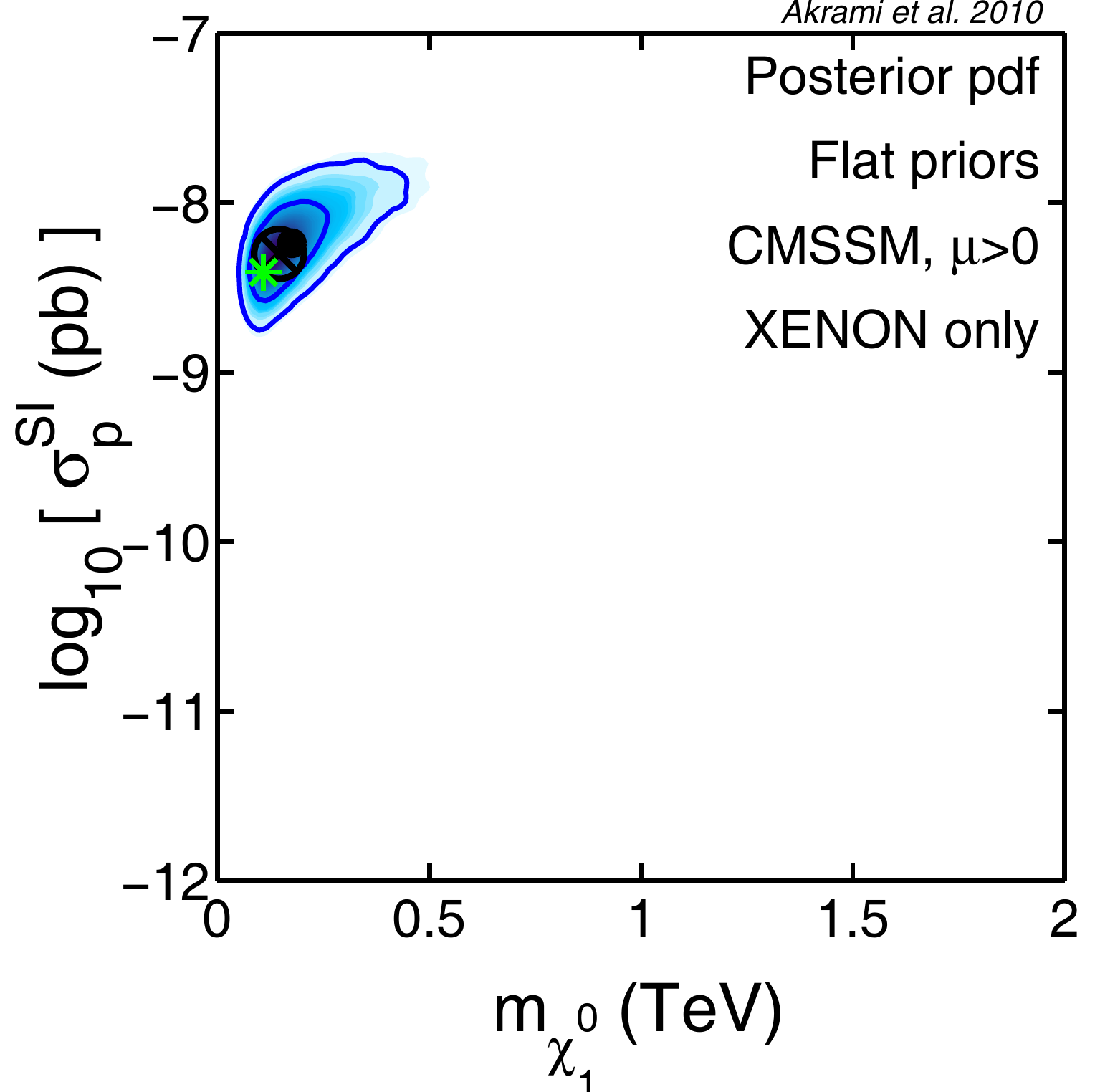}}
\subfigure{\includegraphics[scale=0.23, trim = 40 230 130 123, clip=true]{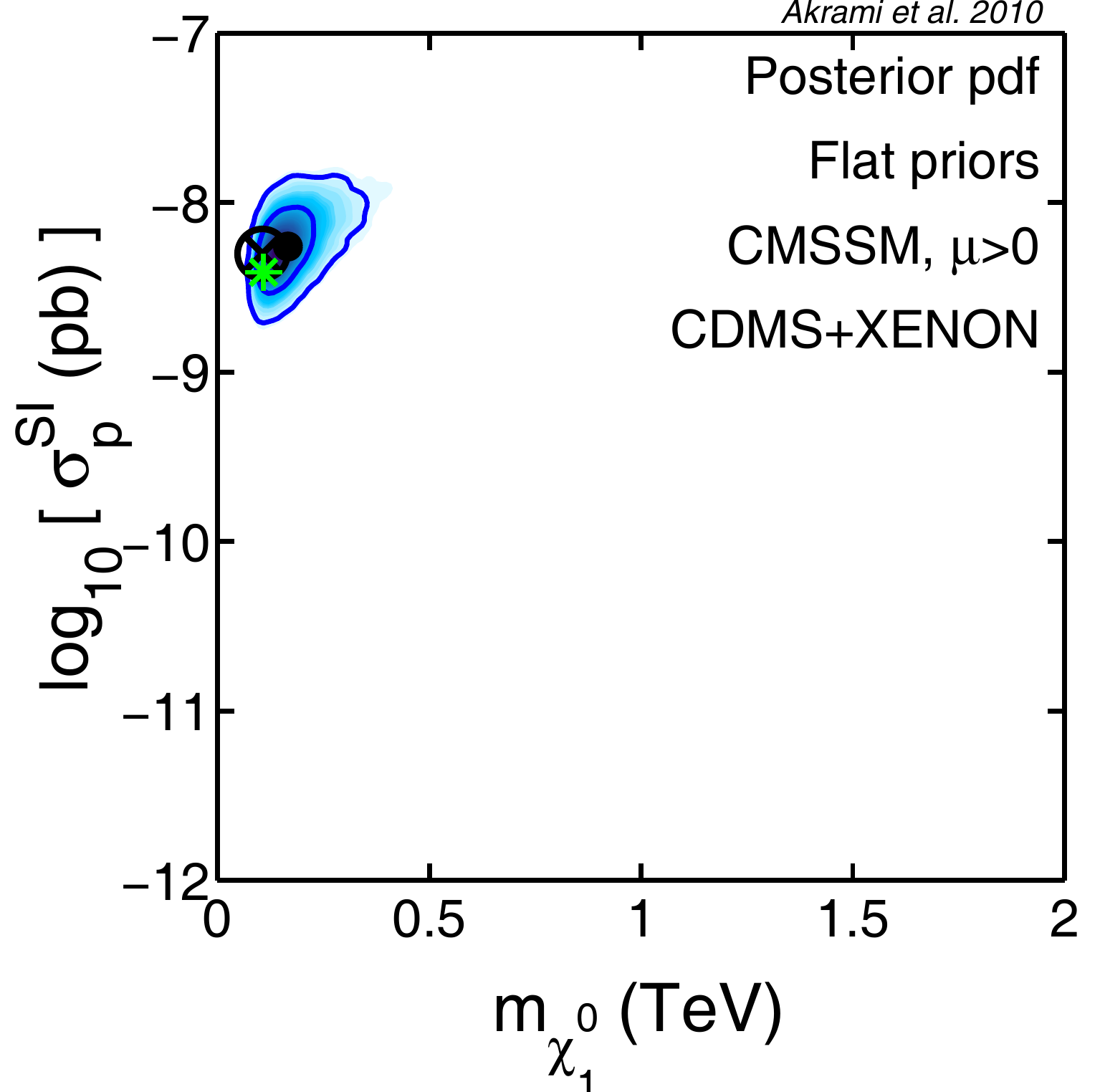}}
\subfigure{\includegraphics[scale=0.23, trim = 40 230 60 123, clip=true]{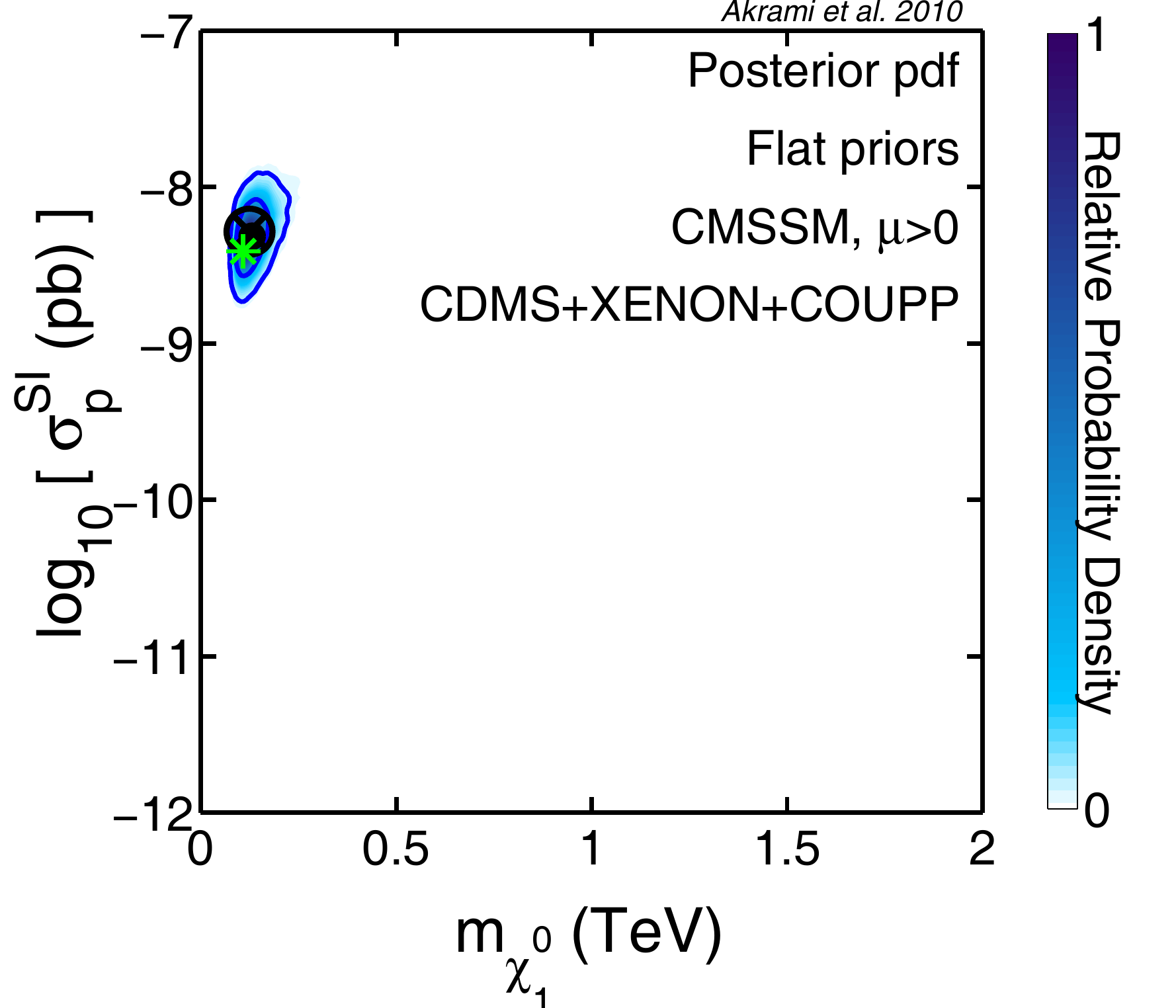}}\\
\subfigure{\includegraphics[scale=0.23, trim = 40 230 130 123, clip=true]{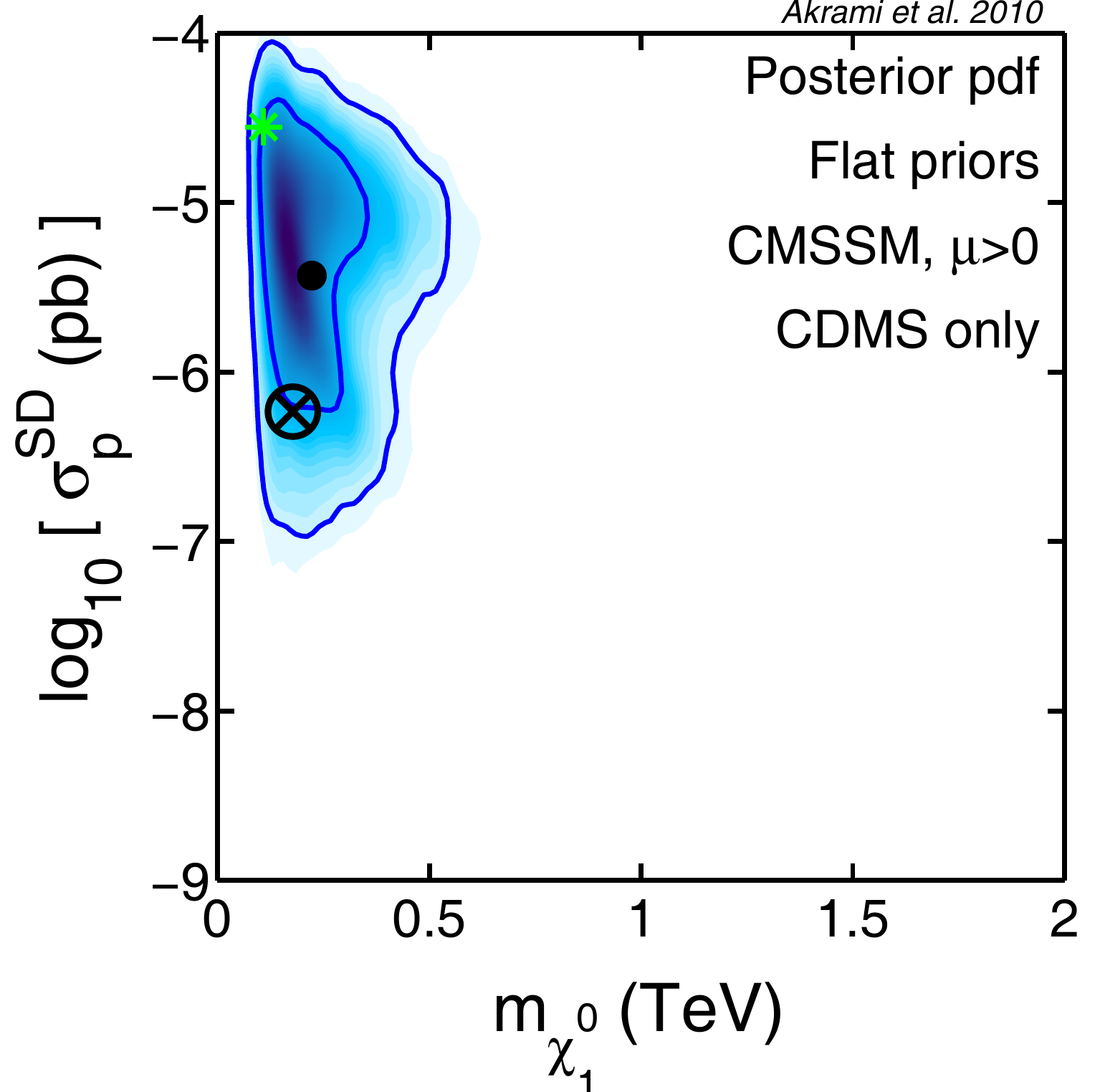}}
\subfigure{\includegraphics[scale=0.23, trim = 40 230 130 123, clip=true]{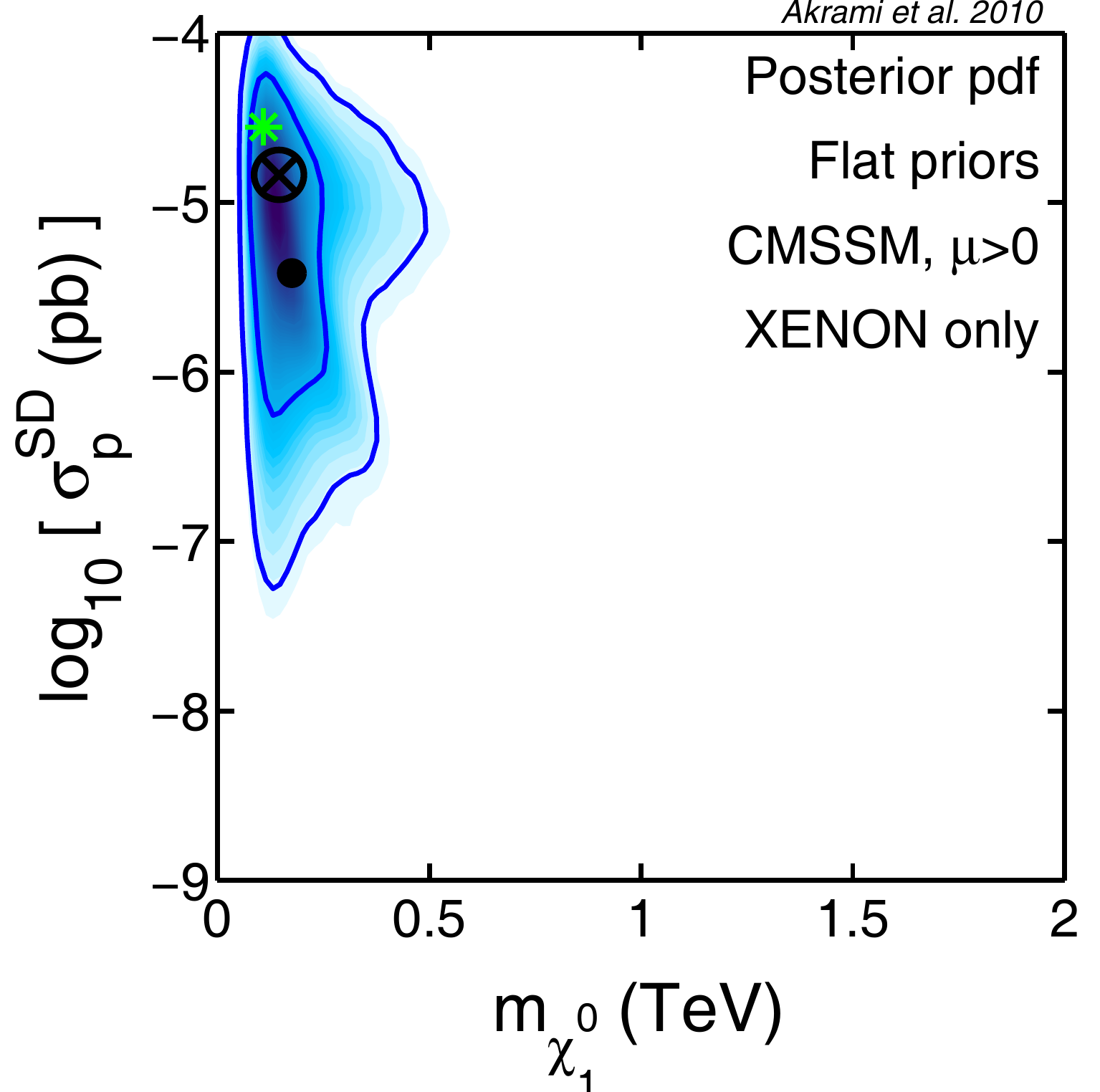}}
\subfigure{\includegraphics[scale=0.23, trim = 40 230 130 123, clip=true]{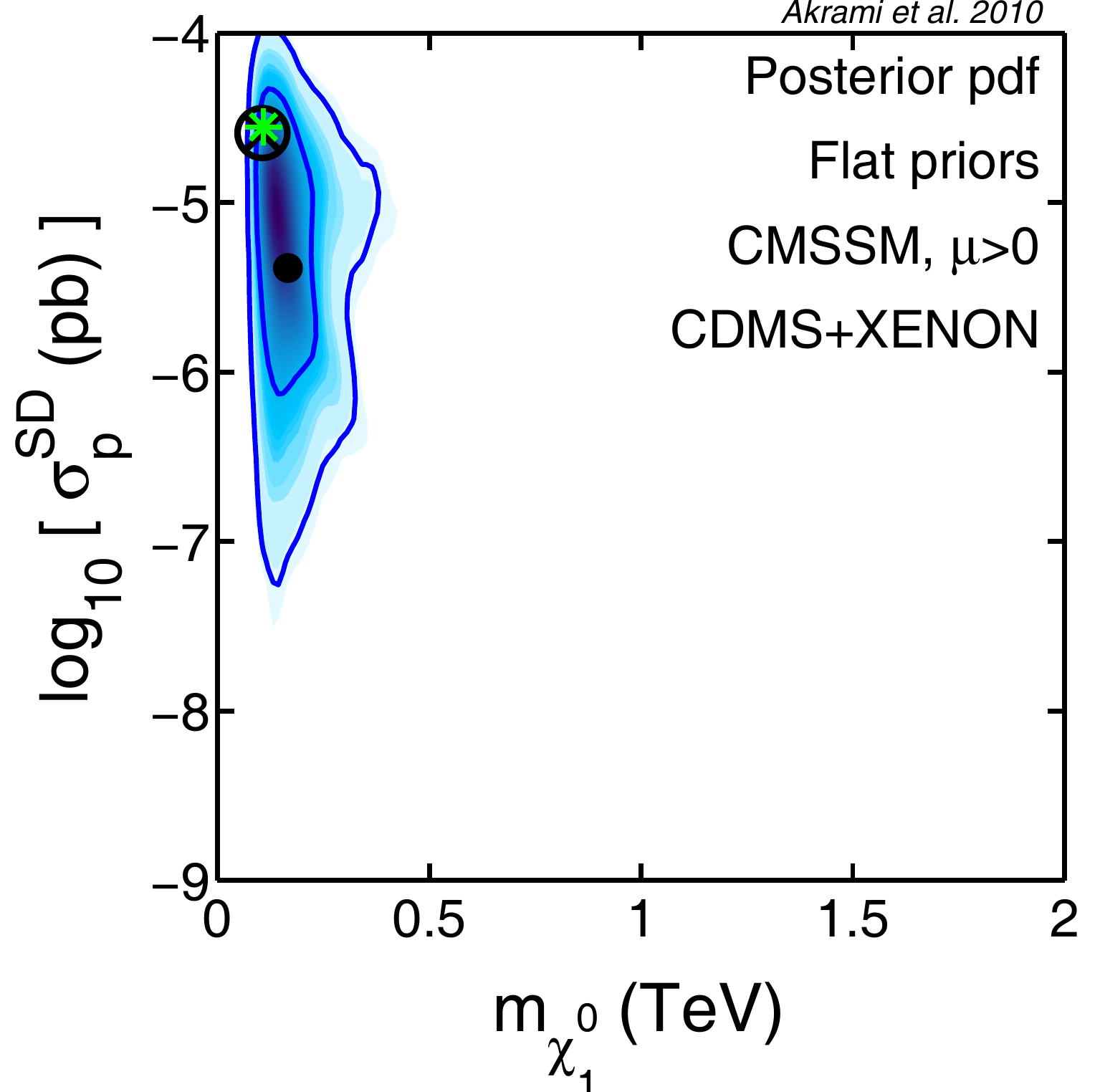}}
\subfigure{\includegraphics[scale=0.23, trim = 40 230 60 123, clip=true]{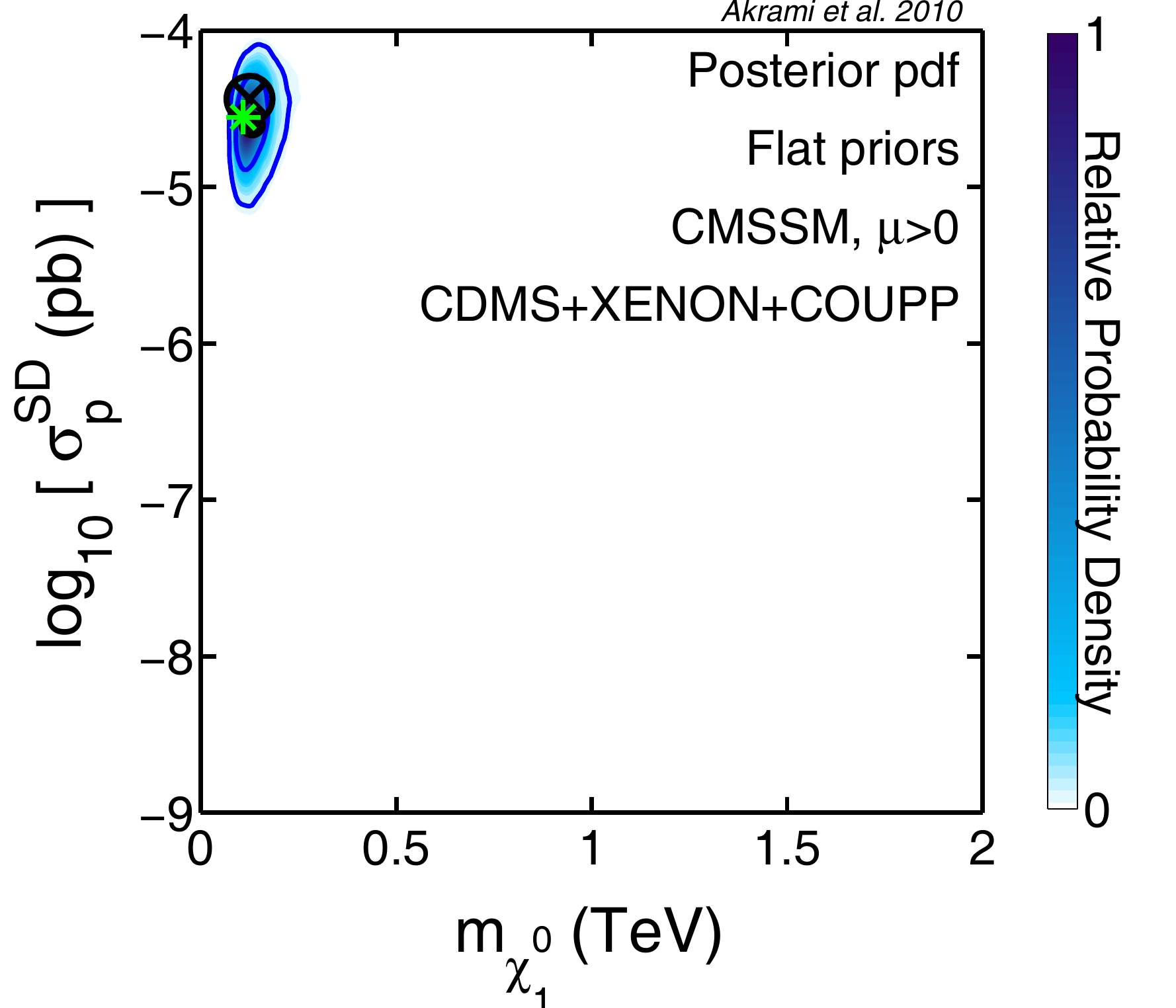}}\\
\subfigure{\includegraphics[scale=0.23, trim = 40 230 130 123, clip=true]{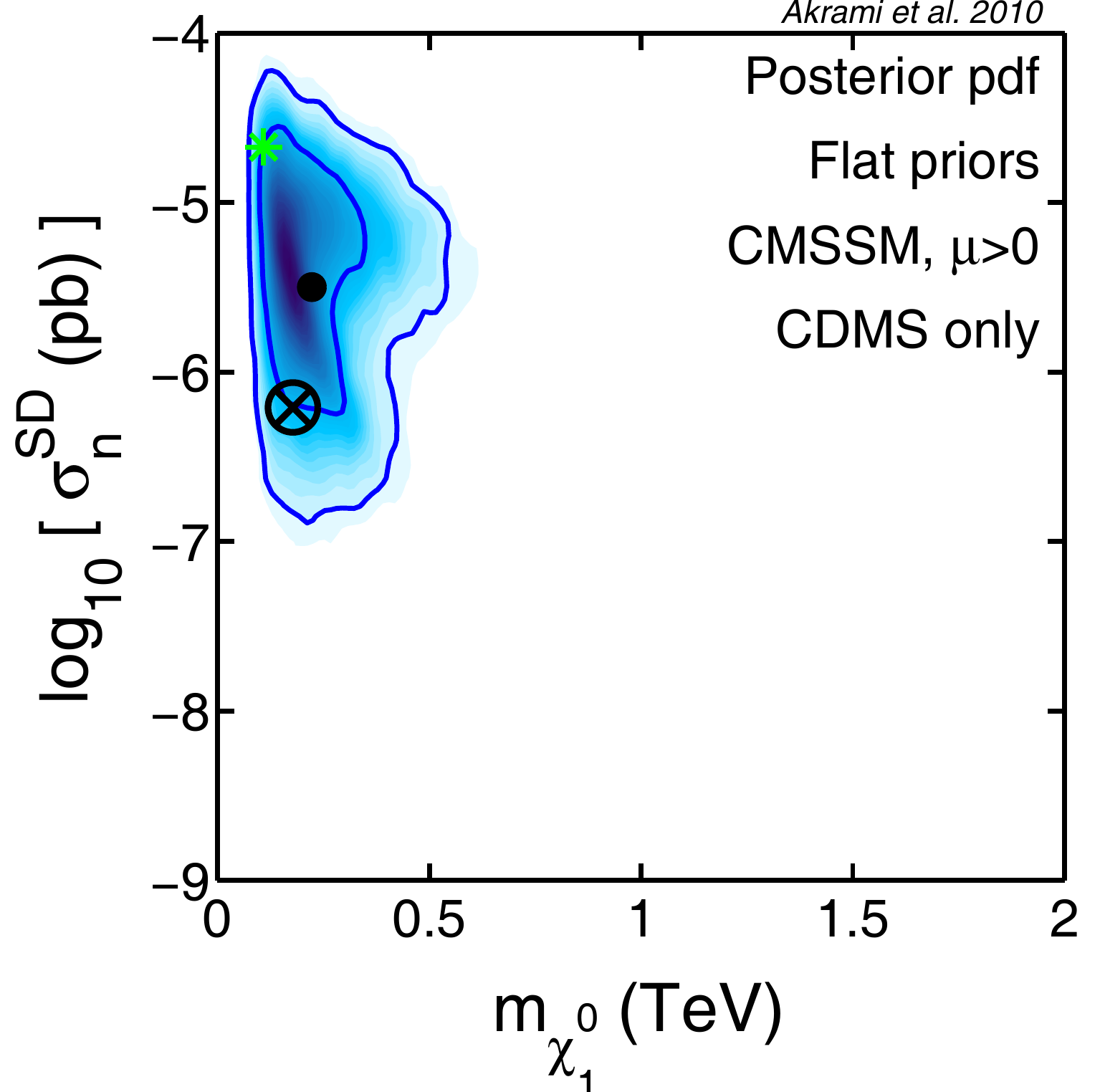}}
\subfigure{\includegraphics[scale=0.23, trim = 40 230 130 123, clip=true]{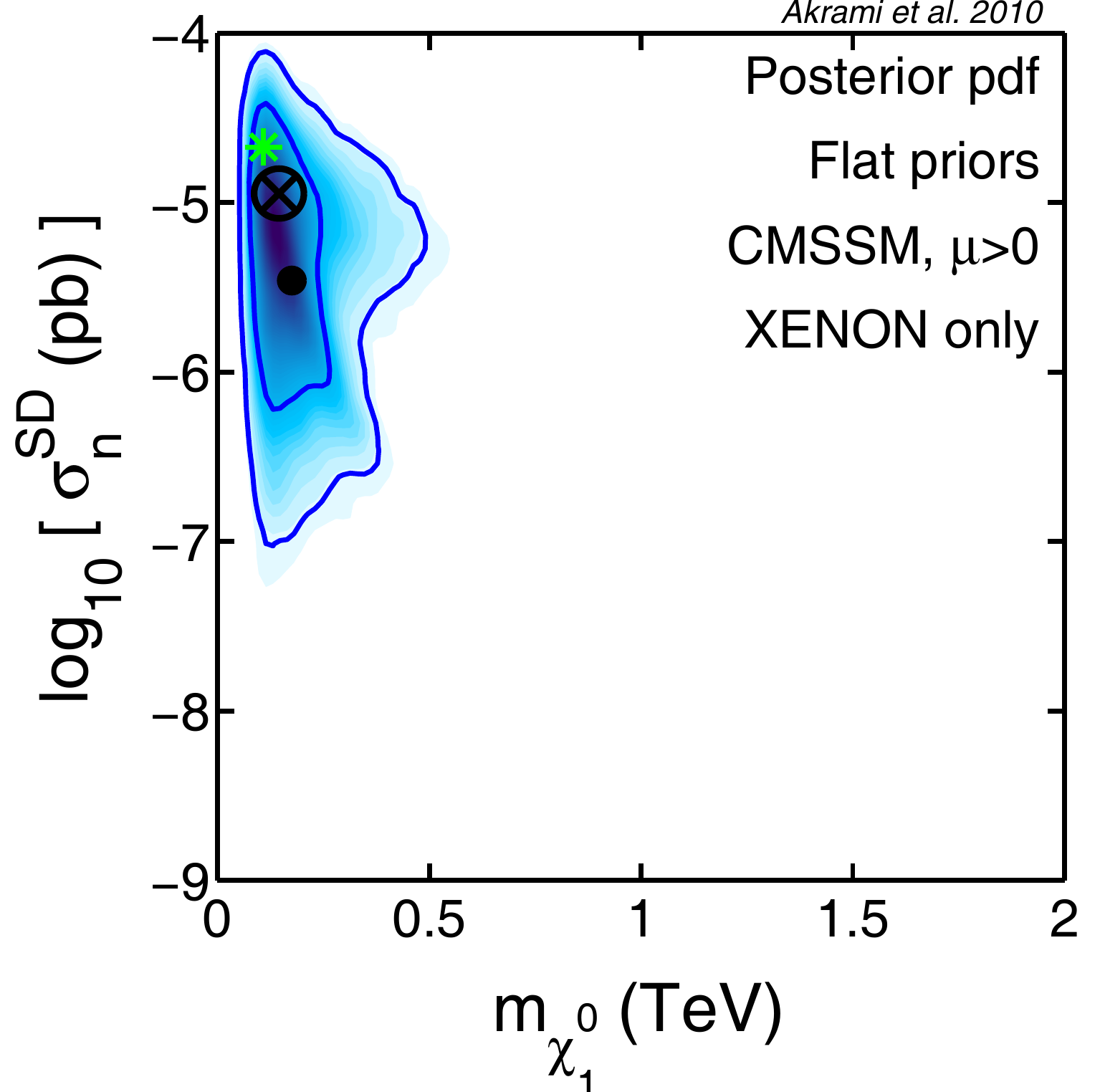}}
\subfigure{\includegraphics[scale=0.23, trim = 40 230 130 123, clip=true]{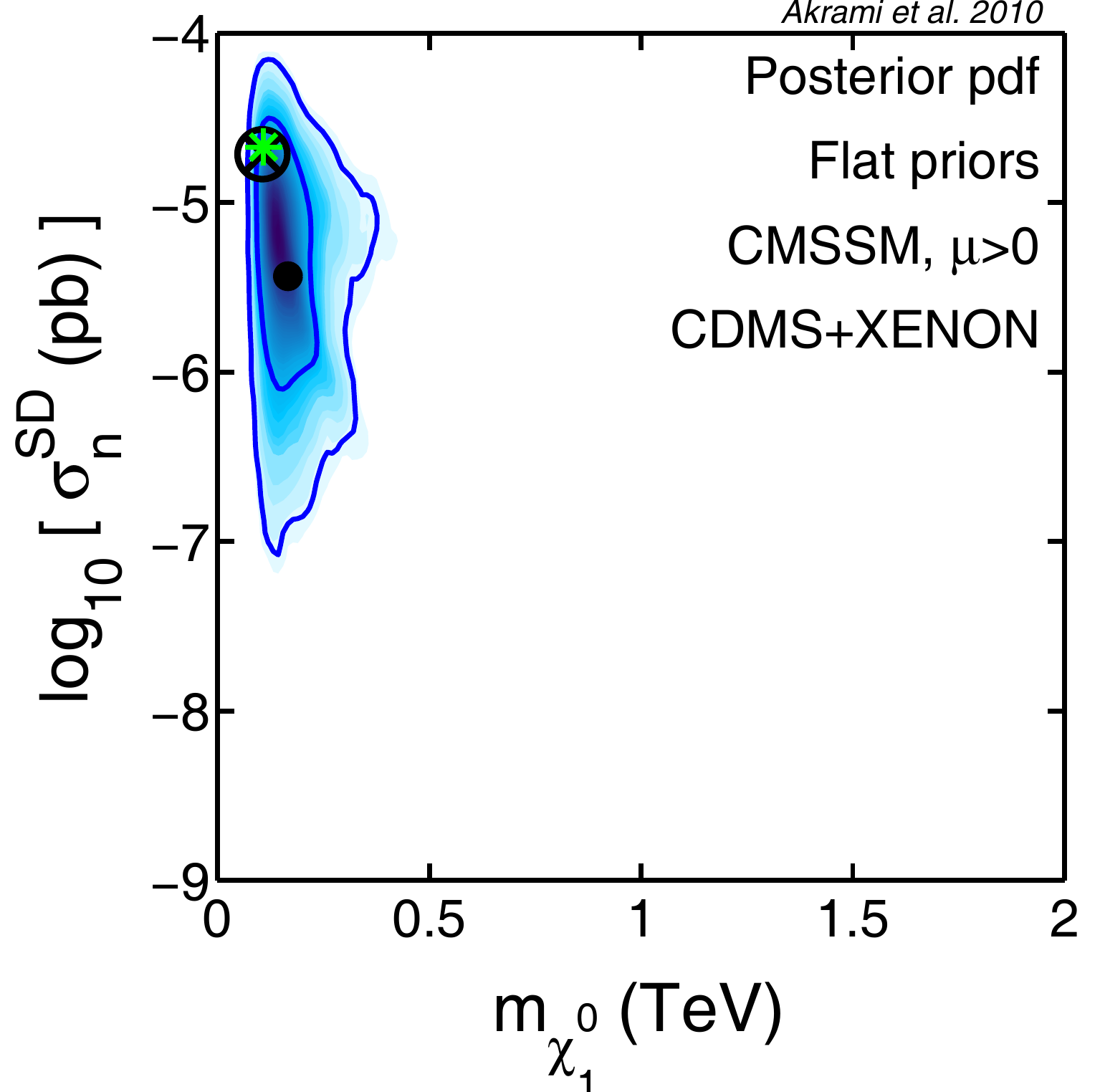}}
\subfigure{\includegraphics[scale=0.23, trim = 40 230 60 123, clip=true]{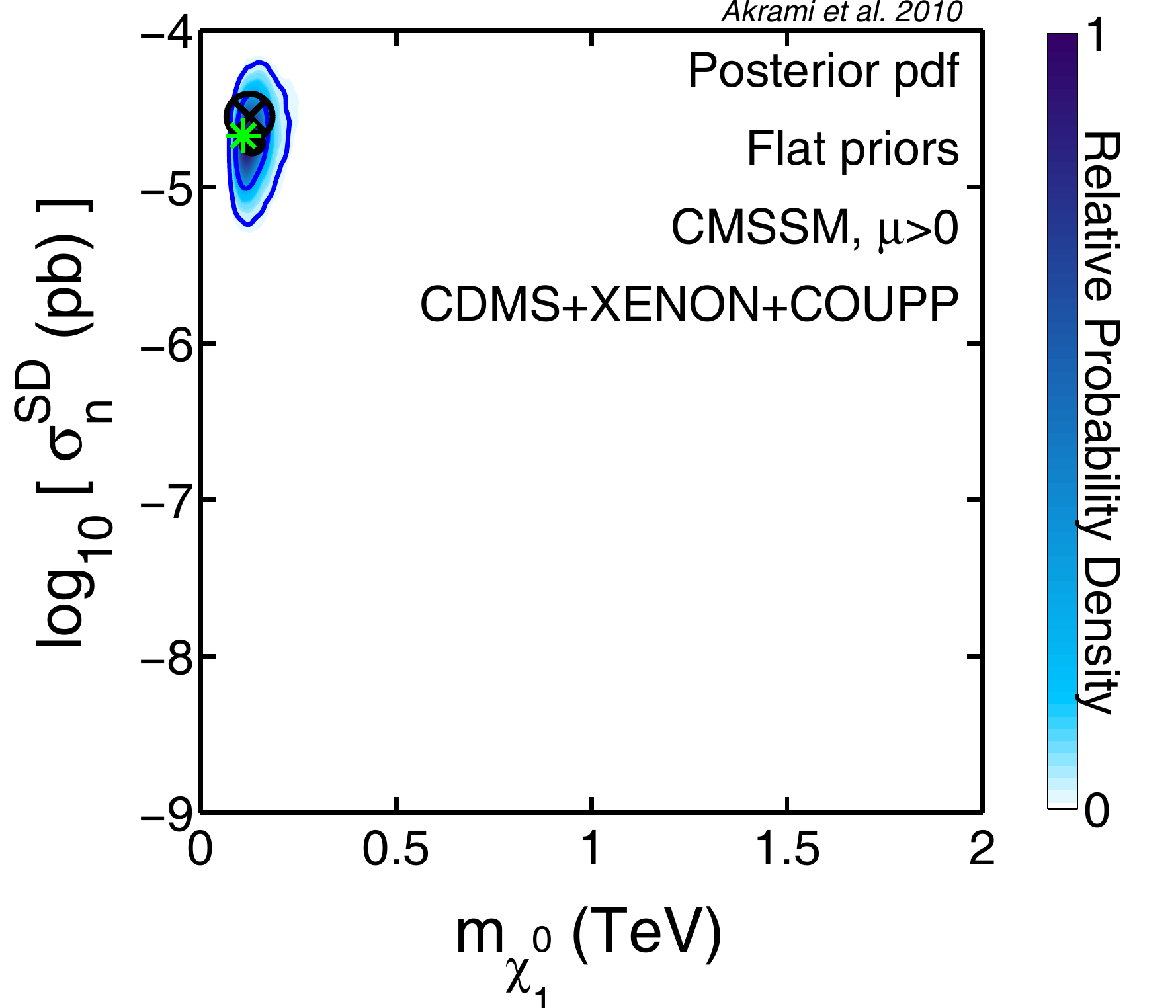}}\\
\caption[aa]{\footnotesize{Two-dimensional marginalised posterior PDFs for the mass and nuclear scattering cross-sections of the lightest neutralino for benchmark 1 when different combinations of direct detection likelihoods are used in our scans. The inner and outer contours in each panel represent $68.3\%$ ($1\sigma$) and $95.4\%$ ($2\sigma$) confidence levels, respectively. Black dots and crosses show the posterior means and best-fit points, respectively, and benchmark values are marked with green stars.}}\label{fig:LHmarg}
\end{figure}

\begin{figure}[t]
\subfigure{\includegraphics[scale=0.23, trim = 40 230 130 123, clip=true]{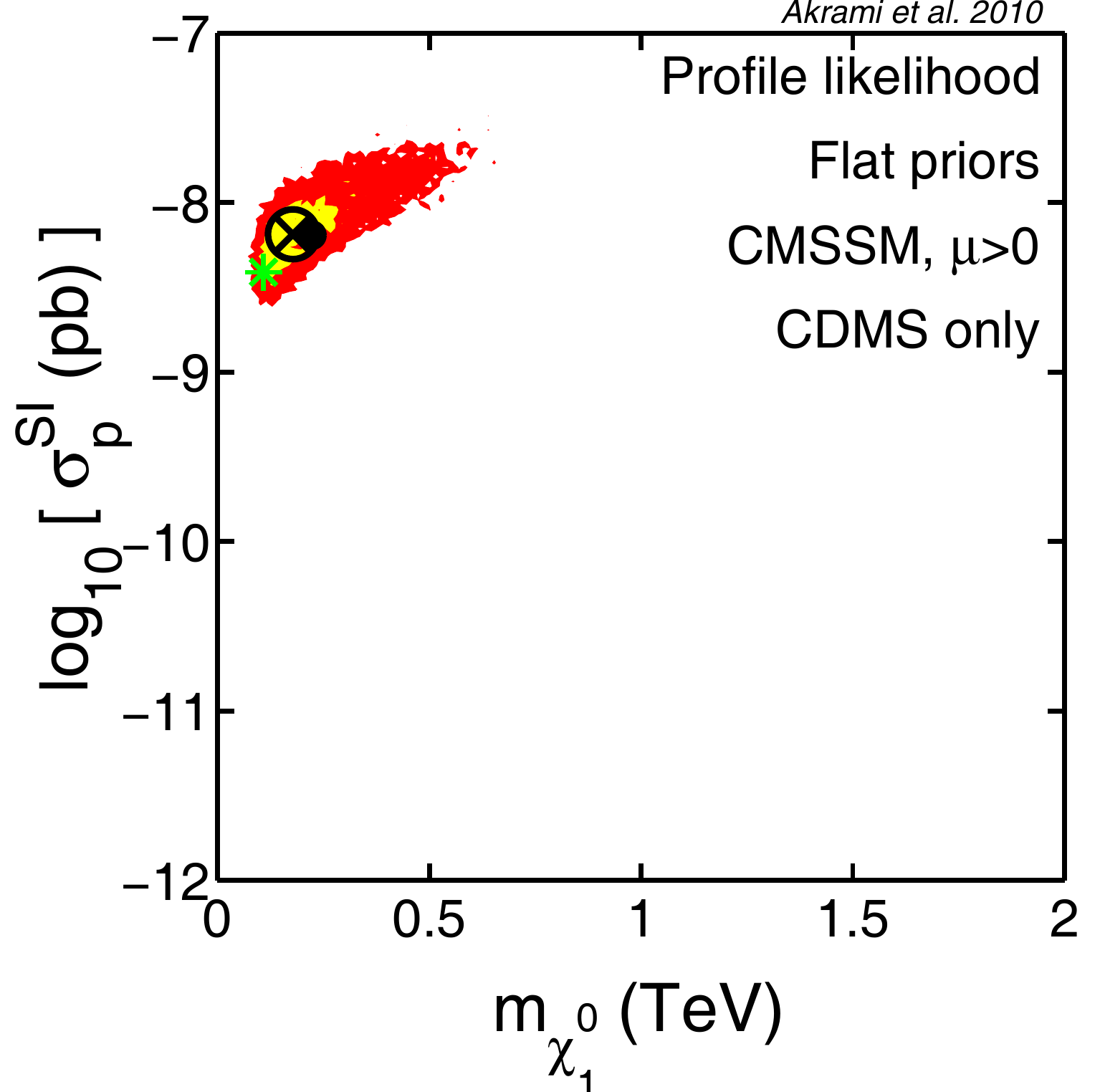}}
\subfigure{\includegraphics[scale=0.23, trim = 40 230 130 123, clip=true]{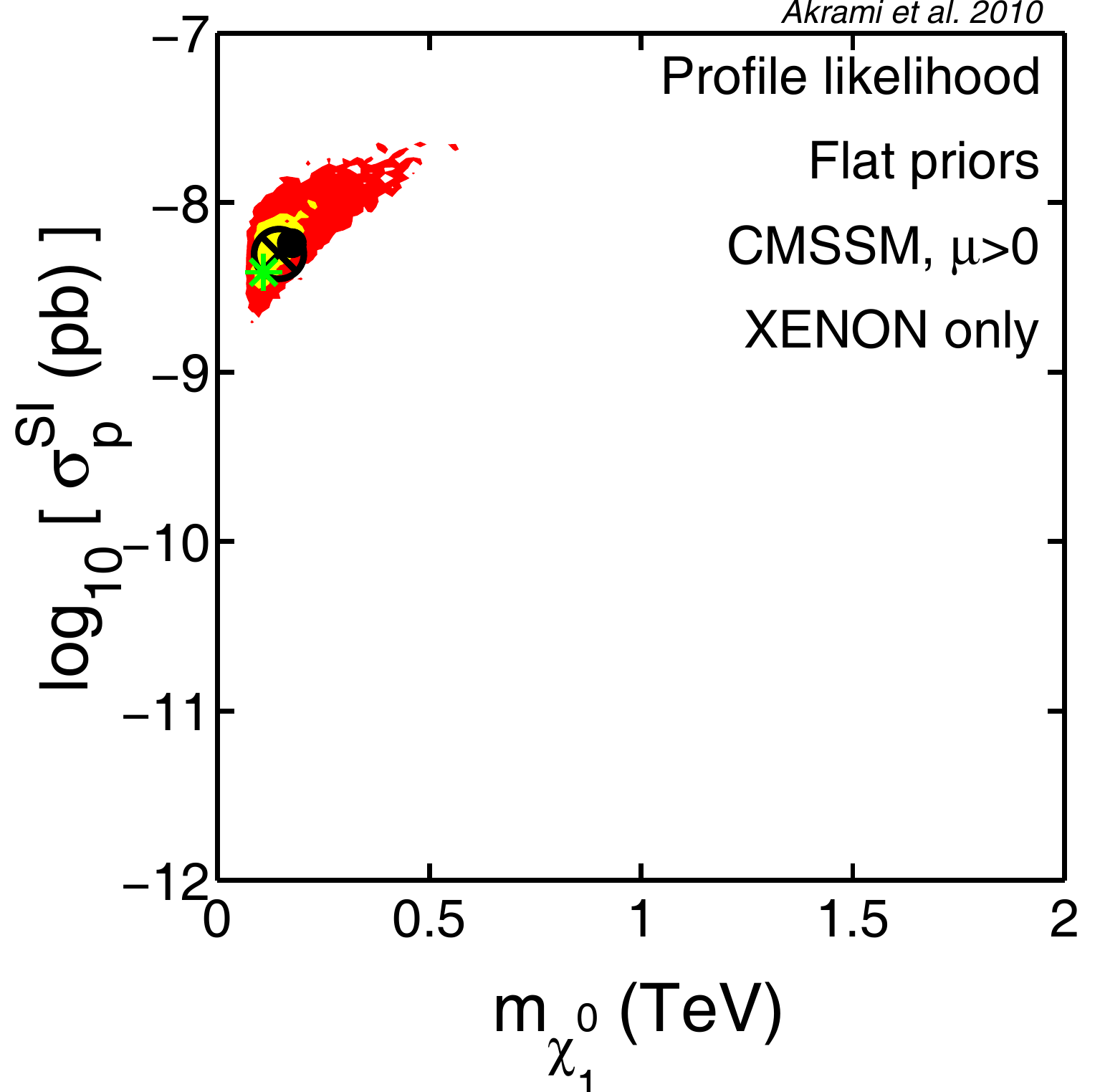}}
\subfigure{\includegraphics[scale=0.23, trim = 40 230 130 123, clip=true]{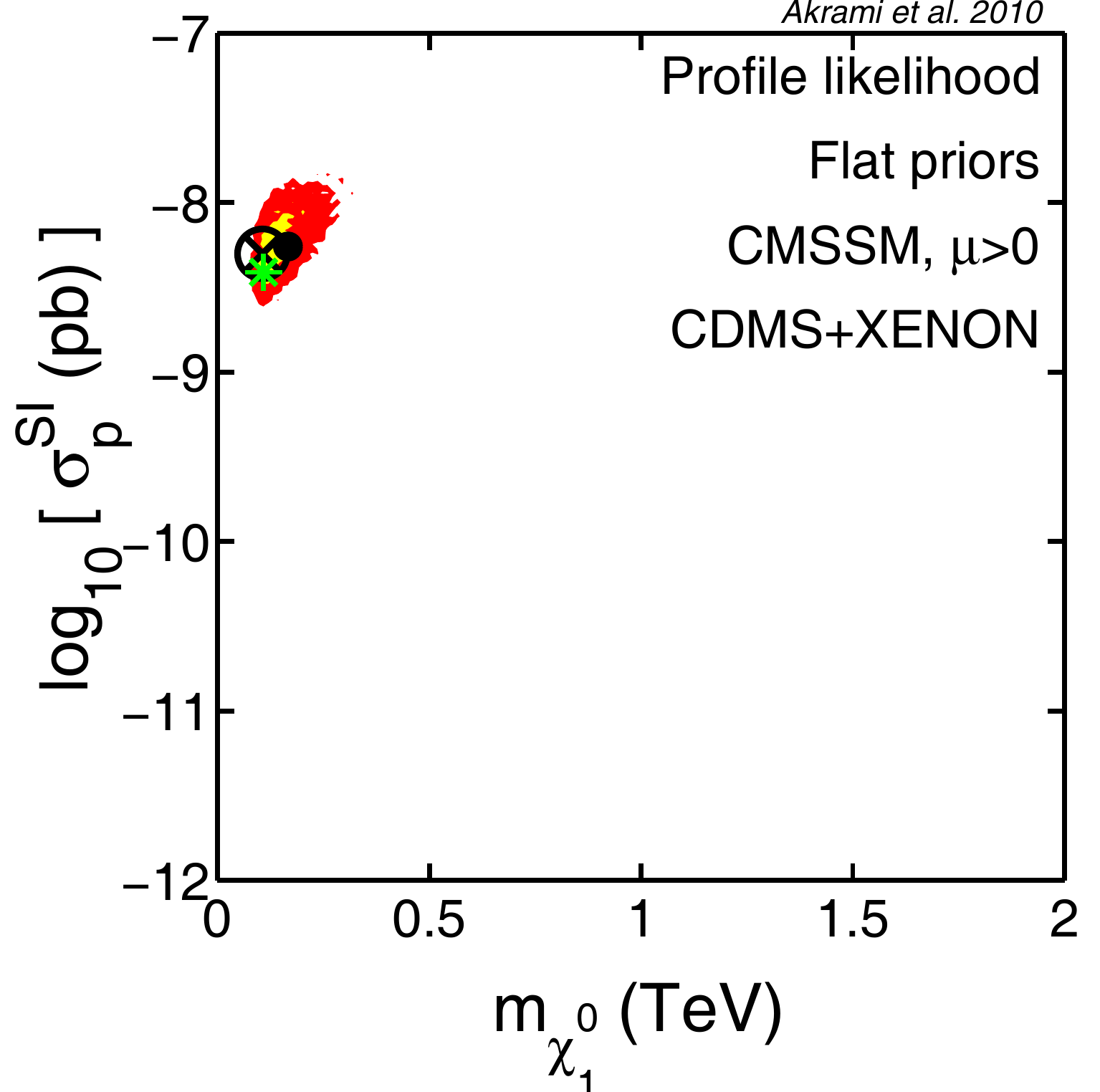}}
\subfigure{\includegraphics[scale=0.23, trim = 40 230 60 123, clip=true]{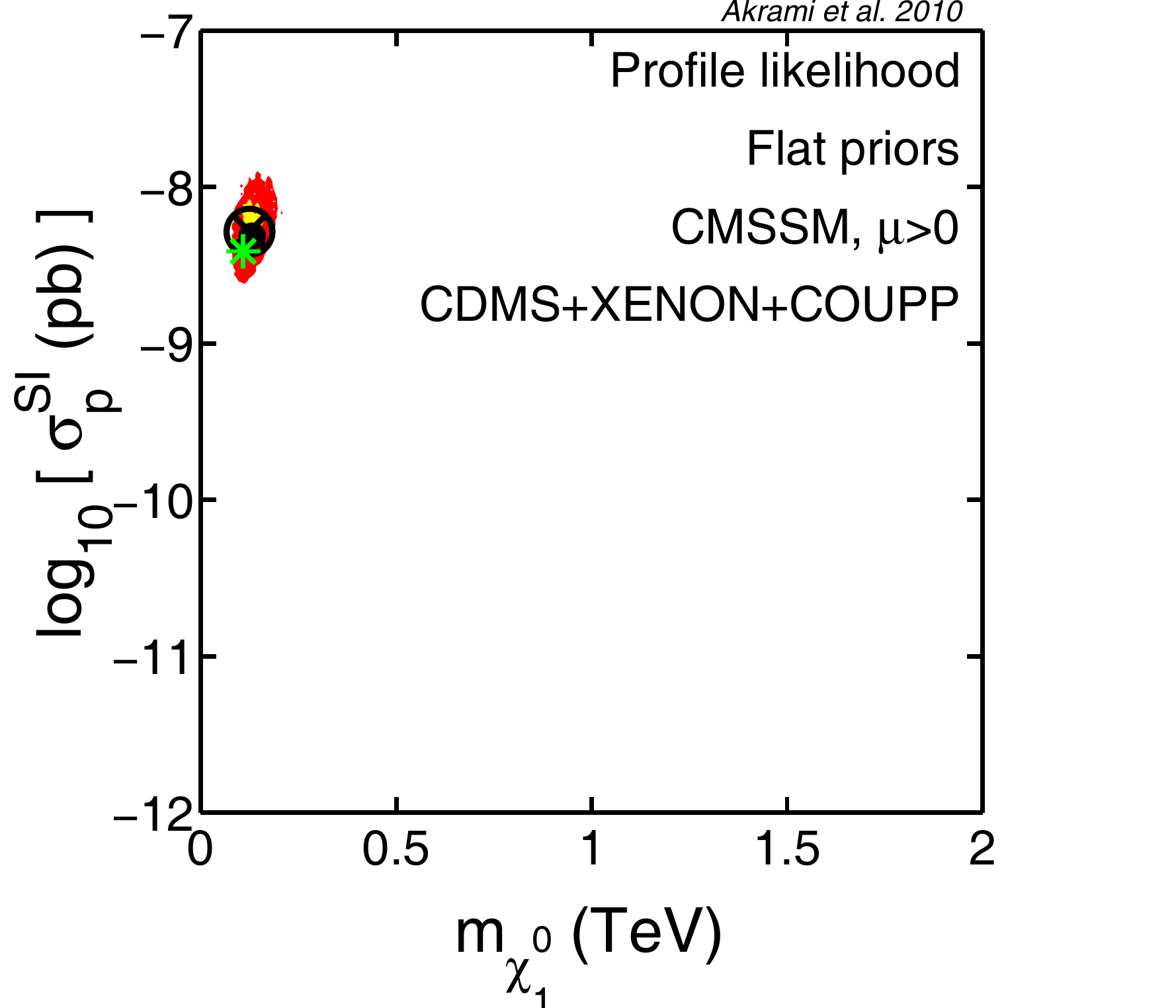}}\\
\subfigure{\includegraphics[scale=0.23, trim = 40 230 130 123, clip=true]{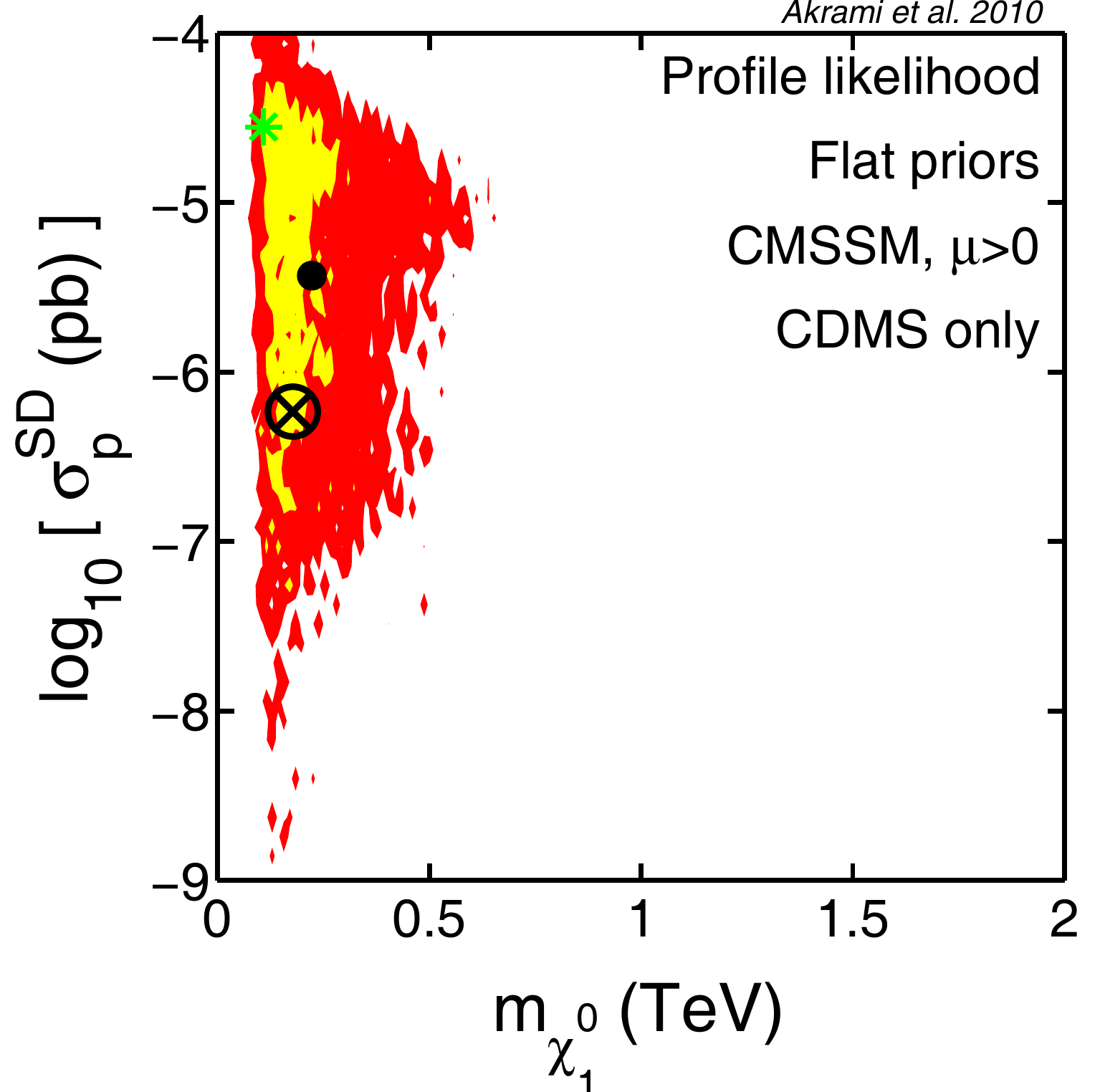}}
\subfigure{\includegraphics[scale=0.23, trim = 40 230 130 123, clip=true]{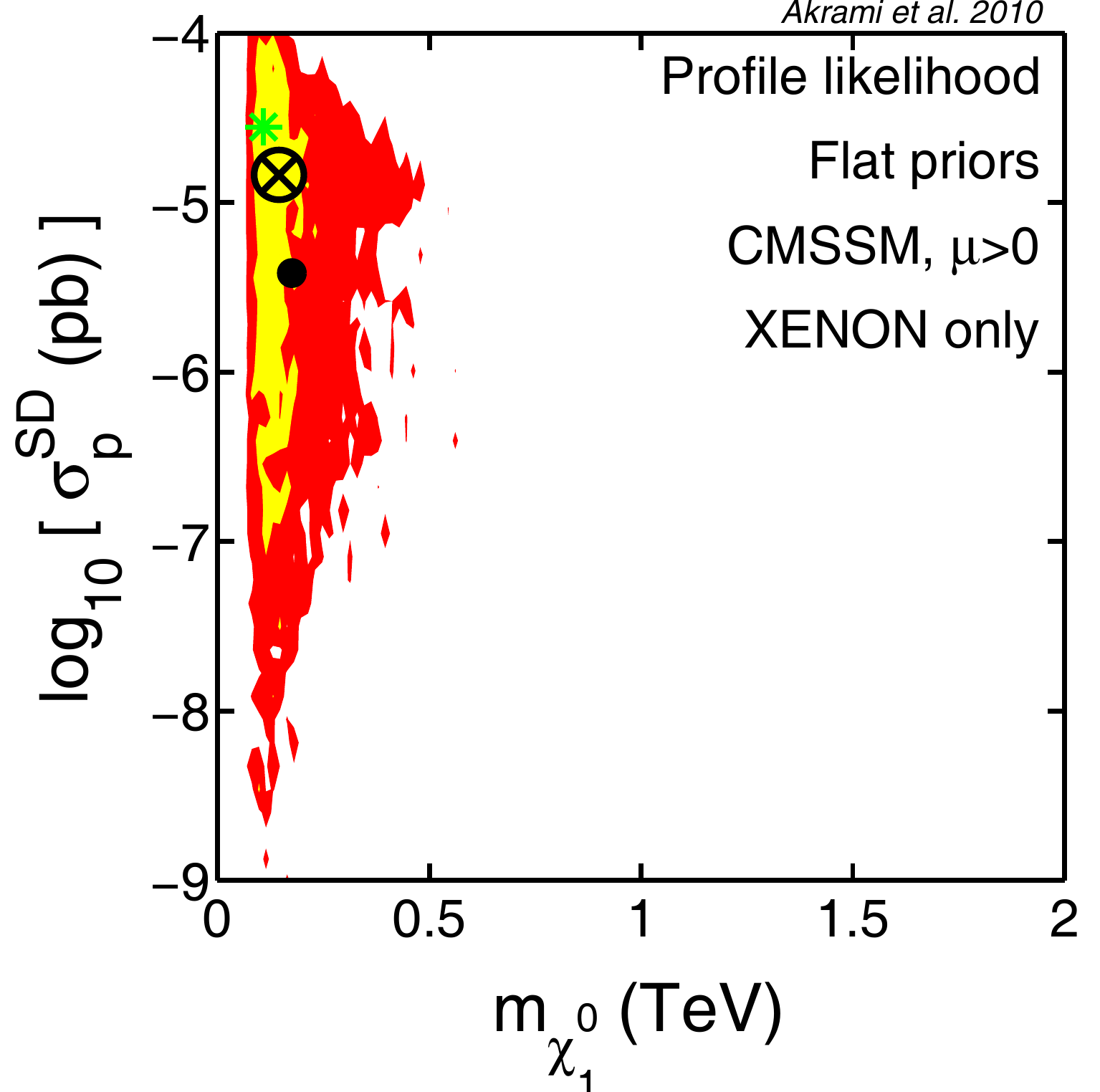}}
\subfigure{\includegraphics[scale=0.23, trim = 40 230 130 123, clip=true]{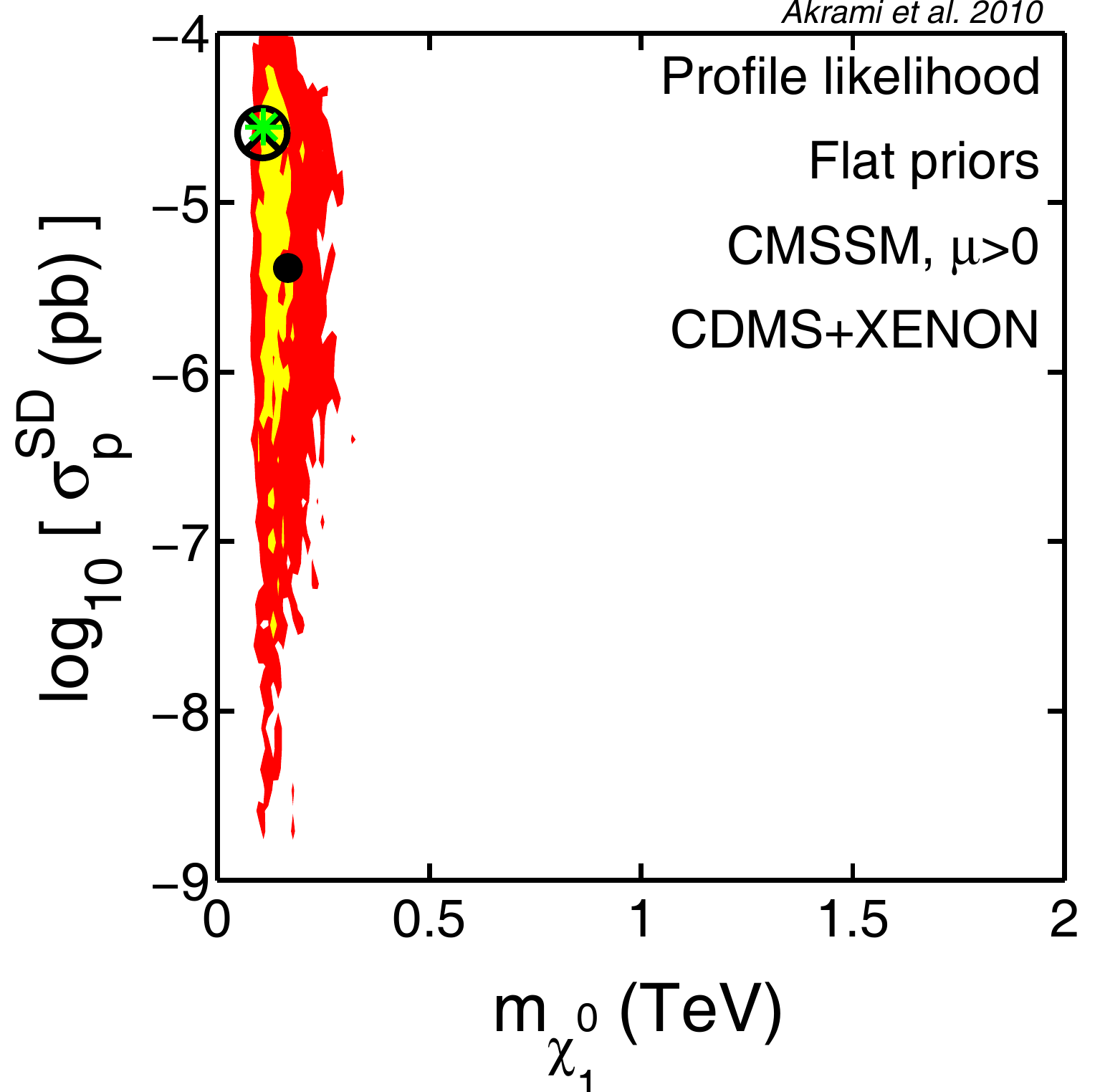}}
\subfigure{\includegraphics[scale=0.23, trim = 40 230 60 123, clip=true]{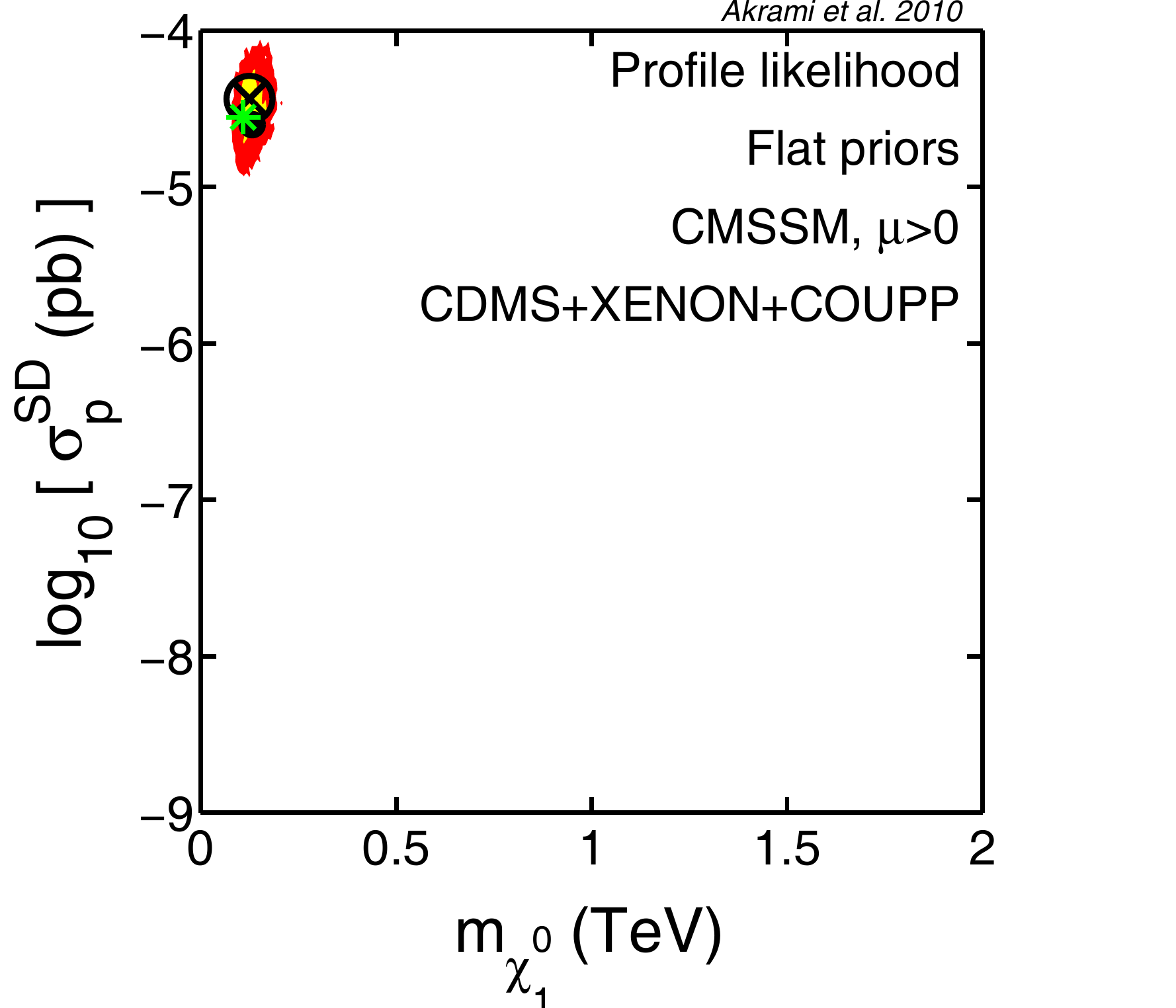}}\\
\subfigure{\includegraphics[scale=0.23, trim = 40 230 130 123, clip=true]{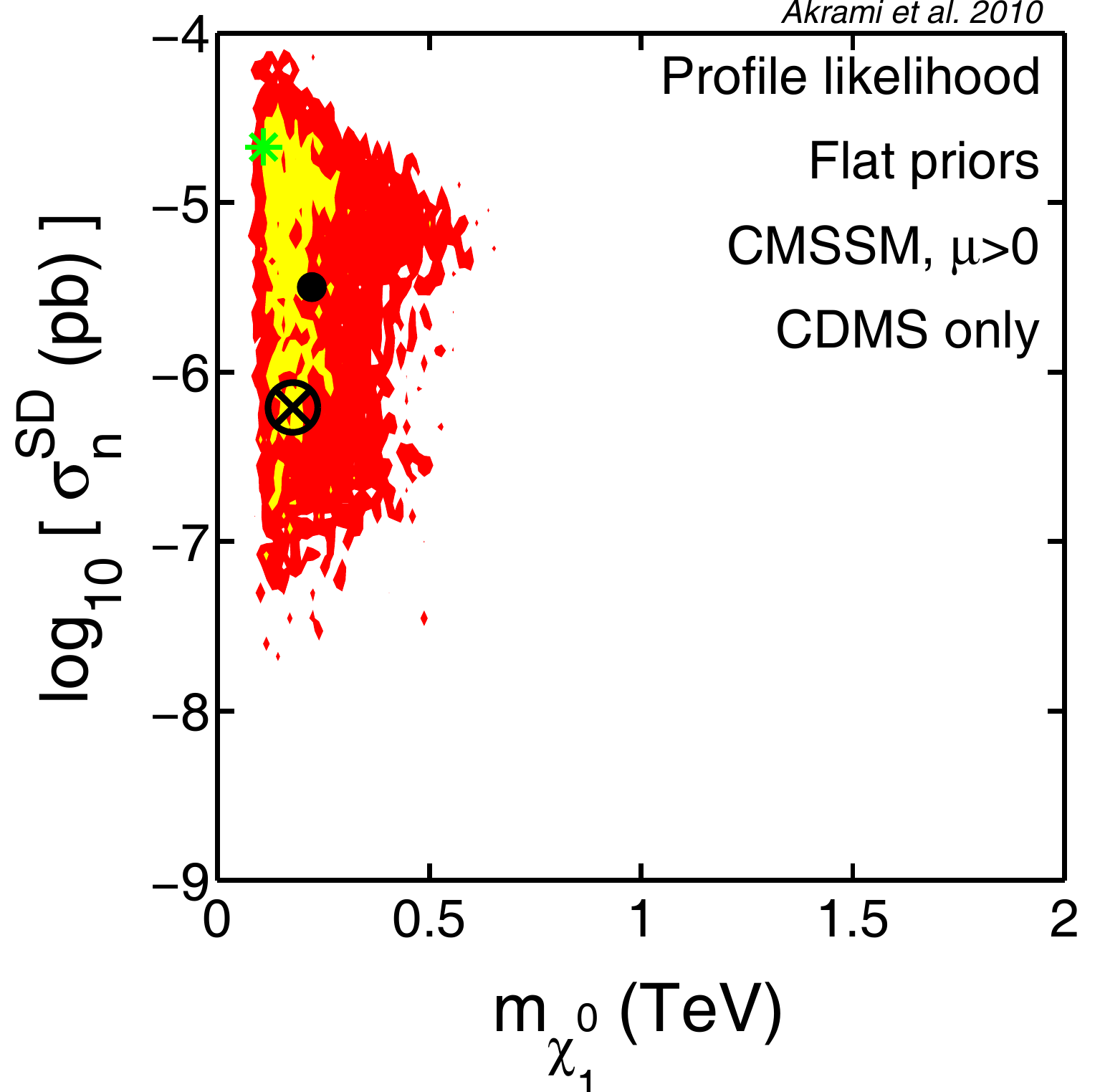}}
\subfigure{\includegraphics[scale=0.23, trim = 40 230 130 123, clip=true]{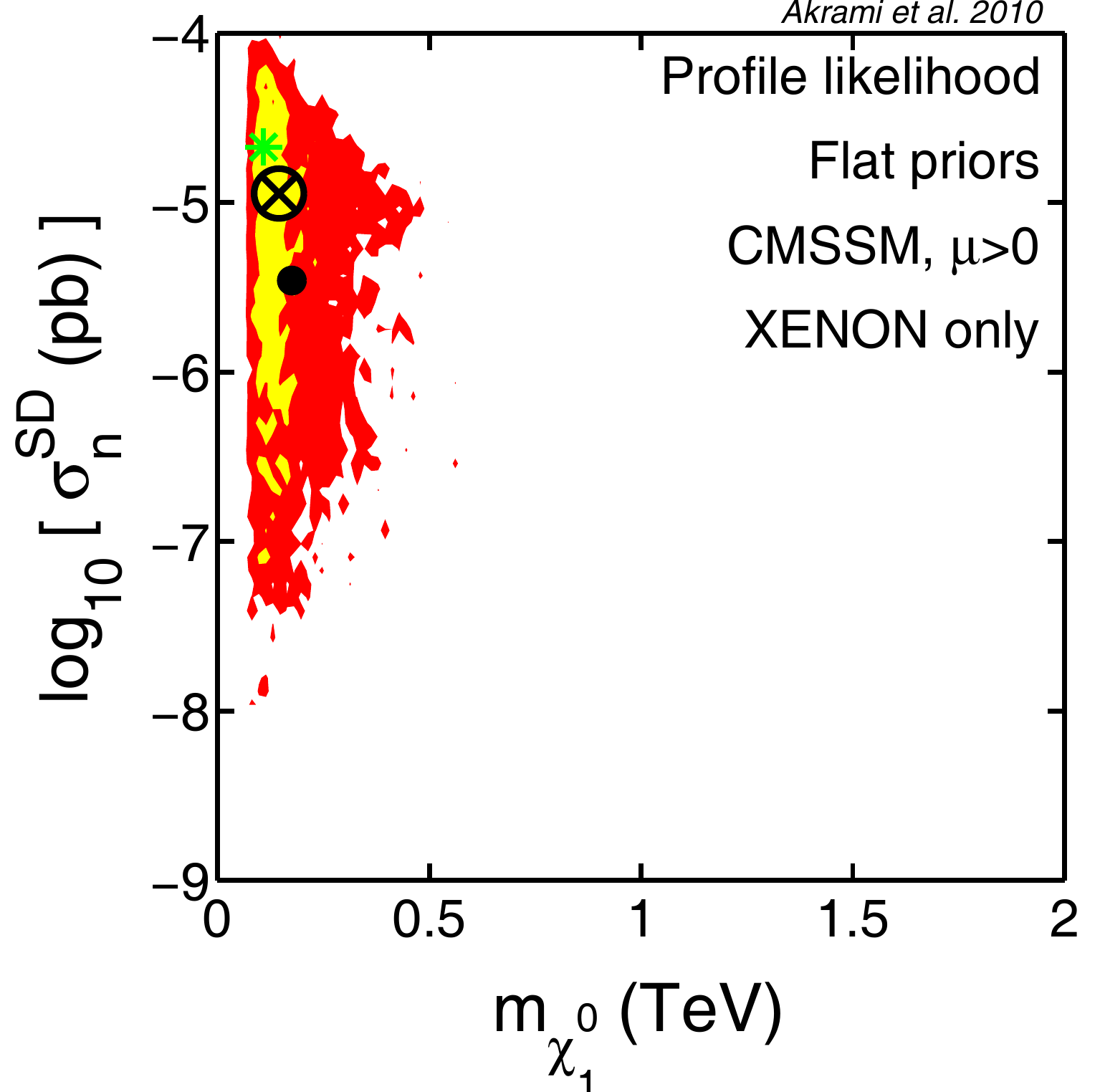}}
\subfigure{\includegraphics[scale=0.23, trim = 40 230 130 123, clip=true]{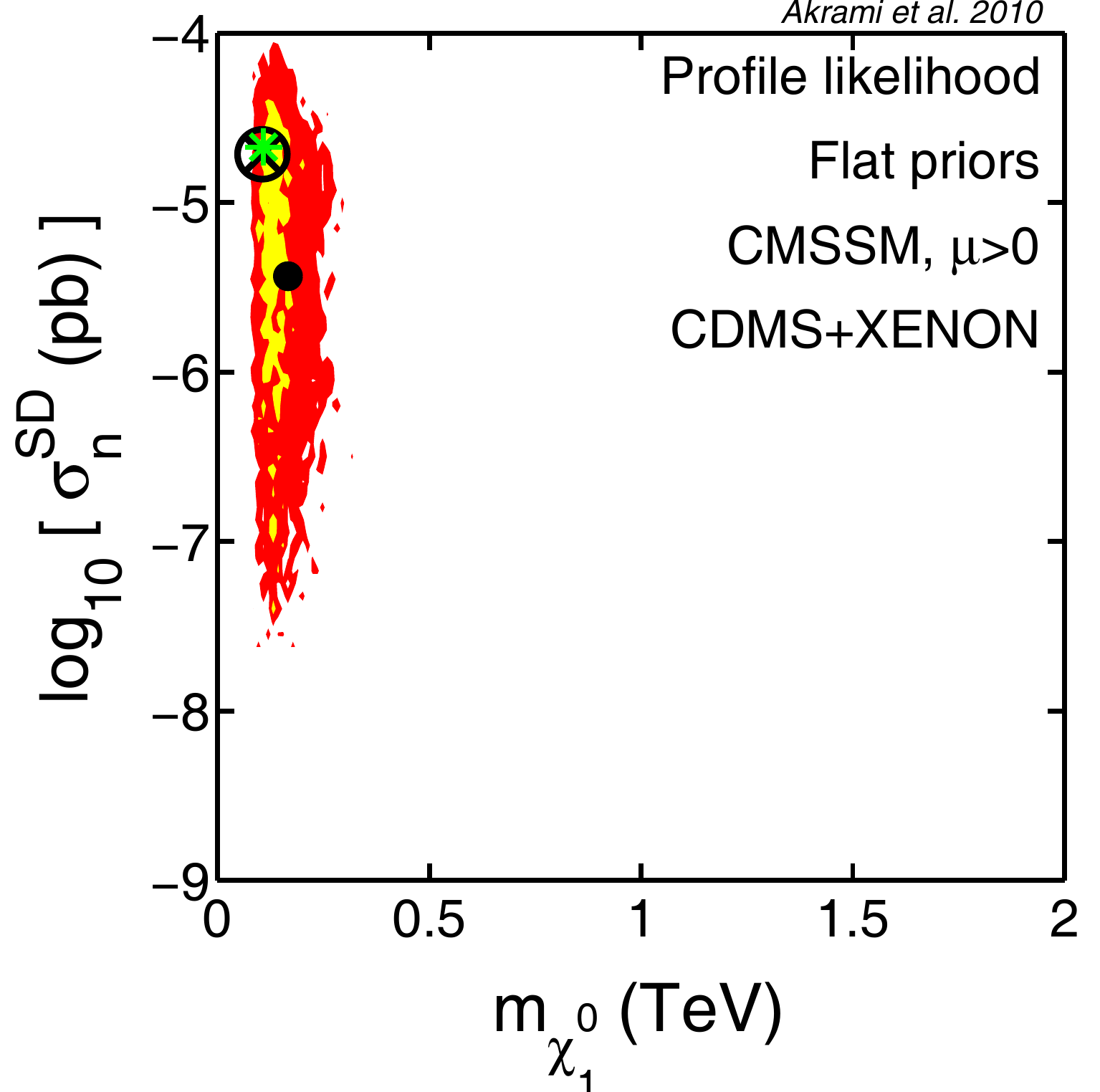}}
\subfigure{\includegraphics[scale=0.23, trim = 40 230 60 123, clip=true]{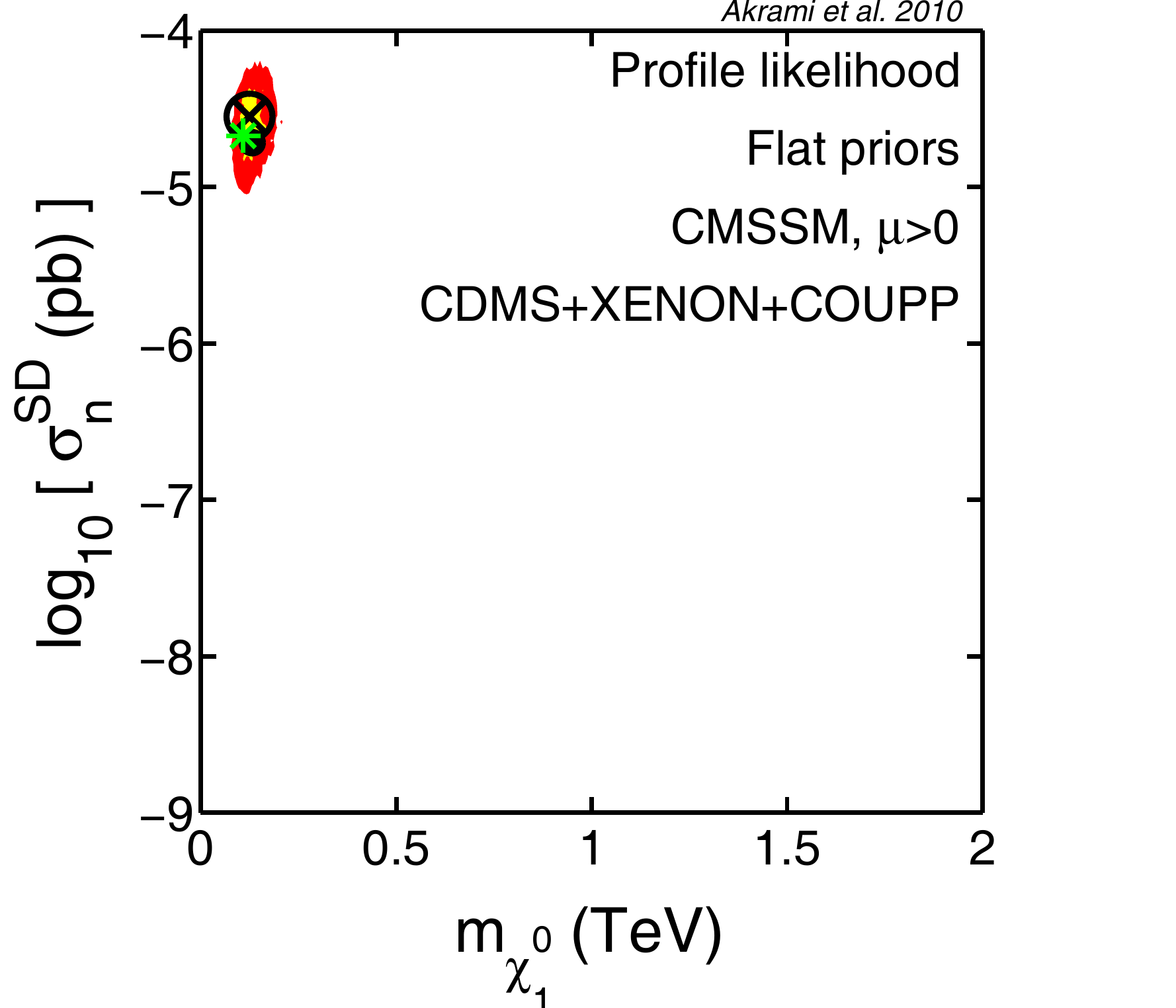}}\\
\caption[aa]{\footnotesize{As in~\fig{fig:LHmarg}, but for two-dimensional profile likelihoods.}}\label{fig:LHprofl}
\end{figure}

\begin{figure}[t]
\subfigure{\includegraphics[scale=0.23, trim = 40 230 130 123, clip=true]{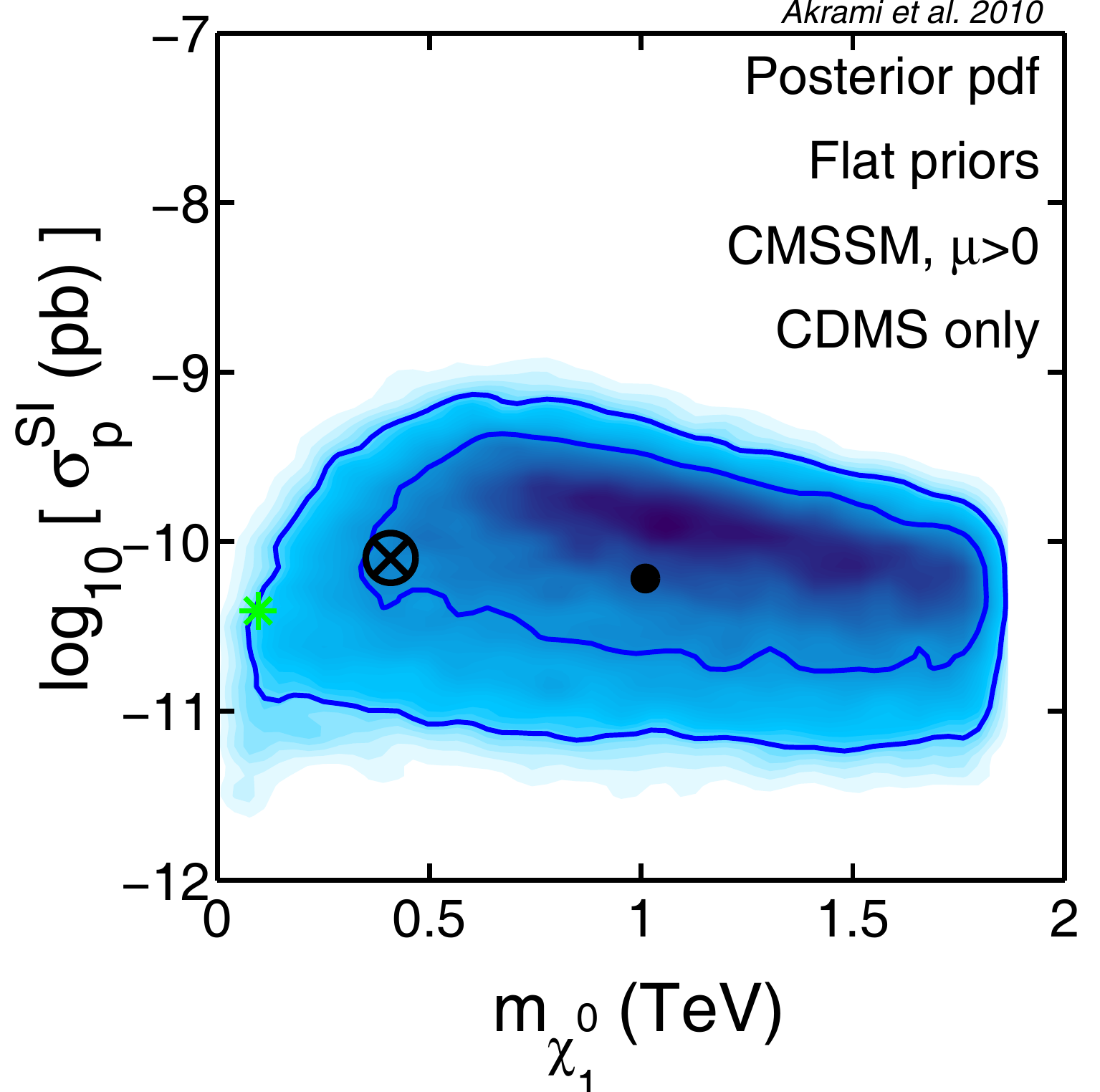}}
\subfigure{\includegraphics[scale=0.23, trim = 40 230 130 123, clip=true]{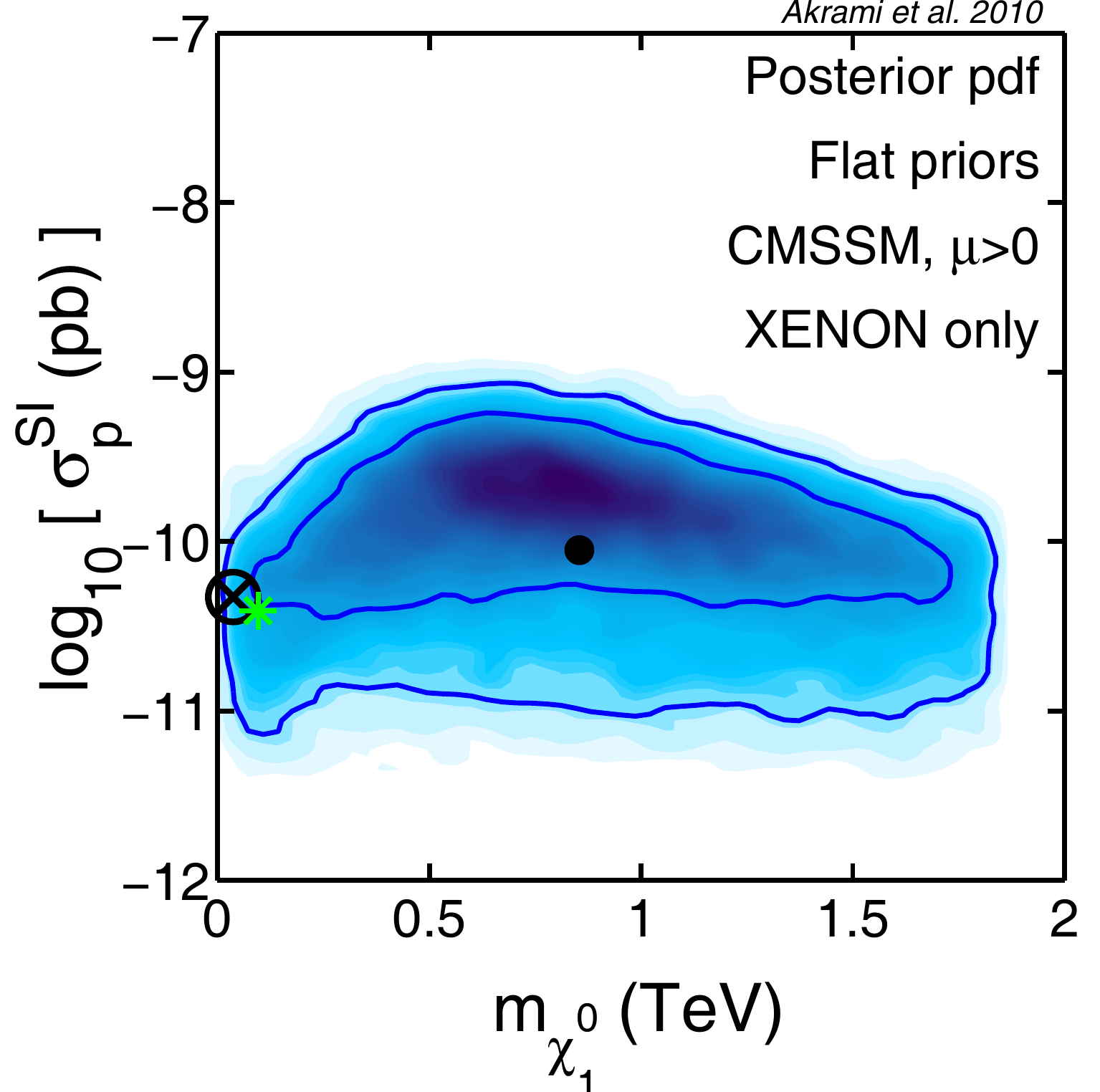}}
\subfigure{\includegraphics[scale=0.23, trim = 40 230 130 123, clip=true]{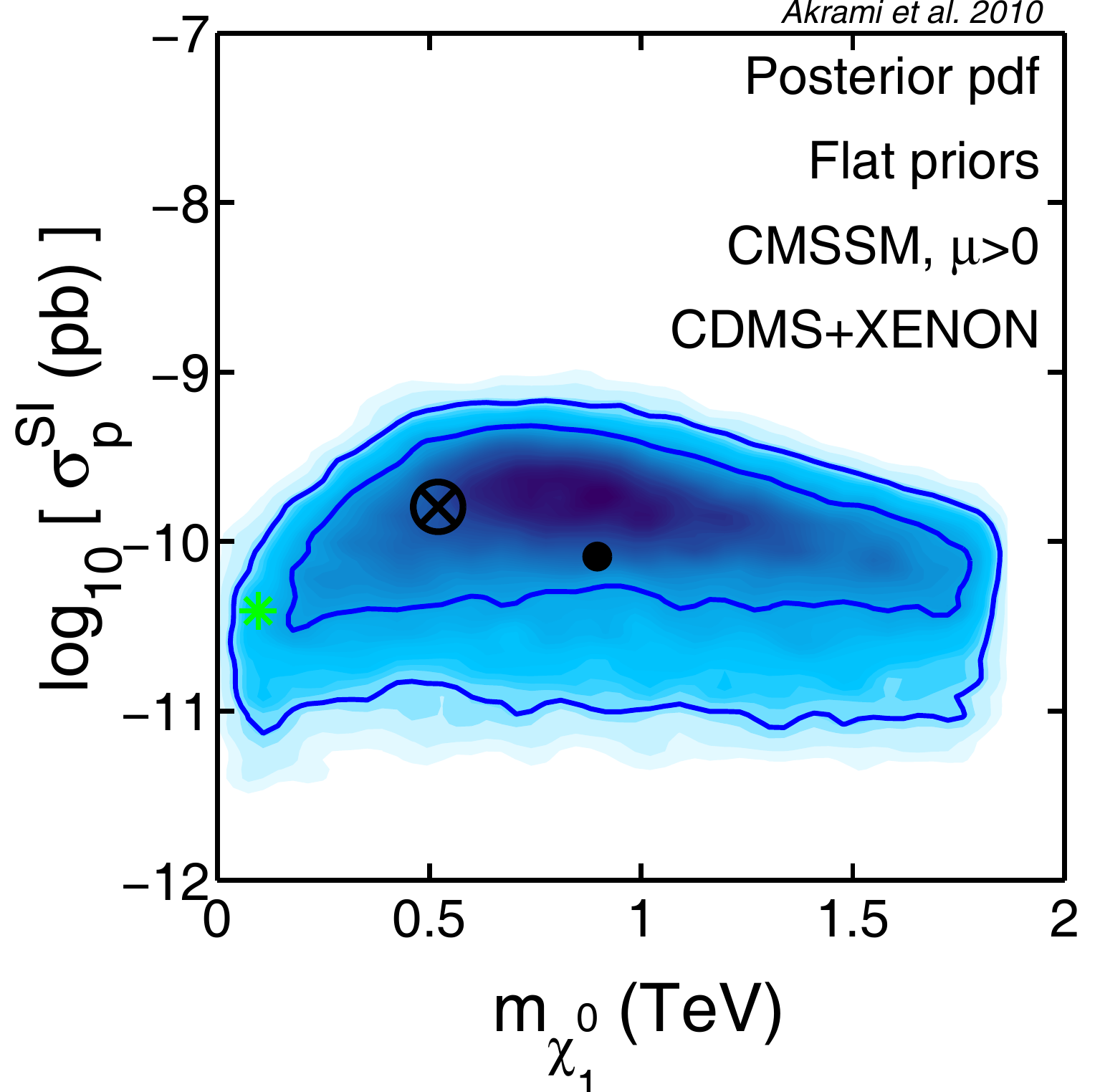}}
\subfigure{\includegraphics[scale=0.23, trim = 40 230 60 123, clip=true]{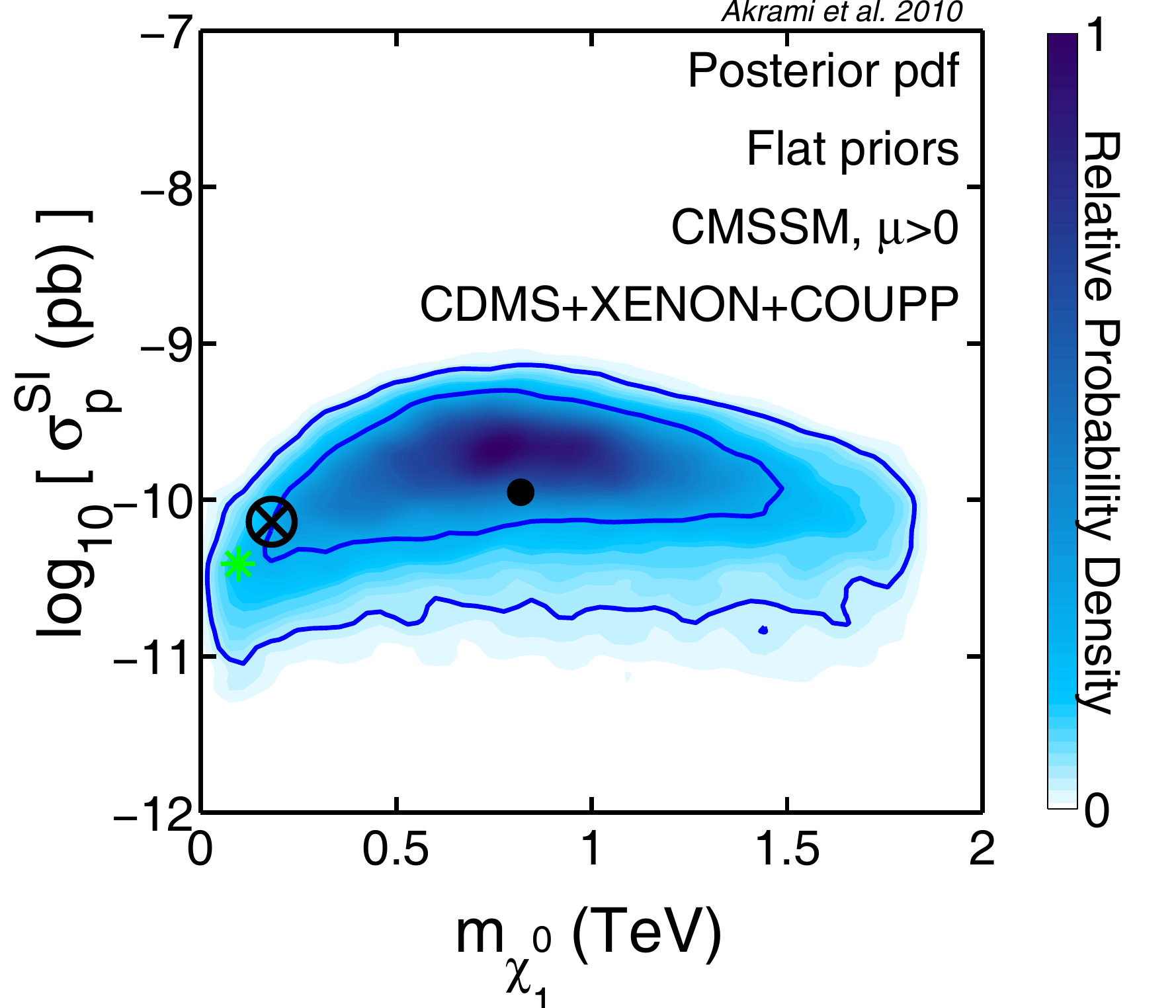}}\\
\subfigure{\includegraphics[scale=0.23, trim = 40 230 130 123, clip=true]{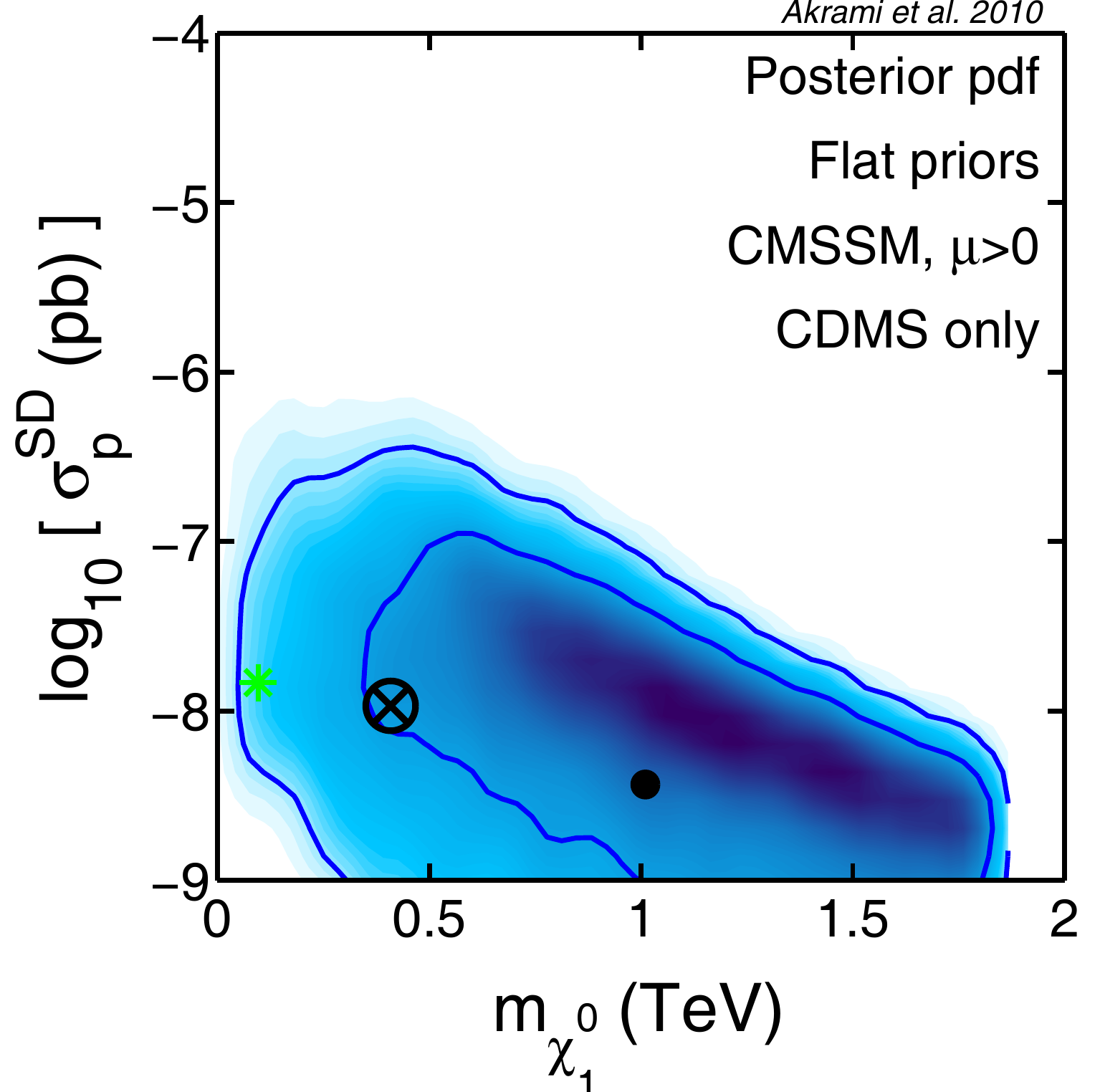}}
\subfigure{\includegraphics[scale=0.23, trim = 40 230 130 123, clip=true]{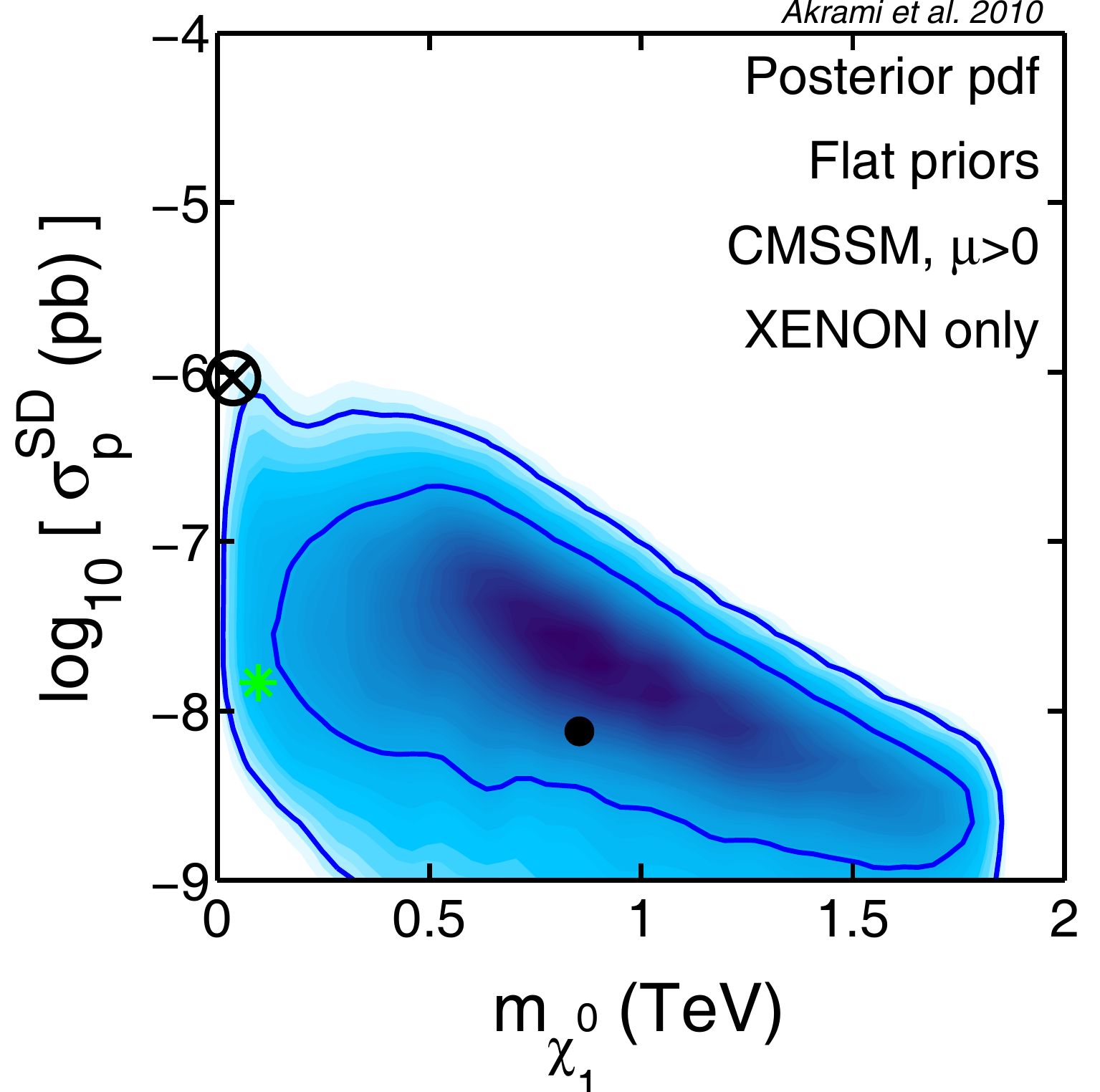}}
\subfigure{\includegraphics[scale=0.23, trim = 40 230 130 123, clip=true]{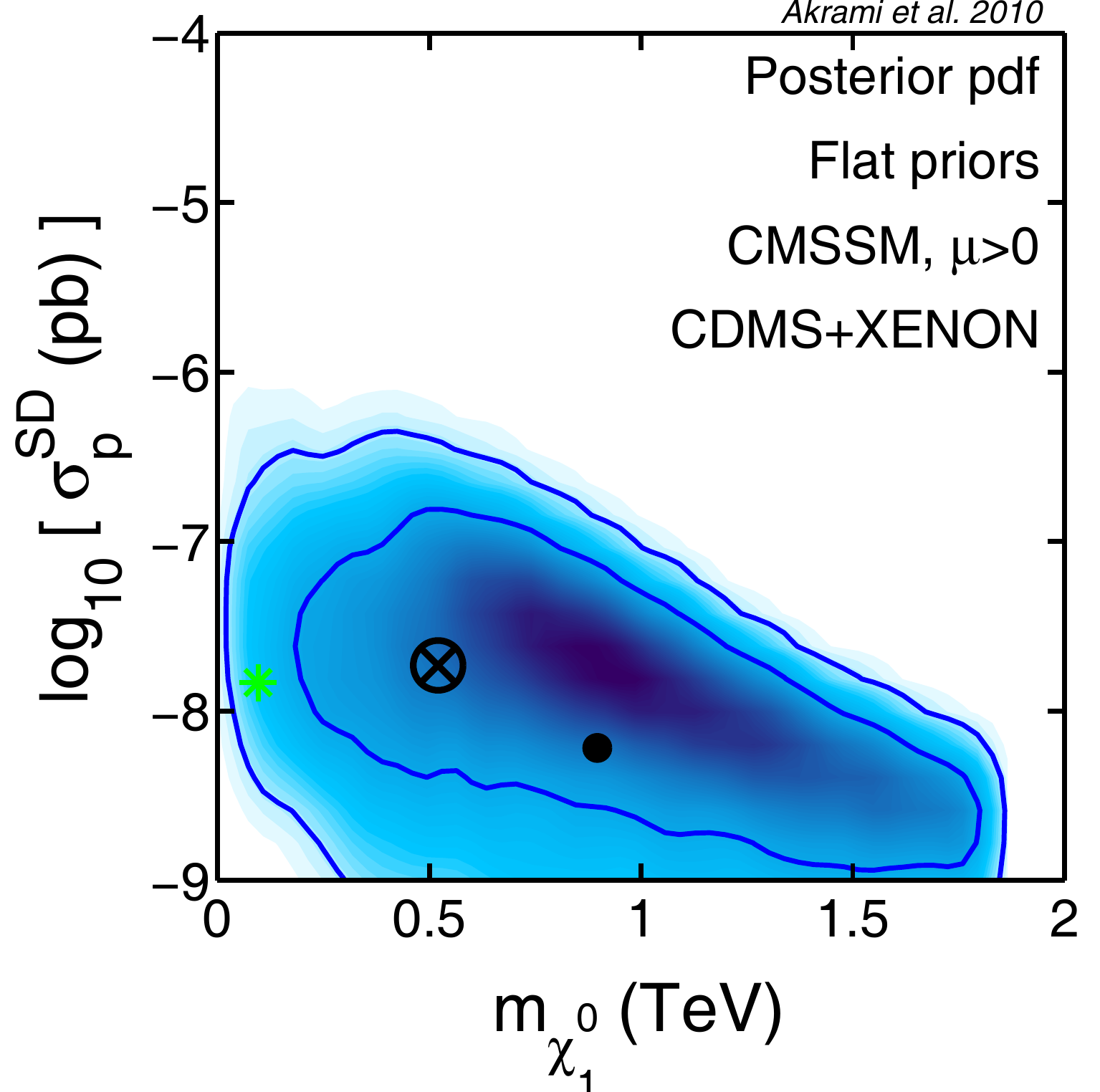}}
\subfigure{\includegraphics[scale=0.23, trim = 40 230 60 123, clip=true]{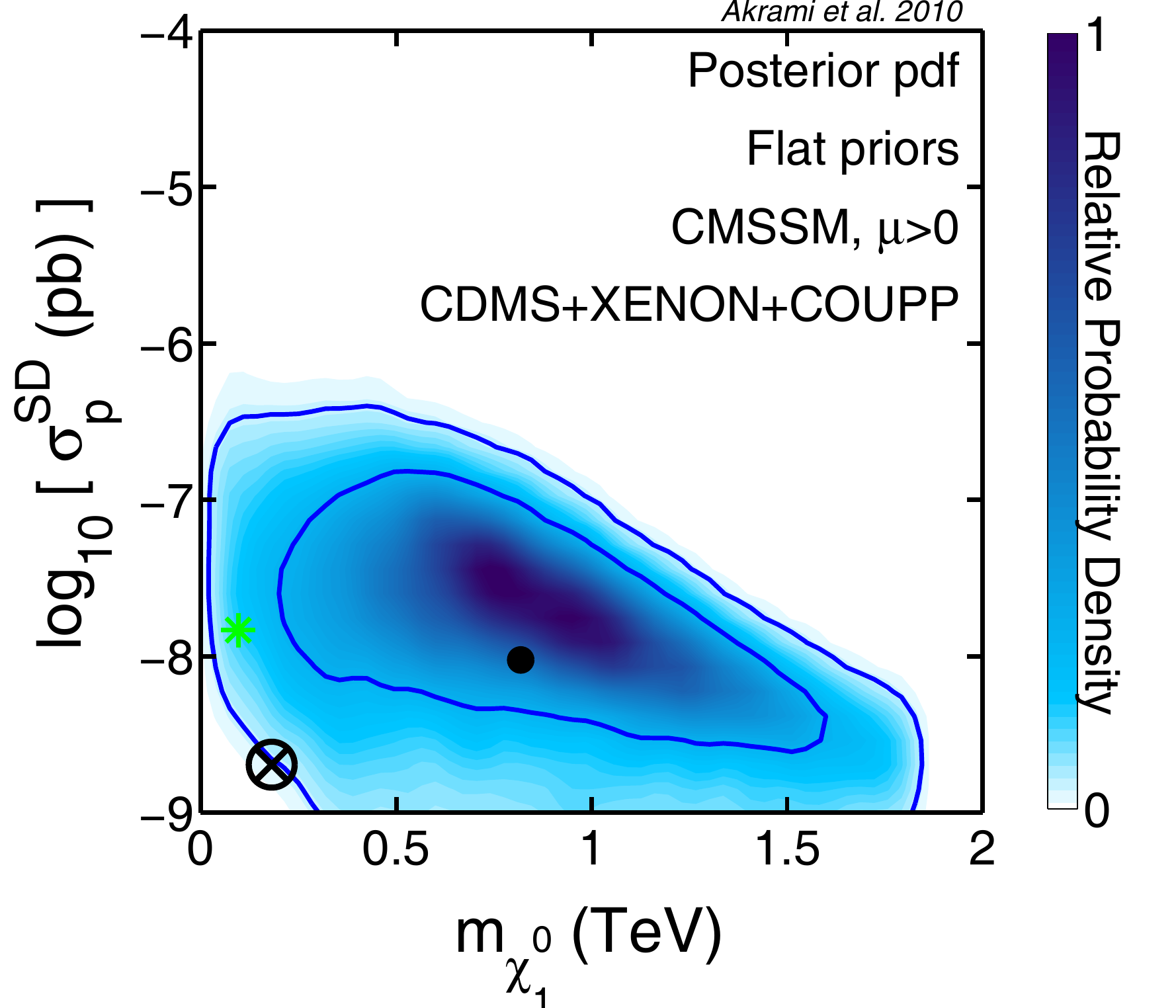}}\\
\subfigure{\includegraphics[scale=0.23, trim = 40 230 130 123, clip=true]{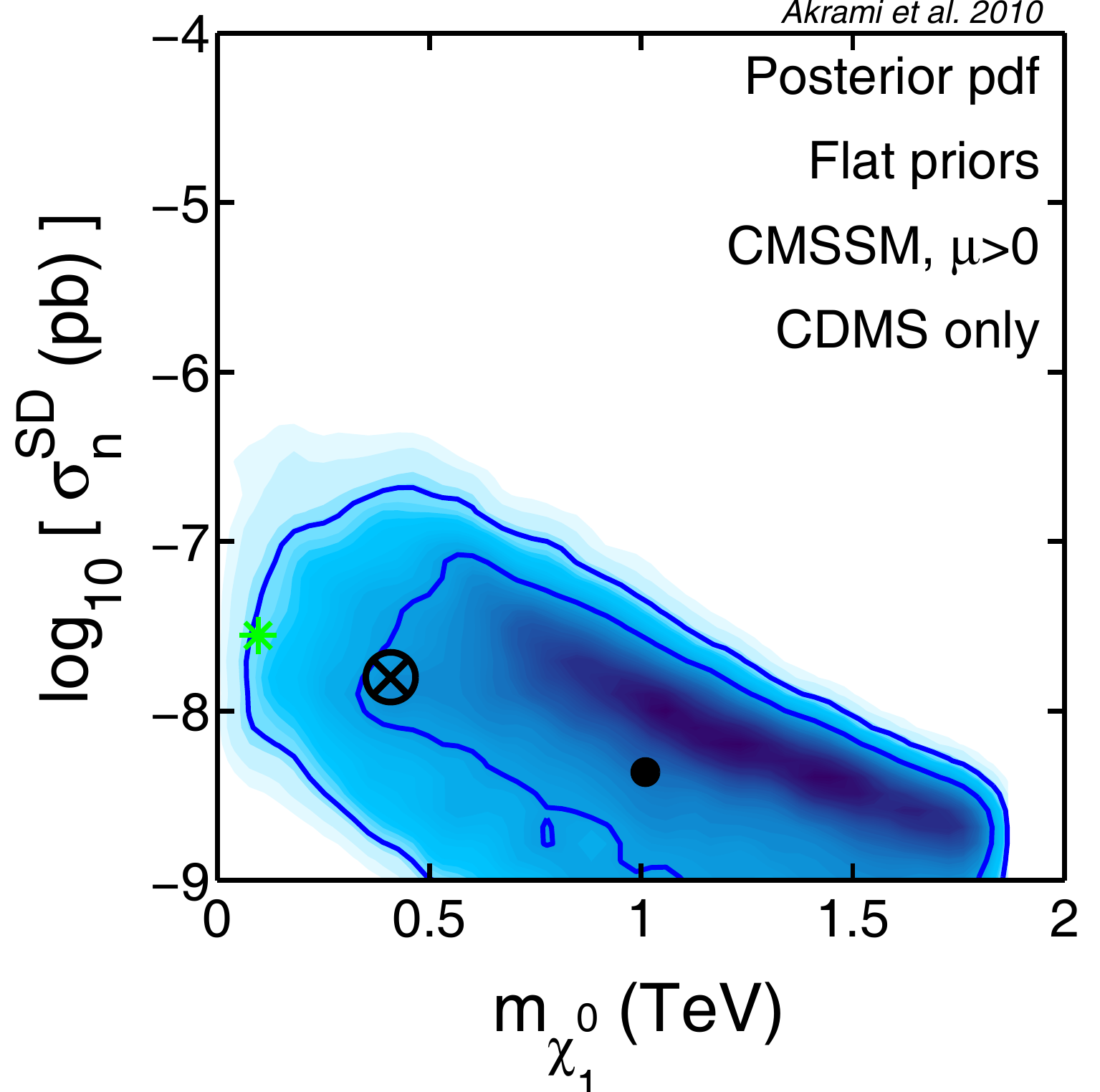}}
\subfigure{\includegraphics[scale=0.23, trim = 40 230 130 123, clip=true]{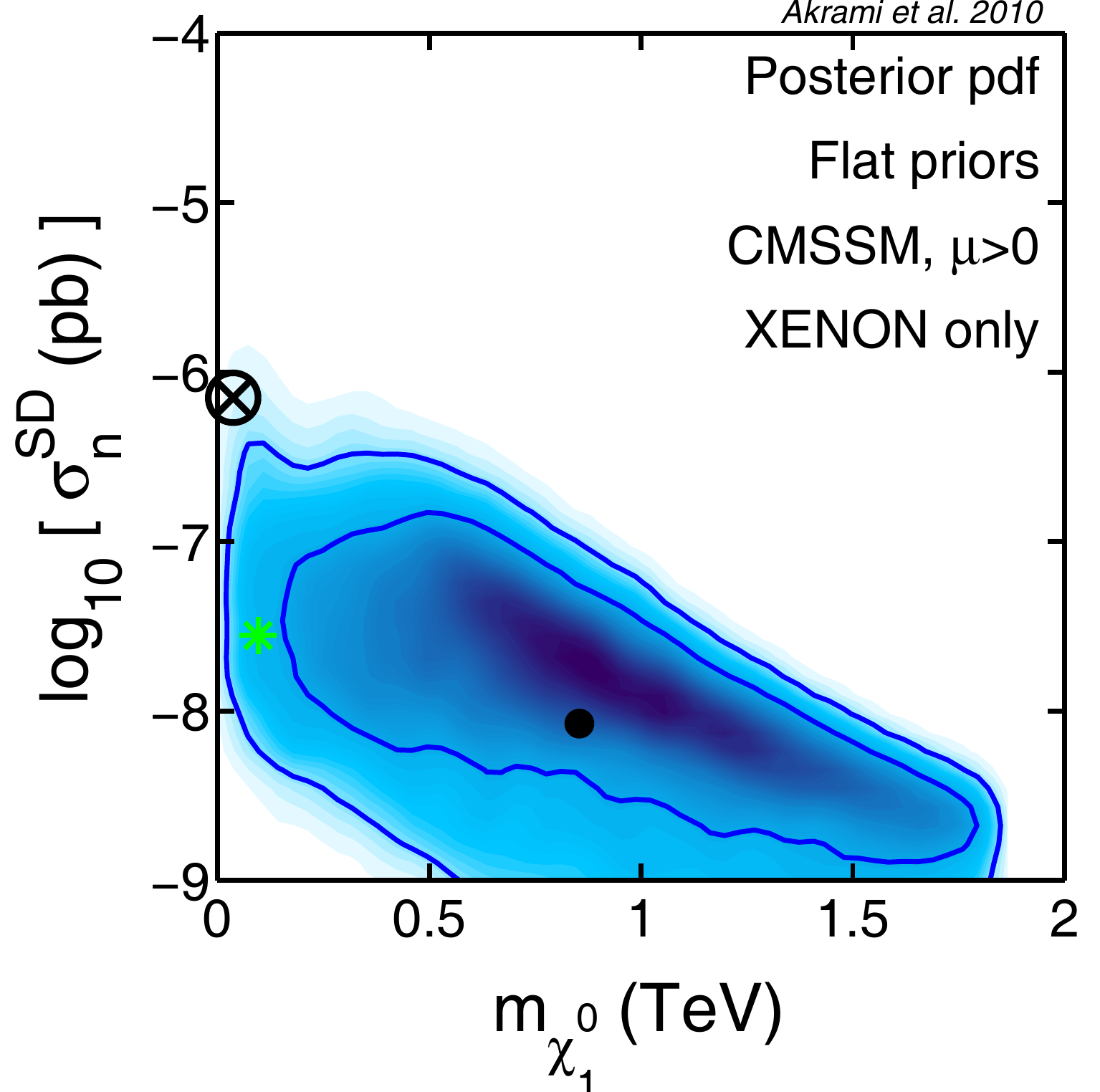}}
\subfigure{\includegraphics[scale=0.23, trim = 40 230 130 123, clip=true]{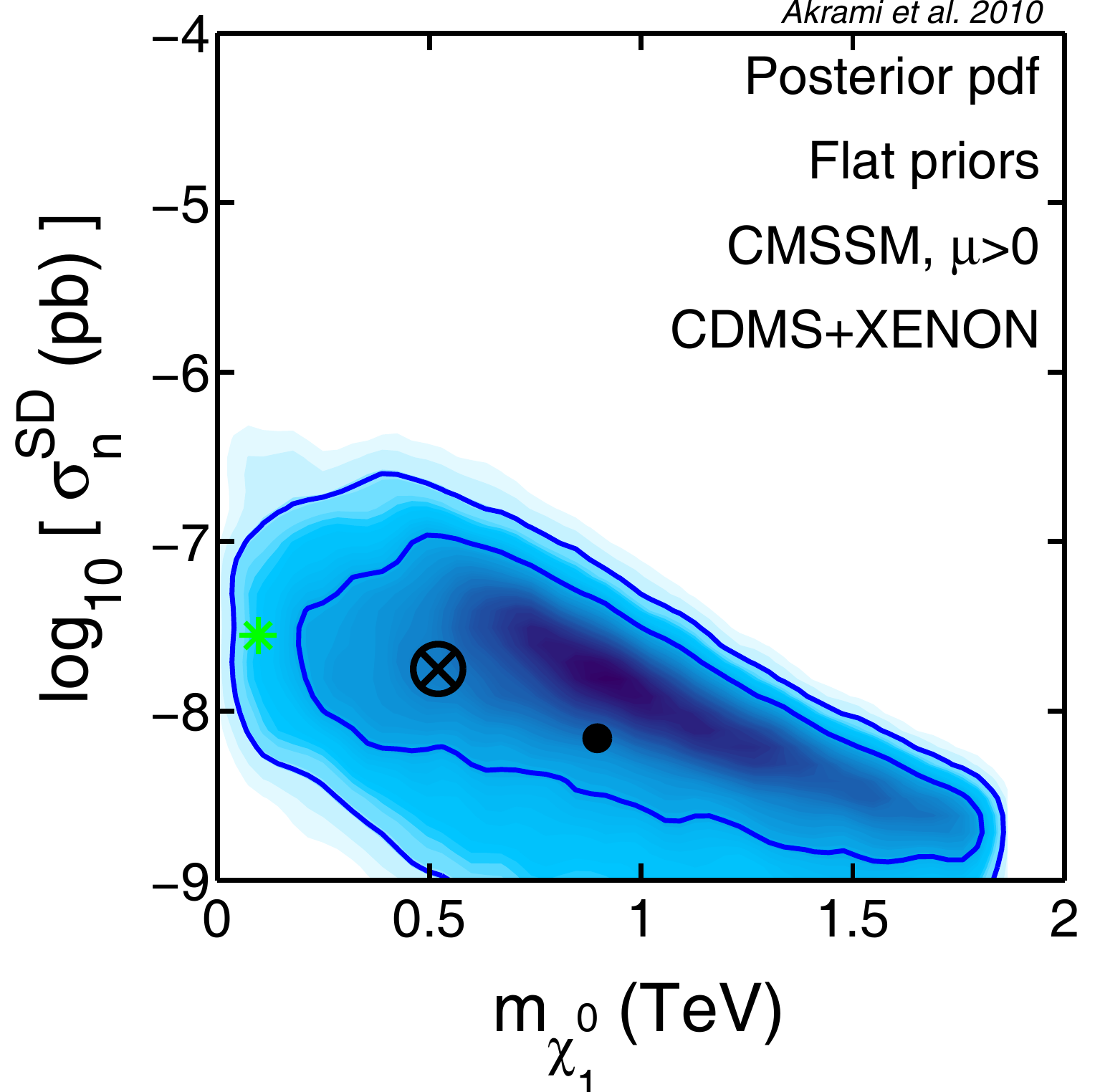}}
\subfigure{\includegraphics[scale=0.23, trim = 40 230 60 123, clip=true]{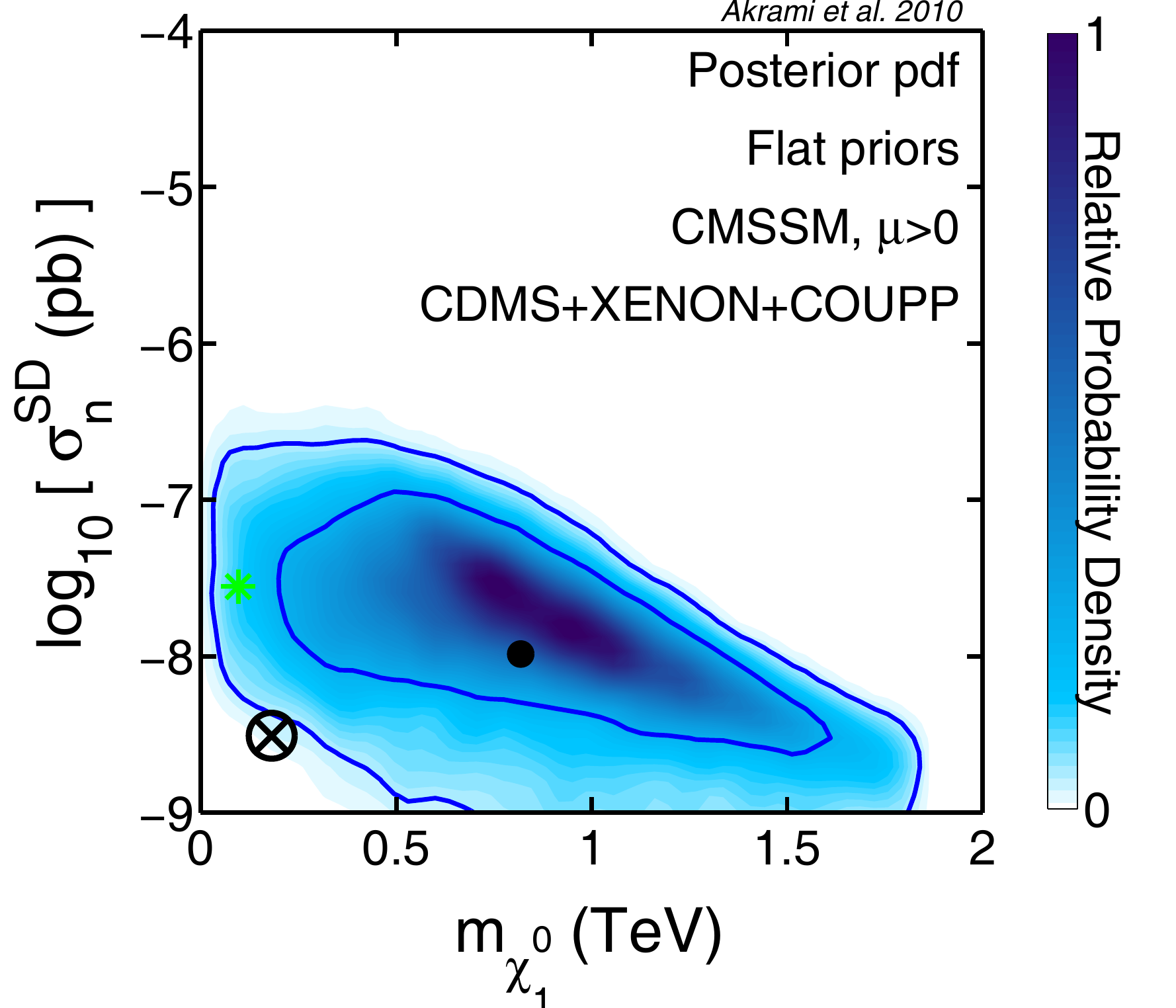}}\\
\caption[aa]{\footnotesize{As in~\fig{fig:LHmarg}, but for benchmark 2.}}\label{fig:LLmarg}
\end{figure}

\begin{figure}[t]
\subfigure{\includegraphics[scale=0.23, trim = 40 230 130 123, clip=true]{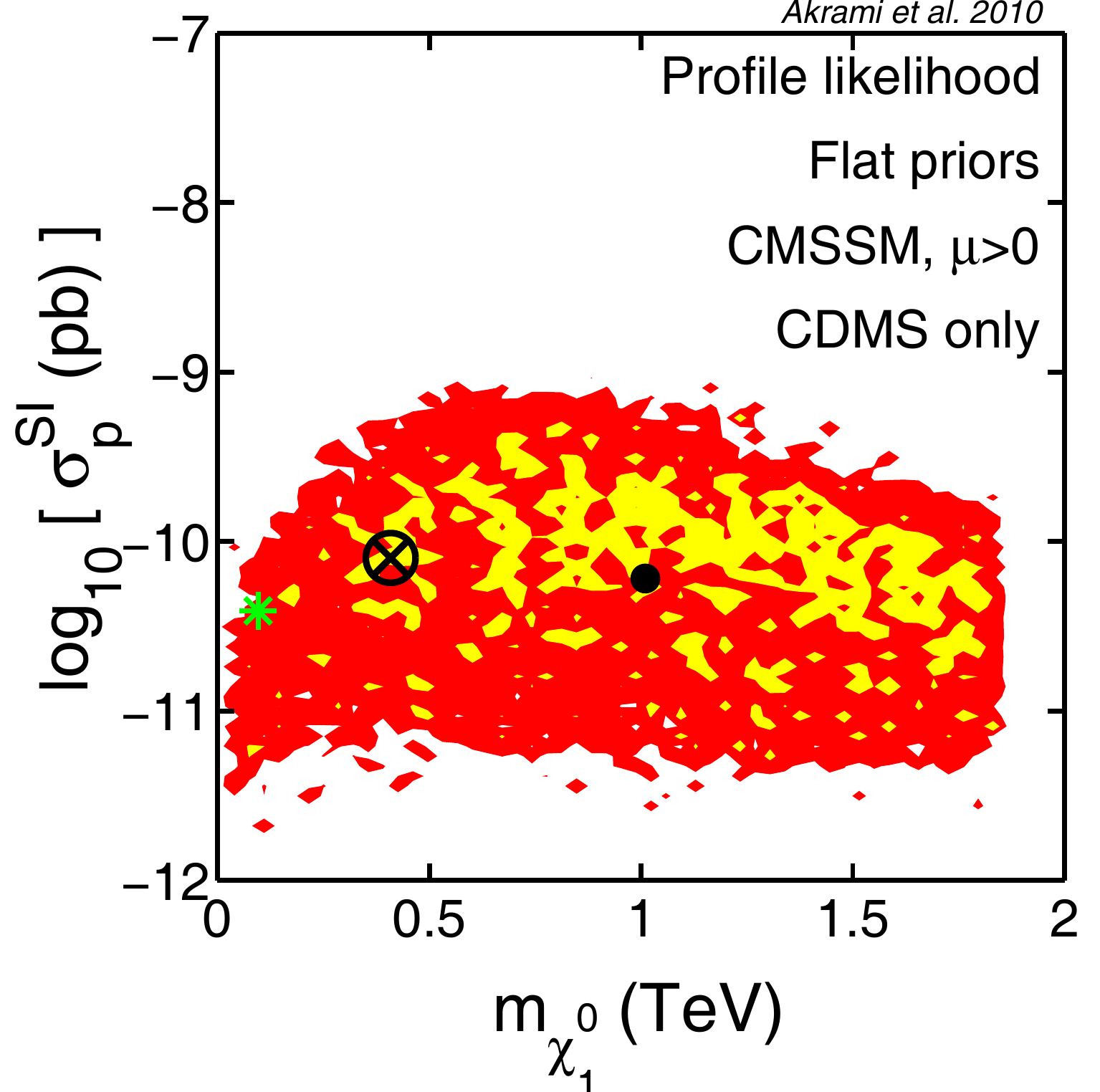}}
\subfigure{\includegraphics[scale=0.23, trim = 40 230 130 123, clip=true]{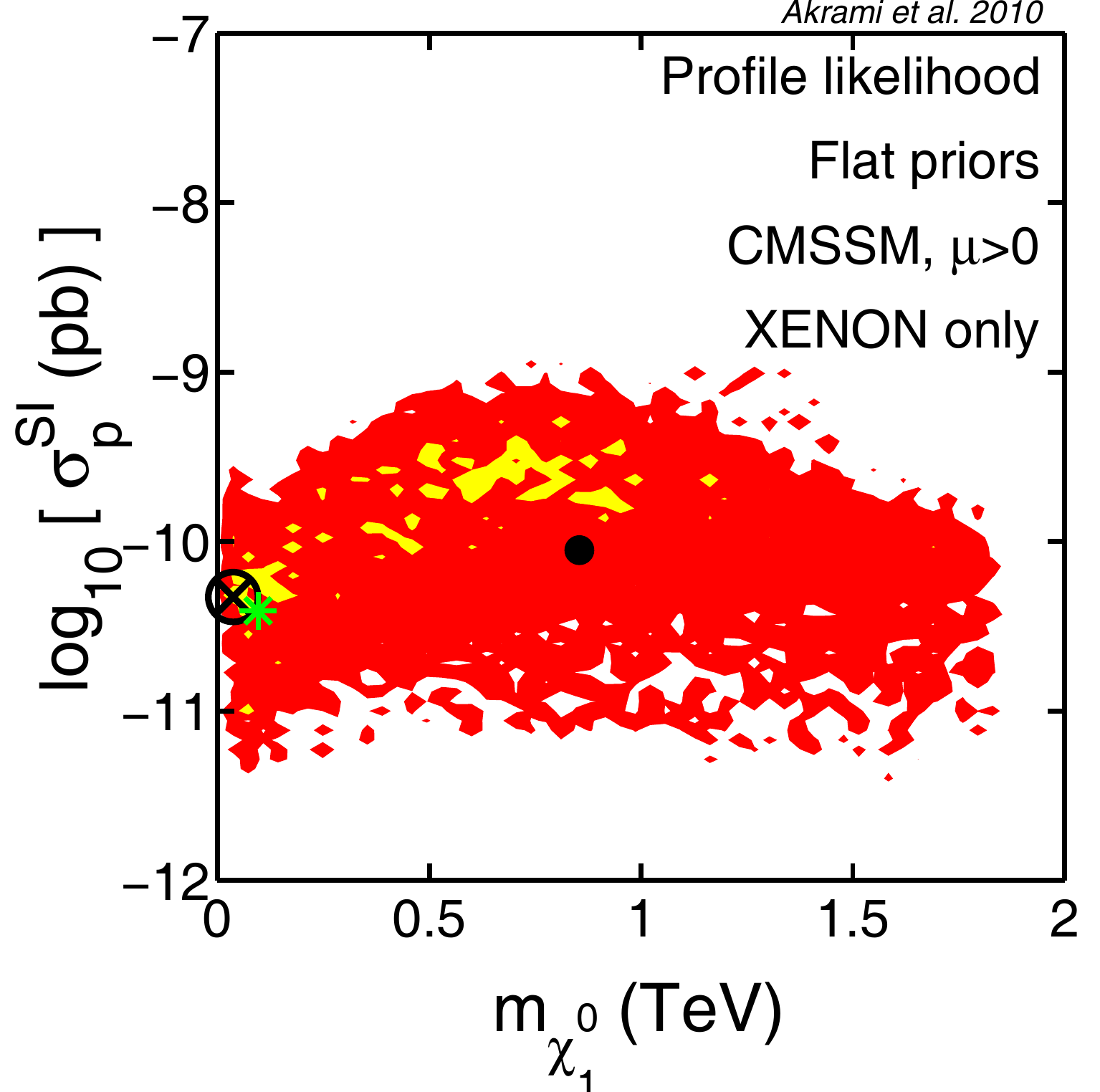}}
\subfigure{\includegraphics[scale=0.23, trim = 40 230 130 123, clip=true]{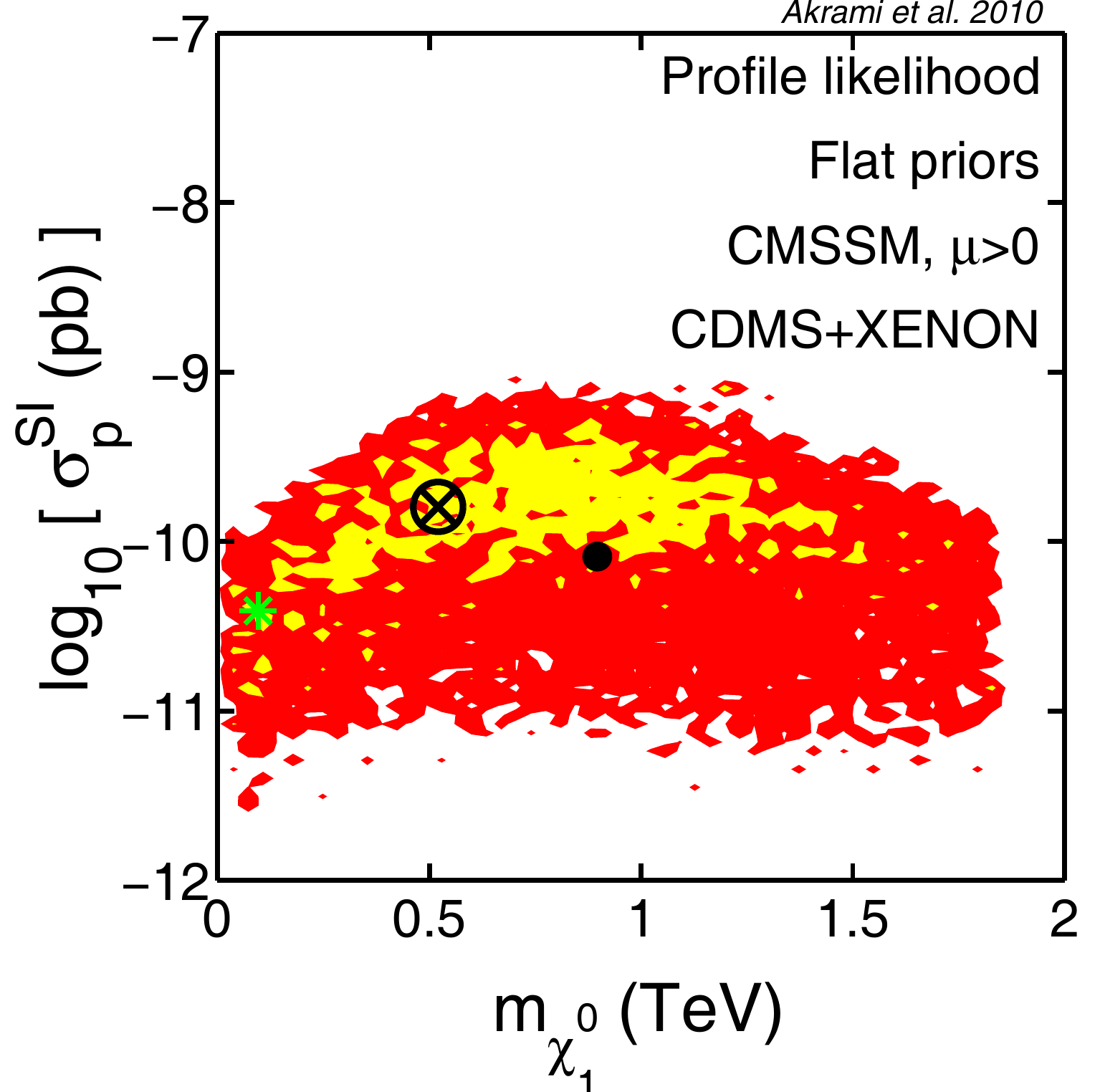}}
\subfigure{\includegraphics[scale=0.23, trim = 40 230 60 123, clip=true]{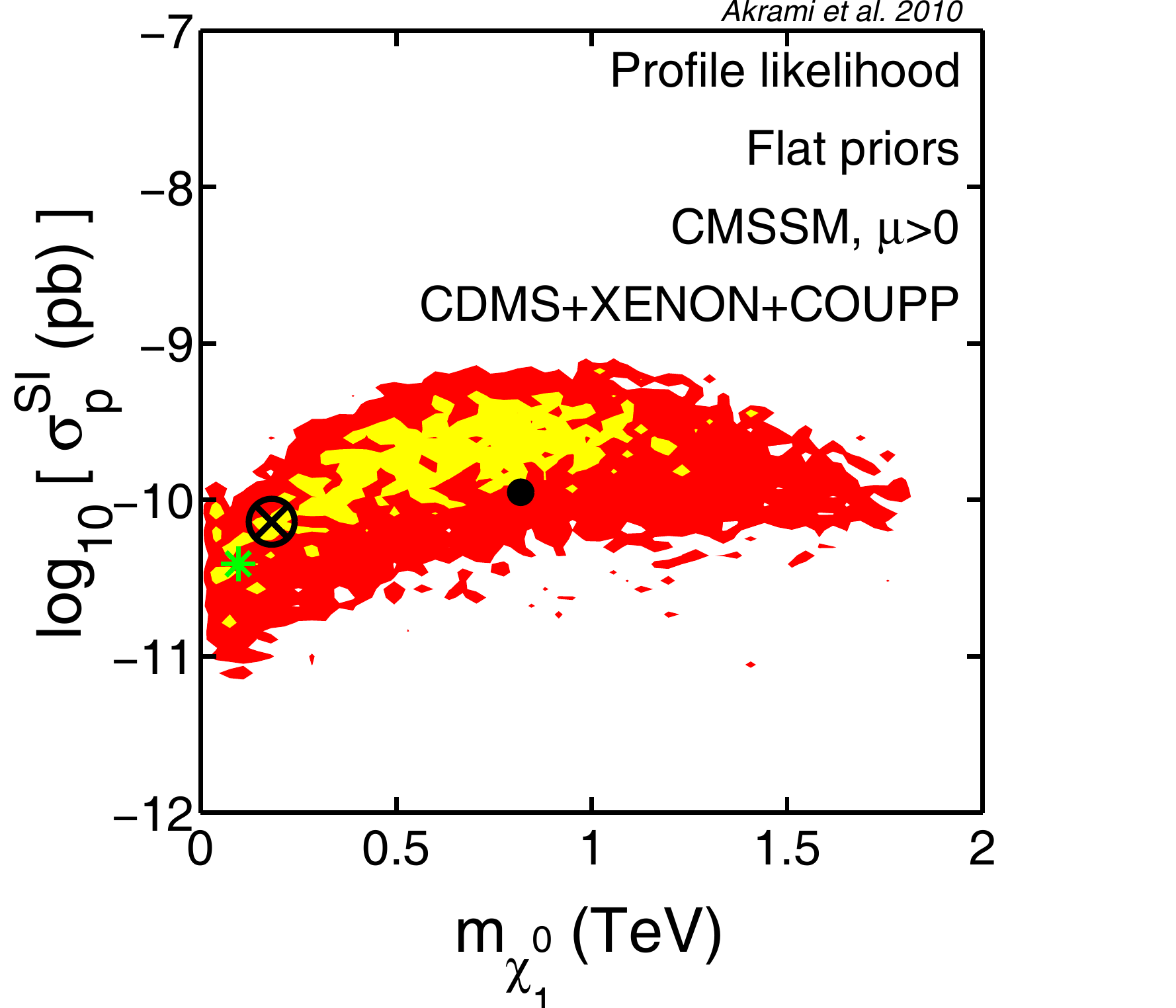}}\\
\subfigure{\includegraphics[scale=0.23, trim = 40 230 130 123, clip=true]{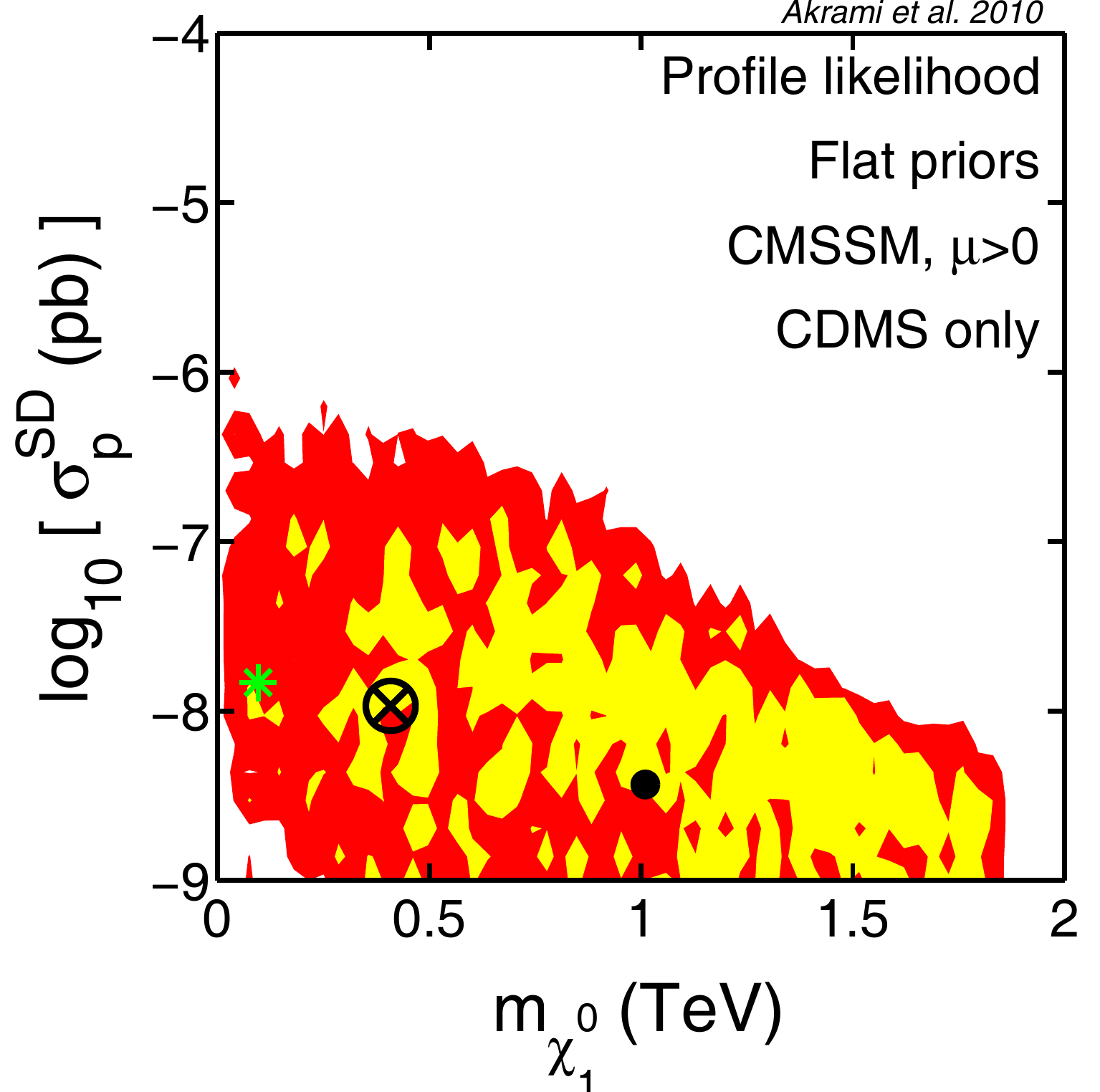}}
\subfigure{\includegraphics[scale=0.23, trim = 40 230 130 123, clip=true]{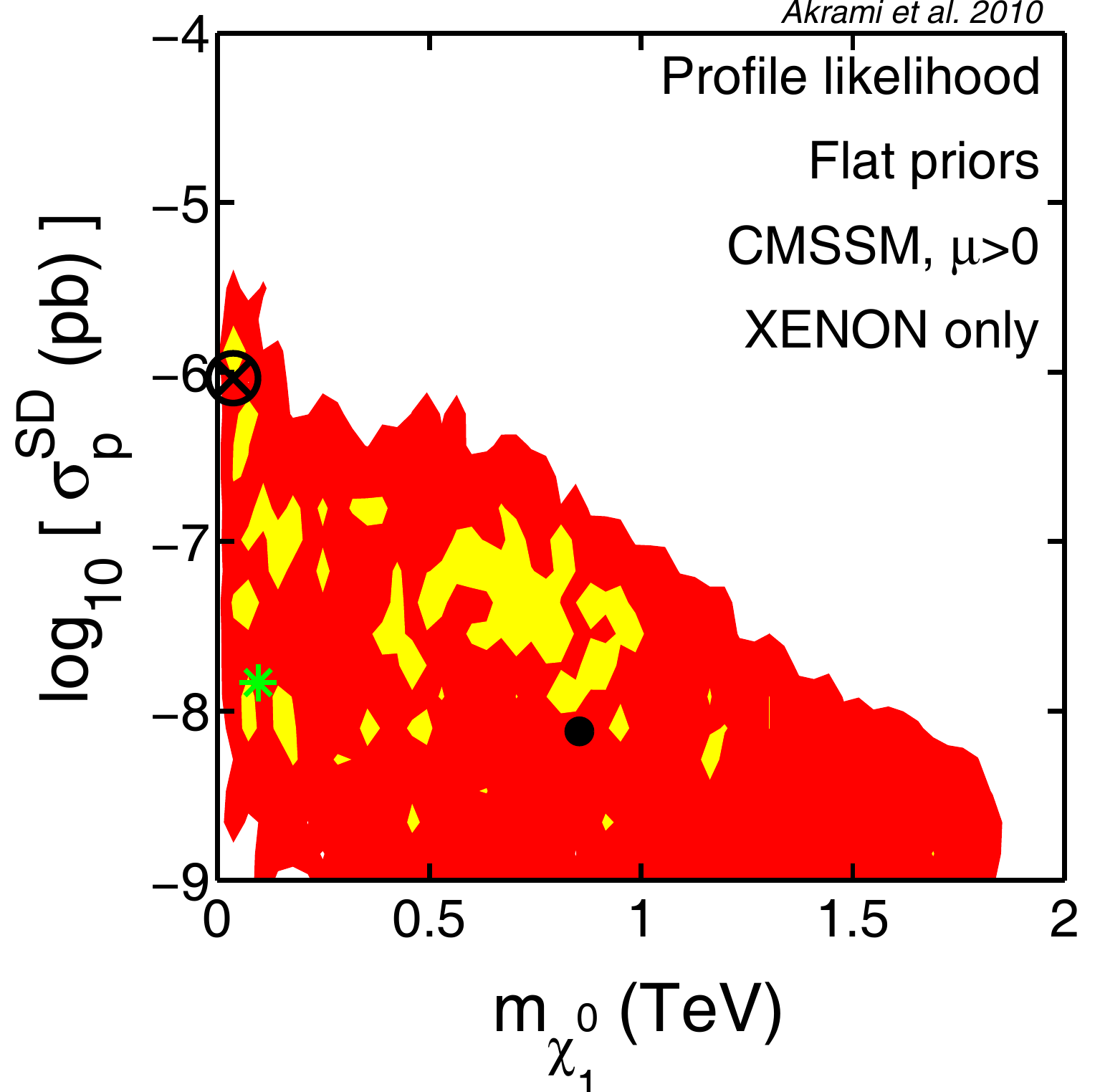}}
\subfigure{\includegraphics[scale=0.23, trim = 40 230 130 123, clip=true]{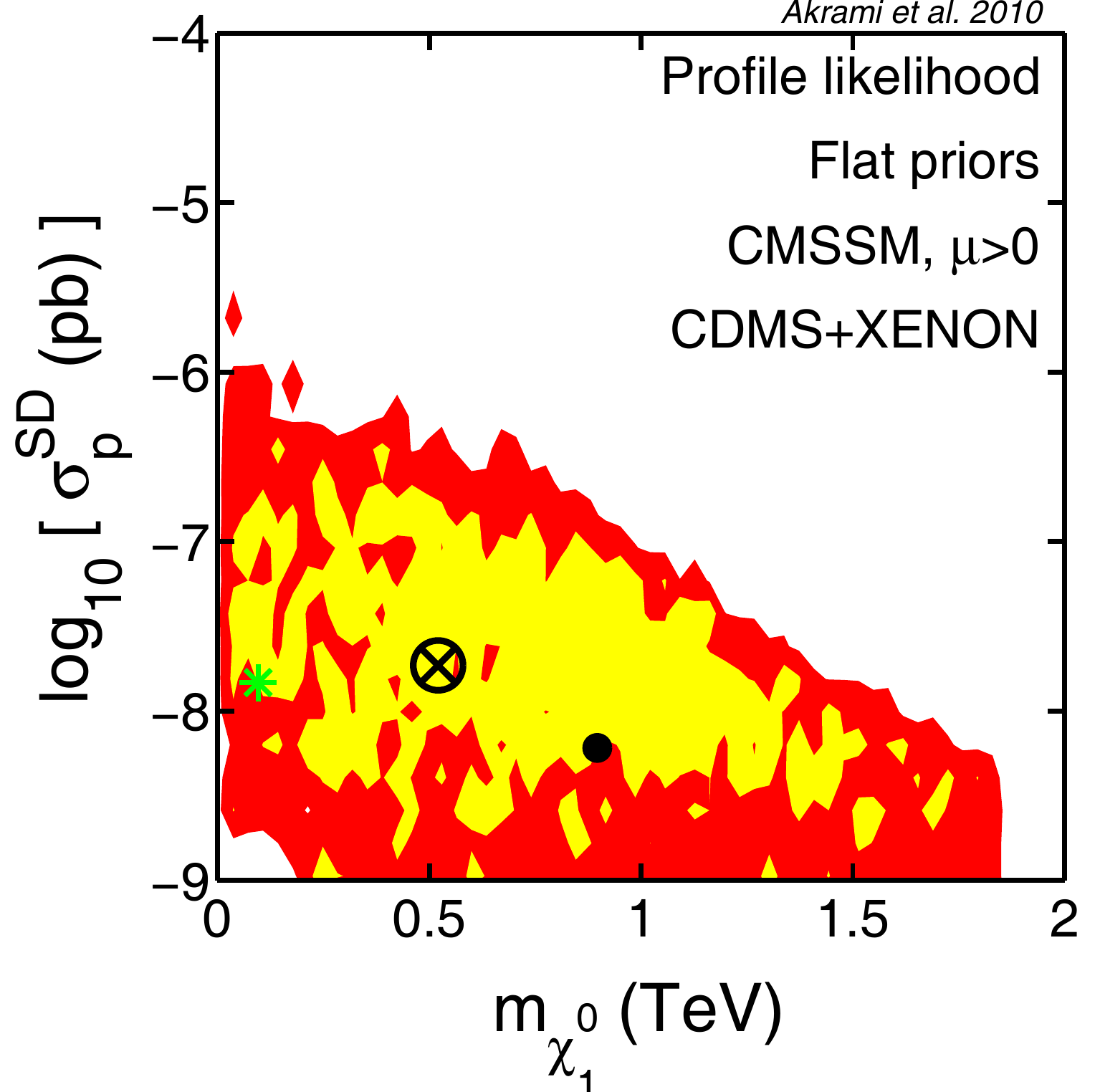}}
\subfigure{\includegraphics[scale=0.23, trim = 40 230 60 123, clip=true]{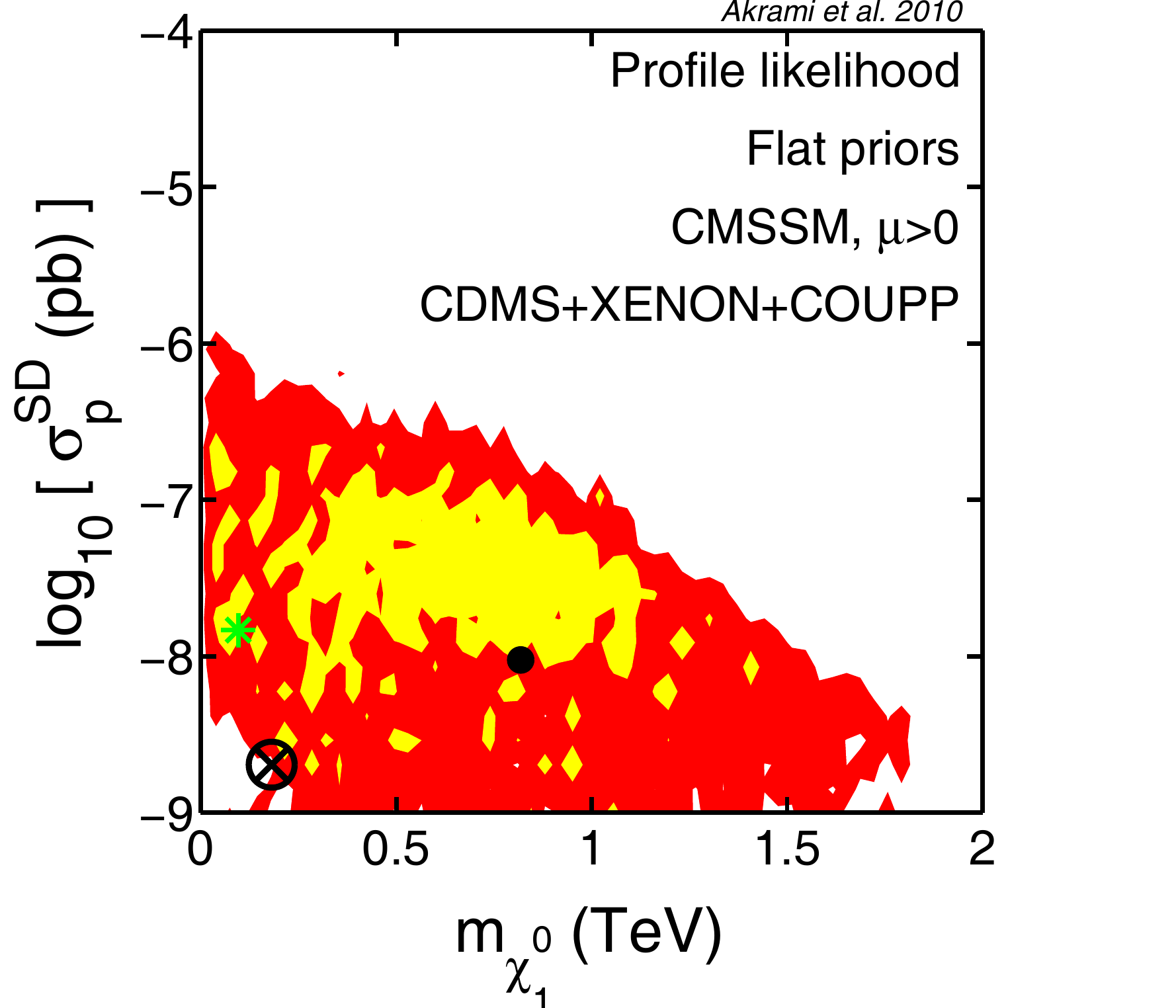}}\\
\subfigure{\includegraphics[scale=0.23, trim = 40 230 130 123, clip=true]{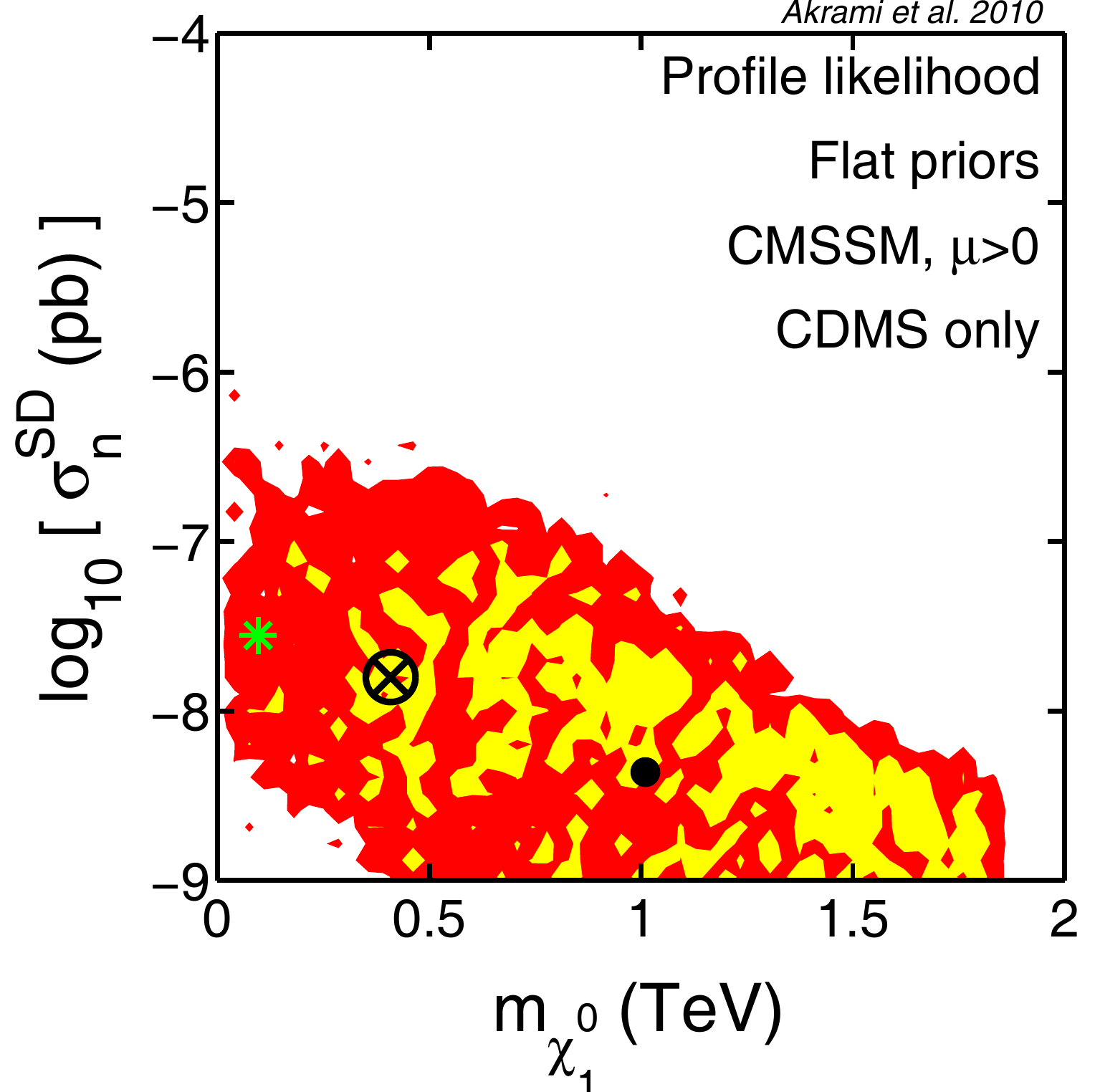}}
\subfigure{\includegraphics[scale=0.23, trim = 40 230 130 123, clip=true]{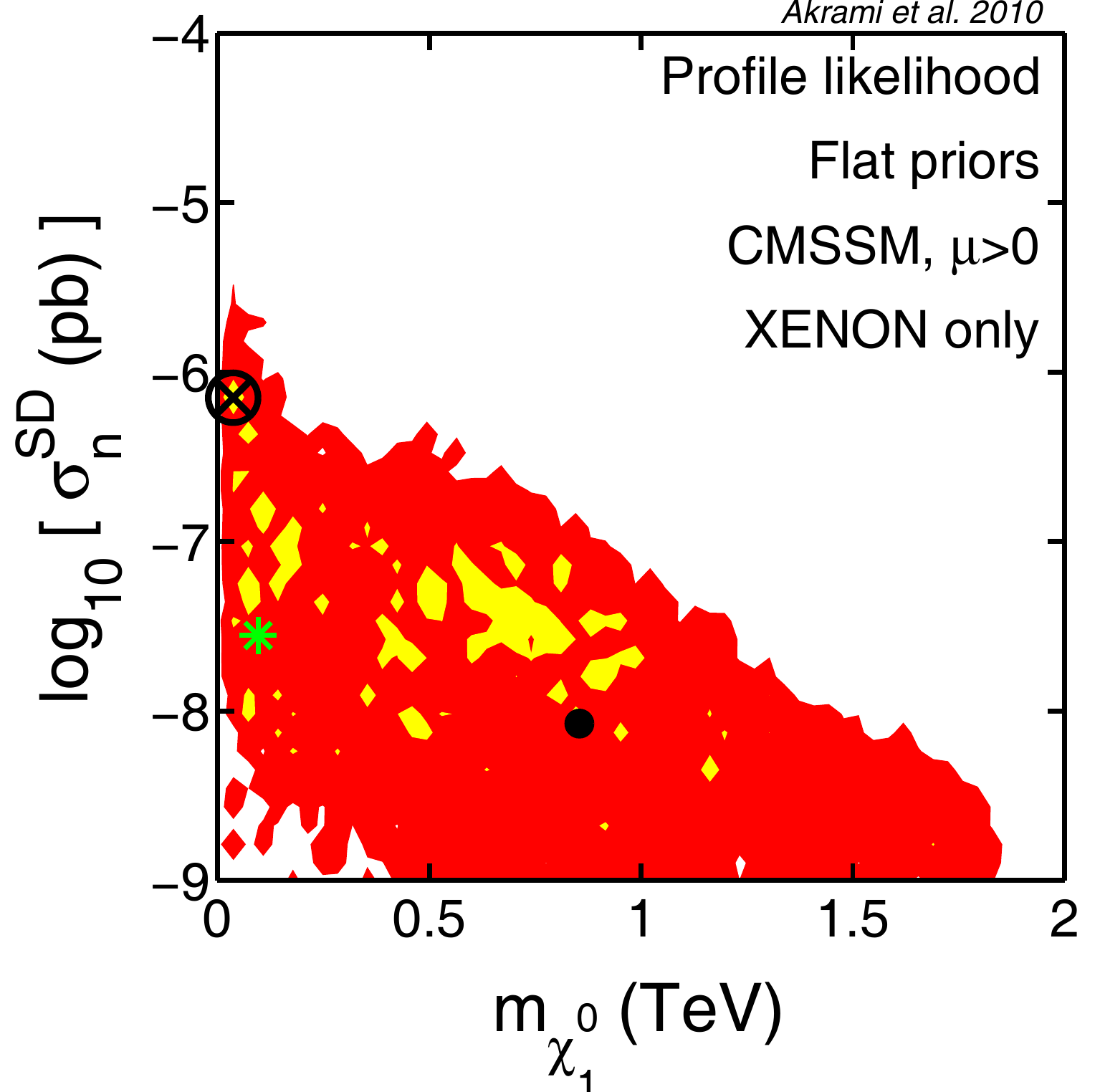}}
\subfigure{\includegraphics[scale=0.23, trim = 40 230 130 123, clip=true]{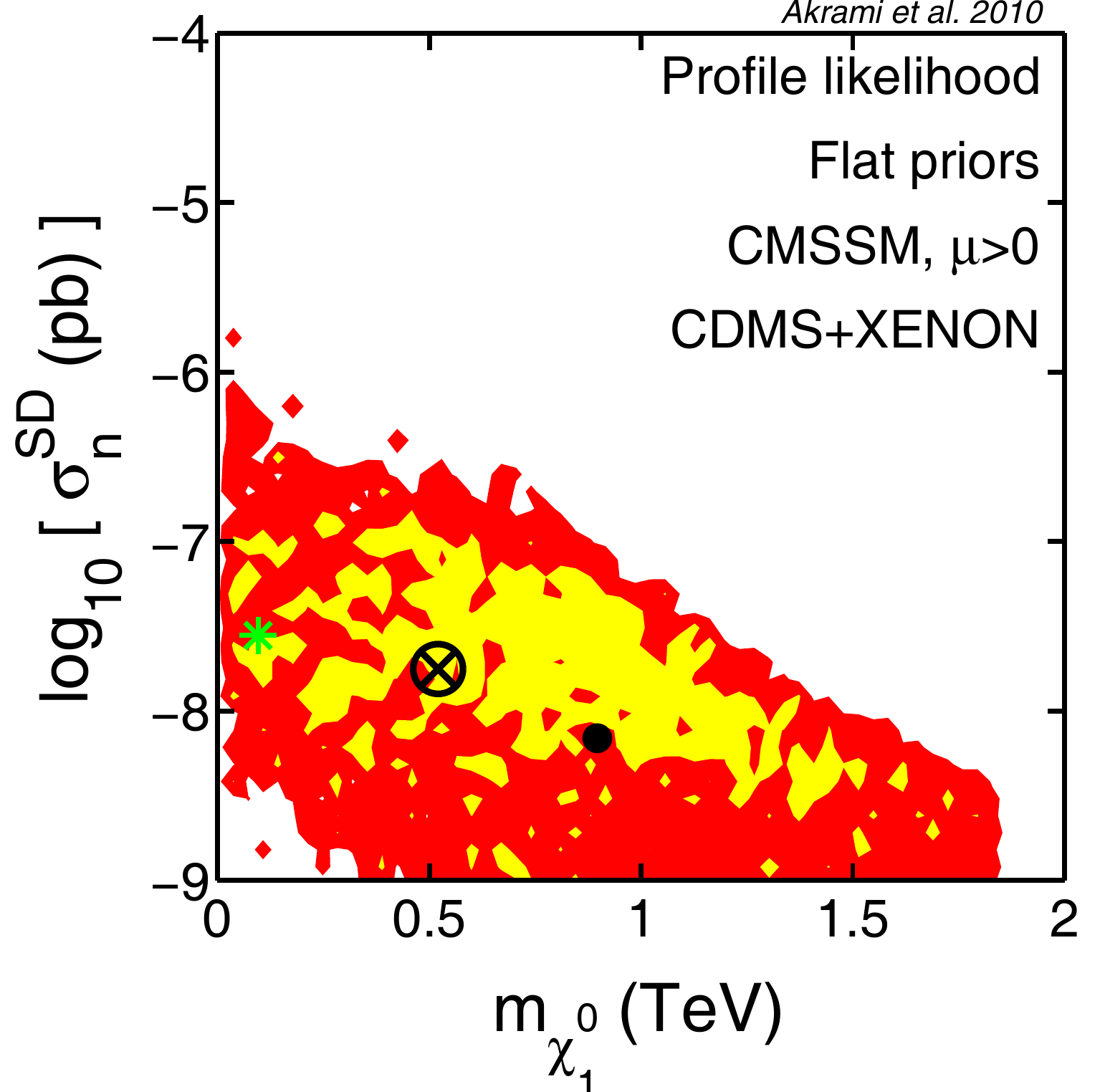}}
\subfigure{\includegraphics[scale=0.23, trim = 40 230 60 123, clip=true]{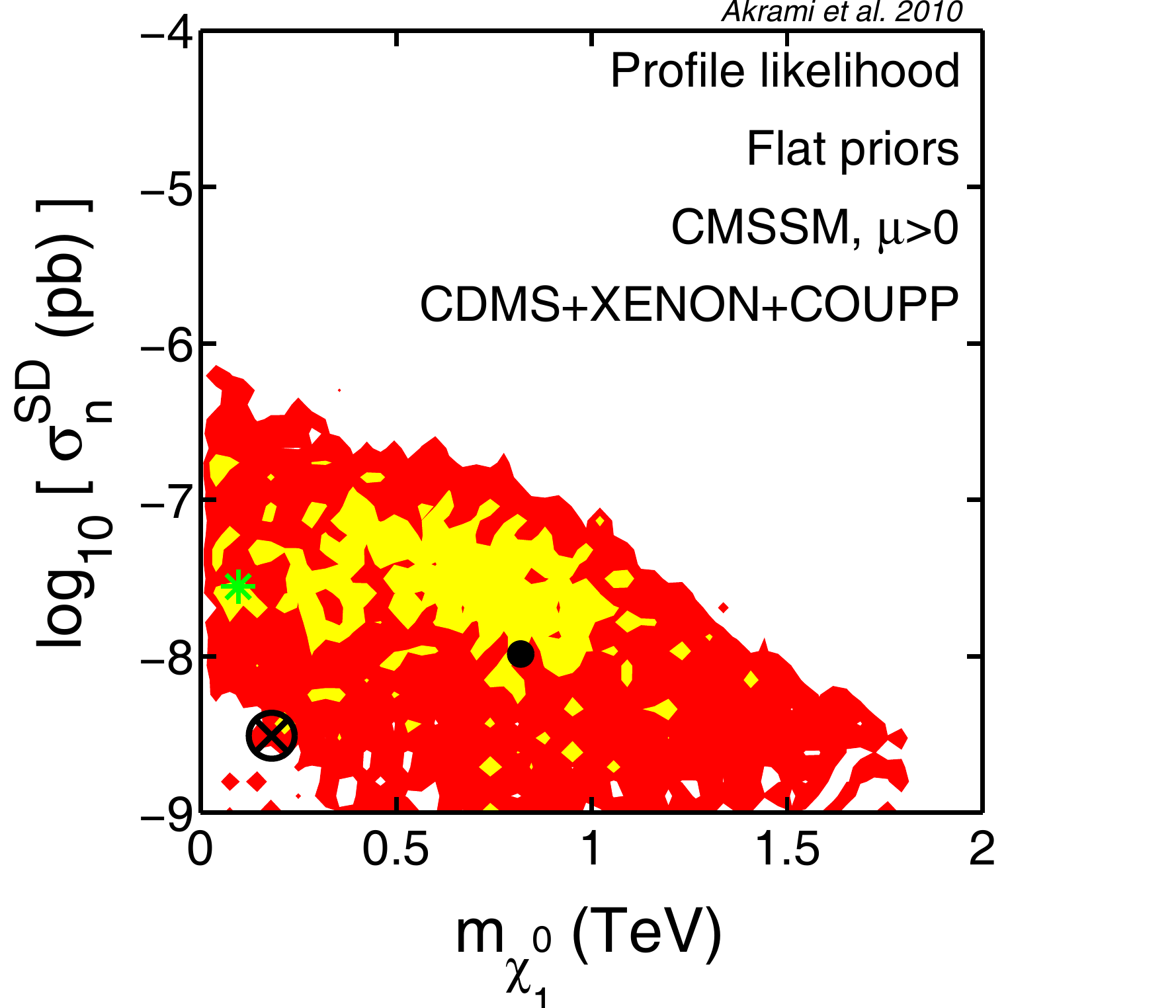}}\\
\caption[aa]{\footnotesize{As in~\fig{fig:LHprofl}, but for benchmark 2.}}\label{fig:LLprofl}
\end{figure}

\begin{figure}[t]
\subfigure{\includegraphics[scale=0.23, trim = 40 230 130 123, clip=true]{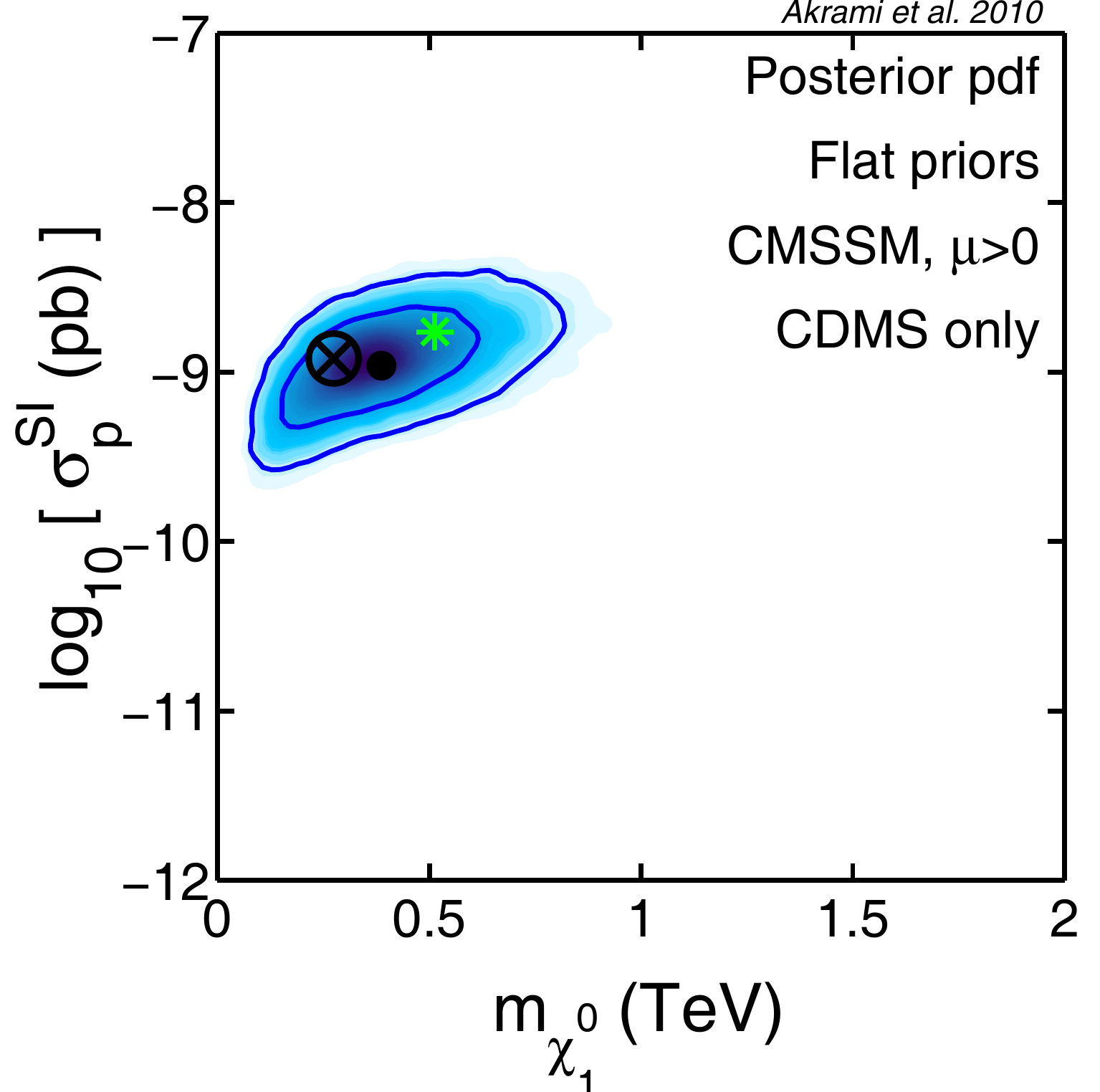}}
\subfigure{\includegraphics[scale=0.23, trim = 40 230 130 123, clip=true]{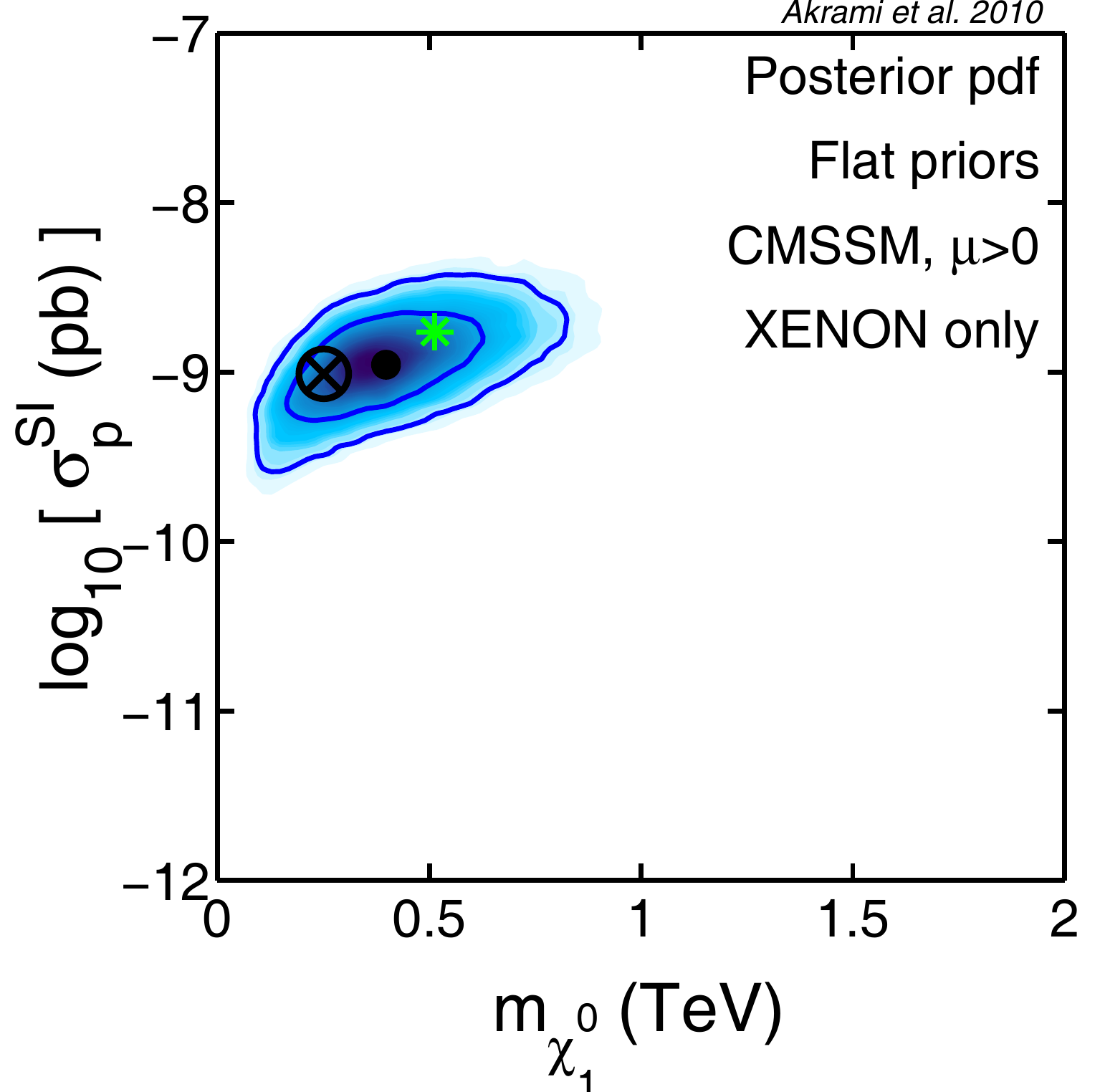}}
\subfigure{\includegraphics[scale=0.23, trim = 40 230 130 123, clip=true]{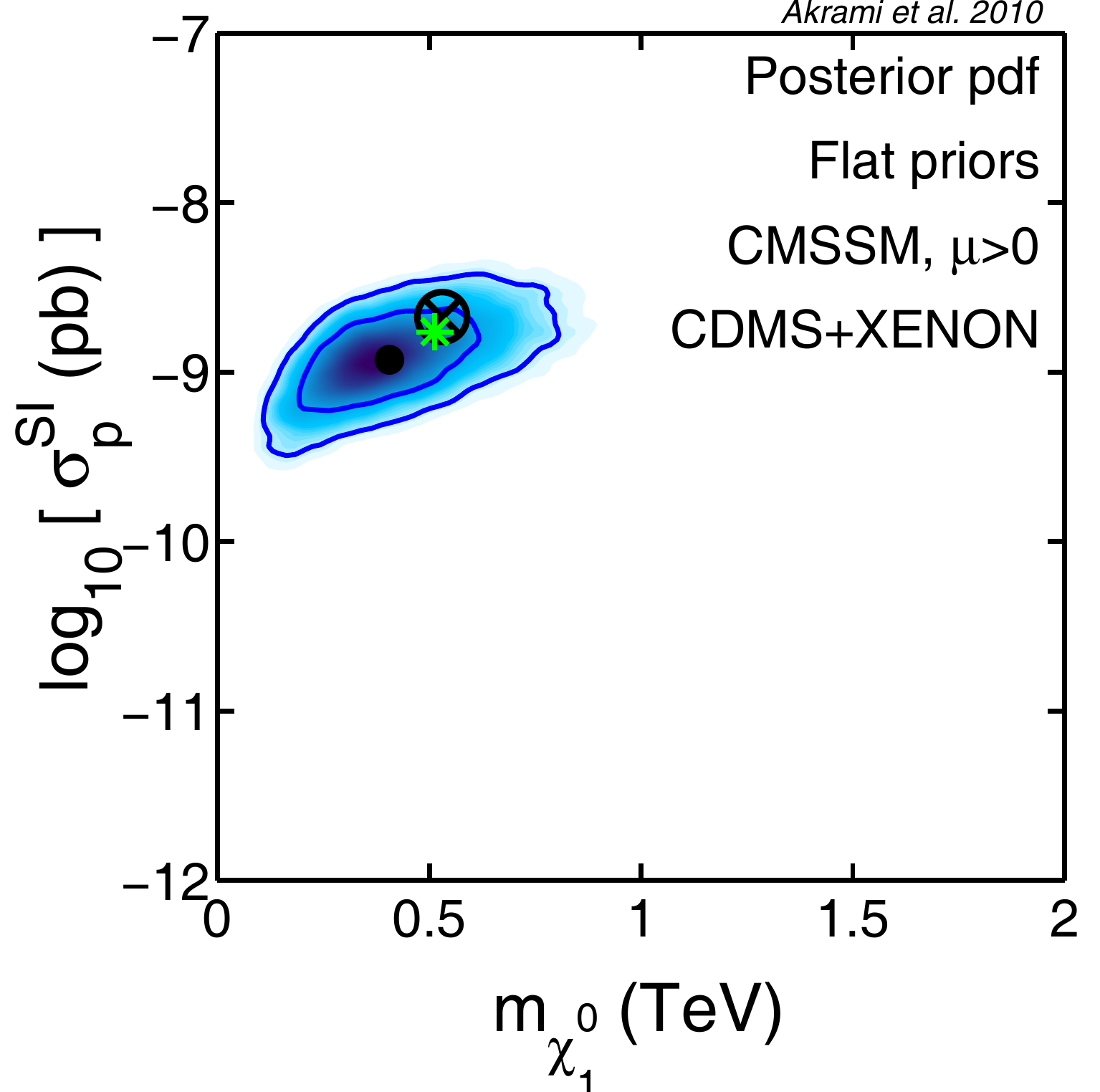}}
\subfigure{\includegraphics[scale=0.23, trim = 40 230 60 123, clip=true]{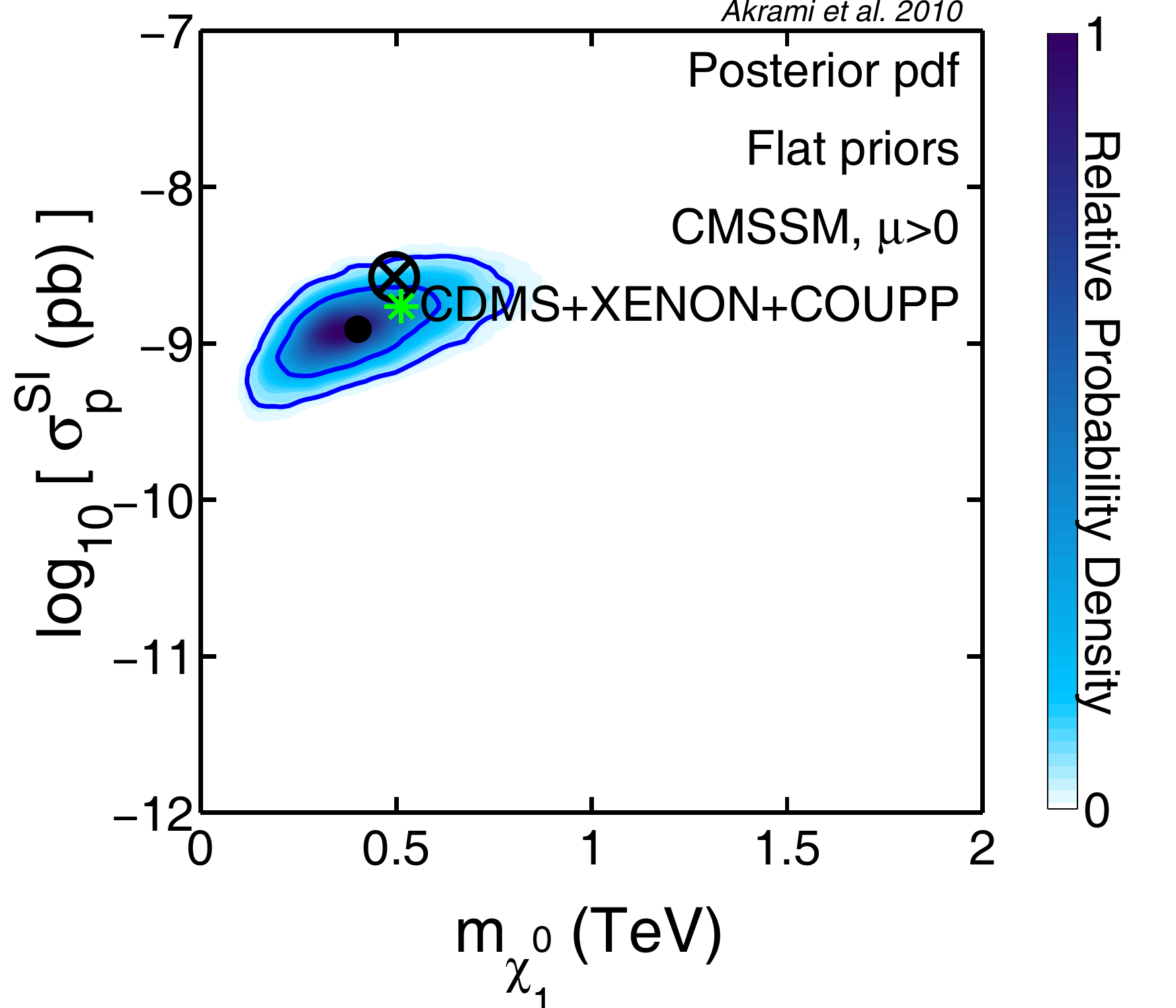}}\\
\subfigure{\includegraphics[scale=0.23, trim = 40 230 130 123, clip=true]{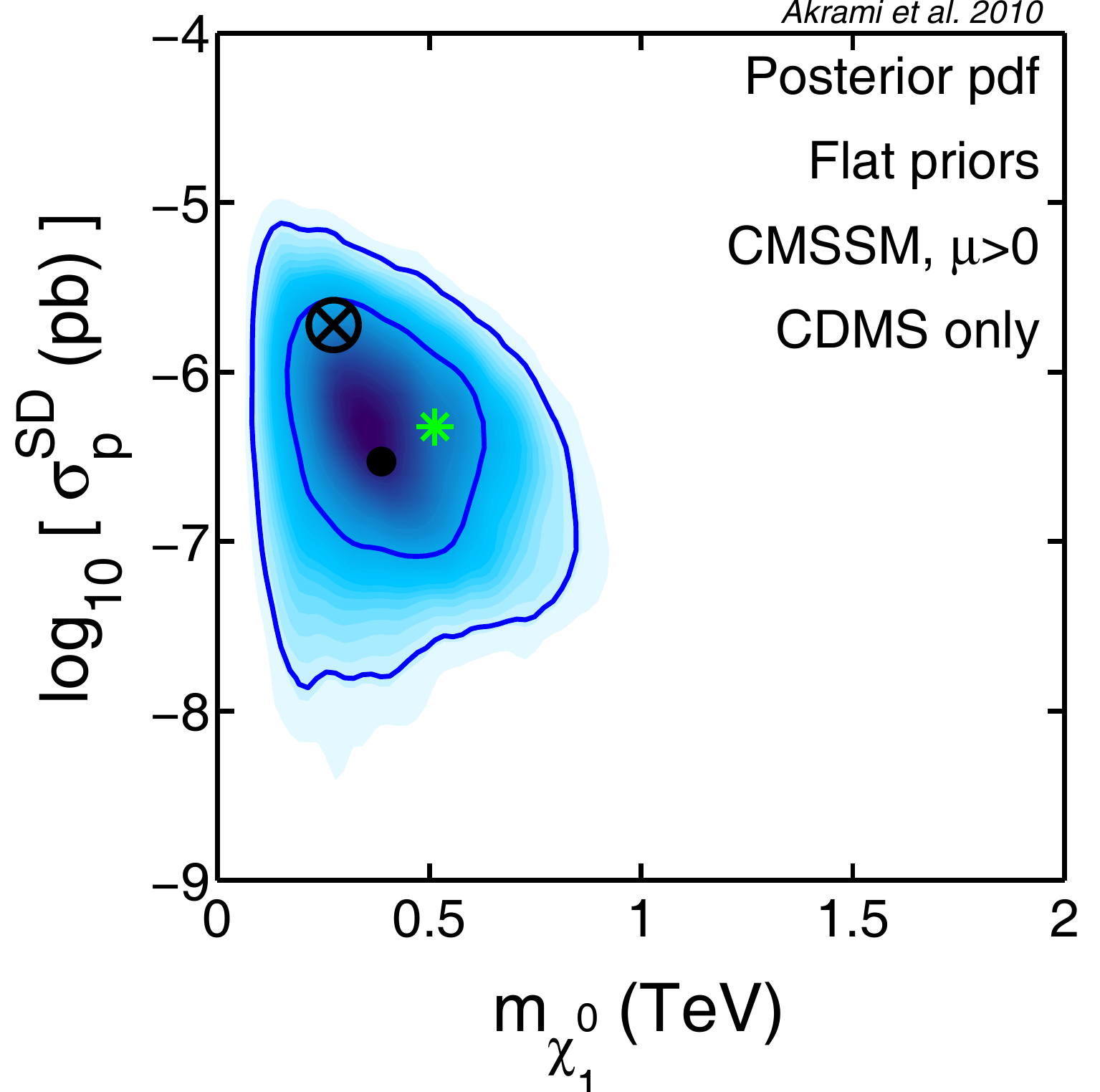}}
\subfigure{\includegraphics[scale=0.23, trim = 40 230 130 123, clip=true]{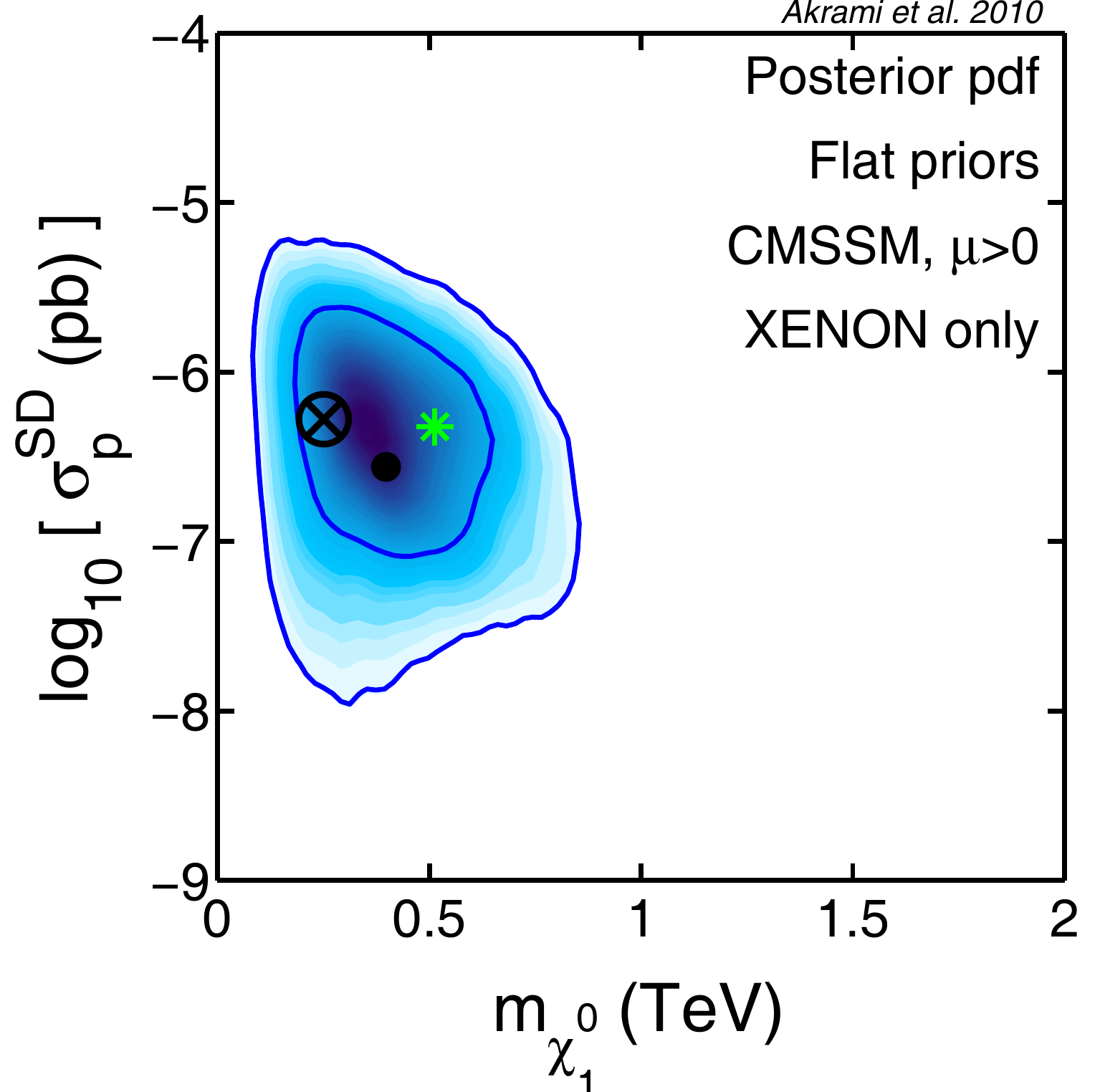}}
\subfigure{\includegraphics[scale=0.23, trim = 40 230 130 123, clip=true]{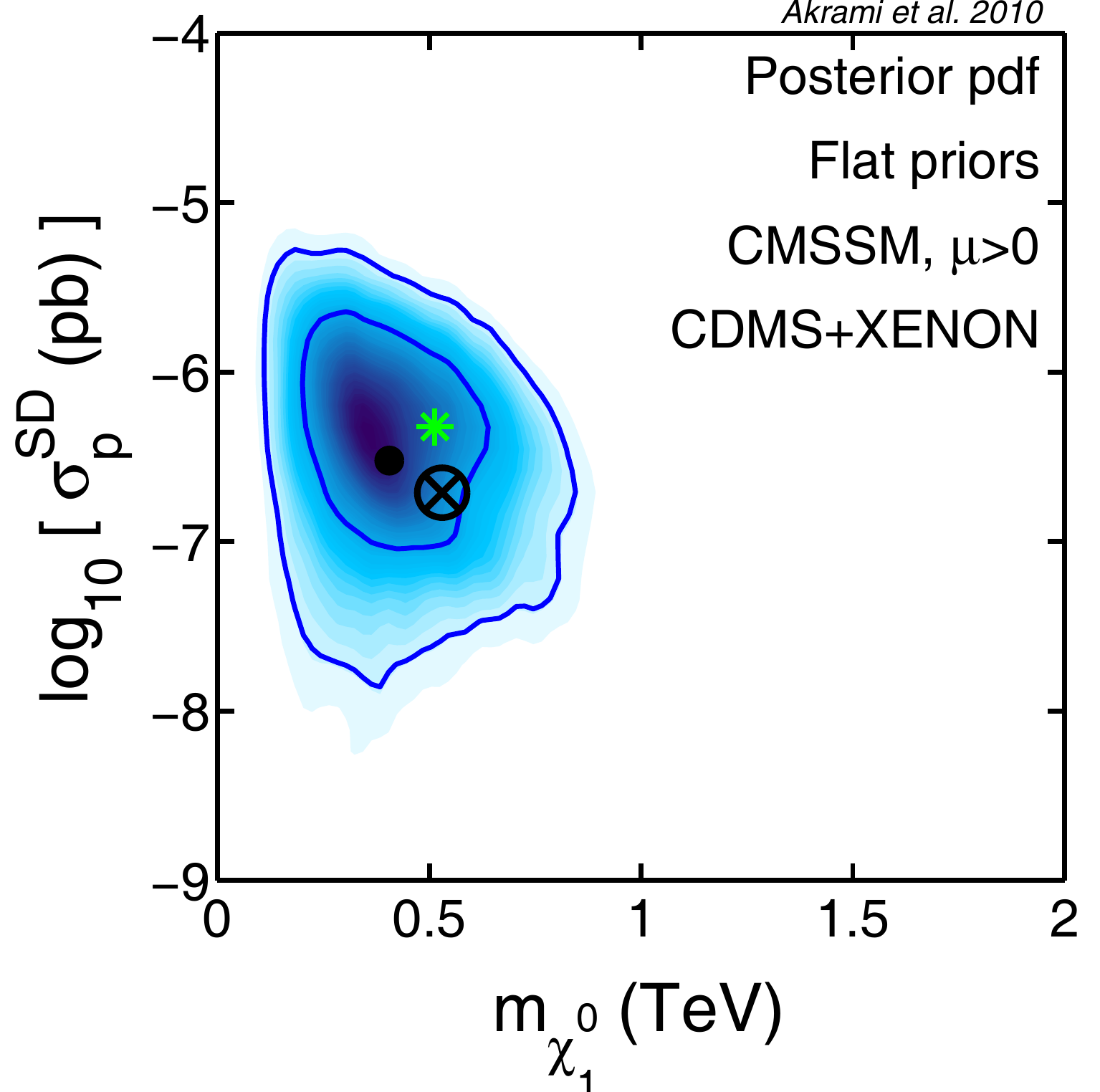}}
\subfigure{\includegraphics[scale=0.23, trim = 40 230 60 123, clip=true]{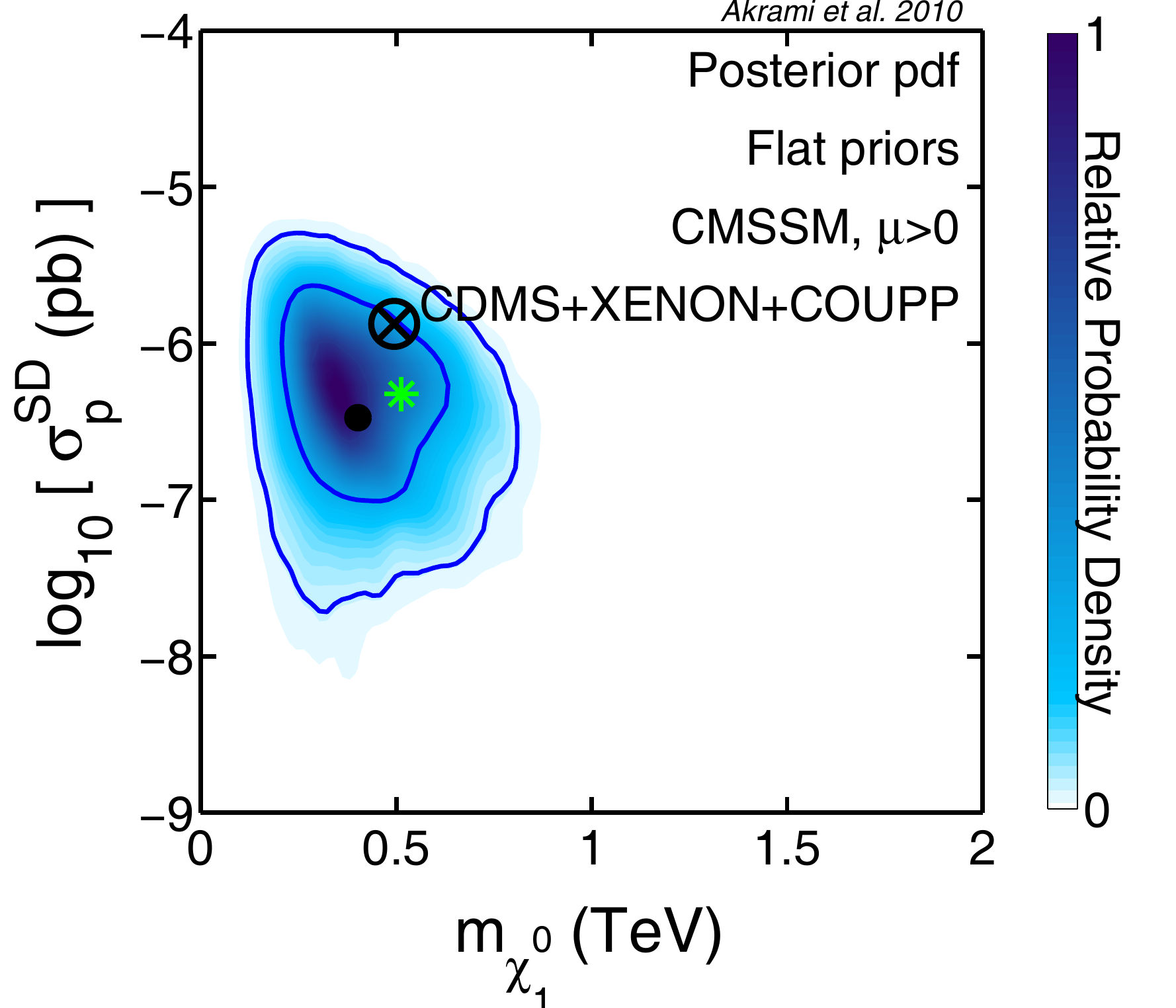}}\\
\subfigure{\includegraphics[scale=0.23, trim = 40 230 130 123, clip=true]{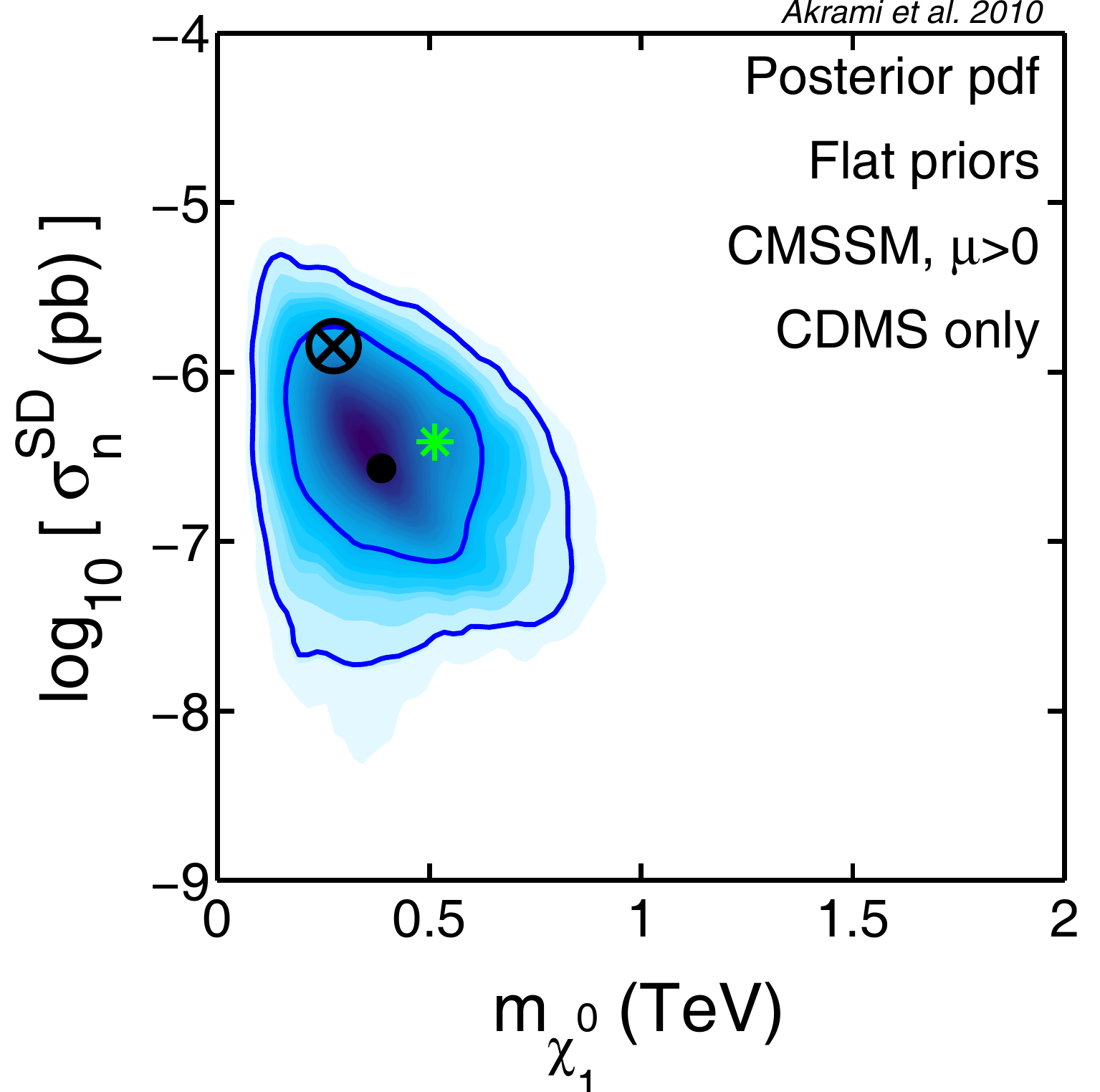}}
\subfigure{\includegraphics[scale=0.23, trim = 40 230 130 123, clip=true]{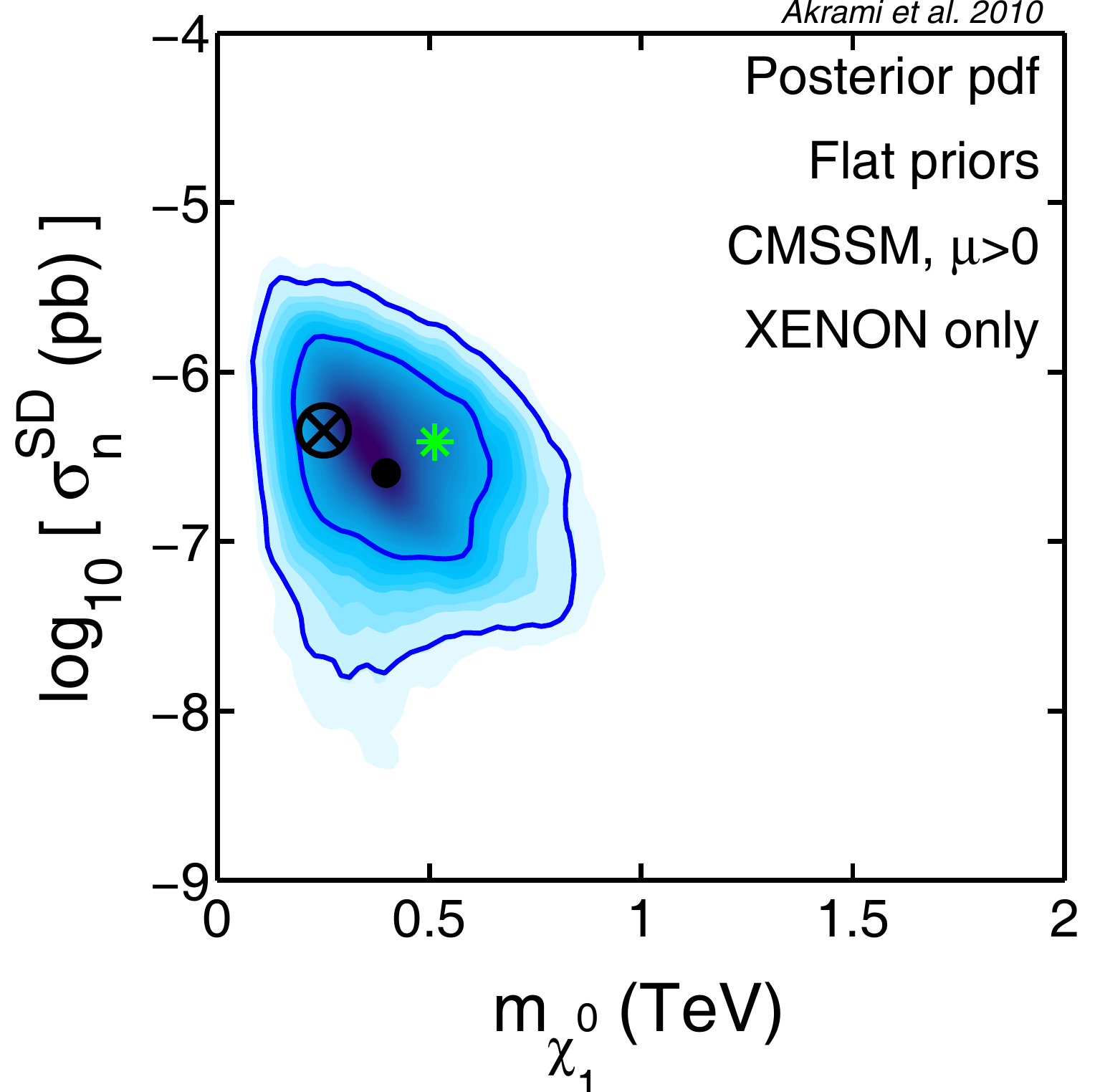}}
\subfigure{\includegraphics[scale=0.23, trim = 40 230 130 123, clip=true]{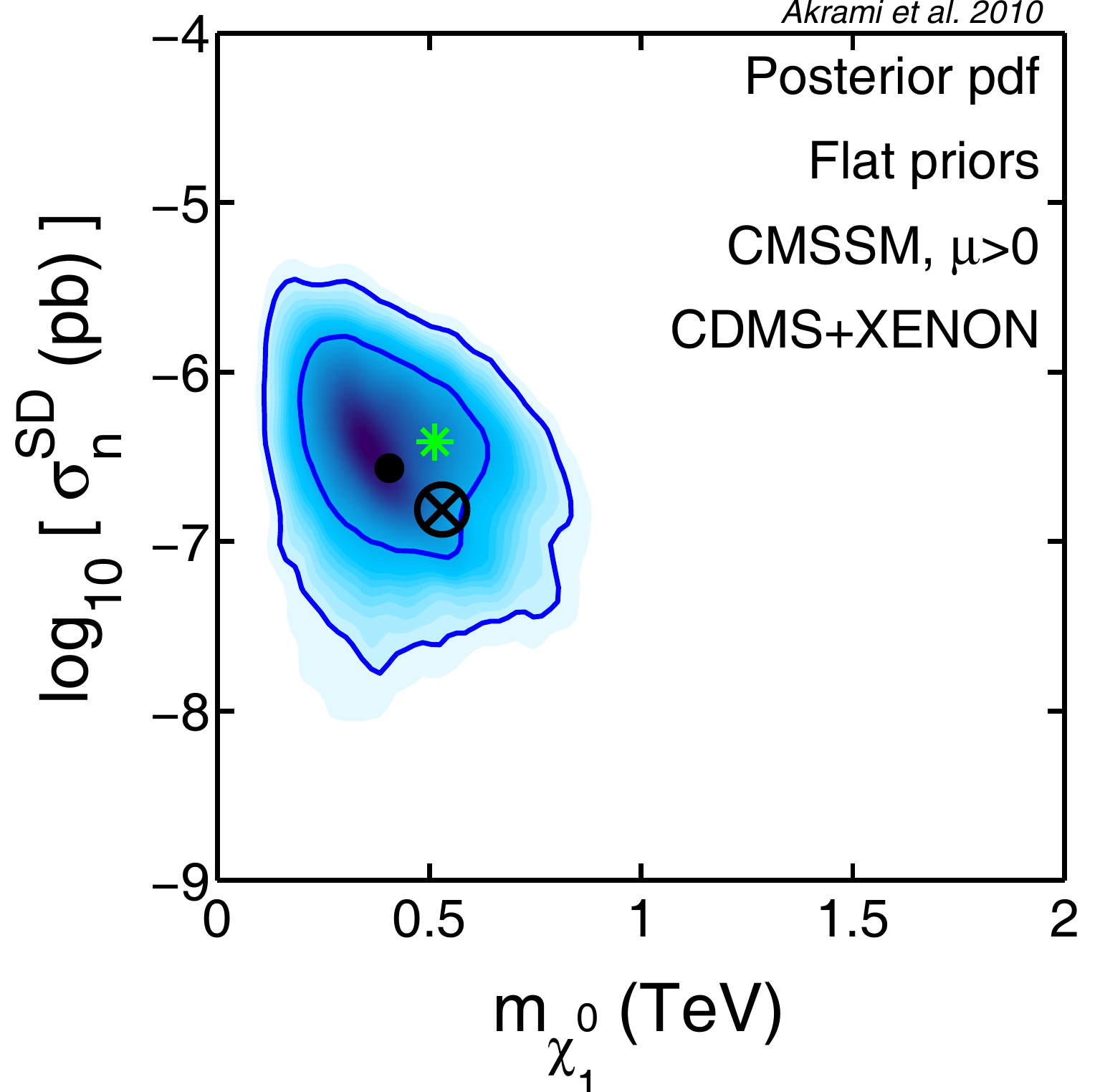}}
\subfigure{\includegraphics[scale=0.23, trim = 40 230 60 123, clip=true]{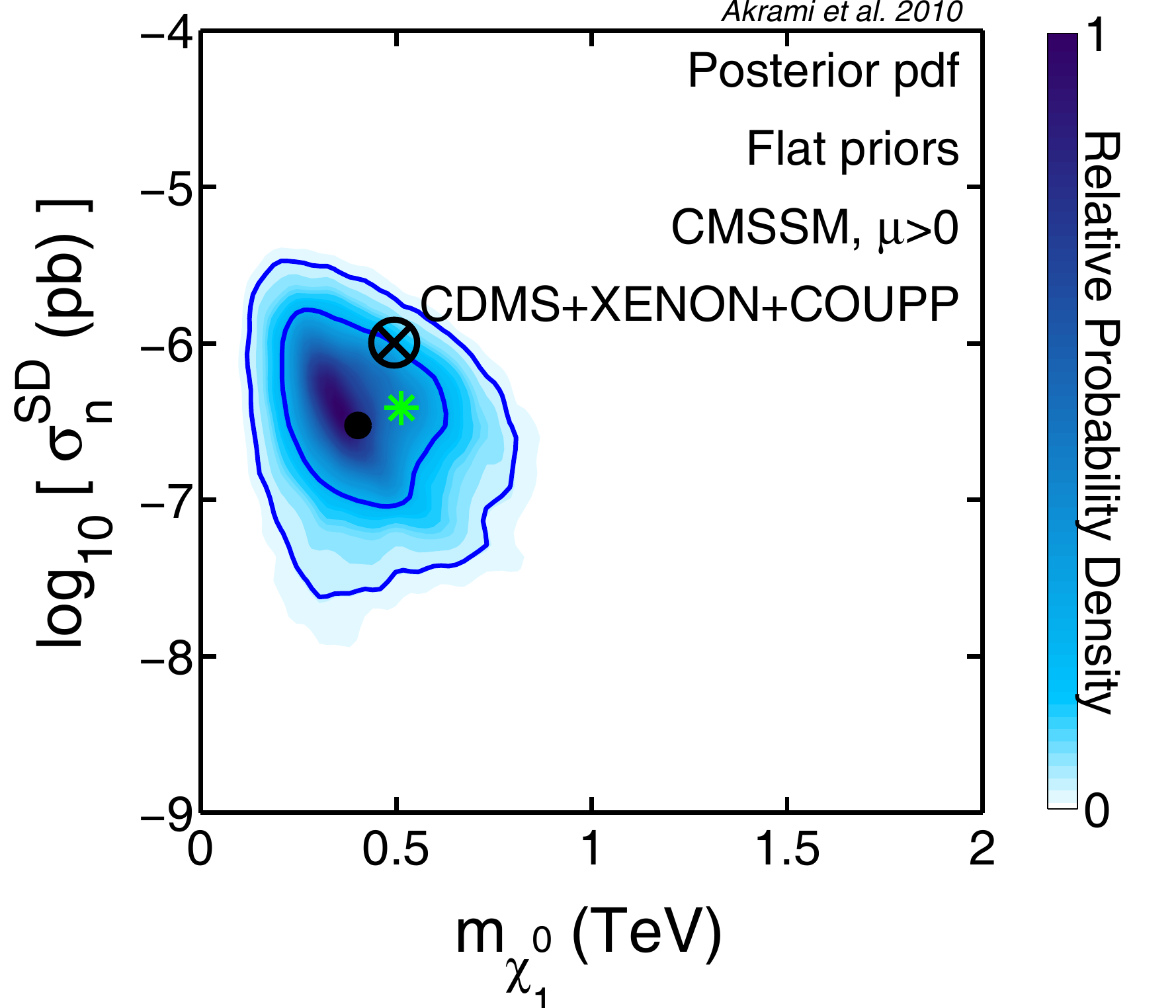}}\\
\caption[aa]{\footnotesize{As in~\figs{fig:LHmarg}{fig:LLmarg}, but for benchmark 3.}}\label{fig:MMmarg}
\end{figure}

\begin{figure}[t]
\subfigure{\includegraphics[scale=0.23, trim = 40 230 130 123, clip=true]{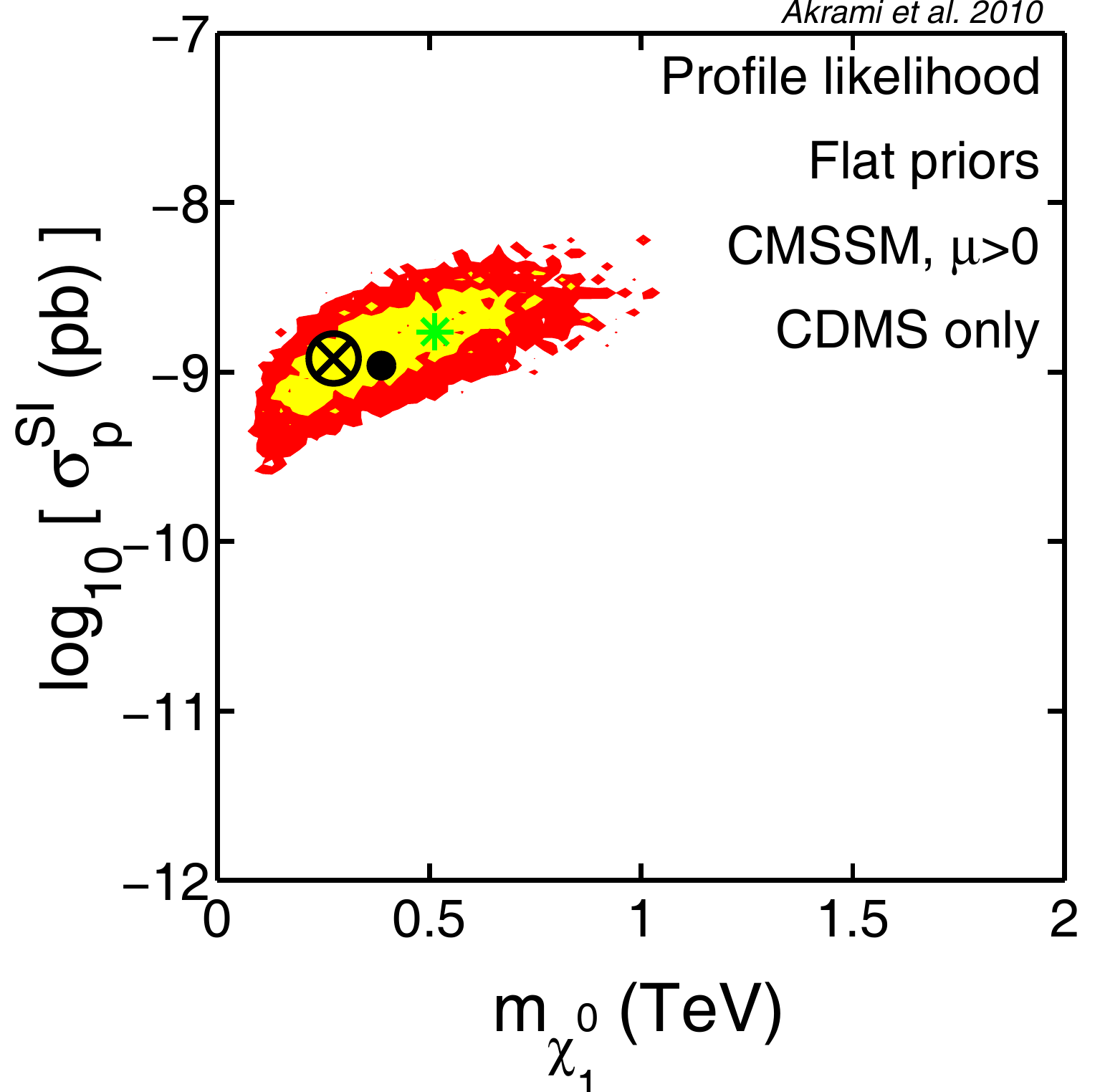}}
\subfigure{\includegraphics[scale=0.23, trim = 40 230 130 123, clip=true]{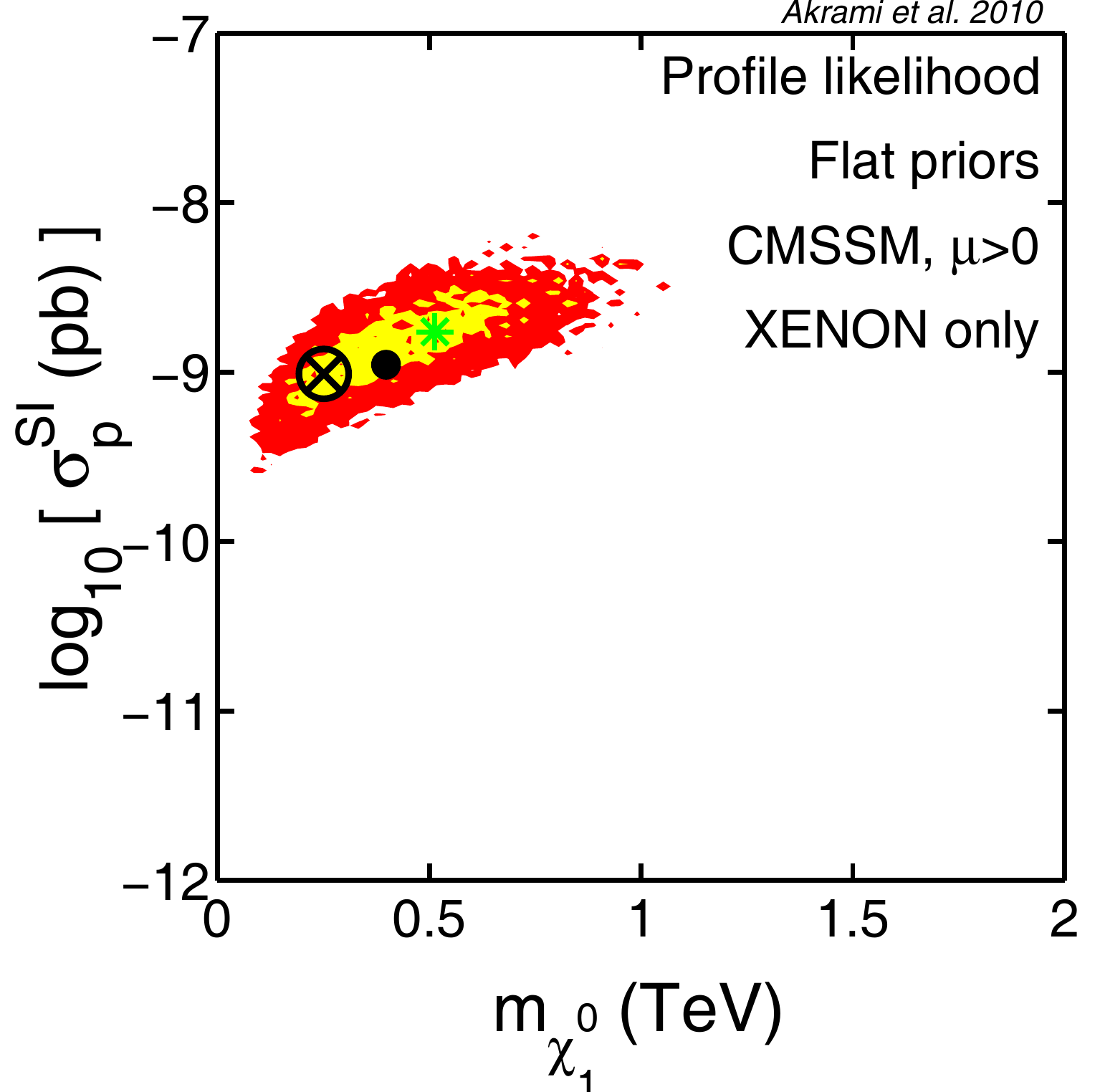}}
\subfigure{\includegraphics[scale=0.23, trim = 40 230 130 123, clip=true]{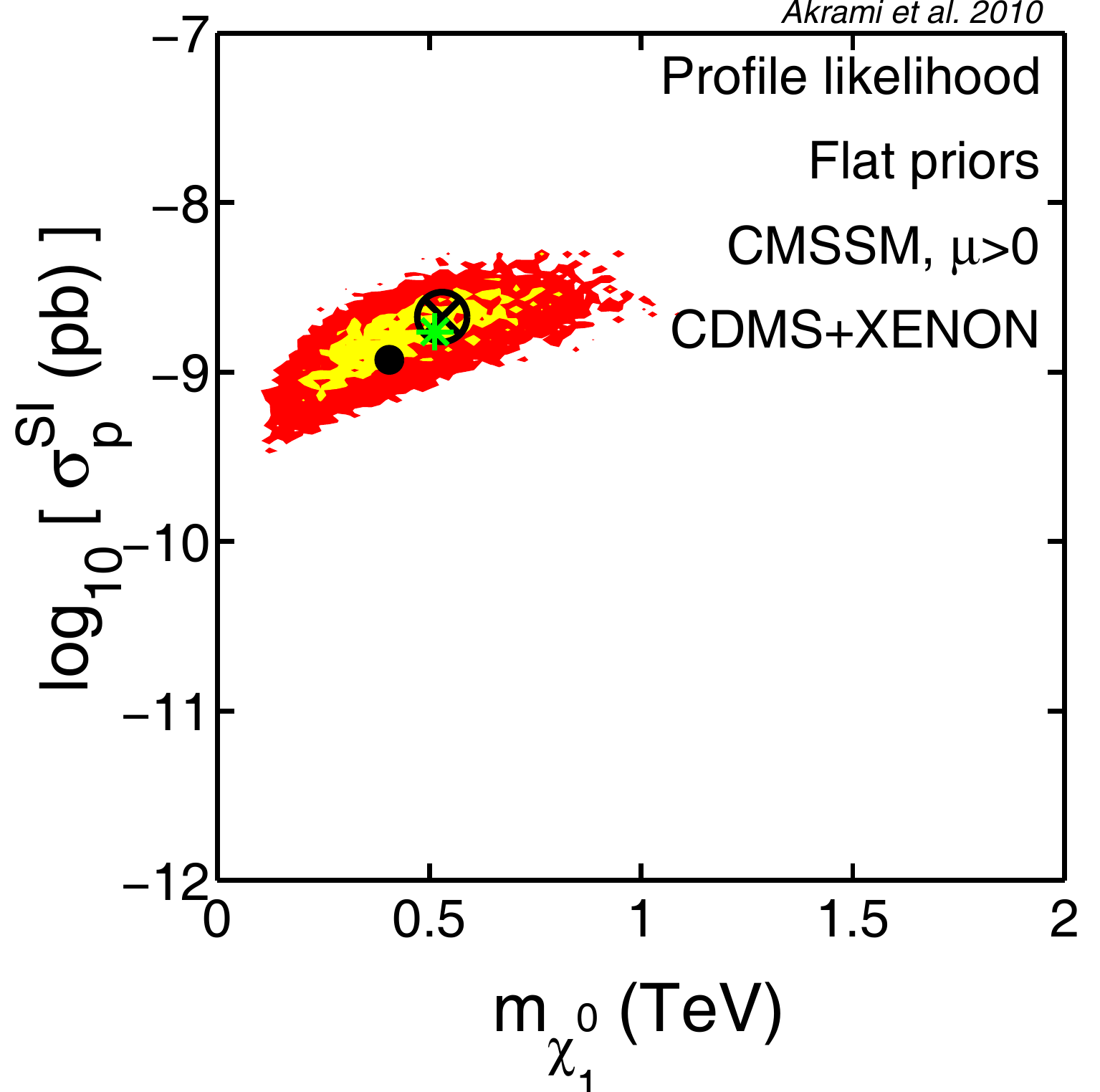}}
\subfigure{\includegraphics[scale=0.23, trim = 40 230 60 123, clip=true]{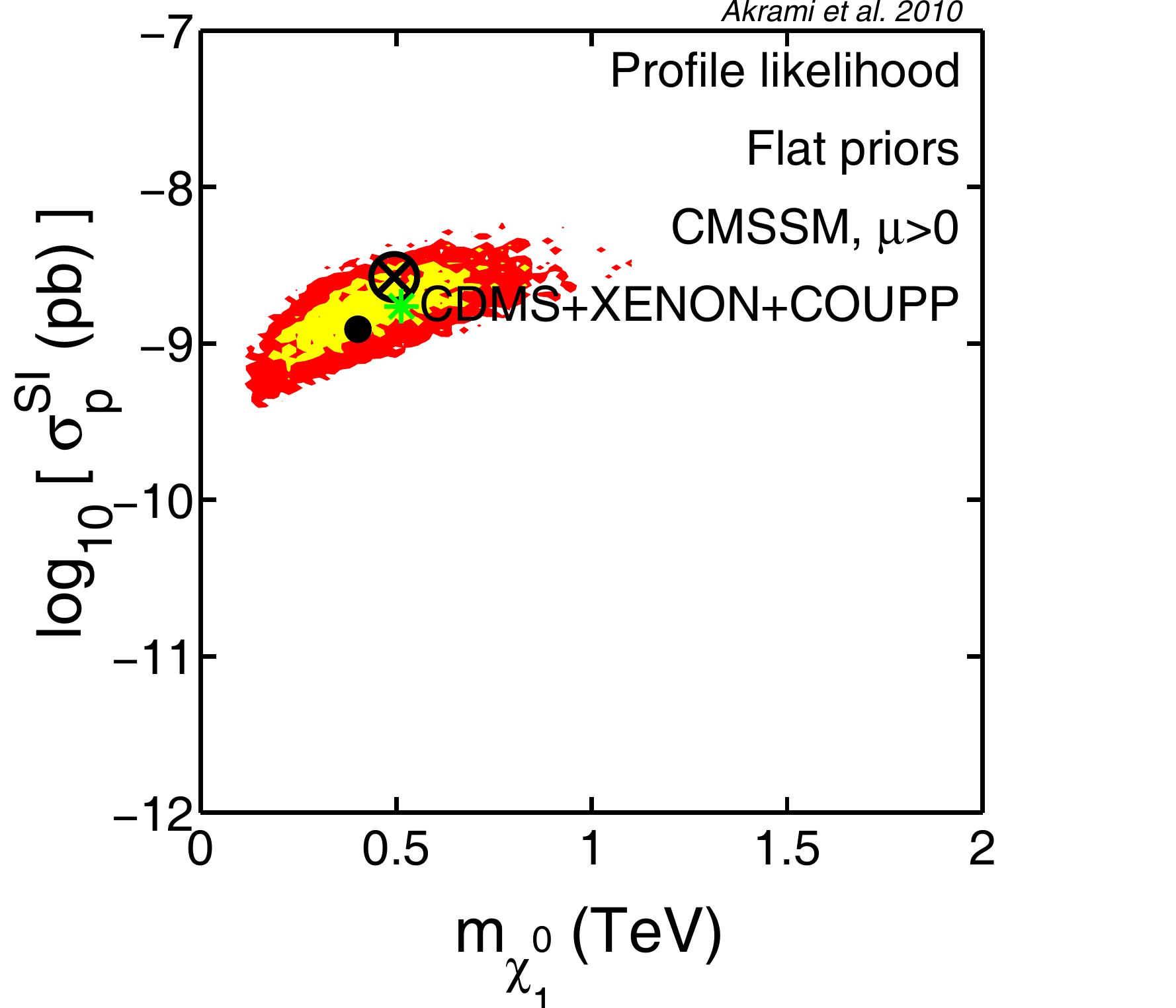}}\\
\subfigure{\includegraphics[scale=0.23, trim = 40 230 130 123, clip=true]{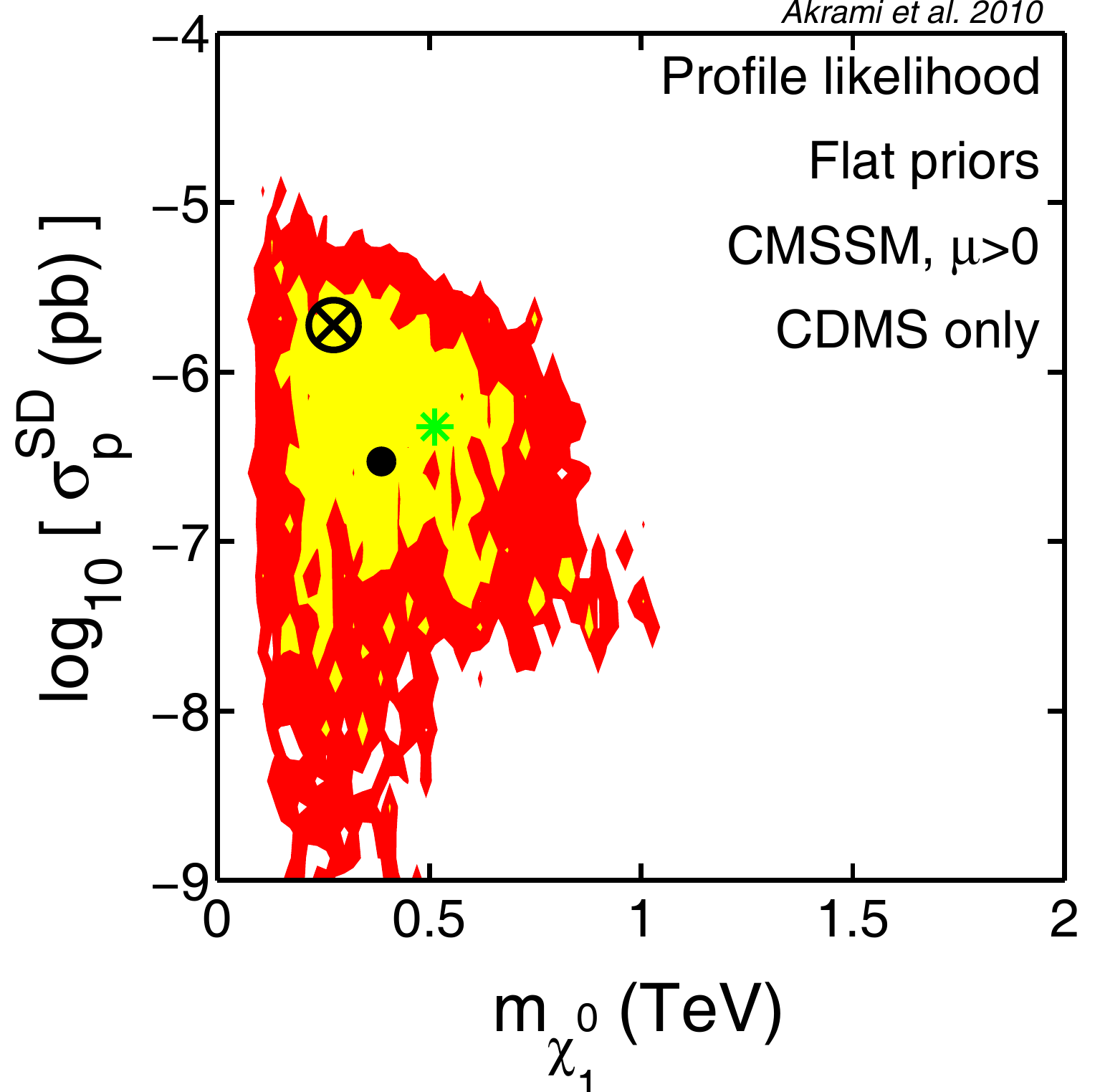}}
\subfigure{\includegraphics[scale=0.23, trim = 40 230 130 123, clip=true]{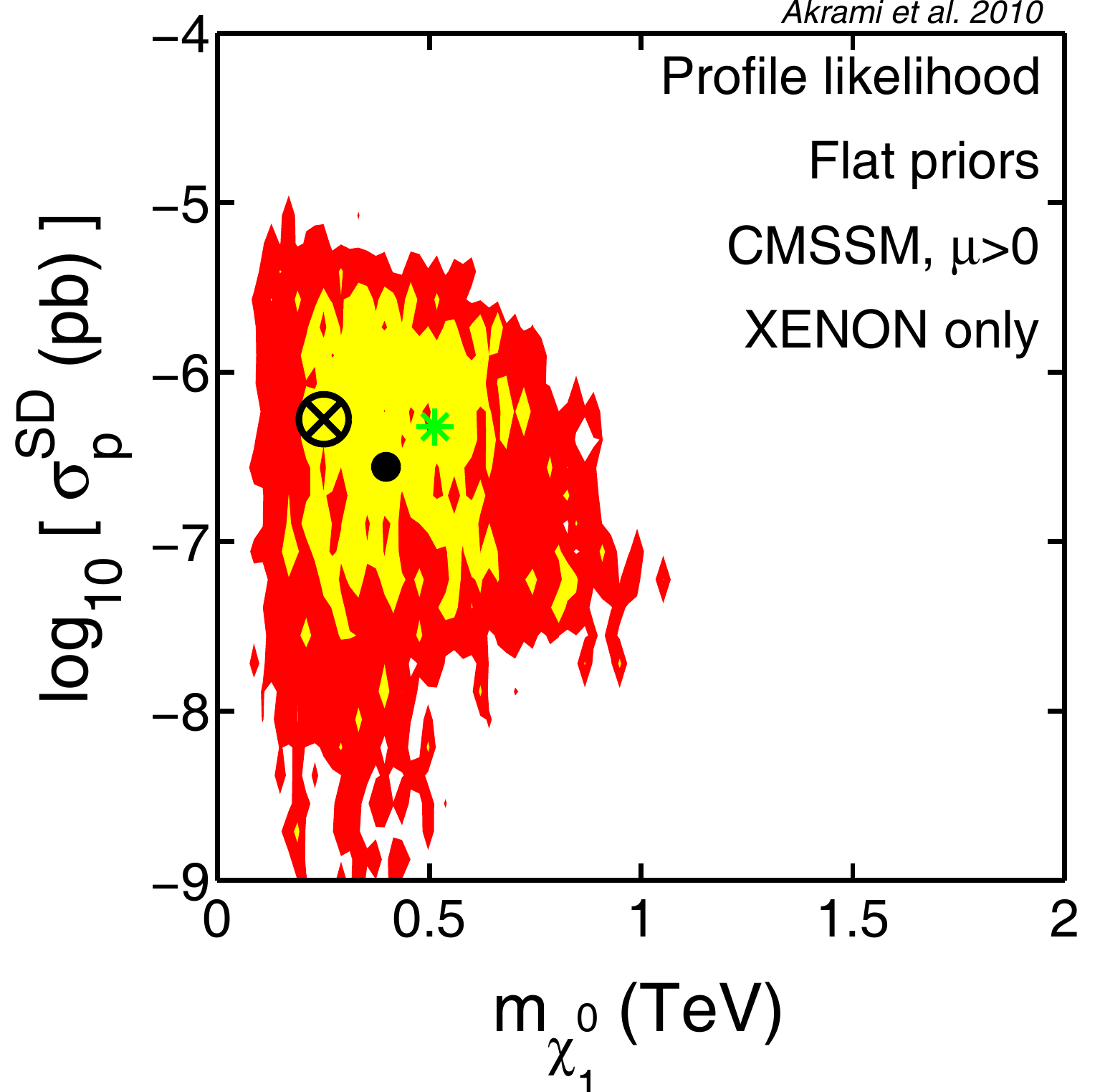}}
\subfigure{\includegraphics[scale=0.23, trim = 40 230 130 123, clip=true]{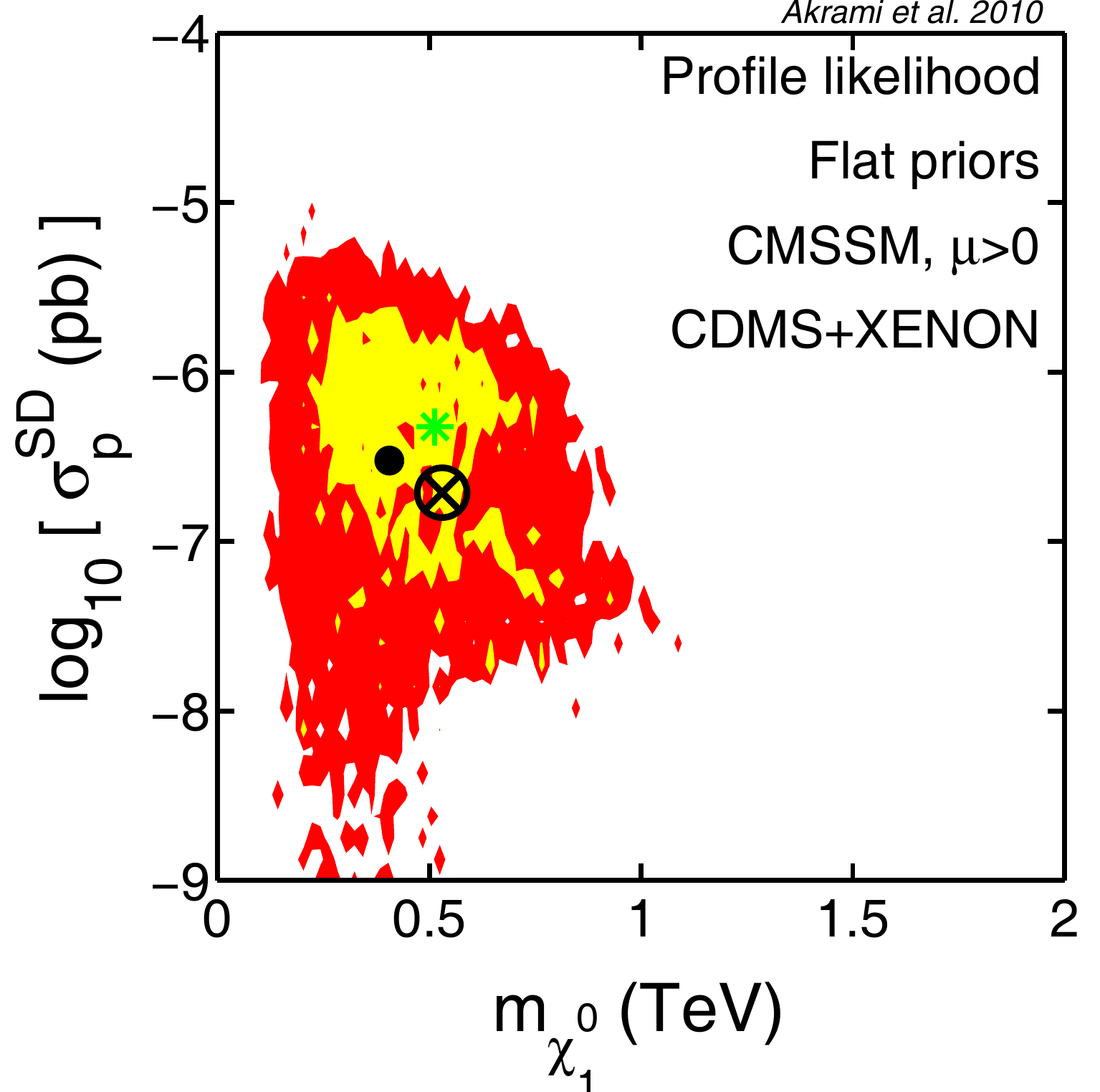}}
\subfigure{\includegraphics[scale=0.23, trim = 40 230 60 123, clip=true]{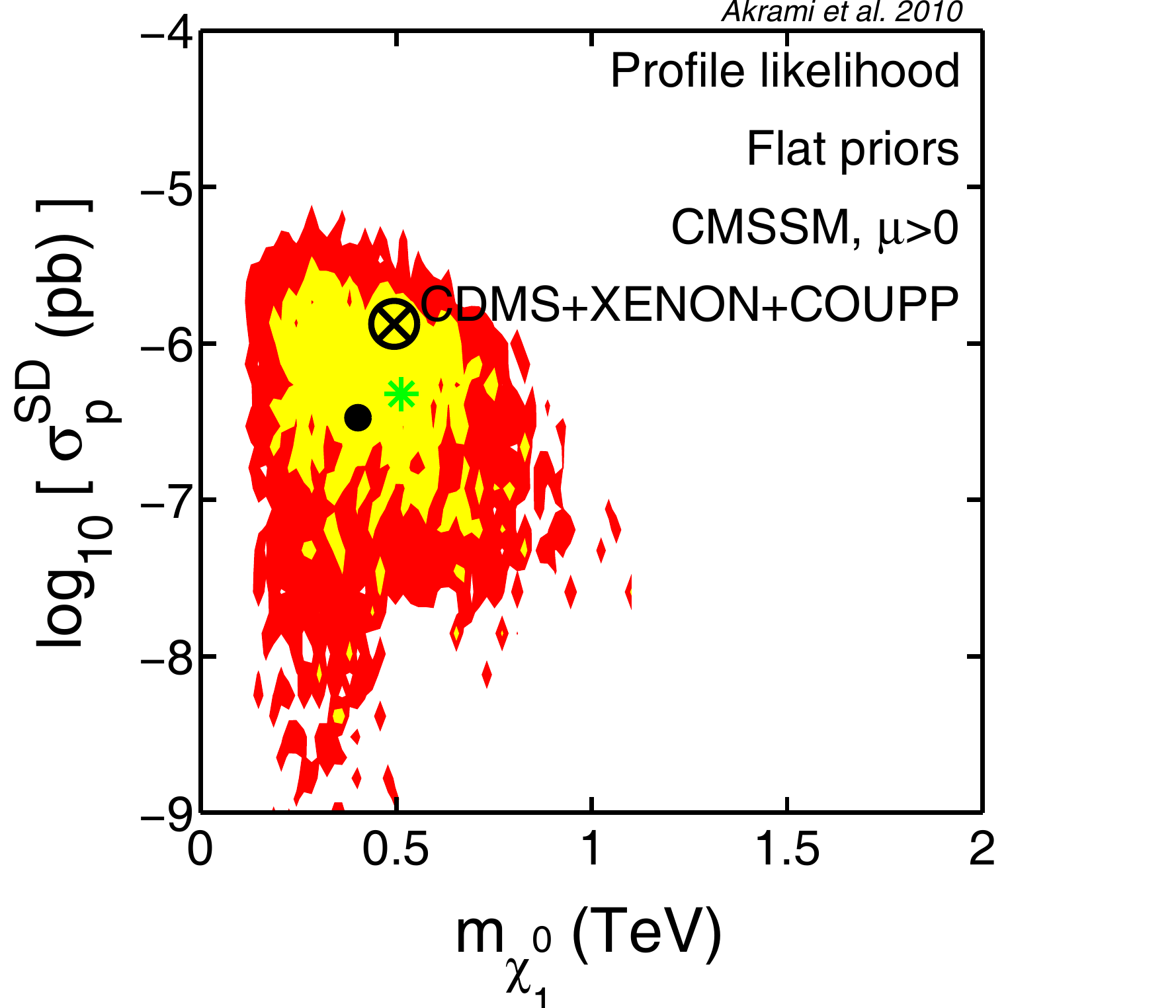}}\\
\subfigure{\includegraphics[scale=0.23, trim = 40 230 130 123, clip=true]{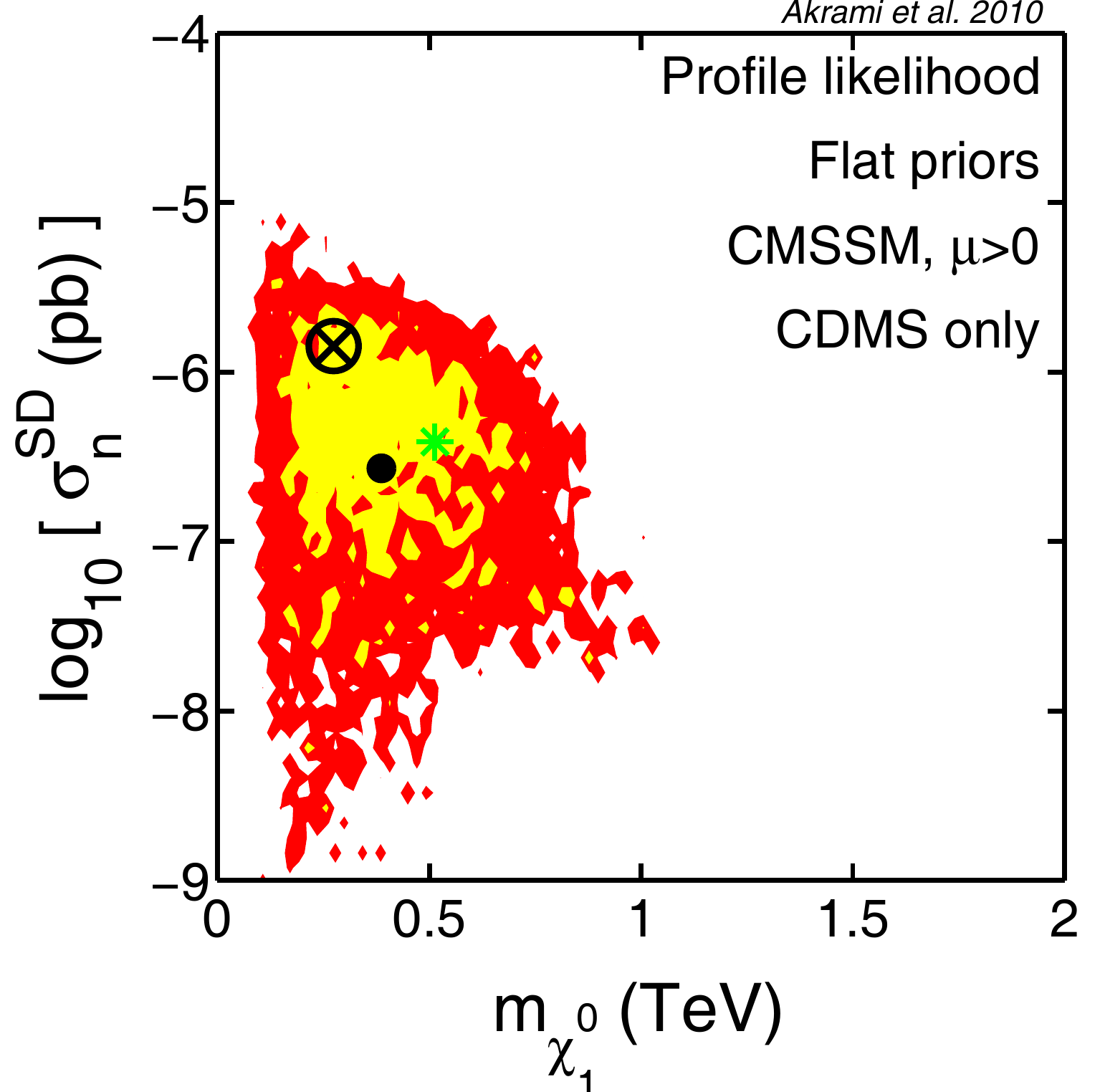}}
\subfigure{\includegraphics[scale=0.23, trim = 40 230 130 123, clip=true]{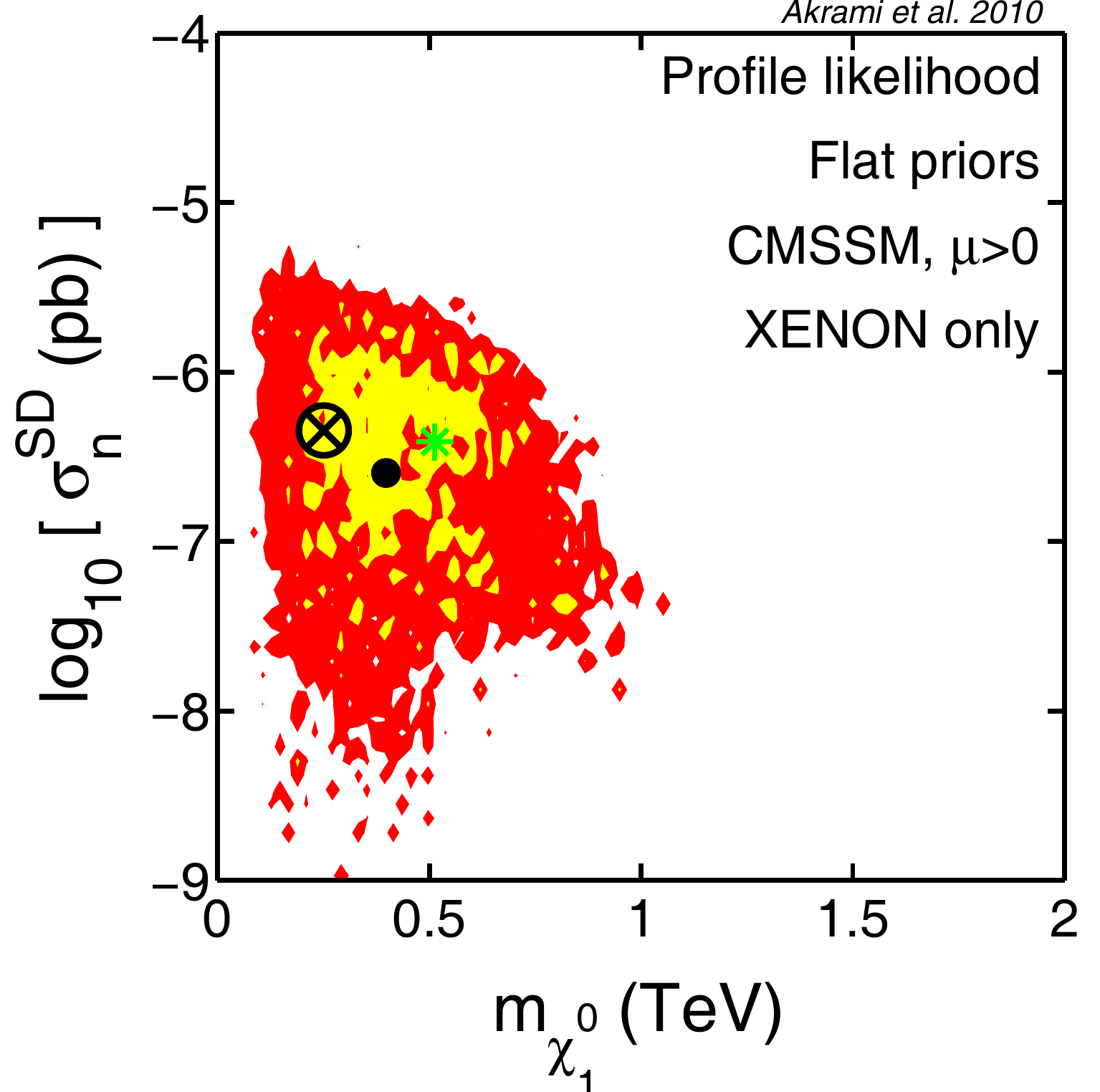}}
\subfigure{\includegraphics[scale=0.23, trim = 40 230 130 123, clip=true]{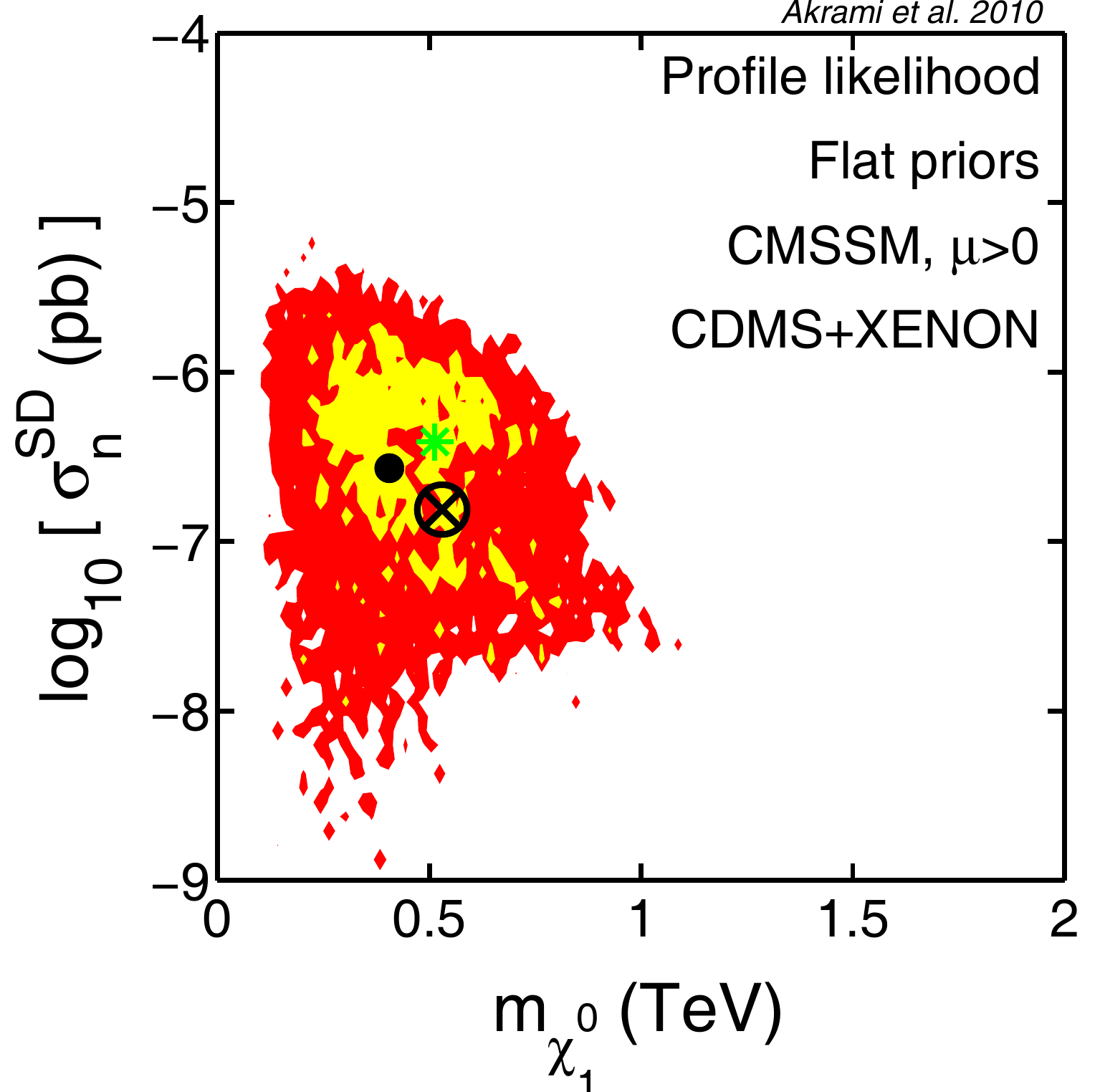}}
\subfigure{\includegraphics[scale=0.23, trim = 40 230 60 123, clip=true]{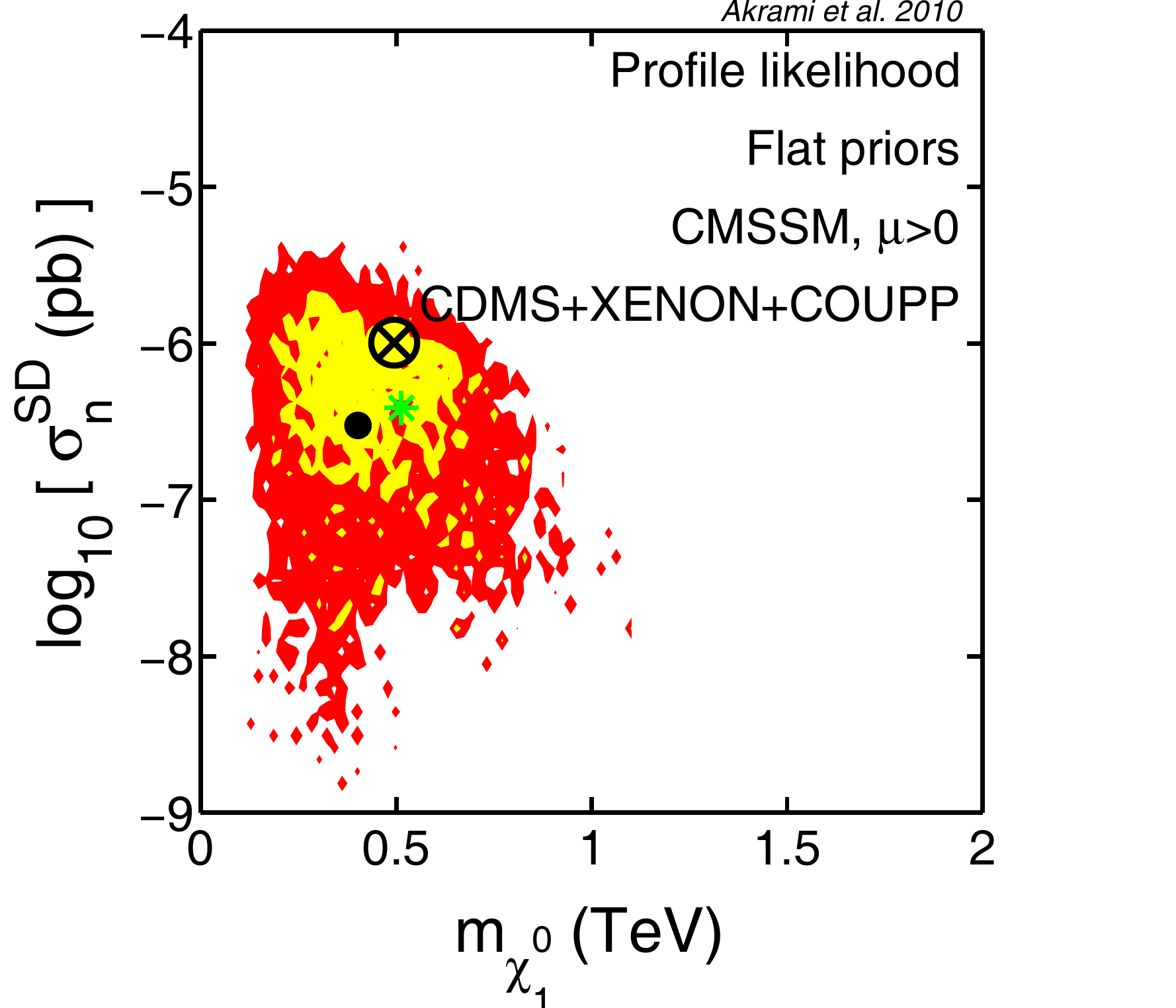}}\\
\caption[aa]{\footnotesize{As in~\figs{fig:LHprofl}{fig:LLprofl}, but for benchmark 3.}}\label{fig:MMprofl}
\end{figure}

\begin{figure}[t]
\subfigure{\includegraphics[scale=0.23, trim = 40 230 130 123, clip=true]{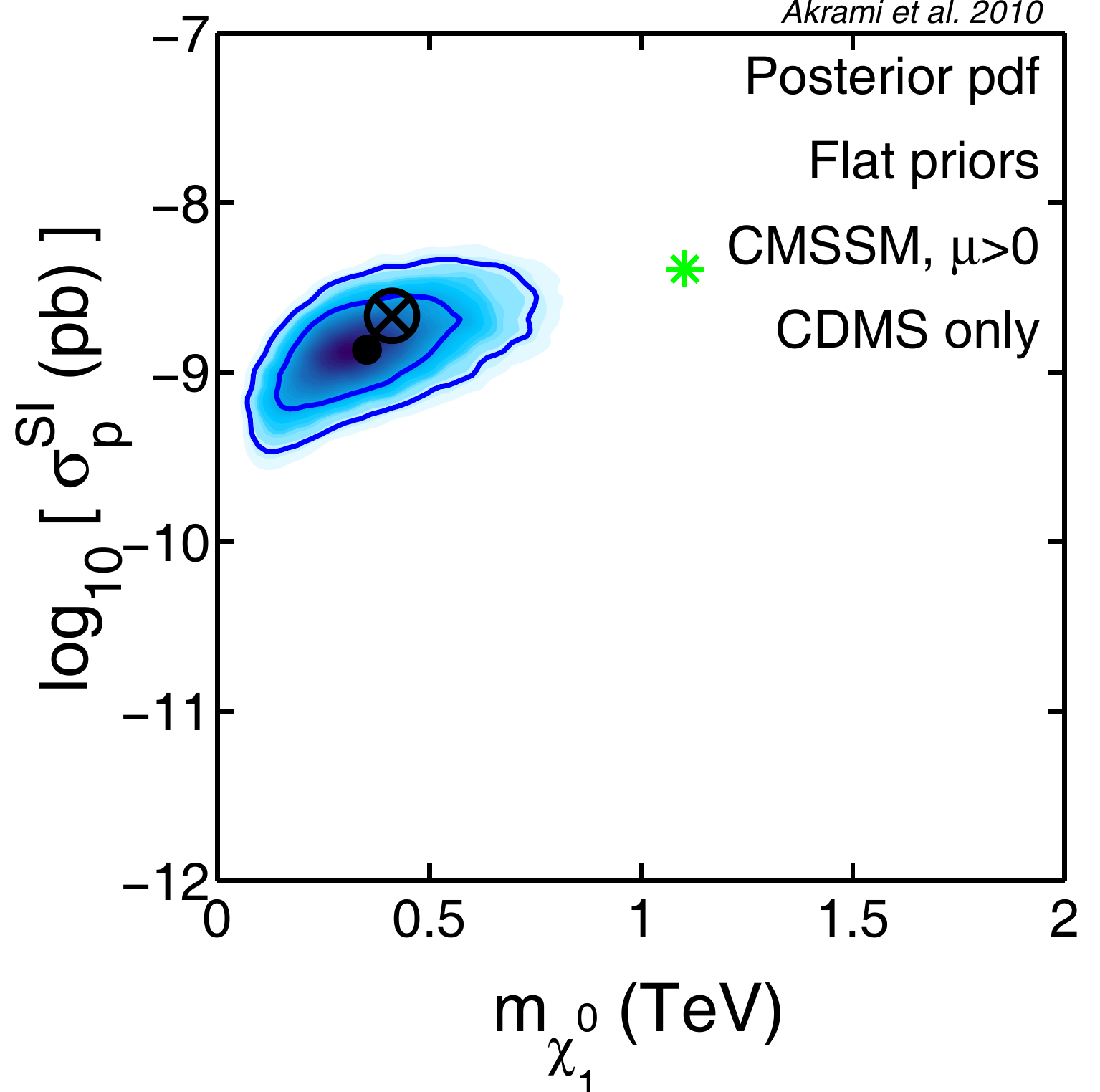}}
\subfigure{\includegraphics[scale=0.23, trim = 40 230 130 123, clip=true]{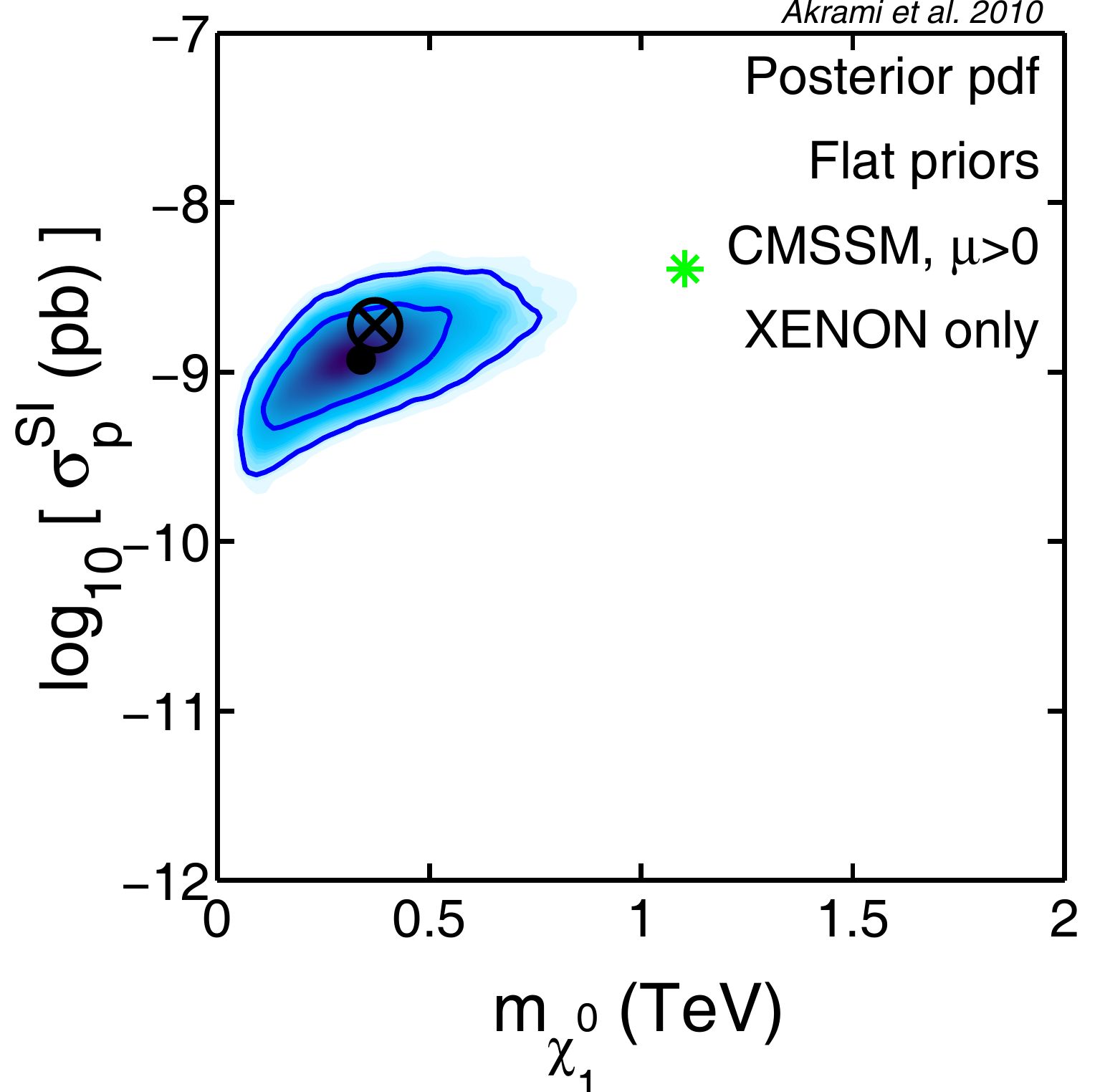}}
\subfigure{\includegraphics[scale=0.23, trim = 40 230 130 123, clip=true]{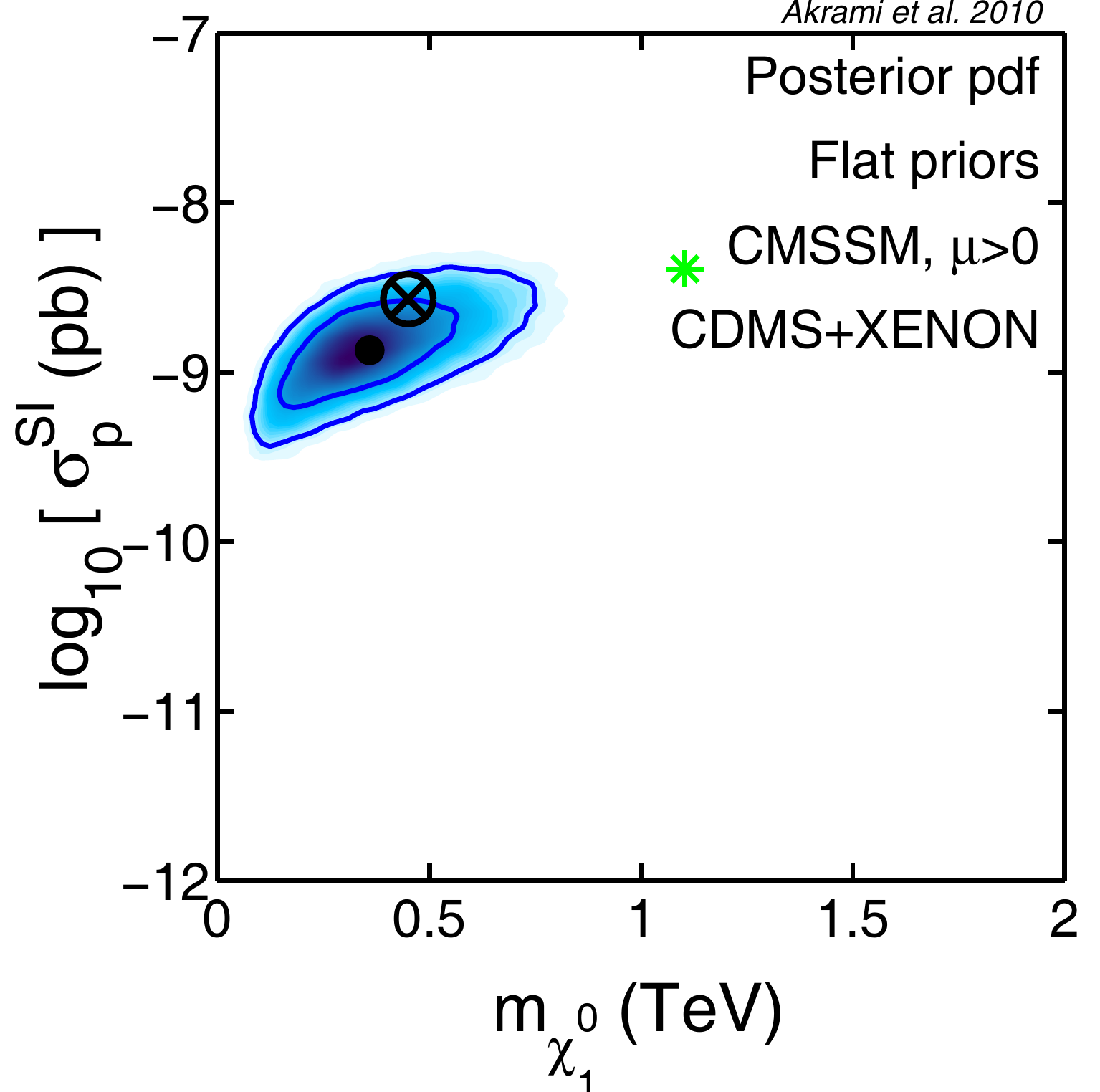}}
\subfigure{\includegraphics[scale=0.23, trim = 40 230 60 123, clip=true]{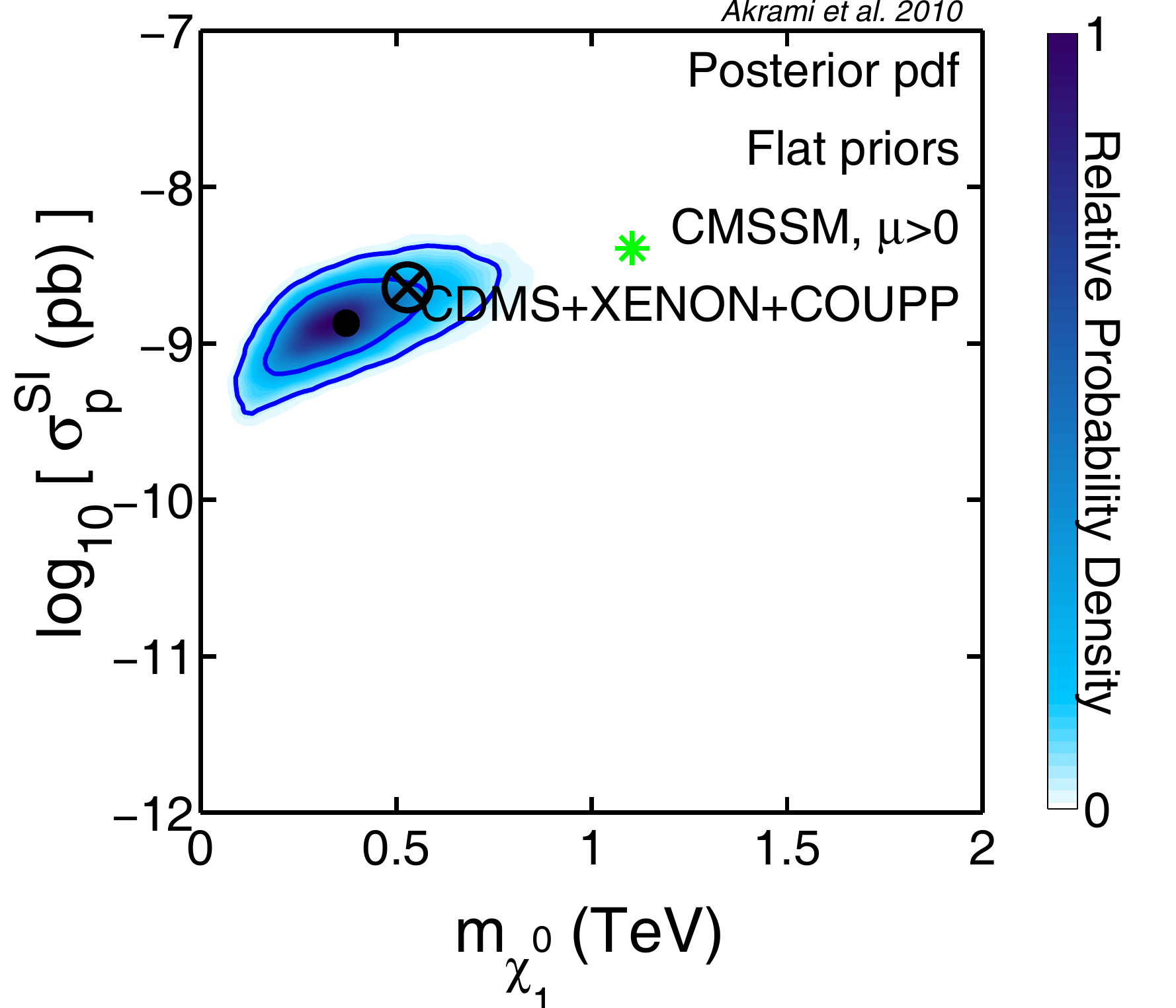}}\\
\subfigure{\includegraphics[scale=0.23, trim = 40 230 130 123, clip=true]{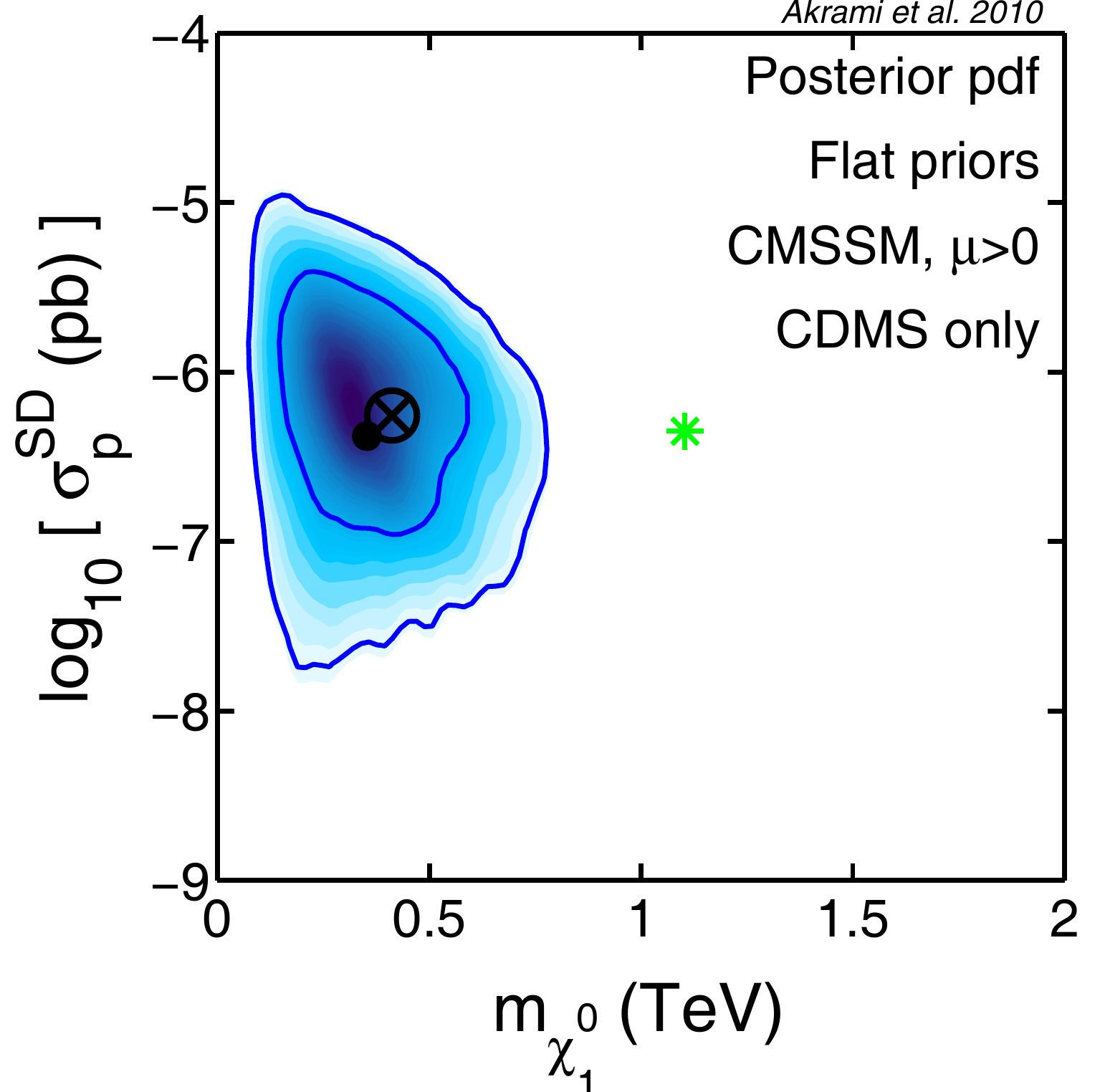}}
\subfigure{\includegraphics[scale=0.23, trim = 40 230 130 123, clip=true]{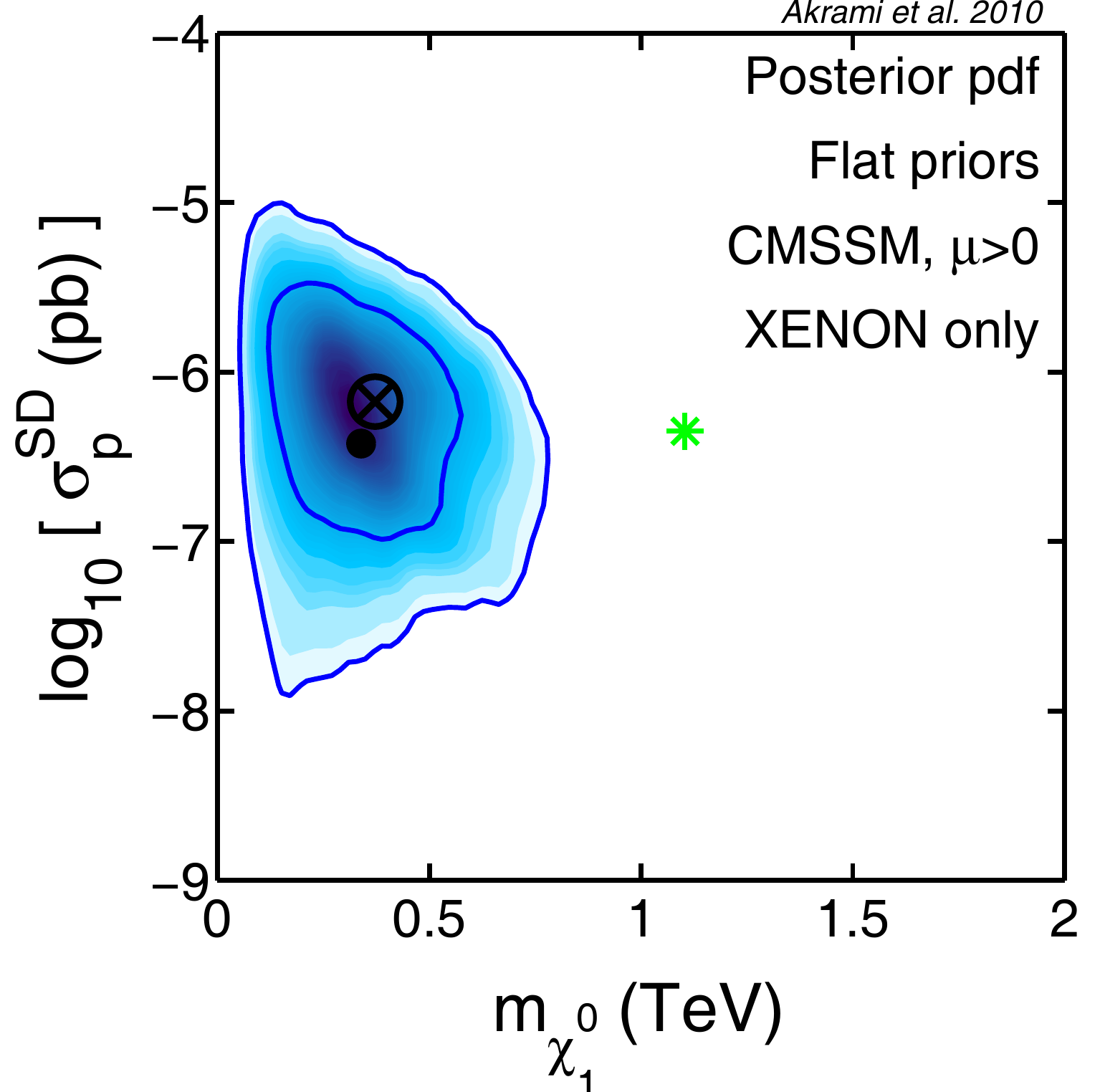}}
\subfigure{\includegraphics[scale=0.23, trim = 40 230 130 123, clip=true]{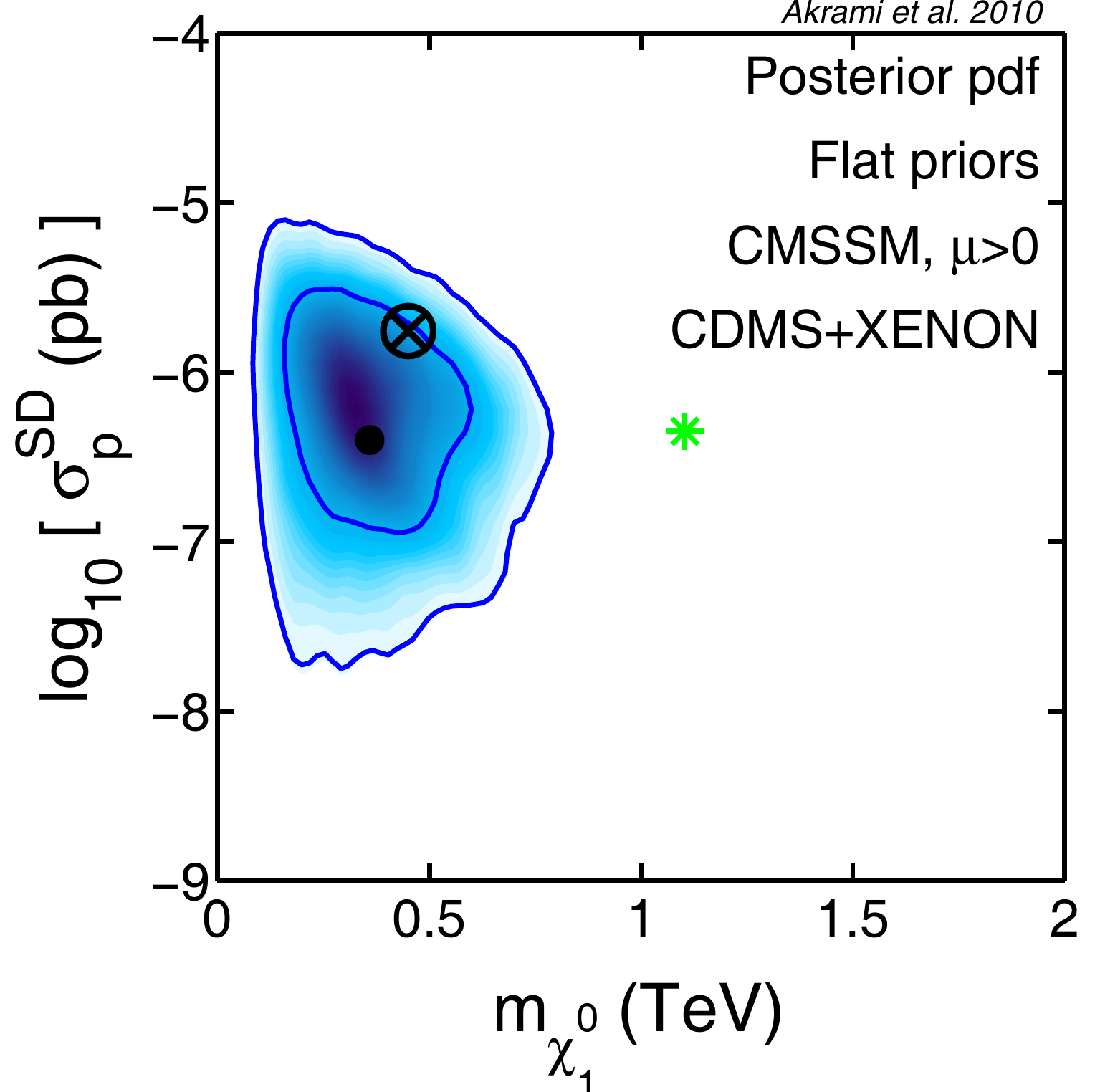}}
\subfigure{\includegraphics[scale=0.23, trim = 40 230 60 123, clip=true]{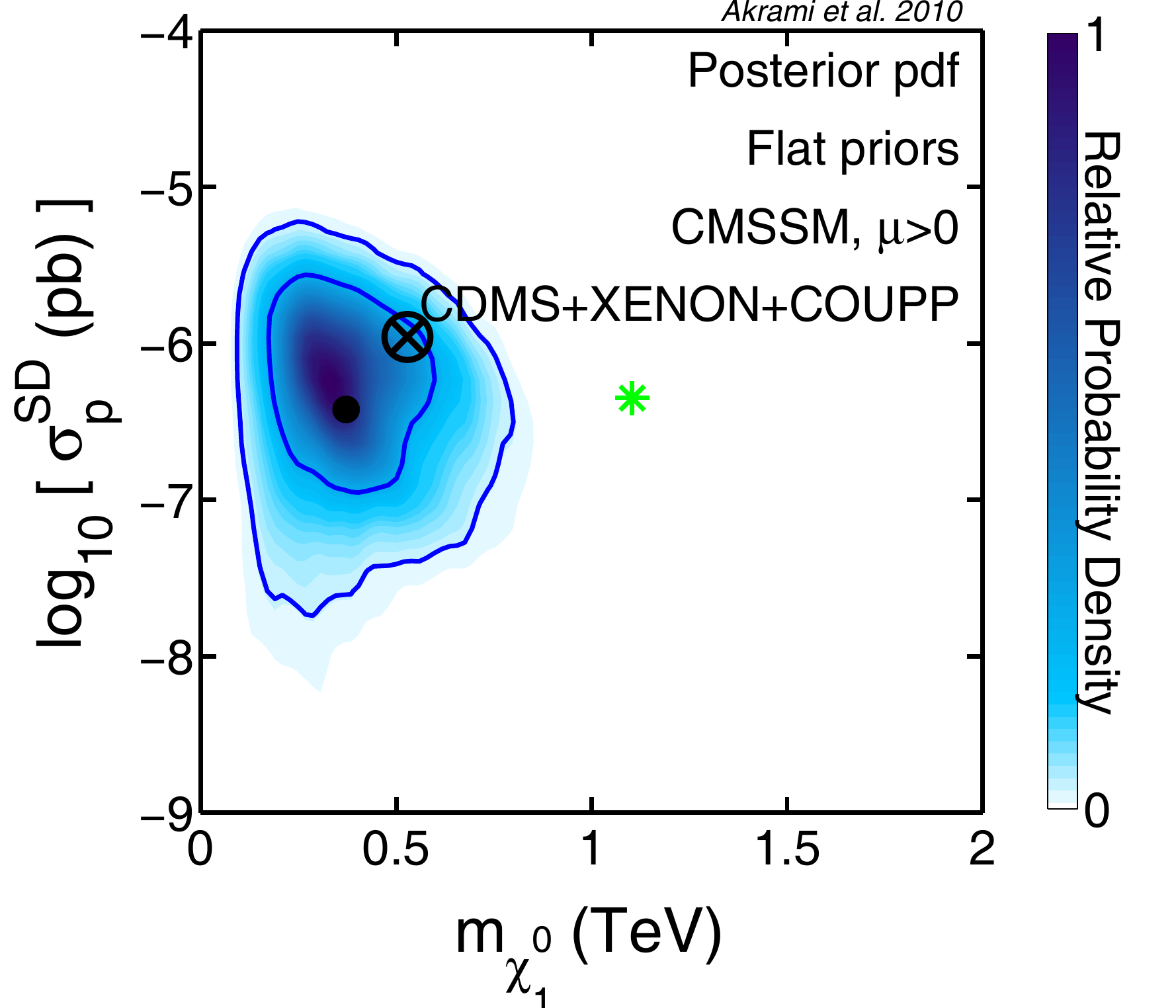}}\\
\subfigure{\includegraphics[scale=0.23, trim = 40 230 130 123, clip=true]{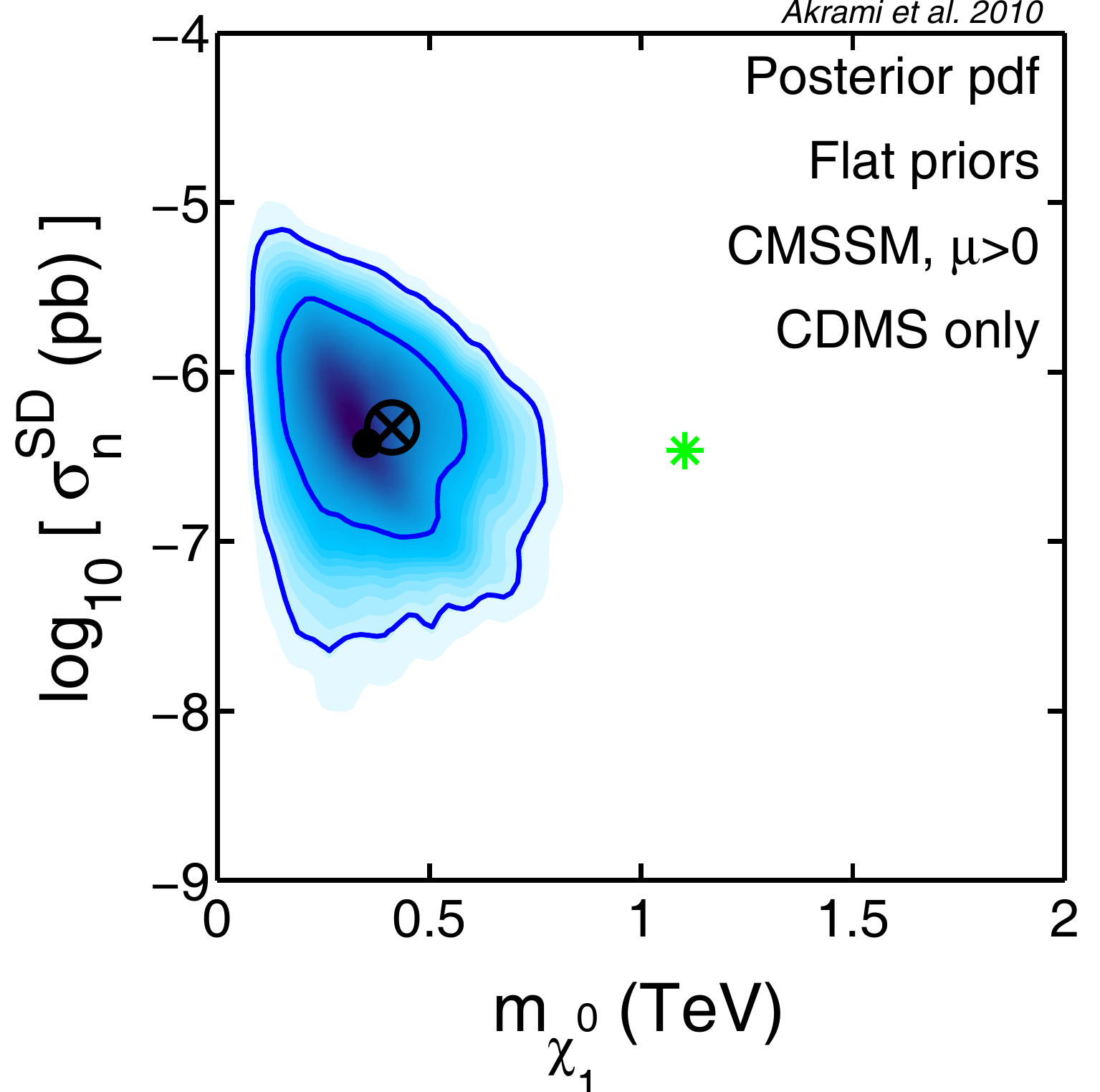}}
\subfigure{\includegraphics[scale=0.23, trim = 40 230 130 123, clip=true]{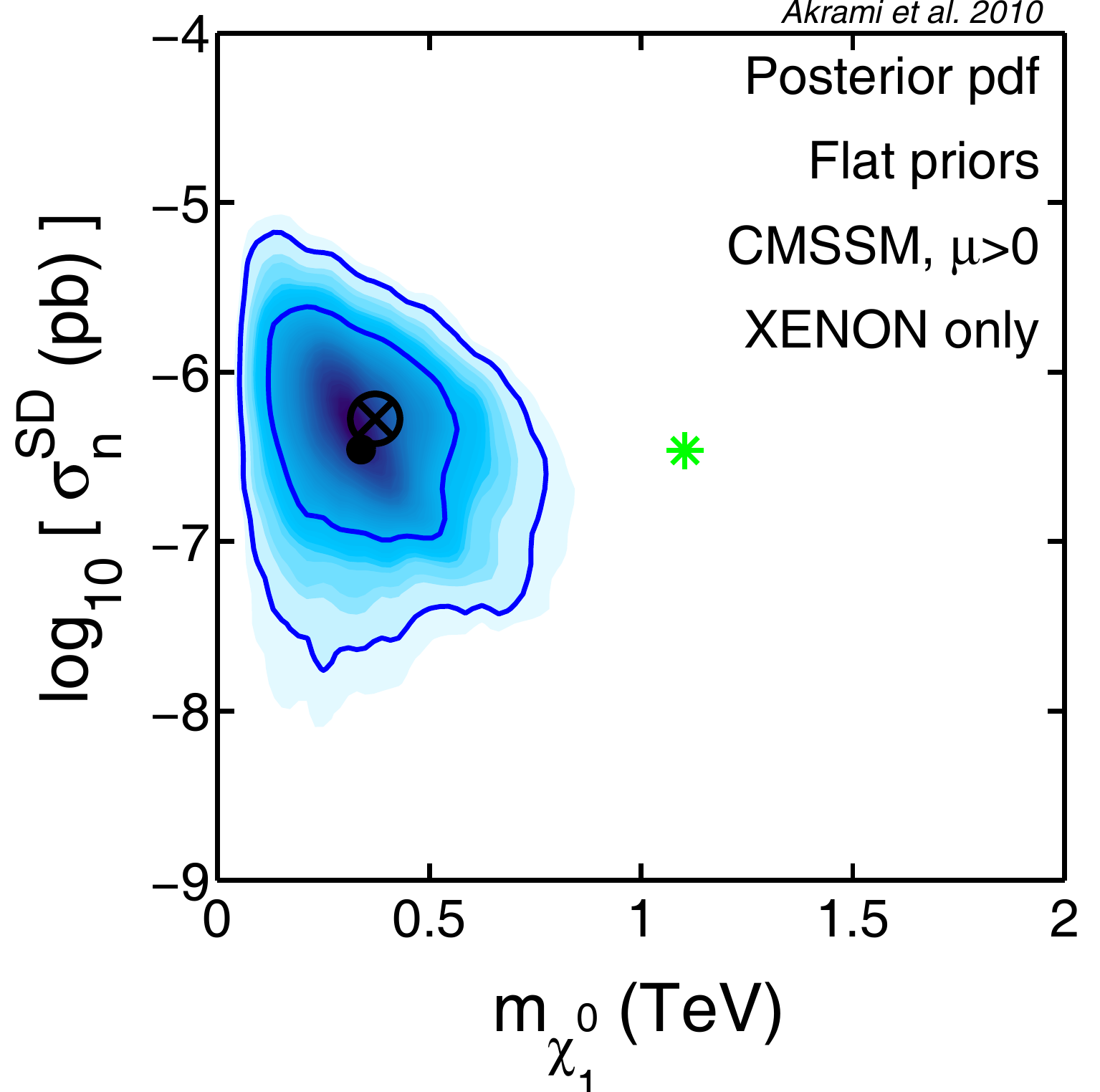}}
\subfigure{\includegraphics[scale=0.23, trim = 40 230 130 123, clip=true]{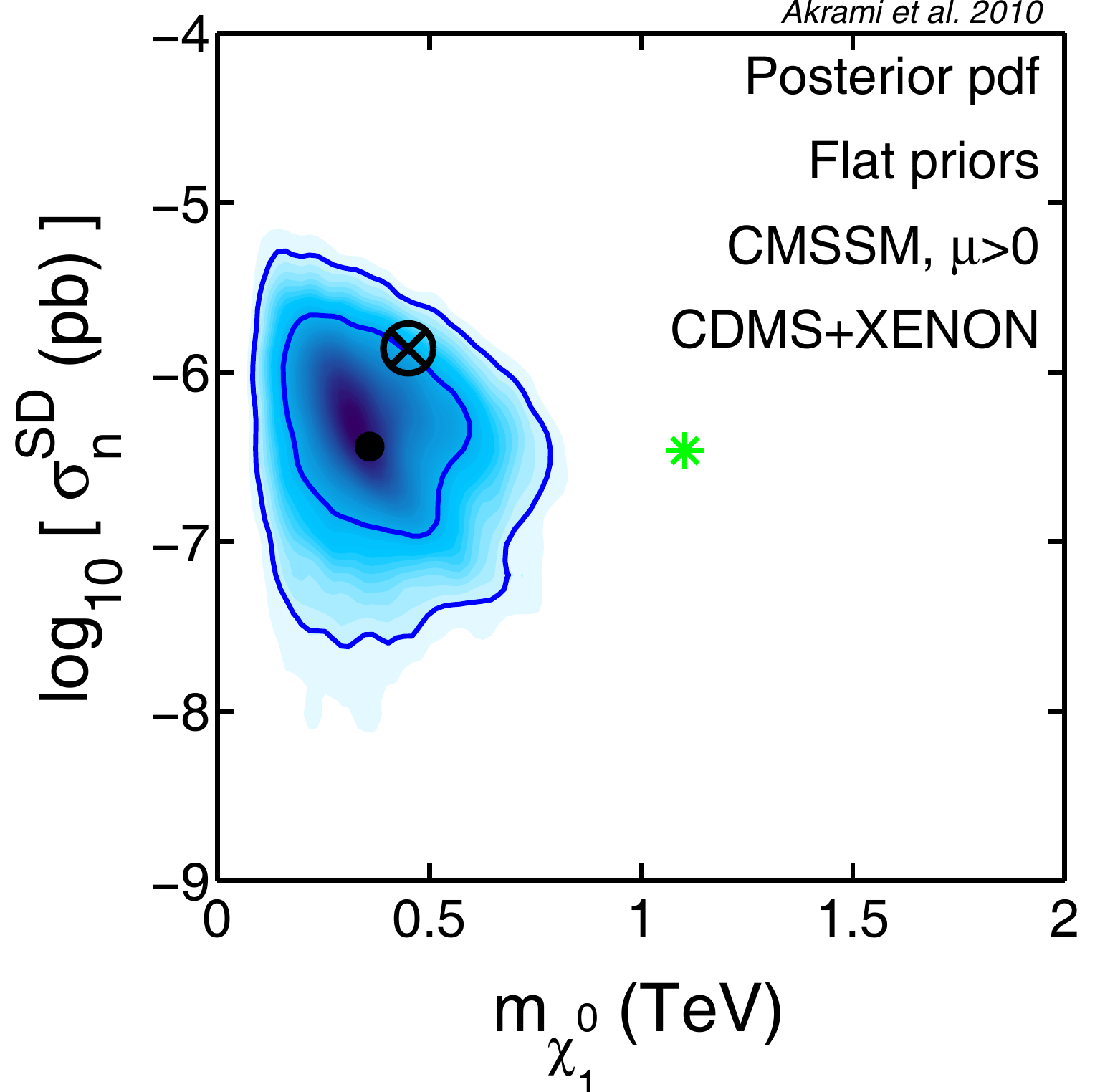}}
\subfigure{\includegraphics[scale=0.23, trim = 40 230 60 123, clip=true]{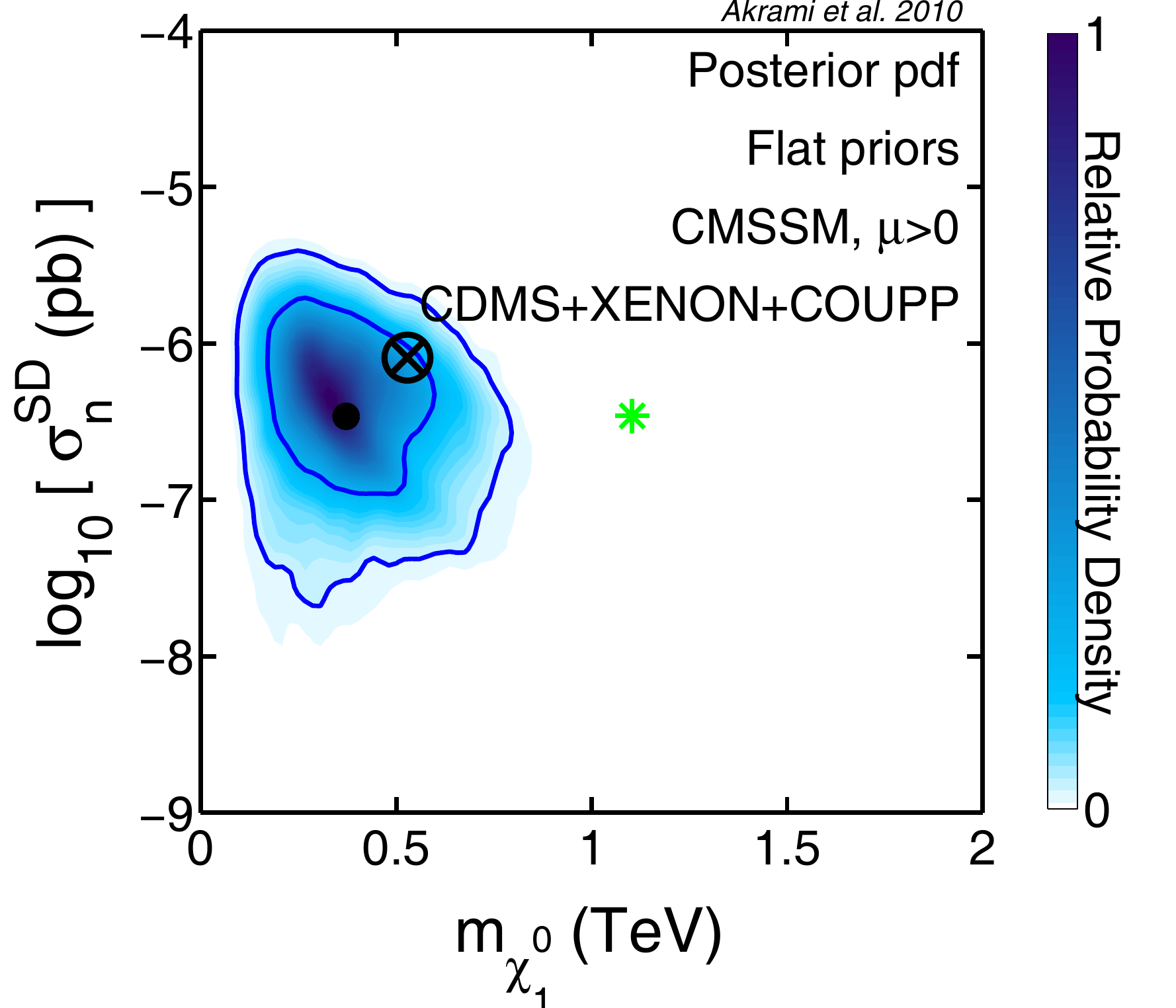}}\\
\caption[aa]{\footnotesize{As in Figs.~\ref{fig:LHmarg},~\ref{fig:LLmarg} and~\ref{fig:MMmarg}, but for benchmark 4.}}\label{fig:HHmarg}
\end{figure}

\begin{figure}[t]
\subfigure{\includegraphics[scale=0.23, trim = 40 230 130 123, clip=true]{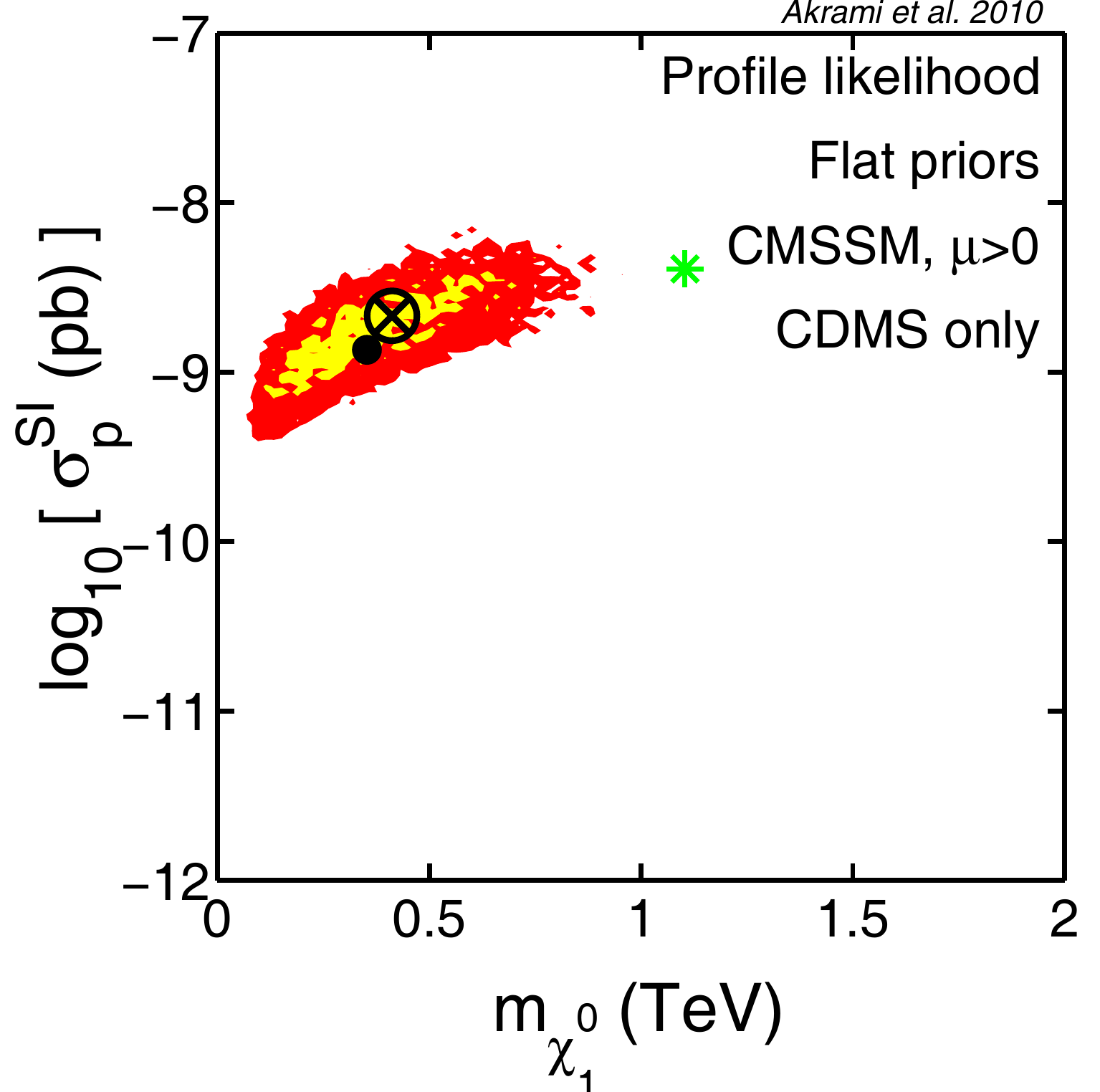}}
\subfigure{\includegraphics[scale=0.23, trim = 40 230 130 123, clip=true]{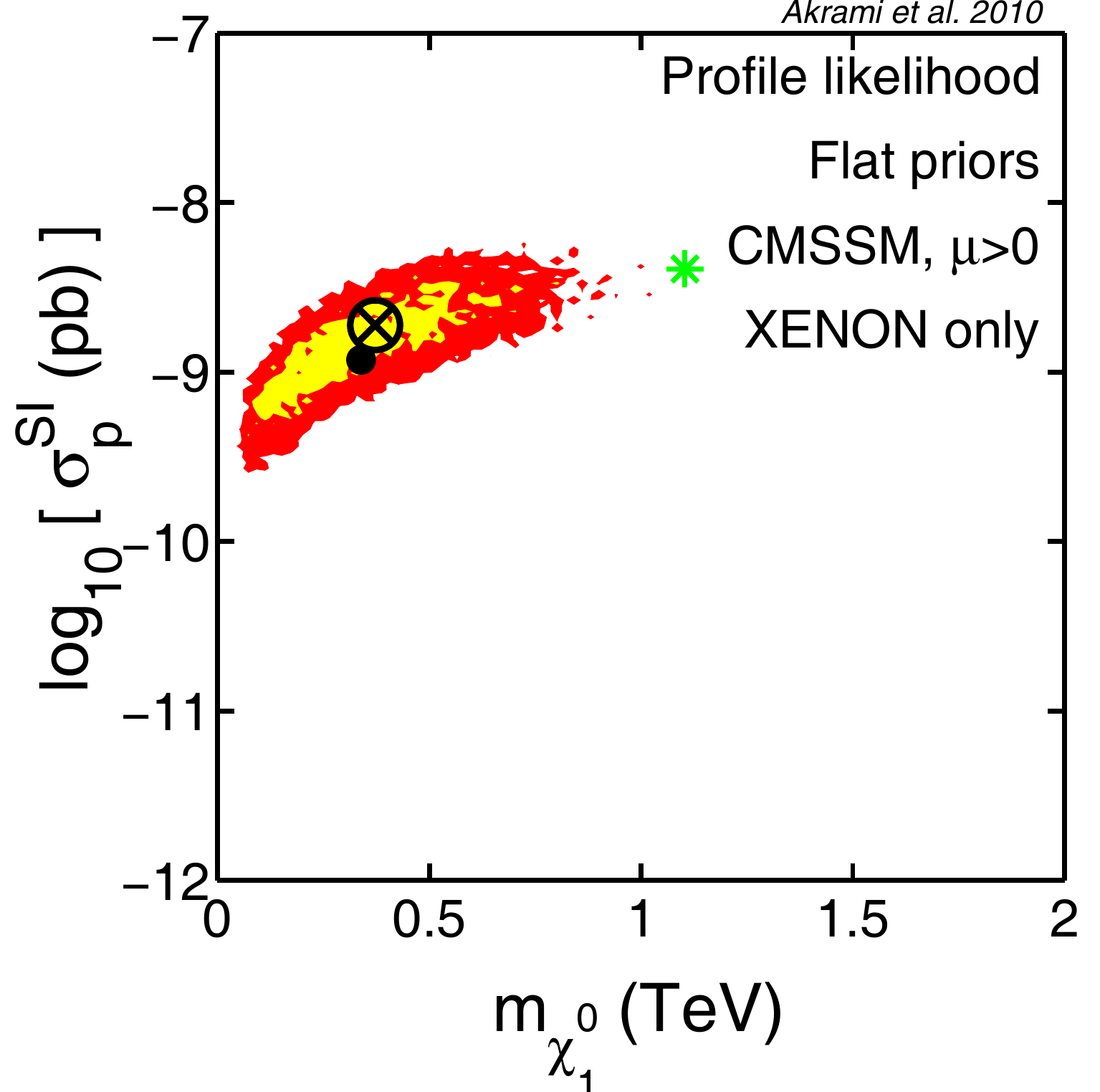}}
\subfigure{\includegraphics[scale=0.23, trim = 40 230 130 123, clip=true]{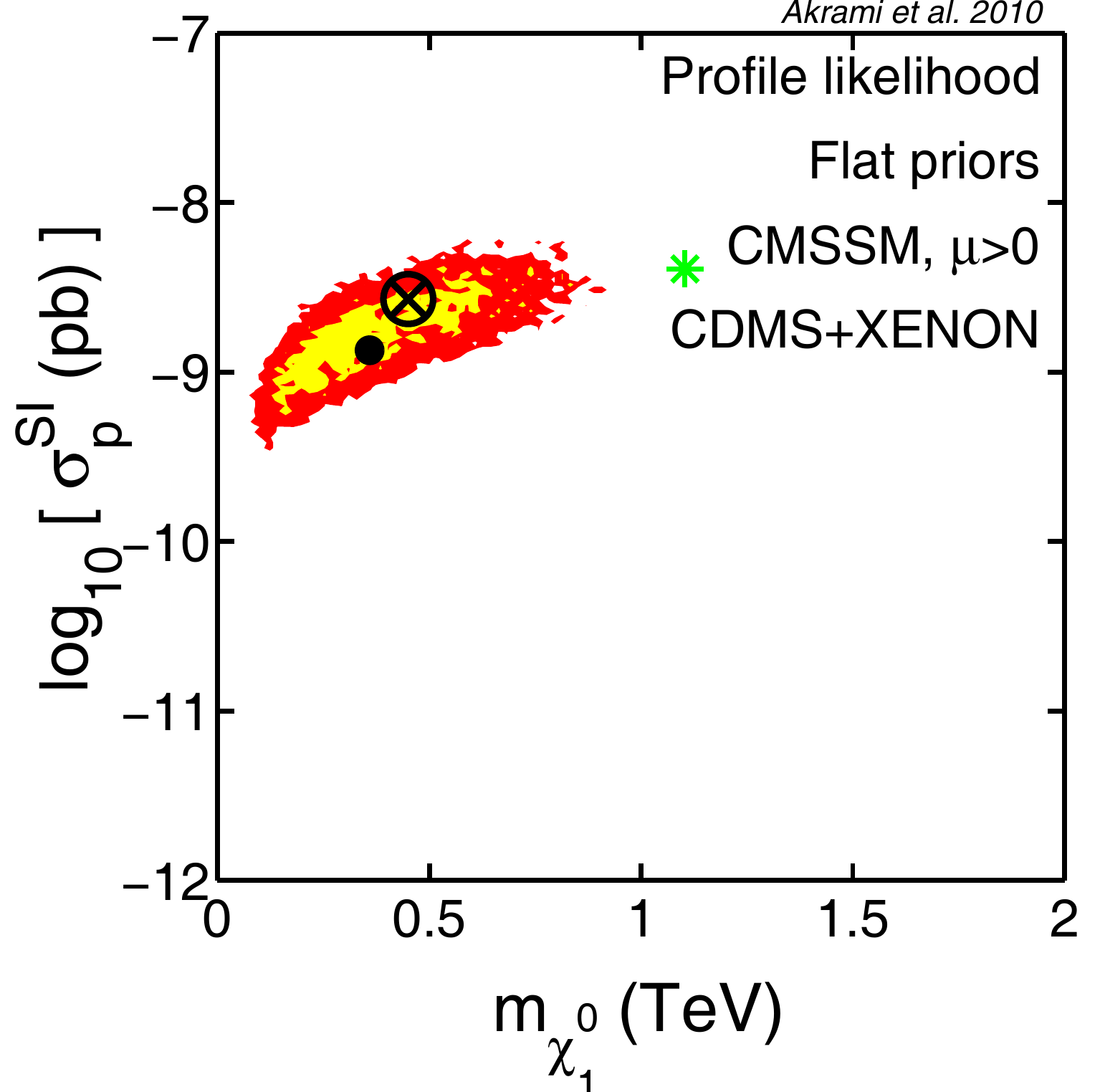}}
\subfigure{\includegraphics[scale=0.23, trim = 40 230 60 123, clip=true]{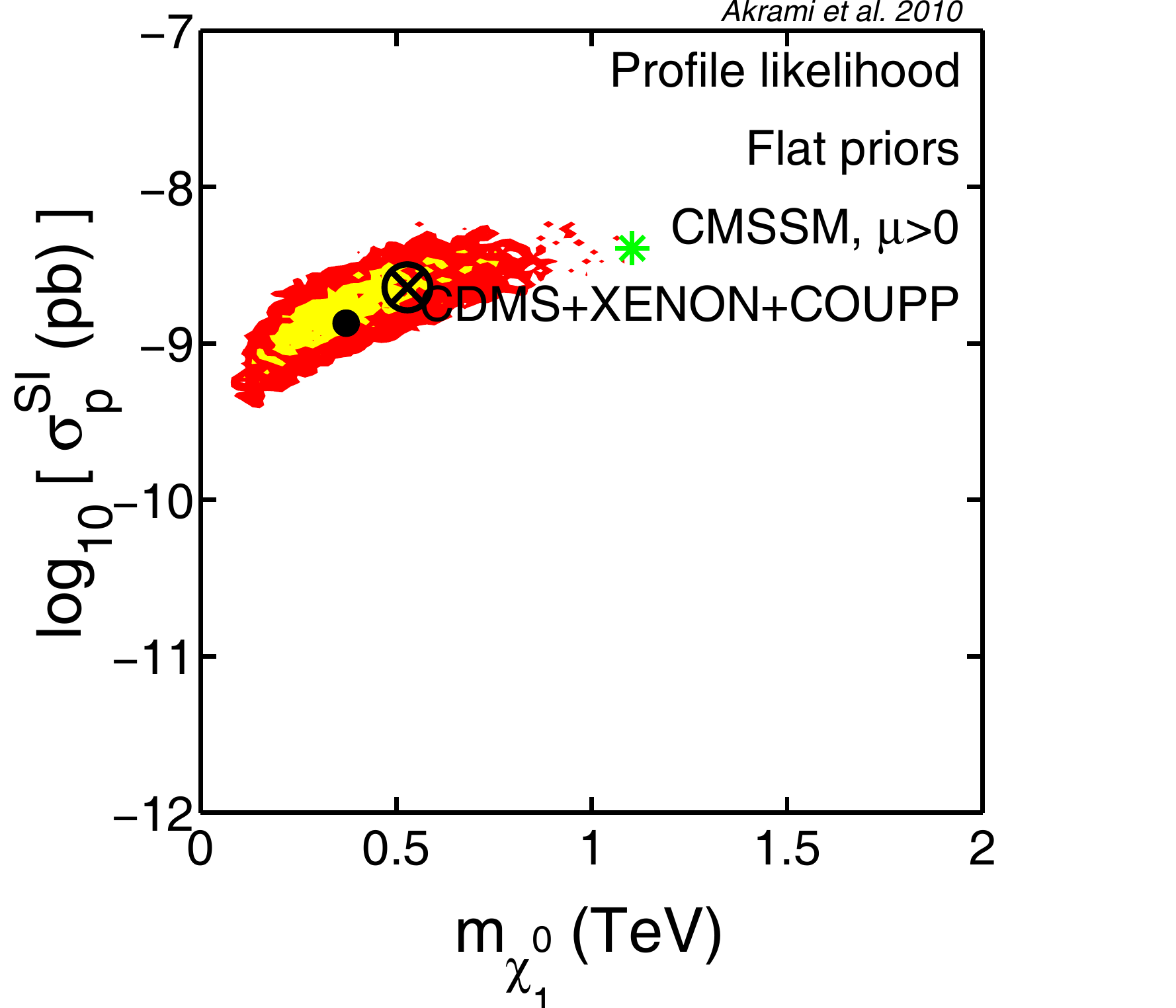}}\\
\subfigure{\includegraphics[scale=0.23, trim = 40 230 130 123, clip=true]{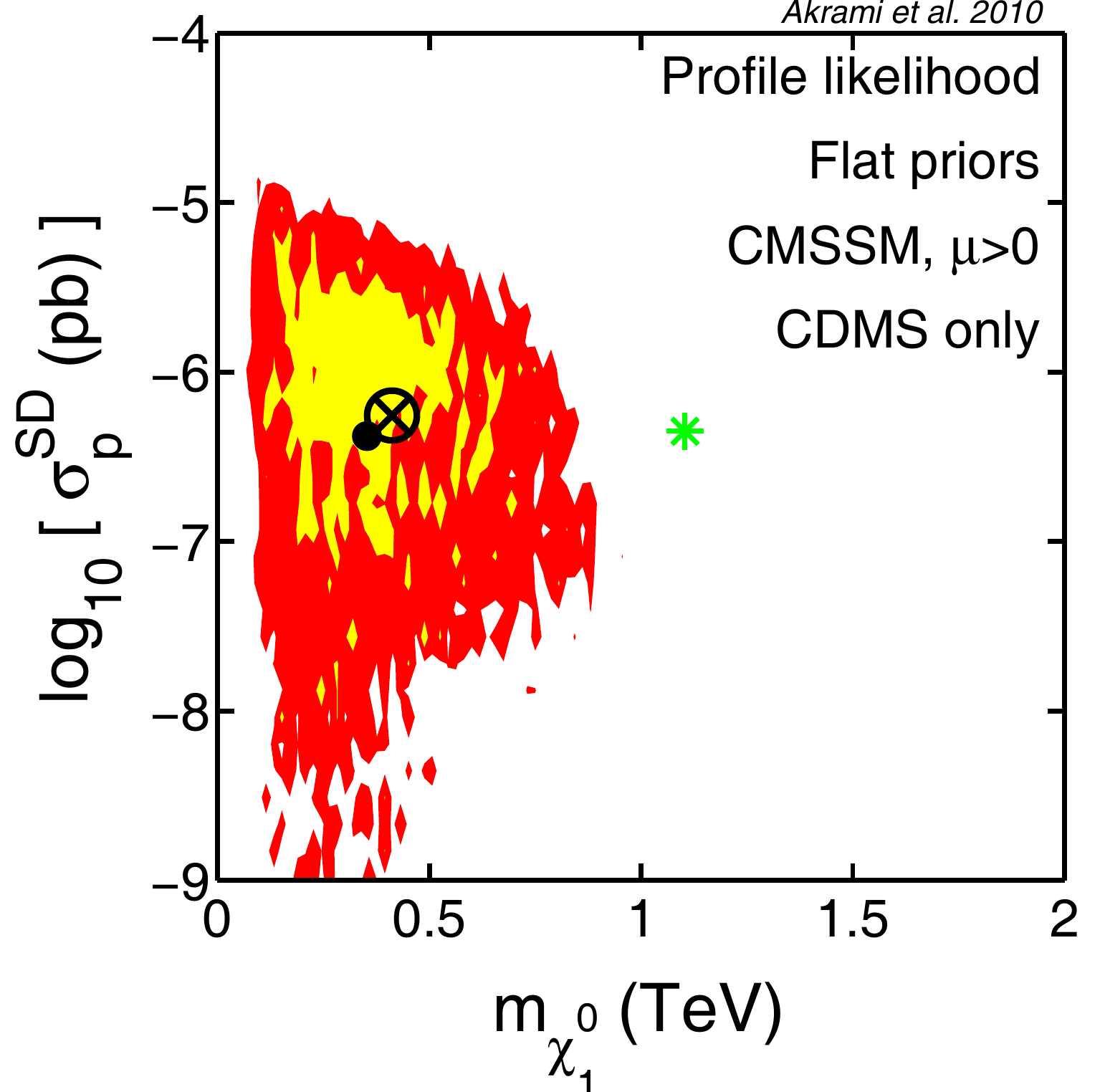}}
\subfigure{\includegraphics[scale=0.23, trim = 40 230 130 123, clip=true]{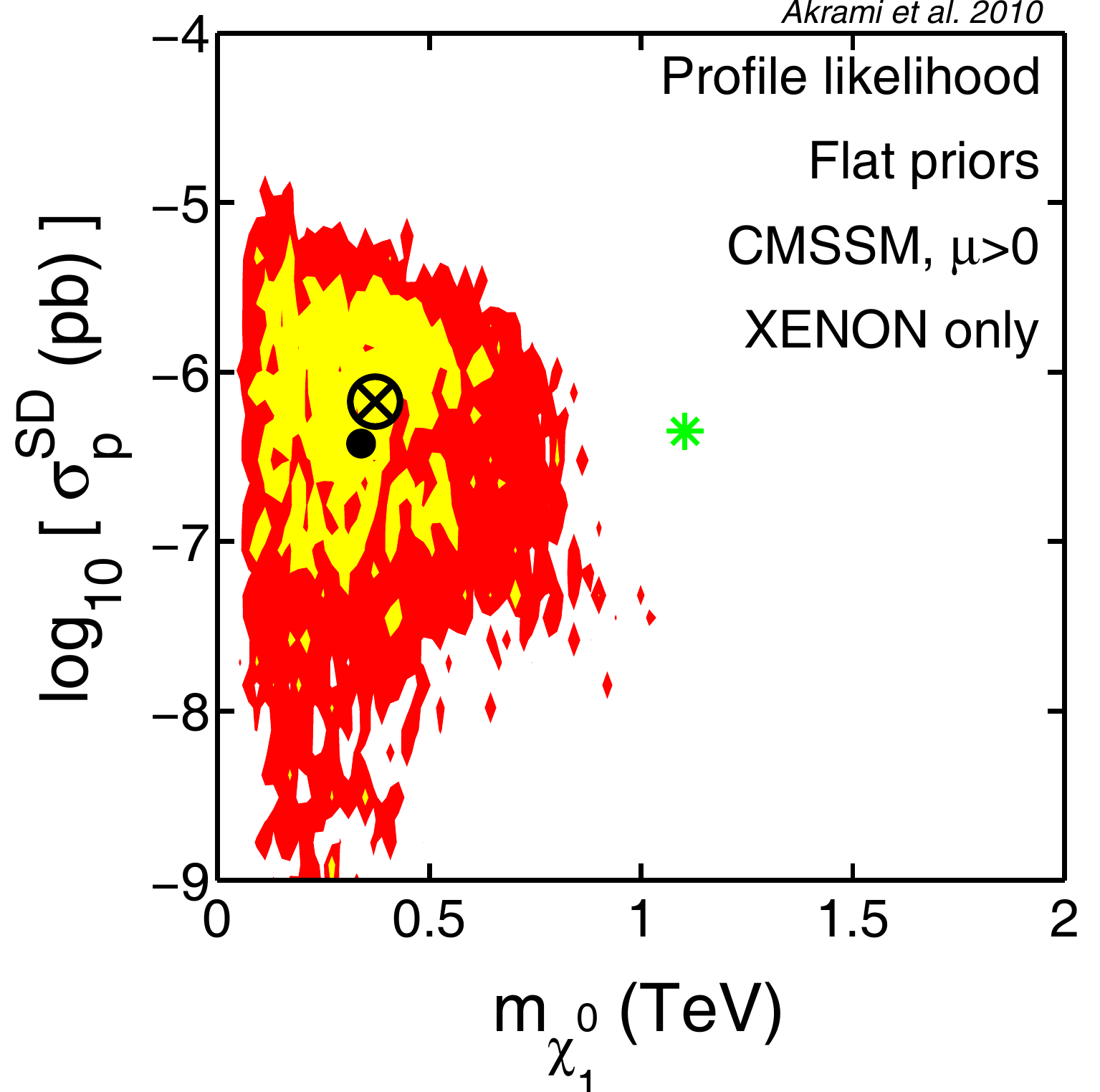}}
\subfigure{\includegraphics[scale=0.23, trim = 40 230 130 123, clip=true]{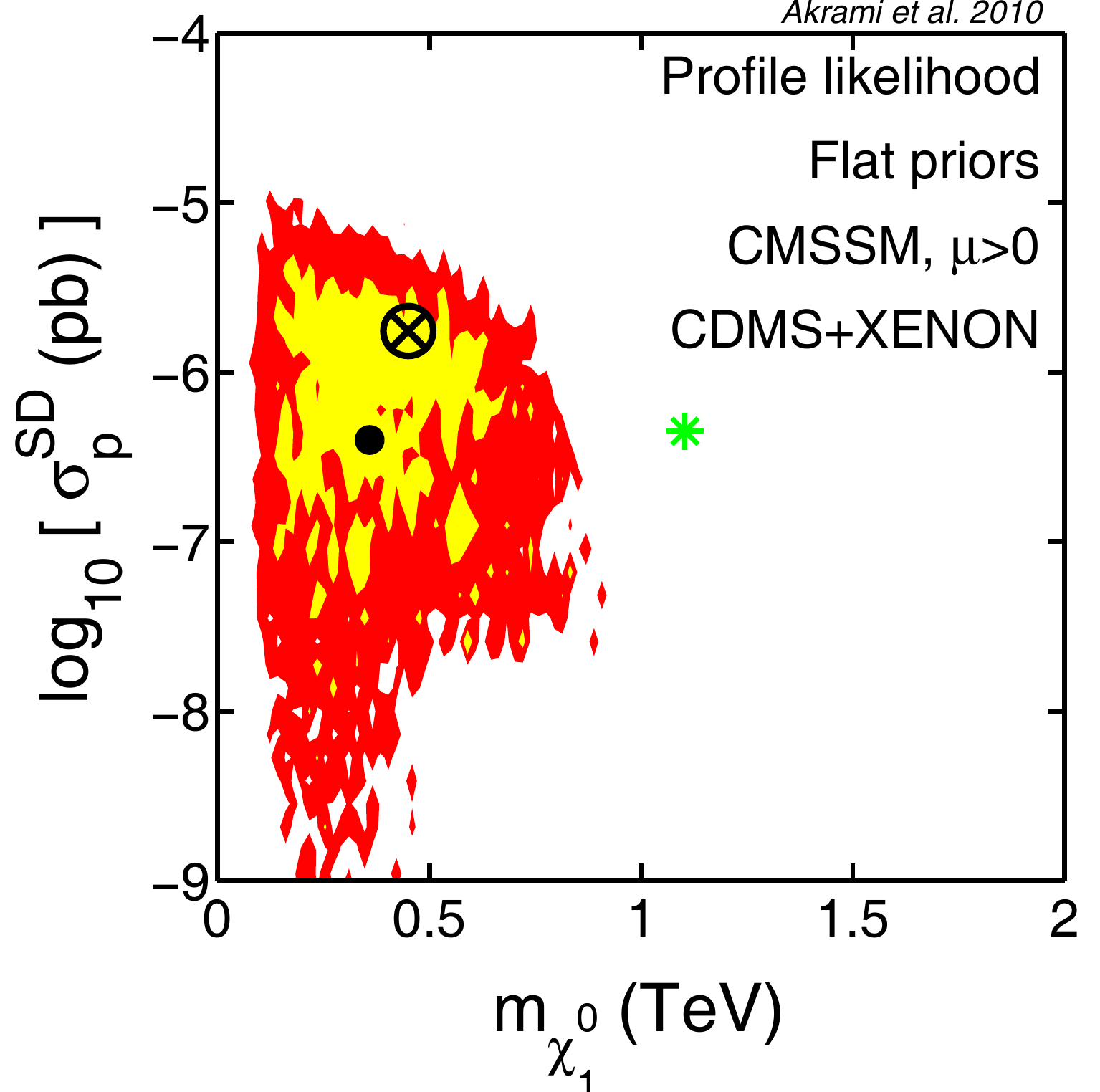}}
\subfigure{\includegraphics[scale=0.23, trim = 40 230 60 123, clip=true]{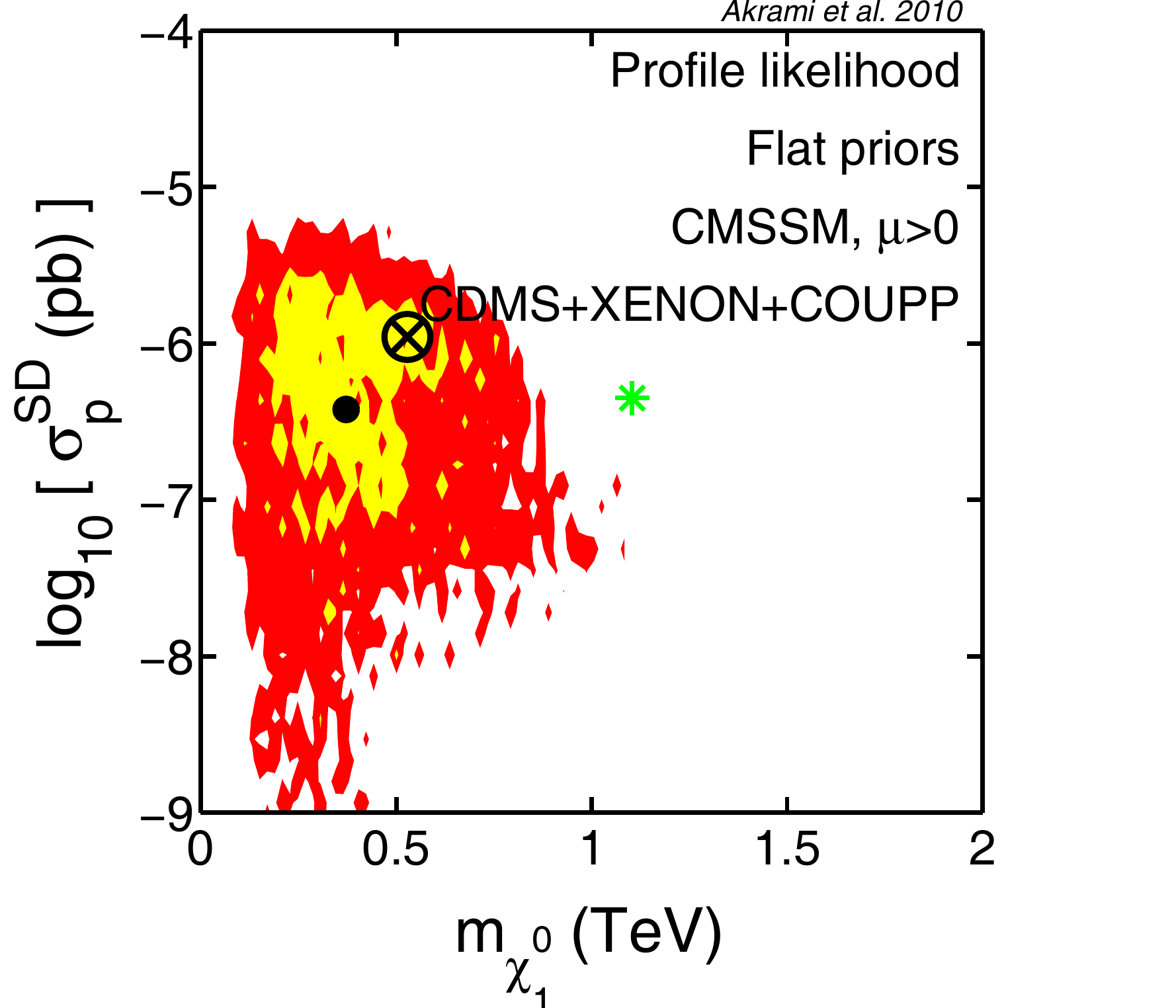}}\\
\subfigure{\includegraphics[scale=0.23, trim = 40 230 130 123, clip=true]{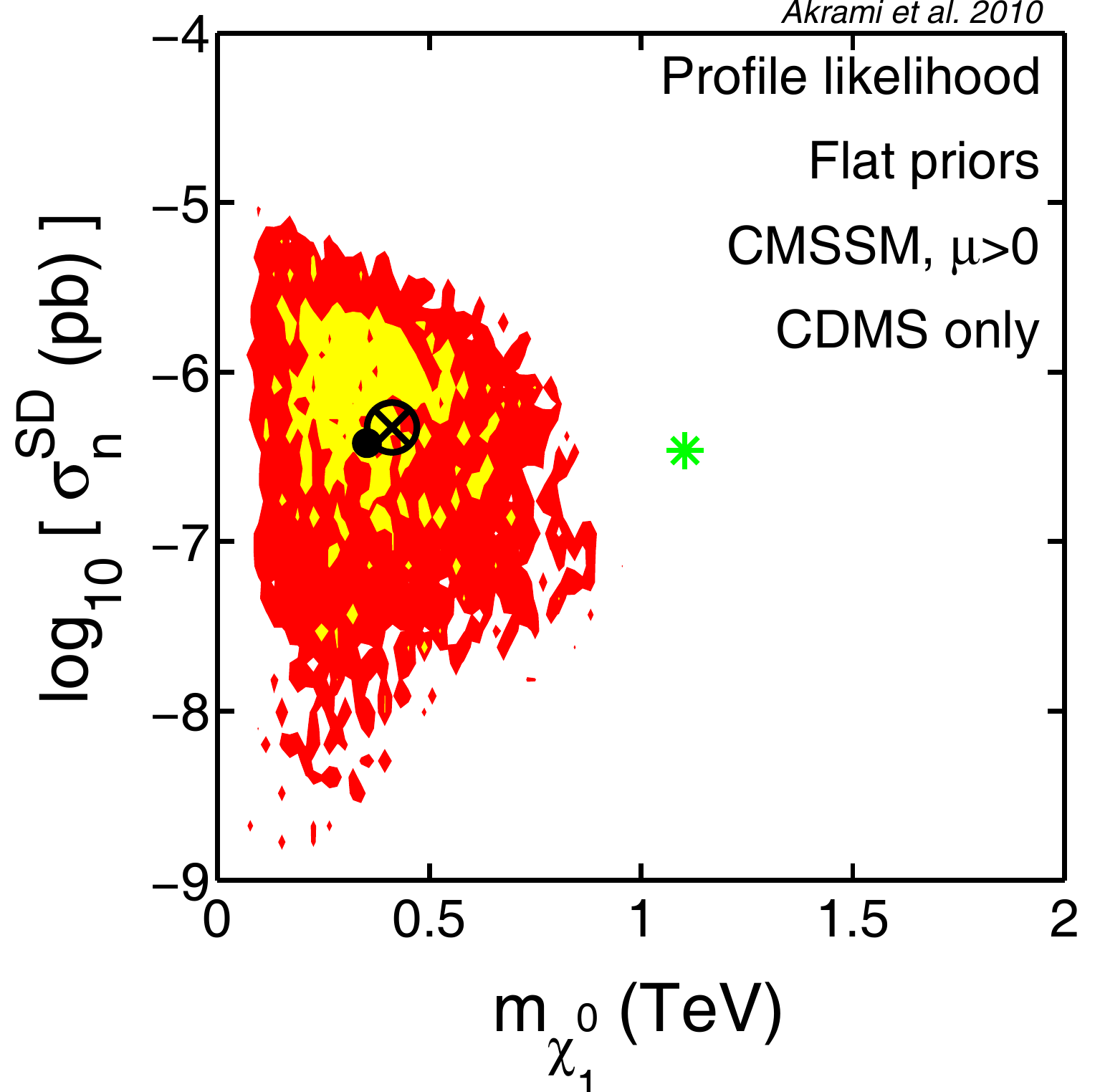}}
\subfigure{\includegraphics[scale=0.23, trim = 40 230 130 123, clip=true]{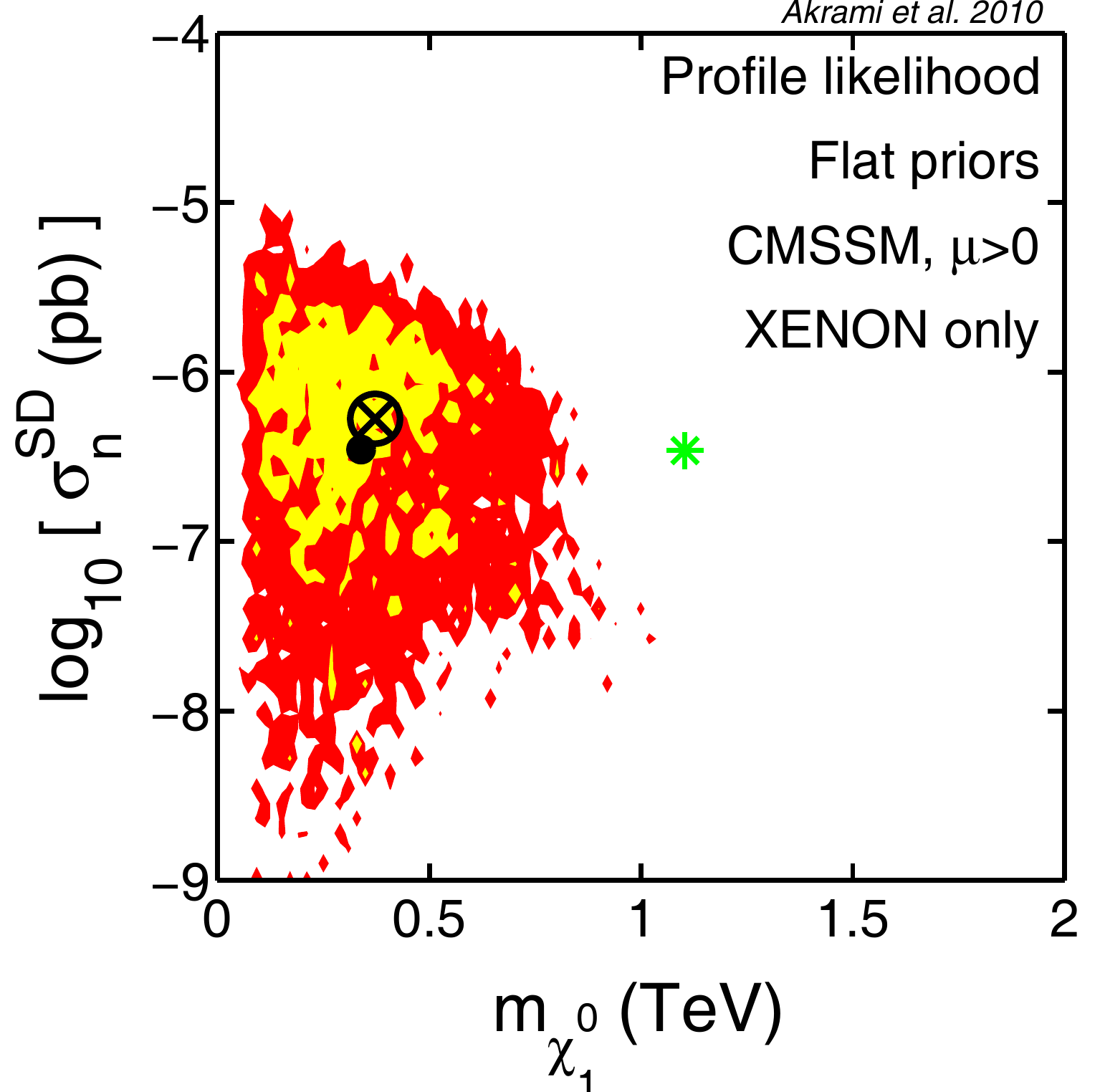}}
\subfigure{\includegraphics[scale=0.23, trim = 40 230 130 123, clip=true]{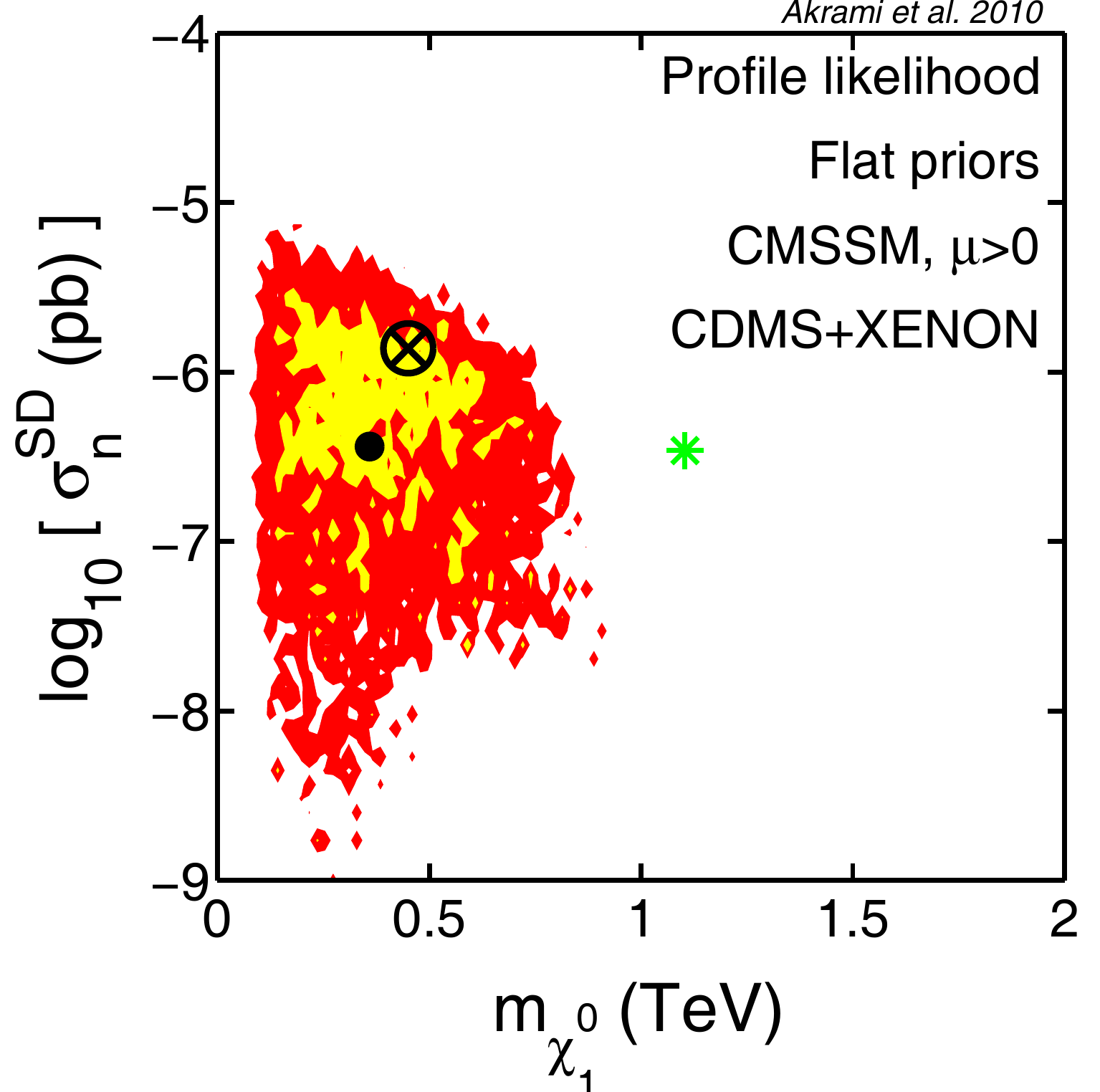}}
\subfigure{\includegraphics[scale=0.23, trim = 40 230 60 123, clip=true]{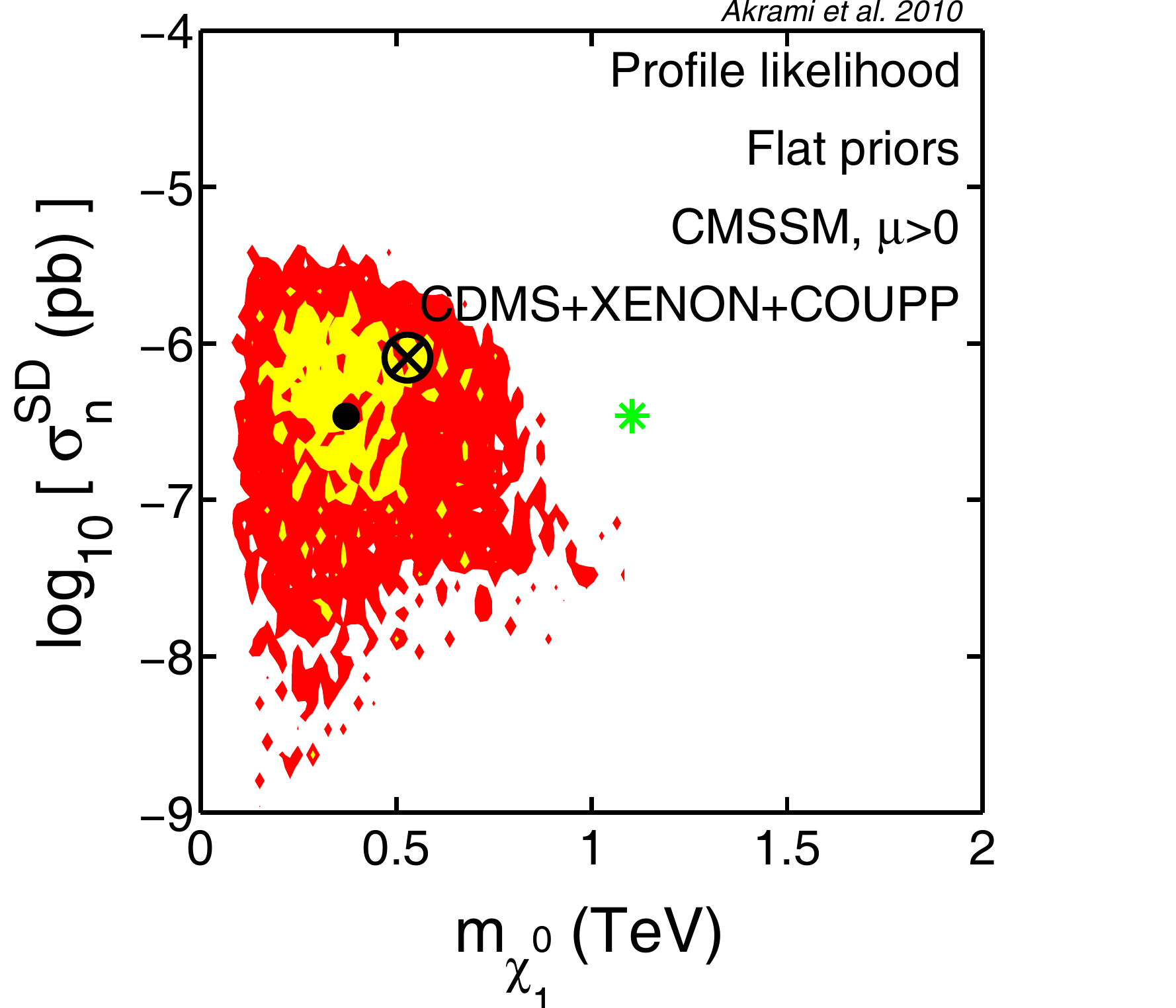}}\\
\caption[aa]{\footnotesize{As in Figs.~\ref{fig:LHprofl},~\ref{fig:LLprofl} and~\ref{fig:MMprofl}, but for benchmark 4.}}\label{fig:HHprofl}
\end{figure}

We then present our main results when synthetic data from the DD experiments CDMS1T, XENON1T and COUPP1T are included. They are shown in Figs~\ref{fig:LHmarg}$-$\ref{fig:HHprofl} for different benchmarks. For each benchmark, we first present two-dimensional marginal posteriors for the three scattering cross-sections $\sigma^{SI}_p$, $\sigma^{SD}_p$ and $\sigma^{SD}_n$ versus the neutralino mass $m_{\tilde\chi^0_1}$. Again, as in~\fig{fig:PriorsOnly}, $1\sigma$ and $2\sigma$ contours are shown in dark and light blue, respectively. Corresponding two-dimensional profile likelihoods are given in separate figures for the same planes of observables. Here $1\sigma$ and $2\sigma$ contours are depicted in yellow and red, and posterior means and best-fit points are denoted by black dots and crosses, respectively. In all plots, we also mark the locations of the benchmark values by green stars.

In each figure, we show the credible/confidence regions for different combinations of likelihoods for DD experiments: the first column in the left corresponds to the case where only CDMS1T is used, the second is for XENON1T only, the third corresponds to a combination of CDMS1T and XENON1T, and the last column shows the regions when all three experiments, CDMS1T, XENON1T and COUPP1T, are taken into account.
Regions for COUPP1T alone were not determined as, without a measurement
of event energies (the recoil spectrum), COUPP1T has almost no ability
to constrain the neutralino mass and provides much weaker overall
constraints than CDMS1T or XENON1T alone.  However, as will be discussed
below, COUPP1T can prove to be very useful in constraining various
quantities when combined with one or both of the other experiments.

\begin{figure}[t]
\subfigure[][\footnotesize{\textbf{Benchmark 1:}}]{\includegraphics[scale=0.23, trim = 40 230 130 110, clip=true]{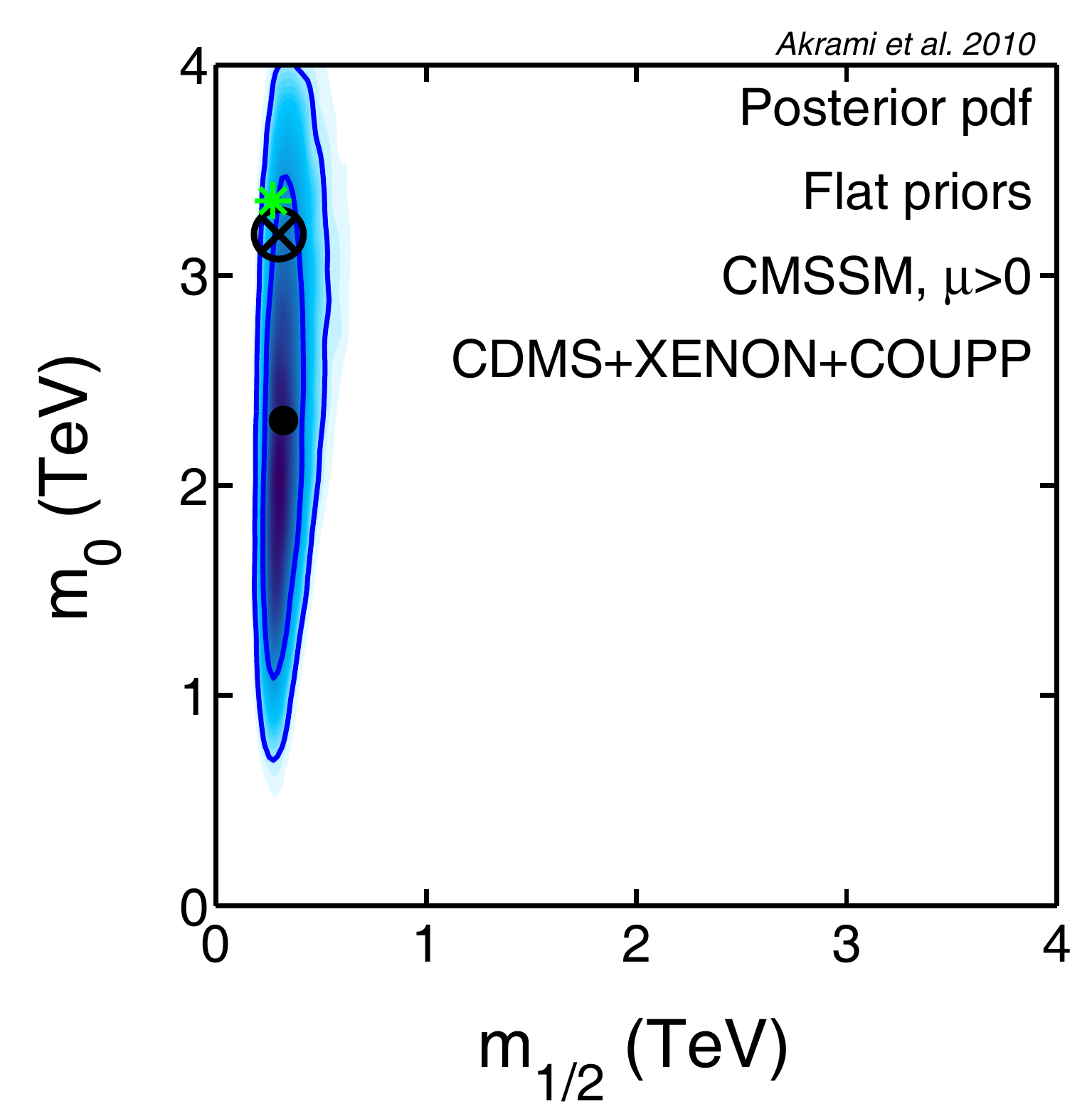}}
\subfigure[][\footnotesize{\textbf{Benchmark 2:}}]{\includegraphics[scale=0.23, trim = 40 230 130 110, clip=true]{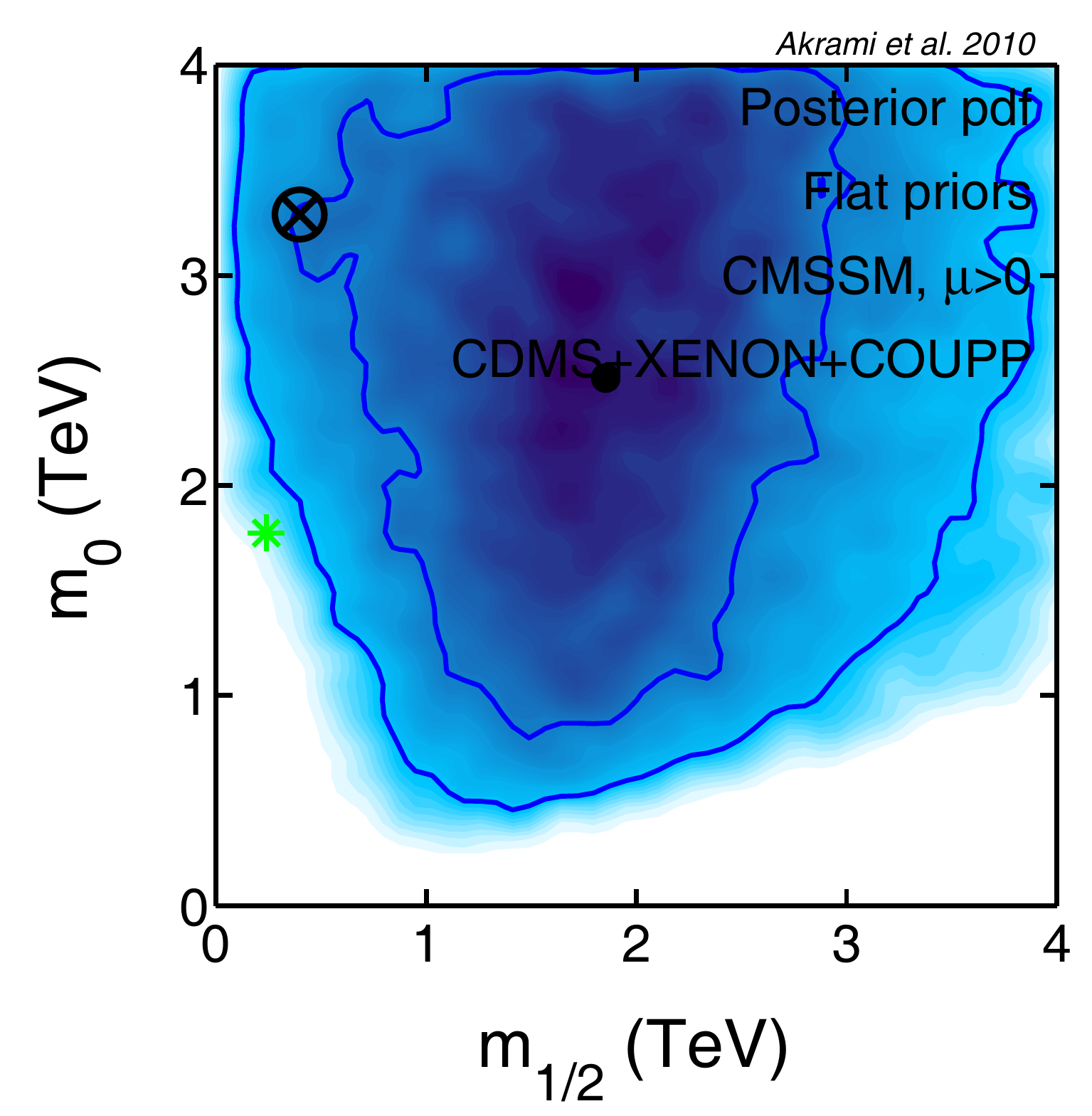}}
\subfigure[][\footnotesize{\textbf{Benchmark 3:}}]{\includegraphics[scale=0.23, trim = 40 230 130 110, clip=true]{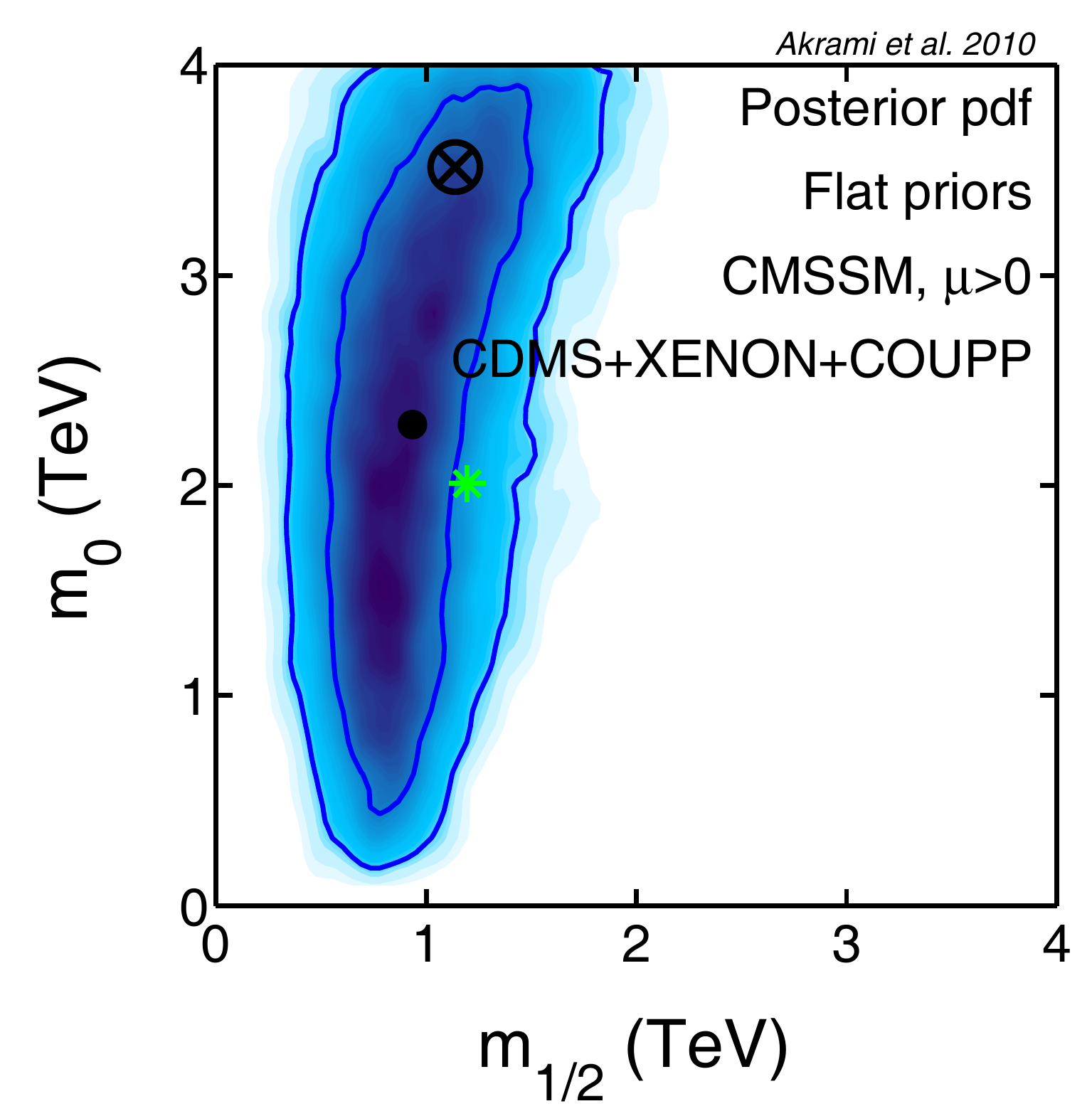}}
\subfigure[][\footnotesize{\textbf{Benchmark 4:}}]{\includegraphics[scale=0.23, trim = 40 230 60 110, clip=true]{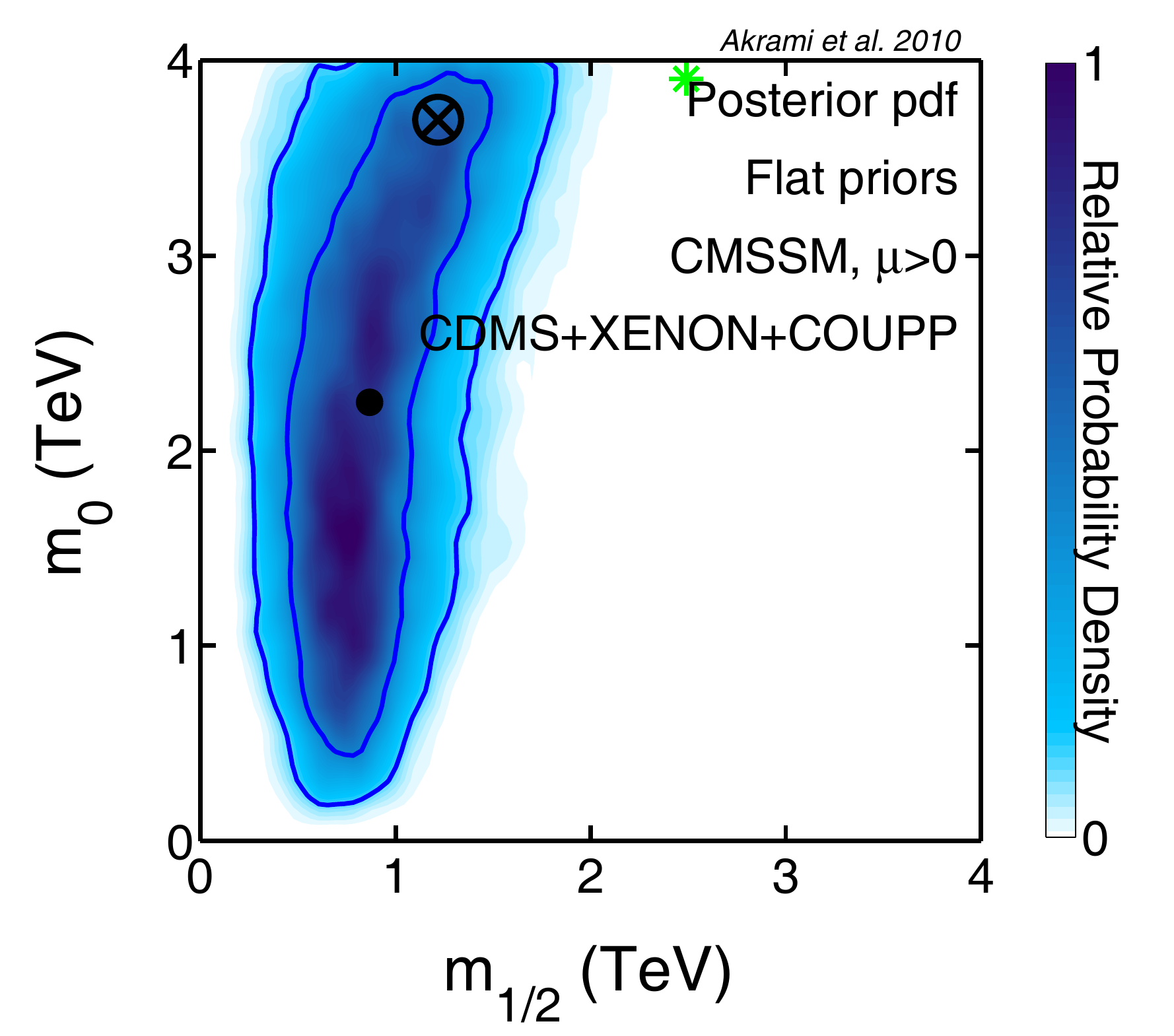}}\\
\subfigure{\includegraphics[scale=0.23, trim = 40 230 130 123, clip=true]{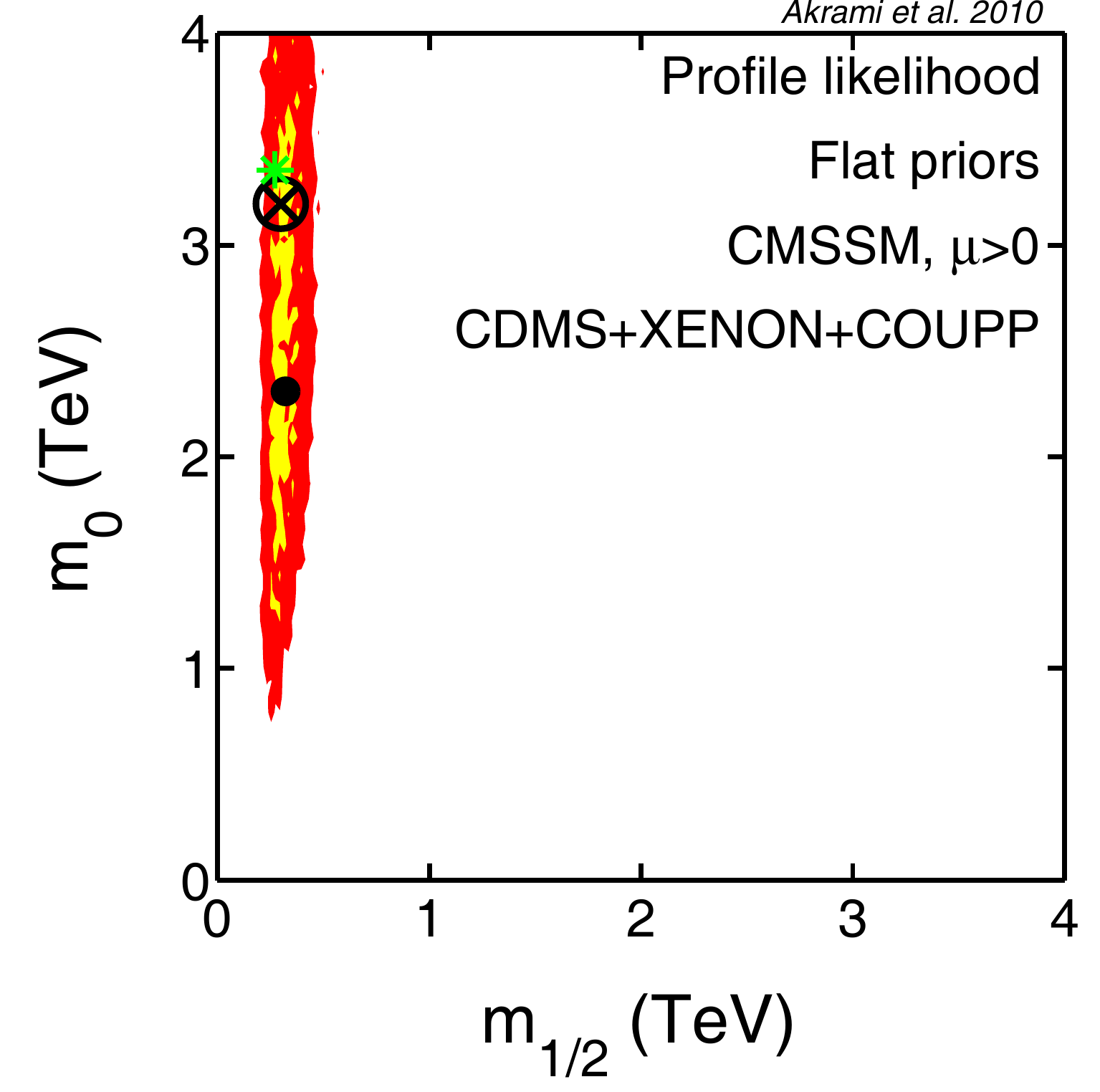}}
\subfigure{\includegraphics[scale=0.23, trim = 40 230 130 123, clip=true]{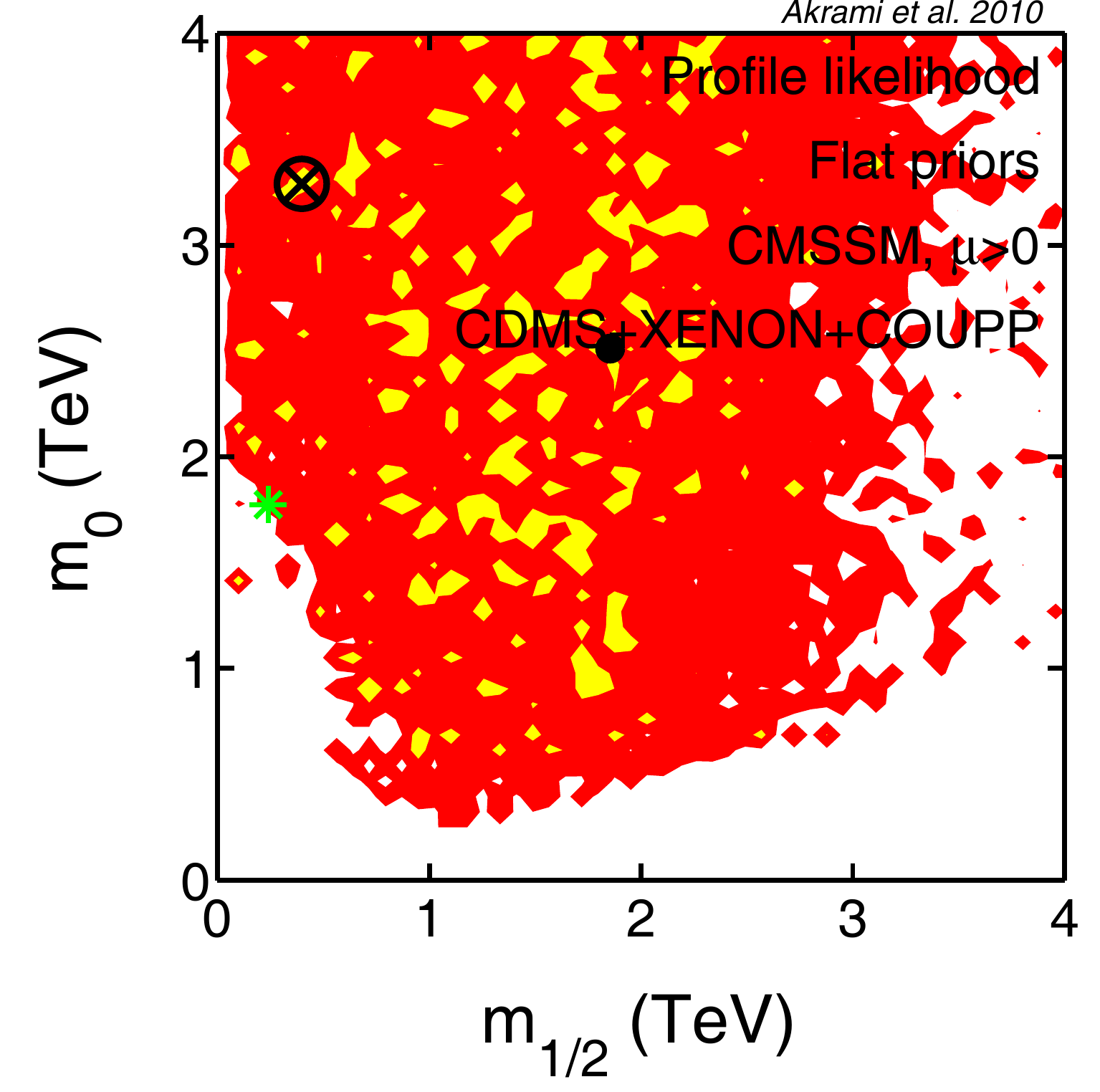}}
\subfigure{\includegraphics[scale=0.23, trim = 40 230 130 123, clip=true]{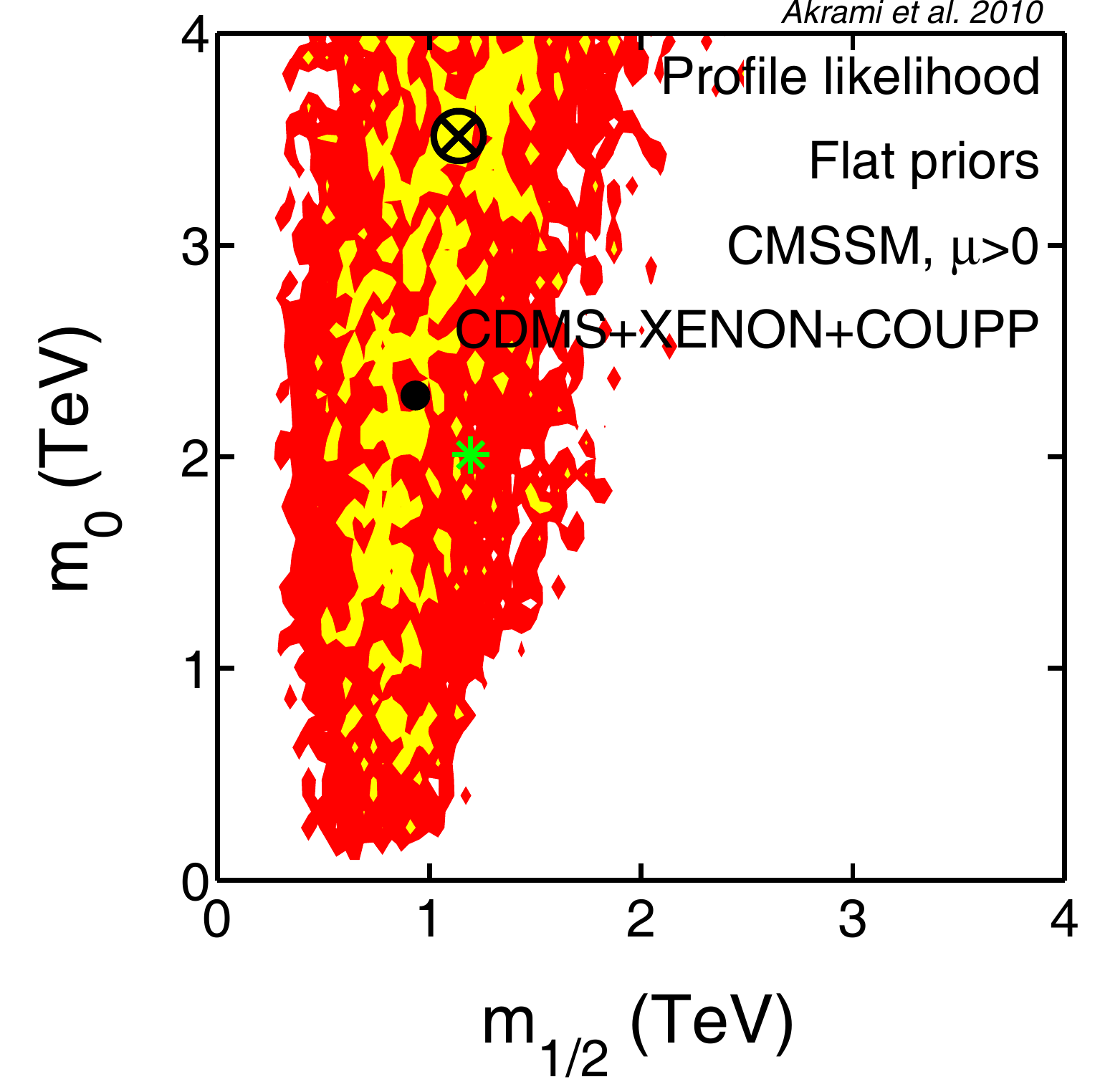}}
\subfigure{\includegraphics[scale=0.23, trim = 40 230 60 123, clip=true]{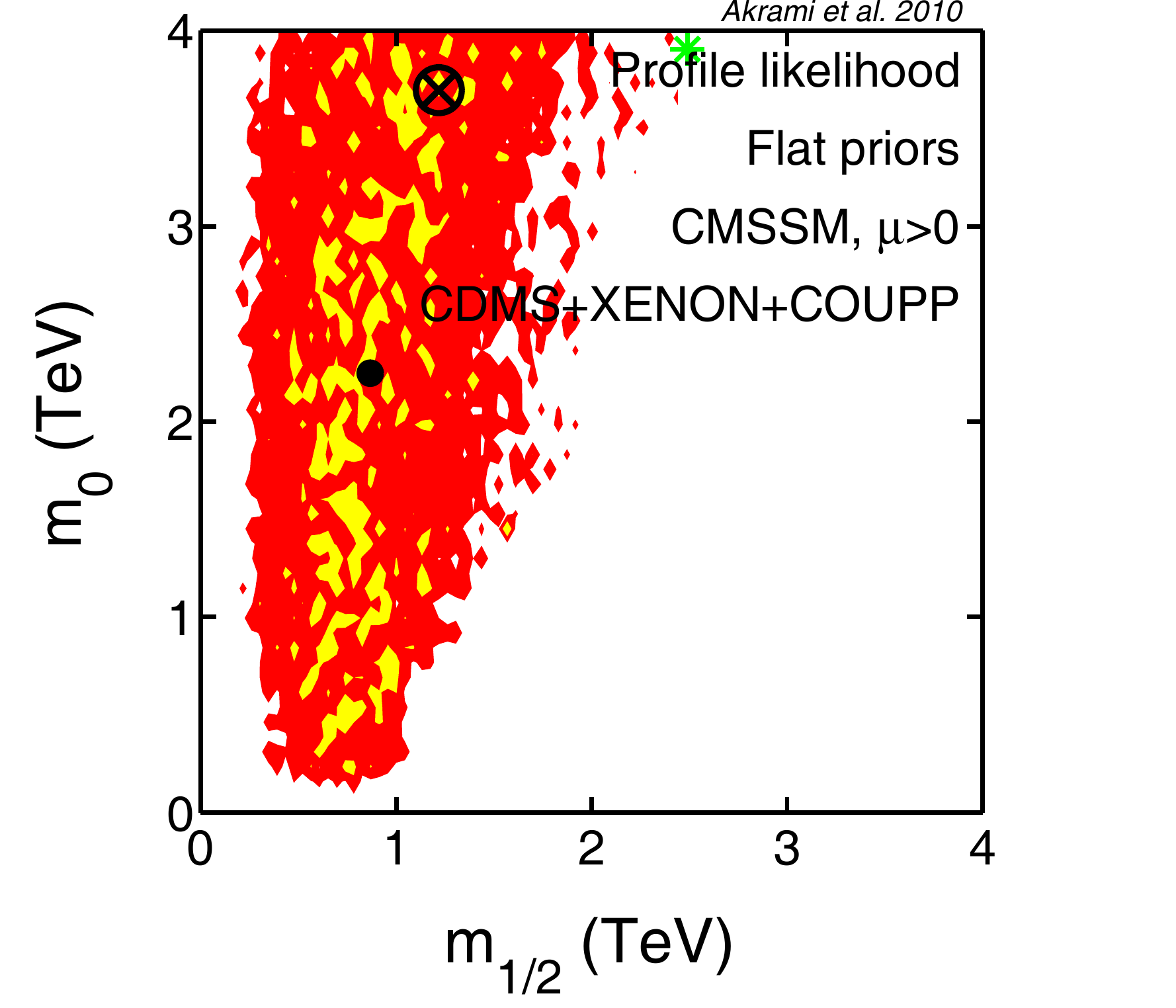}}\\
\subfigure{\includegraphics[scale=0.23, trim = 40 230 130 123, clip=true]{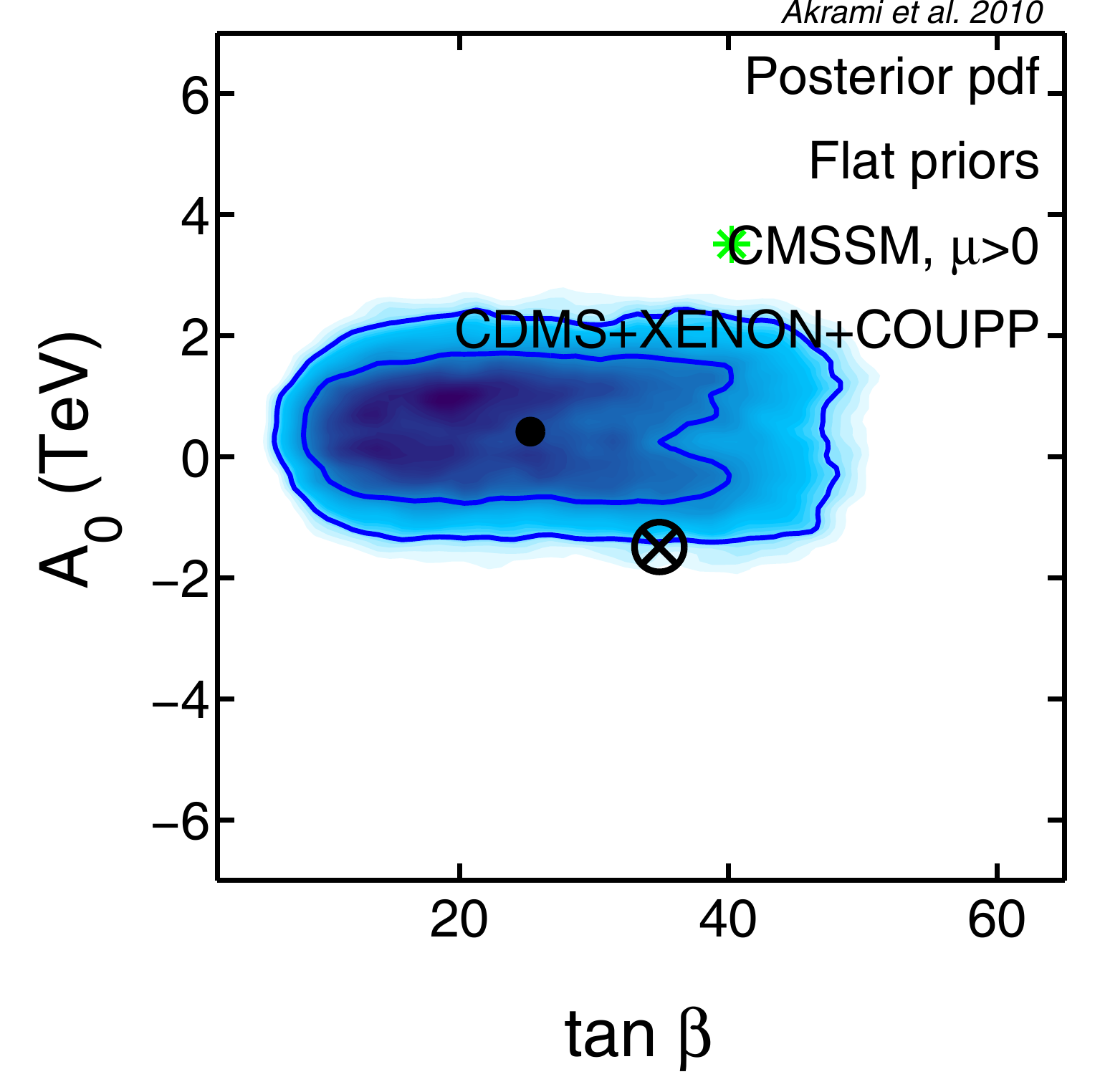}}
\subfigure{\includegraphics[scale=0.23, trim = 40 230 130 123, clip=true]{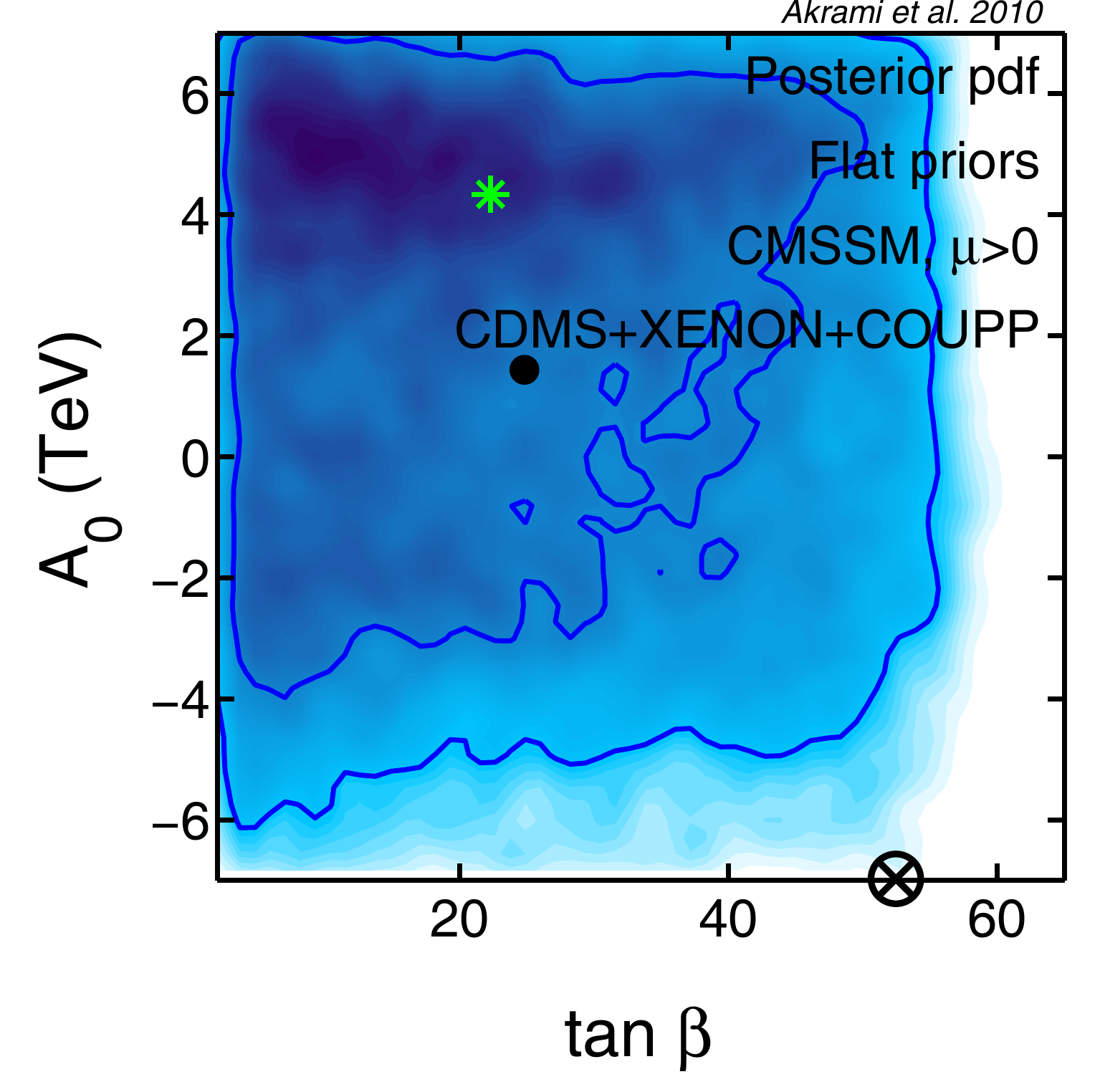}}
\subfigure{\includegraphics[scale=0.23, trim = 40 230 130 123, clip=true]{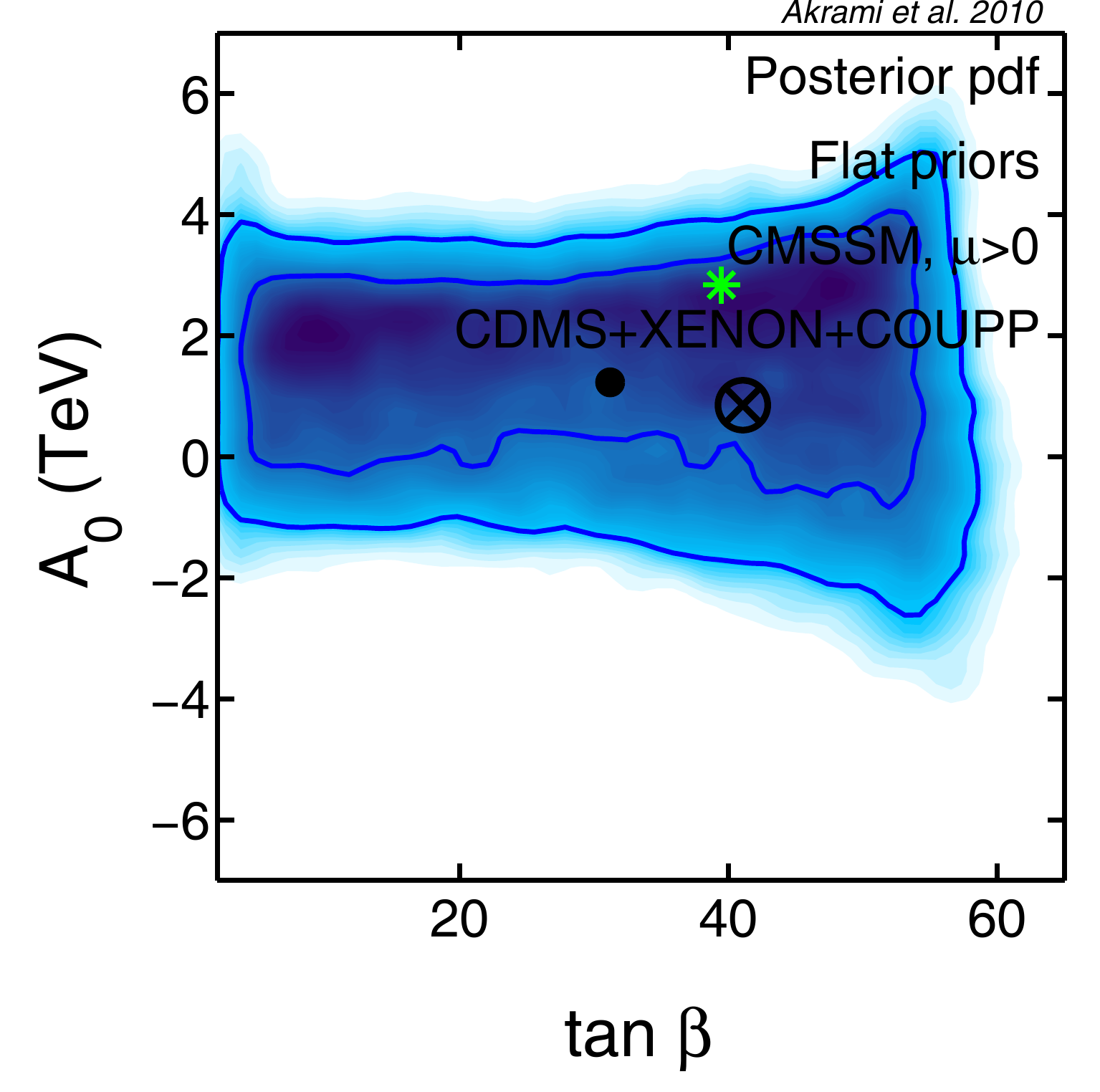}}
\subfigure{\includegraphics[scale=0.23, trim = 40 230 60 123, clip=true]{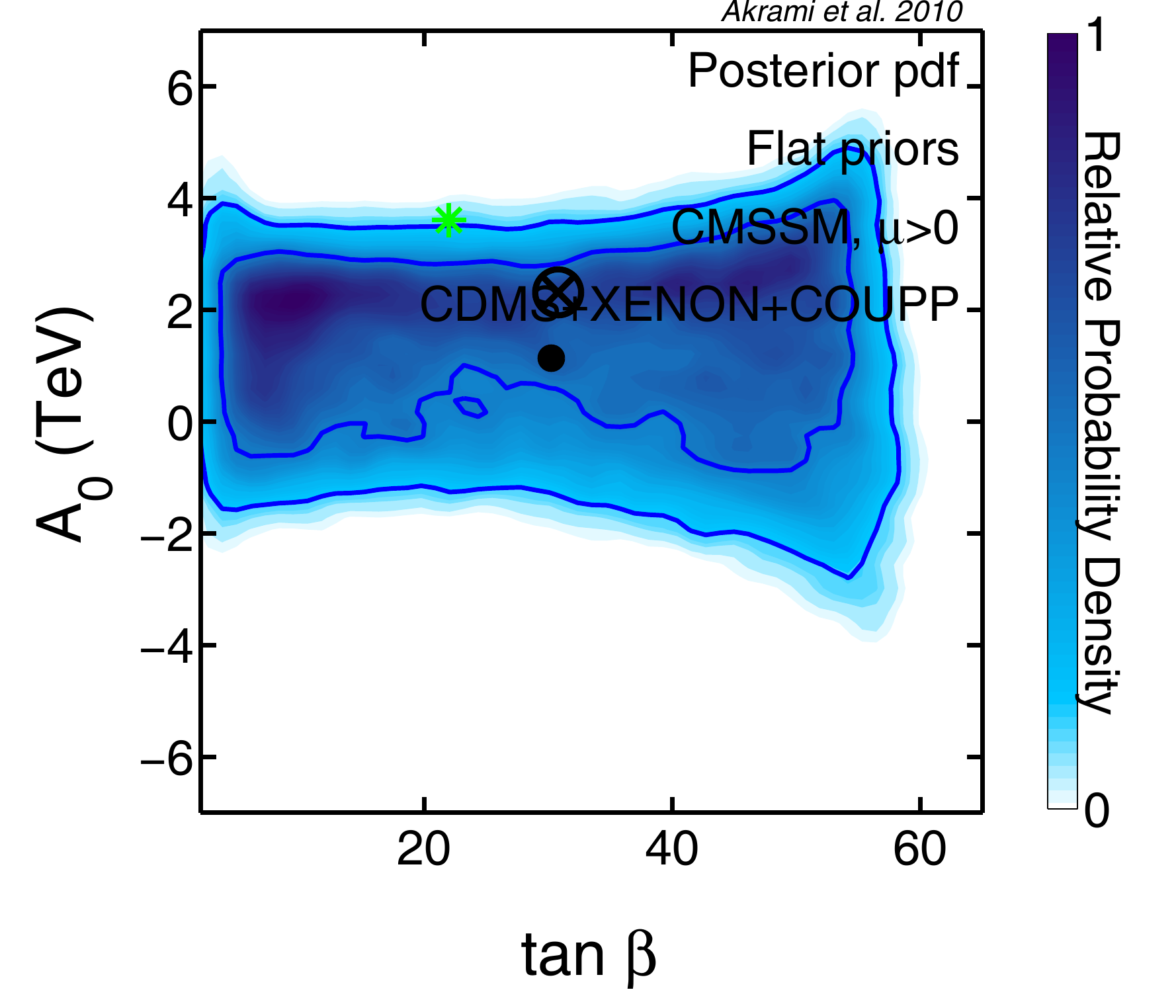}}\\
\subfigure{\includegraphics[scale=0.23, trim = 40 230 130 123, clip=true]{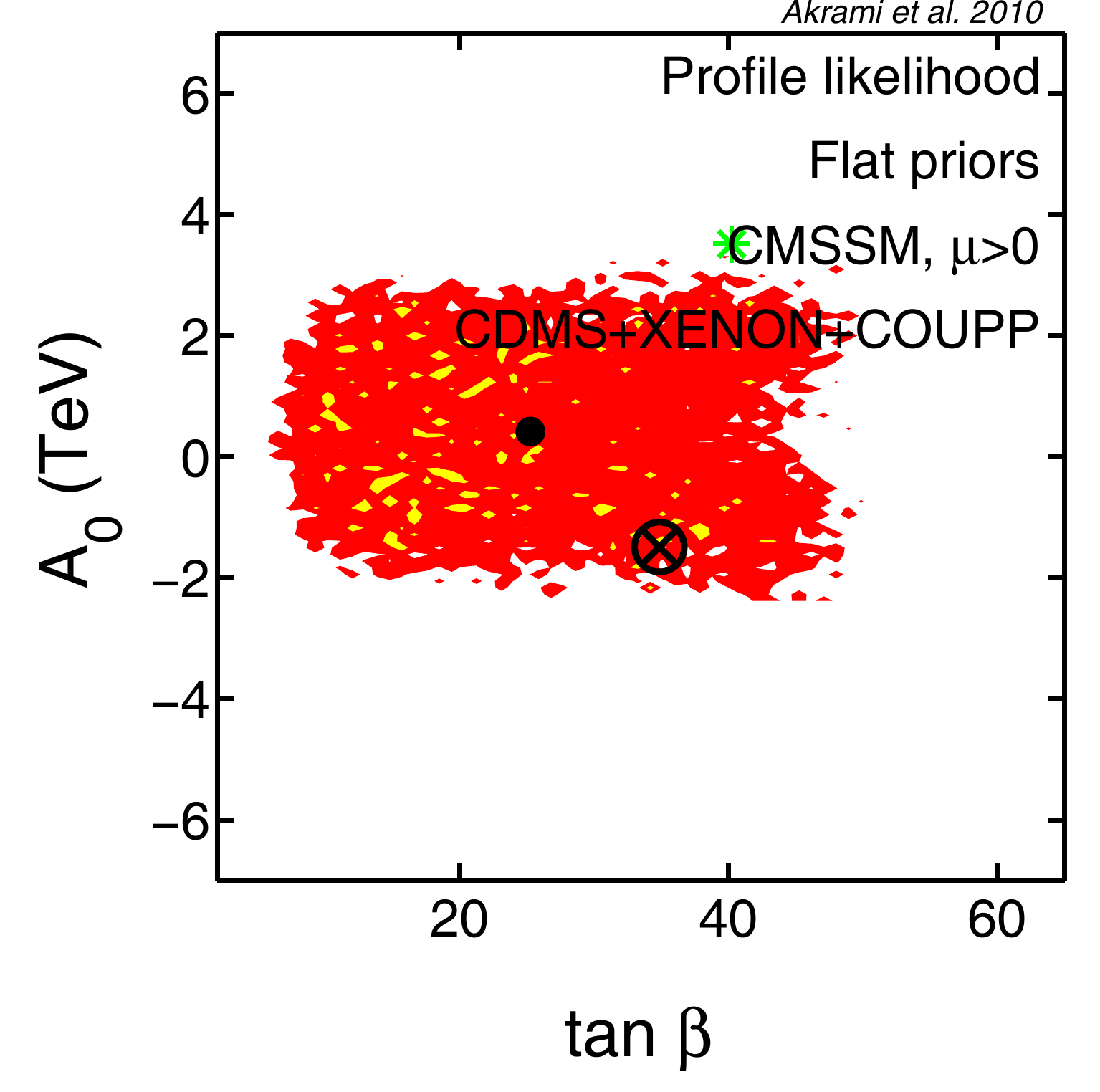}}
\subfigure{\includegraphics[scale=0.23, trim = 40 230 130 123, clip=true]{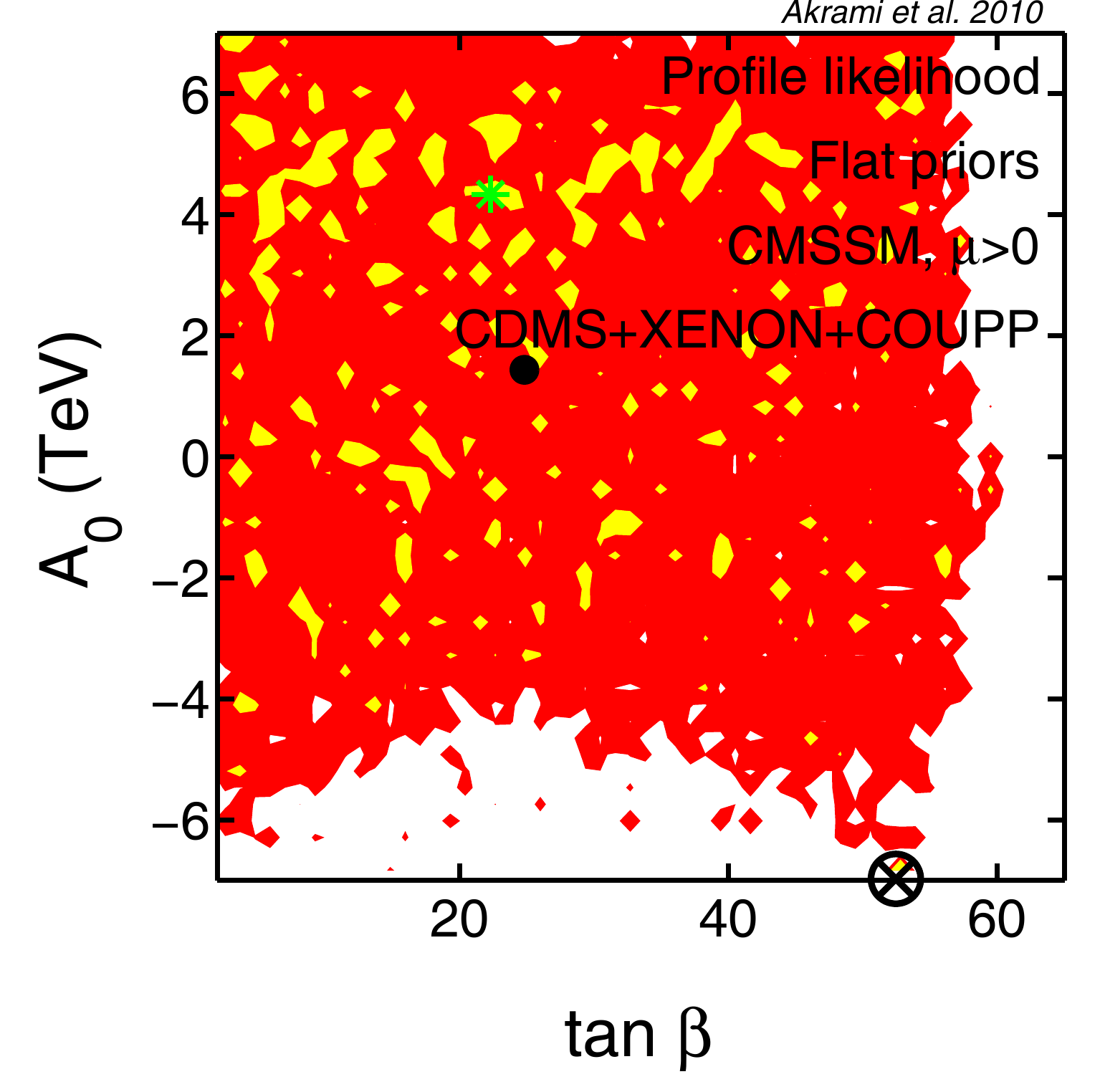}}
\subfigure{\includegraphics[scale=0.23, trim = 40 230 130 123, clip=true]{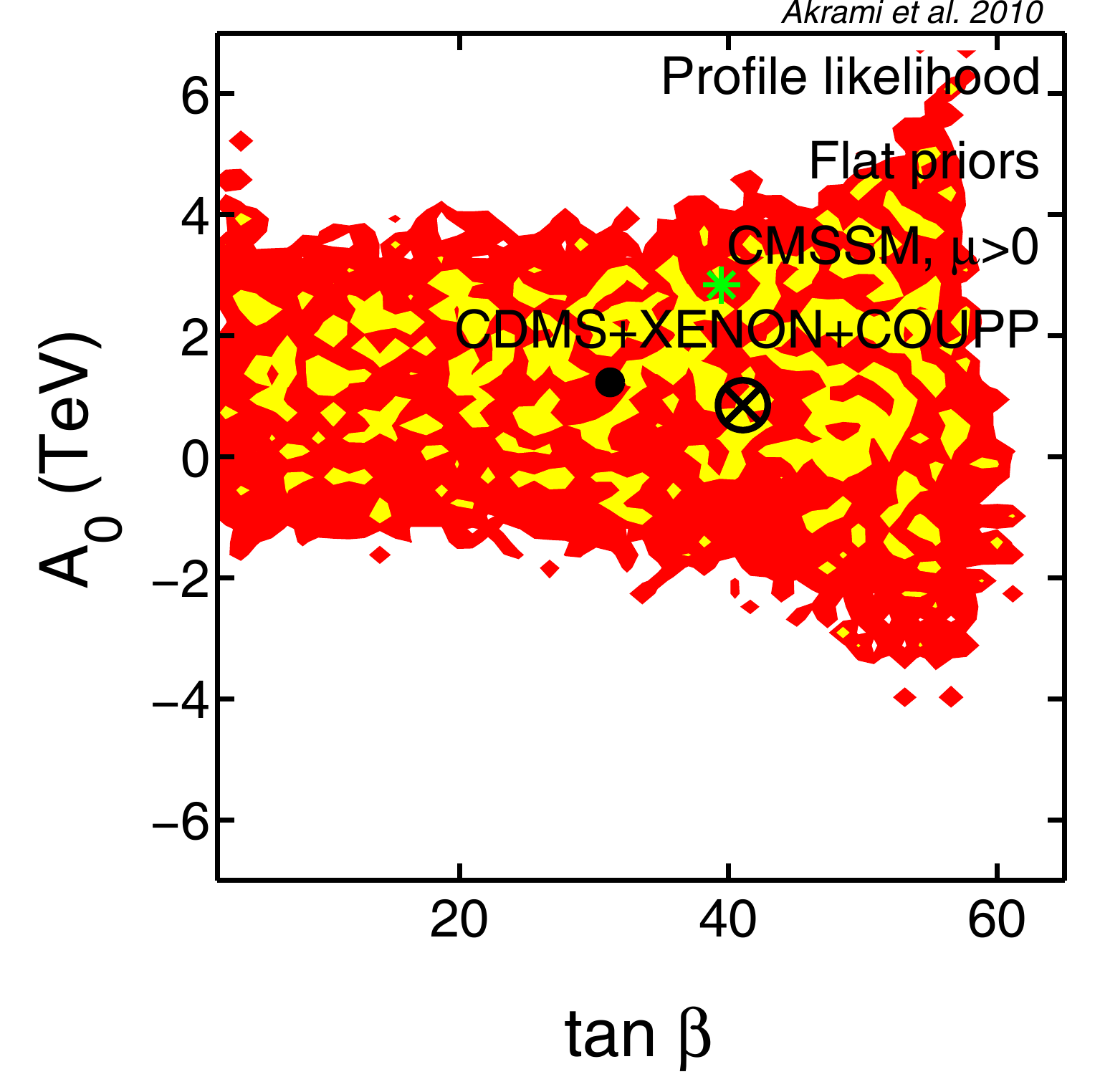}}
\subfigure{\includegraphics[scale=0.23, trim = 40 230 60 123, clip=true]{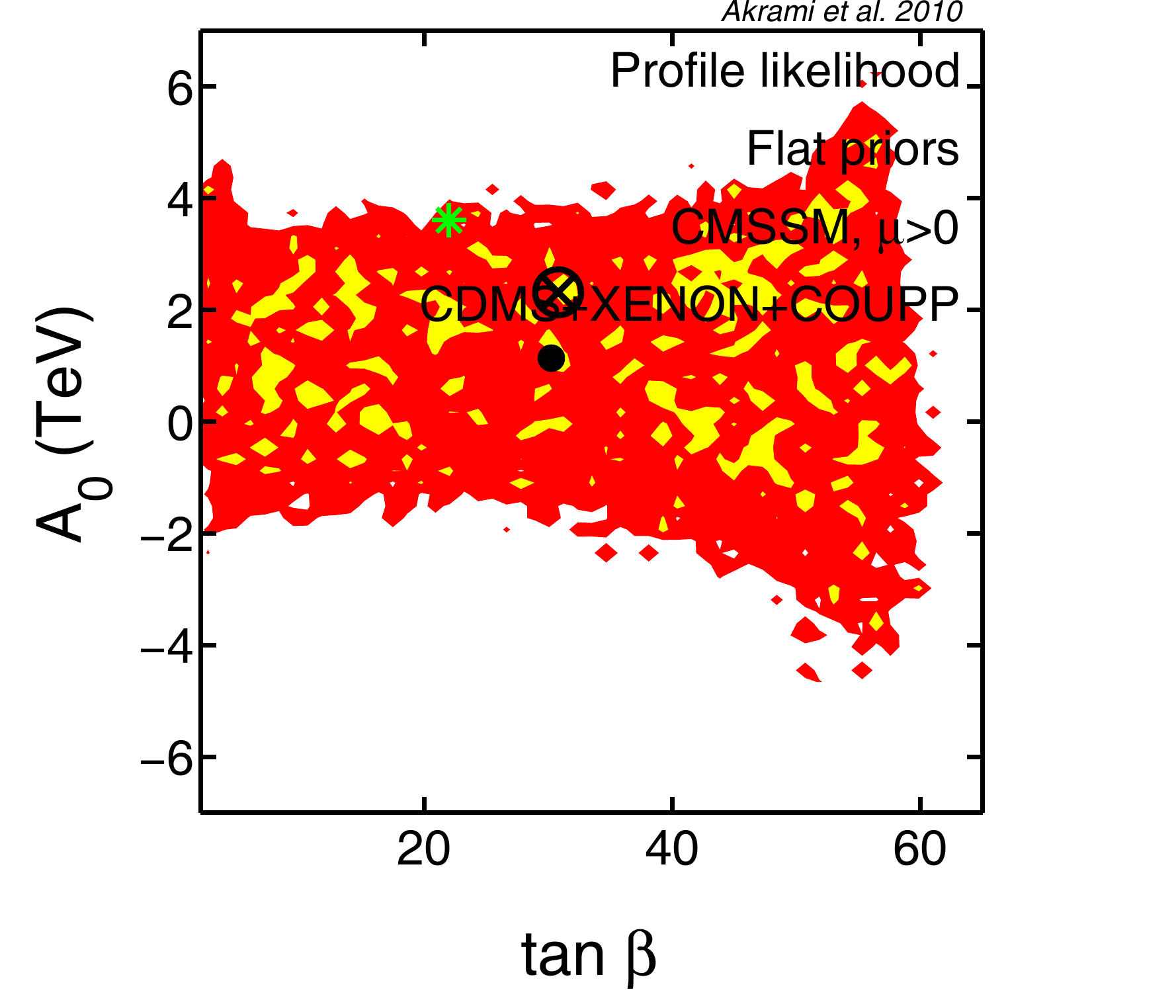}}\\
\caption[aa]{\footnotesize{Two-dimensional marginalised posterior PDFs and profile likelihoods for CMSSM parameters and for all four benchmarks, when the combination of all three direct detection likelihoods is used in the scans. The inner and outer contours in each panel represent $68.3\%$ ($1\sigma$) and $95.4\%$ ($2\sigma$) confidence levels, respectively. Black dots and crosses show the posterior means and best-fit points, respectively, and benchmark values are marked with green stars.}}\label{fig:CMSSMmargprofl}
\end{figure}

In~\fig{fig:CMSSMmargprofl}, we also present two-dimensional marginal posteriors and profile likelihoods for the CMSSM parameters $\mzero$ and $\azero$ versus $\mhalf$ and $\tanb$, respectively, when different benchmarks are used in our reconstruction procedure. Since the focus of this paper is the dark matter mass and cross-sections, we show only the cases where combination of all three DD likelihoods is used in the scans so as to avoid the large number of figures in the paper (for plots with other combinations of DD likelihoods, see the first preprint of the paper available on the \textsf{arXiv}).

In~\figs{fig:LHmarg}{fig:LHprofl} for benchmark 1, with relatively high cross-section and low neutralino mass, all the experiments constrain the combination of $\sigma^{SI}_p$ and $m_{\tilde\chi^0_1}$ quite successfully, even when uncertainties on the halo and cross-section parameters are taken into account. Both the best-fit point and the posterior mean are located very close to the benchmark point in all cases. This is what we would expect from all three experiments when there are many signal events (as in this case), most of which are SI events.

\begin{figure}[t!]
\setcounter{subfigure}{0}
\subfigure[][\footnotesize{\textbf{Benchmark 1:}}]{\includegraphics[scale=0.23, trim = 40 230 130 130, clip=true]{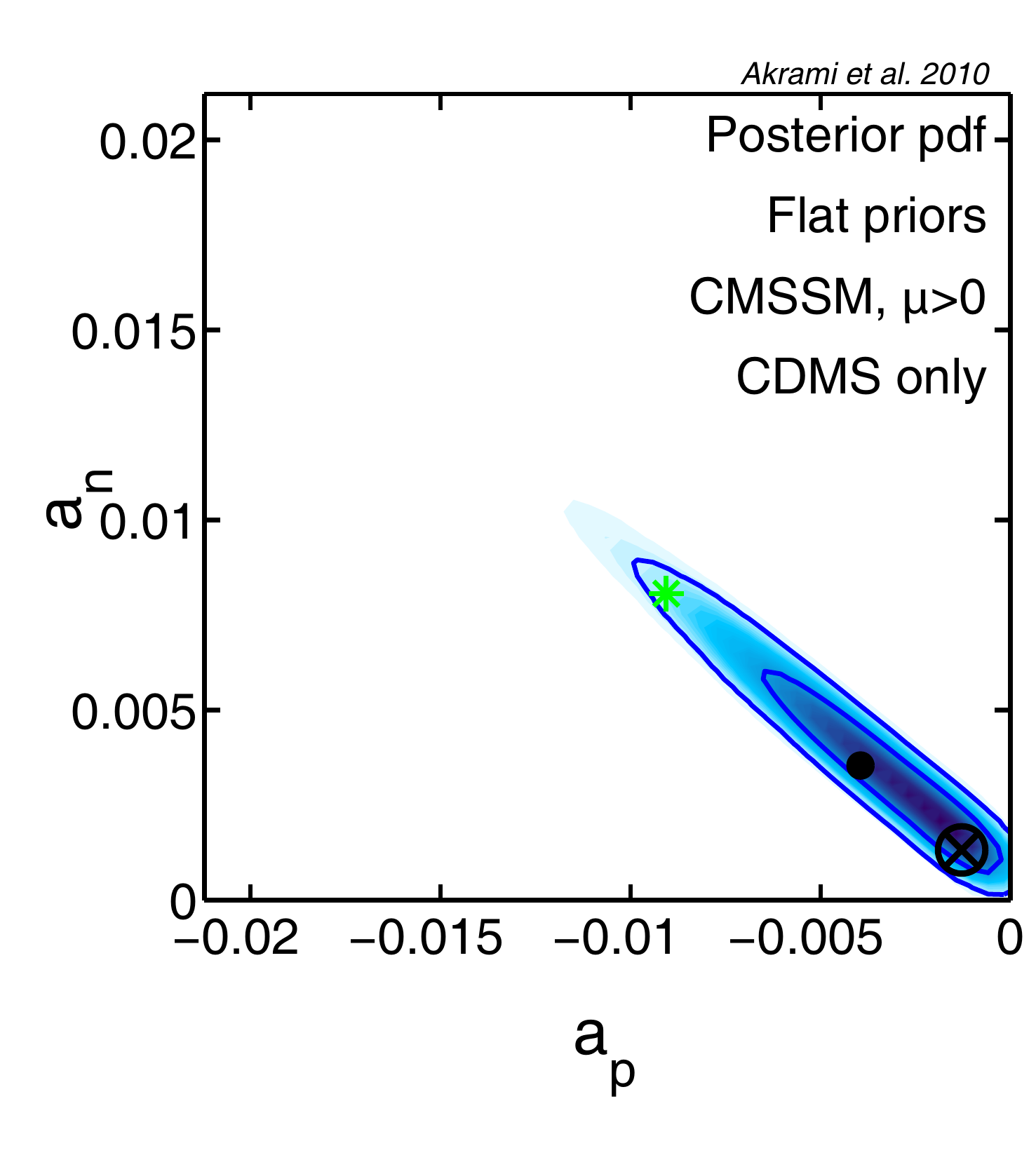}}
\subfigure{\includegraphics[scale=0.23, trim = 40 230 130 130, clip=true]{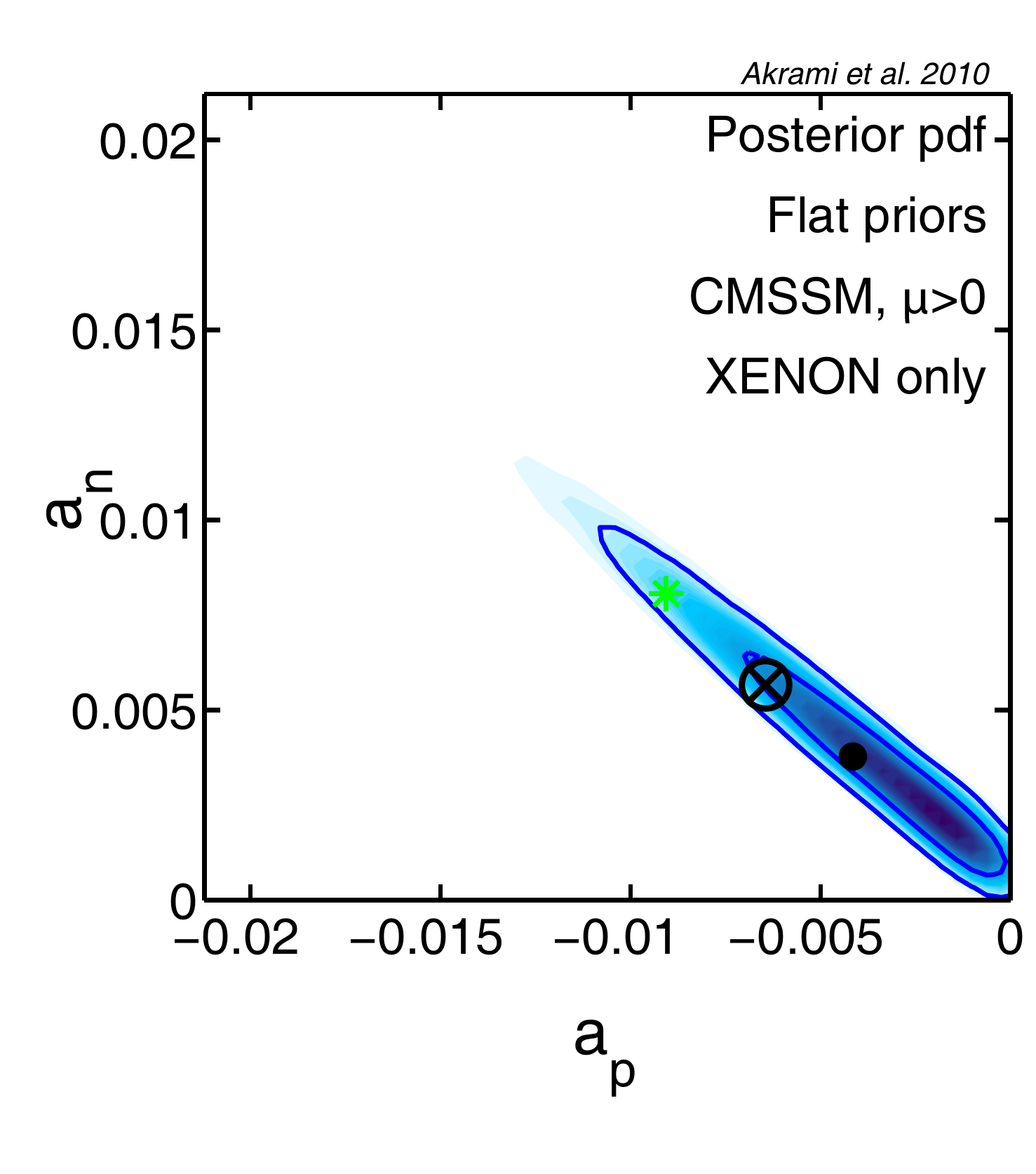}}
\subfigure{\includegraphics[scale=0.23, trim = 40 230 130 130, clip=true]{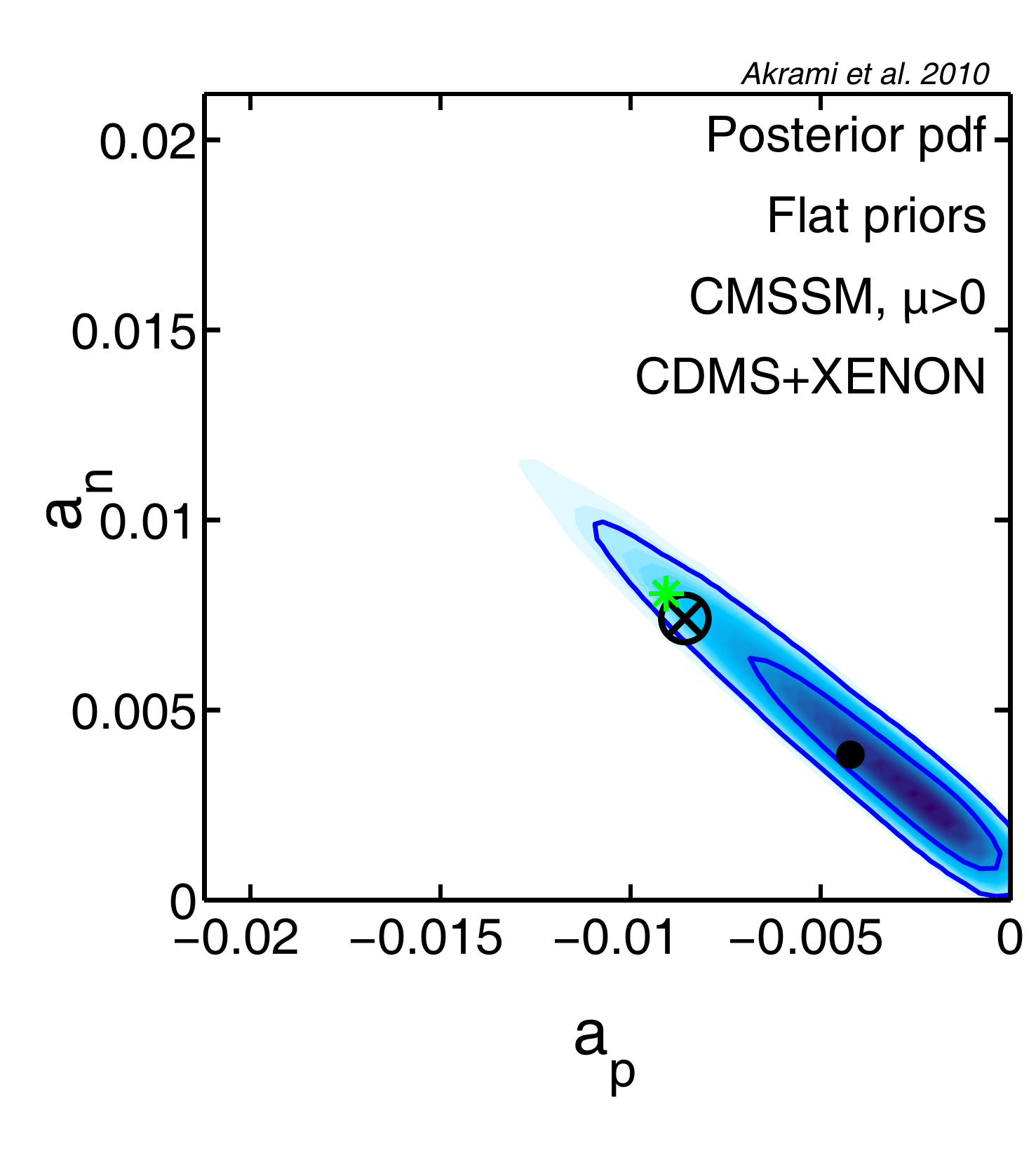}}
\subfigure{\includegraphics[scale=0.23, trim = 40 230 60 130, clip=true]{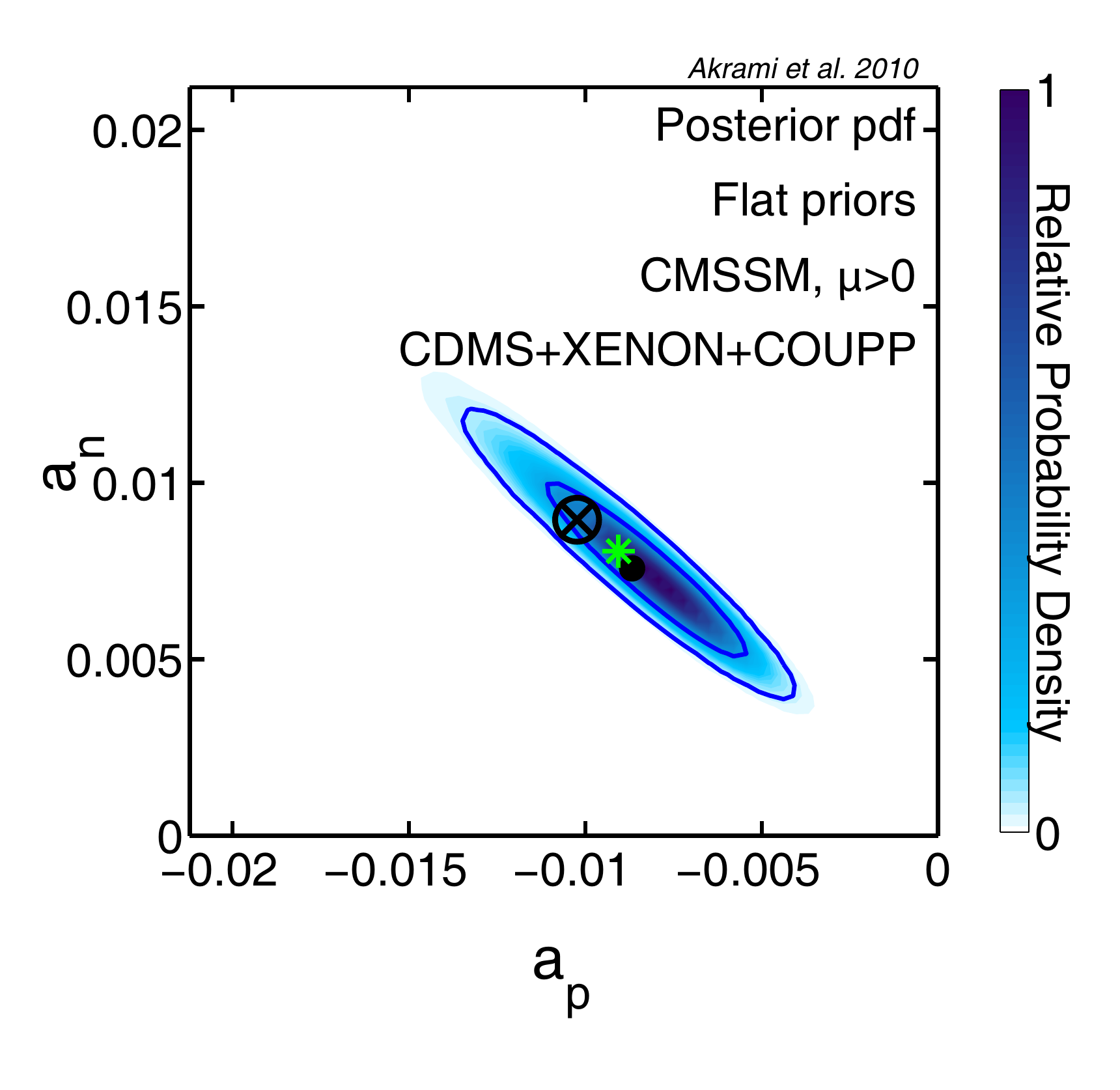}}\\
\setcounter{subfigure}{1}
\subfigure[][\footnotesize{\textbf{Benchmark 2:}}]{\includegraphics[scale=0.23, trim = 40 230 130 130, clip=true]{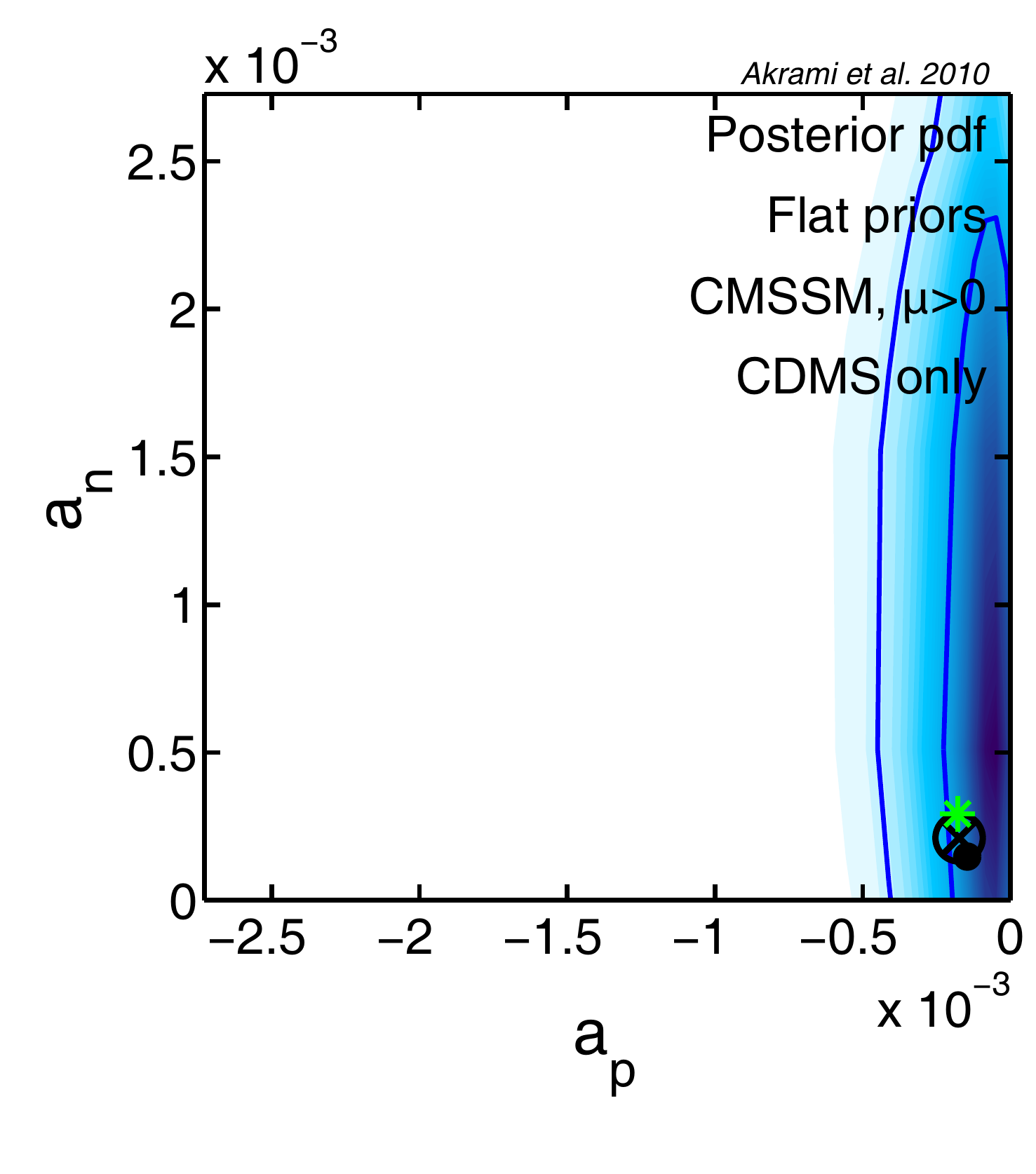}}
\subfigure{\includegraphics[scale=0.23, trim = 40 230 130 130, clip=true]{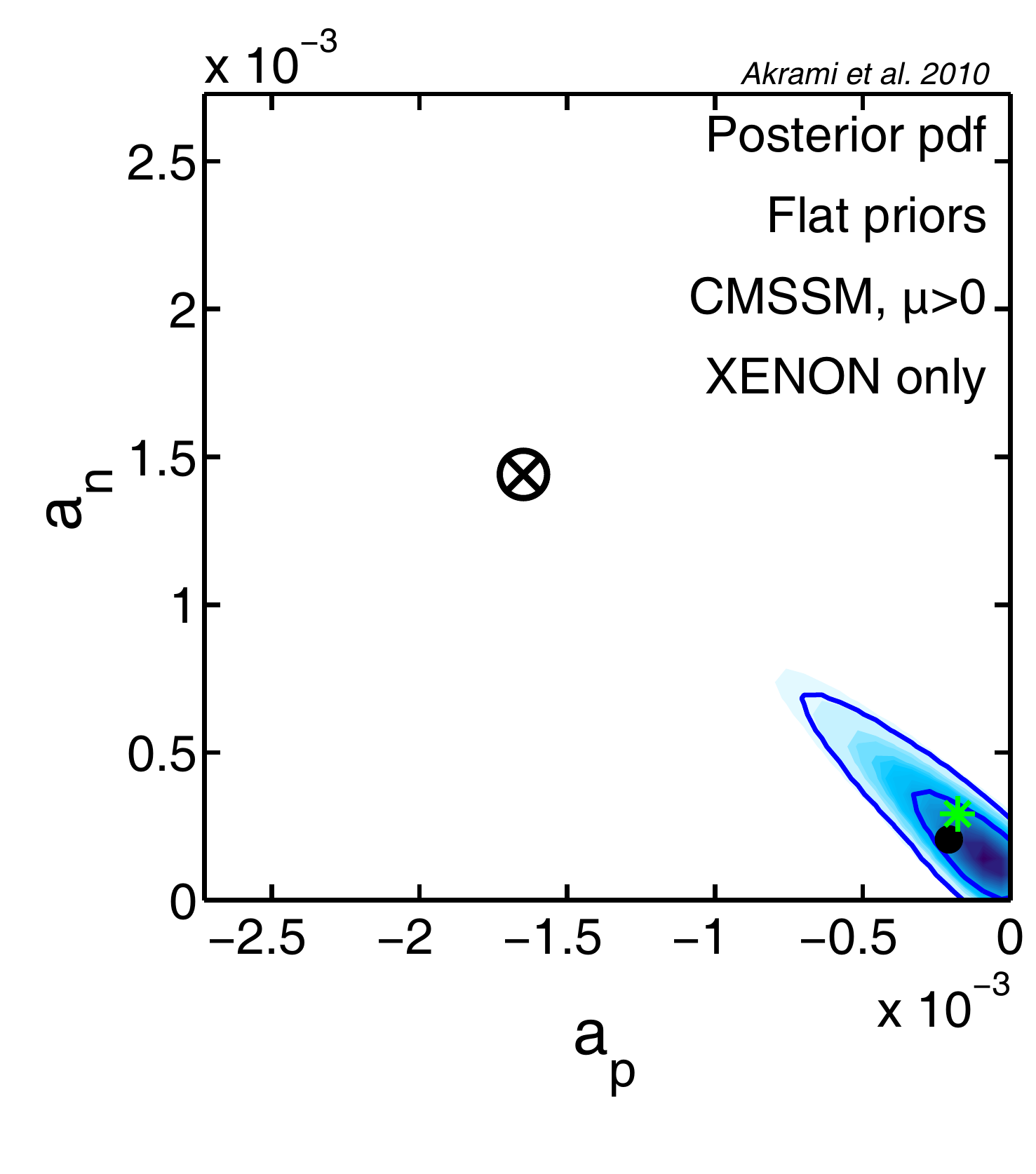}}
\subfigure{\includegraphics[scale=0.23, trim = 40 230 130 130, clip=true]{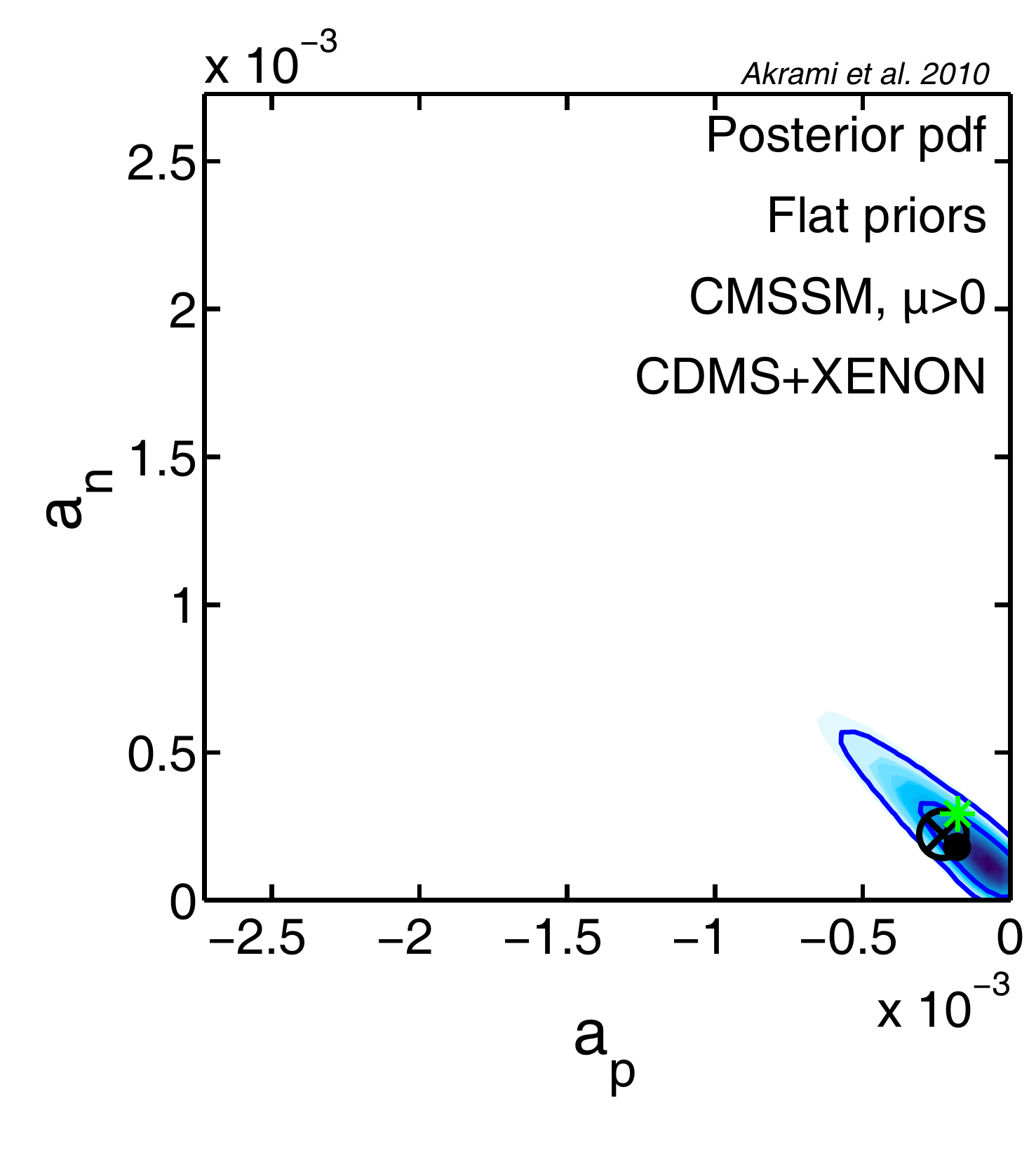}}
\subfigure{\includegraphics[scale=0.23, trim = 40 230 60 130, clip=true]{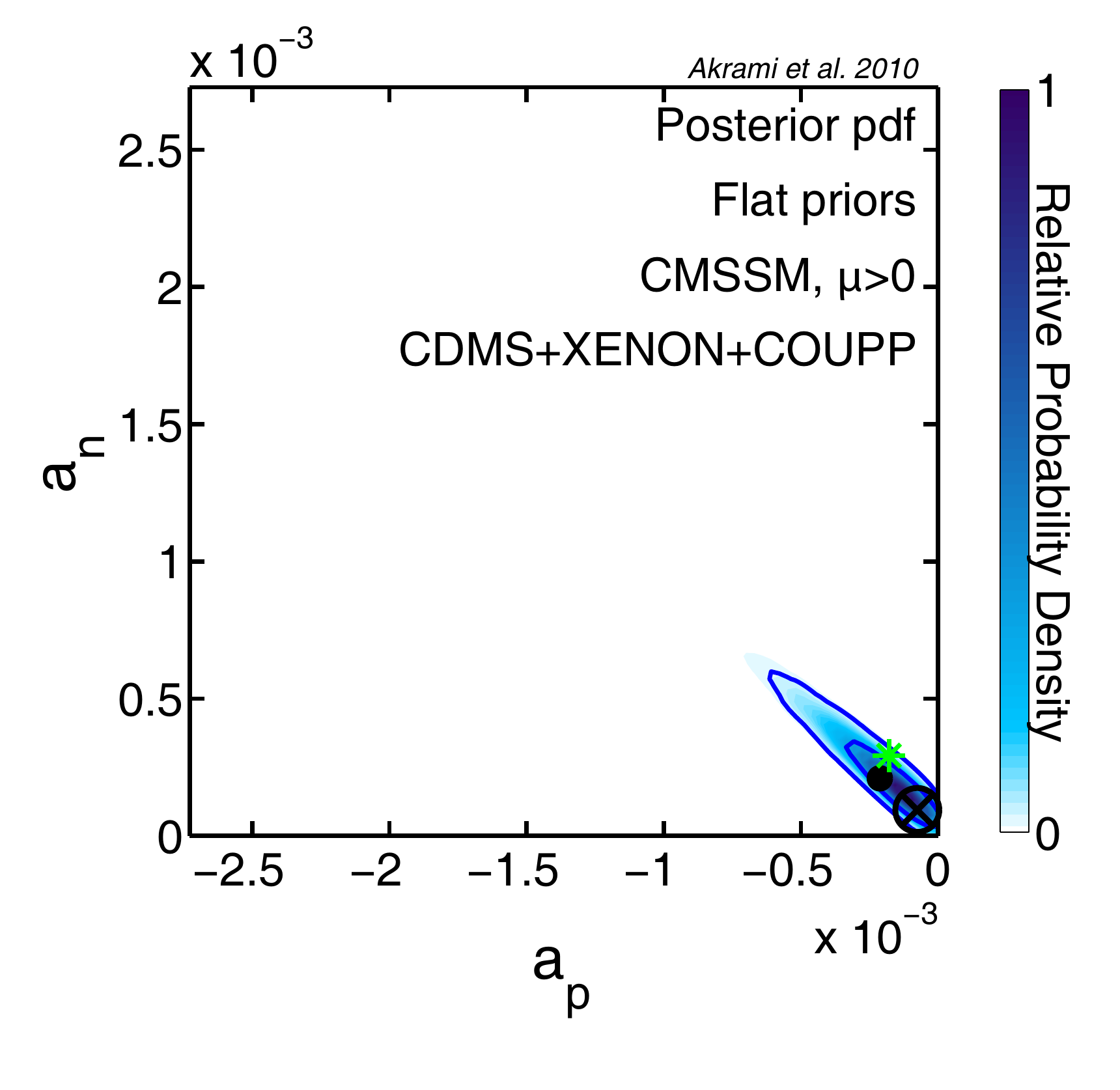}}\\
\setcounter{subfigure}{2}
\subfigure[][\footnotesize{\textbf{Benchmark 3:}}]{\includegraphics[scale=0.23, trim = 40 230 130 130, clip=true]{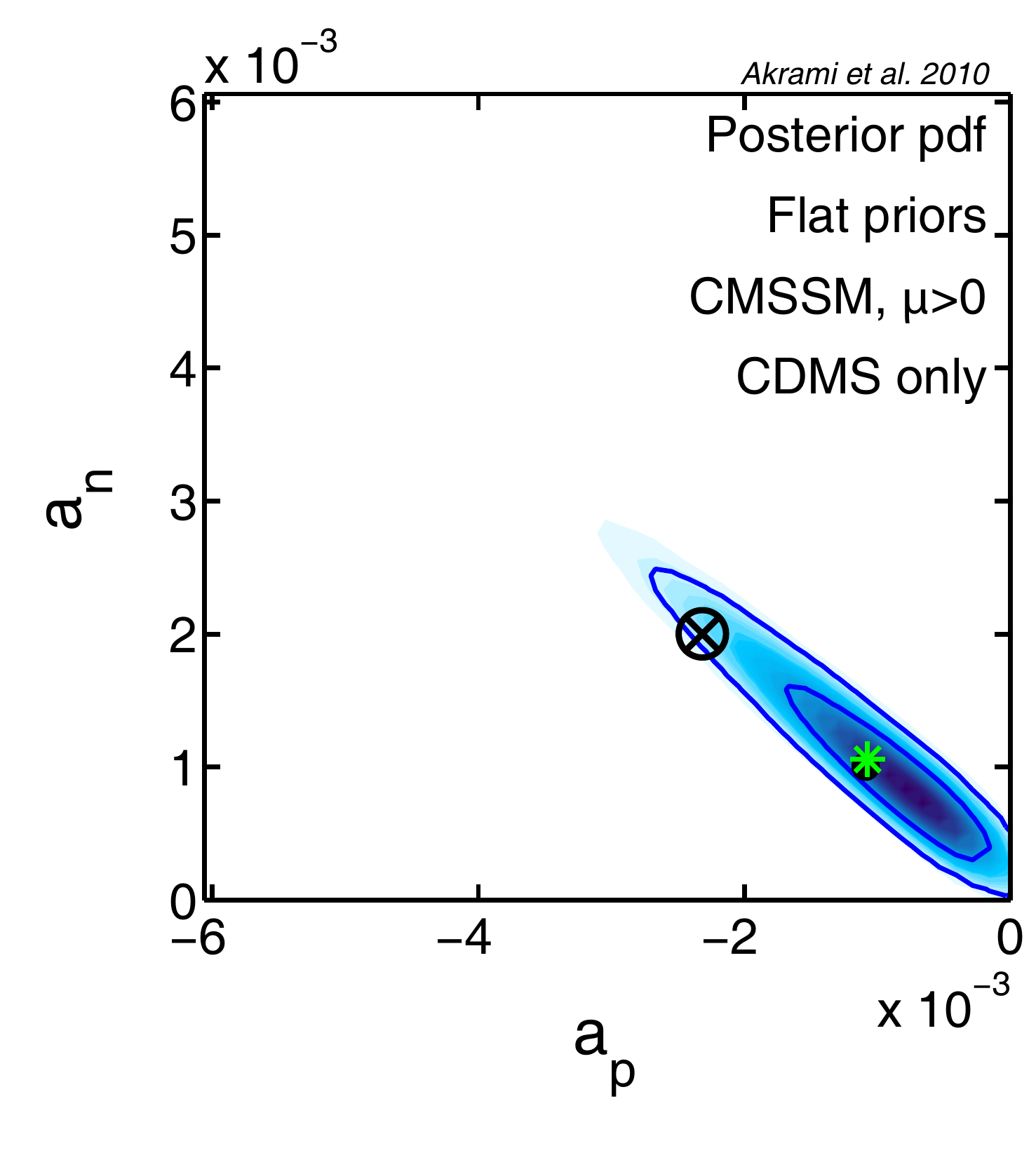}}
\subfigure{\includegraphics[scale=0.23, trim = 40 230 130 130, clip=true]{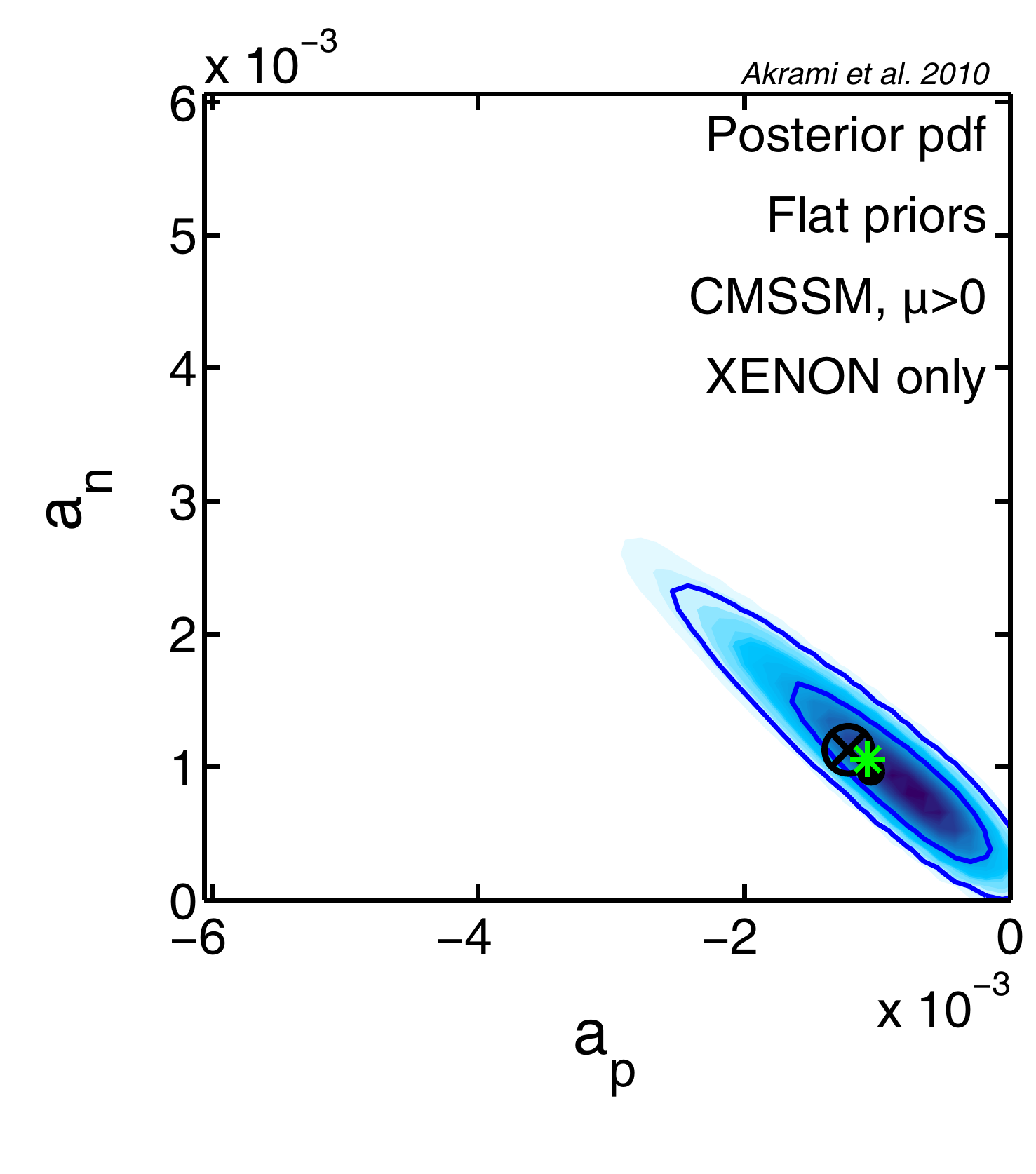}}
\subfigure{\includegraphics[scale=0.23, trim = 40 230 130 130, clip=true]{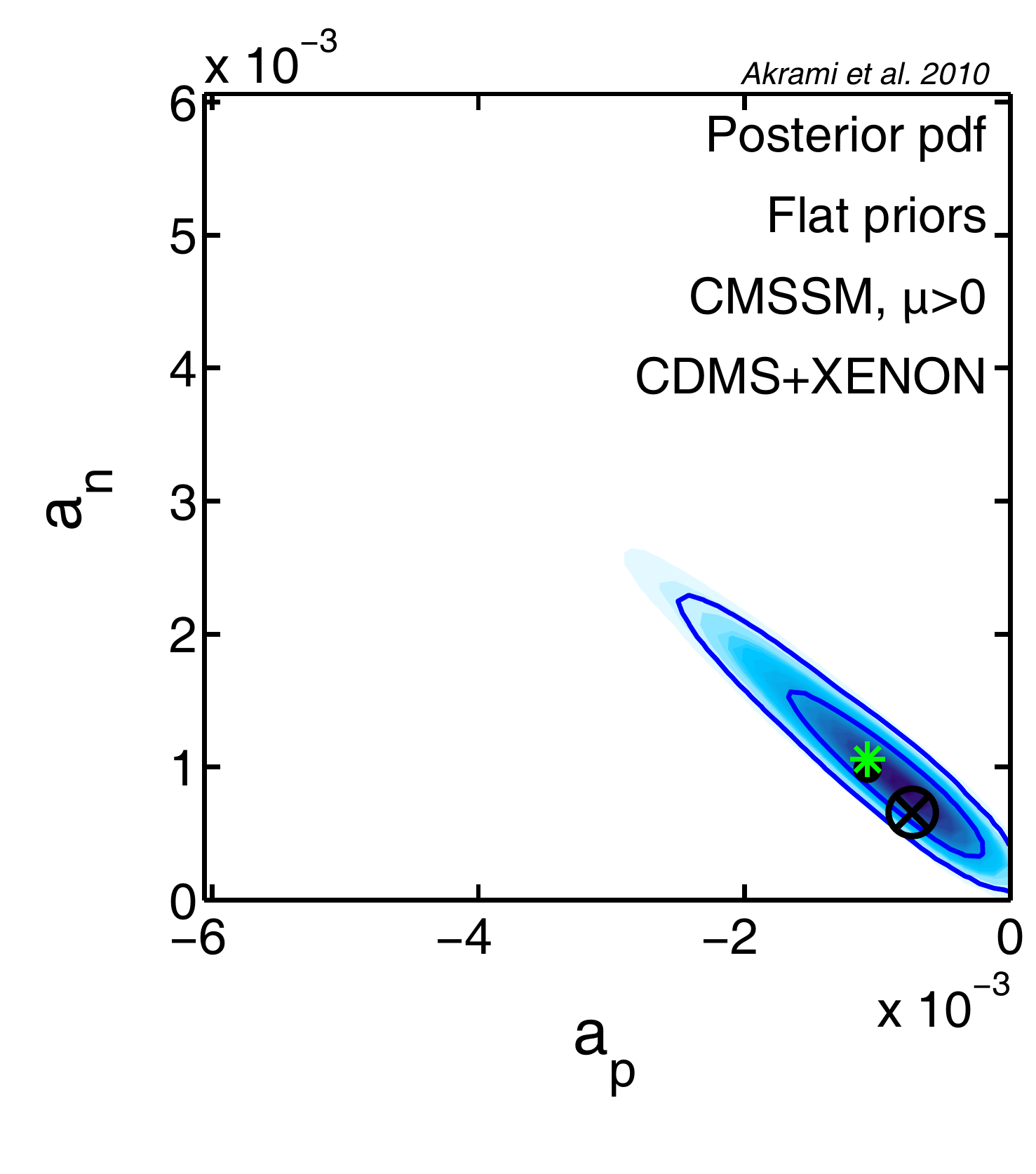}}
\subfigure{\includegraphics[scale=0.23, trim = 40 230 60 130, clip=true]{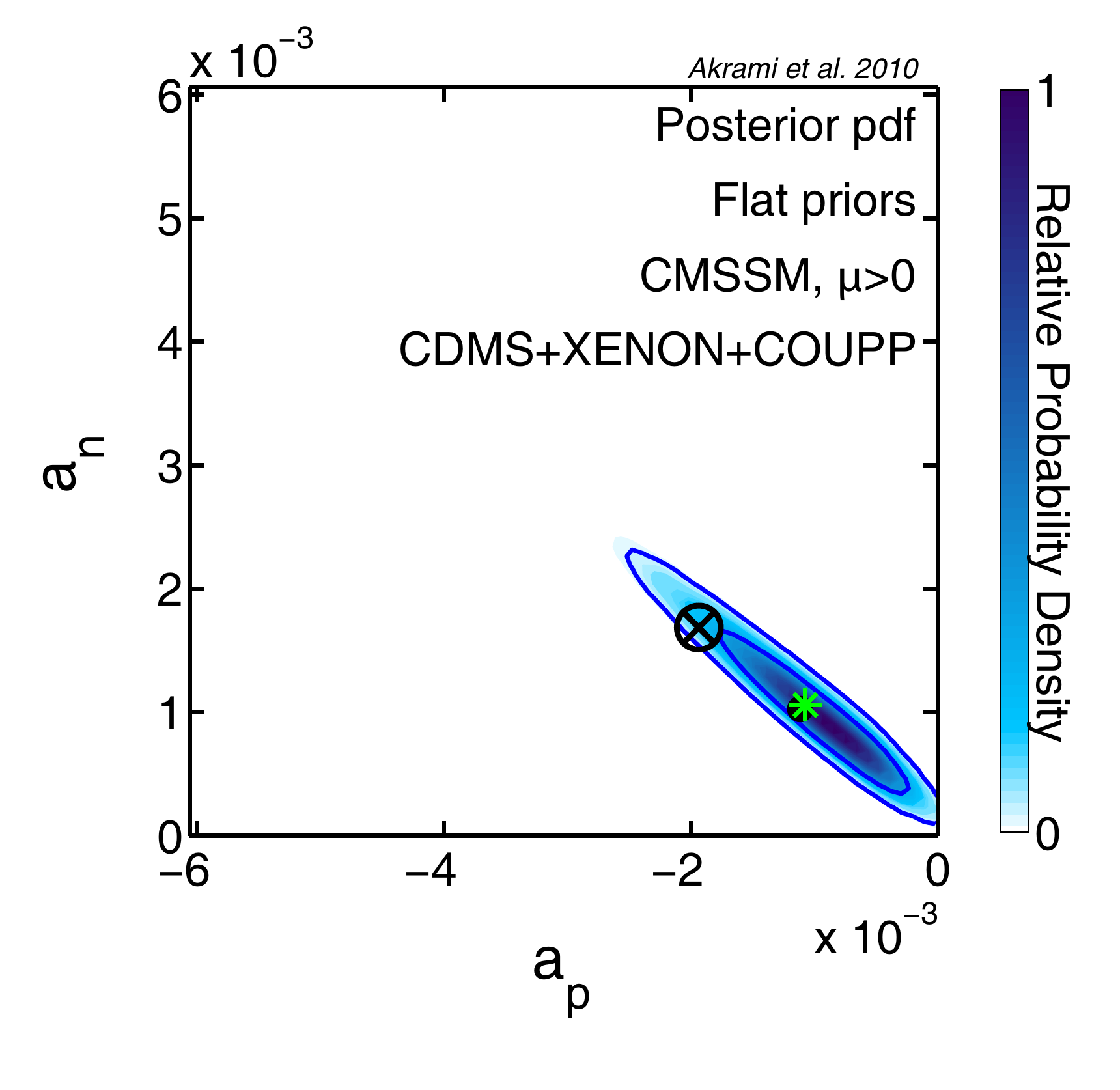}}\\
\setcounter{subfigure}{3}
\subfigure[][\footnotesize{\textbf{Benchmark 4:}}]{\includegraphics[scale=0.23, trim = 40 230 130 130, clip=true]{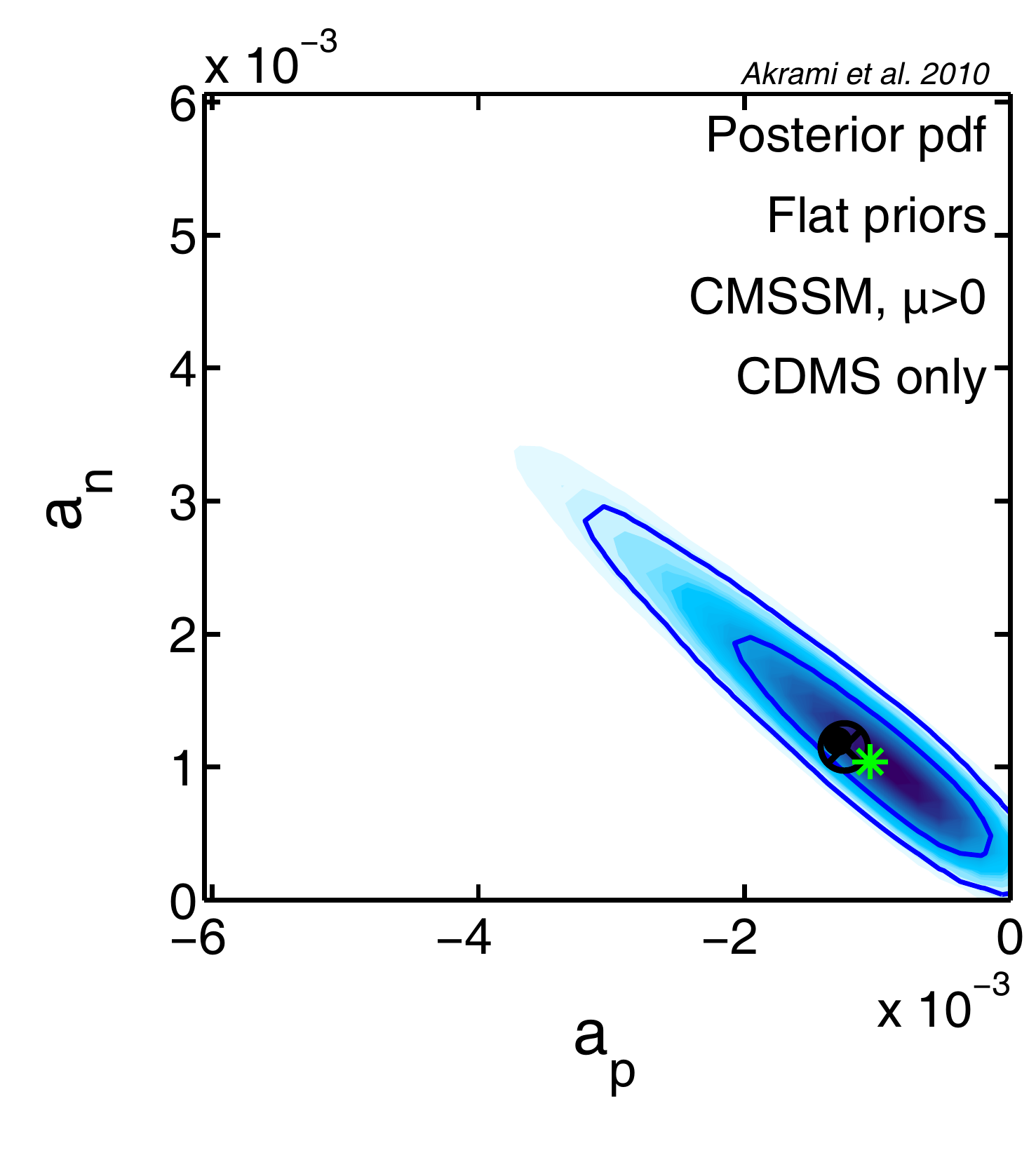}}
\subfigure{\includegraphics[scale=0.23, trim = 40 230 130 130, clip=true]{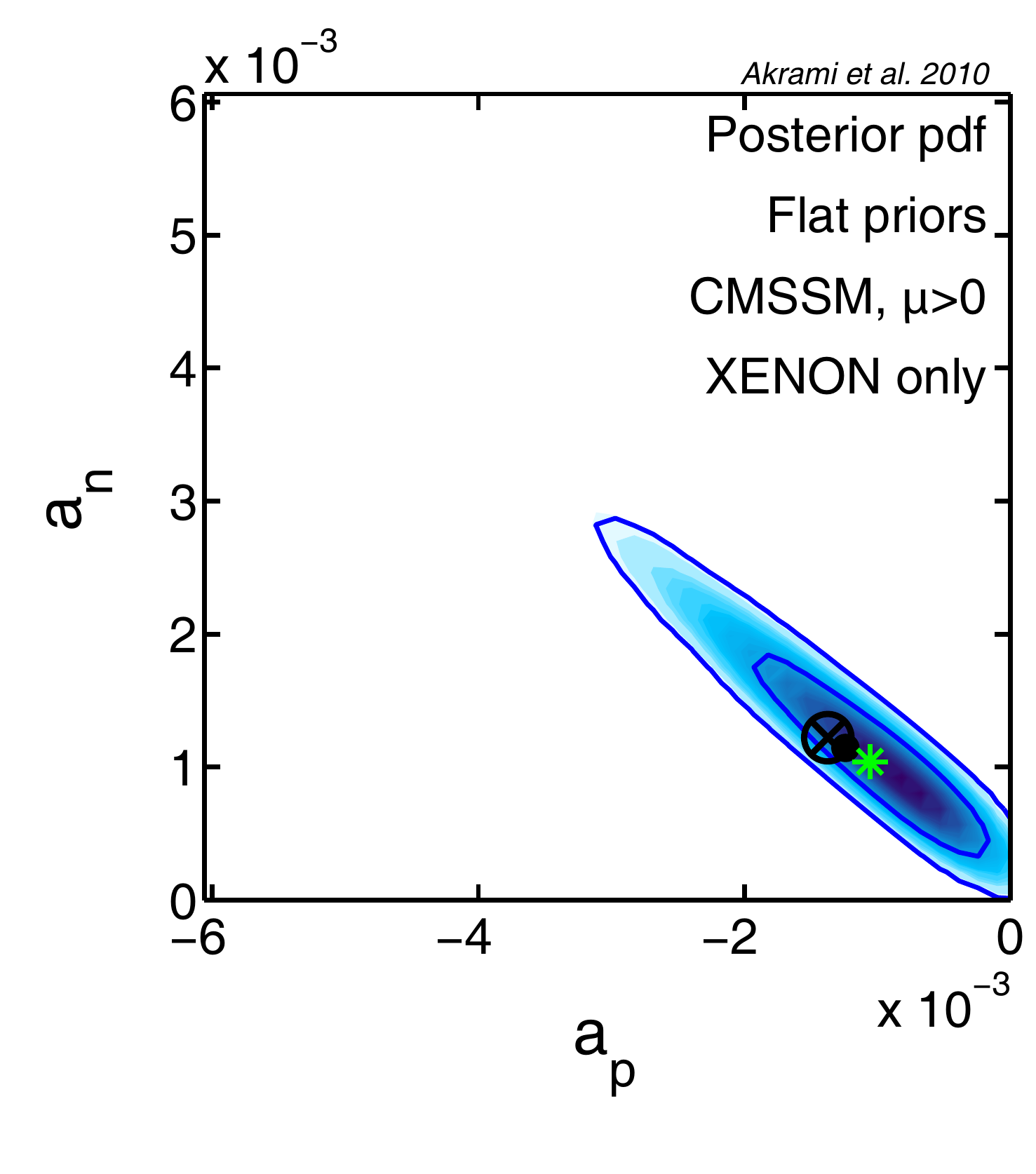}}
\subfigure{\includegraphics[scale=0.23, trim = 40 230 130 130, clip=true]{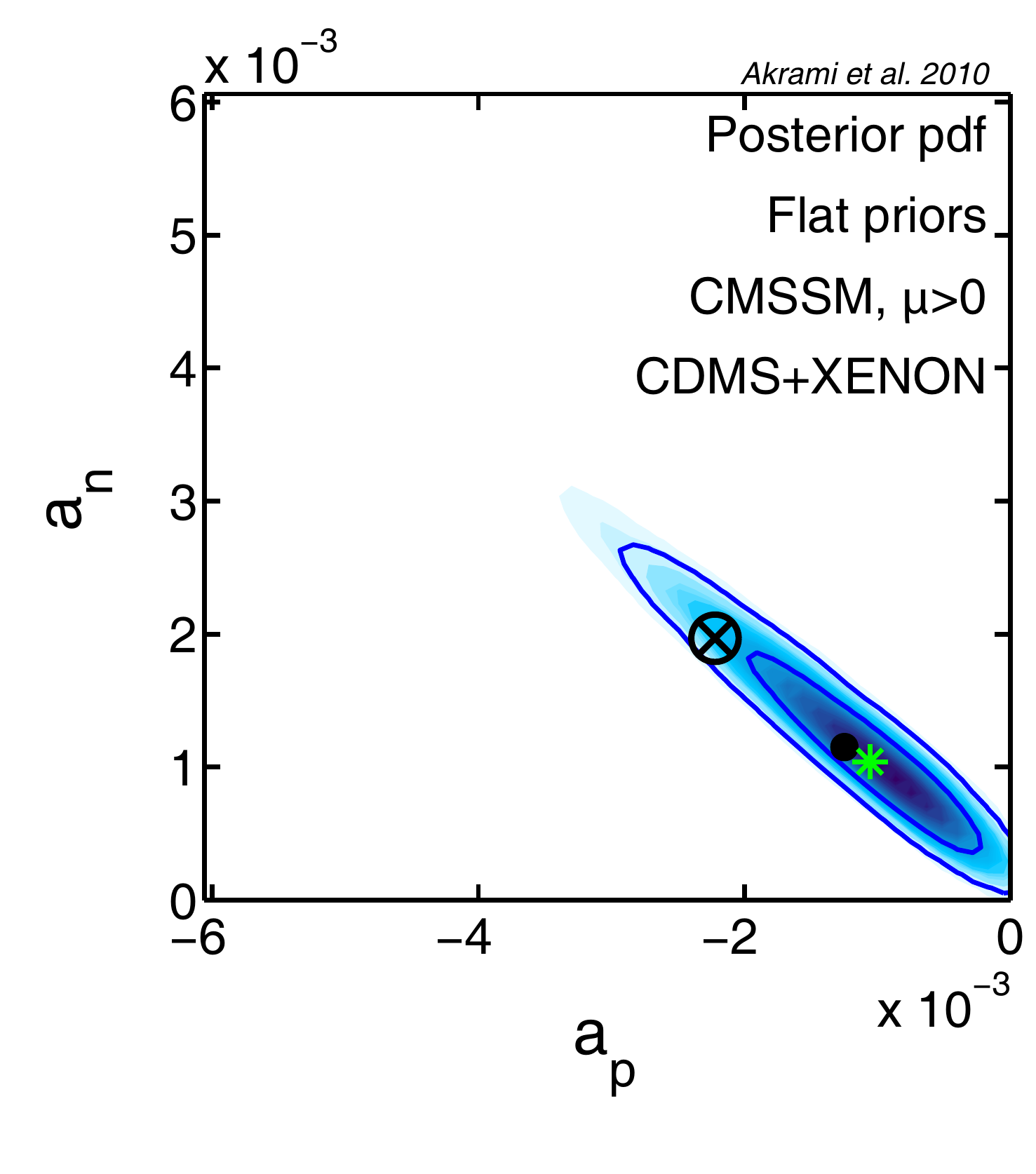}}
\subfigure{\includegraphics[scale=0.23, trim = 40 230 60 130, clip=true]{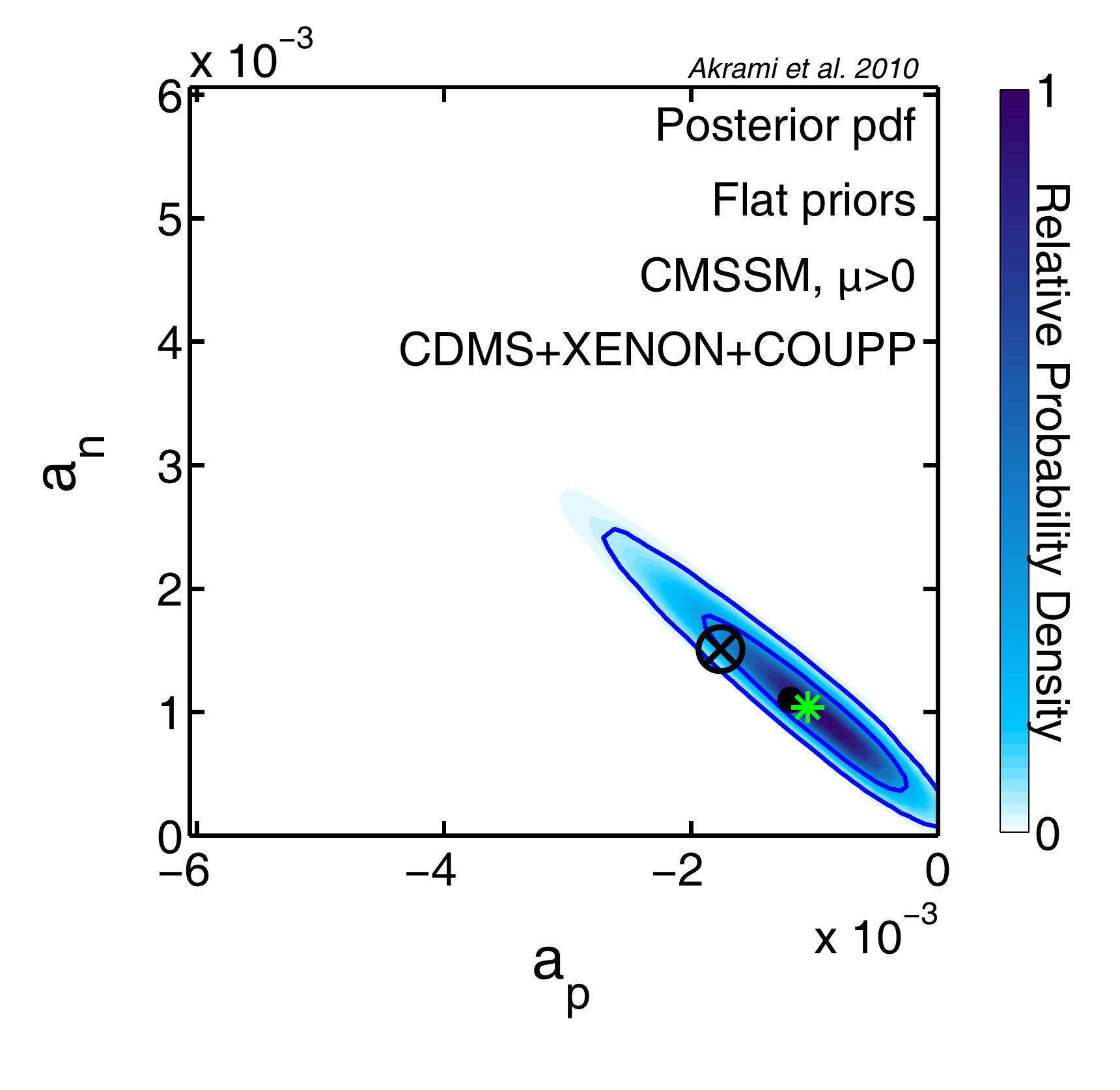}}\\
\caption[aa]{\footnotesize{Two-dimensional marginalised posterior PDFs for the effective spin-dependent neutralino-neutron ($a_n$) and neutralino-proton ($a_p$) couplings, when different combinations of direct detection likelihoods are used in our scans. The inner and outer contours in each panel represent $68.3\%$ ($1\sigma$) and $95.4\%$ ($2\sigma$) confidence levels, respectively. Black dots and crosses show the posterior means and best-fit points, respectively. Benchmarks are marked with green stars.}}\label{fig:anapmarg}
\end{figure}

By looking at the $\sigma^{SD}_p$-$m_{\tilde\chi^0_1}$ and $\sigma^{SD}_n$-$m_{\tilde\chi^0_1}$ planes, we see that the capability of CDMS1T and XENON1T to constrain SD cross-sections is much less than for the SI case. This is to be expected, as these experiments do not have optimal target nuclei for SD detection. The situation is different for COUPP1T (rightmost column of~\figs{fig:LHmarg}{fig:LHprofl}). SD events for COUPP1T constitute a large fraction of the total events, which strongly breaks the degeneracy in the $\sigma^{SD}_p$-$m_{\tilde\chi^0_1}$ and $\sigma^{SD}_N$-$m_{\tilde\chi^0_1}$ planes. This can be seen in the substantial narrowing of contours when data from COUPP1T is included in the fit. To clarify this picture even further, we show in~\figs{fig:anapmarg}{fig:anapprofl} two-dimensional posterior PDFs and profile likelihoods in terms of the two effective couplings of the neutralino to the proton and neutron, $a_p$ and $a_n$ (\eq{eqn:aN}). These two quantities are strongly correlated in the CMSSM. By looking at the panels in the first row we see that for the cases where results from COUPP1T are not included, virtually no constraint is placed upon $a_p$ and $a_n$, and $|a_p|$ and $|a_n|$ can take values as small as zero. With COUPP1T, however, relatively large values of $|a_p|$ and $|a_n|$ are favoured, meaning that in this case, SD interactions are important and COUPP1T has been able to detect them.
Here, CDMS1T and XENON1T are able to constrain both the neutralino mass
and the SI cross-section, but are unable to place any significant
limits on the SD cross-sections due to the statistically insignificant
number of SD events expected (see \reftab{eventsBMs}).  Once the
mass and SI cross-section are approximately determined, which COUPP1T
cannot do alone, a clear excess of COUPP1T events, relative to the
expected number of SI events, becomes apparent: 392 events are observed
in this pseudo-experiment, but only $\sim$270 SI events are expected for
the neutralino mass and SI cross-section determined by CDMS1T and/or
XENON1T.  This $\sim$120 event excess can be attributed to SD
scattering, allowing the SD cross-sections to be constrained.  In this
case, COUPP1T proves to be a quite useful addition to the other
experiments and this situation illustrates how multiple DD experiments
can provide complementary (rather than redundant) results.

The prior effects seen in~\fig{fig:PriorsOnly}, and discussed previously, are strongly compensated for in this case and the credible/confidence regions are dominated by data rather than priors (either real or effective). The $\mzero$-$\mhalf$ plots (first column in the left of~\fig{fig:CMSSMmargprofl}) show that DD data for benchmark 1 with low neutralino mass can strongly constrain $\mhalf$; this is not surprising, given the strong correlation between the neutralino mass and $\mhalf$ in the CMSSM. For the $\azero$-$\tanb$ planes, DD data exclude regions at very high or very low $\azero$. Such data do not put strong constraints upon $\tanb$, except for the case where COUPP1T is added to the combination. Adding COUPP1T also excludes regions with very high or very low $\tanb$ (as shown in~\fig{fig:CMSSMmargprofl}). 

For benchmark 2,~\figs{fig:LLmarg}{fig:LLprofl} and~\fig{fig:CMSSMmargprofl} (second column in the left) show that the credible/confidence regions are strongly dominated by prior effects. For example, comparing panels in the first row of~\fig{fig:LLmarg} for $\sigma^{SI}_p$ versus $m_{\tilde\chi^0_1}$ with the corresponding plot in~\fig{fig:PriorsOnly} shows that including data has only excluded high cross-sections at low neutralino masses. The situation improves slightly when the COUPP1T likelihood is added to the analysis; in this case, parts of the region at low cross-sections and high masses are also excluded. The reason why our DD experiments cannot constrain the model well in this case is the small number of signal events compared to background (\tab{tab:eventsBMs}), giving rather low statistics. The same conclusions can be drawn from the $\mzero$-$\mhalf$ and $\azero$-$\tanb$ plots for benchmark 2 in~\fig{fig:CMSSMmargprofl}. They are again dominated by priors except for very low values of $\azero$ in the $\azero$-$\tanb$ plane, which seem to be excluded. Regions with low values of both $\mzero$ and $\mhalf$, as well as high-$\mhalf$ regions are also somewhat disfavoured. The reason for the unusual $\anSD$-$\apSD$ regions for the CDMS1T-only
case with benchmark 2, shown in \figs{fig:anapmarg}{fig:anapprofl}, is not completely
clear, but we suspect it arises due to this particular pseudo-experimental
result, which had four events at energies of $13.3$, $79.7$, $90.5$, and $95.3$~keV.  The three higher energy events are likely due to a $\sim 2\sigma$ random upward
fluctuation in the number of flat background events (an average of one
flat background event is expected).  These few high-energy events, with
a lack of low-energy events, are not particularly consistent with
either a neutralino spectrum or the exponentially-falling background.
The combination of the few events and unusual energy distribution may
have driven the scans to prefer atypical sets of parameters.
We note that, though these regions are unusual, the true parameters
still lie within the 1$\sigma$ contours.

\figs{fig:MMmarg}{fig:MMprofl} and~\fig{fig:CMSSMmargprofl} (third column in the left) present our results for benchmark 3. These plots show that compared to benchmark 1, which has a much lower neutralino mass (\figs{fig:LHmarg}{fig:LHprofl} and second column of~\fig{fig:CMSSMmargprofl}), the experiments are less constraining in this case. Combining different experiments does not improve the results significantly. This is because the number of total events is largely reduced and most of the events come from SI interactions (\tab{tab:eventsBMs}). The latter point explains why the degeneracy in the two SD couplings is not broken in this case when the COUPP1T likelihood is taken into account.

Finally, similar plots are given in~\figs{fig:HHmarg}{fig:HHprofl} and~\fig{fig:CMSSMmargprofl}~(last column) for benchmark 4. The neutralino mass for this point is relatively large ($\sim$~1 TeV). Our credible/confidence regions in this case are very similar to the previous case for benchmark 3.

One interesting feature of~\figs{fig:HHmarg}{fig:HHprofl} is that contrary to the other benchmarks, the true values for the mass and cross-sections do not fall within the $1\sigma$ and $2\sigma$ regions. From a Bayesian point of view, such a situation can naturally happen in the presence of strong priors that dominate the statistical inference, in the absence of high statistics provided by data. In a frequentist framework, however, where the inference should be independent of priors, any confidence regions and intervals provided by a successful parameter estimation procedure, `must' show ``proper coverage''. The latter simply means that in an ideal case, regions with $95.4\%$ ($2\sigma$) confidence level, for instance, contain the ``true'' parameters in $95.4\%$ of the times that a large number of scans are performed using synthetic data generated from a psudo-experiment. This indicates that the plots in the first row of~\fig{fig:HHprofl}, where $\sigma^{SI}_p$ is plotted against $m_{\tilde\chi^0_1}$ for the two-dimensional profile likelihoods, and the benchmark is outside the $2\sigma$ regions, can be perfectly acceptable as long as such plots do not appear in more than $4.6\%$ of the times. This is not the case however, as is discussed in detail in the companion paper~\cite{Akrami:2010}, where statistical coverage is studied in SUSY parameter estimation when data from DD experiments are used (also see ref.~\cite{Bridges:2010de} for a similar coverage study of the CMSSM, but in a different context). By looking at the first rows of~\fig{fig:HHprofl}, we observe that contours taper off at high masses, and only a few points exist in those regions even though very good fits continue to be found as neutralino masses increase. For neutralinos much heavier than the nuclear targets, as is the case here, DD experiments are unable to provide an upper limit on the neutralino mass and the best-fit contours in the $\sigmapSI$-$m_{\tilde\chi^0_1}$ plane should extend indefinitely along a line of constant $\sigmapSI/m_{\tilde\chi^0_1}$ at high masses. This then suggests that our CMSSM scans, due to some ``sampling effects'', map out only the lower mass end of these expected contours. As discussed in ref.~\cite{Akrami:2010}, we believe that this under-coverage seen in the profile likelihood results, is due to sampling effects caused by strong prior effects that reduce the ability of our scanning algorithm (i.e.~\MN) to explore regions at high masses and cross-sections.~\MN~is a technique optimised for Bayesian statistics and, consequently, any frequentist inference it provides can be indirectly (and in some cases strongly) influenced by the prior effects. The configuration we use for our~\MN~scans (i.e. the particular values of the tolerance parameter and the number of live points) is the same as what is used in, for example,~\cite{Trotta:08093792}. A few scans with a smaller value of the tolerance parameter and a higher number of live points (as suggested in ref.~\cite{Feroz:2011bj}) show that one might obtain better coverage in some cases (for more details, see ref.~\cite{Akrami:2010}).

In order to see what the predictions of our scans are for other types of experiments, in~\figs{fig:IDmarg}{fig:IDprofl} we give two-dimensional marginal posteriors and profile likelihoods for the velocity-averaged neutralino self-annihilation cross-section $\left\langle \sigma v\right\rangle$, as a function of the neutralino mass $m_{\tilde\chi^0_1}$. These plots are commonly used when indirect detection constraints are put upon WIMPs like the lightest neutralino. Every row in the figure corresponds to one benchmark point when different combinations of the DD experiments CDMS1T, XENON1T and COUPP1T are employed. For panels in the first row of each figure, where benchmark 1 (with relatively large scattering cross-sections and low neutralino mass) is studied, we see that our scans predict rather tight contours, especially when COUPP1T is considered. The neutralino mass is highly constrained, as we have seen in previous figures (\figs{fig:LHmarg}{fig:LHprofl}). In addition, the neutralino annihilation and scattering cross-sections are strongly correlated, so because the latter are highly constrained for benchmark 1, the correlation results in relatively tight constraints on the annihilation cross-section. We suspect that a large fraction of the favoured region in this case is the `funnel' region, where one of the neutral Higgs particles in the MSSM has roughly twice the mass of the lightest neutralino, and neutralino annihilation is increased by a mass resonance with that Higgs particle (for a discussion of distinct regions in the CMSSM parameter space and their relevance for indirect detection analysis, see e.g. ref.~\cite{Scott:2009jn}). The funnel region seems to be found also in our scans for benchmark 2 (regions at low neutralino masses in second rows of~\figs{fig:IDmarg}{fig:IDprofl}). This funnel region is often difficult to find when employing the same priors on sparticle masses as we have here (flat priors), so it is interesting that they show up in our scans.

\begin{figure}[t!]
\setcounter{subfigure}{0}
\subfigure[][\footnotesize{\textbf{Benchmark 1:}}]{\includegraphics[scale=0.23, trim = 40 230 130 130, clip=true]{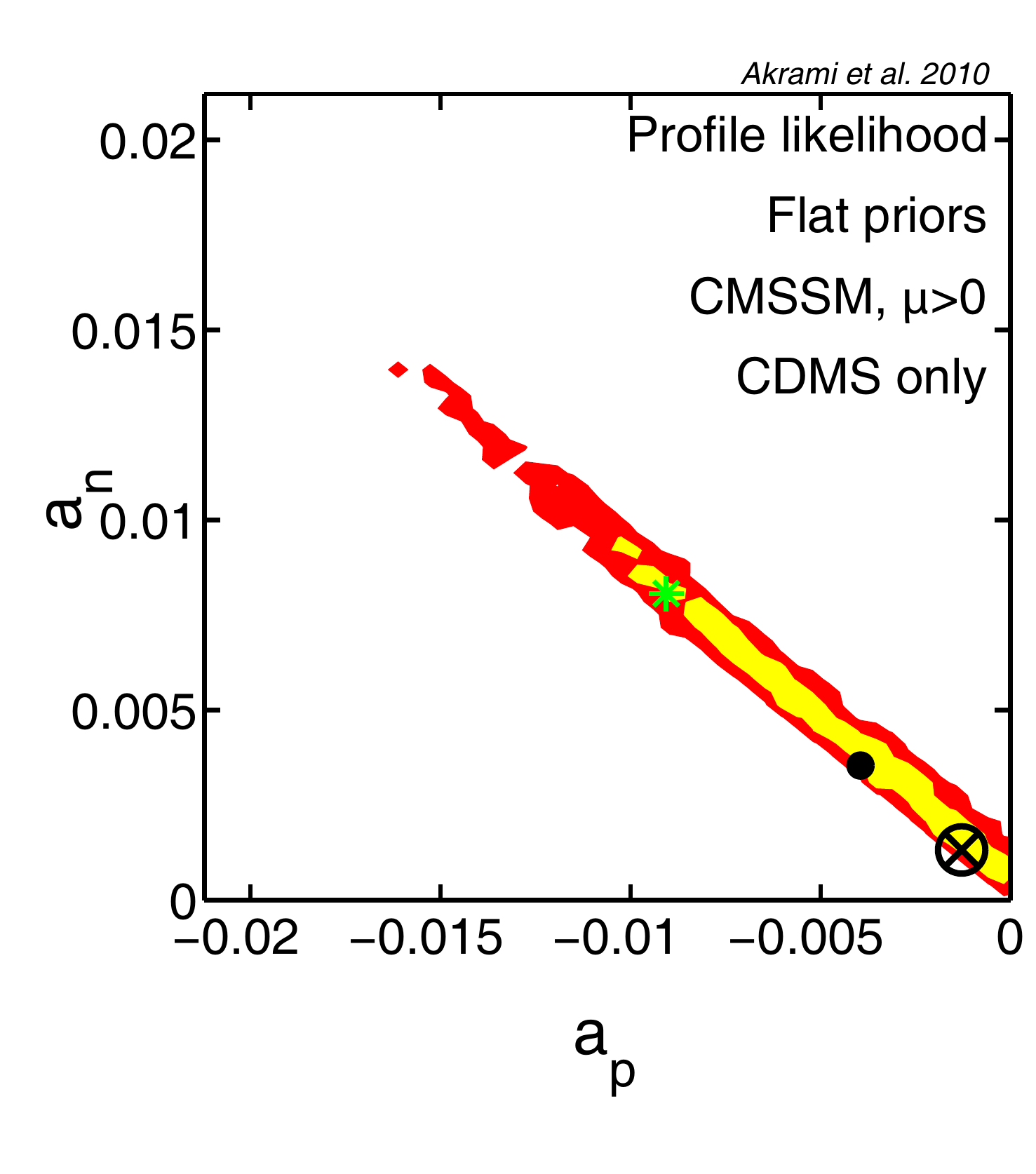}}
\subfigure{\includegraphics[scale=0.23, trim = 40 230 130 130, clip=true]{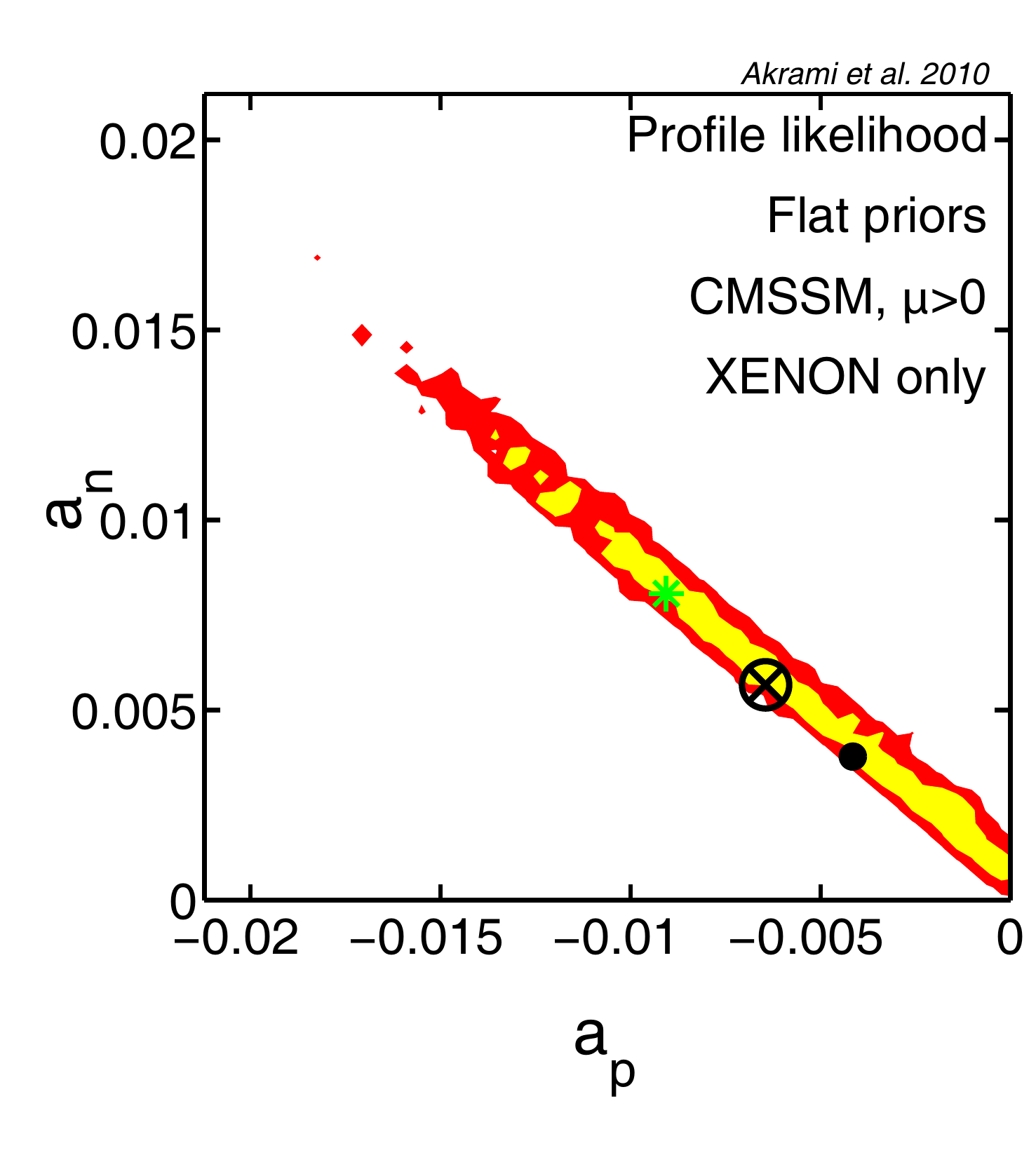}}
\subfigure{\includegraphics[scale=0.23, trim = 40 230 130 130, clip=true]{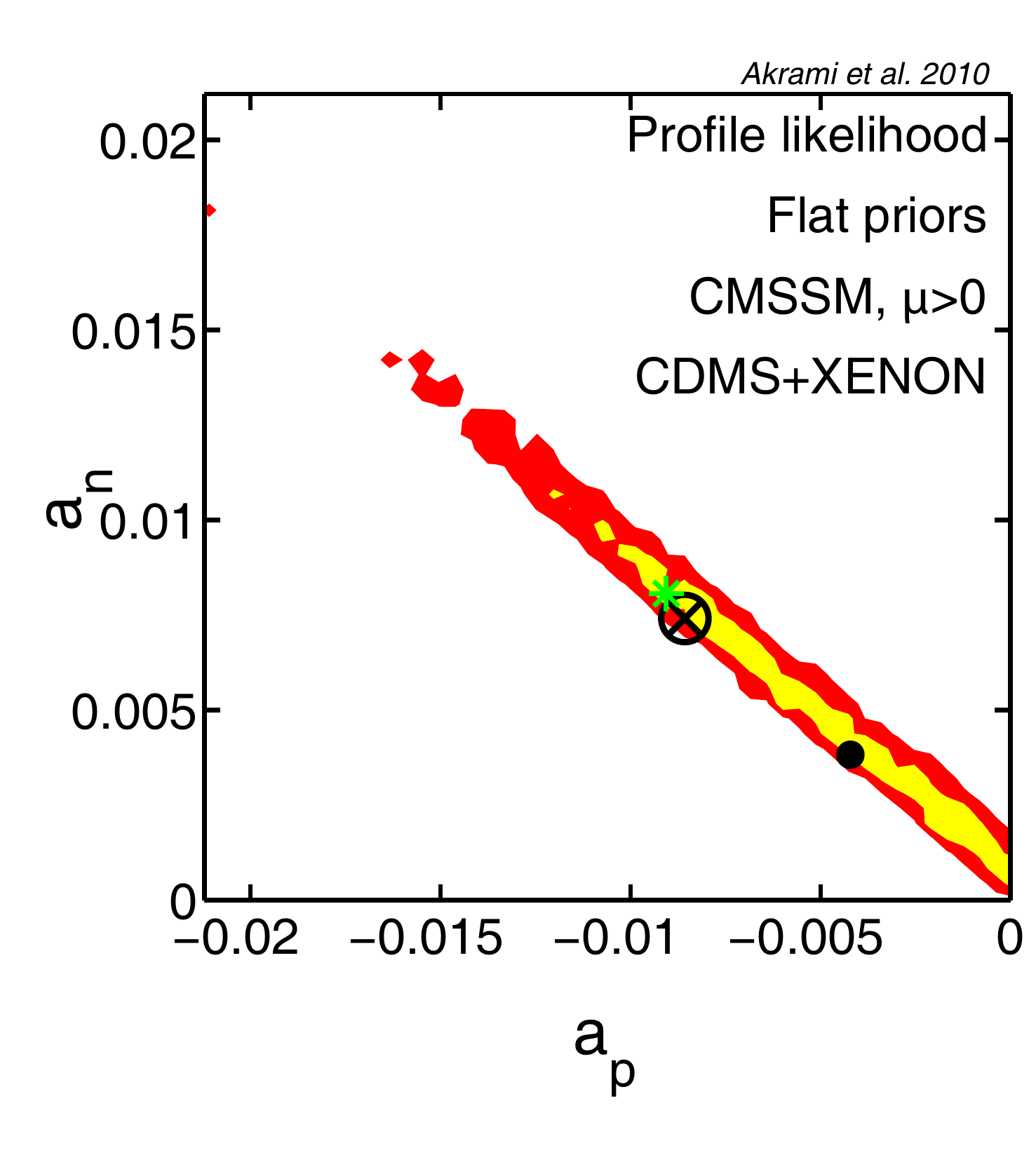}}
\subfigure{\includegraphics[scale=0.23, trim = 40 230 60 130, clip=true]{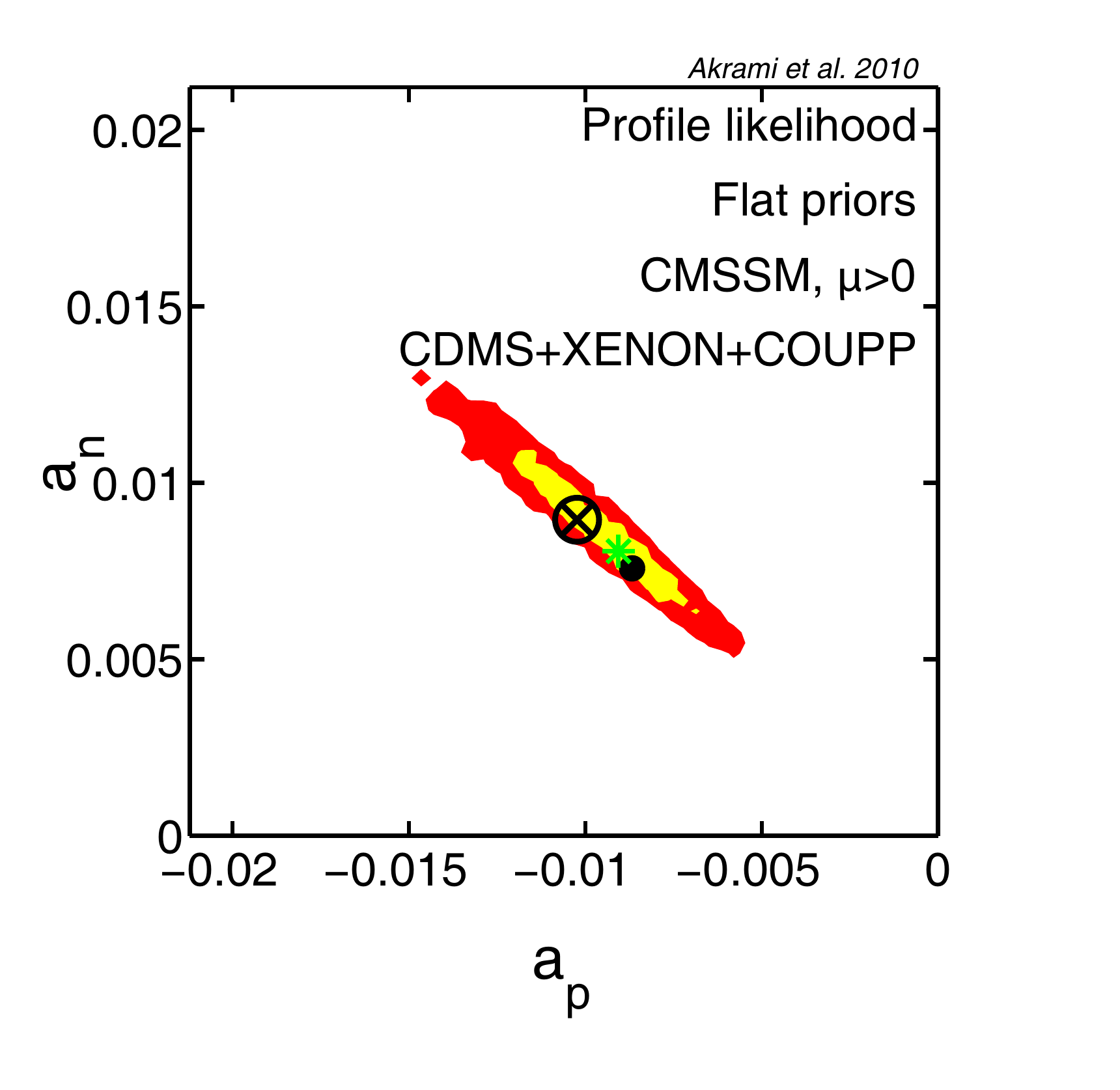}}\\
\setcounter{subfigure}{1}
\subfigure[][\footnotesize{\textbf{Benchmark 2:}}]{\includegraphics[scale=0.23, trim = 40 230 130 130, clip=true]{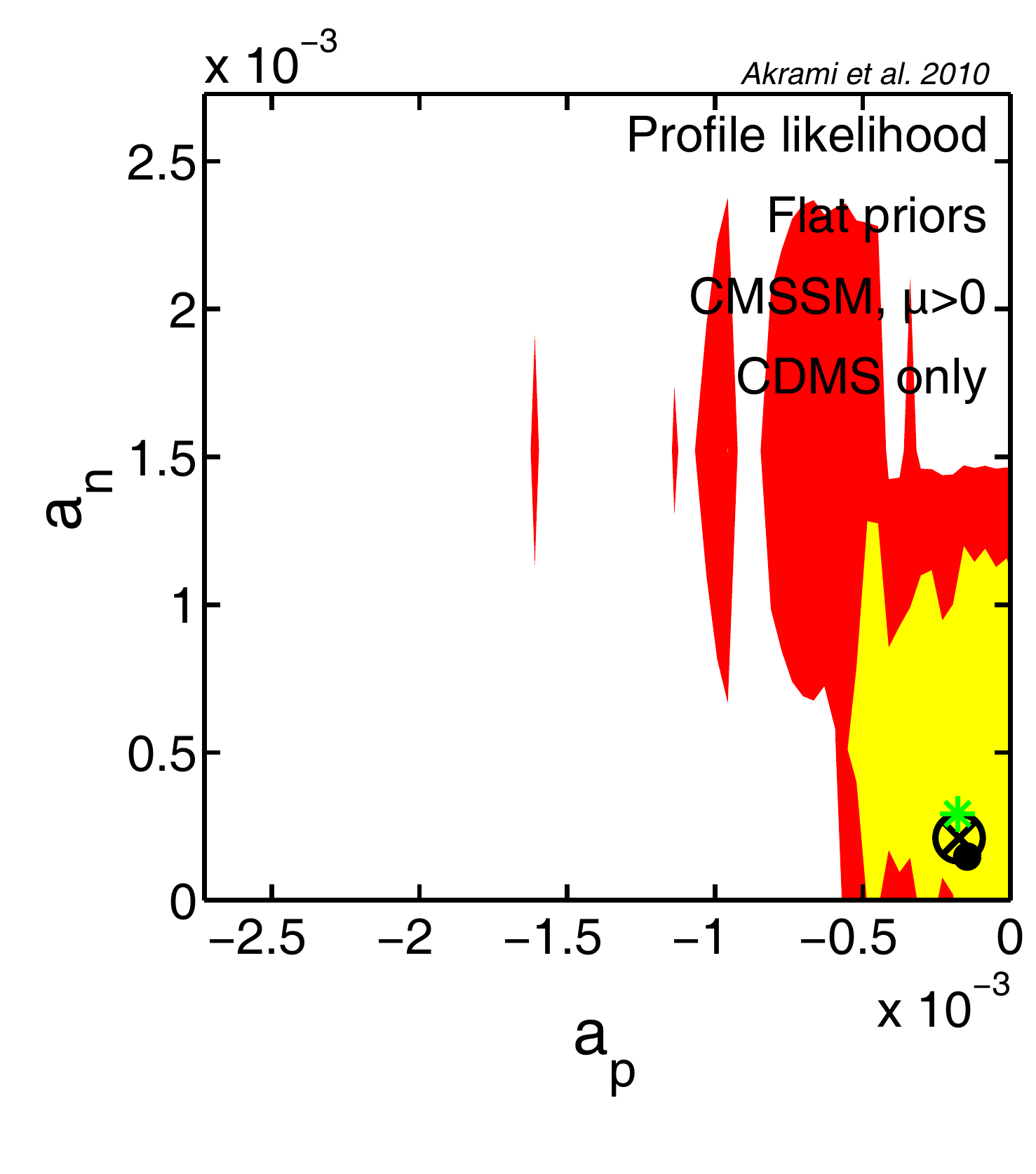}}
\subfigure{\includegraphics[scale=0.23, trim = 40 230 130 130, clip=true]{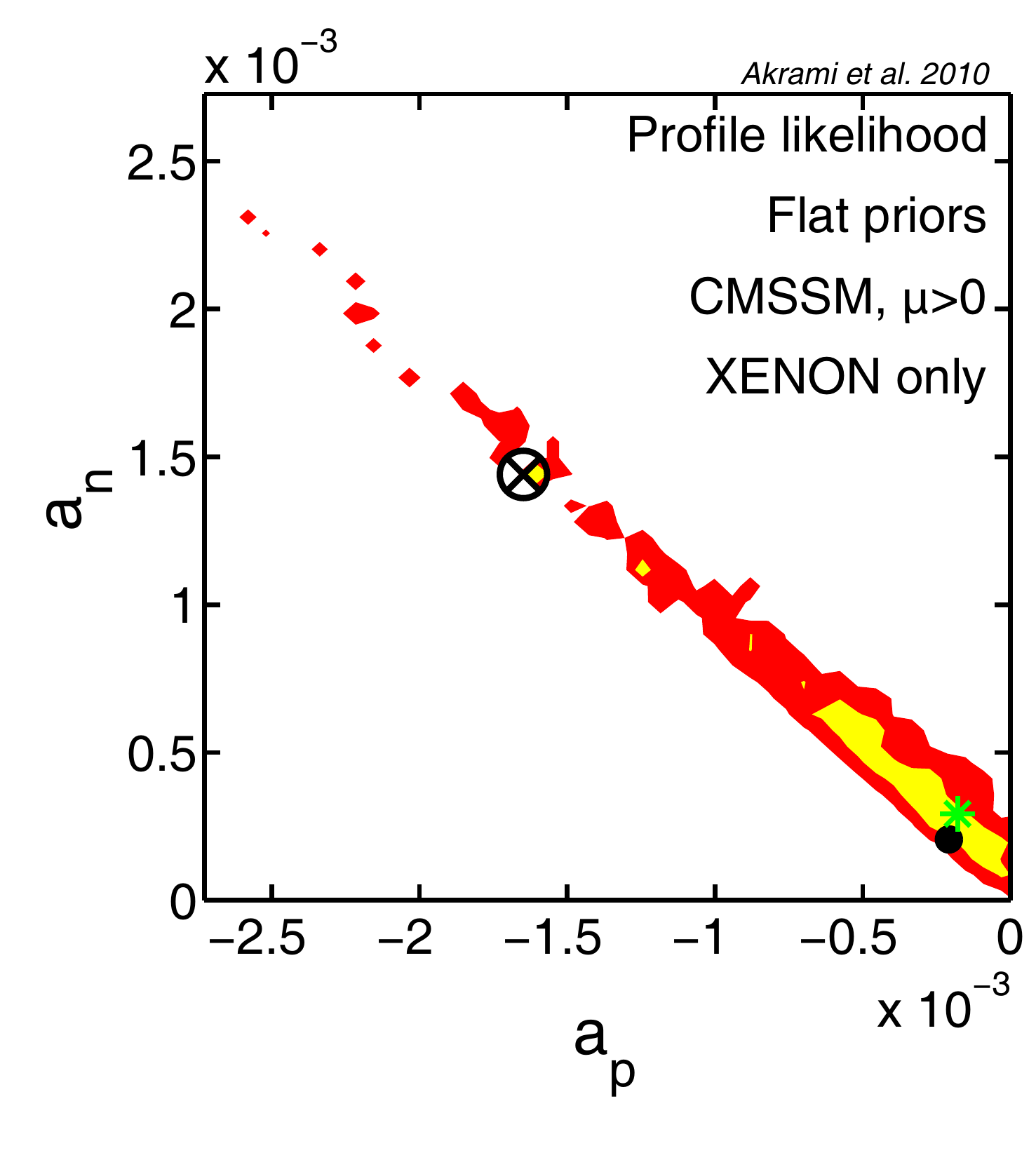}}
\subfigure{\includegraphics[scale=0.23, trim = 40 230 130 130, clip=true]{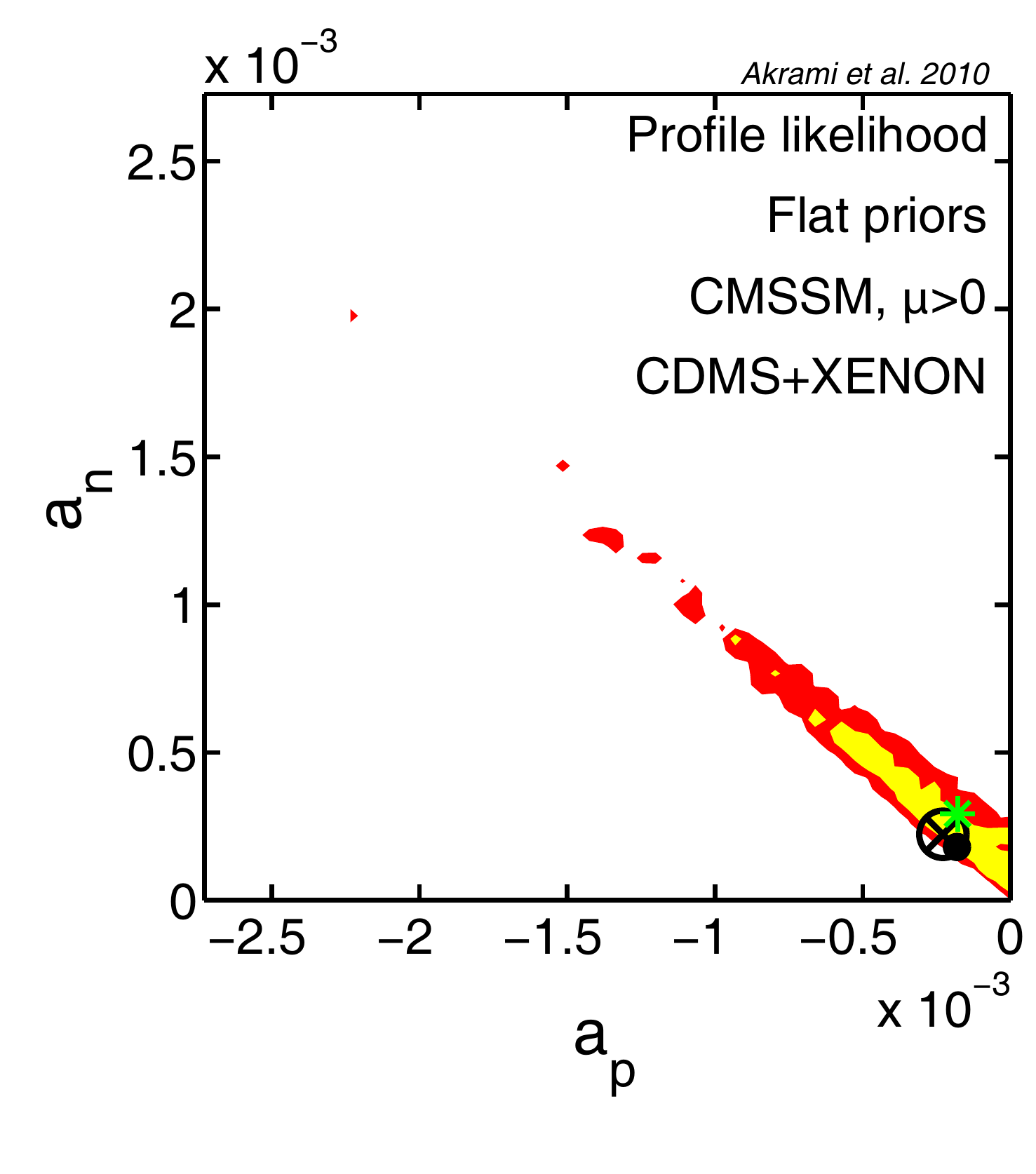}}
\subfigure{\includegraphics[scale=0.23, trim = 40 230 60 130, clip=true]{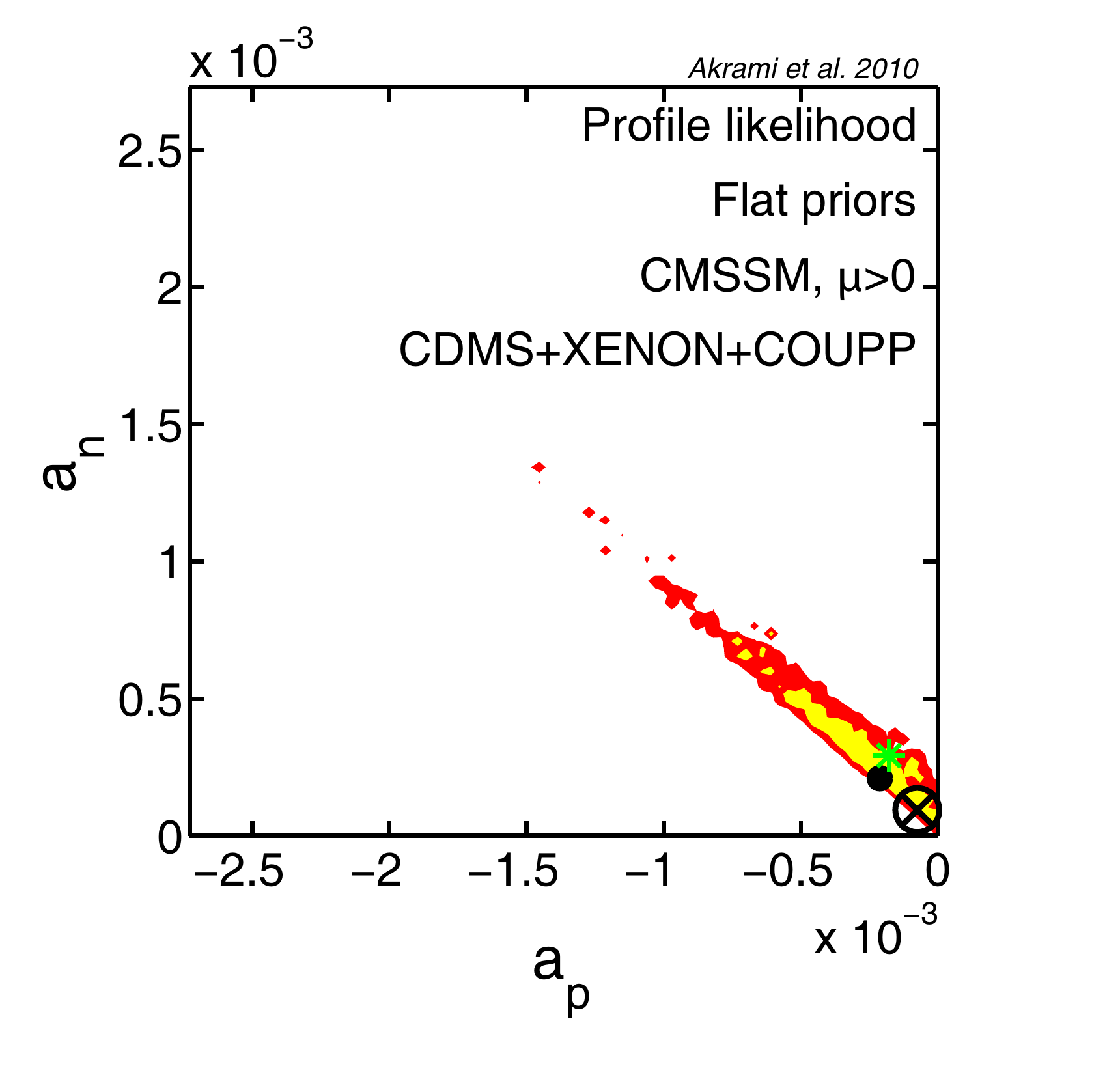}}\\
\setcounter{subfigure}{2}
\subfigure[][\footnotesize{\textbf{Benchmark 3:}}]{\includegraphics[scale=0.23, trim = 40 230 130 130, clip=true]{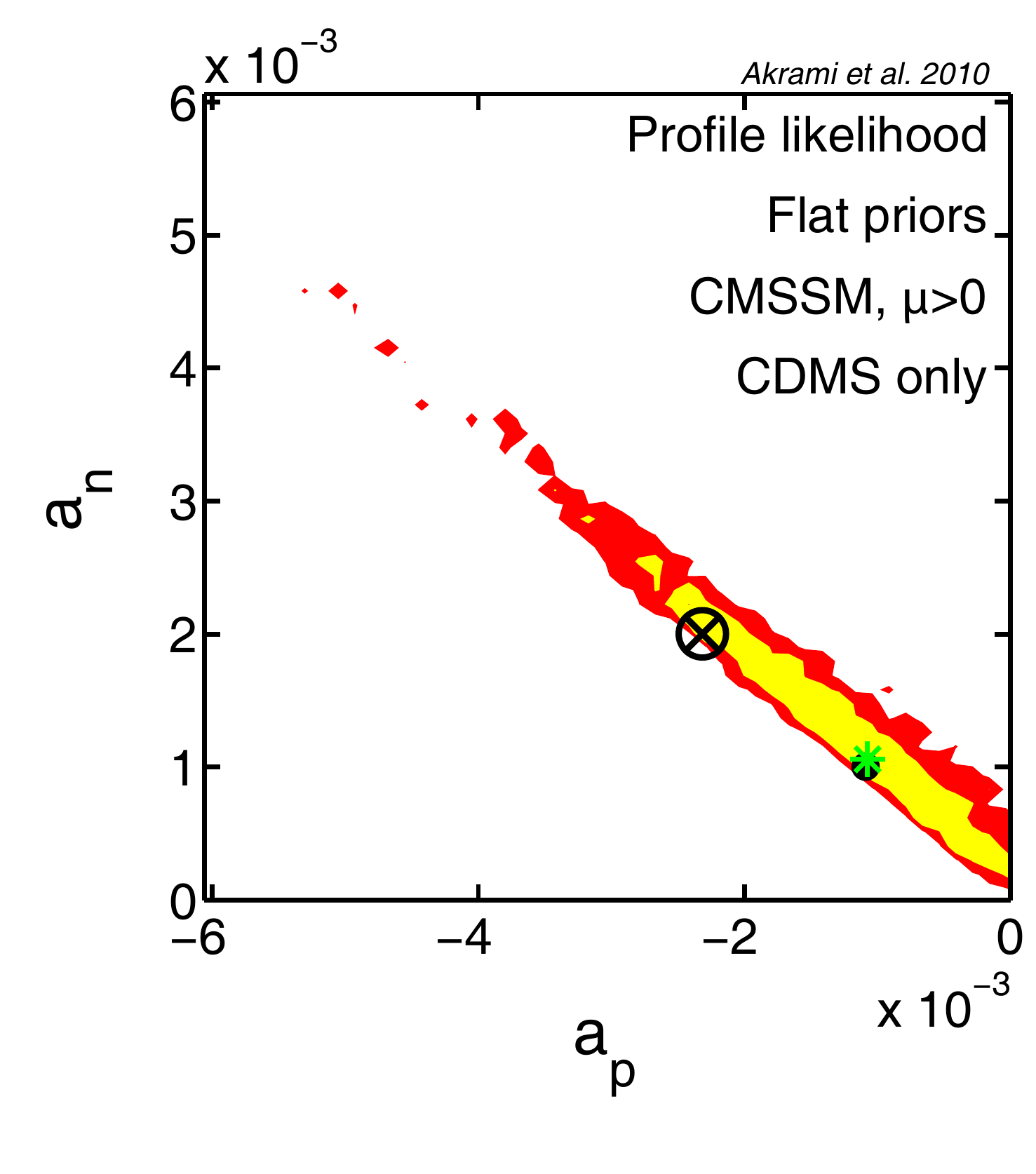}}
\subfigure{\includegraphics[scale=0.23, trim = 40 230 130 130, clip=true]{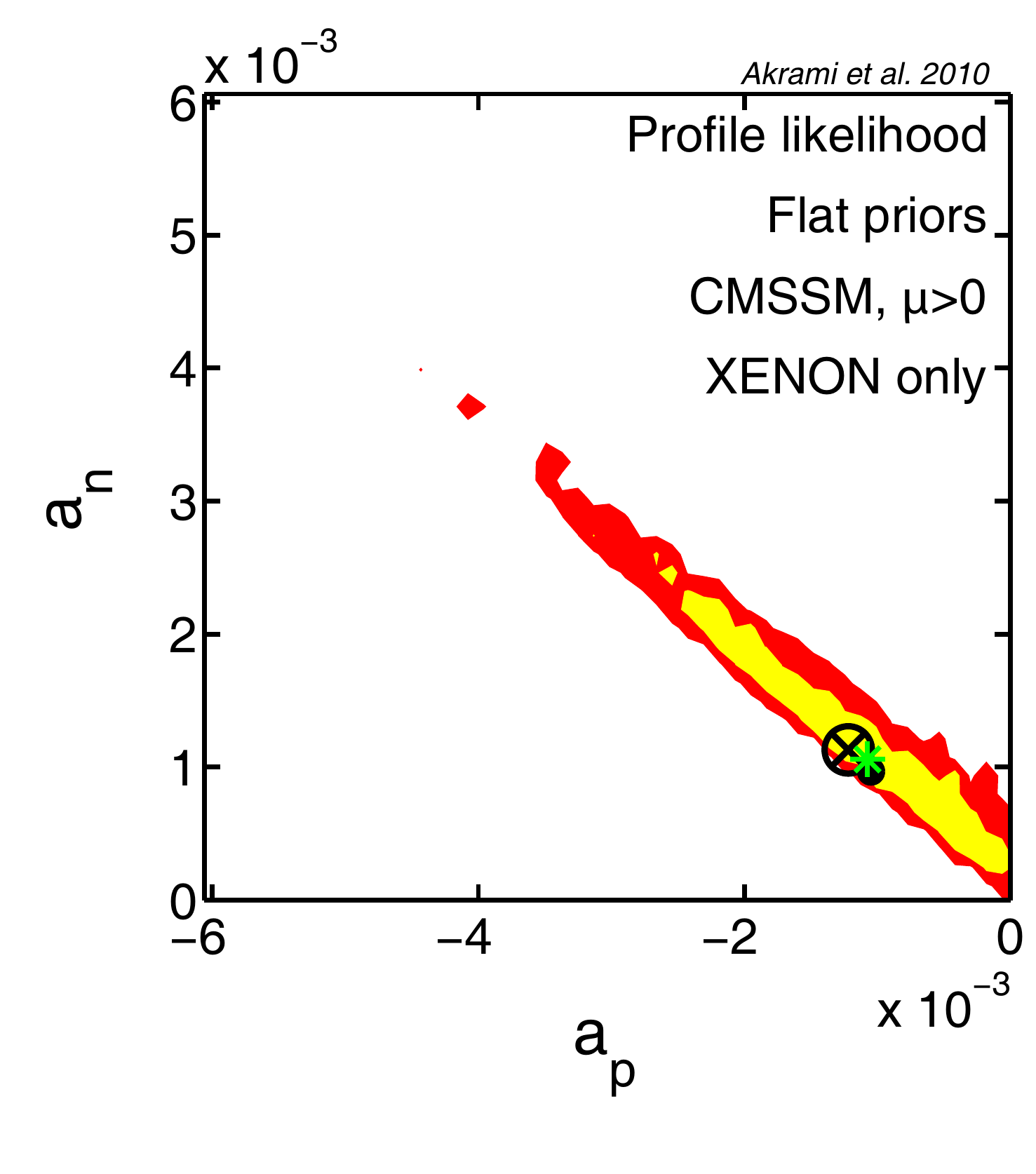}}
\subfigure{\includegraphics[scale=0.23, trim = 40 230 130 130, clip=true]{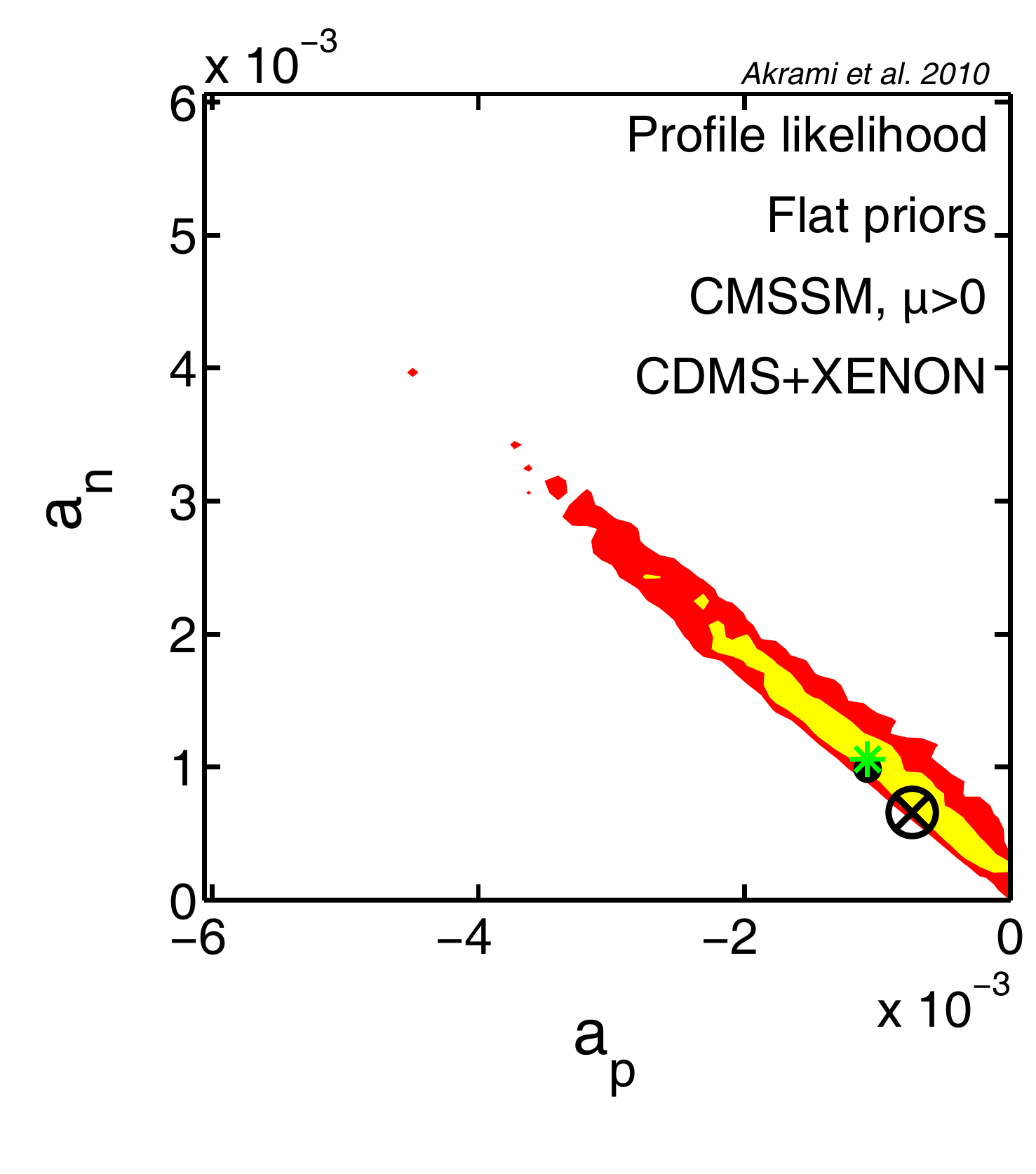}}
\subfigure{\includegraphics[scale=0.23, trim = 40 230 60 130, clip=true]{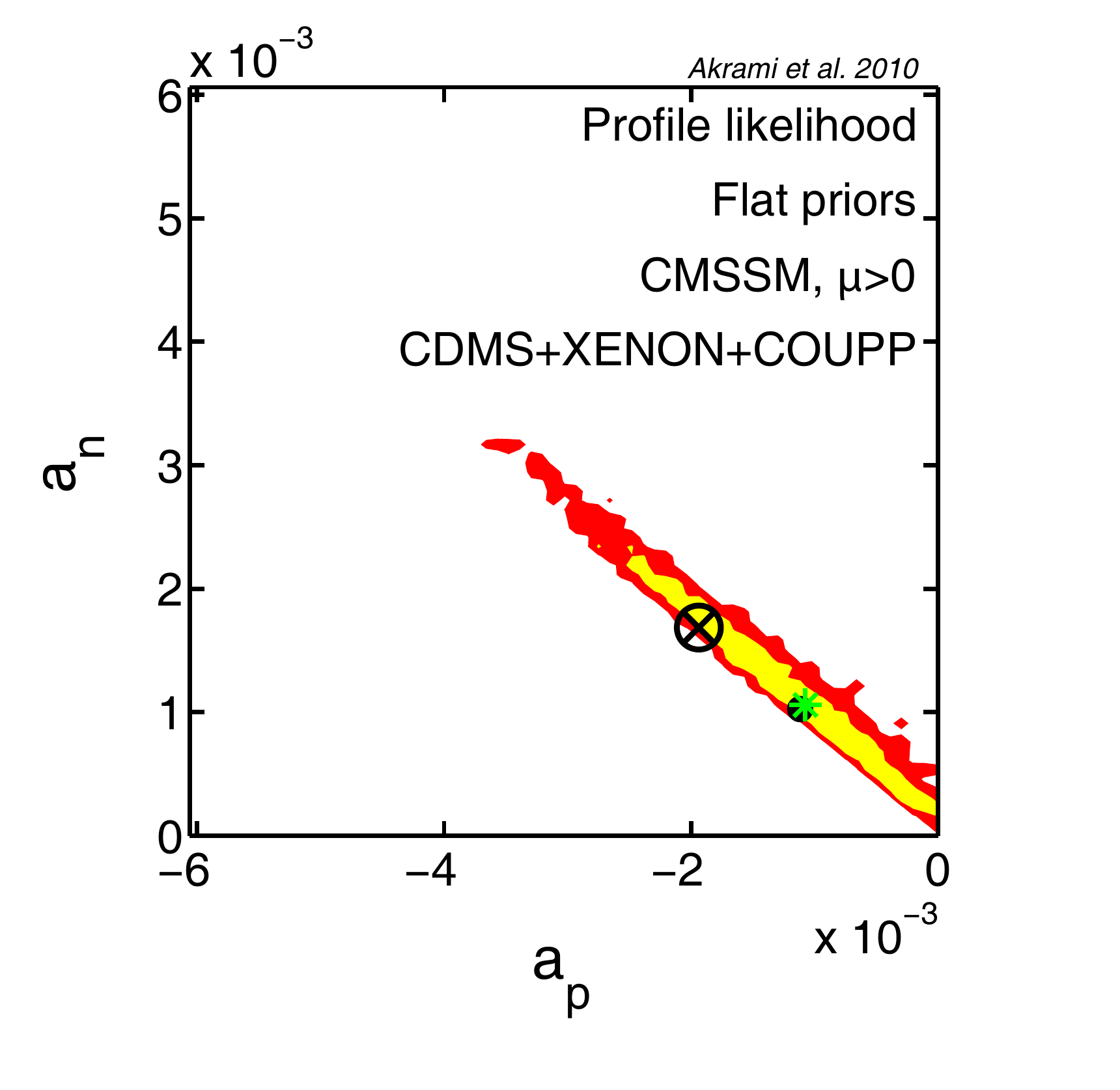}}\\
\setcounter{subfigure}{3}
\subfigure[][\footnotesize{\textbf{Benchmark 4:}}]{\includegraphics[scale=0.23, trim = 40 230 130 130, clip=true]{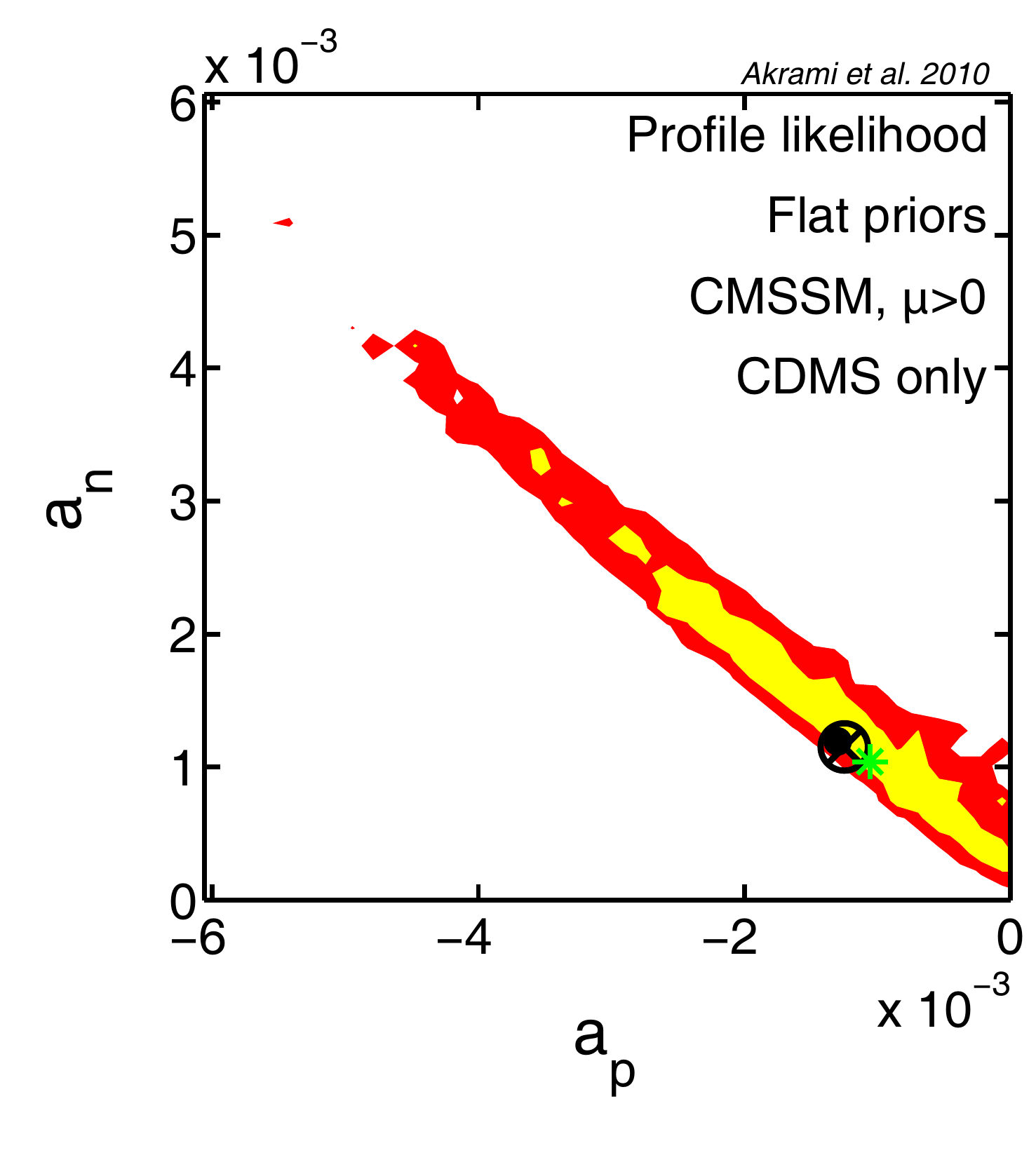}}
\subfigure{\includegraphics[scale=0.23, trim = 40 230 130 130, clip=true]{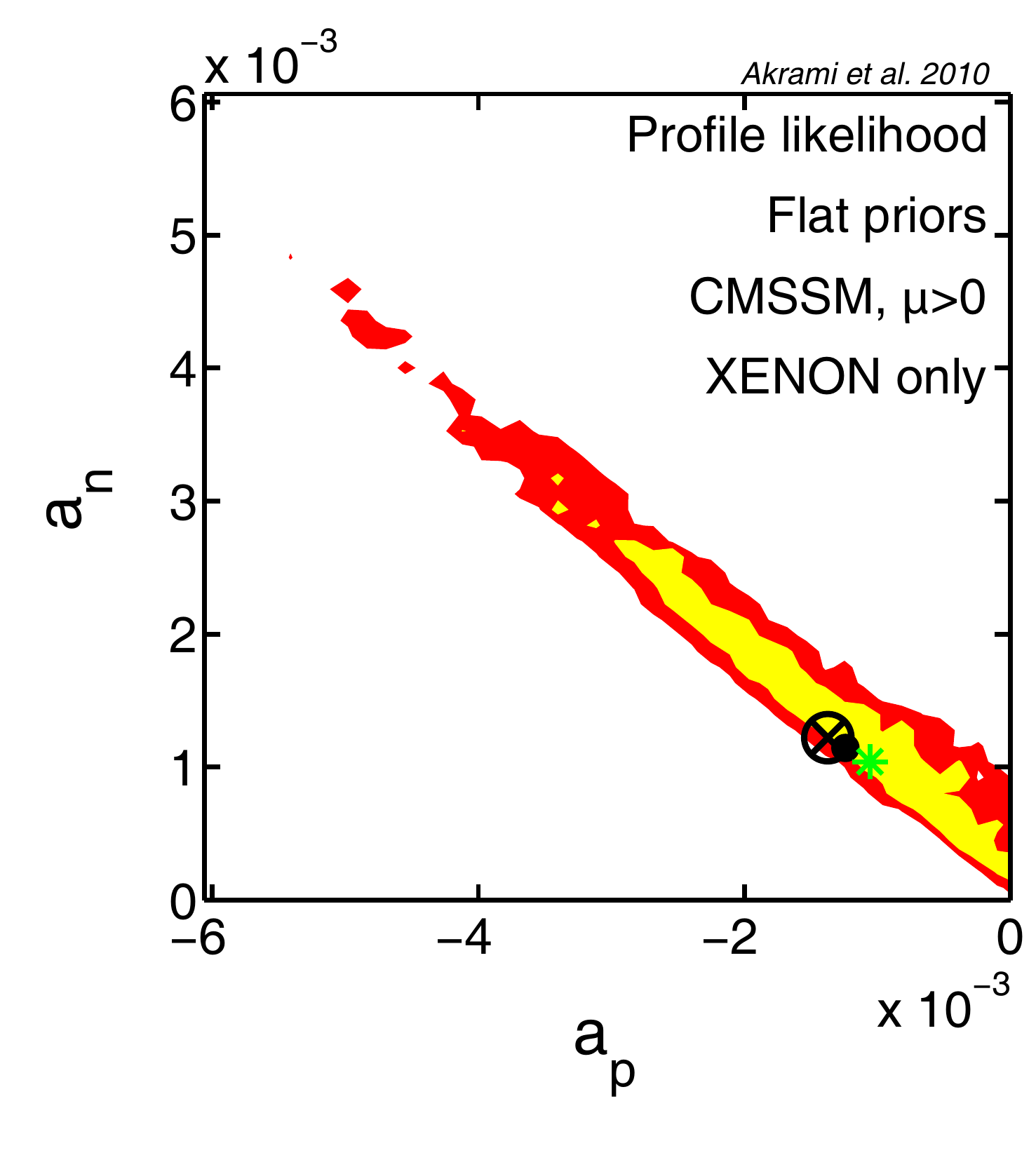}}
\subfigure{\includegraphics[scale=0.23, trim = 40 230 130 130, clip=true]{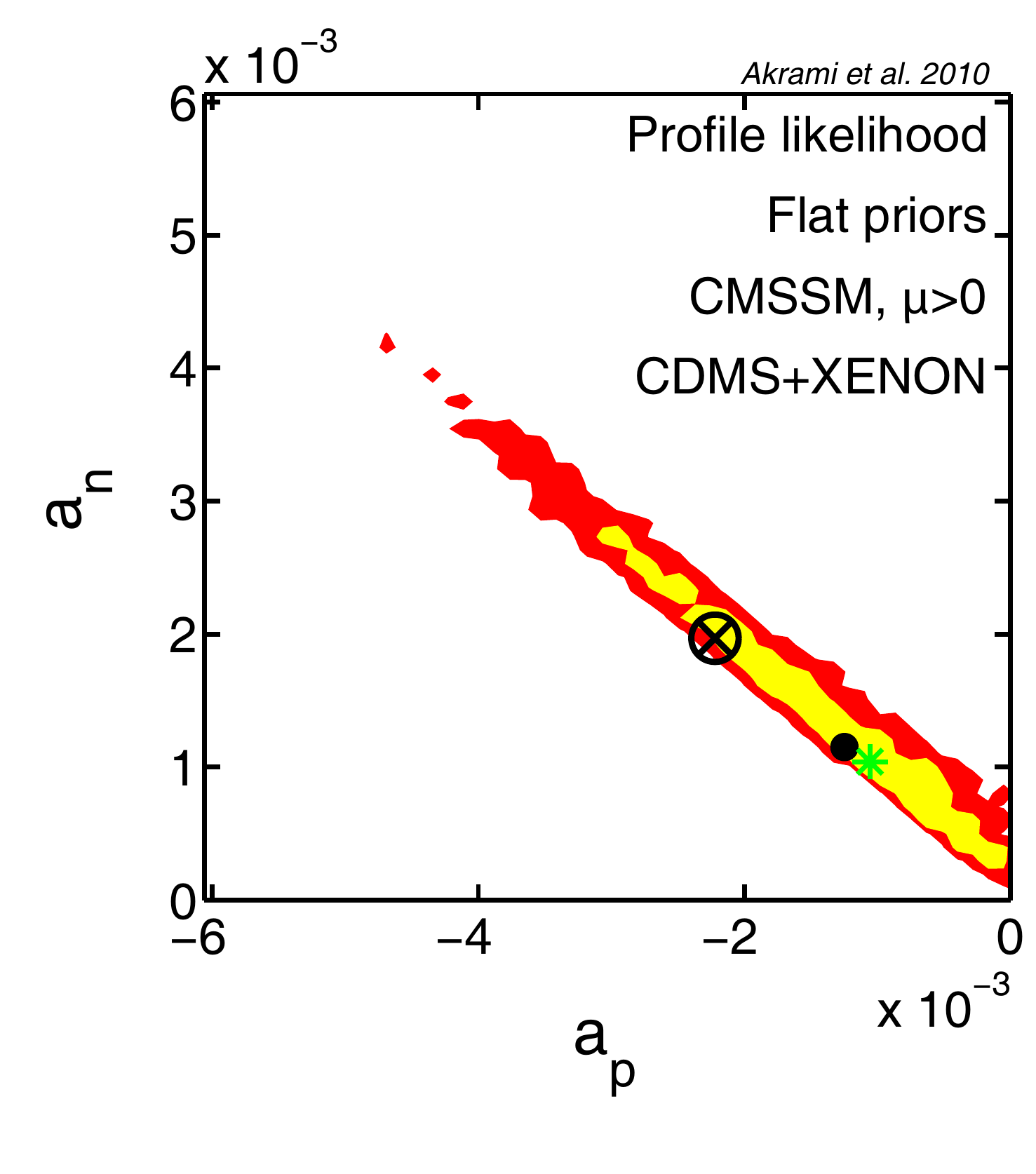}}
\subfigure{\includegraphics[scale=0.23, trim = 40 230 60 130, clip=true]{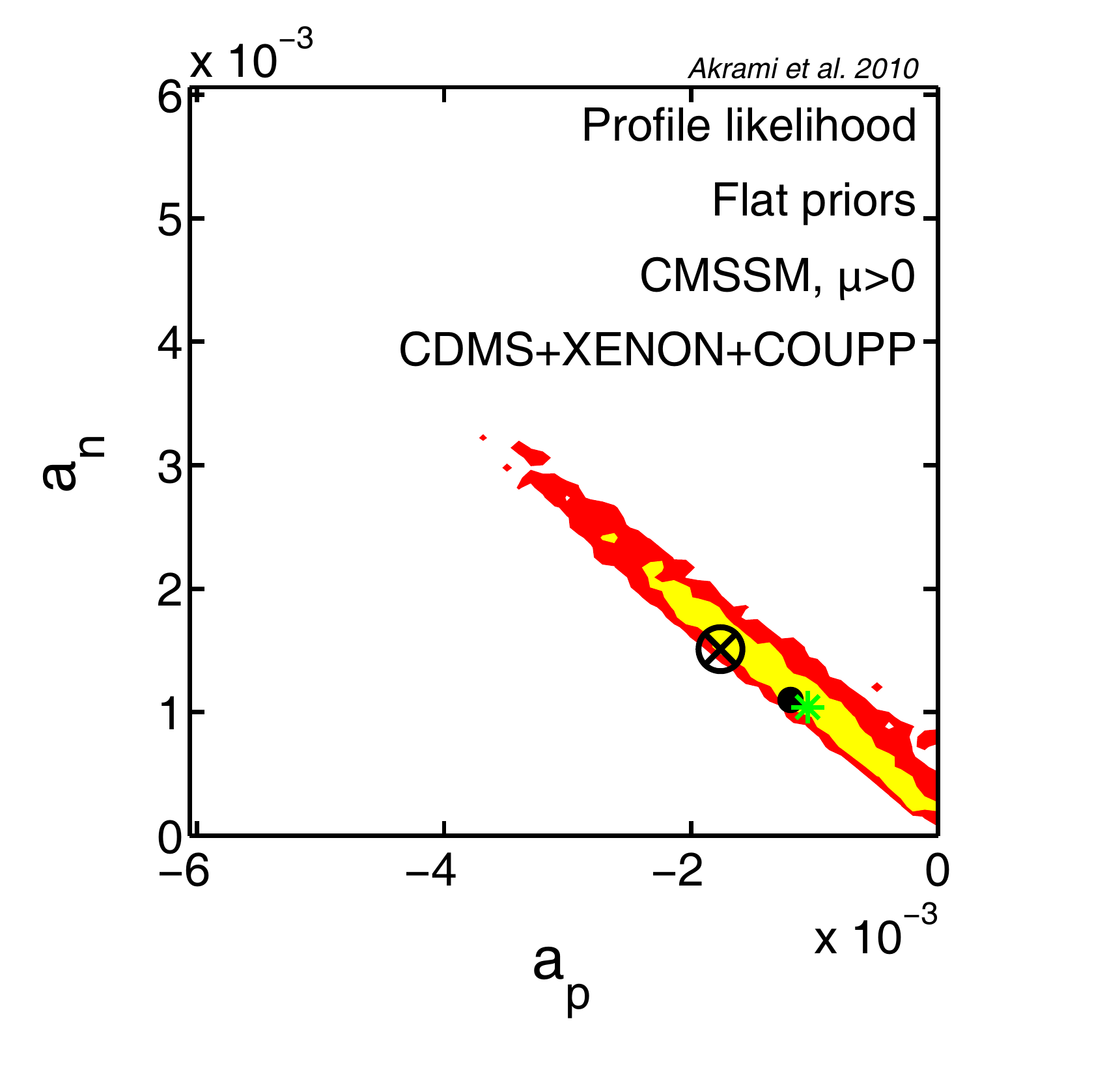}}\\
\caption[aa]{\footnotesize{As in~\fig{fig:anapmarg}, but for two-dimensional profile likelihoods.}}\label{fig:anapprofl}
\end{figure}

\begin{figure}[t!]
\setcounter{subfigure}{0}
\subfigure[\footnotesize{\textbf{Benchmark 1:}}]{\includegraphics[scale=0.23, trim = 40 230 130 100, clip=true]{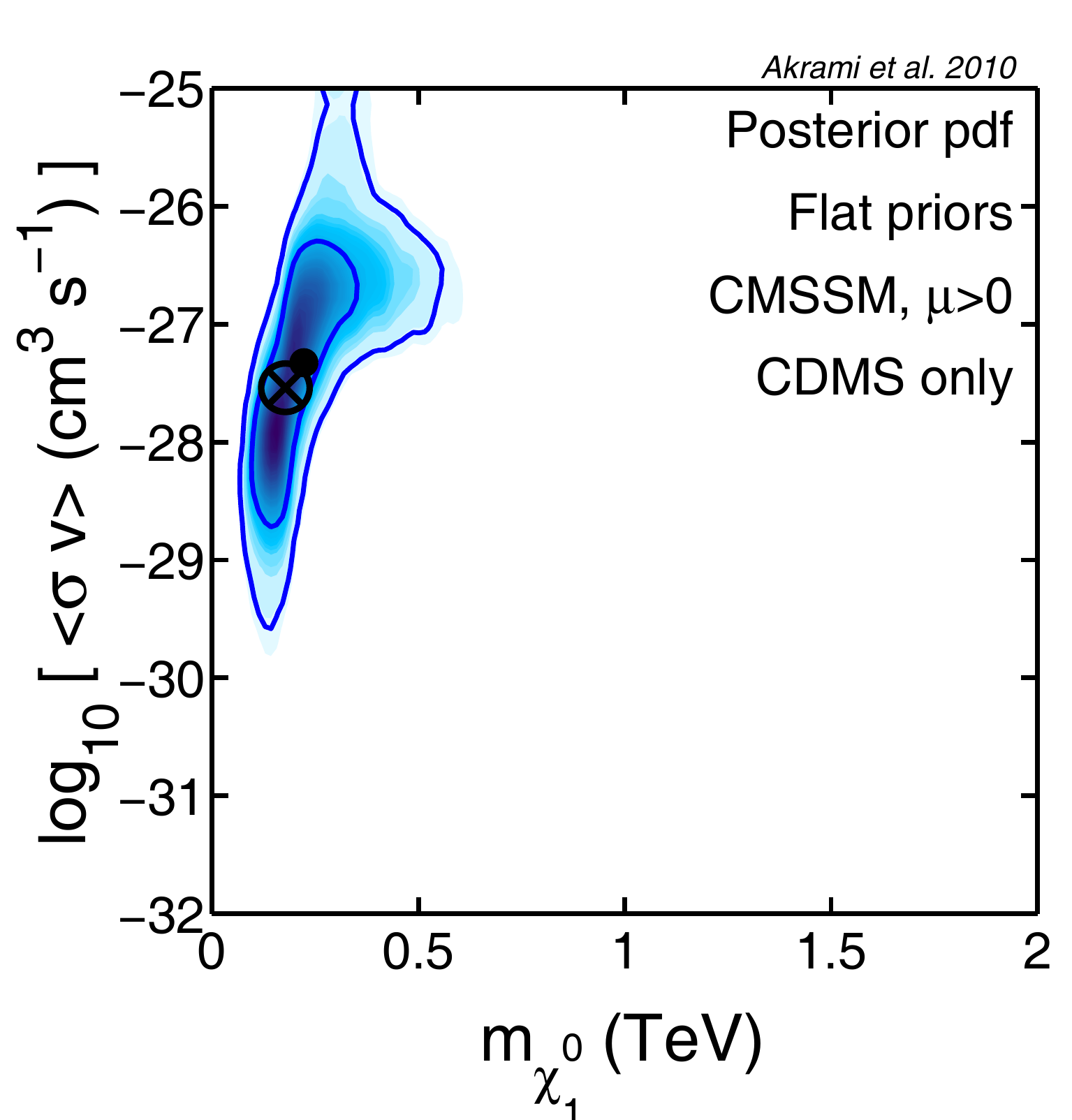}}
\subfigure{\includegraphics[scale=0.23, trim = 40 230 130 100, clip=true]{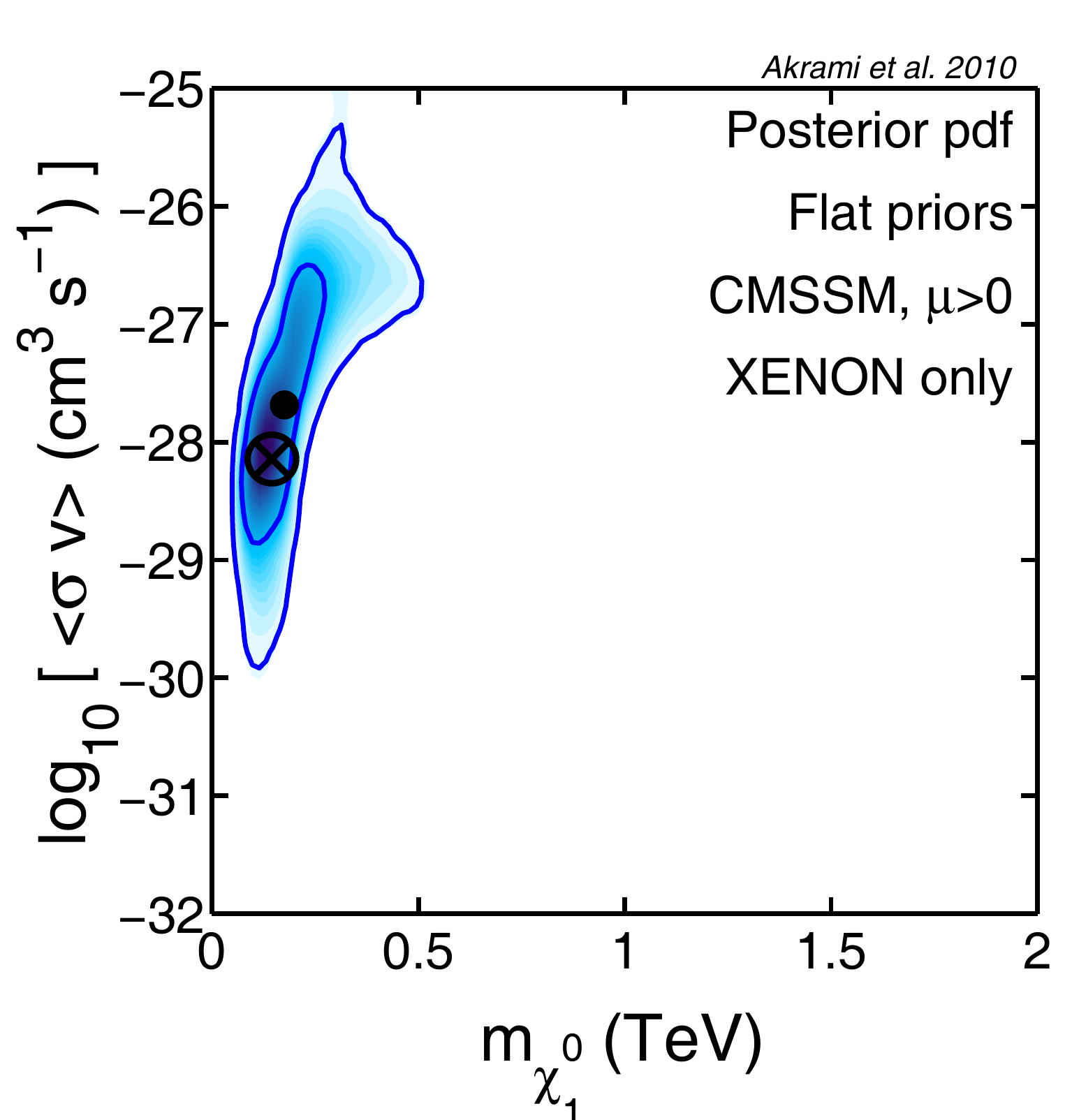}}
\subfigure{\includegraphics[scale=0.23, trim = 40 230 130 100, clip=true]{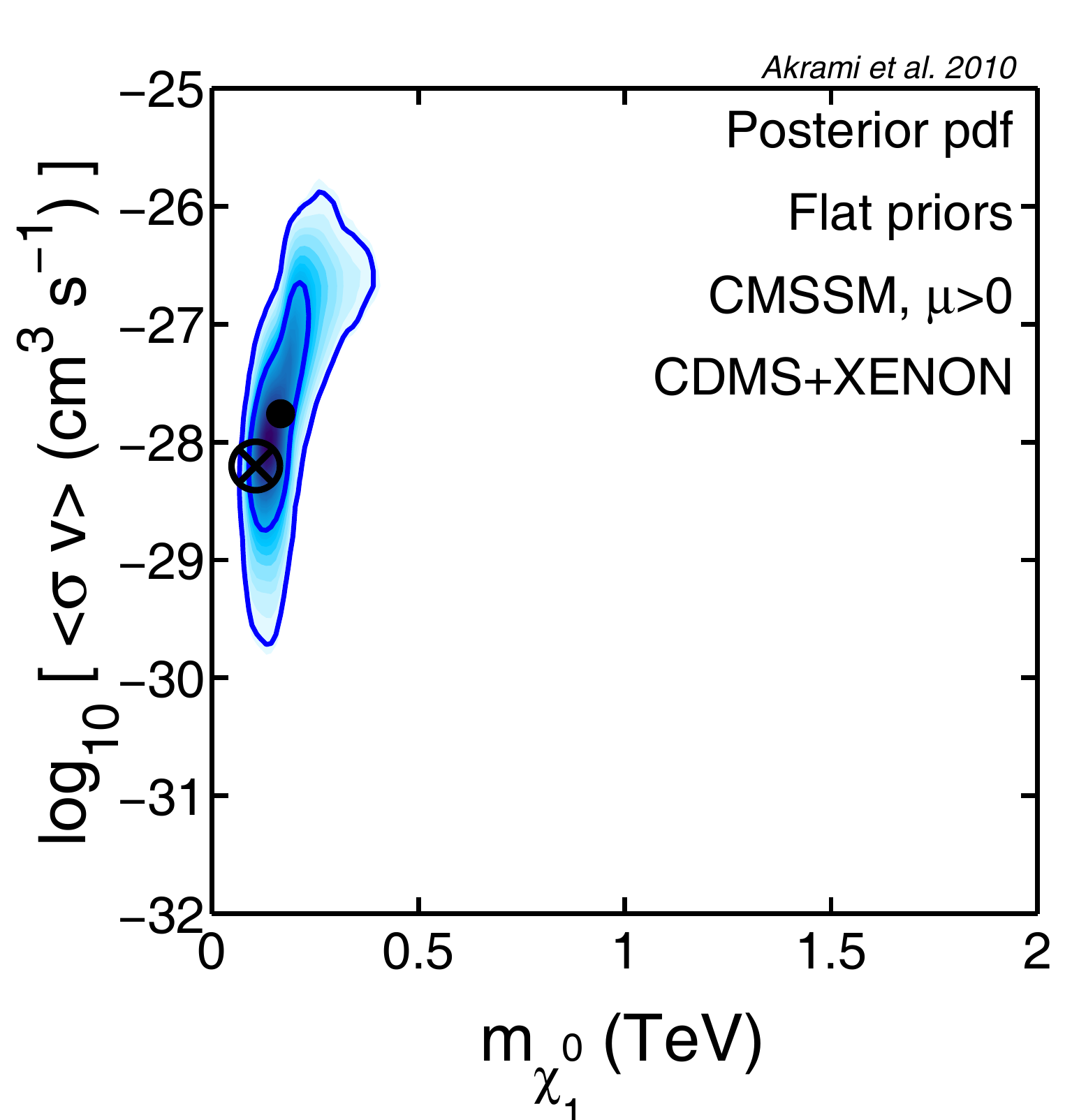}}
\subfigure{\includegraphics[scale=0.23, trim = 40 230 60 100, clip=true]{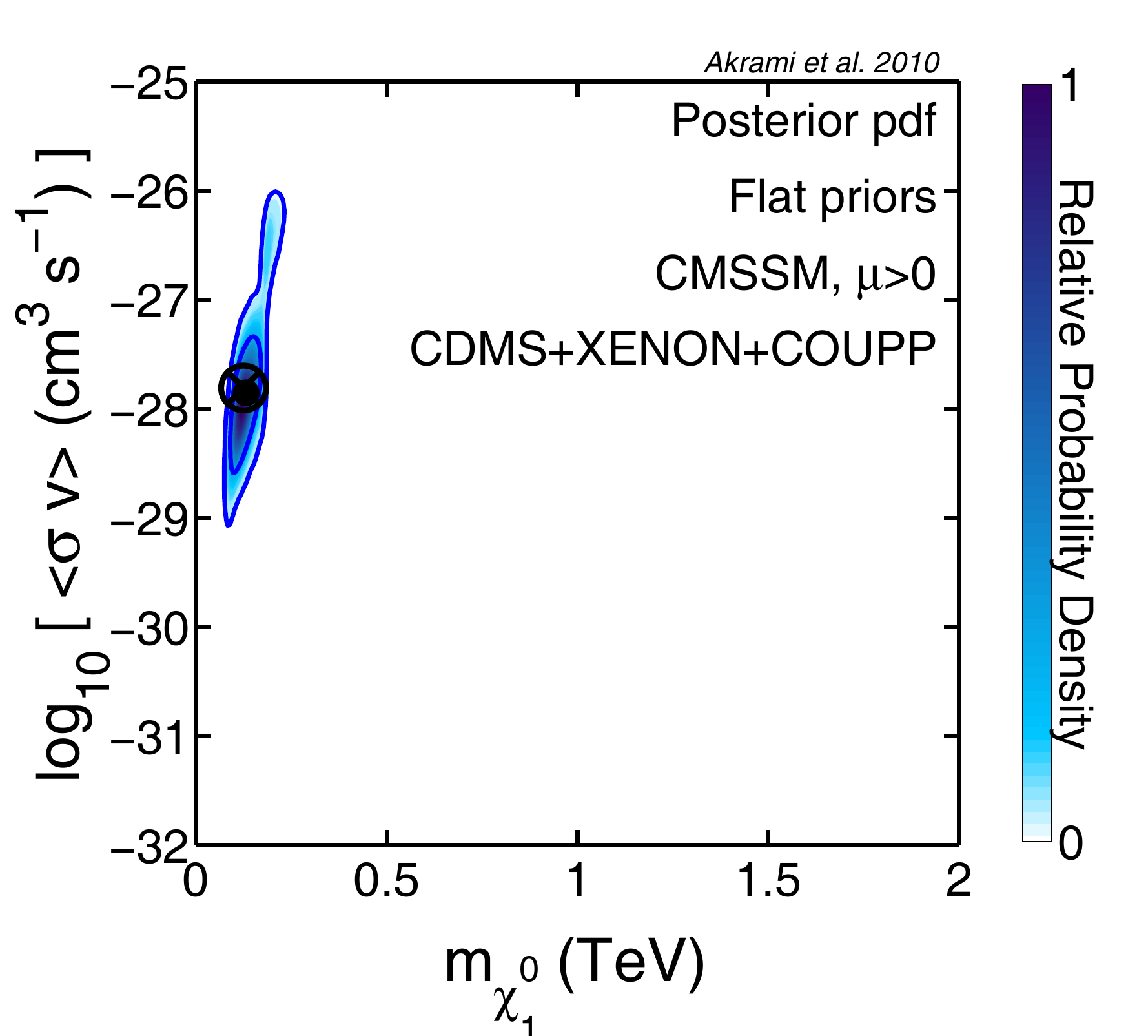}}\\
\setcounter{subfigure}{1}
\subfigure[\footnotesize{\textbf{Benchmark 2:}}]{\includegraphics[scale=0.23, trim = 40 230 130 100, clip=true]{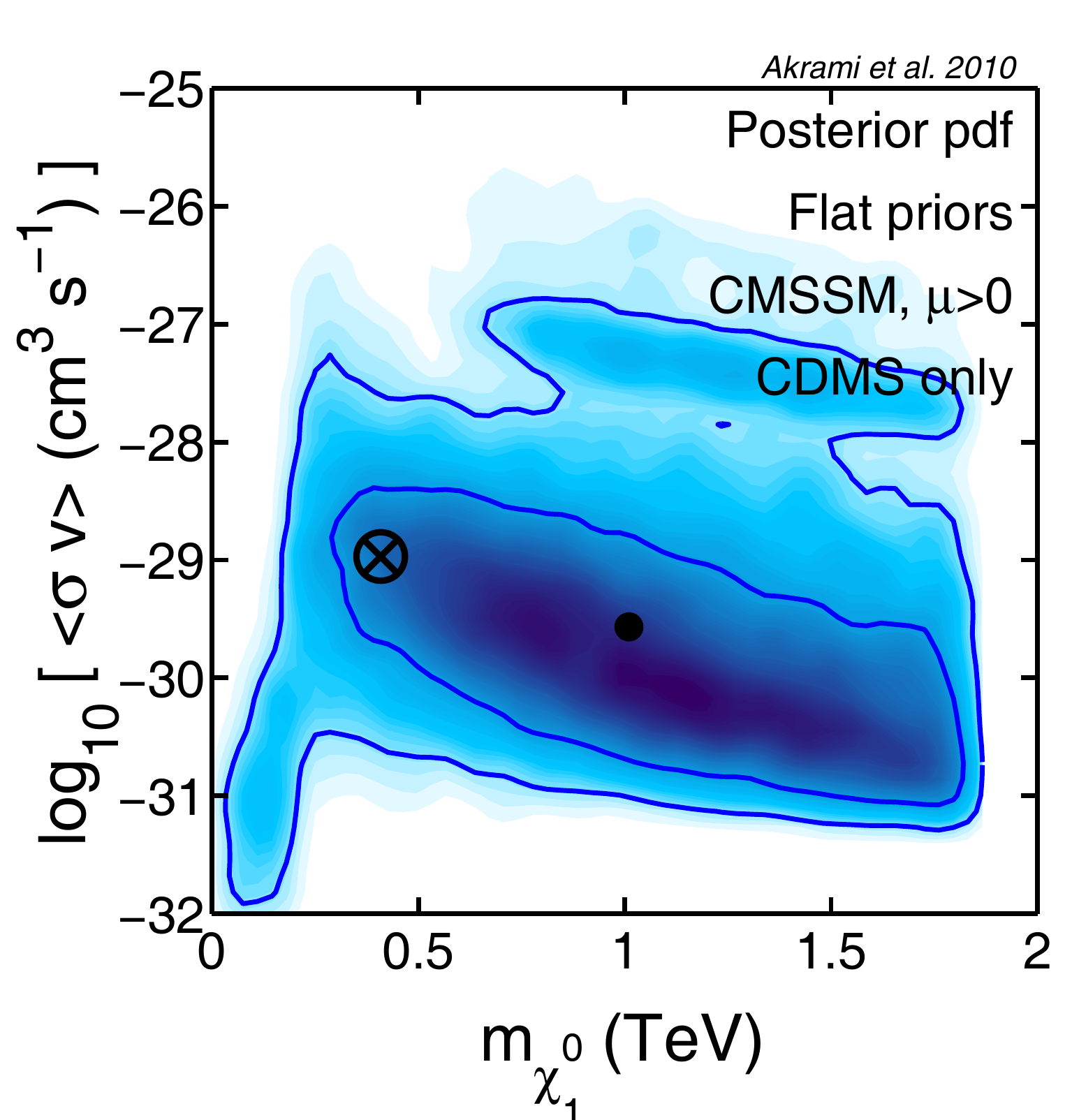}}
\subfigure{\includegraphics[scale=0.23, trim = 40 230 130 100, clip=true]{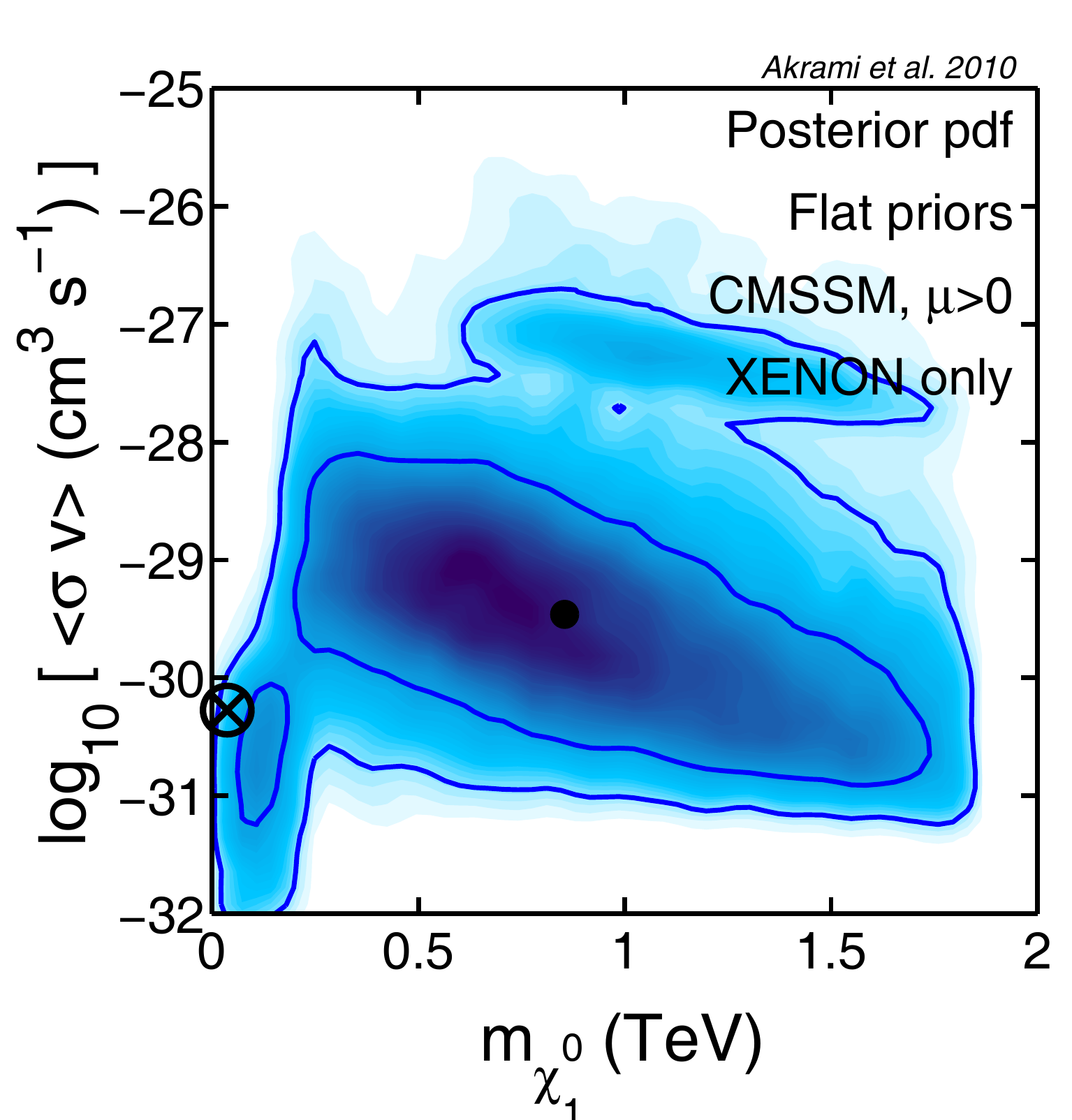}}
\subfigure{\includegraphics[scale=0.23, trim = 40 230 130 100, clip=true]{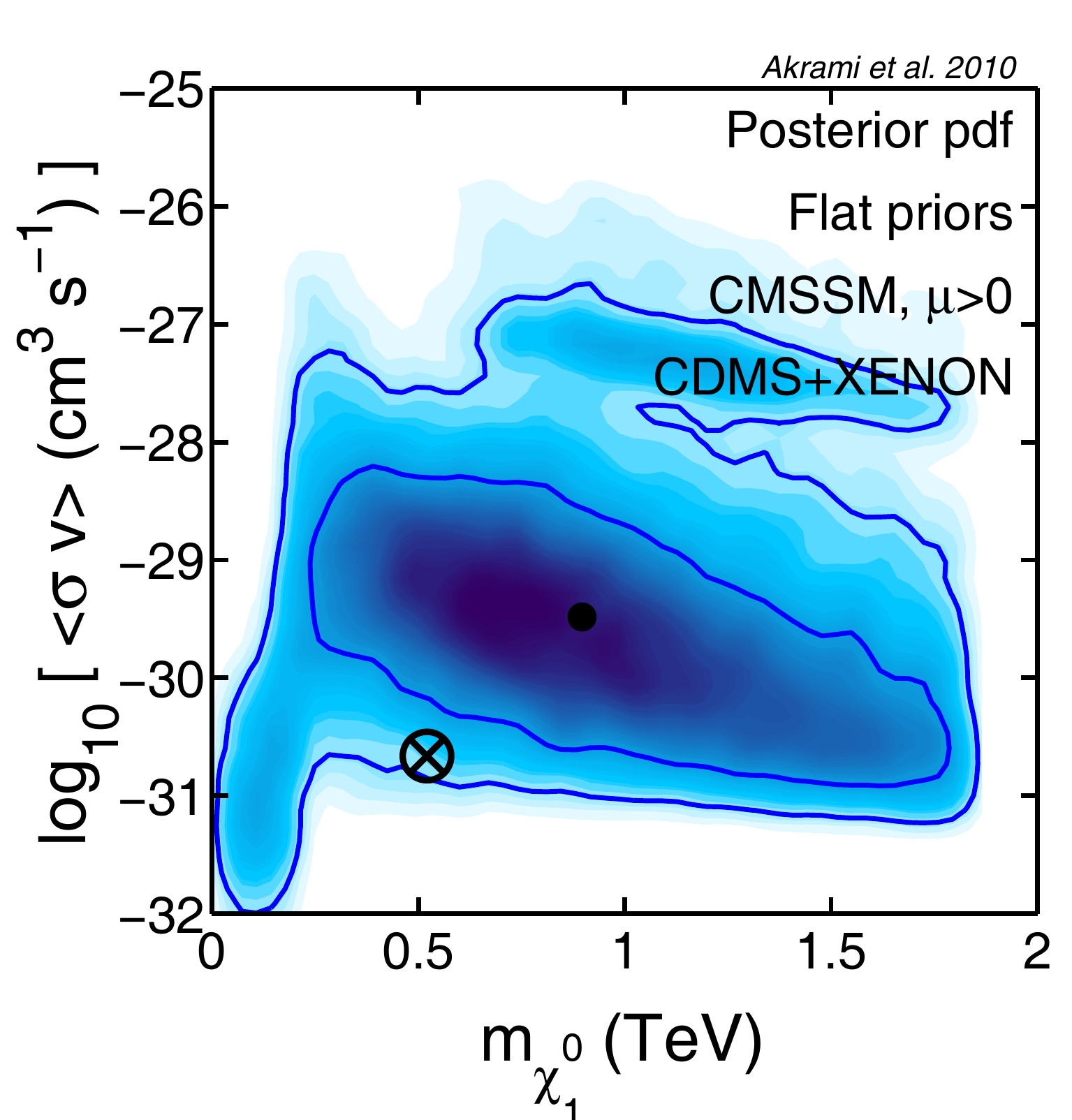}}
\subfigure{\includegraphics[scale=0.23, trim = 40 230 60 100, clip=true]{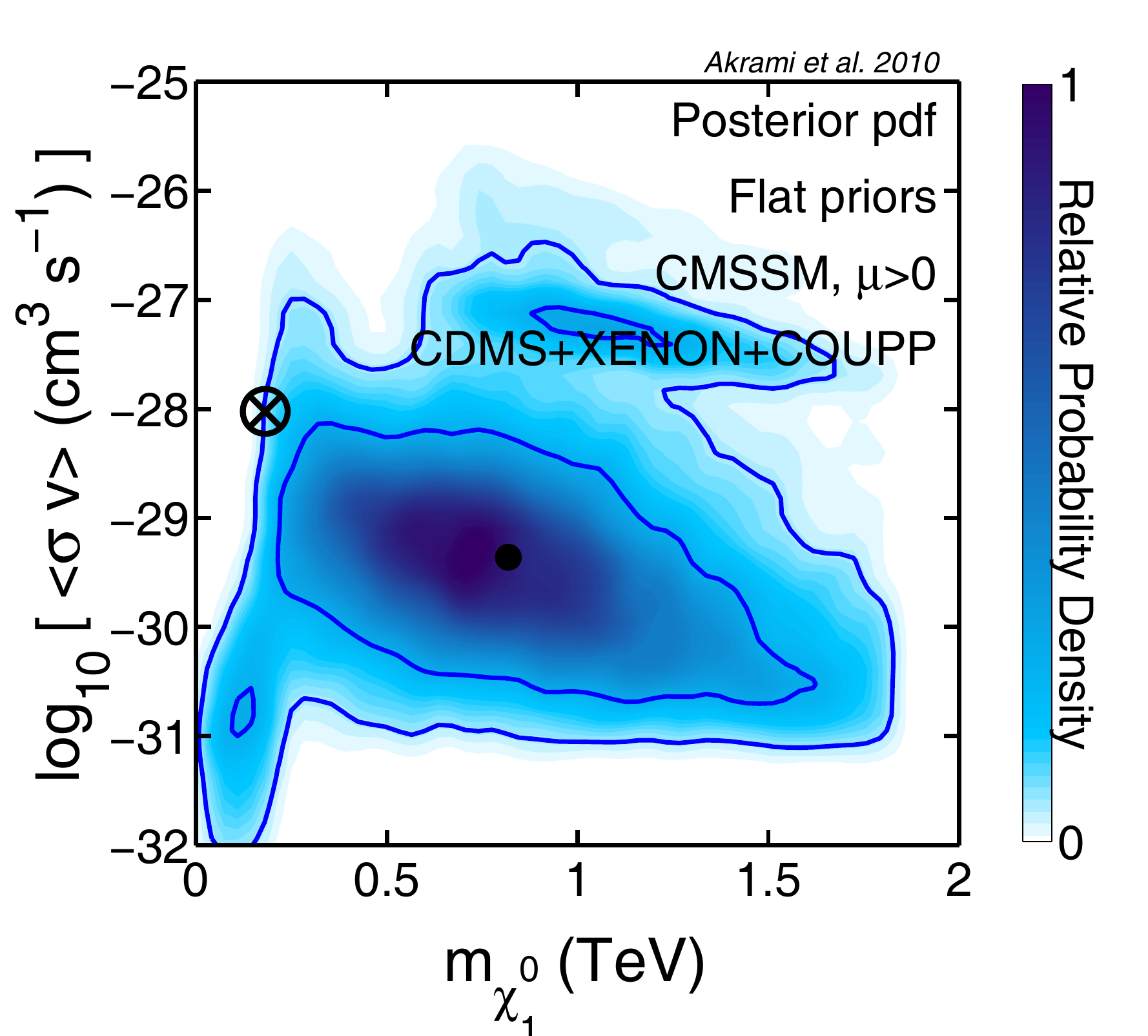}}\\
\setcounter{subfigure}{2}
\subfigure[\footnotesize{\textbf{Benchmark 3:}}]{\includegraphics[scale=0.23, trim = 40 230 130 100, clip=true]{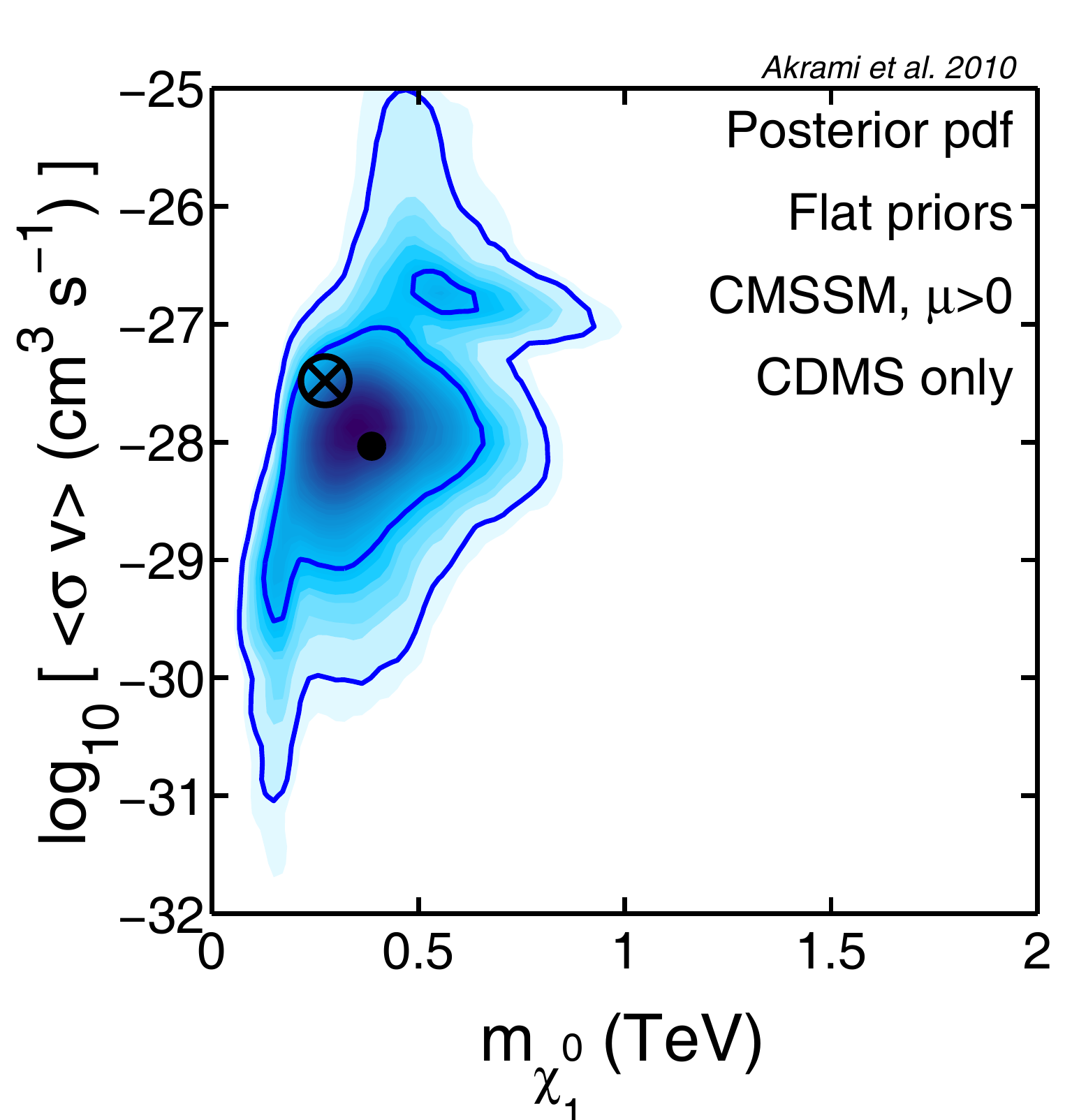}}
\subfigure{\includegraphics[scale=0.23, trim = 40 230 130 100, clip=true]{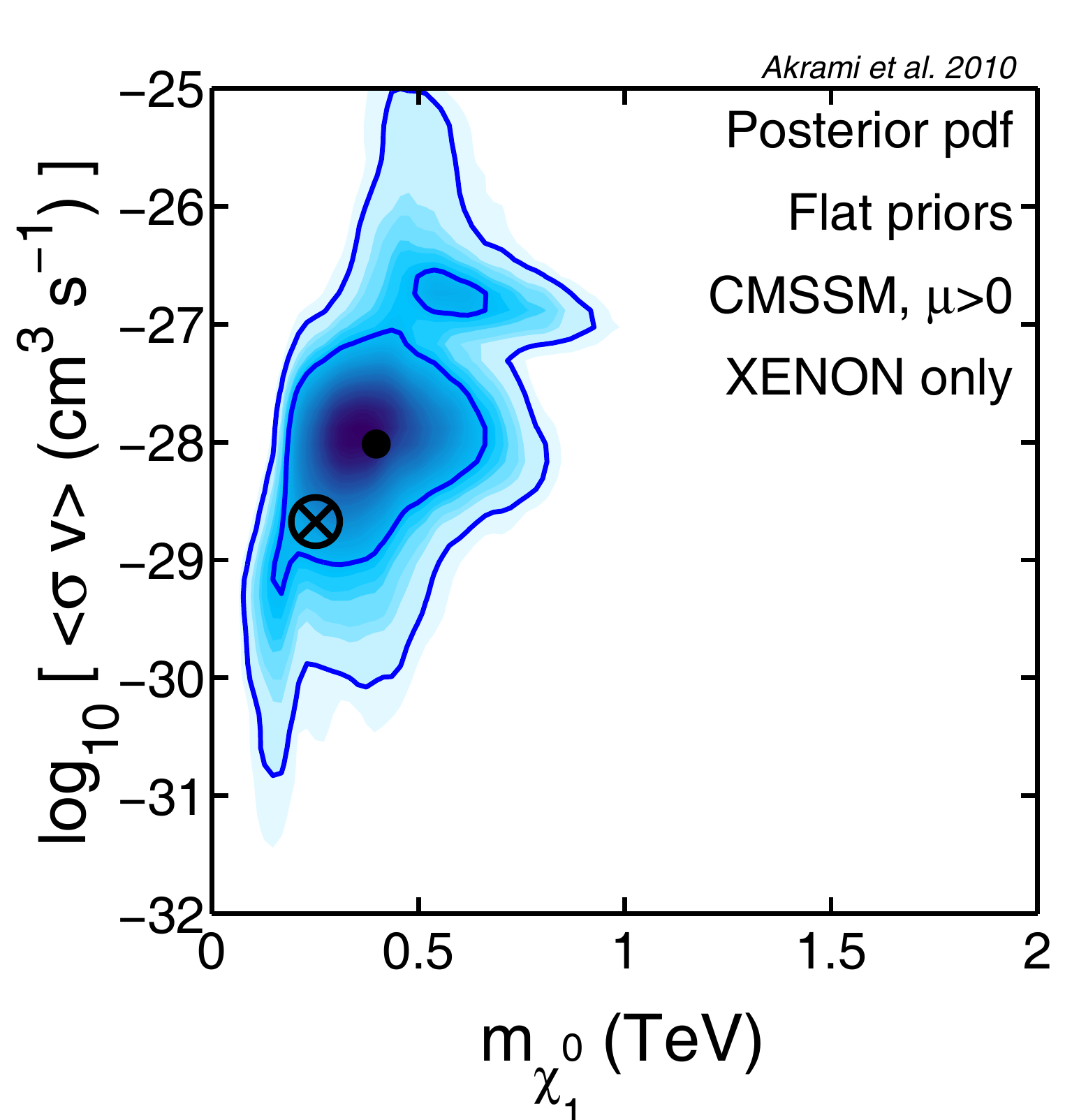}}
\subfigure{\includegraphics[scale=0.23, trim = 40 230 130 100, clip=true]{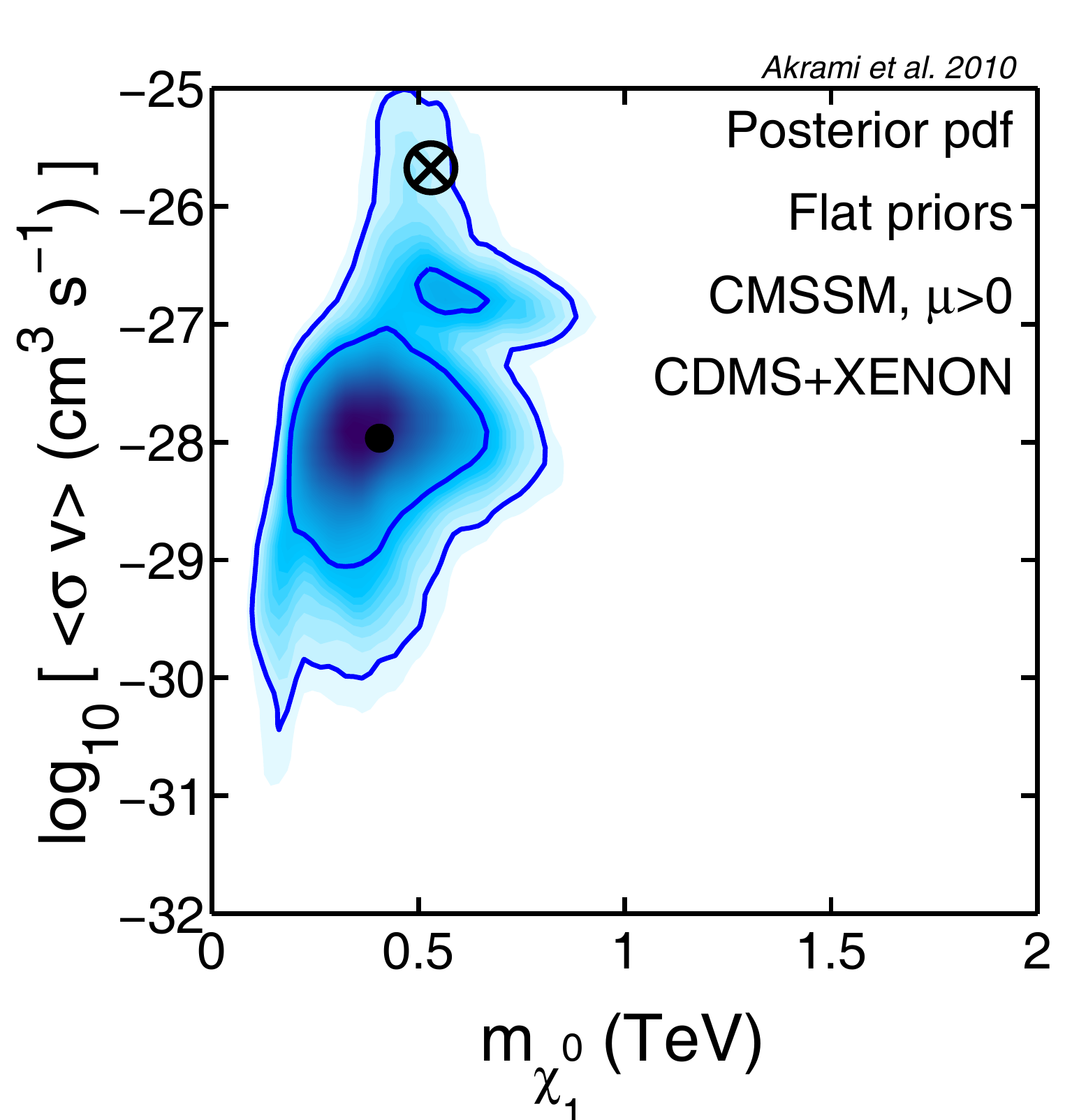}}
\subfigure{\includegraphics[scale=0.23, trim = 40 230 60 100, clip=true]{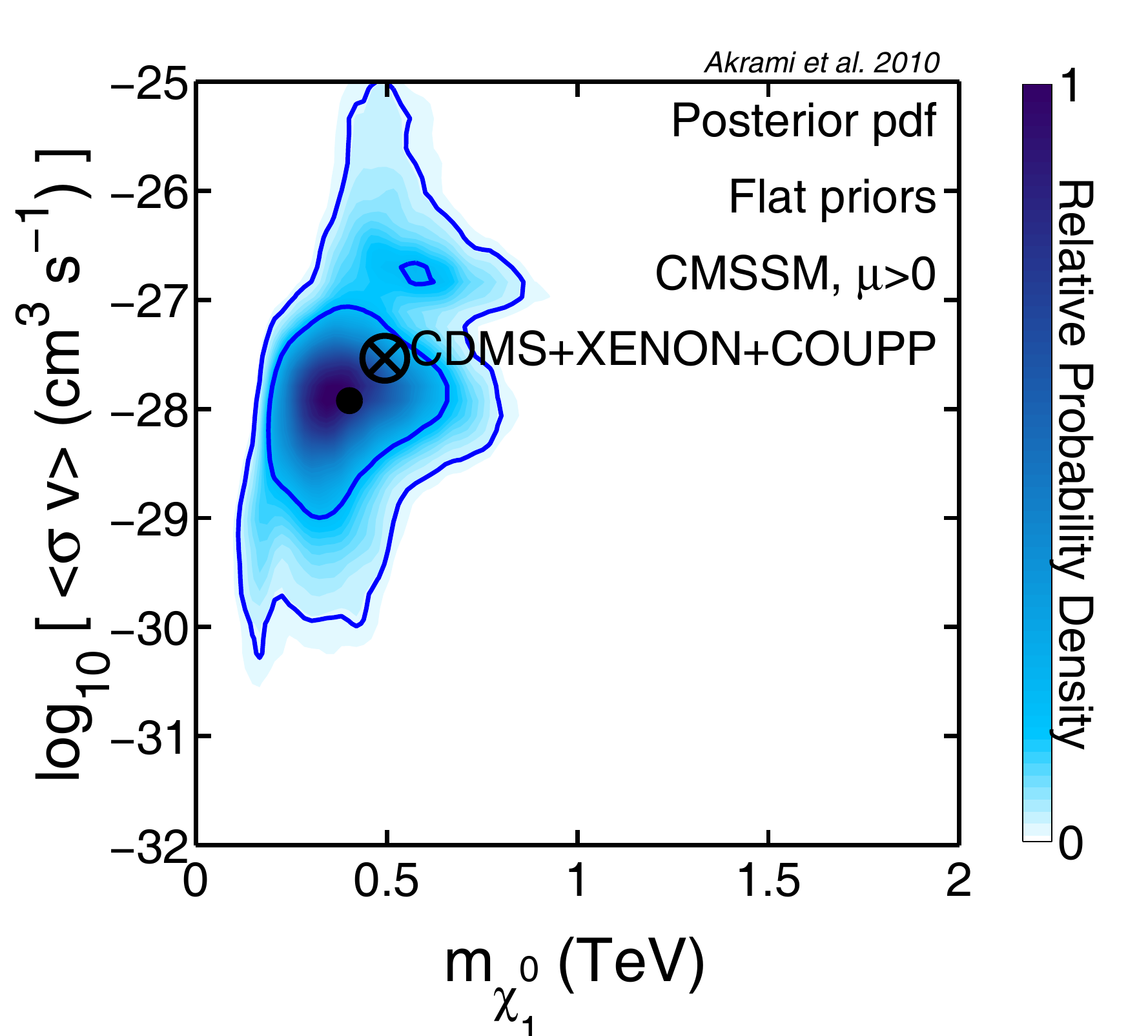}}\\
\setcounter{subfigure}{3}
\subfigure[\footnotesize{\textbf{Benchmark 4:}}]{\includegraphics[scale=0.23, trim = 40 230 130 100, clip=true]{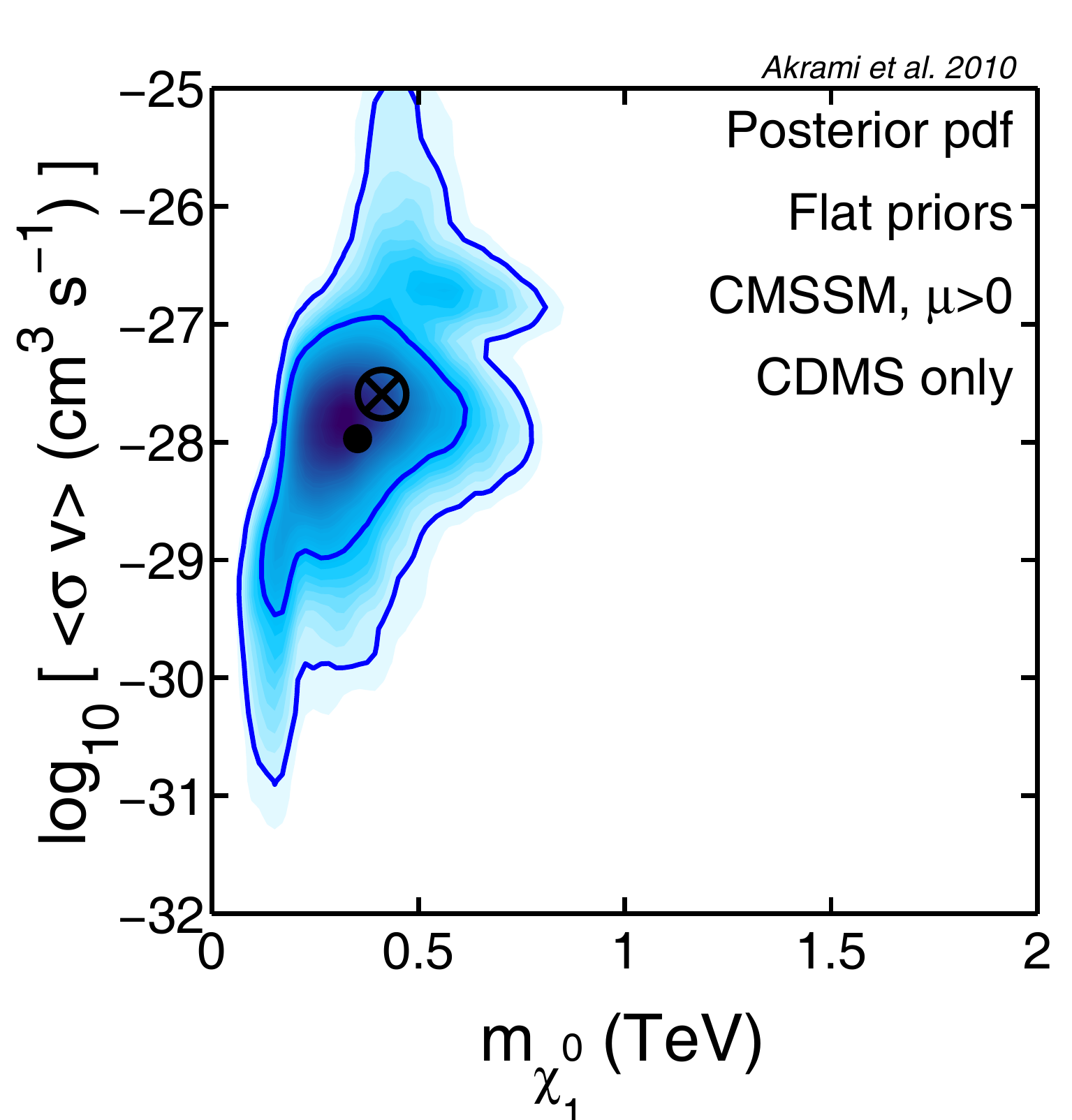}}
\subfigure{\includegraphics[scale=0.23, trim = 40 230 130 100, clip=true]{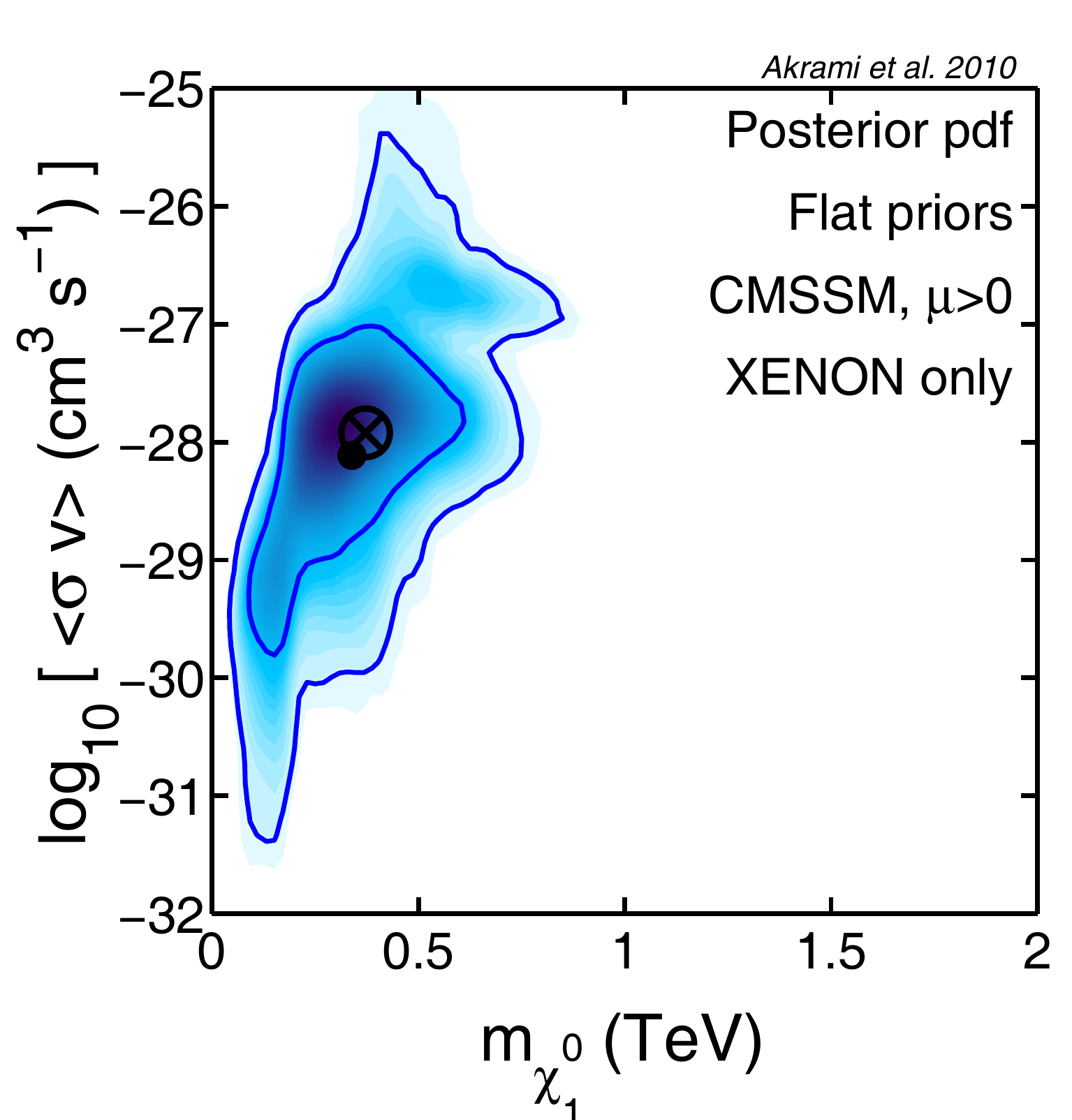}}
\subfigure{\includegraphics[scale=0.23, trim = 40 230 130 100, clip=true]{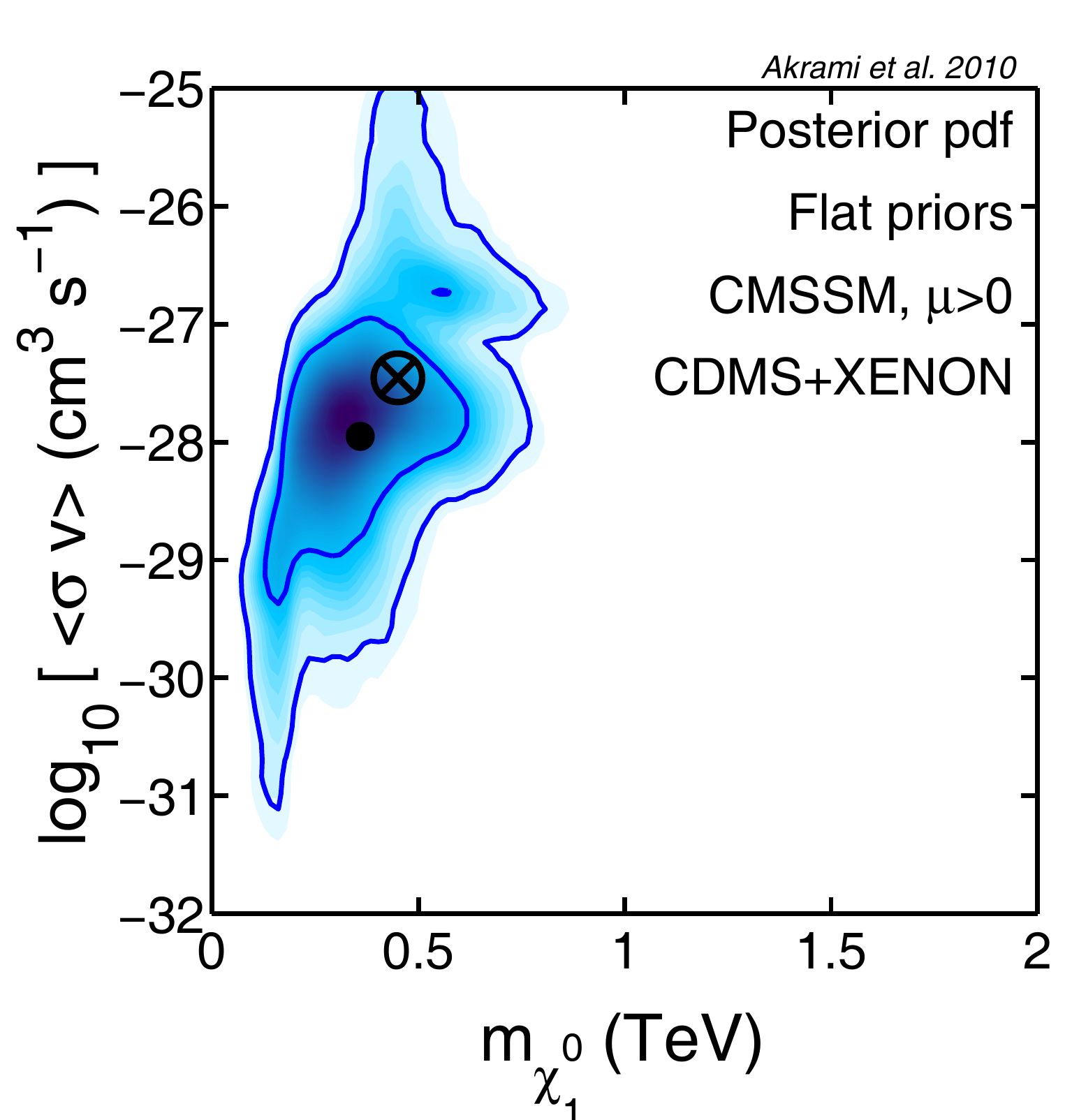}}
\subfigure{\includegraphics[scale=0.23, trim = 40 230 60 100, clip=true]{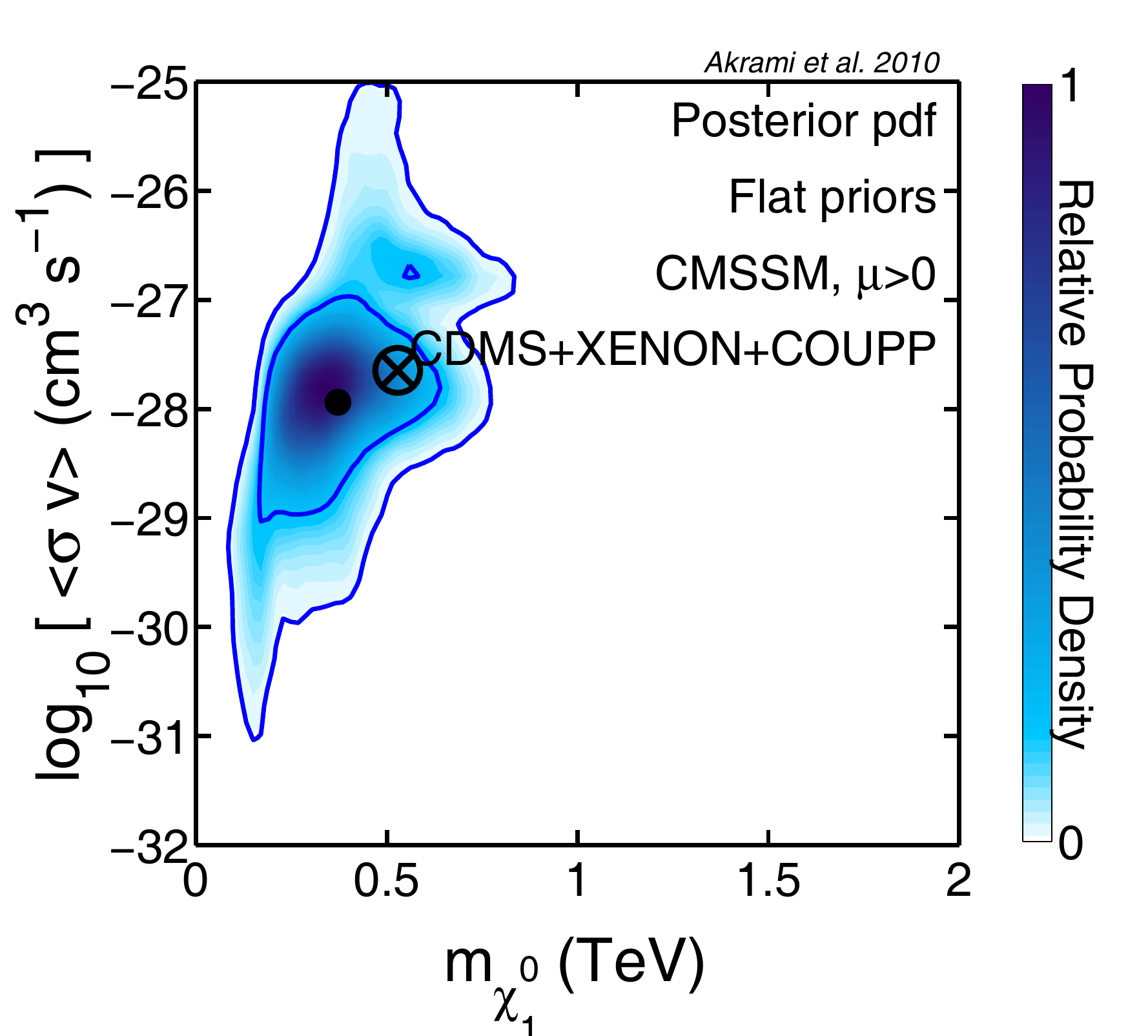}}\\
\caption[aa]{\footnotesize{Two-dimensional marginalised posterior PDFs for the velocity-averaged neutralino self-annihilation cross-section $\left\langle \sigma v\right\rangle$ as a function of the neutralino mass $m_{\tilde\chi^0_1}$, when different combinations of direct detection likelihoods are used in our scans. The inner and outer contours in each panel represent $68.3\%$ ($1\sigma$) and $95.4\%$ ($2\sigma$) confidence levels, respectively. Black dots and crosses show the posterior means and best-fit points, respectively.}}\label{fig:IDmarg}
\end{figure}

\begin{figure}[t!]
\setcounter{subfigure}{0}
\subfigure[][\footnotesize{\textbf{Benchmark 1:}}]{\includegraphics[scale=0.23, trim = 40 230 130 100, clip=true]{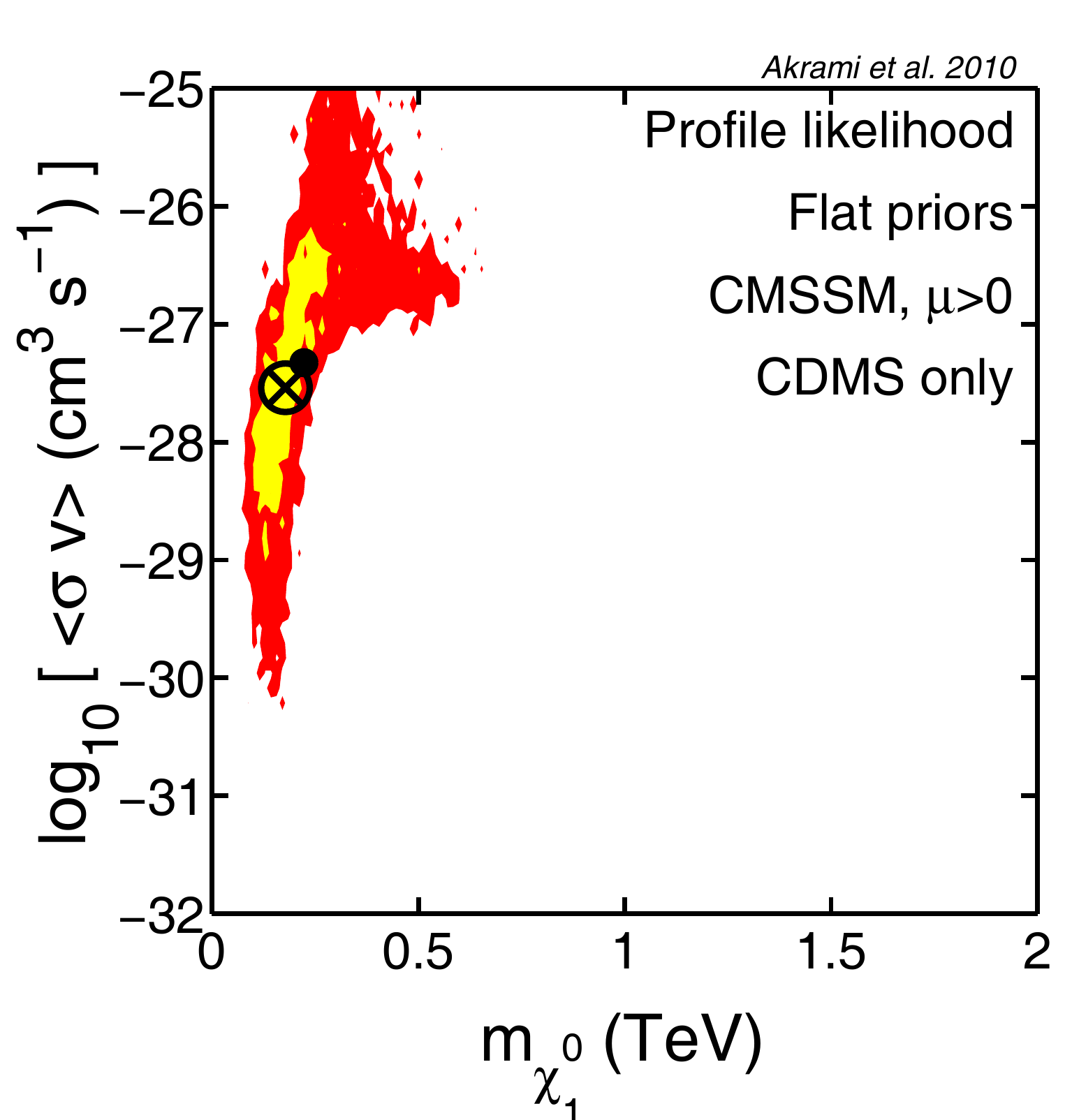}}
\subfigure{\includegraphics[scale=0.23, trim = 40 230 130 100, clip=true]{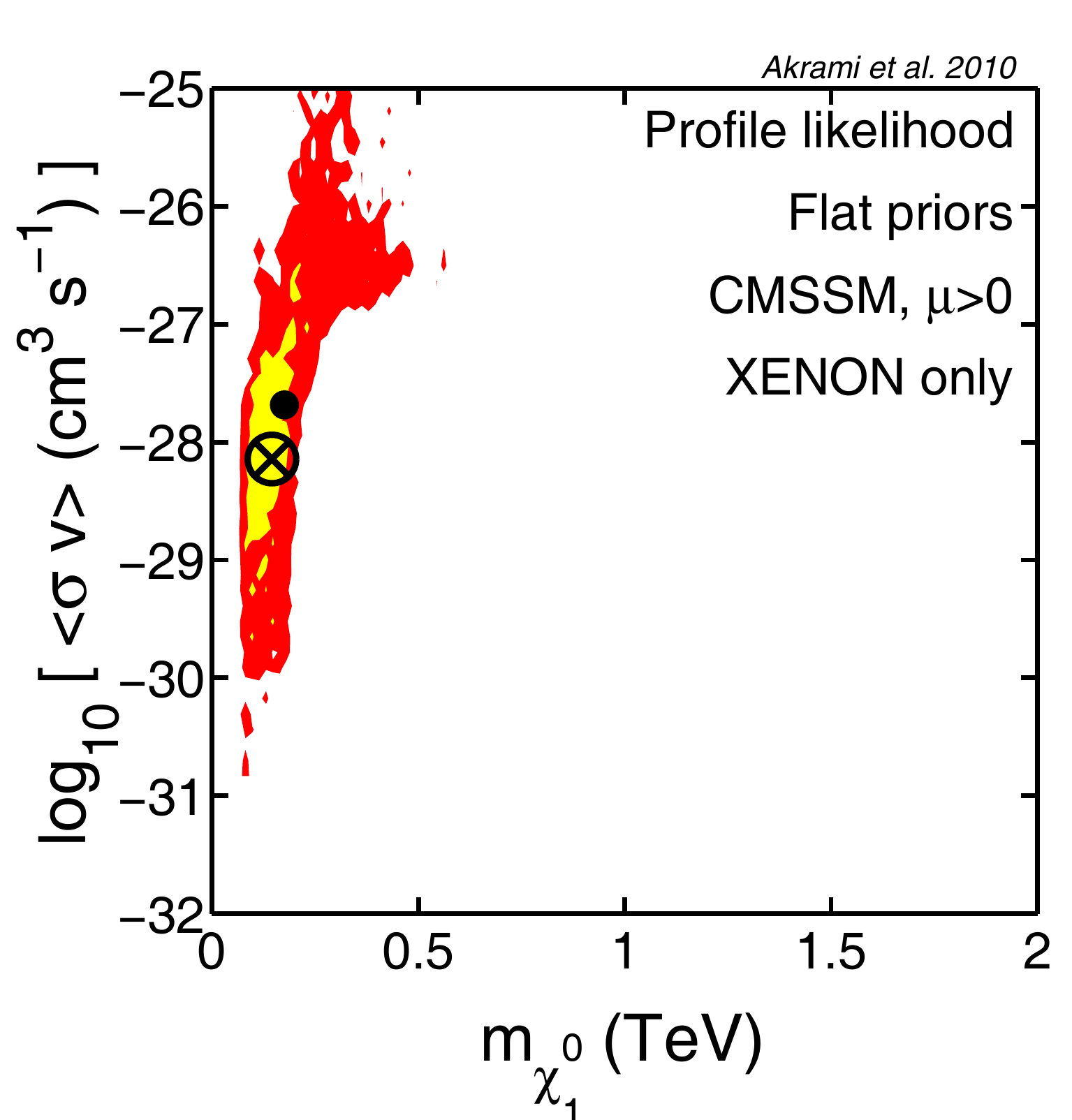}}
\subfigure{\includegraphics[scale=0.23, trim = 40 230 130 100, clip=true]{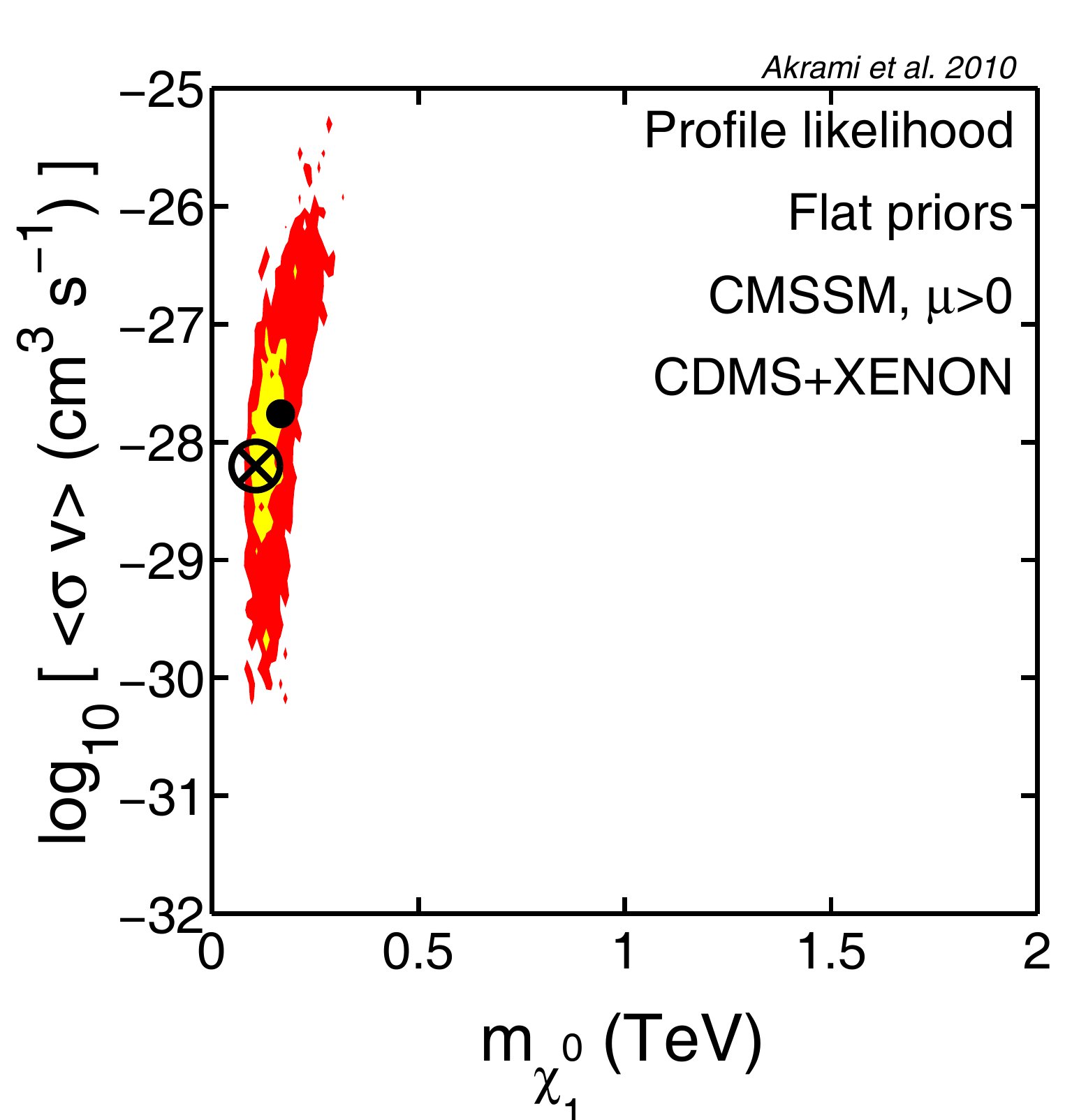}}
\subfigure{\includegraphics[scale=0.23, trim = 40 230 60 100, clip=true]{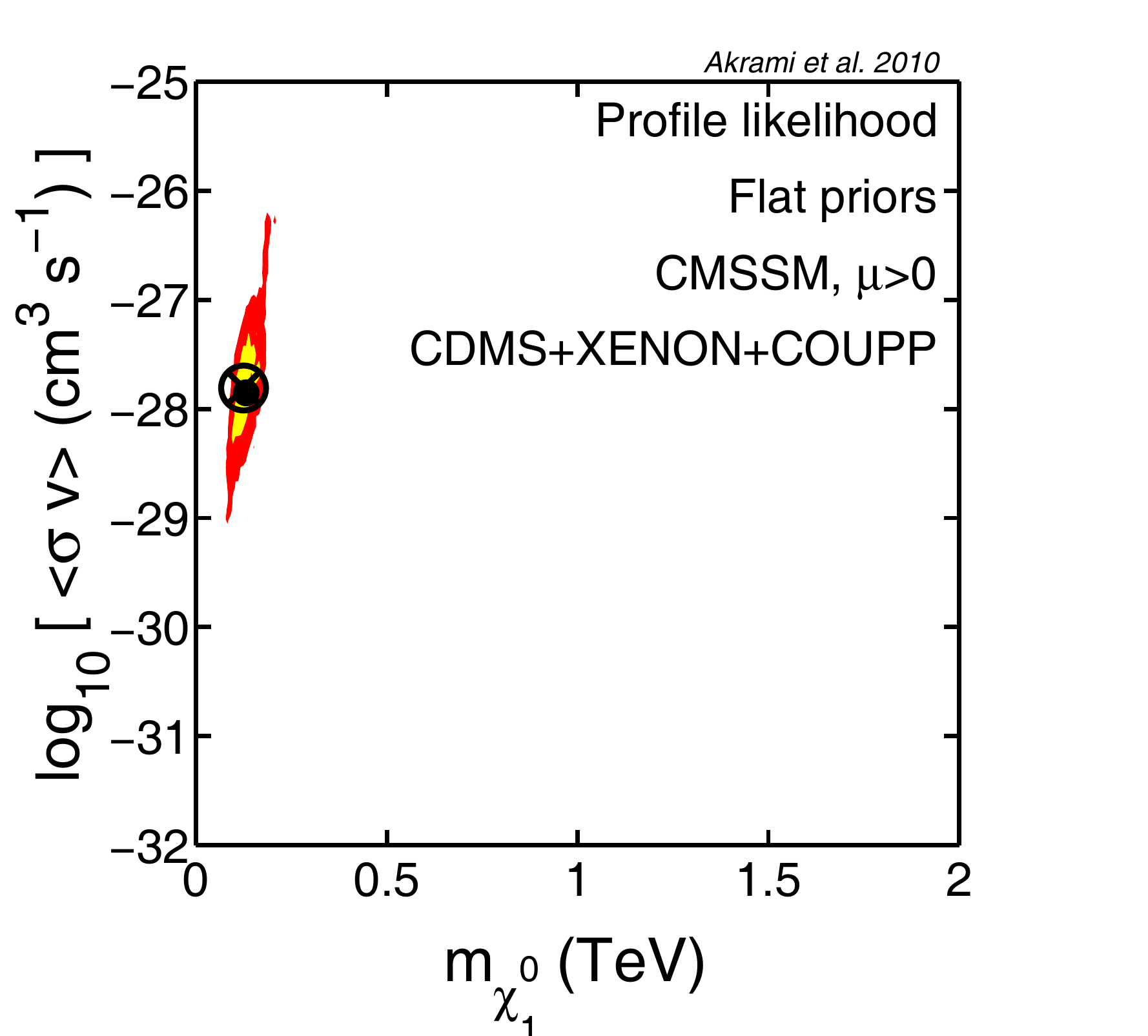}}\\
\setcounter{subfigure}{1}
\subfigure[][\footnotesize{\textbf{Benchmark 2:}}]{\includegraphics[scale=0.23, trim = 40 230 130 100, clip=true]{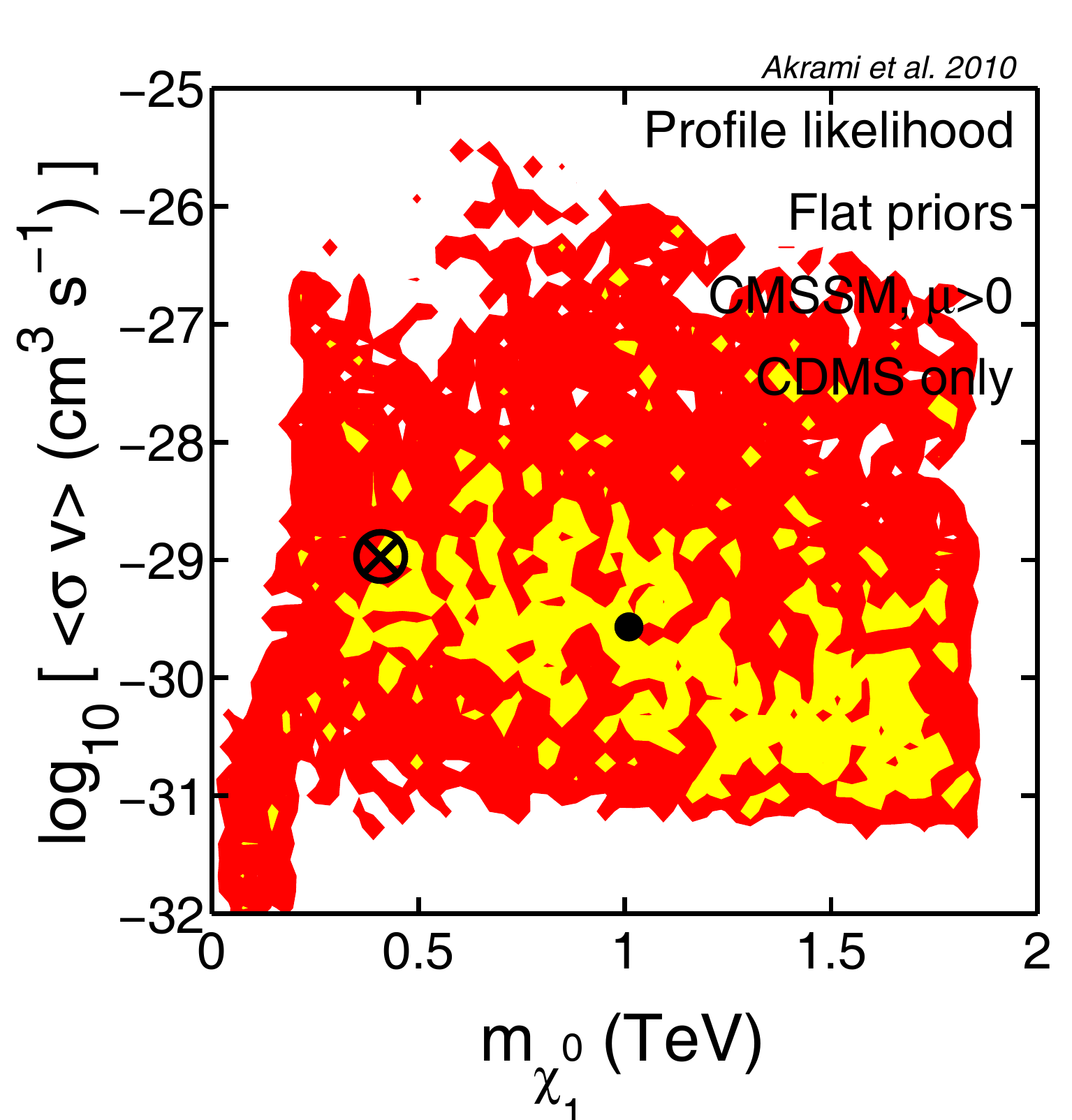}}
\subfigure{\includegraphics[scale=0.23, trim = 40 230 130 100, clip=true]{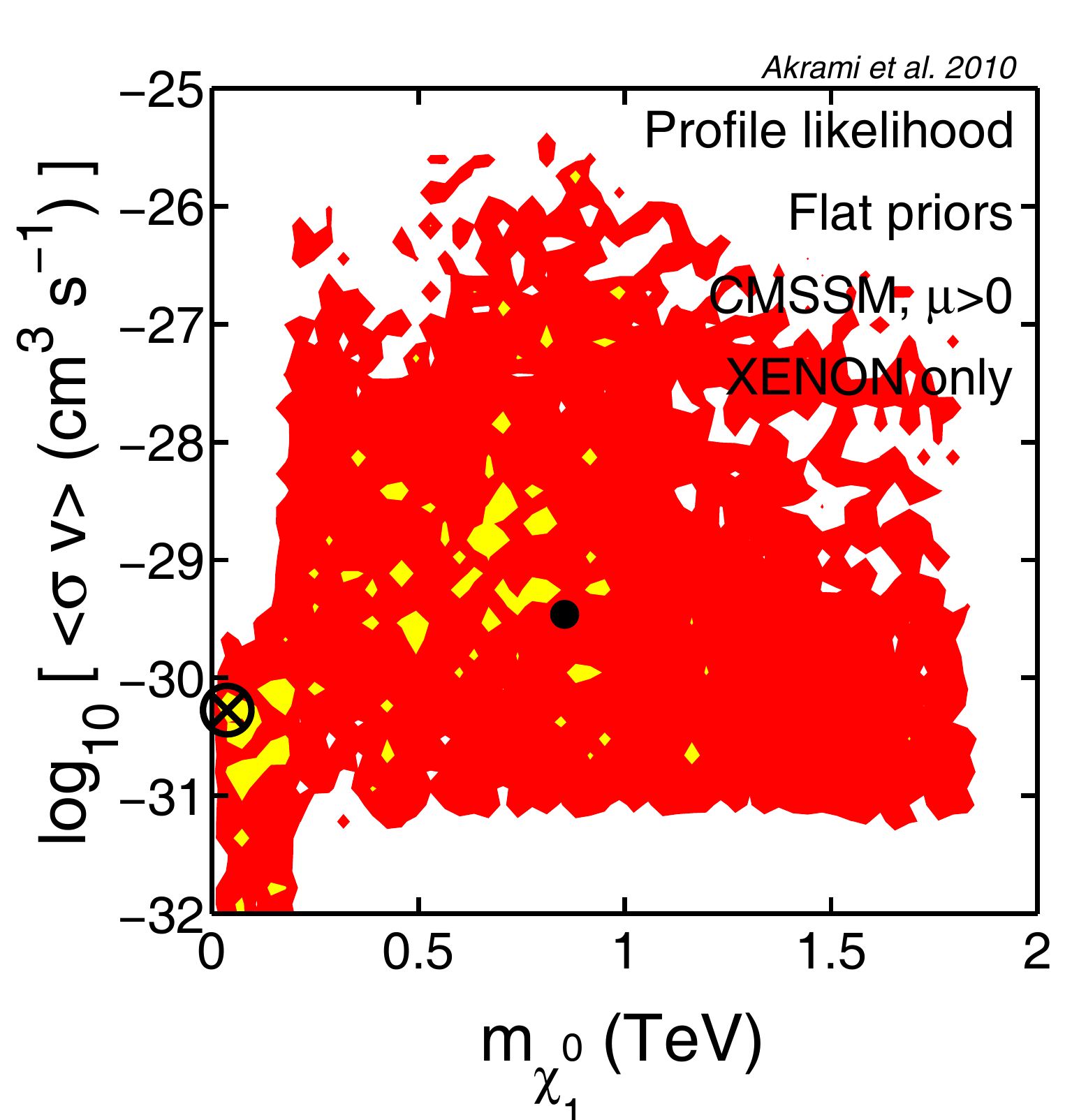}}
\subfigure{\includegraphics[scale=0.23, trim = 40 230 130 100, clip=true]{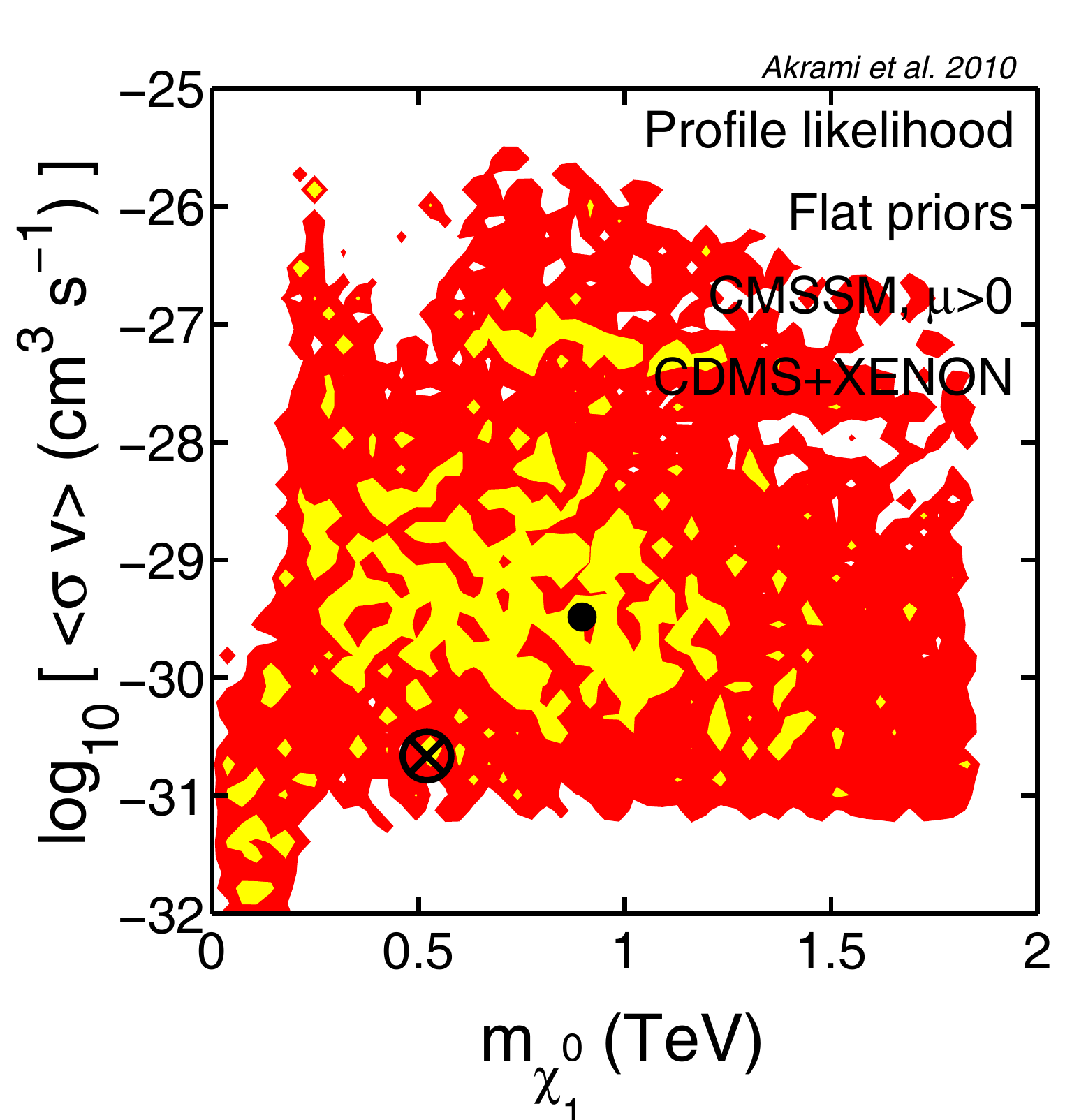}}
\subfigure{\includegraphics[scale=0.23, trim = 40 230 60 100, clip=true]{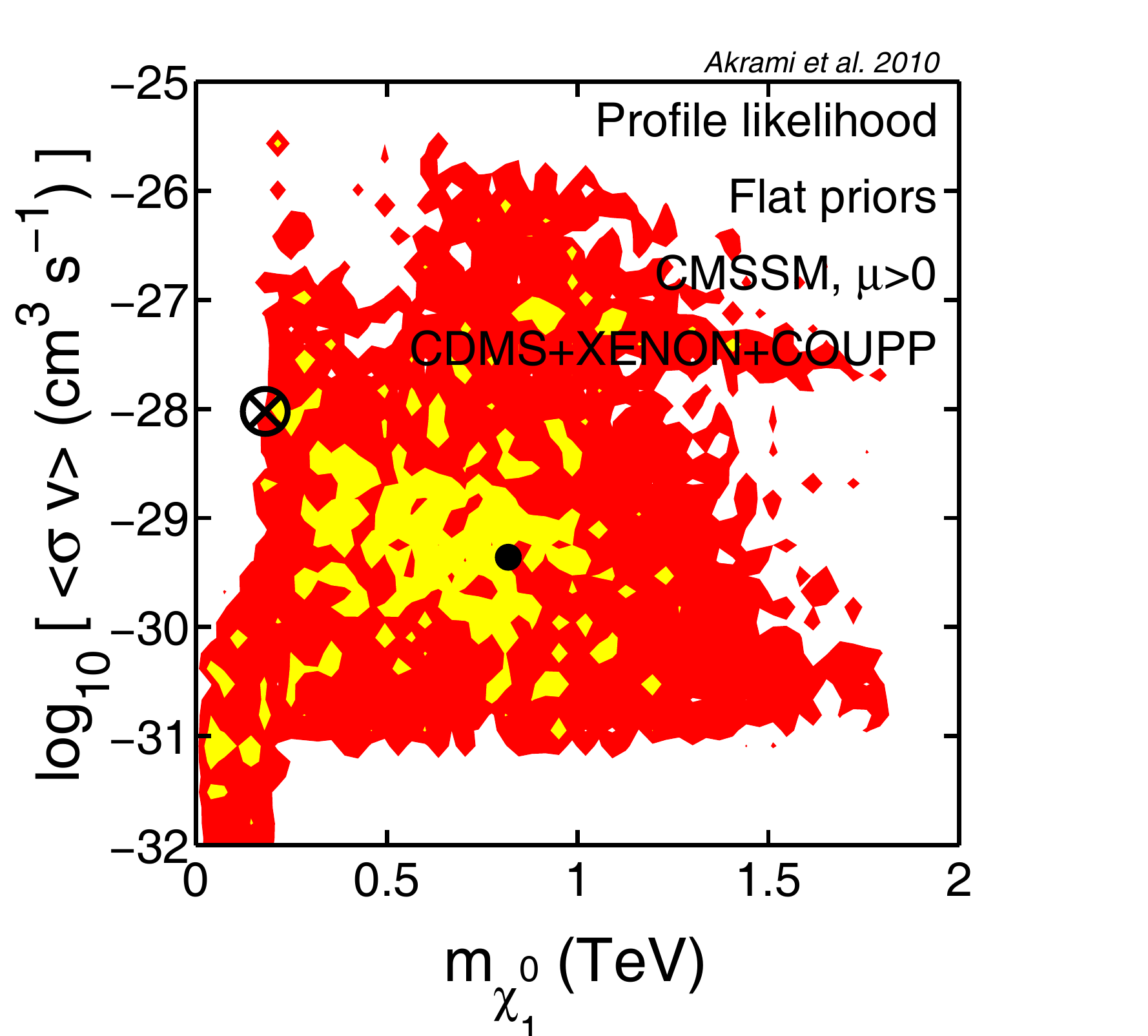}}\\
\setcounter{subfigure}{2}
\subfigure[][\footnotesize{\textbf{Benchmark 3:}}]{\includegraphics[scale=0.23, trim = 40 230 130 100, clip=true]{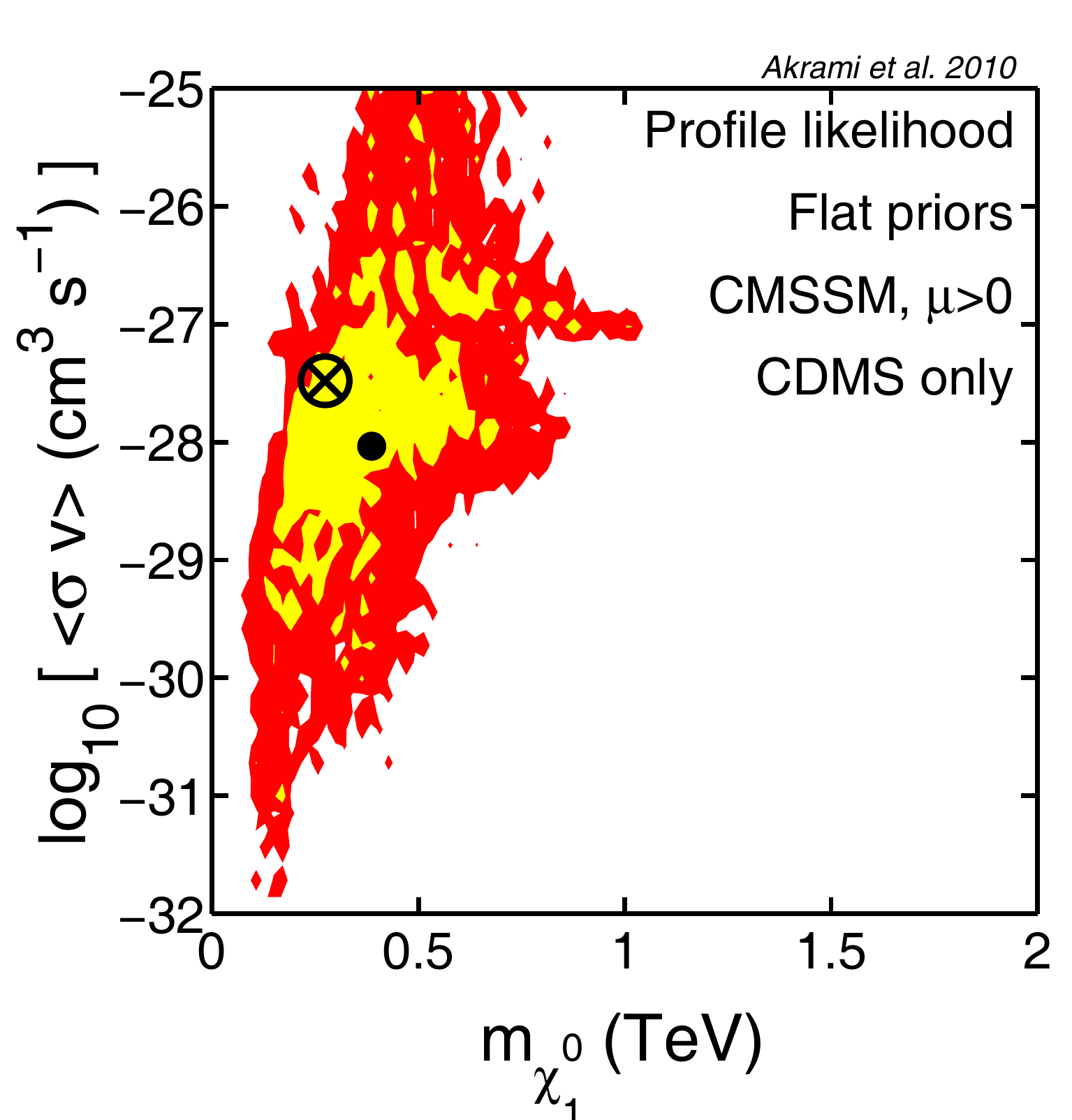}}
\subfigure{\includegraphics[scale=0.23, trim = 40 230 130 100, clip=true]{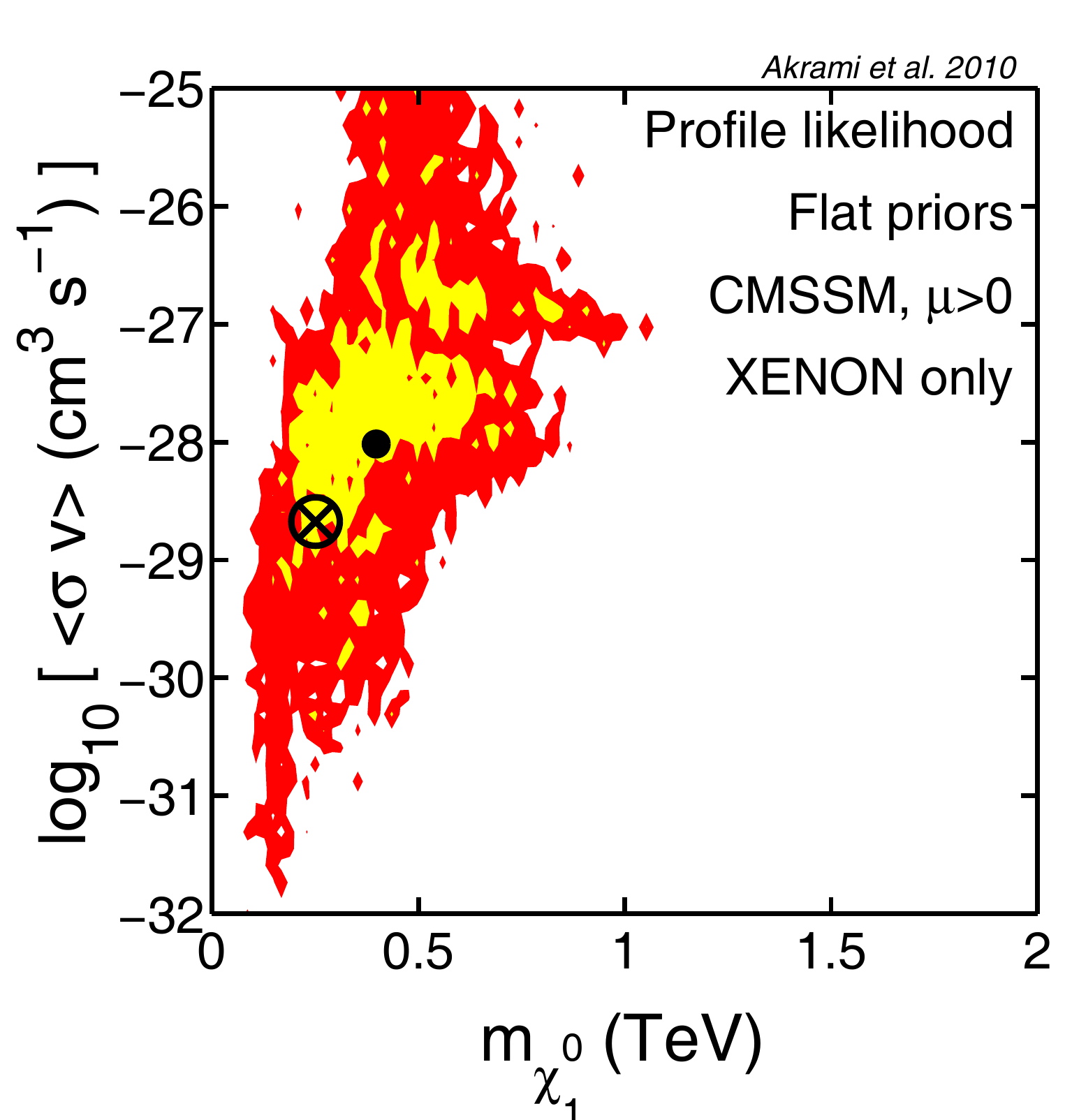}}
\subfigure{\includegraphics[scale=0.23, trim = 40 230 130 100, clip=true]{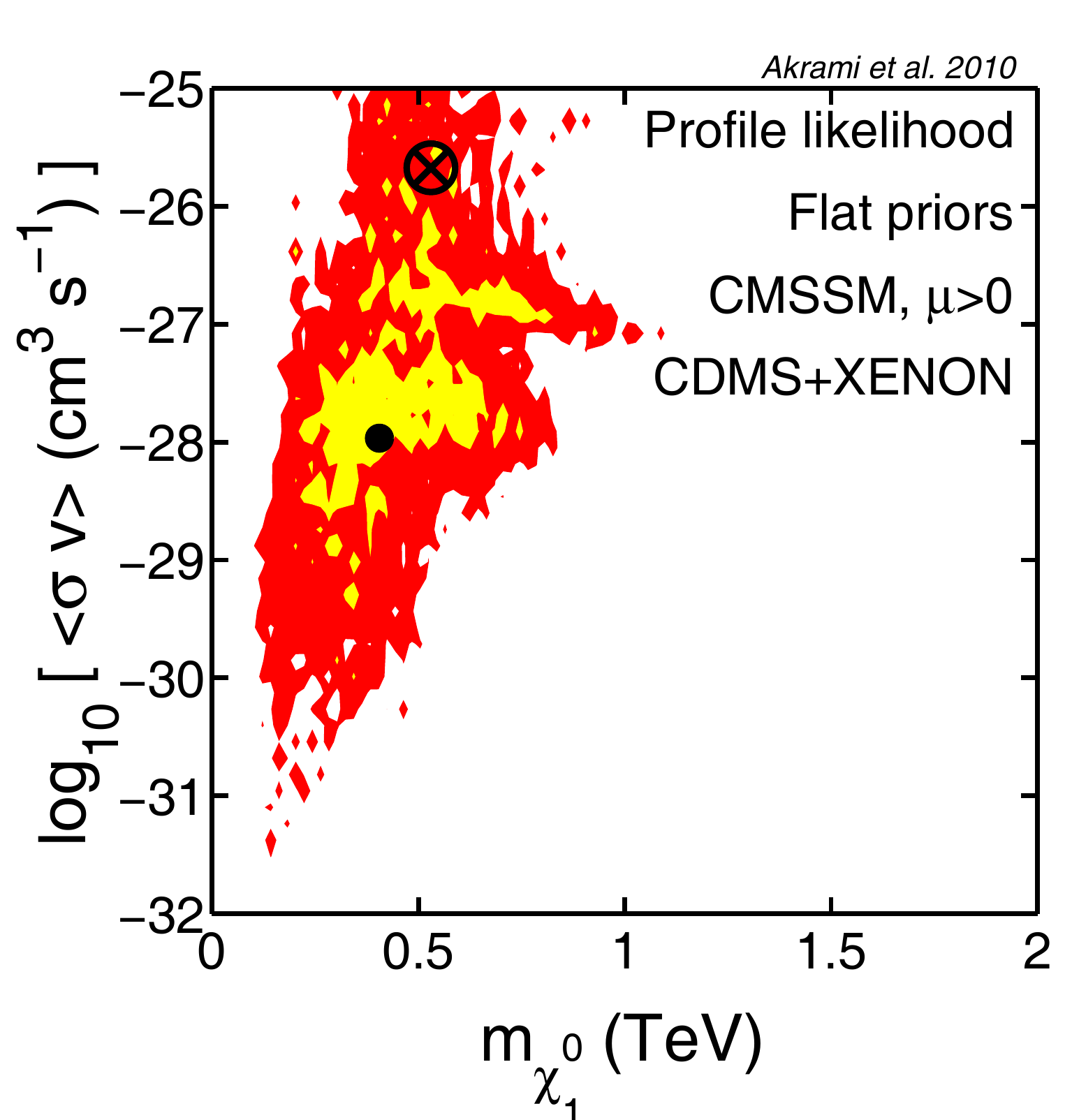}}
\subfigure{\includegraphics[scale=0.23, trim = 40 230 60 100, clip=true]{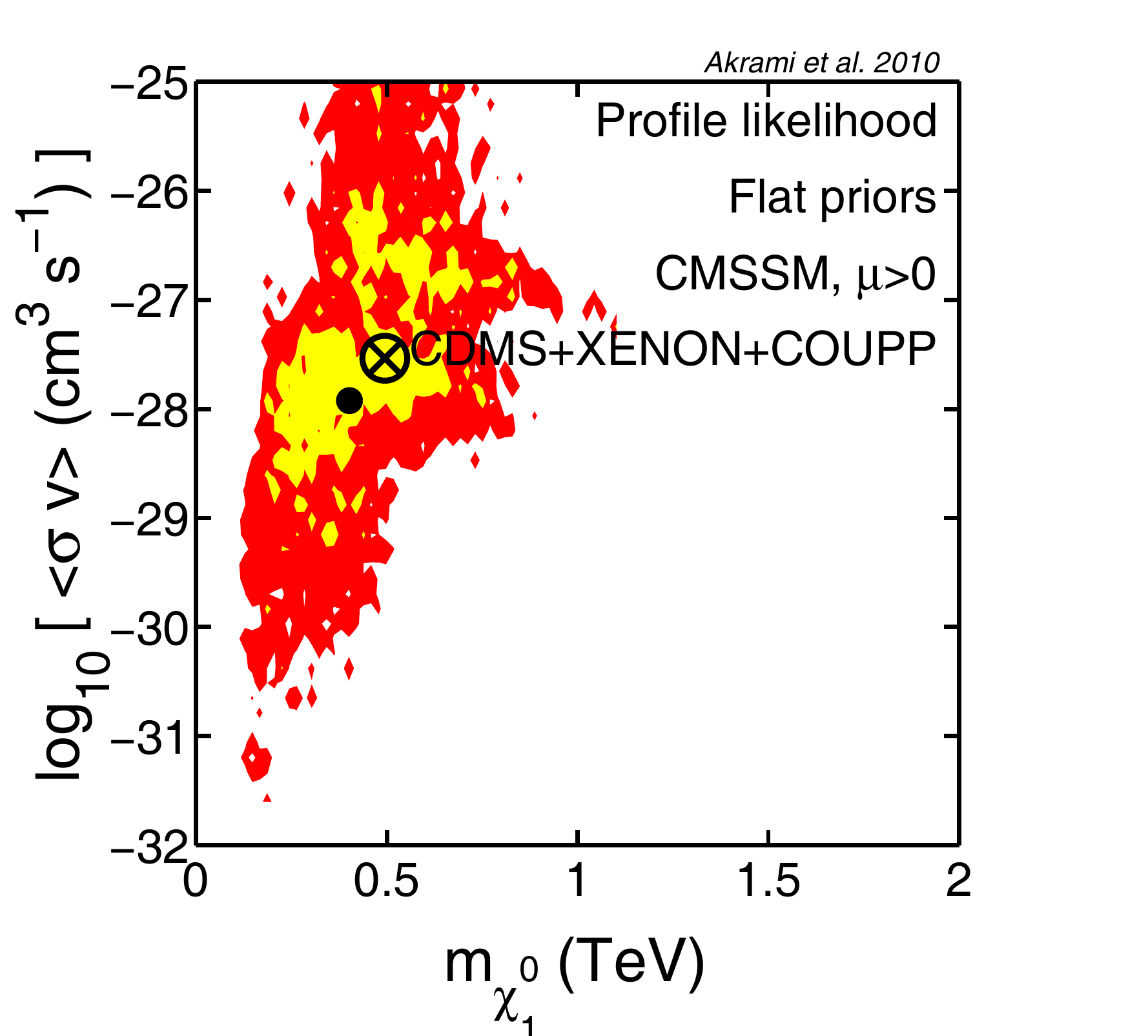}}\\
\setcounter{subfigure}{3}
\subfigure[][\footnotesize{\textbf{Benchmark 4:}}]{\includegraphics[scale=0.23, trim = 40 230 130 100, clip=true]{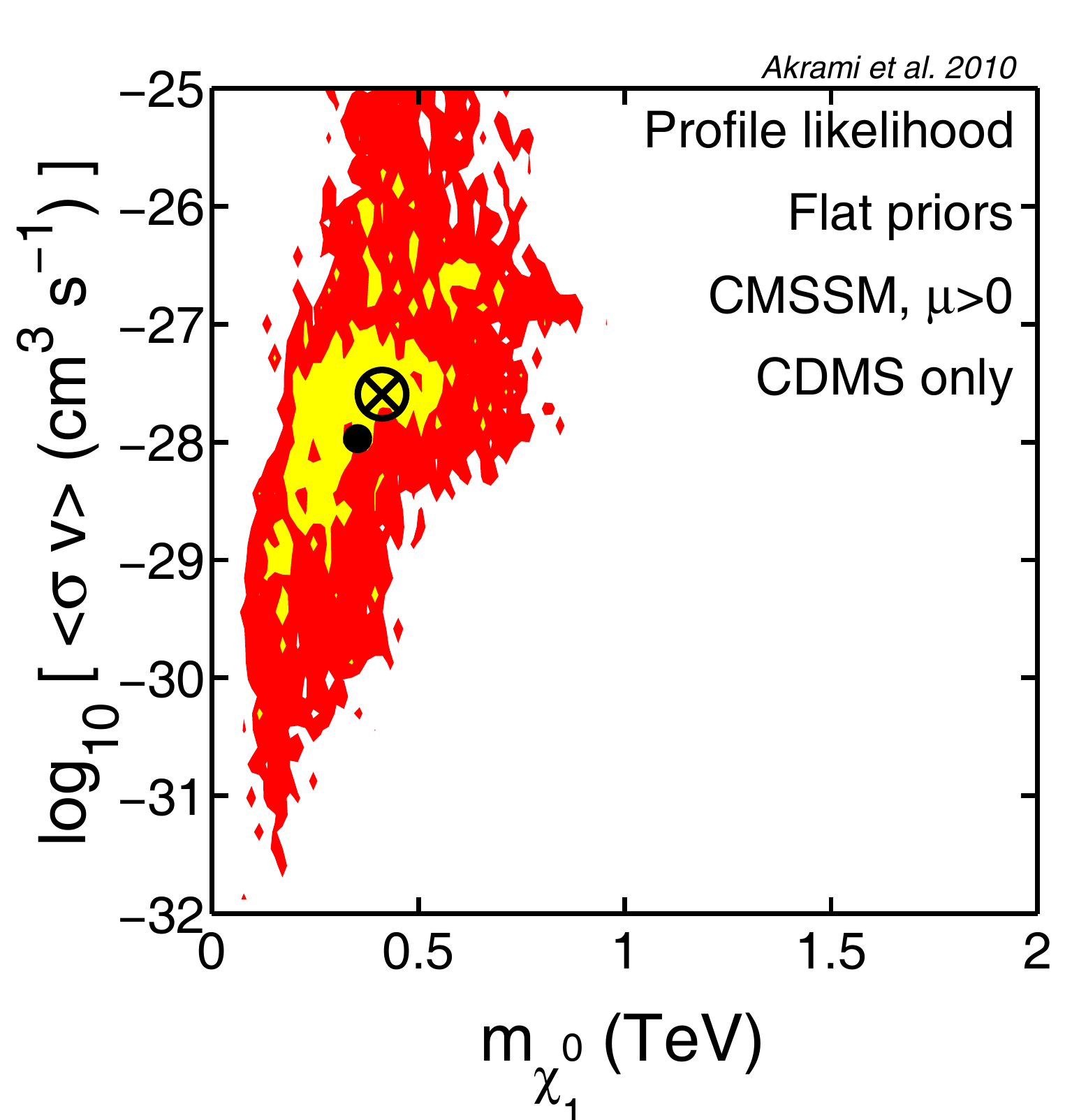}}
\subfigure{\includegraphics[scale=0.23, trim = 40 230 130 100, clip=true]{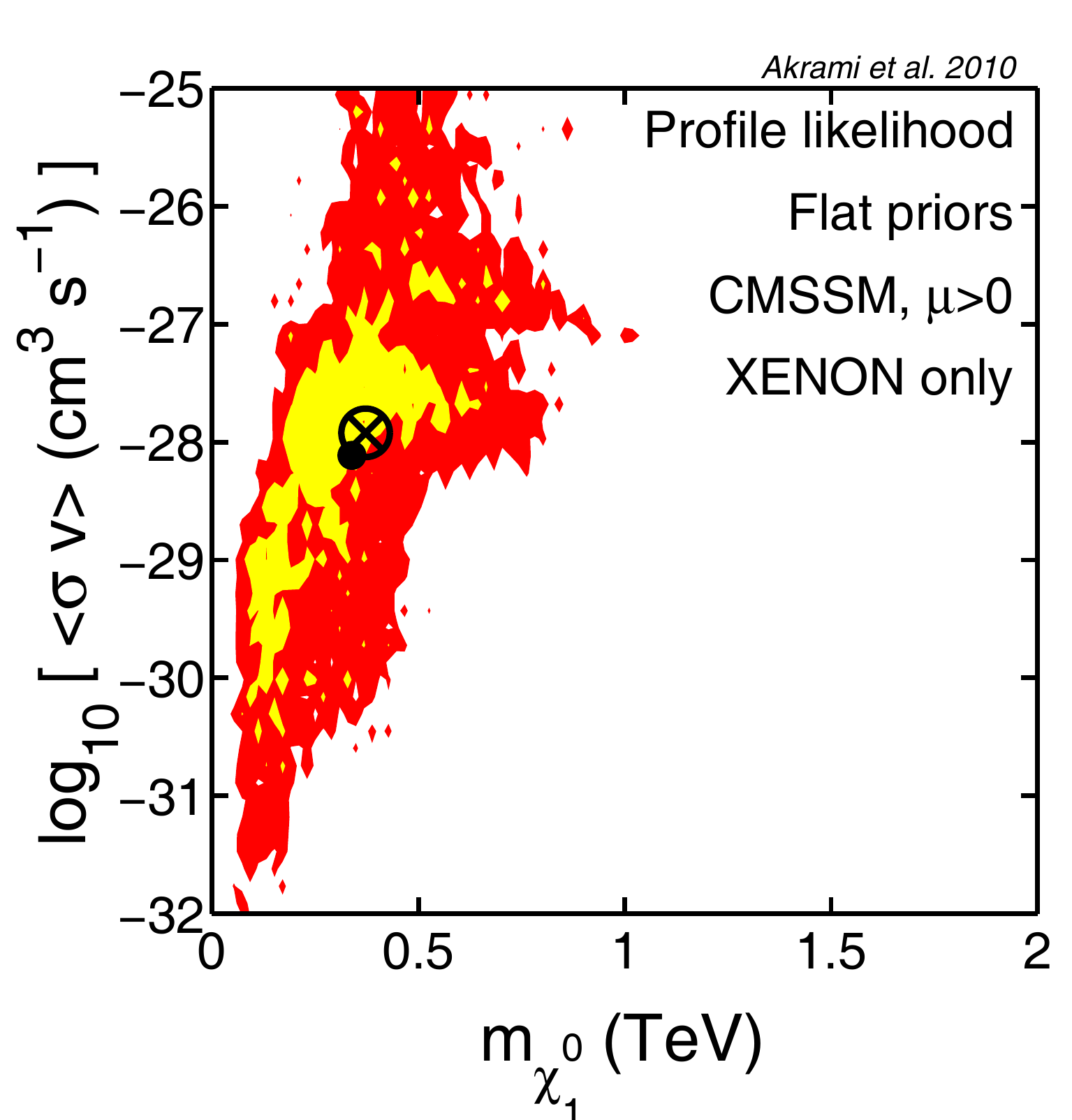}}
\subfigure{\includegraphics[scale=0.23, trim = 40 230 130 100, clip=true]{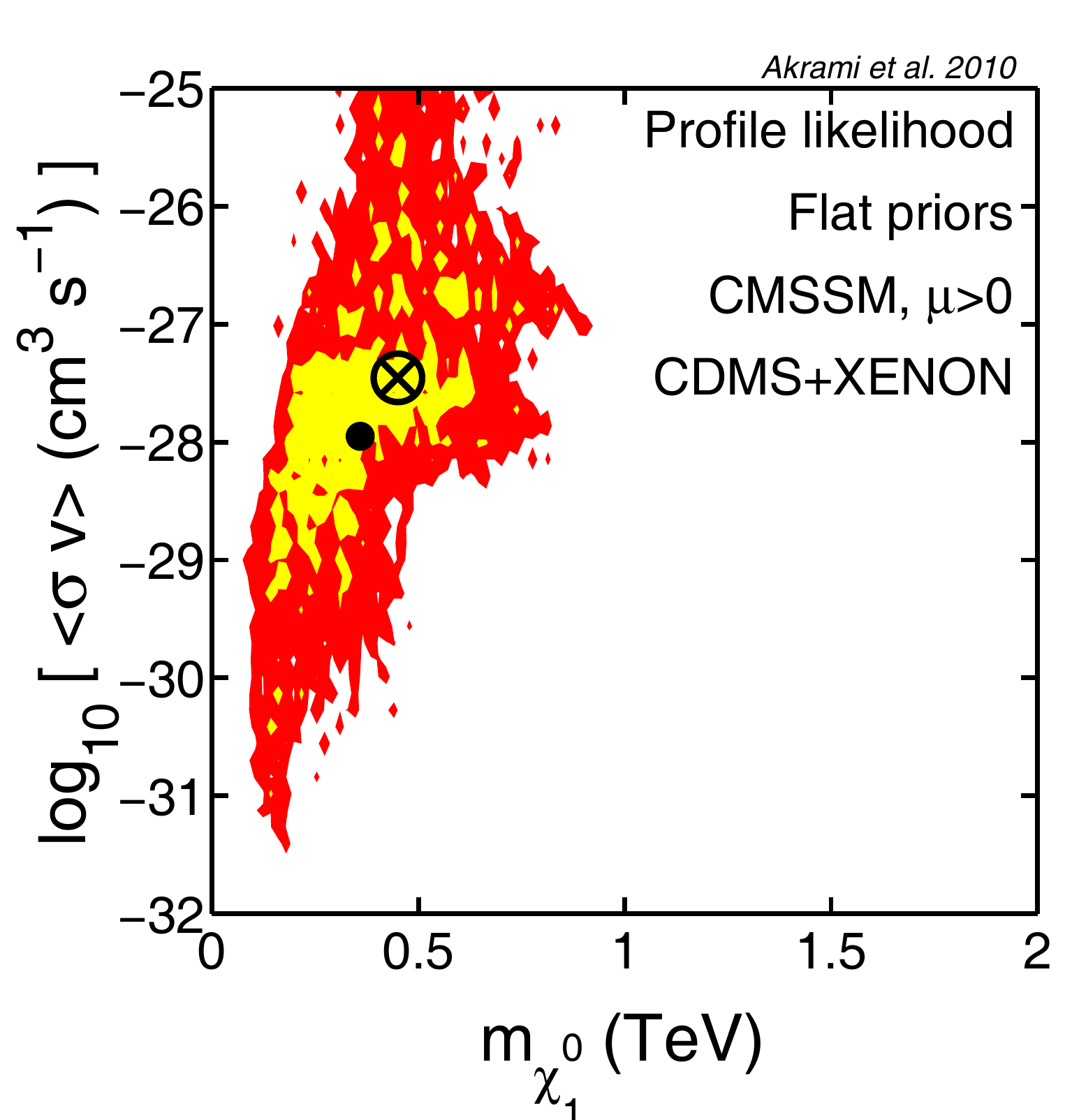}}
\subfigure{\includegraphics[scale=0.23, trim = 40 230 60 100, clip=true]{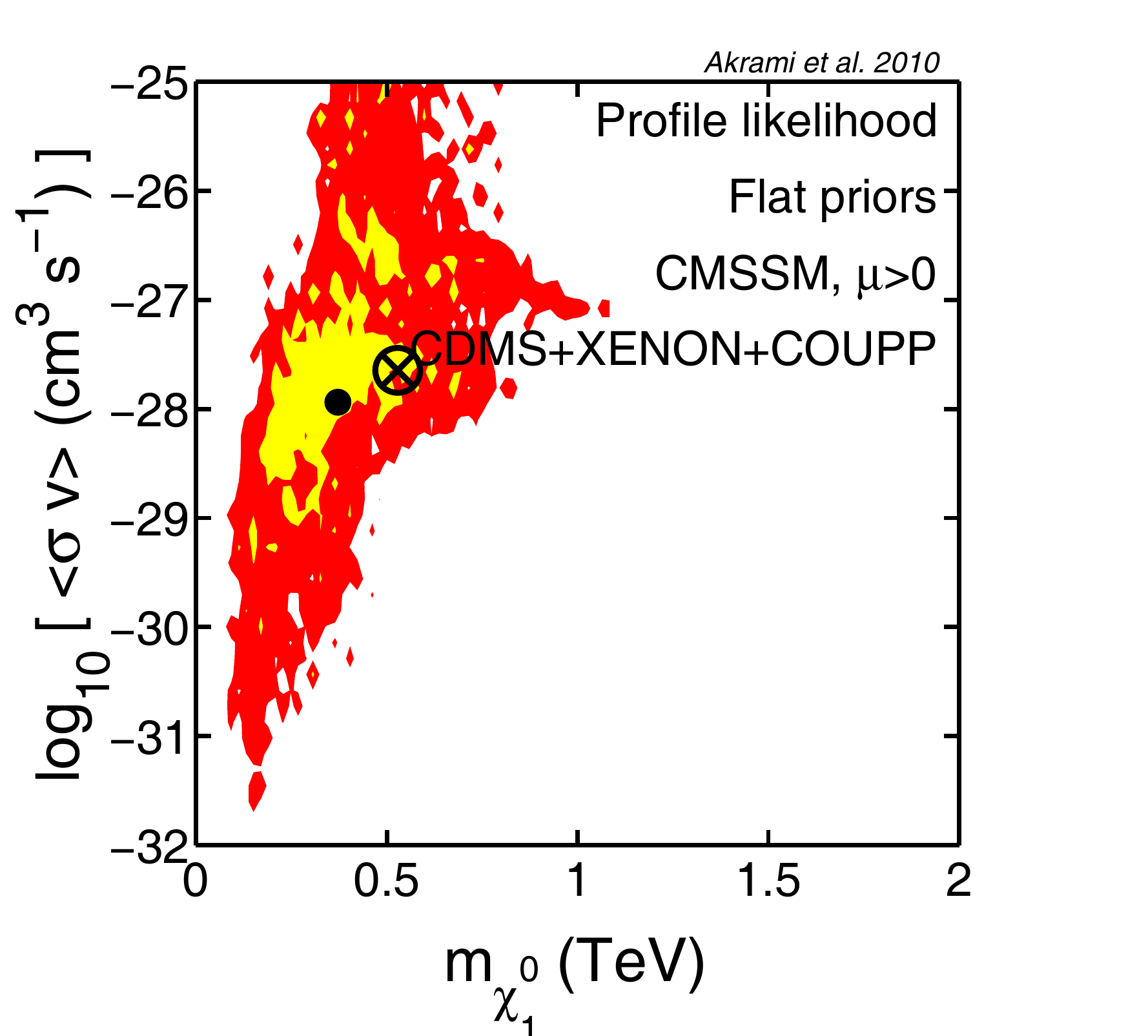}}\\
\caption[aa]{\footnotesize{As in~\fig{fig:IDmarg}, but for two-dimensional profile likelihoods.}}\label{fig:IDprofl}
\end{figure}

\afterpage{\clearpage}

We finally show in~\tab{tab:DDSensitivity} experimental sensitivities (detection prospects) for the different benchmarks.  The table gives the fraction of the time that our ton-scale DD experiments are expected to detect a dark matter signal at the 3 or 5$\sigma$ level based only on the number of events observed. We consider WIMPs to be detected if the observed number of events greatly exceeds that expected due to backgrounds alone. This happens when the observed number has a p-value smaller than $2.7 \times 10^{-3}$ and $5.7 \times 10^{-7}$ (corresponding to the 3$\sigma$ and 5$\sigma$ detection, respectively) for a Poisson distribution based on an average number of background events $\mu_B$. The average background for the individual experiments is $\mu_B = 2.0$ and this corresponds to observing 8 (13) or more events at the 3$\sigma$ (5$\sigma$) level detection. We also give in the last row the combined results of all three experiments. This provides greater statistical power than each experiment alone. For an average background $\mu_B = 6.0$ in this case, a 3$\sigma$ (5$\sigma$) detection corresponds to observing 15 (22) or more total events. Although we do not include event energies in our appraisal of the detection capability of the experiments, this could however moderately improve the detection prospects. Our results show that benchmarks 1, 3 and 4 will be detected in almost all cases while for benchmark 2 the best case would be when all three experiments are combined, with 27.5\% and 0.94\% detection probabilities at the 3 or 5$\sigma$ level, respectively.

\begin{table}[t]
  \begin{center}
  \addtolength{\tabcolsep}{0.5em}
  \begin{tabular}{lcccc}
    \toprule
    \textbf{Experiment} & \textbf{BM1} & \textbf{BM2} & \textbf{BM3} & \textbf{BM4} \\
    \toprule
    CDMS1T     & 1.000 (1.000) & 0.023 (0.00006)
               & 0.998 (0.931) & 1.000 (0.975)    \\
    XENON1T    & 1.000 (1.000) & 0.057 (0.00034)
               & 1.000 (0.996) & 1.000 (0.999)    \\
    COUPP1T    & 1.000 (1.000) & 0.133 (0.00202)
               & 1.000 (1.000) & 1.000 (1.000)    \\
    (Combined) & 1.000 (1.000) & 0.275 (0.00940)
               & 1.000 (1.000) & 1.000 (1.000)    \\
    \bottomrule
  \end{tabular}
  \end{center}
  \caption[Experimental sensitivities]{\footnotesize{
    Fraction of the time that the direct detection experiments are
    expected to detect a dark matter signal at the 3$\sigma$
    (5$\sigma$) level for each benchmark model based only on the number
    of observed events.
    Dark matter is considered to be detected at the 3$\sigma$
    (5$\sigma$) level if the observed number of events greatly
    exceeds that expected due to backgrounds alone,
    specifically if the observed number of events has a p-value
    smaller than $2.7 \times 10^{-3}$ ($5.7 \times 10^{-7}$) for
    a Poisson distribution based on an average number of background
    events $\mu_B$.
    For the average background $\mu_B = 2.0$ assumed for each of the
    individual experiments, this corresponds to observing 8 (13) or
    more events.
    The last row represents the combined results of all three
    experiments, which provides greater statistical power than
    each experiment alone.  A 3$\sigma$ (5$\sigma$) detection in
    this case corresponds to observing 15 (22) or more total events
    for an average background $\mu_B = 6.0$.
    The use of event energies in addition to the number of events
    could give a moderate improvement in the detection prospects.
    }}
  \label{tab:DDSensitivity}
\end{table}

\subsection{Effects of halo and cross-section nuisances} \label{sec:confreg}

\begin{figure}[t]
\setcounter{subfigure}{0}
\subfigure[][\scriptsize{\textbf{CS/H Nuis:~~}}]{\includegraphics[scale=0.23, trim = 40 230 130 100, clip=true]{figs/MM_CDMSXENONCOUPP_wRefPoint/MM_CDMSXENONCOUPP_wRefPoint_2D_marg_15}}
\subfigure[][\scriptsize{\textbf{CS Nuis Only:}}]{\includegraphics[scale=0.23, trim = 40 230 130 100, clip=true]{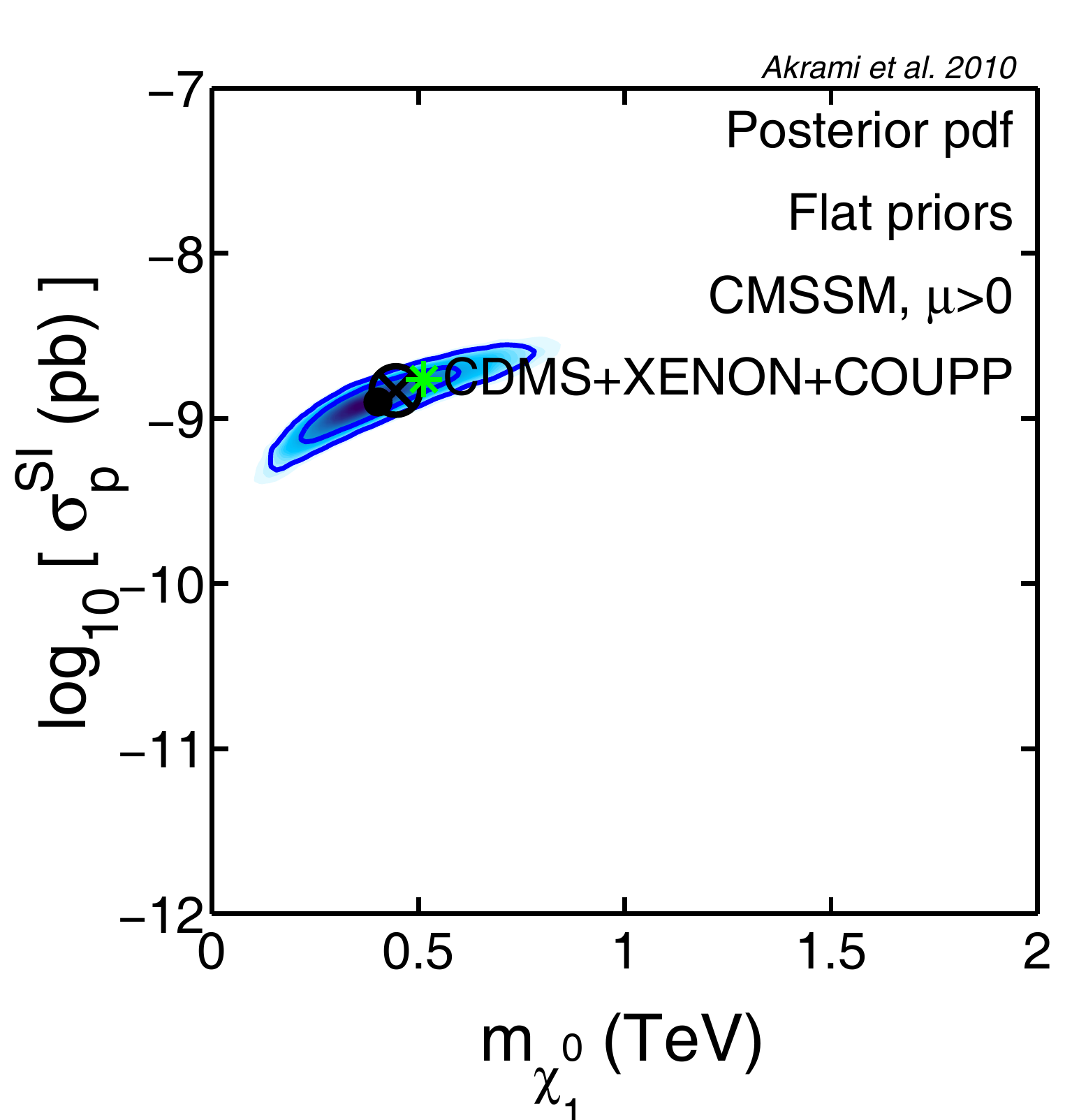}}
\subfigure[][\scriptsize{\textbf{H Nuis Only:~}}]{\includegraphics[scale=0.23, trim = 40 230 130 100, clip=true]{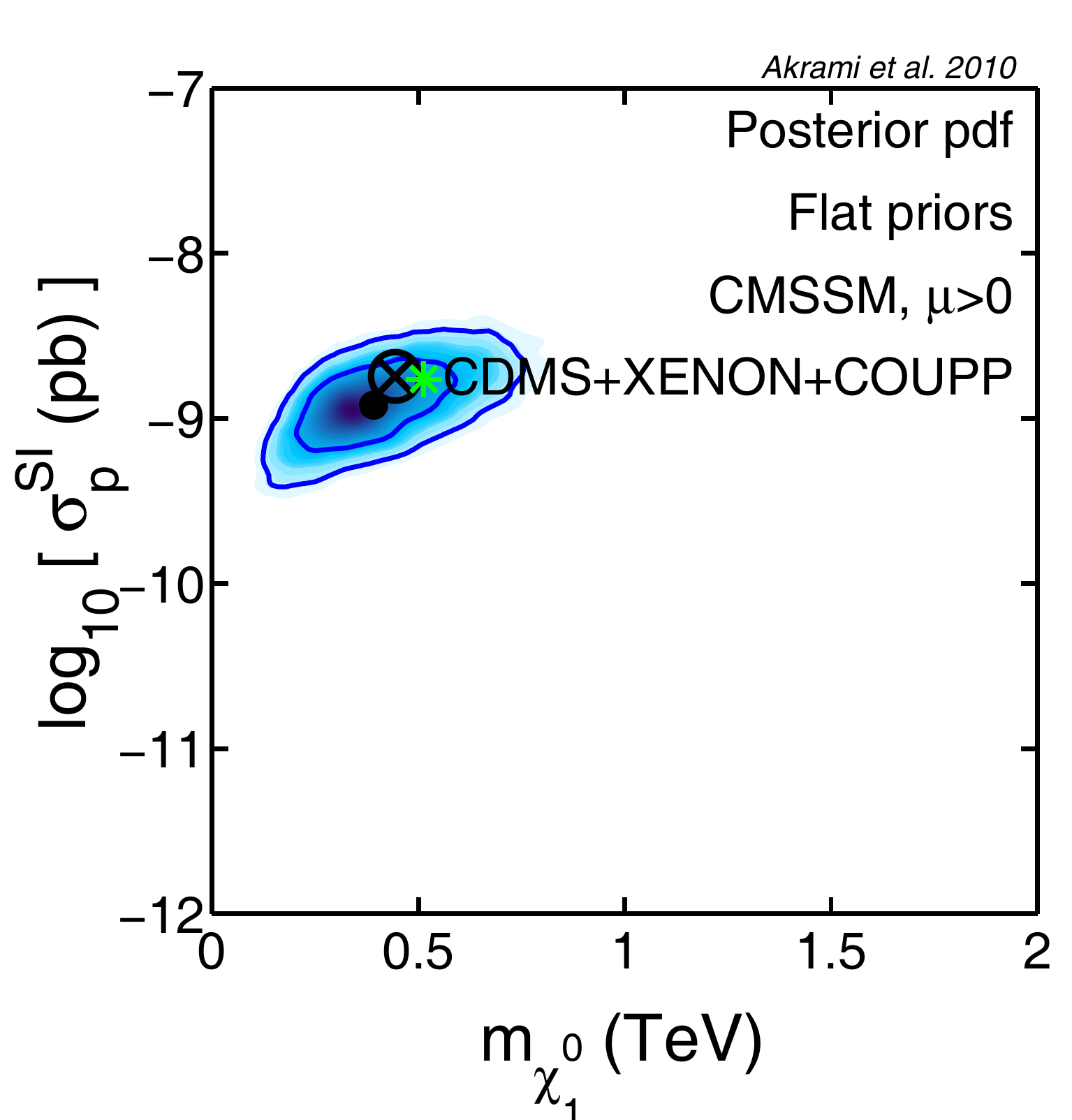}}
\subfigure[][\scriptsize{\textbf{No CS/H  Nuis:~~}}]{\includegraphics[scale=0.23, trim = 40 230 60 100, clip=true]{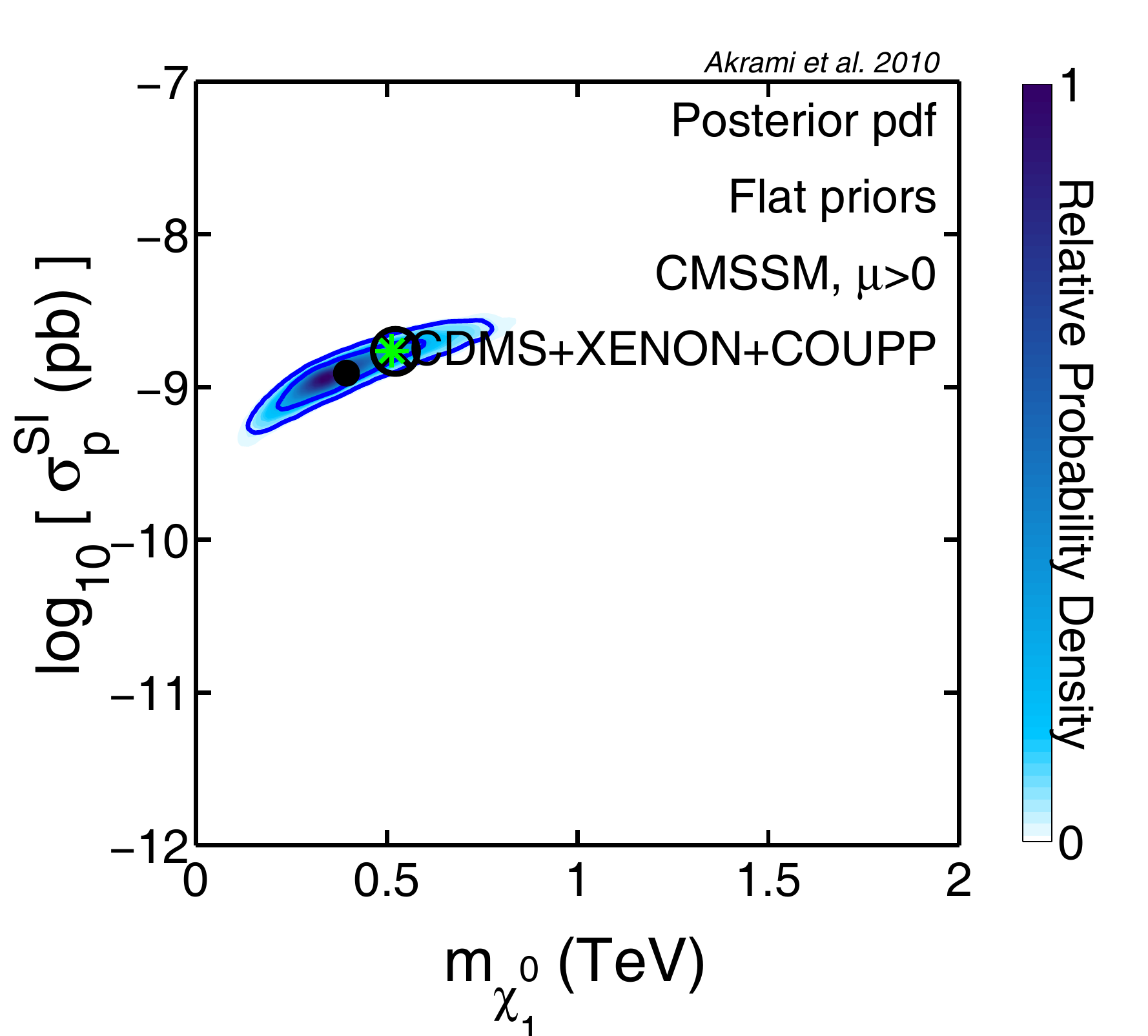}}\\
\subfigure{\includegraphics[scale=0.23, trim = 40 230 130 123, clip=true]{figs/MM_CDMSXENONCOUPP_wRefPoint/MM_CDMSXENONCOUPP_wRefPoint_2D_marg_16}}
\subfigure{\includegraphics[scale=0.23, trim = 40 230 130 123, clip=true]{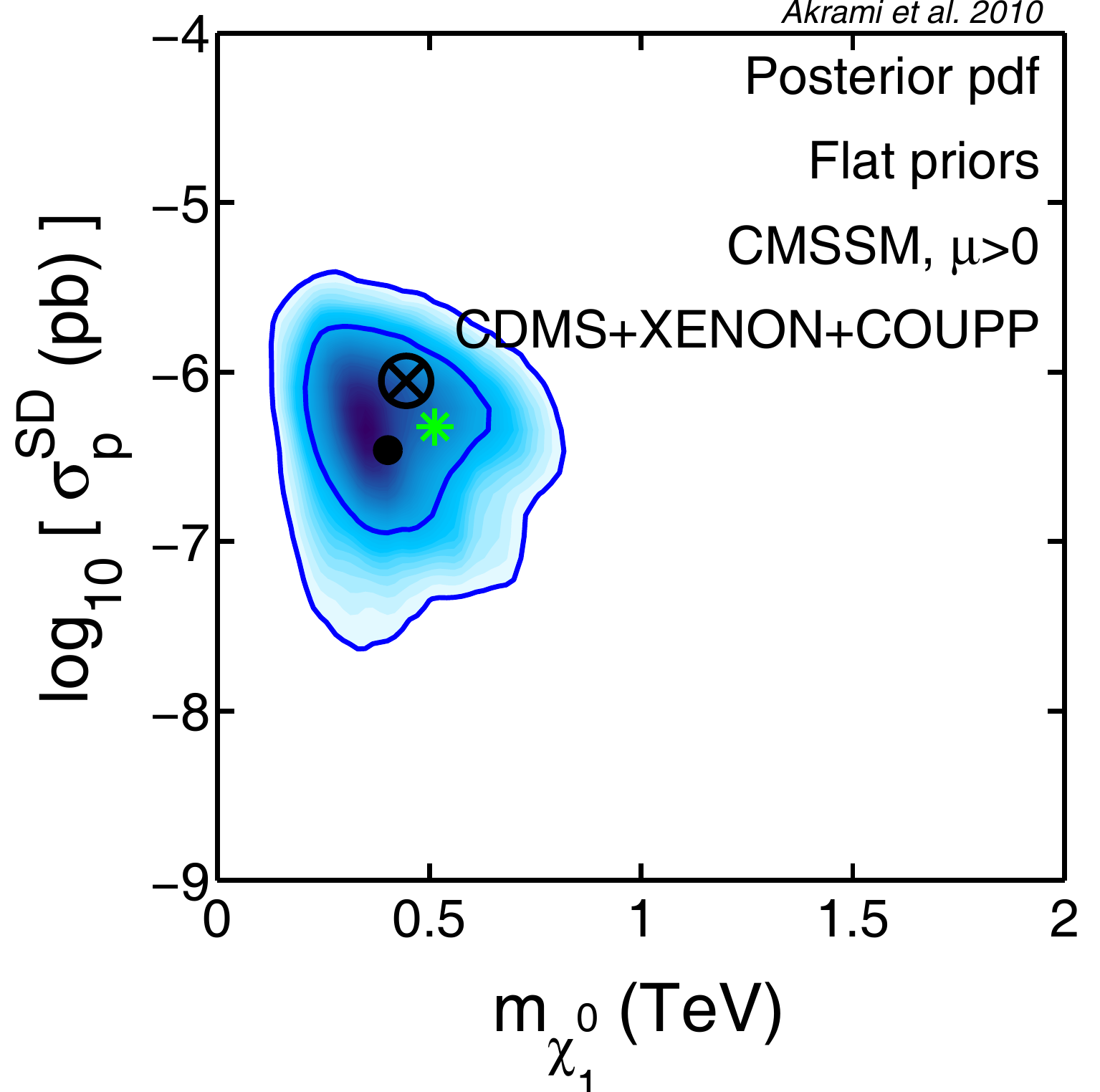}}
\subfigure{\includegraphics[scale=0.23, trim = 40 230 130 123, clip=true]{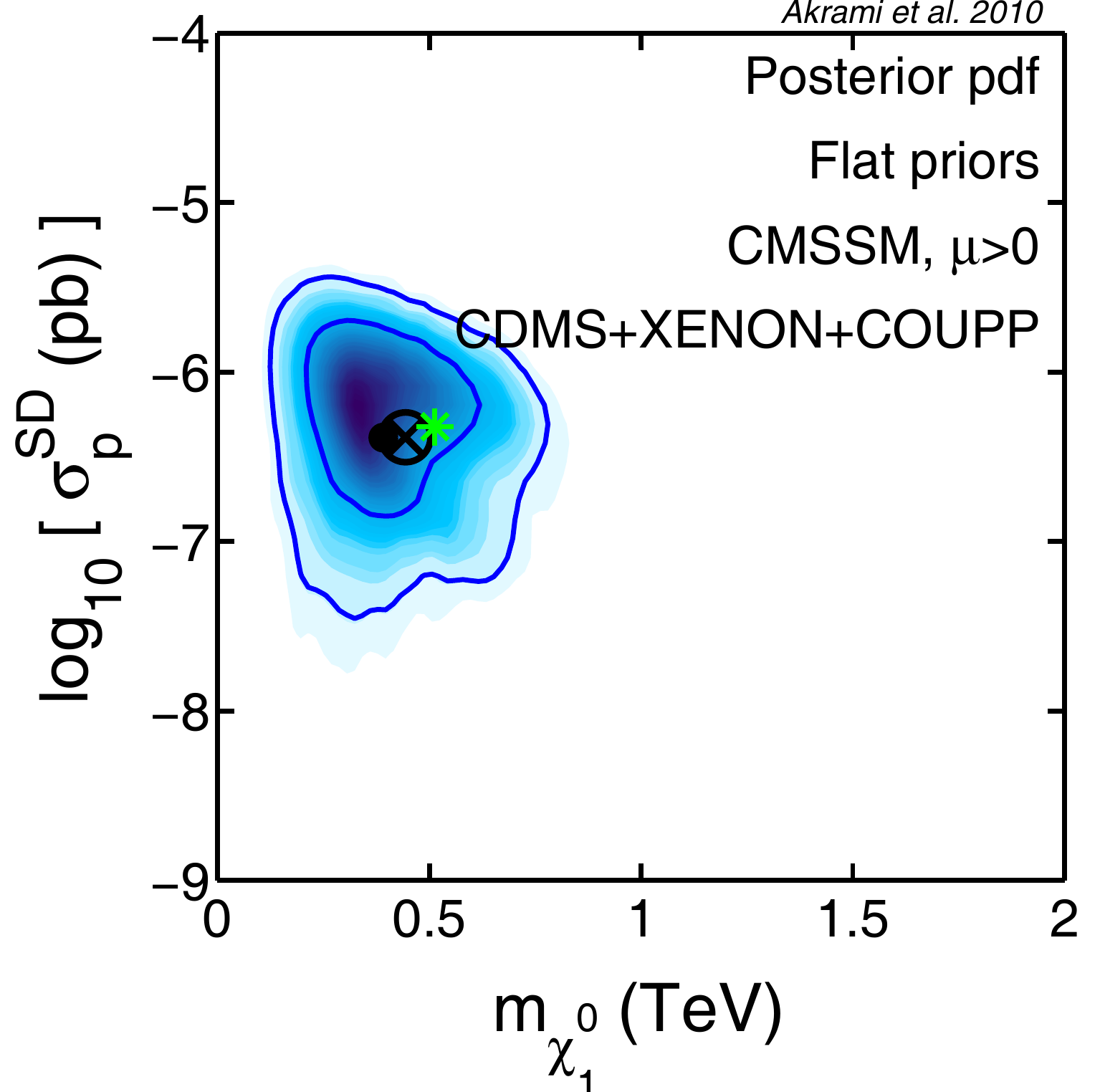}}
\subfigure{\includegraphics[scale=0.23, trim = 40 230 60 123, clip=true]{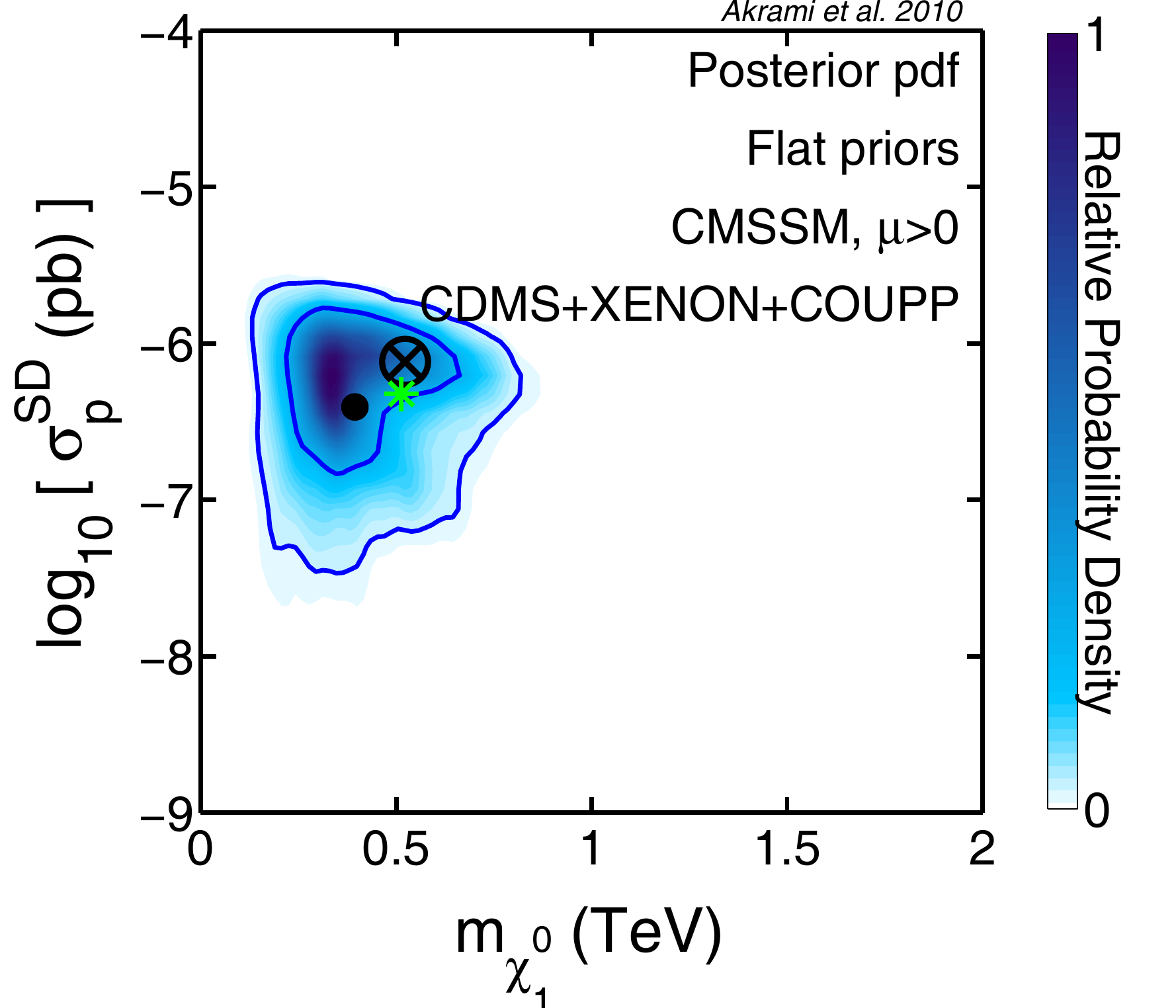}}\\
\subfigure{\includegraphics[scale=0.23, trim = 40 230 130 123, clip=true]{figs/MM_CDMSXENONCOUPP_wRefPoint/MM_CDMSXENONCOUPP_wRefPoint_2D_marg_17}}
\subfigure{\includegraphics[scale=0.23, trim = 40 230 130 123, clip=true]{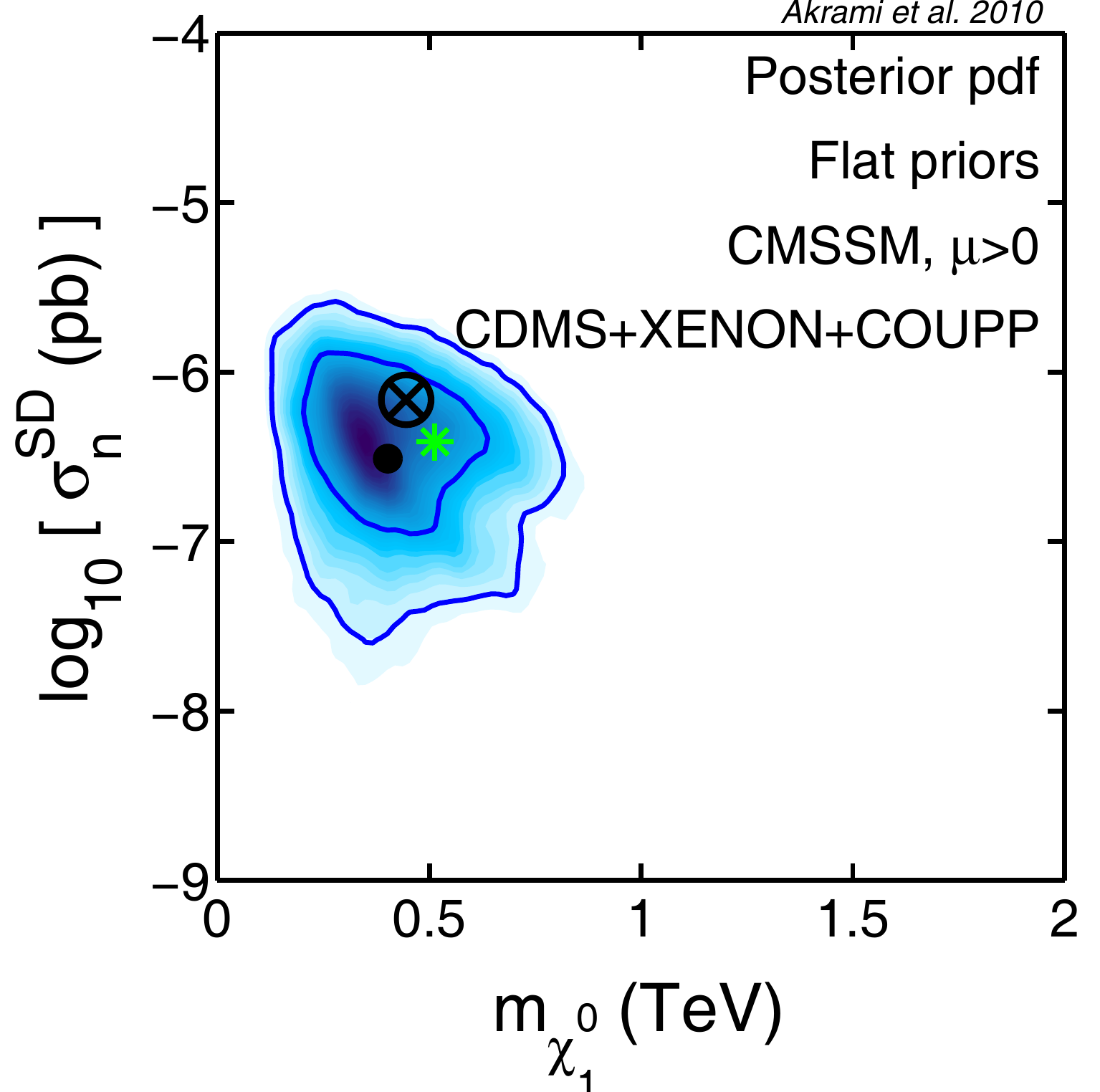}}
\subfigure{\includegraphics[scale=0.23, trim = 40 230 130 123, clip=true]{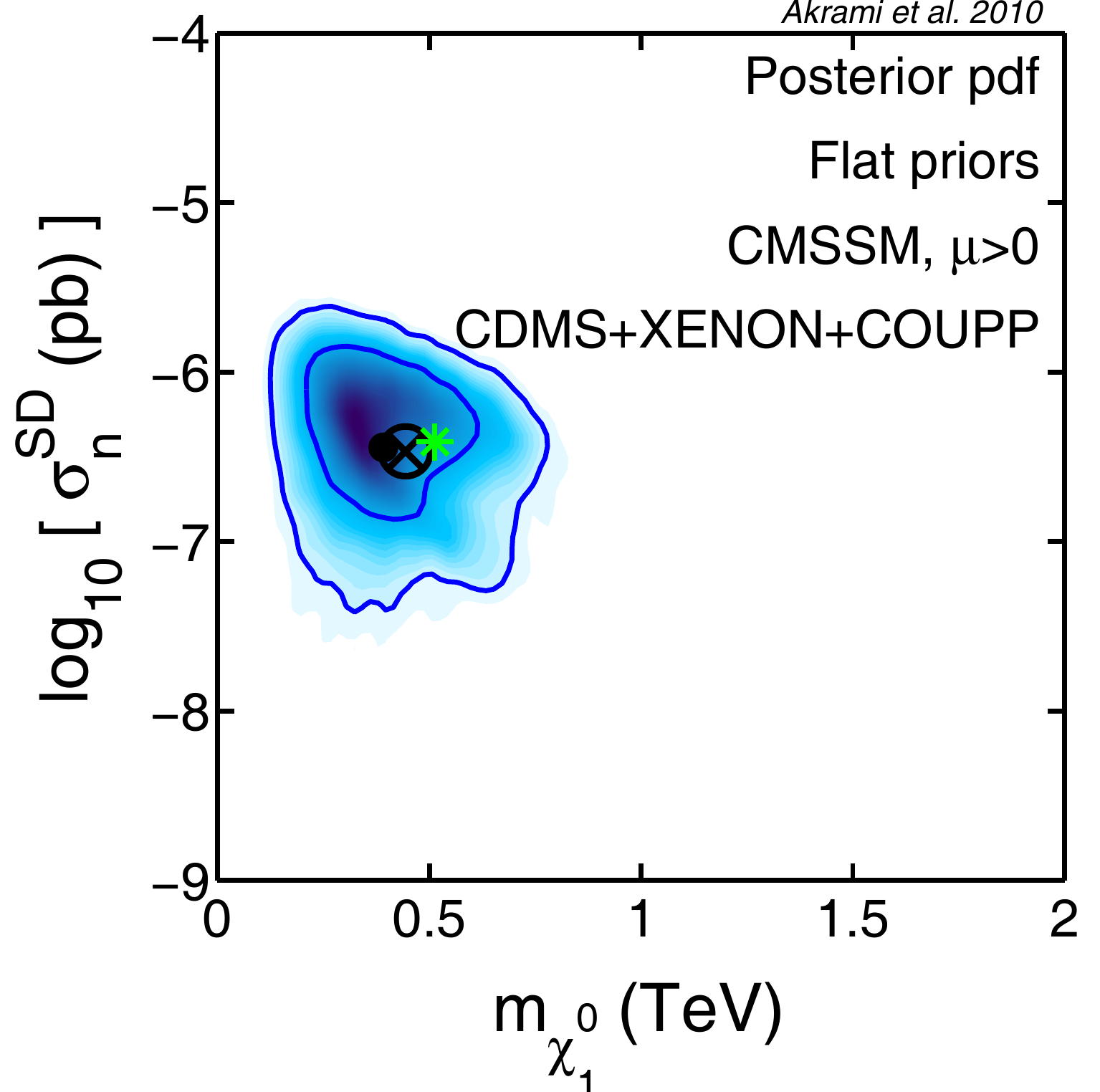}}
\subfigure{\includegraphics[scale=0.23, trim = 40 230 60 123, clip=true]{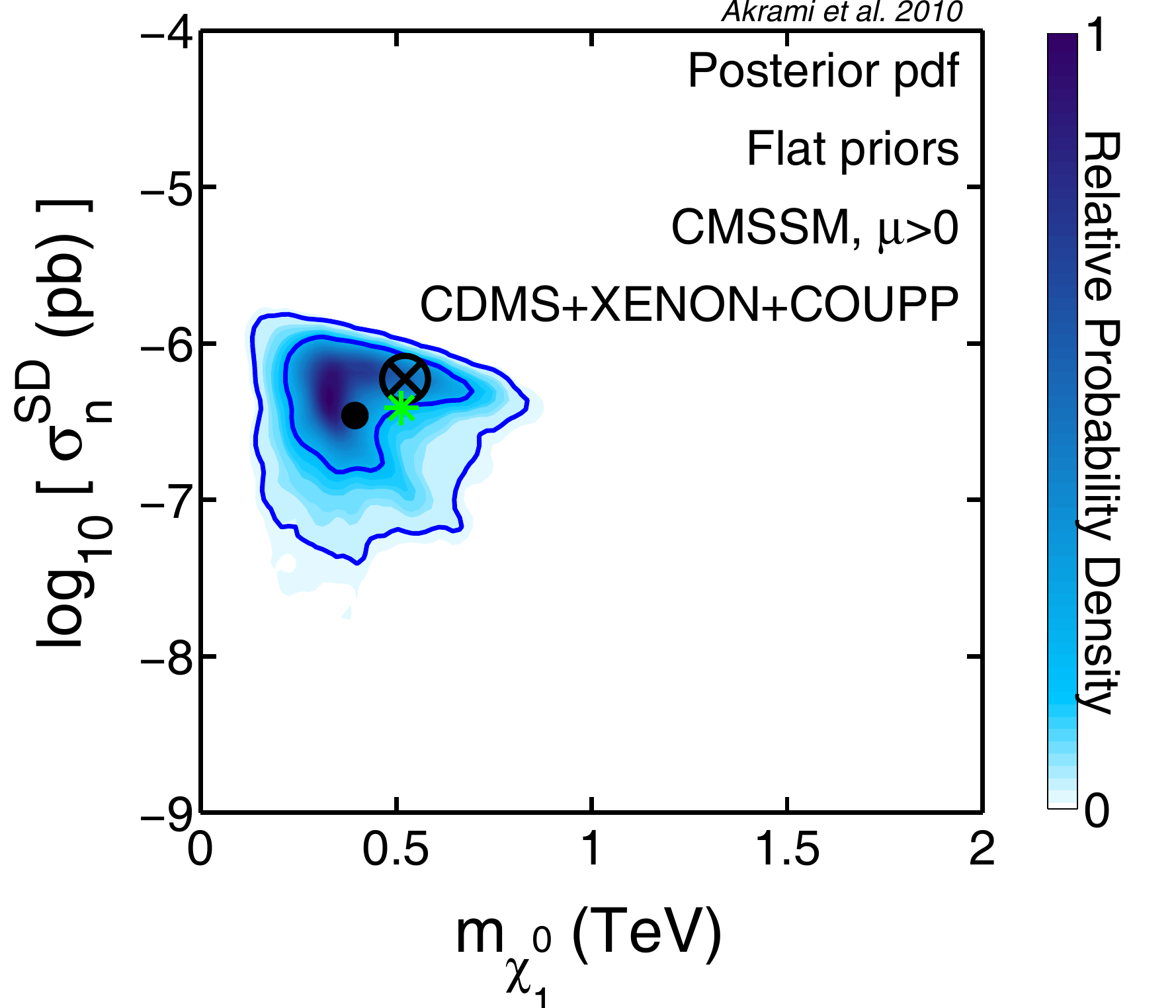}}\\
\caption[aa]{\footnotesize{Two-dimensional marginalised posterior PDFs for the mass and nuclear scattering cross-sections of the lightest neutralino, for benchmark 3 when all, some or none of the halo and cross-section nuisance parameters vary along with the CMSSM parameters in our scans. CS and H in the subcaptions stand for cross-section and halo nuisances, respectively. Again, the inner and outer contours in each panel represent $68.3\%$ ($1\sigma$) and $95.4\%$ ($2\sigma$) confidence levels, respectively. Black dots and crosses show the posterior means and best-fit points, respectively, and benchmark values are marked with green stars.}}\label{fig:MMNuiscompmarg}
\end{figure}

In our analysis, we included uncertainties on the halo parameters $\rhoDM$, $\vrot$, $\vmp$ and $\vesc$, as well as on the neutralino-nucleon coupling parameters $\Deltaps$, $\sigma_0$ and $\SigmapiN$, by treating them as nuisance parameters. It is interesting to know how large these nuisance effects are, and how strongly they impact our scanning results. In~\figs{fig:MMNuiscompmarg}{fig:MMNuiscompprofl} we show the results of some scans for benchmark 3 with data from all three experiments, CDMS1T, XENON1T and COUPP1T, and where none or only some of these nuisances are included (for brevity purposes, we again only show the results in terms of the mass and scattering cross-sections of the neutralino, and refer the reader to the first preprint of the paper on the \textsf{arXiv} for similar plots in the CMSSM planes).

By comparing panels in the first rows (the $\sigma^{SI}_p$-$m_{\tilde\chi^0_1}$ planes) we see that when the nuisance parameters are allowed to vary, the favoured region becomes wider in cross-section. This effect is mostly caused by the halo nuisances (compare the third and fourth panels in the first rows), and fixing those whilst only varying the cross-section parameters (second panels) does not change the size and shape of the confidence/credible region significantly, compared to the case where all nuisances are fixed (fourth panels).
The latter behaviour is expected: changing the values of $\sigma_0$ and
$\SigmapiN$ do not affect the value of $\sigmapSI$ that best fits the
experimental results, but it does affect which CMSSM models yield that
SI cross-section.
An interesting feature in the case where all nuisance parameters are fixed is the sharp cut-off in the SD cross-sections $\sigma^{SD}_p$ and $\sigma^{SD}_n$ at high values (fourth panels in the second and third rows of~\figs{fig:MMNuiscompmarg}{fig:MMNuiscompprofl}); this can be understood from the correlation between SD and SI cross-sections.

\begin{figure}[t]
\setcounter{subfigure}{0}
\subfigure[][\scriptsize{\textbf{CS/H Nuis:~~}}]{\includegraphics[scale=0.23, trim = 40 230 130 100, clip=true]{figs/MM_CDMSXENONCOUPP_wRefPoint/MM_CDMSXENONCOUPP_wRefPoint_2D_profl_15}}
\subfigure[][\scriptsize{\textbf{CS Nuis Only:}}]{\includegraphics[scale=0.23, trim = 40 230 130 100, clip=true]{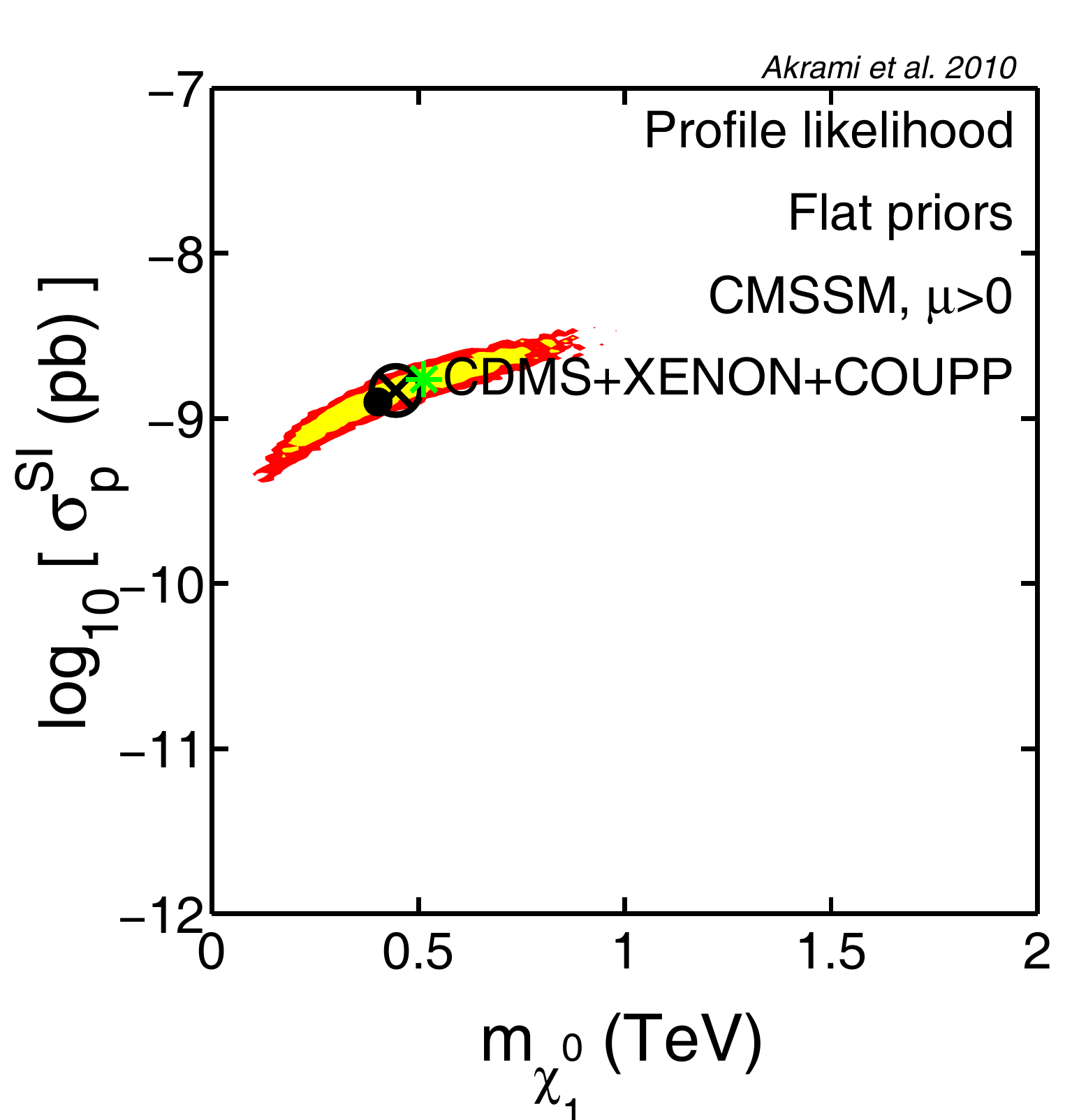}}
\subfigure[][\scriptsize{\textbf{H Nuis Only:~}}]{\includegraphics[scale=0.23, trim = 40 230 130 100, clip=true]{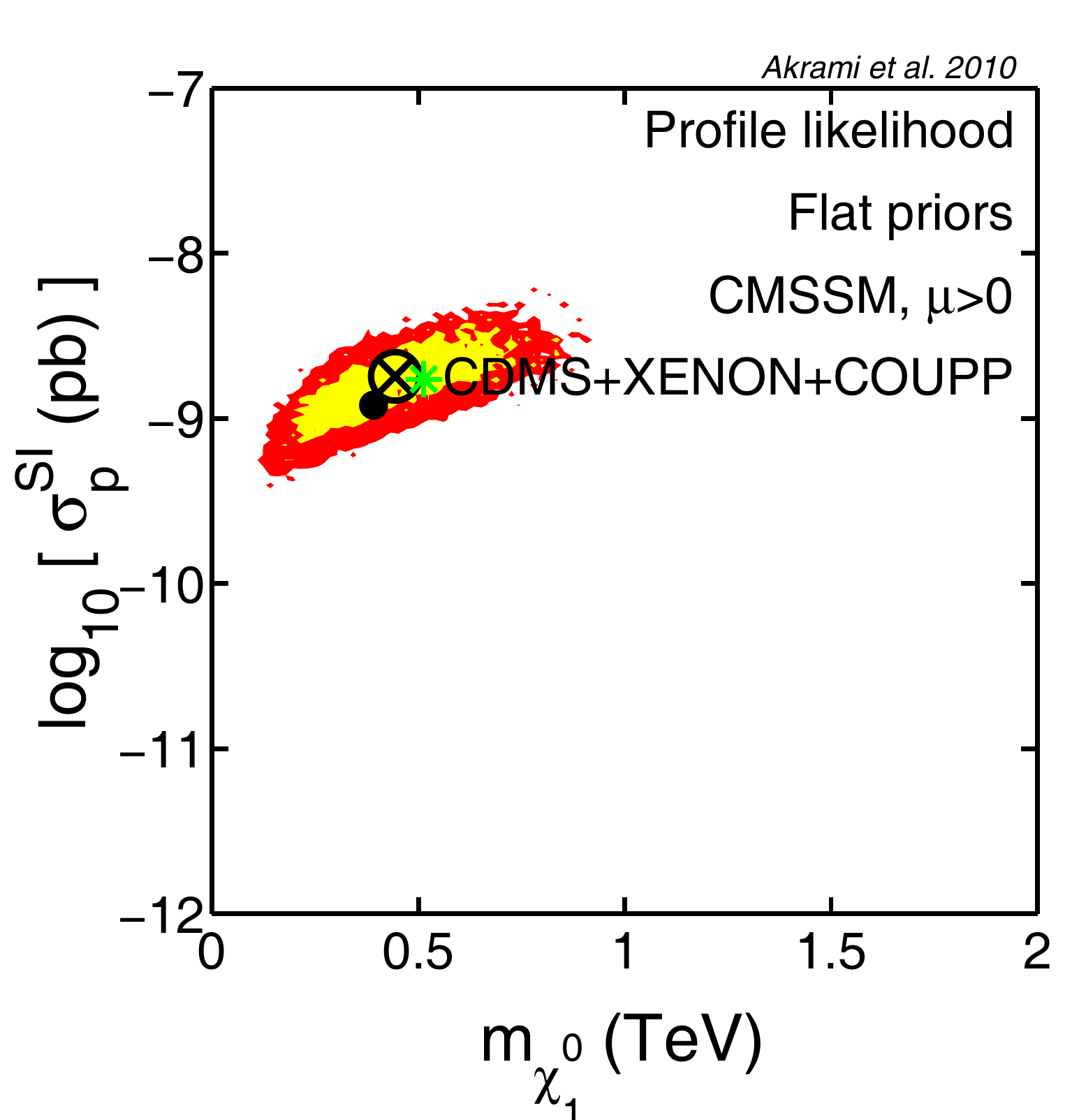}}
\subfigure[][\scriptsize{\textbf{No CS/H  Nuis:~~}}]{\includegraphics[scale=0.23, trim = 40 230 60 100, clip=true]{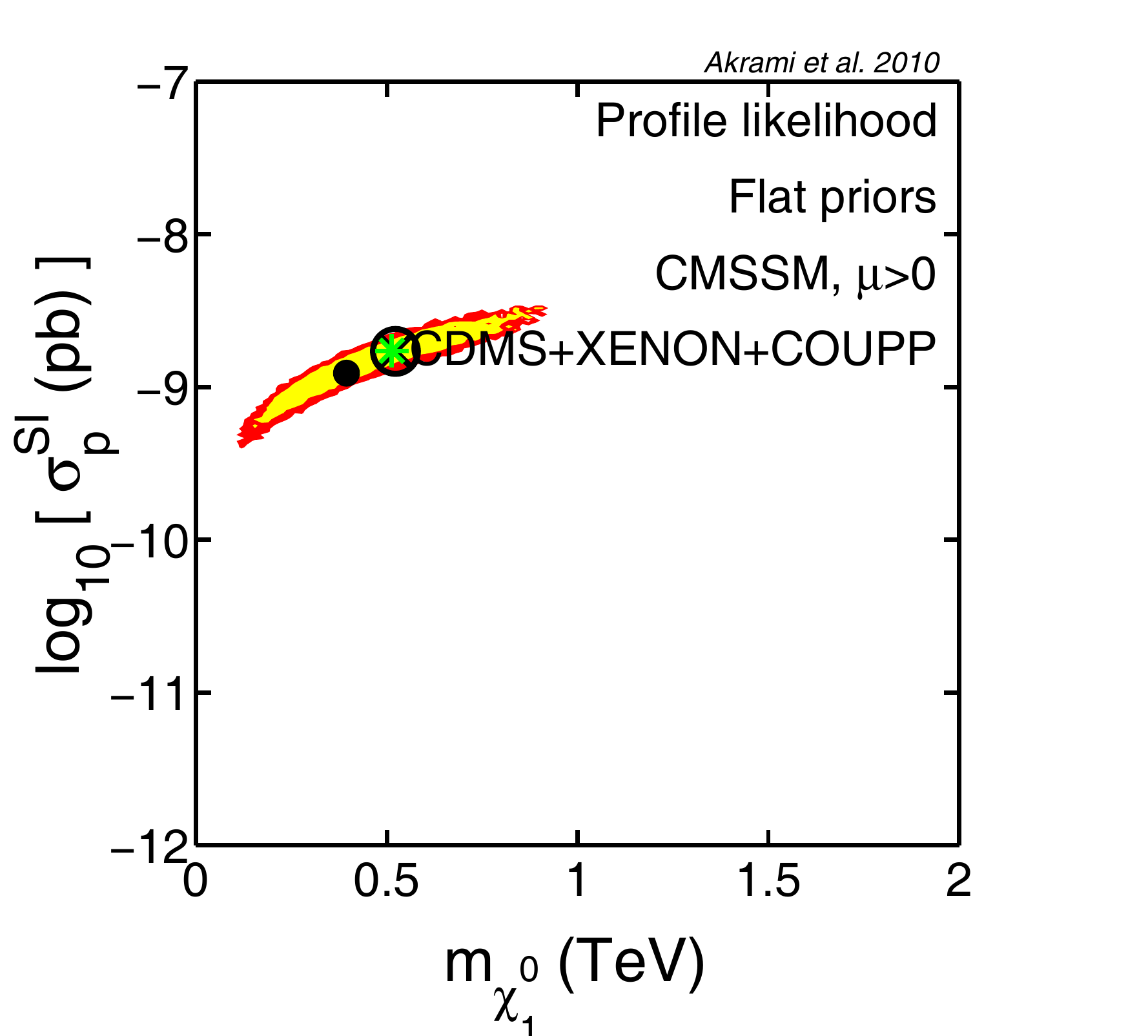}}\\
\subfigure{\includegraphics[scale=0.23, trim = 40 230 130 123, clip=true]{figs/MM_CDMSXENONCOUPP_wRefPoint/MM_CDMSXENONCOUPP_wRefPoint_2D_profl_16}}
\subfigure{\includegraphics[scale=0.23, trim = 40 230 130 123, clip=true]{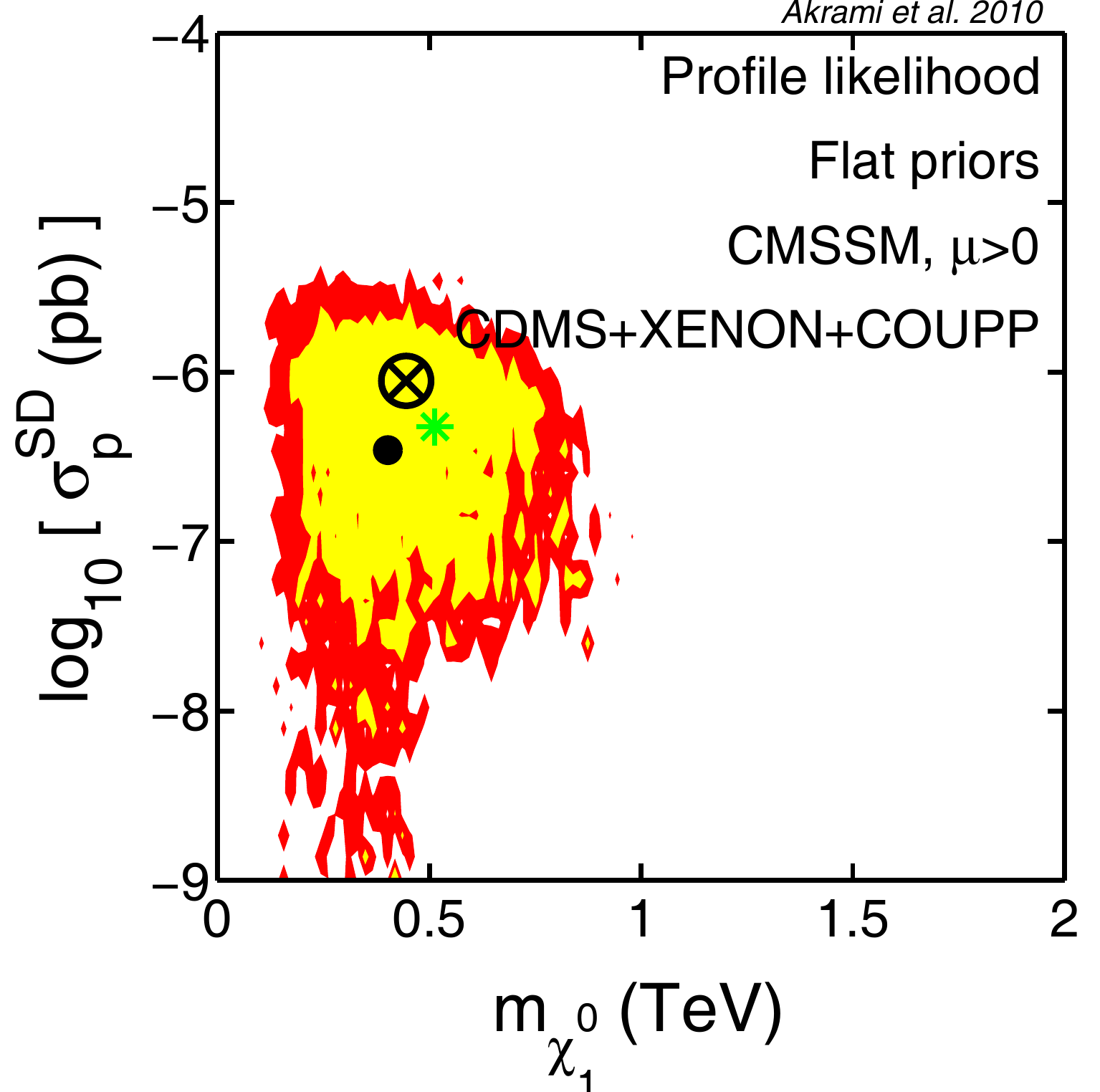}}
\subfigure{\includegraphics[scale=0.23, trim = 40 230 130 123, clip=true]{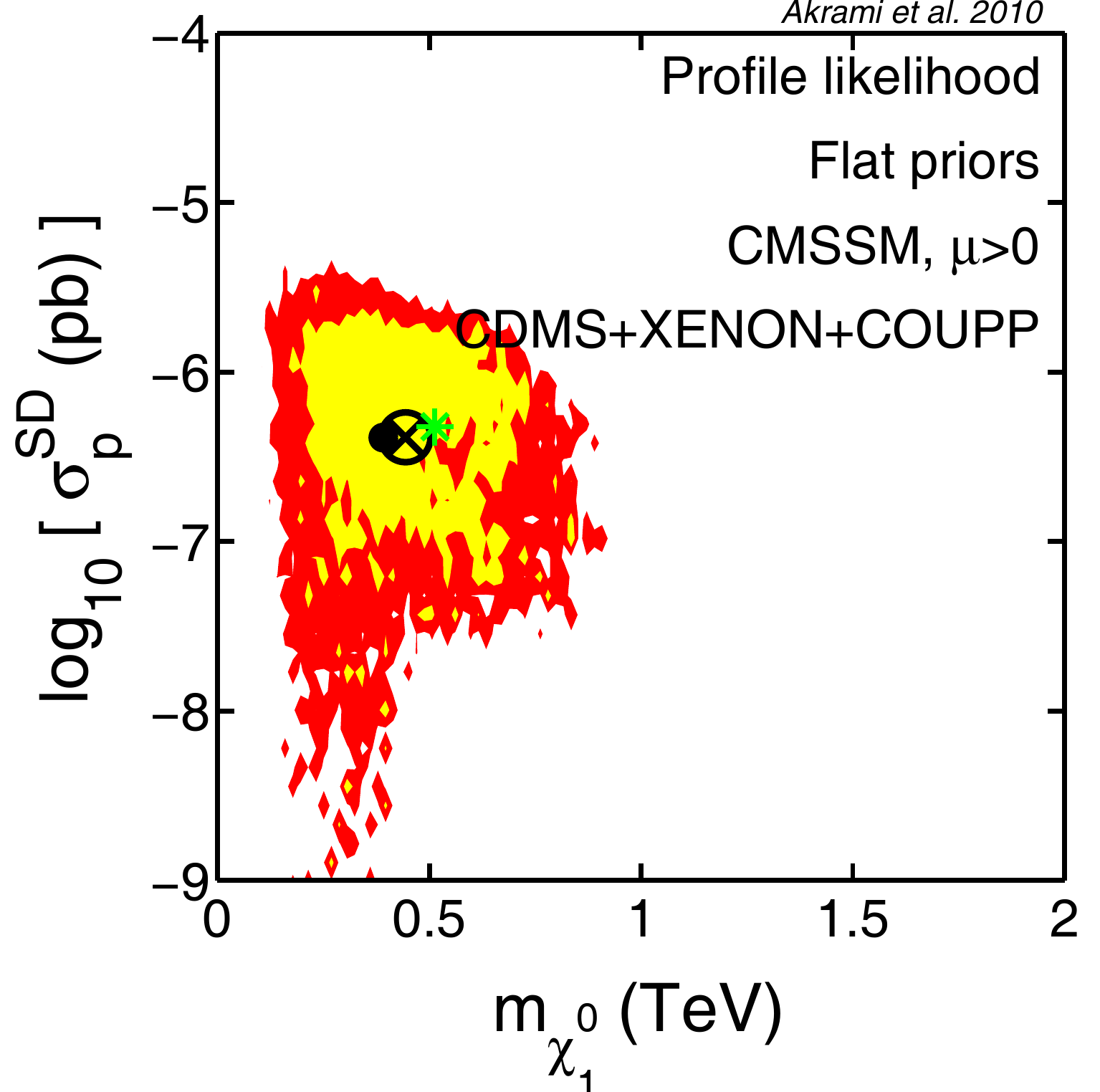}}
\subfigure{\includegraphics[scale=0.23, trim = 40 230 60 123, clip=true]{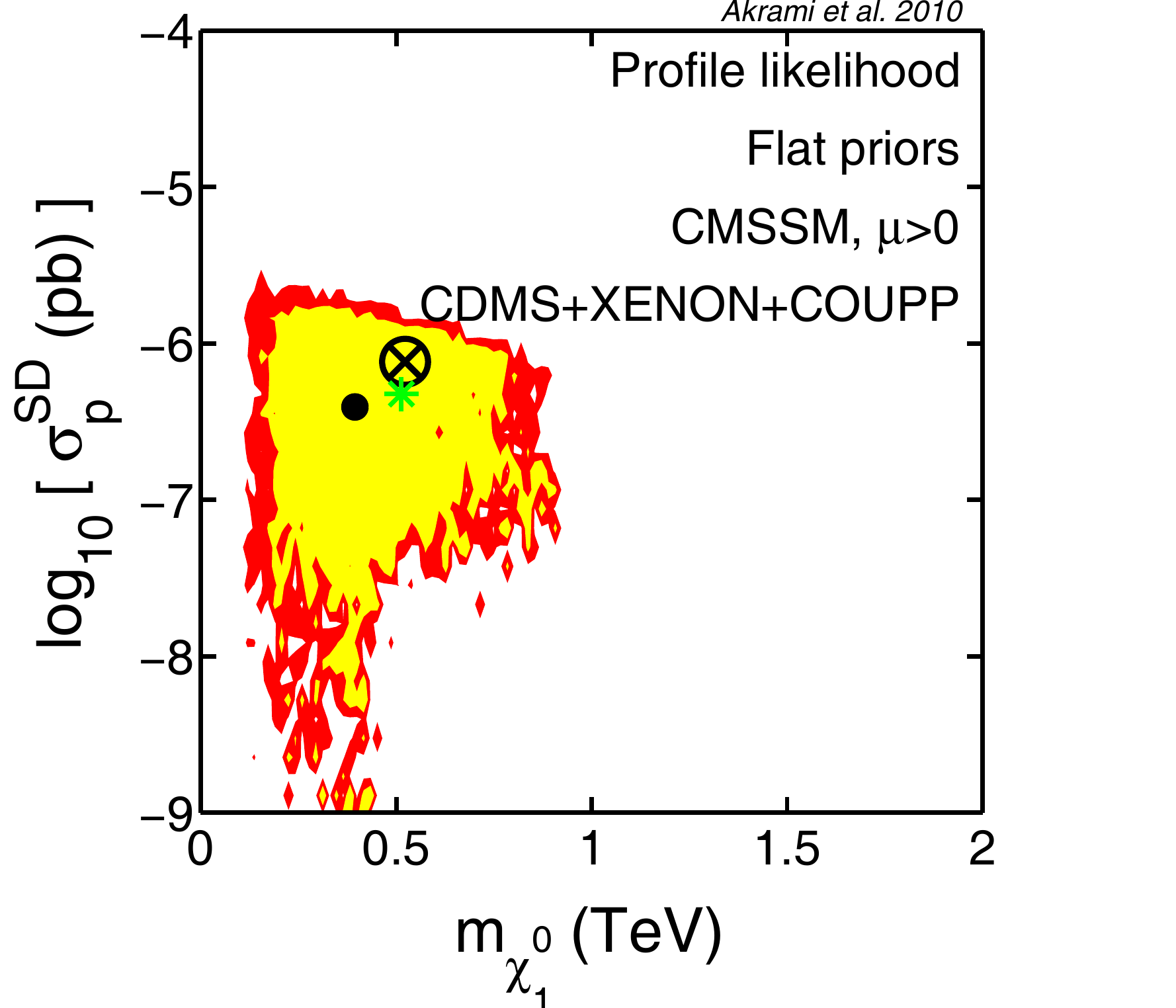}}\\
\subfigure{\includegraphics[scale=0.23, trim = 40 230 130 123, clip=true]{figs/MM_CDMSXENONCOUPP_wRefPoint/MM_CDMSXENONCOUPP_wRefPoint_2D_profl_17}}
\subfigure{\includegraphics[scale=0.23, trim = 40 230 130 123, clip=true]{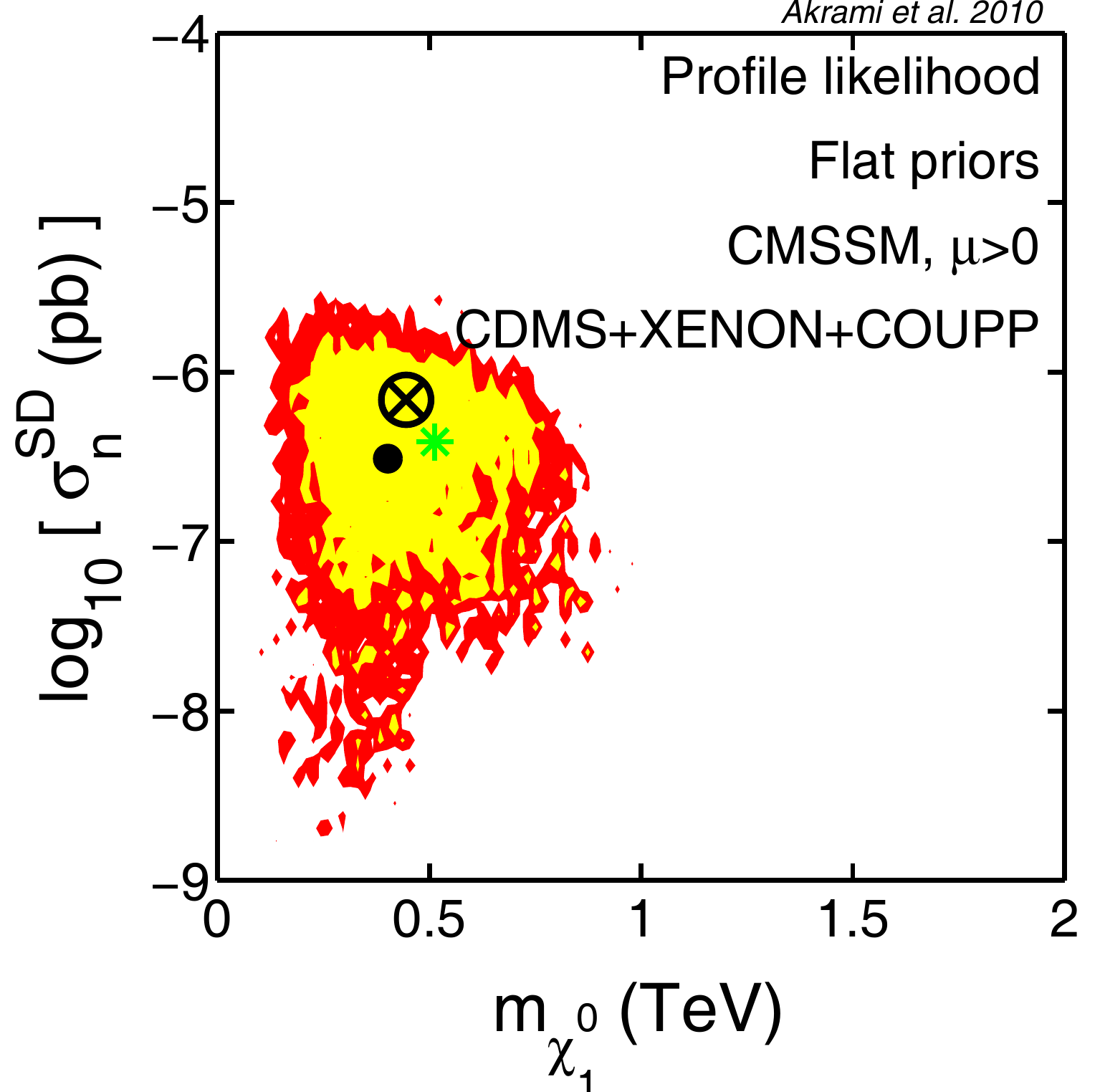}}
\subfigure{\includegraphics[scale=0.23, trim = 40 230 130 123, clip=true]{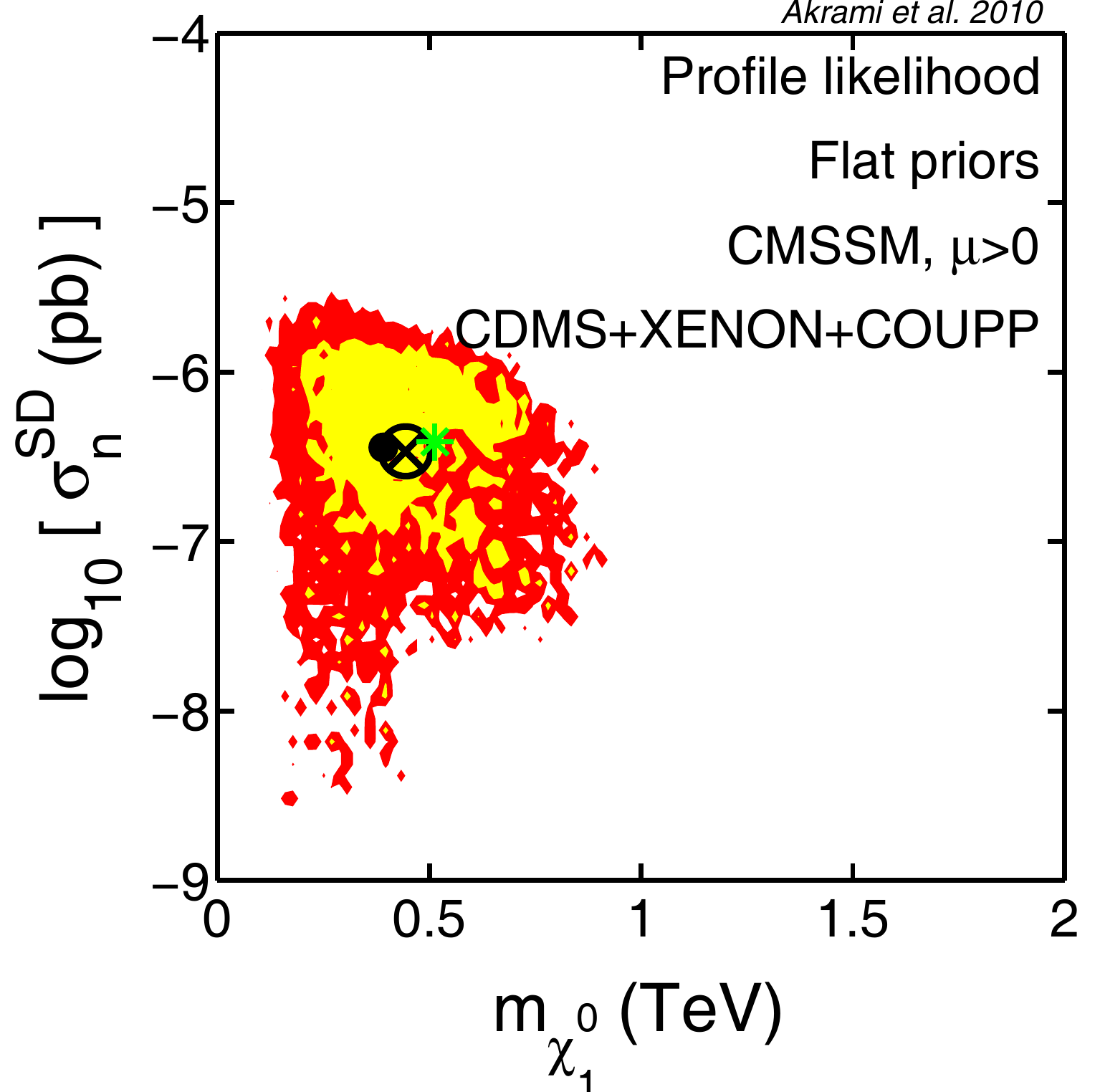}}
\subfigure{\includegraphics[scale=0.23, trim = 40 230 60 123, clip=true]{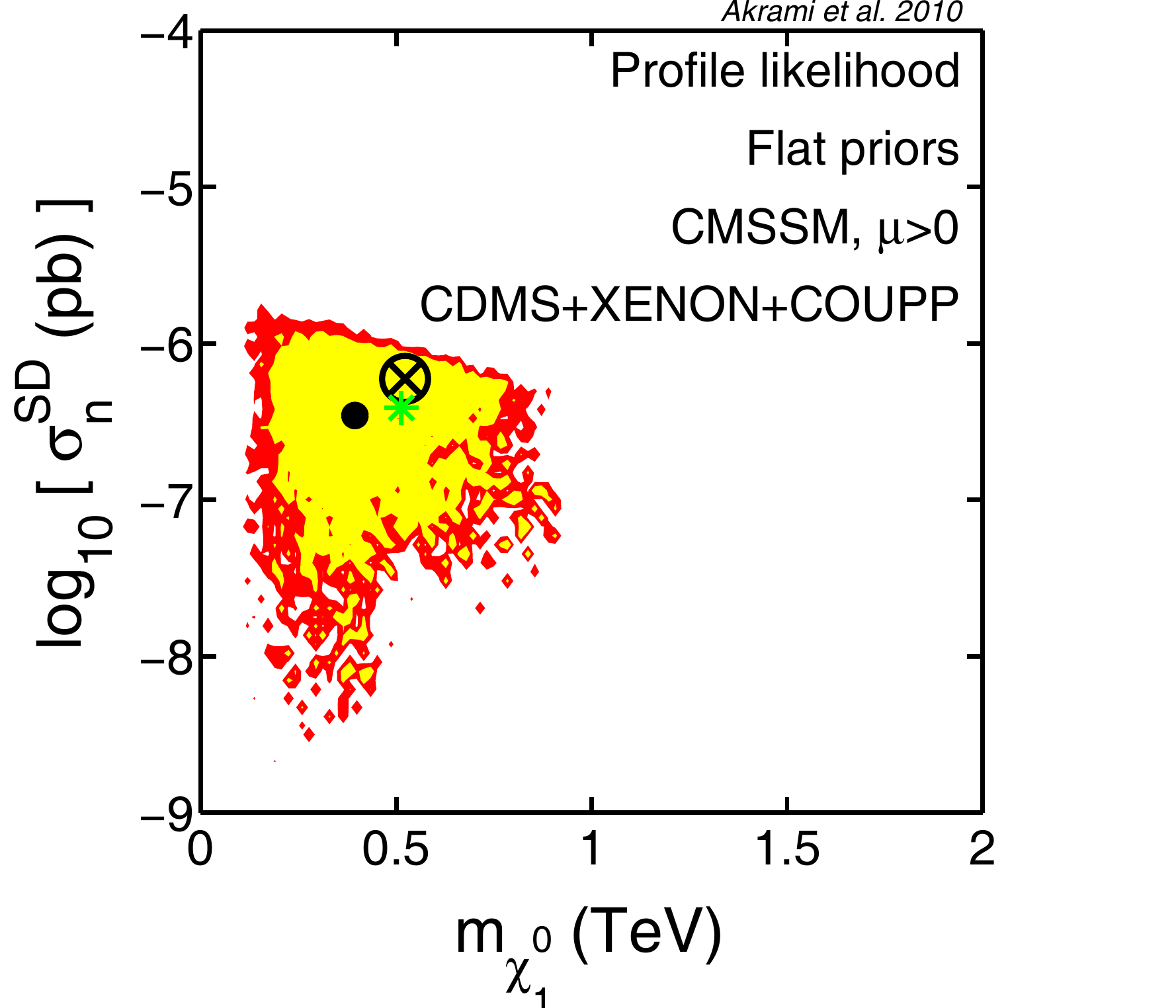}}\\
\caption[aa]{\footnotesize{As in~\fig{fig:MMNuiscompmarg}, but for two-dimensional profile likelihoods.}}\label{fig:MMNuiscompprofl}
\end{figure}

\begin{figure}[t]
\setcounter{subfigure}{0}
\begin{center}
\subfigure[][\scriptsize{\textbf{CS/H Nuis:~~}}]{\includegraphics[scale=0.225, trim = 40 230 130 100, clip=true]{figs/LH_CDMSXENONCOUPP_wRefPoint/LH_CDMSXENONCOUPP_wRefPoint_2D_marg_15}}
\subfigure[][\scriptsize{\textbf{No CS/H  Nuis:}}]{\includegraphics[scale=0.225, trim = 40 230 130 100, clip=true]{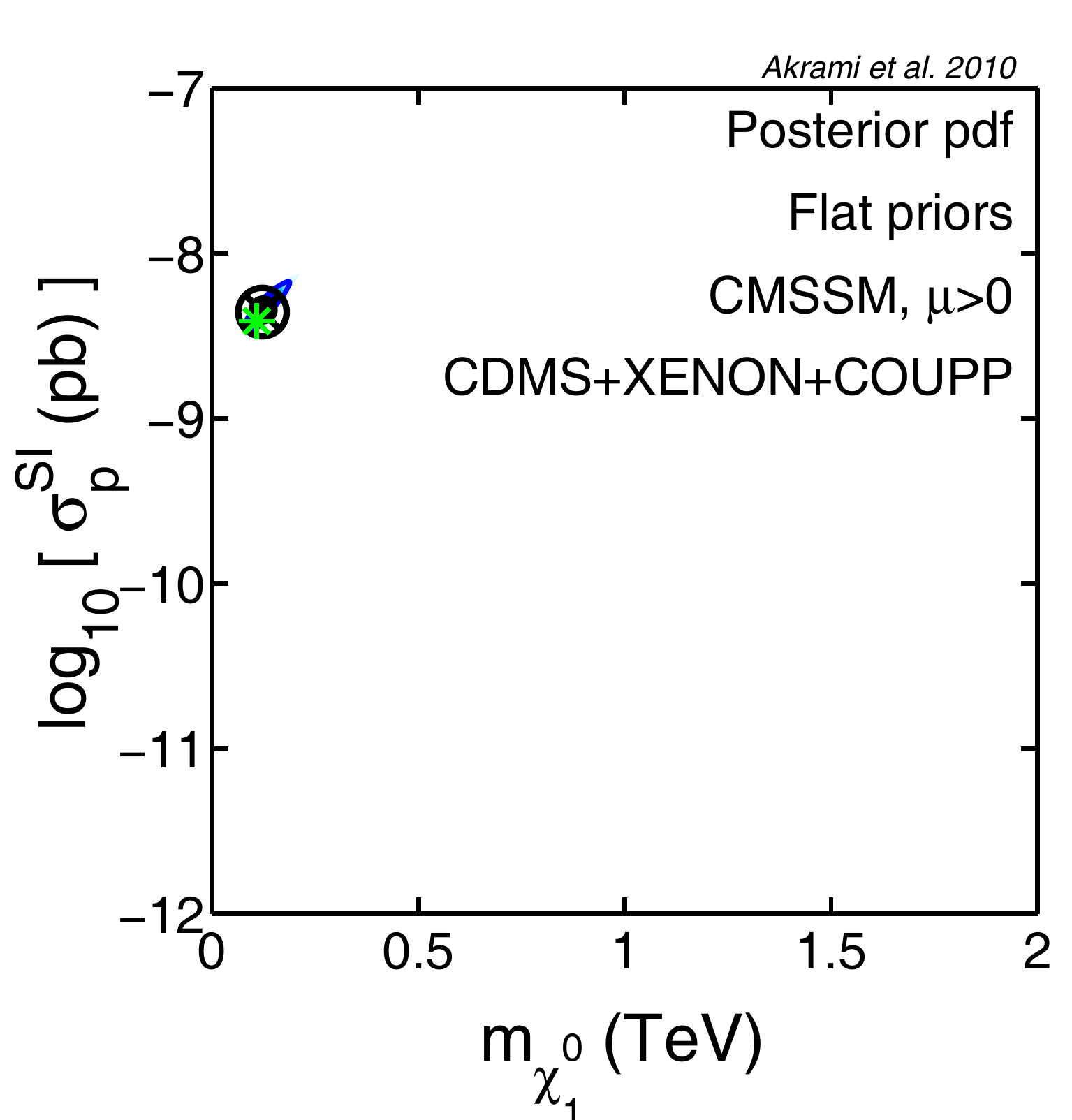}}
\subfigure[][\scriptsize{\textbf{CS/H Nuis:~~}}]{\includegraphics[scale=0.225, trim = 40 230 130 100, clip=true]{figs/LH_CDMSXENONCOUPP_wRefPoint/LH_CDMSXENONCOUPP_wRefPoint_2D_profl_15}}
\subfigure[][\scriptsize{\textbf{No CS/H  Nuis:~~~}}]{\includegraphics[scale=0.225, trim = 40 230 60 100, clip=true]{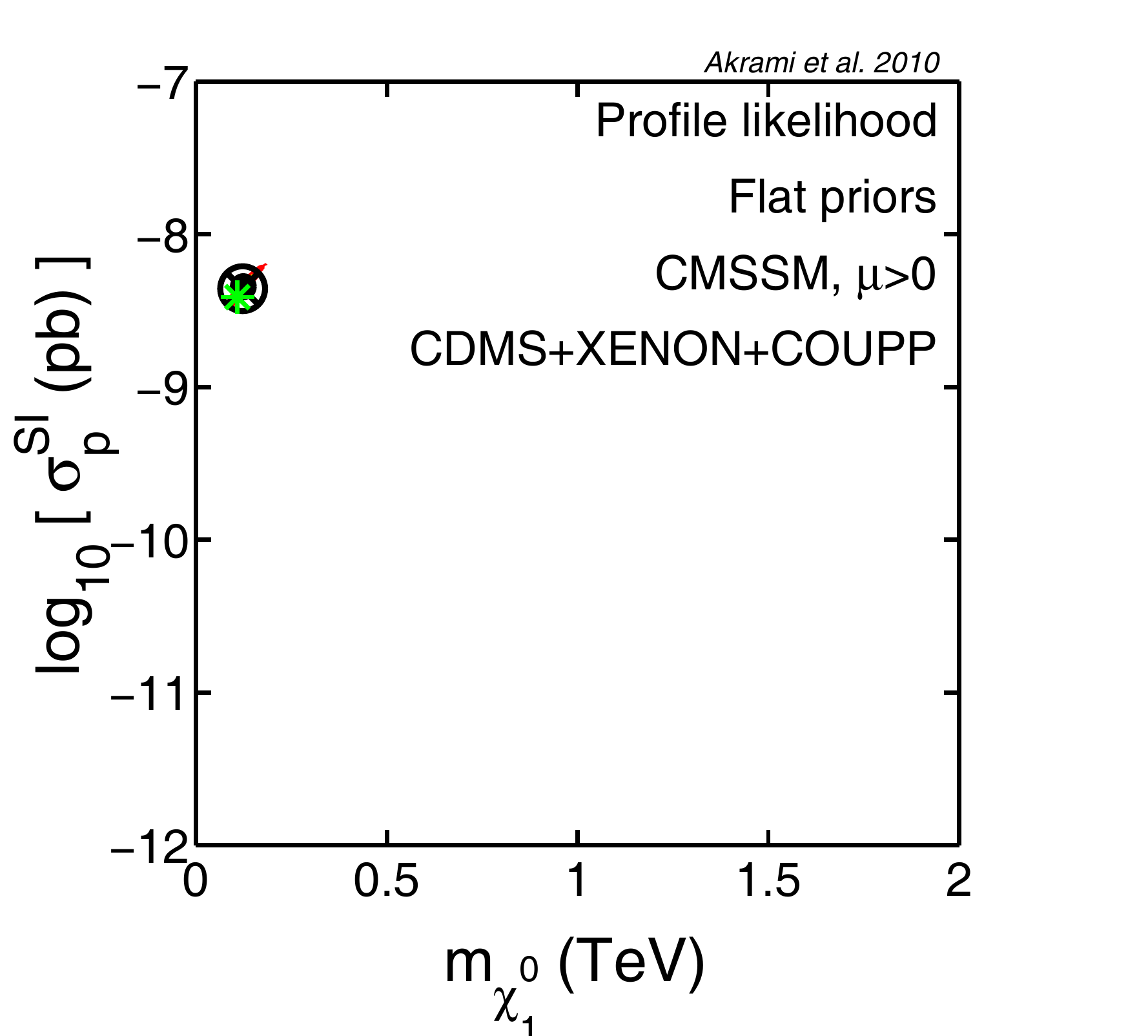}}\\
\subfigure{\includegraphics[scale=0.225, trim = 40 230 130 123, clip=true]{figs/LH_CDMSXENONCOUPP_wRefPoint/LH_CDMSXENONCOUPP_wRefPoint_2D_marg_16}}
\subfigure{\includegraphics[scale=0.225, trim = 40 230 130 123, clip=true]{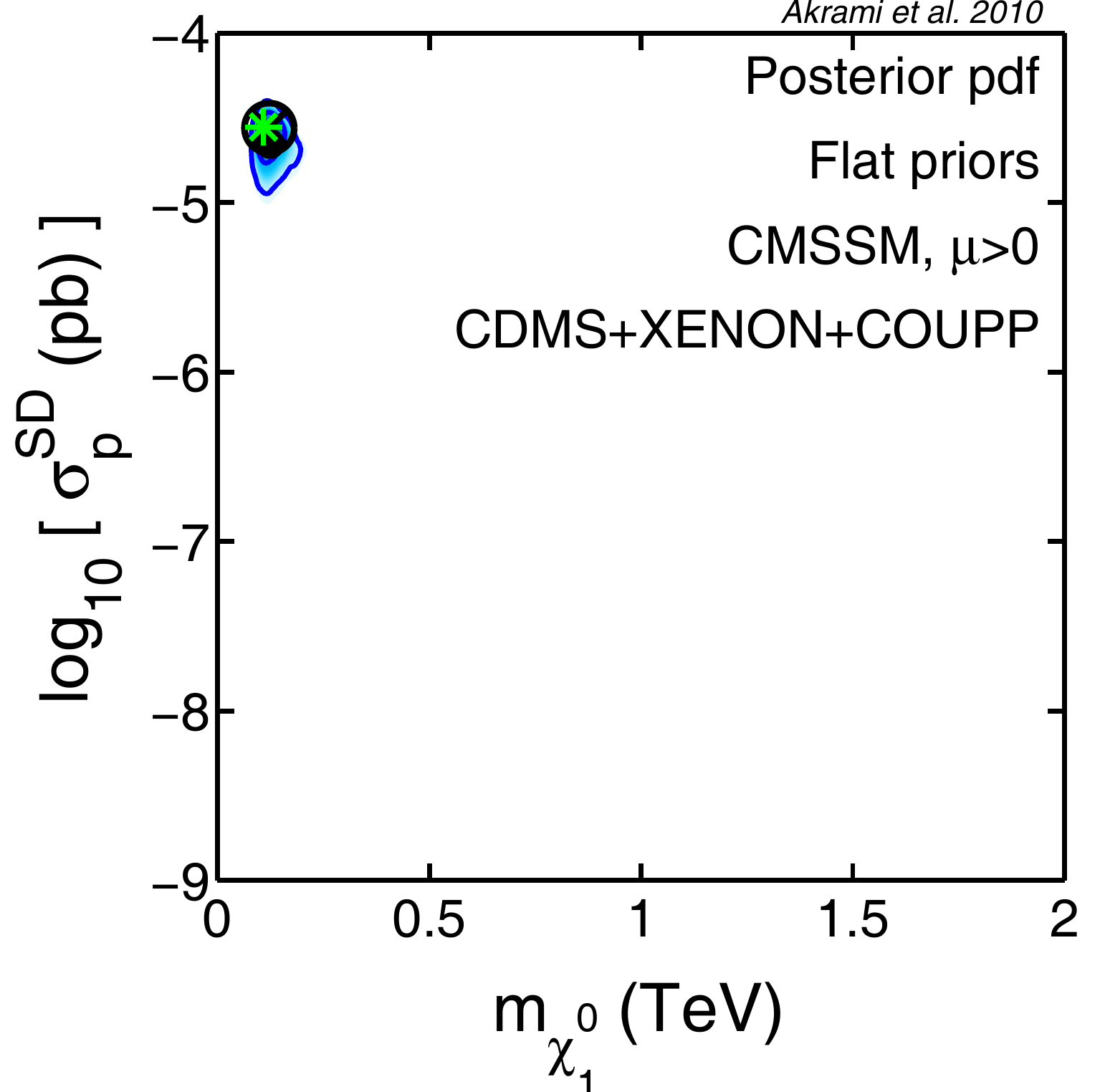}}
\subfigure{\includegraphics[scale=0.225, trim = 40 230 130 123, clip=true]{figs/LH_CDMSXENONCOUPP_wRefPoint/LH_CDMSXENONCOUPP_wRefPoint_2D_profl_16}}
\subfigure{\includegraphics[scale=0.225, trim = 40 230 60 123, clip=true]{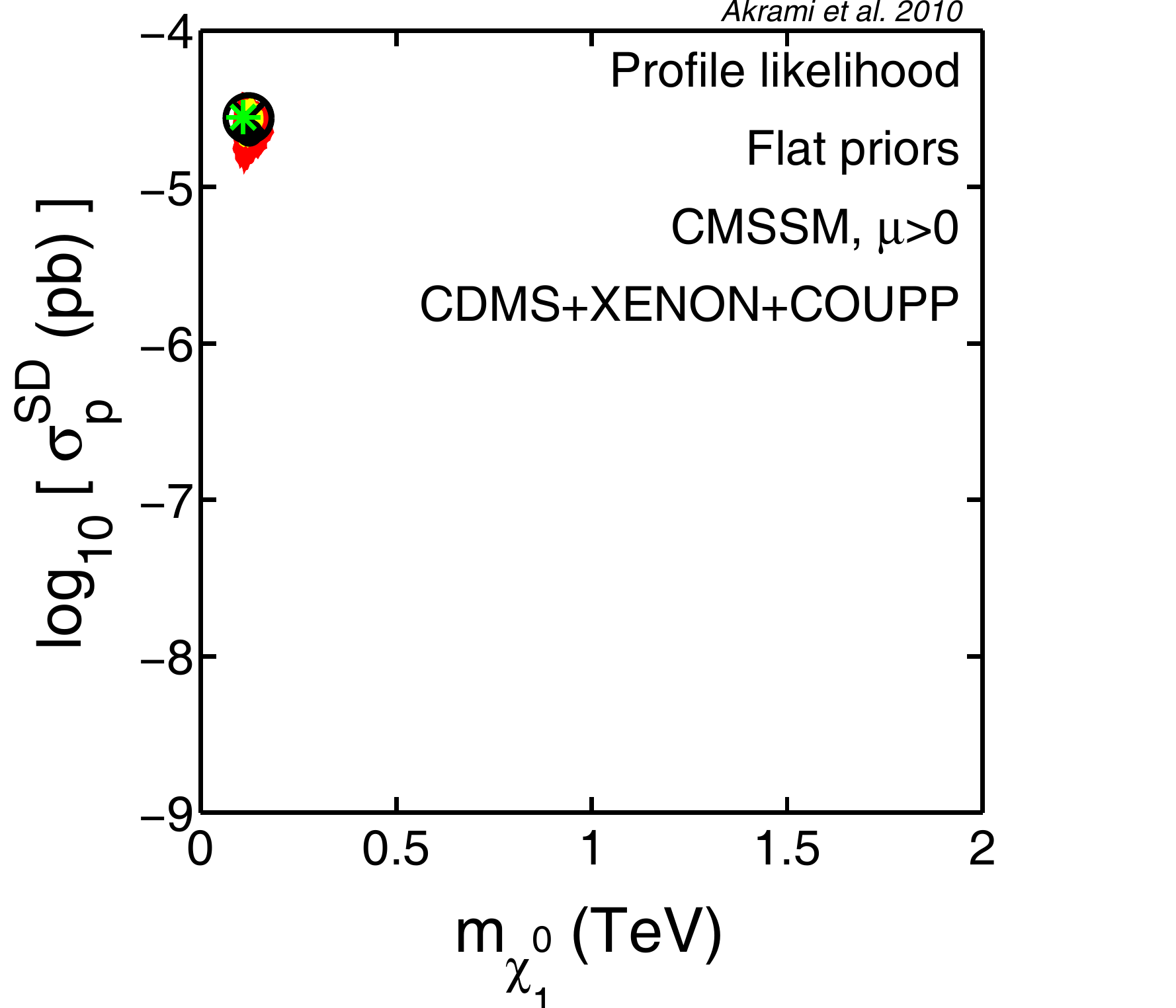}}\\
\subfigure{\includegraphics[scale=0.225, trim = 40 230 130 123, clip=true]{figs/LH_CDMSXENONCOUPP_wRefPoint/LH_CDMSXENONCOUPP_wRefPoint_2D_marg_17}}
\subfigure{\includegraphics[scale=0.225, trim = 40 230 130 123, clip=true]{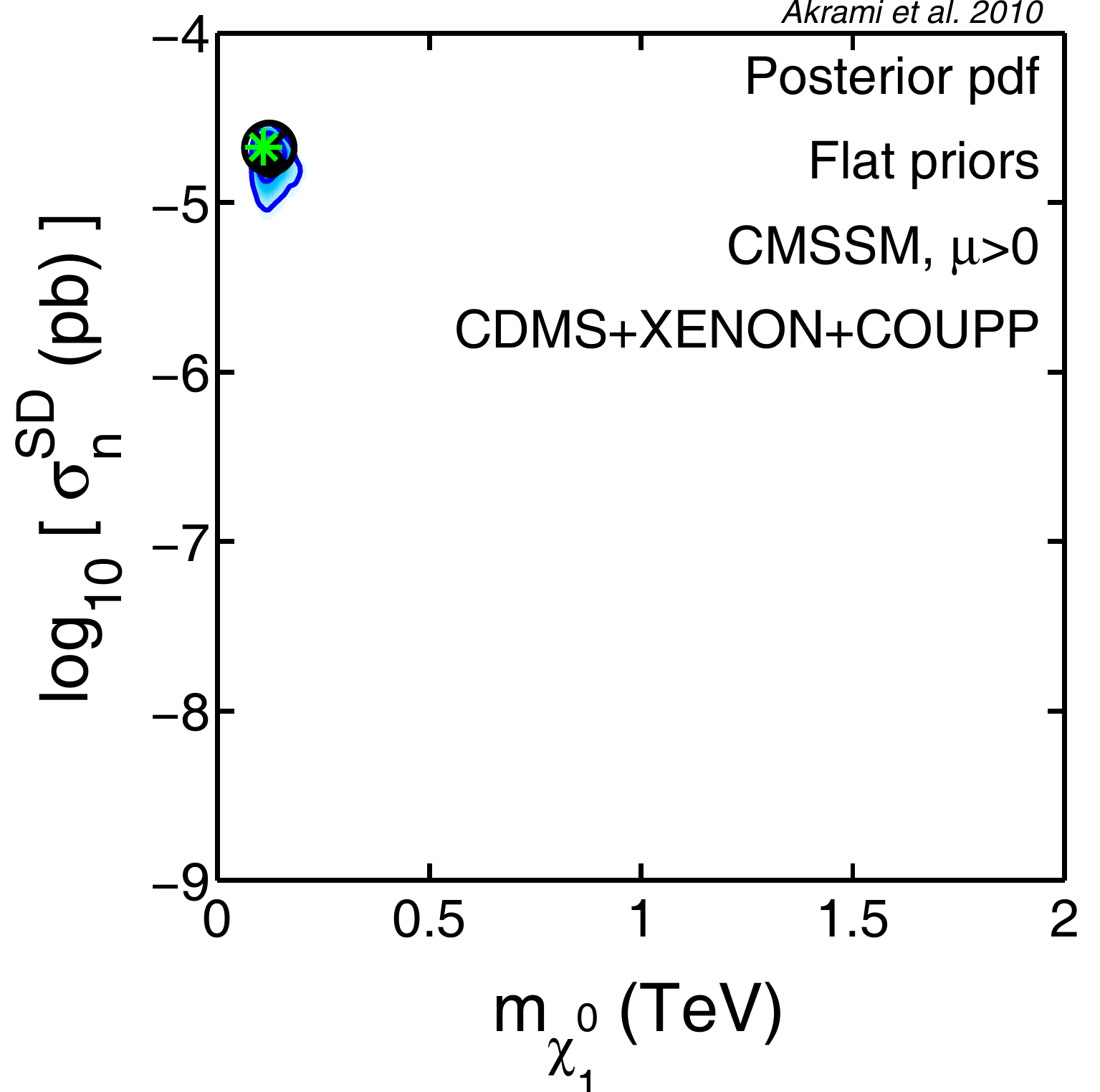}}
\subfigure{\includegraphics[scale=0.225, trim = 40 230 130 123, clip=true]{figs/LH_CDMSXENONCOUPP_wRefPoint/LH_CDMSXENONCOUPP_wRefPoint_2D_profl_17}}
\subfigure{\includegraphics[scale=0.225, trim = 40 230 60 123, clip=true]{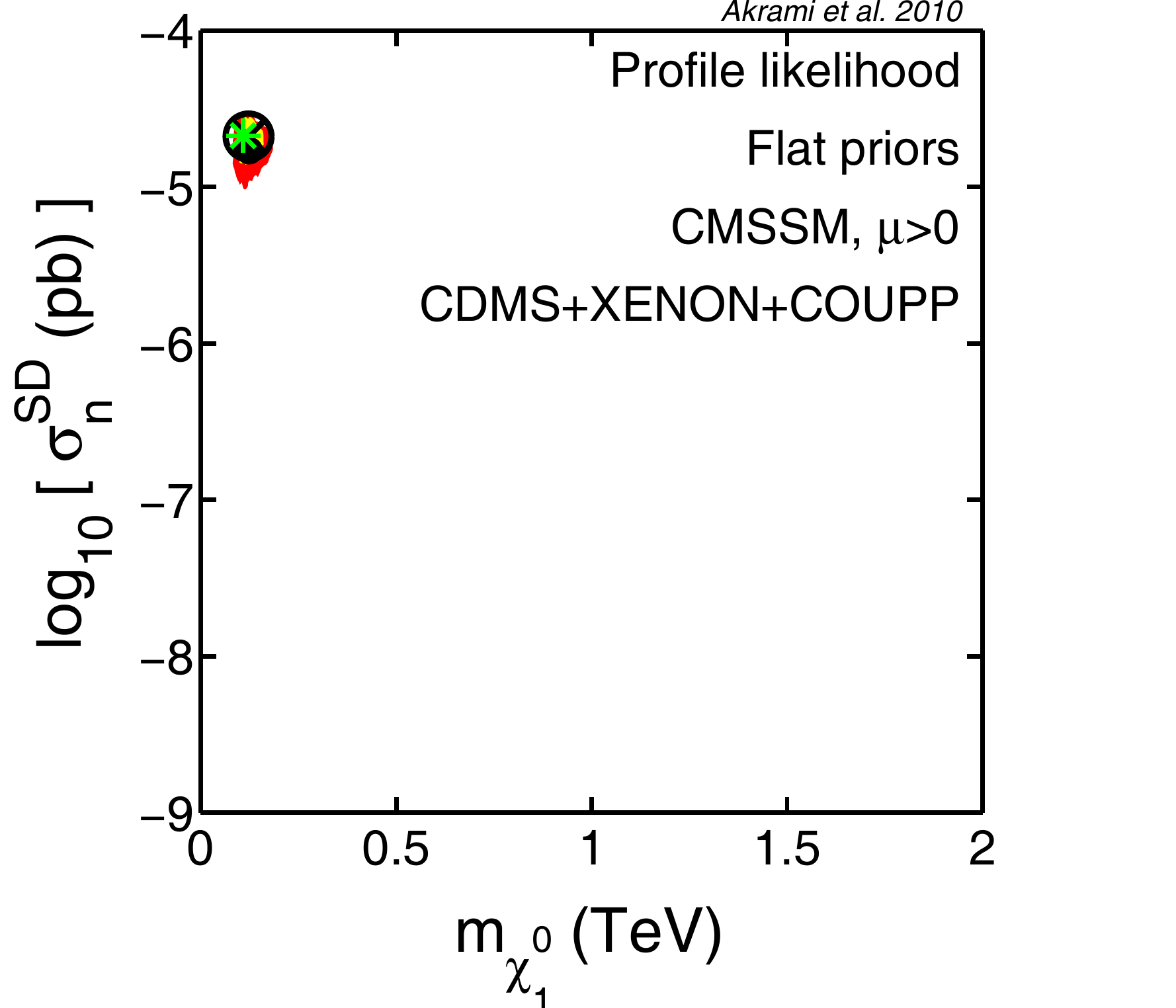}}\\
\subfigure{\includegraphics[scale=0.225, trim = 40 230 130 123, clip=true]{figs/LH_CDMSXENONCOUPP_wRefPoint/LH_CDMSXENONCOUPP_wRefPoint_2D_marg_1}}
\subfigure{\includegraphics[scale=0.225, trim = 40 230 130 123, clip=true]{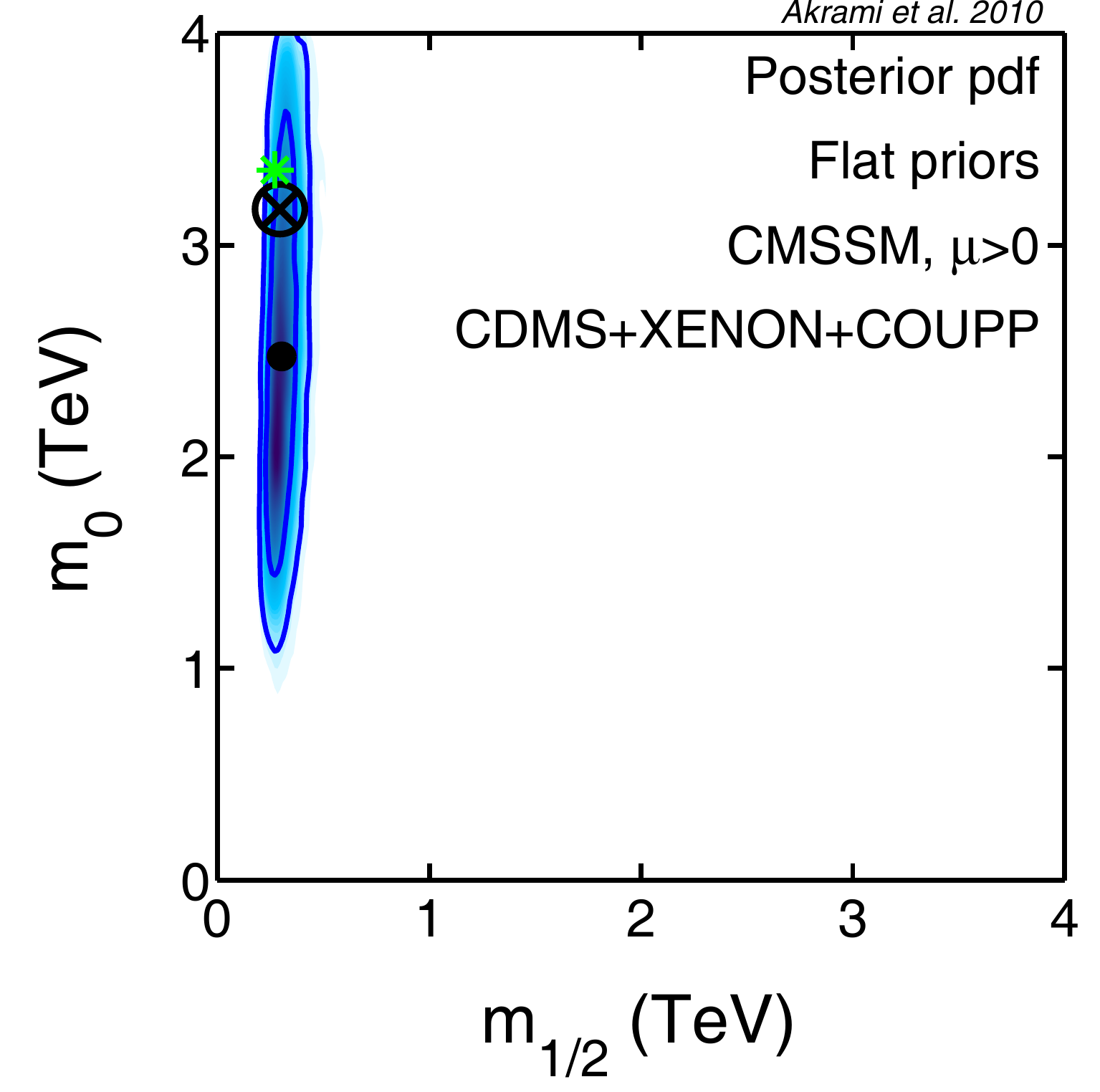}}
\subfigure{\includegraphics[scale=0.225, trim = 40 230 130 123, clip=true]{figs/LH_CDMSXENONCOUPP_wRefPoint/LH_CDMSXENONCOUPP_wRefPoint_2D_profl_1}}
\subfigure{\includegraphics[scale=0.225, trim = 40 230 60 123, clip=true]{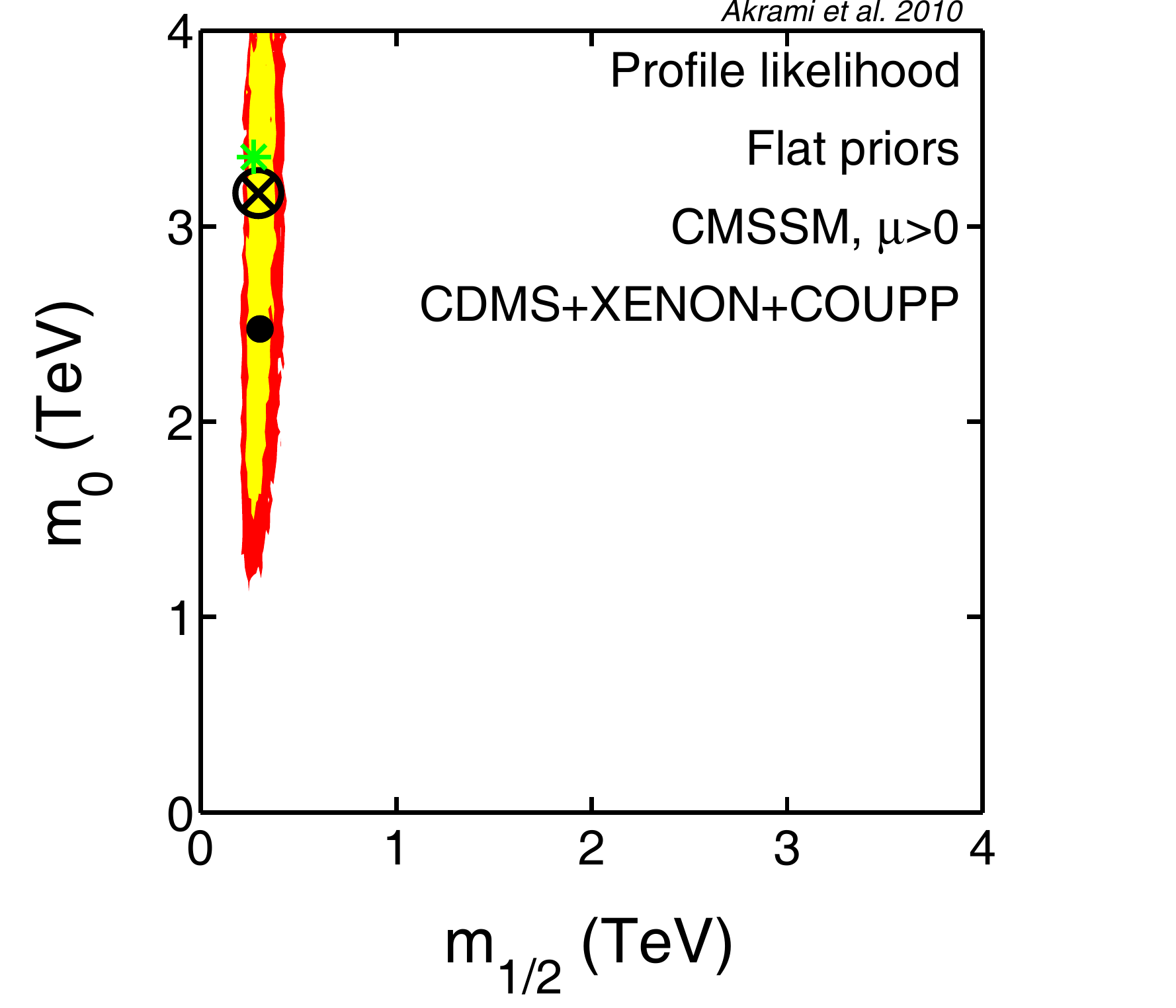}}\\
\subfigure{\includegraphics[scale=0.225, trim = 40 230 130 123, clip=true]{figs/LH_CDMSXENONCOUPP_wRefPoint/LH_CDMSXENONCOUPP_wRefPoint_2D_marg_6}}
\subfigure{\includegraphics[scale=0.225, trim = 40 230 130 123, clip=true]{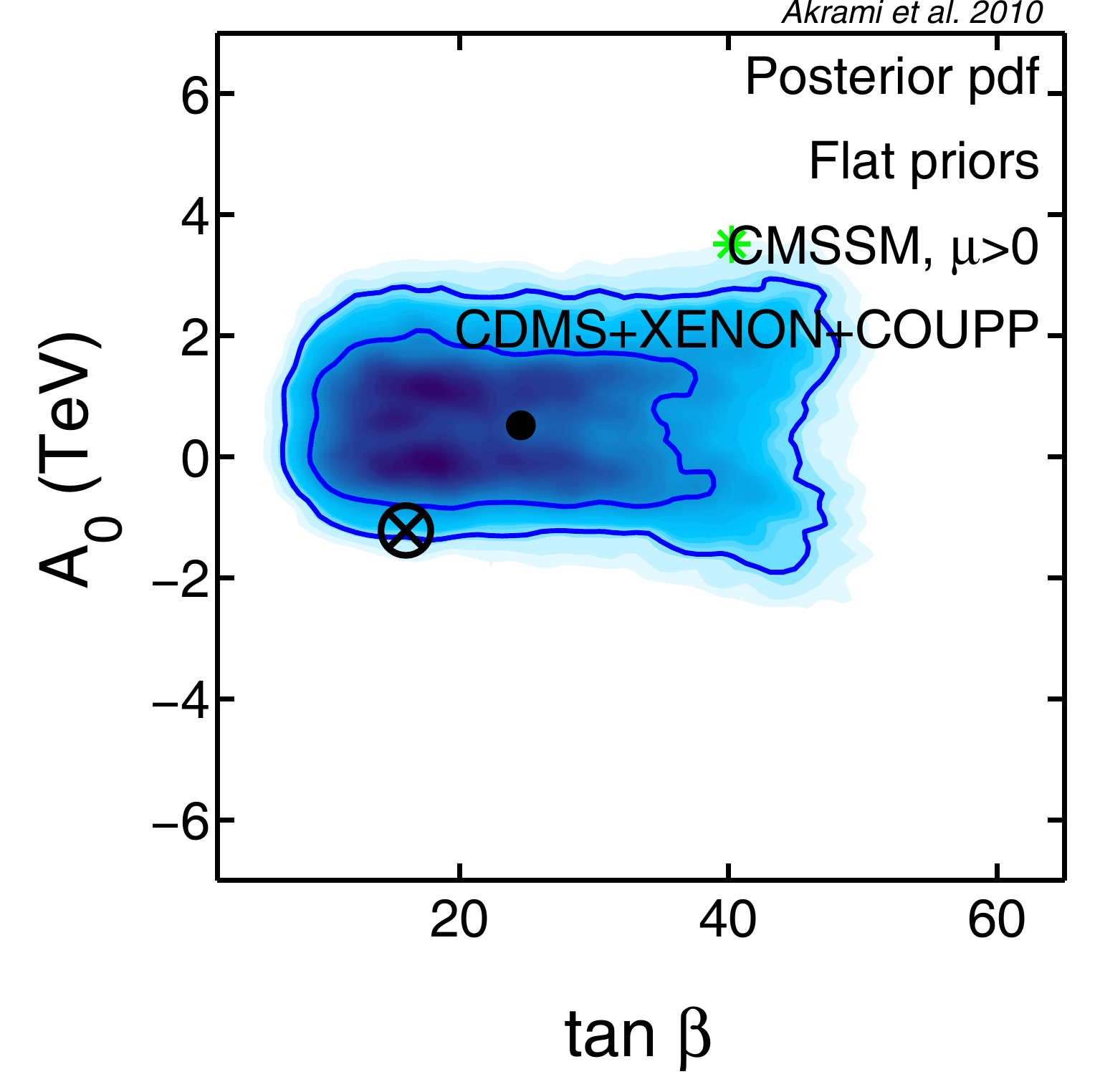}}
\subfigure{\includegraphics[scale=0.225, trim = 40 230 130 123, clip=true]{figs/LH_CDMSXENONCOUPP_wRefPoint/LH_CDMSXENONCOUPP_wRefPoint_2D_profl_6}}
\subfigure{\includegraphics[scale=0.225, trim = 40 230 60 123, clip=true]{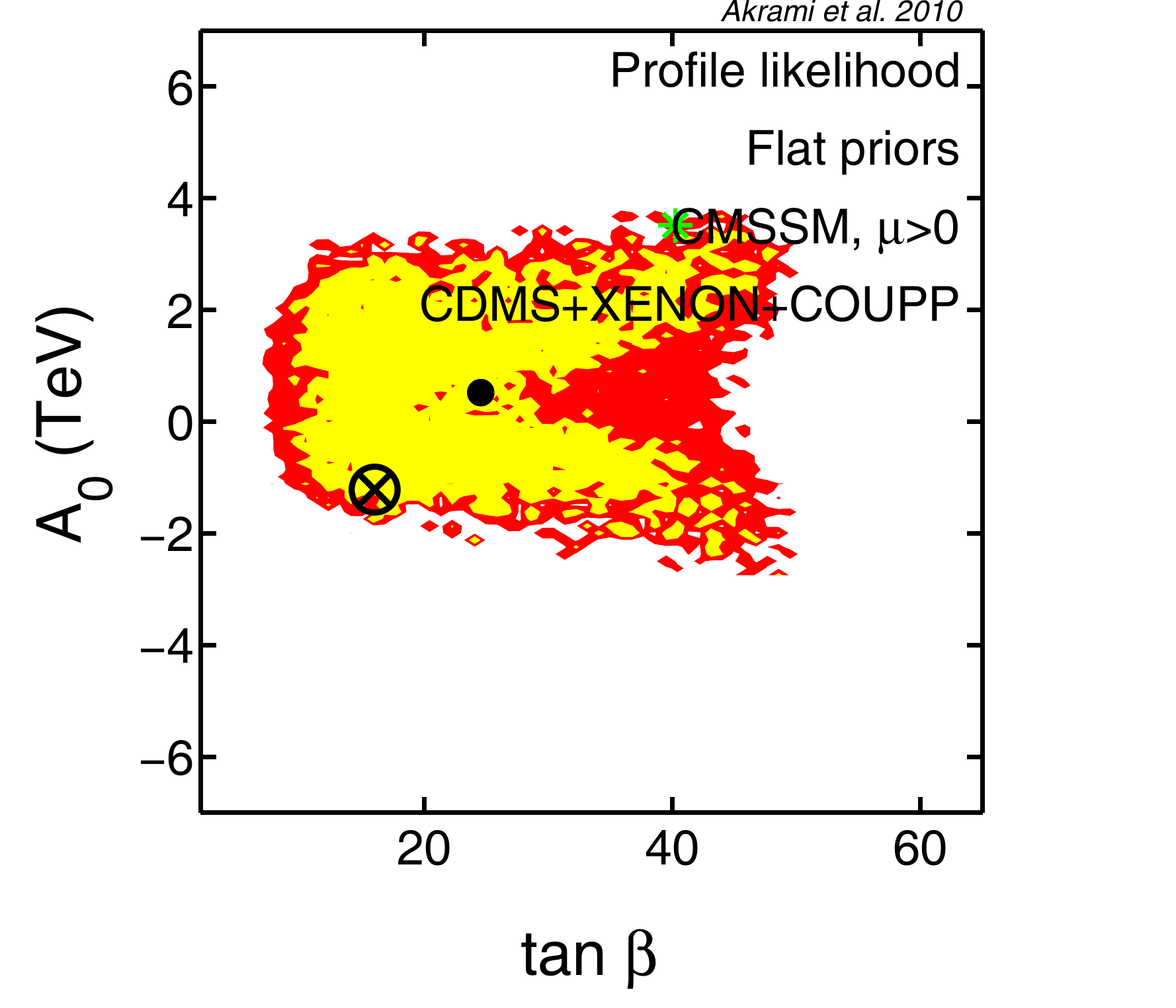}}\\
\caption[aa]{\footnotesize{Two-dimensional marginalised posterior PDFs and profile likelihoods for the mass and nuclear scattering cross-sections of the lightest neutralino, as well as the CMSSM parameters, for benchmark 1 when all or none of the halo and cross-section nuisance parameters vary along with the CMSSM parameters in our scans. CS and H in the subcaptions stand for cross-section and halo nuisances, respectively. The difference between this figure and Figs.~\protect\ref{fig:MMNuiscompmarg} and \protect\ref{fig:MMNuiscompprofl} is the choice of benchmark point; here, we consider a case where direct detection data strongly constrain the parameters (benchmark 1), whereas in Figs.~\protect\ref{fig:MMNuiscompmarg} and \protect\ref{fig:MMNuiscompprofl} we considered a case where constraints were weaker (benchmark 3). Again, the inner and outer contours in each panel represent $68.3\%$ ($1\sigma$) and $95.4\%$ ($2\sigma$) confidence levels, respectively. Black dots and crosses show the posterior means and best-fit points, respectively, and benchmark values are marked with green stars.}}\label{fig:LHNuiscomp}
\end{center}
\end{figure}

An interesting case to look at is when our DD experiments highly constrain the CMSSM parameter space (i.e.~benchmark 1), and when the likelihoods from all three experiments are used. Fig.~\ref{fig:LHNuiscomp} compares the results for two cases: (i) when all the halo and cross-section nuisances vary and (ii) when all the halo and cross-section nuisances are fixed to their mean values. This figure clearly shows that including COUPP1T in the latter case allows us to reconstruct the true values of the scattering cross-sections and neutralino mass almost perfectly, with uncertainties coming essentially only from our ignorance about the nuisance parameters. Fig.~\ref{fig:LHNuiscomp} also shows how well one can constrain the CMSSM parameters in two cases (i) and (ii). Although these measurements impose strong constraints on the values of $\mhalf$, still wide ranges of values for $\mzero$, $\azero$ and $\tanb$ are equally consistent with the data. This therefore indicates that other types of experiments, such as indirect detection or accelerator searches are required to break these degeneracies and pin down the actual values of the model parameters.

\begin{figure}[t]
\subfigure[][\scriptsize{\textbf{Mean values:}}]{\includegraphics[scale=0.23, trim = 40 230 130 100, clip=true]{figs/MM_CDMSXENONCOUPP_wRefPoint/MM_CDMSXENONCOUPP_wRefPoint_2D_marg_15}}
\subfigure[][\scriptsize{\textbf{Random values:}}]{\includegraphics[scale=0.23, trim = 40 230 130 100, clip=true]{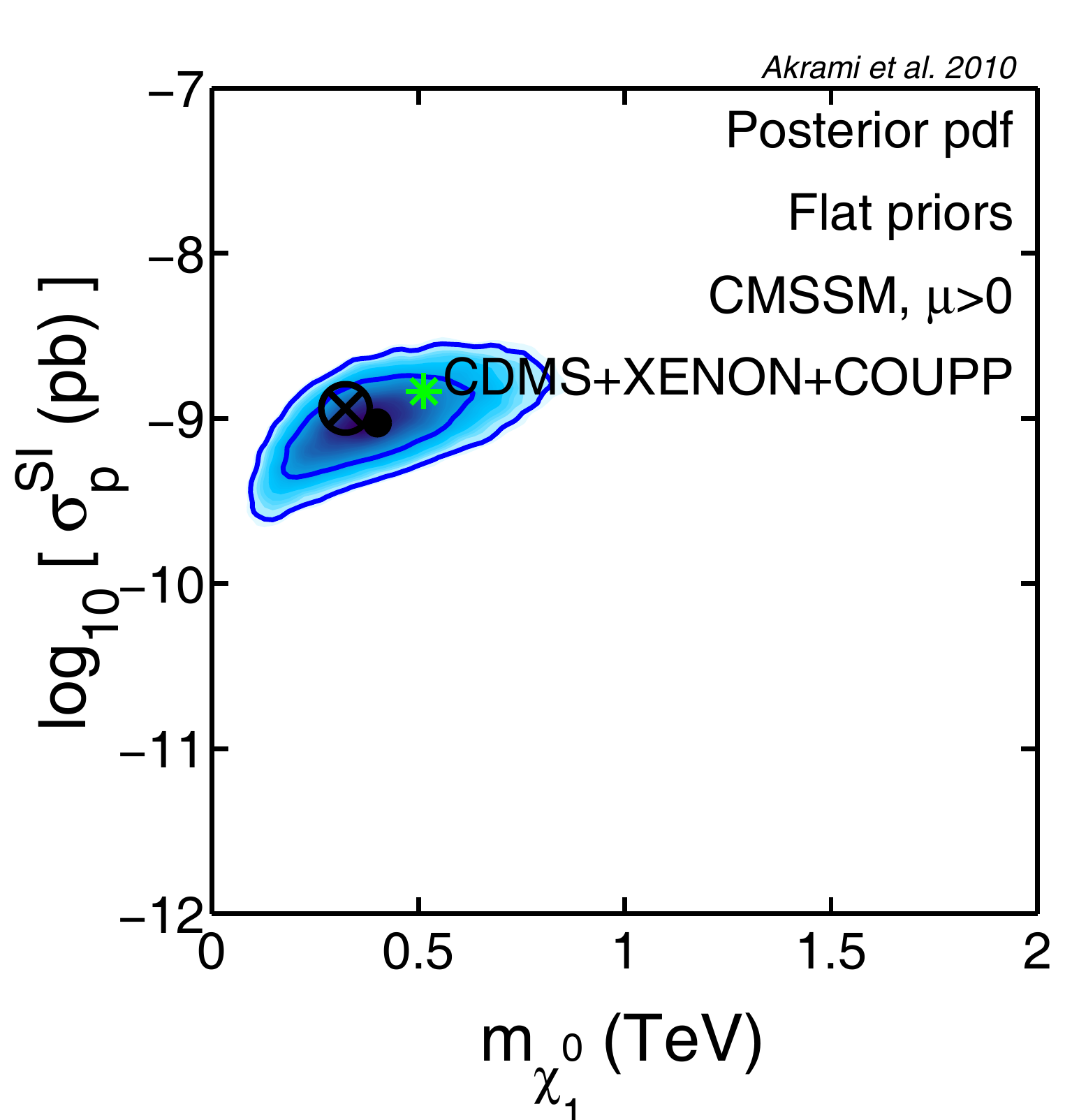}}
\subfigure[][\scriptsize{\textbf{Mean values:~~}}]{\includegraphics[scale=0.23, trim = 40 230 130 100, clip=true]{figs/MM_CDMSXENONCOUPP_wRefPoint/MM_CDMSXENONCOUPP_wRefPoint_2D_profl_15}}
\subfigure[][\scriptsize{\textbf{Random values:}}]{\includegraphics[scale=0.23, trim = 40 230 130 100, clip=true]{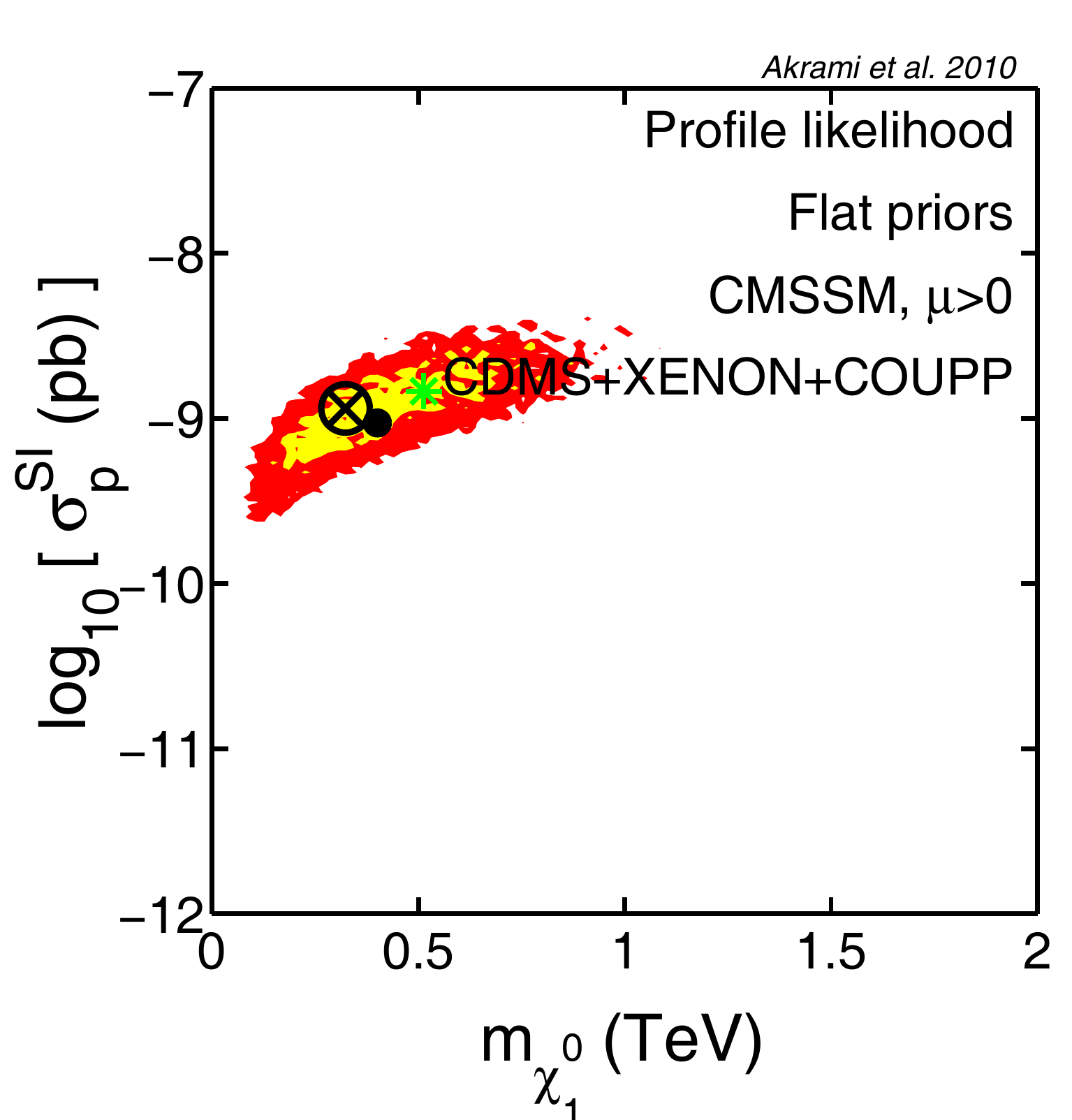}}\\
\subfigure{\includegraphics[scale=0.23, trim = 40 230 130 123, clip=true]{figs/MM_CDMSXENONCOUPP_wRefPoint/MM_CDMSXENONCOUPP_wRefPoint_2D_marg_16}}
\subfigure{\includegraphics[scale=0.23, trim = 40 230 130 123, clip=true]{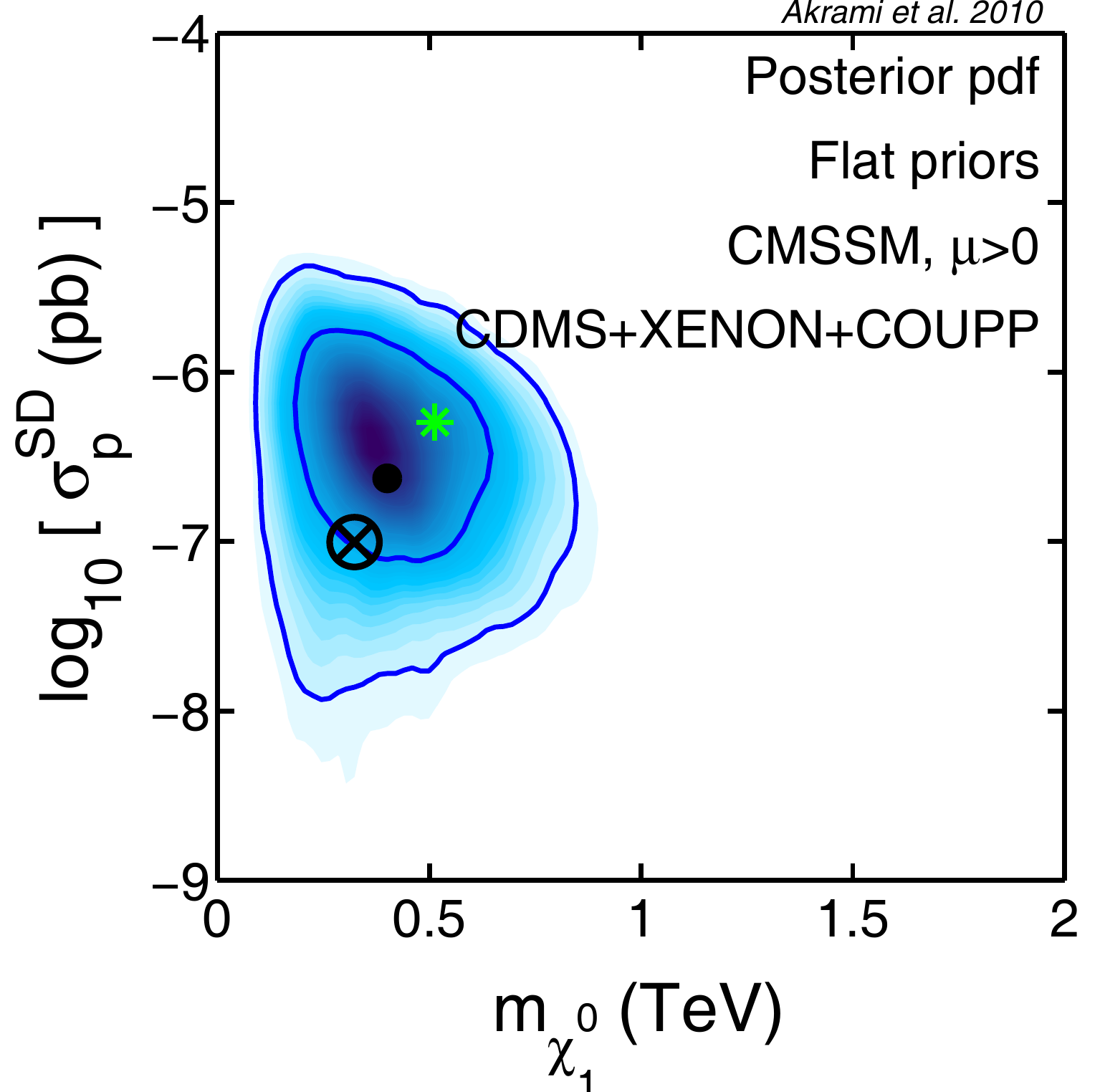}}
\subfigure{\includegraphics[scale=0.23, trim = 40 230 130 123, clip=true]{figs/MM_CDMSXENONCOUPP_wRefPoint/MM_CDMSXENONCOUPP_wRefPoint_2D_profl_16}}
\subfigure{\includegraphics[scale=0.23, trim = 40 230 130 123, clip=true]{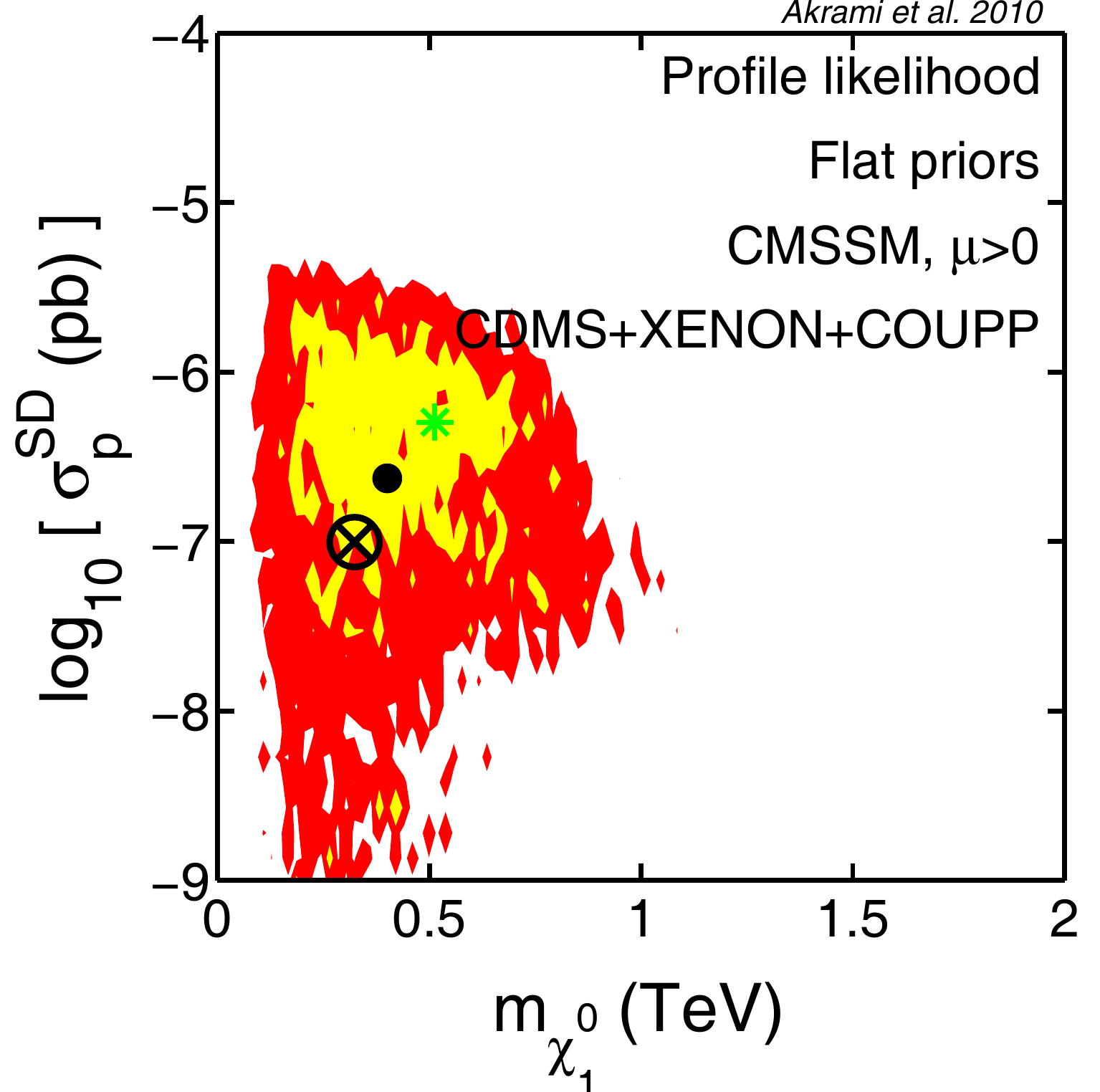}}\\
\subfigure{\includegraphics[scale=0.23, trim = 40 230 130 123, clip=true]{figs/MM_CDMSXENONCOUPP_wRefPoint/MM_CDMSXENONCOUPP_wRefPoint_2D_marg_17}}
\subfigure{\includegraphics[scale=0.23, trim = 40 230 130 123, clip=true]{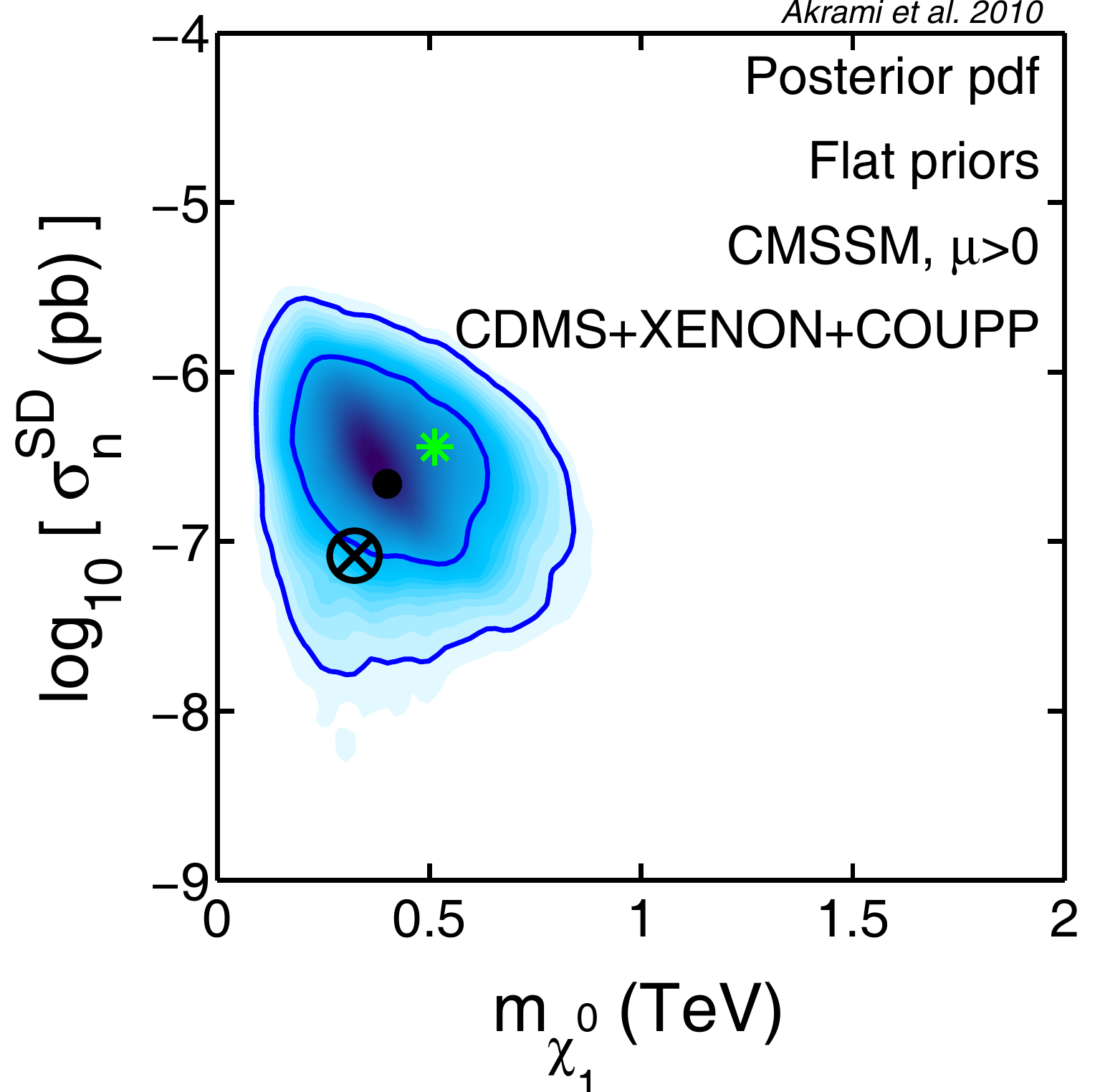}}
\subfigure{\includegraphics[scale=0.23, trim = 40 230 130 123, clip=true]{figs/MM_CDMSXENONCOUPP_wRefPoint/MM_CDMSXENONCOUPP_wRefPoint_2D_profl_17}}
\subfigure{\includegraphics[scale=0.23, trim = 40 230 130 123, clip=true]{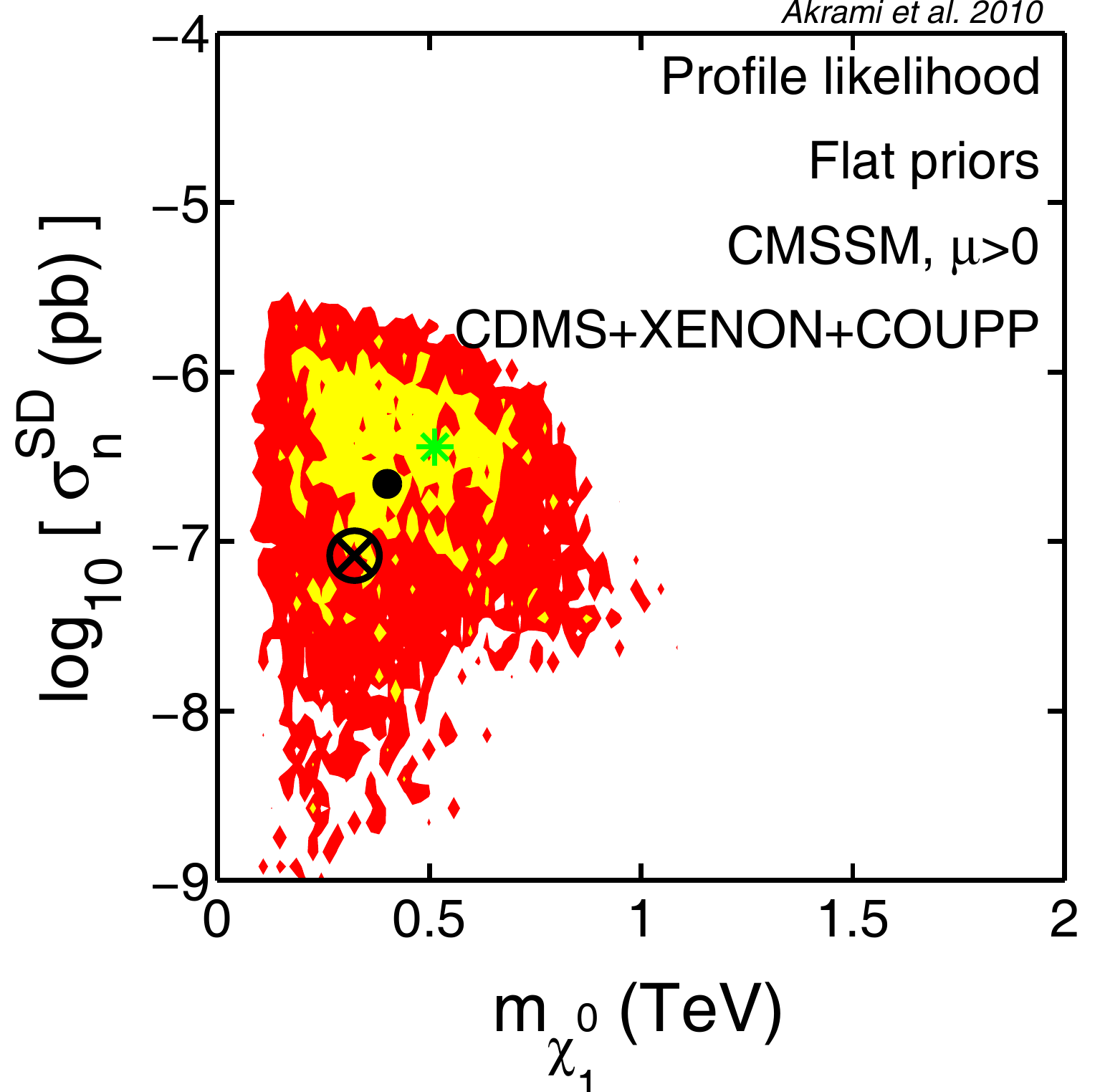}}\\
\caption[aa]{\footnotesize{Comparison of two-dimensional marginalised posterior PDFs and profile likelihoods for two different choices of the `true' halo and cross-section nuisance parameters. All plots are for benchmark 3. First and third columns show the results when the nuisances are set to their mean values when generating synthetic data. Second and fourth columns show the same, but for randomly chosen values of the nuisances.}}\label{fig:MMRandNuis}
\end{figure}

Throughout the paper, in generating synthetic DD data at each benchmark point, we fixed the halo and cross-section nuisance parameters at their mean values given in~\tab{tab:DDParameters}. We believe that our results would not change significantly if we chose other benchmark values for these parameters within e.g.~their $1$ or $2\sigma$ ranges and generated the simulated data using those new benchmark values of nuisances; this is essentially the whole idea behind marginalisation. To demonstrate this, we scanned over the CMSSM parameter space assuming a set of randomly generated values for the nuisances: $\Deltaps = -0.124$, $\sigma_0 = 33.0$~(MeV), $\SigmapiN = 57.6$~(MeV), $\rhoDM = 0.308$~(GeV/cm$^3$), $\vrot = 203.6$~(km/s), $\vmp = 238.7$~(km/s), $\vesc = 538.2$~(km/s). In other words, for each benchmark point (with CMSSM parameters given in~\tab{tab:BMs}), instead of the mean values of~\tab{tab:DDParameters}, we used the above values for nuisance parameters, and then generated the synthetic data to be used in the scans. Likelihood distributions used in these scans had the same forms (with the same mean values) as before (i.e. introduced in~\tab{tab:DDParameters}). We show the results in~\fig{fig:MMRandNuis}, where the two-dimensional posterior PDFs and profile likelihoods are compared to the case where the nuisances are fixed to their mean values. As expected, credible/confidence regions are not affected strongly; they are only slightly larger in the case where the true values of the nuisances differ from the means of their distributions.

\afterpage{\clearpage}

\section{Summary and conclusions} \label{sec:concl}

We have studied the ability of future ton-scale dark matter direct detection experiments to put constraints upon the parameter space of the CMSSM. The experiments we considered were ton-scale extrapolations of CDMS, XENON10/100 and COUPP. We assumed 1000~kg-years of raw exposure for all three experiments, with efficiencies, energy resolutions and energy thresholds similar to their current versions. In our construction of the likelihoods, we included the number of events seen by the experiments, as well as the observed event energies for CDMS 1-ton and XENON 1-ton. We modelled different background events with a background spectrum containing a flat and an exponentially-falling component.

We generated synthetic data for some benchmark points with particular nuclear-scattering cross-sections and neutralino masses. We scanned over the CMSSM parameter space and reconstructed the marginalised posterior PDFs and profile likelihoods. These gave us Bayesian credible and frequentist confidence regions for different combinations of parameters and observables. We have also taken into account uncertainties in different halo model parameters and neutralino-quark couplings, by considering them as nuisance parameters and marginalising over them.

Our results show that by combining all three experiments, one can successfully determine the WIMP mass and cross-sections for significantly low-mass and high-cross-section neutralinos. This ability lessens substantially when WIMPs have much higher masses and lower cross-sections. We have shown that COUPP can be very helpful for breaking degeneracies between different parameters when a large fraction of its signal event rate comes from SD interactions. We also observed that uncertainties in the halo model have a reasonably large effect on the constructed credible/confidence regions, whereas cross-section uncertainties do not change the results significantly.

Finally, our results imply that even in the best-case scenario where the neutralino has relatively low mass and high cross-sections, direct detection experiments would not constrain all CMSSM parameters strongly; the tightest constraint would be on the gaugino mass parameter, which is strongly correlated with the neutralino mass. This indicates that if indeed SUSY is realised in Nature, other types of experiments, such as indirect detection or accelerator searches, are needed in order to break degeneracies between the remaining parameters.  This illustrates the high degree of complementarity in different searches for SUSY; our results show great promise for future direct detection experiments in characterising SUSY together with other searches, if it is discovered.





\acknowledgments{We thank Lars Bergstr\"om, Roberto Ruiz de Austri, Subir Sarkar and Roberto Trotta for helpful discussions.  We also thank Gianfranco Bertone and the other authors of ref.~\cite{BertoneDD} for sharing their preliminary results. We are grateful to the Swedish Research Council (VR) for financial support.  PS is supported by the Lorne Trottier Chair in Astrophysics and an Institute for Particle Physics Theory Fellowship.  JC is a Royal Swedish Academy of Sciences Research Fellow supported by a grant from the Knut and Alice Wallenberg Foundation.}


\end{document}